\documentstyle[epsfig]{dissertation}
\title{A STUDY OF \\NUCLEON SPIN STRUCTURE\\
FROM QUANTUM CHROMODYNAMICS}
\author{Jonathan Andrew Osborne}
\date{2000}
\department{Department of Physics}
\acknowledgements{I will always be grateful to my relatives
and friends for supporting me through difficult times
and understanding my need to discuss physics at the most
inappropriate times.  My cousin, Tom, who was always there with a 
beer when I needed a break.  My father, who always pushed me
to do my best.  My mother, who always made me feel I could do it.
  
Over the past eight years, I have benefitted greatly
from discussions with fellow students.  It is only
through heated debate that one completely understands
the true nature of physics, and only through teaching and
explaining that one can fully assimilate new material.
Thanks in particular to Michael Hartl, Eduardo Ponton,
William Linch, and C. K. Chow for many 
useful discussions on the subject matter
of this dissertation (as well as many long, 
enjoyable digressions).

Over the years, I have
had the great pleasure of working with and studying under 
several exceptional physicists and mathematicians.  
Their guidance has helped me to understand physics in 
a simple way which is difficult to present in textbooks.
My undergraduate 
advisor, Prof. Ivan Kramer, once told me to keep on 
pushing and never give up.  I have remembered that sentiment 
more than a few times during the last few months during the 
preparation of this manuscript.  Prof. Thomas Cohen, 
who always has a simple explanation for complicated 
mathematical results, has helped 
me to see physics from another perspective.

Of course, the largest contributions to this 
work come from my advisor, mentor, and collaborator
Xiangdong Ji.  His enthusiasm for the subject
and confidence in his insights have always 
motivated me to prove him wrong - which is (almost!) impossible.
At the same time, his confidence in {\it my}
insights and understanding drives me to work harder to 
prove him right - which is nearly as difficult.
I have been told many times that one can hear both of
our laughs from the other side of the building. 
His influence has touched nearly every part of this 
dissertation, from finding irreducible representations
of $SO(3,1)$ to proving all-orders factorization for DVCS.
Without his patient understanding of my many faults,
this dissertation would not have been possible.

In this revised manuscipt, I must also pay homage
to Loretta Robinette, the beloved matriarch of the Theoretical
Quarks, Hadrons, and Nuclei group at the University of Maryland.
Time and again, she has done far more than I expected 
to help me with the preparation of this corrected version.}

\advisor{Associate Professor Xiangdong Ji}
\committee{Associate Professor Xiangdong Ji, Chair\\
Professor Manoj Banerjee\\Associate Professor Elizabeth Beise\\
Professor Thomas Cohen\\
Professor William Walters}
\begin{document}
\makefrontmatter

\section{Introduction}

In the beginning, God created the Heavens and the Earth.  After that,
it has been the goal of physicists to understand the laws they obey.  
The dream of physics is a single set of 
laws which describe all phenomena from the largest galaxies
to the smallest atoms and beyond.  Recently, physics has come to 
a plateau of understanding in terms of which there are only a few 
physical processes which cannot be explained quantitatively.  
The physics of the very large 
is governed by Einstein's General Relativity \cite{GR}.  This theory 
has had tremendous success in accurately predicting the orbits
of the planets in our solar system, the bending of light by our Sun, and 
the behavior of radiation emitted by distant galaxies.  The 
physics of the very small is thought to be described by the 
Standard Model.  First formulated in its present form 
in 1973\footnote{The `electroweak' sector was formulated
in pieces throughout the 1960's.  The most notable contributions are from
S. L. Glashow \cite{G}, S. Weinberg \cite{W}, and
A. Salam \cite{S}.  The `strong' sector was not written in its present form 
until 1973, as discussed below.}, this model is well-known
for its extraordinarily accurate predictions of the 
properties of many subatomic particles, the excitation spectra of atoms,
and the existence and theory of radioactive decay.  Perhaps the most
famous example of a precision calculation within the 
Standard Model is the calculation of the anomalous magnetic moment
of the electron to 11 decimal places \cite{alpha4}.  
This quantity was measured to 7 parts in a trillion \cite{measprec}
as a test of the Standard Model, making it the
most precisely measured quantity in the history of the 
human race.  Agreement between experiment and theory to this degree of precision
is unheard of in any other branch of physics, making
this aspect of the Standard Model the most precisely tested 
theory in the history of physics.

The Standard Model (SM) consists of three parts.  These can be thought of
as Electricity and Magnetism, the Weak Interactions, and the Strong
Interactions.  Electricity and Magnetism in the SM is
a relativistic quantum version of the familiar classical 
Electrodynamics embodied by the Maxwell Equations and Lorentz force
law.  This is the sector of the SM which is largely
responsible for all of the macroscopic objects we see around us : 
computers, light bulbs, cars, etc.  It is also responsible for almost 
every branch of science.  It is the driving force behind chemistry,
metallurgy, condensed matter physics, biology, etc.  Well understood as a classical
theory, its marriage with quantum mechanics led to 
the first realistic quantum field theory.
Quantum field theories have dramatic implications 
on the structure of matter in general,
some of which were studied for the first time 
in the context of quantum electrodynamics (QED).
In particular, QED employs photons not only as 
the physical quanta of light, but also as carriers of
the electromagnetic force.  In QED, charged 
particles interact by exchanging massless spin-1
photons.  This idea of forces being {\it mediated} 
by vector bosons\footnote{To highlight the difference
between two basic particle types, particles 
which obey Bose-Einstein statistics are called bosons
and those which obey Fermi-Dirac statistics 
are called fermions.  This behavior finds its
origins in the particle's {\it spin},
or intrinsic angular momentum. 
Particles with integer spin are invariably bosons
and those with half-integer spin are fermions.}
plays a fundamental role 
in the other two sectors of the SM.  Indeed, one 
can understand many aspects of these elusive
forces simply by exploiting an analogy to QED. 

The Weak Interactions manifest themselves mainly in certain forms
of radioactive decay.  The processes of beta decay, electron capture,
and positron emission first studied in the 
late 1890's \footnote{True matter radiation 
was first discovered by H. Becquerel \cite{Becque}
in 1896.  Originally discovered in samples of uranium ore,
his `uranic rays' were not understood as properties 
of individual atoms until M. Curie \cite{Curie}.
For a comprehensive history of the 
discovery of radiation, see \cite{inward}.}
all take place via the Weak Interaction.  Their
discovery and subsequent study provided the first evidence
of the existence of this sector of the SM. 
However, it was not understood that these processes required the
introduction of a new force until the discovery of the neutron
in 1932.  Before this monumentous discovery, it was believed
that beta decay was simply the emission of one of the `nuclear electrons'
needed to get the observed nuclear charge and mass right in the 
contemporary model of the nucleus.  The discovery of the neutron
forced scientists to reconsider the process that leads to 
beta decay in the nucleus. 
More recently, many other
manifestations of this interaction have been seen experimentally, most
notably the decay of muons and pions and the existence of the
so-called `strange' particles.
However, this force has so many intricate features and 
is so well-hidden in ordinary matter that it was 
not completed for more than twenty
years after these discoveries.
Its completion  at the end of the 1960's
won Glashow \cite{G}, Weinberg \cite{W}, and 
Salam \cite{S} the Nobel Prize in Physics in 1979.  
G. `t Hooft and M. Velman's 1972 proof that the 
theory is internally consistent and complete \cite{thooft}
won a second Nobel Prize in Physics in 1999.  
The experimental discovery by C. Rubbia and S. van der Meer 
of the W- and Z-bosons \cite{WZexp} proposed to 
mediate the weak interactions won them the Nobel Prize in 1984.

In this work, we will mainly be concerned with the final sector of the
SM.  The Strong Interactions are responsible for the structure of nuclei
and some of their decay modes.  Alpha decay and
the rare decays of spontaneous fission and proton and neutron emission
are mediated by this force.  It was alpha decay that first permitted
the discovery of the nucleus itself.  In 1909, 
H. Geiger and E. Marsden performed an experiment in which
they found that approximately one in eight thousand alpha particles scattered off
of a gold foil are scattered at an angle greater than 90$^\circ$
\cite{Geig}.  The contemporary model of the atom, due to J. J. Thompson \cite{jelly},
involved many negatively charged electrons embedded in a jelly-like
medium of smeared-out positive charge.  Since electrons are many thousands
of times lighter than alpha particles, this model did not provide any 
mechanism for the observed back-scatter.  

E. Rutherford proposed 
that atoms are composed of a very
heavy positively charged hard
core, or nucleus, surrounded by Thompson's
negatively charged electrons.  This core was estimated to be approximately 
one hundred thousand times smaller than the atom \cite{rut}.
He argued that the electric
field of this heavy nucleus was responsible for the back-scattering 
observed in recent experiments, rather than the light electrons or the
diffuse jelly.  This had the unattractive consequence that the vast 
majority of the volume of an atom is empty space.  However, 
future experiments proved the correctness of this model.  Since the
attraction of electrons to the nucleus is accomplished by the 
electromagnetic field, it seemed the only thing left to study was 
the nucleus itself.  

Radioactivity became physicists' `window on the nucleus'
when it was discovered that 
almost all radioactive processes are nuclear in origin.
Physicists were beginning to understand that the nucleus
itself is a strongly interacting bound state.  Rutherford's experiments
in 1919 showed that one could obtain oxygen by bombarding nitrogen
with alpha particles \cite{trans}.  Through years of these experiments
and intense theoretical work, it became clear that nuclei are composed
of two heavy particles, collectively called nucleons.  

The positively charged protons responsible for the 
electric charge of nuclei were 
discovered first partly because they exist independently 
of other nucleons.  These objects are the nuclei of the lightest isotope 
of hydrogen and were called `positive electrons' by Rutherford until
it became clear that they are really quite different from electrons.
Neutrons were the last constituent of `ordinary' 
matter to be discovered since they do not carry an electric
charge and decay into protons when isolated from nuclei.  Long
thought to be an extremely tightly bound system of a proton and an
electron, neutrons were not given separate identities 
until 1932.  In that year, 
J. Chadwick showed
that the neutral radiation resulting from the bombardment of beryllium
with alpha particles was not simply electromagnetic radiation,
as widely believed, but was in fact a new kind of particle which he called
the neutron \cite{chad}.  He received the Nobel Prize for this 
discovery in 1935.  The neutron solved many of the 
paradoxes the new quantum theory brought to nuclear physics.  At last,
the nucleus was seen as a sensible quantum system.  All that was
needed for a complete theory was an interaction responsible for the formation and stability
of the bound states observed.  

Once it was understood that the nucleus is made of positively 
charged protons and neutral neutrons, it became apparent that the 
electromagnetic repulsion inherent in nuclei must be overcome
by a tremendously strong force.  Called
the Strong Force for obvious reasons, this new interaction was proposed
to involve only protons and neutrons.
Neutron, proton, and alpha\footnote{$\alpha$-particles were long known to be 
$^4$He nuclei.}
scattering data on various nuclei gave
scientists of the time a good amount of information on
the excitation spectra, but contemporary models were mainly 
empirical in origin.  A modest degree of success was obtained
by these early models, most notably the composite nucleus model of
N. Bohr \cite{bohrcomp} and the celebrated nuclear shell model 
\cite{nucshell}, but there was much left to be desired.

A partial solution to the problem of the origins of the nuclear
force came with the discovery of a new triplet of particles
and the exploitation of an approximate symmetry seen in nuclear
interactions.  It was proposed by H. Yukawa in 1934 \cite{yuk} that if this
new force is indeed fundamental, a description
based on the newly formulated theory of relativistic
quantum electrodynamics may
lead to new insights.  He argued that since the long-range nature of the 
electromagnetic field is attributed to the  
masslessness of the photon,
a short-range nuclear force may be 
generated by massive particle exchange.  Obtaining the range of the
strong force from scattering experiments, he found that the mass
of this new particle, called a meson, 
should be about two hundred times the mass of the electron.

In reality, there are five particles in Yukawa's predicted mass
range.  Only three of them have the right properties to 
be candidates for the meson required
in his theory.  The other two appear as decay products of
{\it these} three,\footnote{well, two of them...} 
and have nothing whatsoever to do with the strong
nuclear force.\footnote{These objects, the $\mu^\pm$, are actually
heavier copies of electrons.  They provide the first evidence of the 
second generation of fermions.}  This created a 
great deal of confusion in the scientific
community for about ten years, but in 1947 C. F. Powell and collaborators
were finally able to show conclusive evidence testifying to the existence
of two charged strongly interacting bosons with masses approximately 
270 times
the electron mass \cite{pion}.  For his prediction of their existence,
Yukawa received the Nobel 
Prize in 1949.  A third meson in this mass range was discovered
in 1948, this one with no electric charge.  The existence of this
triplet forced physicists to take another look at the
structure of the strong interactions. 
 
For many years, it had been a known fact
that to a very good approximation the strong force does
not distinguish between protons and neutrons.  In 
fact, if we take Yukawa's meson exchange force seriously, 
exchanges of charged mesons actually {\it change} neutrons {\it into}
protons and vice versa.  This new symmetry led physicists to
consider protons and neutrons as different manifestations of the
{\it same particle}.  In these models, protons and 
neutrons are considered as components of a 
two-dimensional vector in a new space.  This internal space
is called `isospin space' since it is mathematically identical
to ordinary quantum-mechanical spin.  The mesons are considered
a vector in isospin space and work to rotate the nucleon states.
This symmetry is actually a very good approximation to 
the true form of strong interaction physics, and had been
useful in classifying particles and relating certain strong
processes to each other. 
It also became instrumental in revealing to us the true nature of
the strong interaction.

Along with the 1950's came a new generation of accelerators
in which one could liberate larger and larger amounts of energy,
which nature used to form exotic new particles.  At this point,
it was expected that all of the fundamental particles were known
and these higher energies would only reveal more detail about the
structure of the strong interactions and the behavior of nuclear 
excited states.  However, hundreds of new 
strongly interacting particles had been
discovered by 1960.  These new objects were classified according to their
charge, spin, decay properties, and isospin in an effort to make some
sense of this `zoo' of particles.  The situation here is analogous to 
Mendeleev's arrangement of elements on his 
periodic table.  Just as his arrangement led
to a greater understanding of chemistry and eventually the atom,
this arrangement of particles led to the creation of the SM.
In order to accommodate all these new particles, the word `meson' was
extended to mean `integer spin strongly interacting particle' and
the mesons of Yukawa's theory were called $\pi$-mesons, or pions.
The word `baryon' was introduced for strongly interacting fermions
and `hadron' as a blanket term for all strongly 
interacting particles.\footnote{Nuclei are 
not included in these designations.  Although the
$^{14}$N nucleus is indeed a boson which participates in the strong
interaction, it is never referred to as a meson.  The terms `meson' and 
`baryon' will be defined more clearly below.} 

One of the most striking patterns in the particle zoo emerges
when we arrange them according to the length of time it
takes for them to decay.  This exercise reveals three distinct 
time scales, showing 
physicists clear evidence of three different forces.
The shortest decay times are on the order of 10$^{-21}$ to 10$^{-24}$ 
seconds, and involve only strongly interacting particles.  These 
are attributed to the strong force.  
The mid-range
lies on the order of 10$^{-17}$ to 10$^{-19}$ seconds and is associated 
with the electromagnetic interaction.  These decays are signaled by the 
appearance of photons in the final state.  The longest lived particles
have half lives on the order 
of 10$^{-10}$ to 10$^{-8}$ 
seconds.\footnote{Two notable exceptions 
are the neutron and the $\mu^\pm$, both of which
have uncharacteristically long lifetimes due to phase space suppression.}  These
decays are attributed to the weak interaction.  Weak decays
are signaled by the existence of leptons (like electrons and muons)
in the final state, as well as missing energy attributed to the
elusive neutrinos postulated to conserve energy and momentum 
in the weak decay of the neutron.

Many of the observed 
decays fit nicely into this scheme and physicists were given confidence in
this three force model.  However, a small number of mesons and 
baryons were found to behave very strangely.  These hadrons
participate in decays that look very much like strong decays,
but they live entirely too long and certain symmetries
seen to a very high degree of accuracy in all true 
strong interactions were badly broken by these decays.  For example
the heavy K-mesons, with a mass of about a thousand electrons, were
found to have the competing decay modes  
\begin{eqnarray}
K^+&\rightarrow& \mu^+ \; + \;\nu_\mu\nonumber\\
   &\rightarrow& \pi^+ \; + \;\pi^0    \;\; .
\end{eqnarray}
The first mode is unambiguously a weak mode signaled by the 
neutrino $\nu_\mu$.  The second appears to be a strong mode since
all of the objects participating are strongly interacting particles.
If this is the case, it should dominate the decay and cause the
half-life of the $K^+$ to be on the order of 
the characteristic time scale of the strong interactions.  In reality,
the half-life of the $K^+$ is known to be $\sim 10^{-8}\,s$, characteristic
of a weak decay.  

A closer inspection of the second decay mode
reveals that it does not respect the isospin symmetry found to
such a good approximation in the strong interactions.  This
is easy to see since the K-mesons (kaons) have isospin-1/2 while pions
have isospin-1.  The final state can only have isospin-2,
while the initial state has isospin-1/2.  All of these
strange new decays were found to violate isospin.  Physicists of the time had
to concede that we were seeing a new manifestation of the weak interaction.
It was postulated that these particles posses a new quantum 
number (strangeness) 
which must be conserved by
the strong interactions, but may be changed by one unit
via the weak interaction.  Mesons could be found only with strangeness
0 and $\pm 1$, but there were baryons with 
strangeness $\pm 2$ and $\pm 3$ as well.
A baryon with strangeness $S=2$ has to decay first to a baryon with
$S=1$, then to a baryon with $S=0$.  

\begin{figure}
\label{fig1}
\epsfig{figure=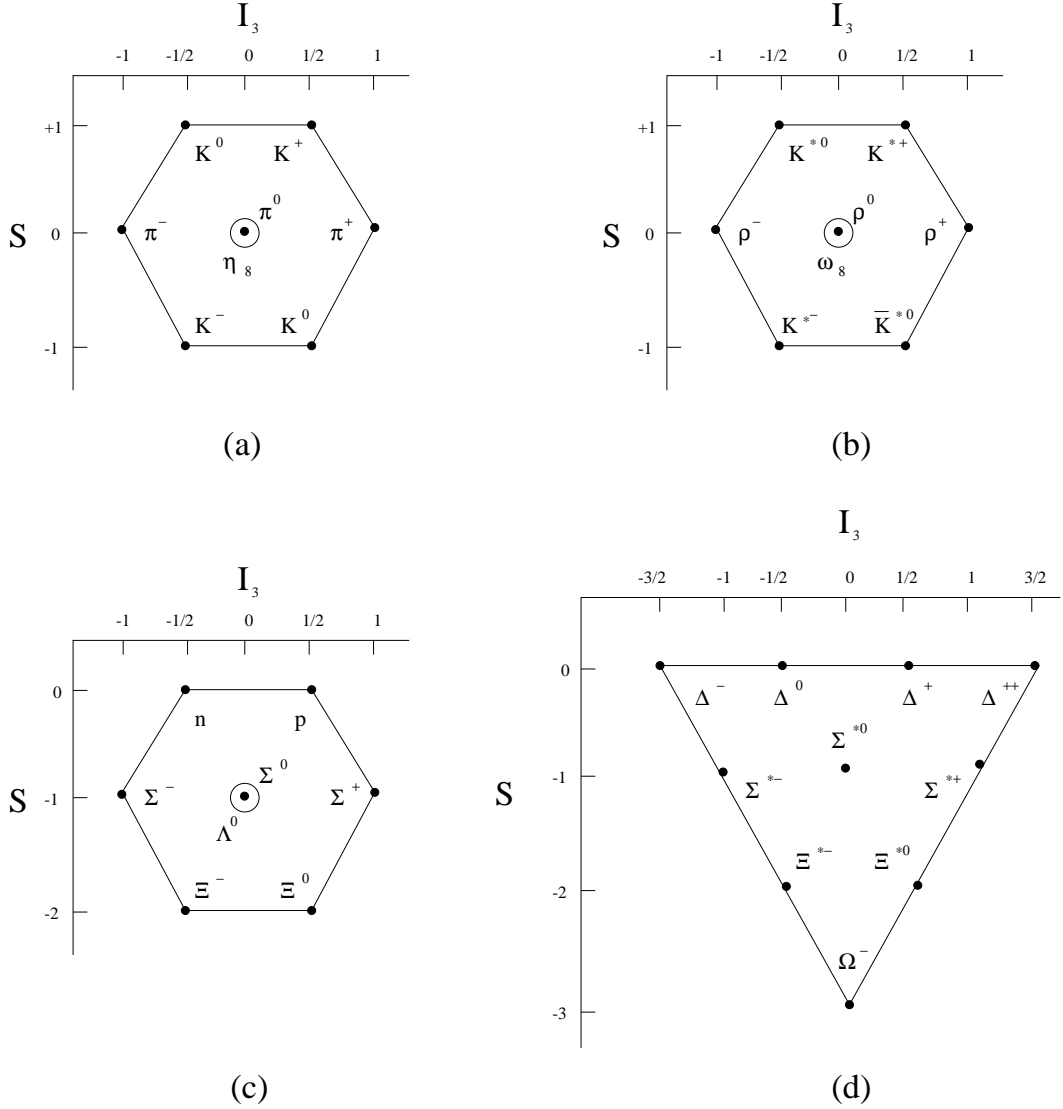,height=5.75in}
\caption{Gell-Mann's Eightfold Way for (a) spin-0 and
(b) spin-1 mesons.  These mesons form the {\bf 8} of
${\bf 3}\times{\bf\overline 3}$.  If $SU(3)$ were exact,
each multiplet would be degenerate and would not
mix with the {\bf 1} of ${\bf 3}\times{\bf\overline 3}$.
However, since in reality $SU(3)$ is badly broken,
the mathematical ideals $\eta_8$ and $\omega_8$ mix with the
singlets $\eta_0$ and $\omega_0$ to form the physical 
states $\eta$, $\eta'$, $\omega$, and $\phi$.
(c) and (d) show the baryon spin-1/2 octet and spin-3/2 
decouplet, respectively.  In each octet there are
two particles in the center : one with isospin-0 and one
with isospin-1.  The fact that they lie on top of each other
is due exclusively to the fact that I have chosen to suppress 
$I^2$.}
\end{figure}

Since the strong interaction is postulated to conserve strangeness
and (approximately) isospin, classification of its states in terms 
of these quantum numbers can give us insight into its true form. 
This observation led physicists to plot 
the particles of a given spin on axes 
labeled by these two quantum numbers.\footnote{There 
are in fact two quantum numbers associated with isospin, as 
shown in Appendix \ref{sun}.  In this plot, 
the isospin casimir $I^2$ has been suppressed and
only the 3-component of isospin, $I_3$, is considered.}
Figure 0.1 shows this
arrangement for the lowest-lying spin-0, -1/2, -1, and -3/2 hadrons.  
The mass-splitting within these multiplets is small compared with the
energy required to jump to the next group of hadrons with the same spin, so to
a first approximation these objects can be considered degenerate.
This is much like the approximation of isospin symmetry 
which proved so fruitful in the past.  In fact, isospin multiplets
(particles lying along a horizontal line in Fig. 0.1) are very
close to degenerate.  One can think of each jump away from the 
$S=0$ axis as a shift of about 350 electron masses, while the
splitting along the horizontal axis is closer to 20 $m_e$.\footnote{
In contrast, the jump to the next multiplet of spin-1/2
baryons is approximately 1000 $m_e$.} 
If the analogy with isospin were exact, we would assign the spin-1/2 
baryons to the fundamental representation of some internal 8-dimensional 
space.  However, this procedure would not allow us to understand the
reason why isospin is a much better symmetry than this new group.
Ideally we would like to place these particles in representations
of some symmetry group which not only allows us to account for all
our new low-lying states in a natural way, \footnote{This is not easy
in a group with an 8-dimensional fundamental representation since the 
spin-3/2 baryons form a 10-dimensional representation.}
but also contains
our isospin symmetry as a subgroup.  This means that we can 
break the full symmetry group in such a way that isospin remains
unbroken.  

Our ideal was achieved in 1962 by M. Gell-Mann's Eightfold
Way \cite{gellman}.  Gell-Mann found that the observed hadron 
spectrum could be placed into representations of the 
special unitary group in three dimensions, $SU(3)$.  Many of the
mathematical properties of this group are discussed 
in Appendix \ref{sun}.  Here, it
is important only to know that there are two inequivalent 
fundamental representations of $SU(3)$ from which one can
build all other representations.  Calling these two 
representations ${\bf 3}$ and ${\bf\overline 3}$, we find that the
mesons of spin-0 and -1 form representations of 
${\bf 3\times\overline 3}$ and the baryons of spin-1/2 and -3/2 
form representations of ${\bf 3\times 3\times 3}$.  If we
postulate the existence of objects in the fundamental representations
of this $SU(3)$, as M. Gell-Mann and G. Zweig did in 1964
\cite{qmod}, say quarks for ${\bf 3}$ and antiquarks for 
${\bf\overline 3}$, the last sentence says that mesons are
mathematically equivalent to bound states of one quark and 
one antiquark and baryons are mathematically equivalent to 
bound states of three quarks.  At this point, the idea
of quarks and antiquarks is merely a mathematical construct.
In order to suggest the actual existence of these particles
we would have to propose a full theory of them explaining,
in particular, why we haven't seen them in the lab.

Even if they don't actually
exist in nature, quarks provide a nice way to organize the 
known particles.  If we name the quarks in the fundamental
representation `up', `down', and `strange' ($u$, $d$, and $s$), we find that the
new strangeness quantum number is simply the number of 
strange quarks in the hadron (with a minus sign from convention).  
Thus a baryon with $S=-2$ is composed of two strange quarks and
one other quark.  If we suppose that the strange quark is 
heavier than the other two, breaking the $SU(3)$ invariance of the
theory, the multiplets break down exactly into the isospin 
multiplets we observe.  The residual isospin symmetry is nothing
but the remaining symmetry between up and down quarks.
The electromagnetic charge assignments of all known particles
can be predicted from the quark content if $d$ and 
$s$ quarks are given a charge of $-1/3$ 
and $u$ quarks are given +2/3, in units of the proton charge.  
Conservation of strangeness is easily understood
in this model if we postulate that the 
strong and electromagnetic 
interactions cannot change the type of quark, 
but the weak interaction can.  In fact, almost all of these 
strangeness changing weak decays can be understood in terms 
of the basic weak decay of an $s$ quark into a
$u$ quark and a negatively charged state (like a $\pi^-$).  Other
weak decays (like that of the neutron) are
understood as a change from $d$ to $u$ or vice versa.  
This idea 
completely changed the
shape of the weak interactions in the years to come.
Spin assignments can be understood by
postulating that quarks (and antiquarks) have spin-1/2:  Since
baryons are representations of ${\bf 3\times 3\times 3}$, 
they may have spin-1/2 or -3/2.\footnote{Higher spins can also
be obtained in this model by adding orbital contributions; these are
always integer, so do not change the fact that the state will 
be a fermion.}  Mesons will have integer spin since they are
representations 
of ${\bf 3\times\overline 3}$.\footnote{Herein 
lies the true meaning of the terms {\it baryon} and
{\it meson}.  A baryon is a three-quark bound 
state while a meson is a bound state of 
a quark and an antiquark.} 
Hence we see that if
quarks actually exist as particles, many of our problems 
will be solved.  Much in the way that the discovery of protons,
neutrons, and electrons reduced the number of fundamental particles
from the eighty or ninety different atoms plus photons and other 
radiation to four, the discovery of quarks would reduce the 
hundreds of strongly interacting particles
to three fundamental objects.

However nice Gell-Mann and Zweig's 
theory of quarks looks at first sight, it 
has some major problems.  One problem is that no mechanism is given
for binding the quarks into hadrons.  Many models were proposed
in analogy with QED to solve this problem, but no acceptable 
solution could be found.  The new force would have to be extremely 
strong to account for the fact that hadrons appeared to be fundamental
to the experimentalists of the time and no free quarks had
been seen in the lab.  In fact, no particle of fractional charge
had ever been seen in any experiment anywhere.
The fact that 
it was necessary to give quarks fractional charges to account for the
observed hadron charges was greatly ridiculed at the time.

Even more devastating is the problem of quantum statistics.  
Suppose, for example, we wish to construct the wavefunctions of the 
lowest-lying spin-3/2 baryons assuming the existence of quarks.
Since we consider ground states, it is natural to assume that 
our wavefunctions are not orbitally excited (i.e. 
the orbital angular momentum of
our states is zero).  The only way to 
obtain spin-3/2 from three quarks without the help
of orbital angular momentum is to align 
all of their spins, making this part of the
wavefunction fully symmetric under quark 
exchange.  Fermi-Dirac statistics then 
demands antisymmetry of the final 
$SU(3)$ sector.  However, the {\bf 10} of
$SU(3)$ is a fully symmetric representation.  
Hence the observed spectrum 
seems to require us to disallow the Pauli exclusion 
principle\footnote{Alternatively,
we could concede that the spin-3/2 baryons are 
indeed orbitally excited.  However, we
would then be forced to explain why we see no 
lower-energy $SU(3)$ singlet state which is
not orbitally excited.} \cite{Pauli} and be satisfied with a 
quantum state which is {\it symmetric}
under the exchange of identical fermions.\footnote{
Even without taking the limit of exact 
$SU(3)$ symmetry, this problem exists.
Consider the $\Omega$ baryon with its spin aligned along the $z$-axis.
Experimentally, the $\Omega$ is seen to have $S=-3$; in the quark model, 
it is viewed as a bound state of three $s$ 
quarks.  In the absence of orbital excitations,
all of our quarks must have their spins aligned along the $z$-axis
to account for the observed spin-3/2 state.  Hence we have
three identical fermions in the same quantum state - 
an explicit violation of Pauli's
exclusion principle without reference to $SU(3)$.} 
In the non-relativistic theory, this principle
had been made independently of the rest of quantum mechanics and
could be argued away as something that is not valid in this
arena.  However, it had been shown by Dirac that a relativistic
theory in which fermions do not obey the Pauli principle 
does not make sense.
It was this argument that seemed like the 
devastating blow to the quark model.  Quarks could not exist as 
real particles because they did not obey the basic requirements
of a sensible quantum theory.  

In reality, this blow was not as devastating as it seemed.
In 1964 it was argued by O. W. Greenberg,
M. Y. Han and Y. Nambu \cite{wally} that 
an extension of the simple quark model could solve this problem.
If one endowed quarks with one other quantum direction in which they
could be antisymmetrized, a sensible quantum theory could be constructed.
There would have to be at least three directions in this
new internal space since each quark in a baryon has to be 
associated with a different direction.  Furthermore, since this
new quantum number had not manifested itself in experiment, all 
observable objects would have to have zero charge.  This requires
the number of internal directions to be exactly three, since otherwise
three quarks could not form a singlet.  Since the quarks must
always be antisymmetrized in this internal direction the rest
of their wavefunction will be symmetric, exactly as observed.
Therefore, at the expense of introducing a new 
quantum number, we have disarmed the most devastating
argument against the esthetically pleasing quark model.  Of course, 
this postulate seems ludicrous at first glance.  
If the esthetics of the quark model were not so
great, this theory may never have been advanced.  However, the
tremendous beauty of the quark model makes us want to 
interpret it physically 
rather than concede that it is nothing but mathematics.
However, we cannot simply say that the quantum number is invisible
and go home.  It must have observable consequences.  Furthermore, our
theory still lacks the mechanism required to bind quarks inside
hadrons.  
 
These problems persisted for many years since no believable candidate
for a theory of dynamical quarks appeared.  At the time, 
people were still trying to make sense of quantum field theory itself,
so there wasn't even a believable playing field on which to make
a model of the strong interactions.  In his book on 
current algebra, Gell-Mann suggested that quantum field theory
be used only as a means to,
``...construct a mathematical theory of the strongly interacting
particles, which may or may not have anything to do with reality,
find suitable algebraic relations that hold in that model, 
postulate their validity, and then throw away the model.''\cite{curr}.
This was the
attitude of the time.  Most of the activity in the field in the
late 1960's was in the area of current algebra.  Current algebra is
a method advanced by Gell-Mann which allows one to derive
experimentally verifiable relationships between observables 
from the symmetries underlying a theory without using the
full formalism of quantum field theory.
While the agreement between theory
and experiment during this time was impressive, 
it was only Gell-Mann's $SU(3)$ that was being tested.  

Real progress in the strong interactions continued in 1969 when 
E. D. Bloom et al. measured 
high energy electron-proton deep-inelastic
scattering at SLAC for the first time
\cite{disexp}.  Their data showed definitively that the proton
behaved at high energy as though it was made of pointlike constituents.
This prompted R. P. Feynman to advance his parton picture
of deep-inelastic scattering (DIS) \cite{parton},
in which he assumes that 
hadrons are made of pointlike 
{\it partons}.\footnote{He is careful not to use `quarks' in his model.}
This model has the advantage of describing DIS without assuming much about
the nature of partons or their interactions.
Further experiments at SLAC showed that the 
partons probed in DIS have spin-1/2.  It seemed that a revival
of the quark model was inevitable.  

Another result from the SLAC data had an even more profound effect on
the future of the strong interactions.  
At large angles, certain functions which parameterize the scattering
were found to {\it scale} with energy.  Rather than depending separately
on the energy and momentum transfer to the proton, they depend only
on a dimensionless ratio involving these two quantities.
This behavior is predicted in
the current algebra of Gell-Mann, as shown by 
C. G. Callan, D. J. Gross and J. D. Bjorken \cite{scaling}, 
and in the operator product expansion techniques of 
Wilson \cite{ope}, which will be discussed 
in detail in Section \ref{opedis}.  However,
the details of renormalization in quantum field theories
usually destroy this property.  Intending to use the 
experimental results to finally put these nonsensical
theories to rest, Gross attempted to
prove that there exist no interacting quantum field theories 
which admit scaling.  In a paper with Callan, he showed that in
order for a field theory to scale
it must have a property known as `asymptotic freedom', that is
it must behave like a free theory in the region of interest
(the asymptotically large energy region in which scaling
is predicted)\cite{CGasymp}.  Shortly thereafter, Gross and 
S. Coleman were able to show that the only kind of theory
known which could be asymptotically free is a non-abelian
gauge theory \cite{coleman}.  
In order to complete his proof
that quantum field theory cannot describe the 
strong interactions, Gross and his student F. Wilczek 
tackled the task of calculating directly whether or not these
theories could admit asymptotic freedom.  Much to their
surprise, the calculation revealed that non-abelian
gauge theories can indeed be asymptotically 
free \cite{asymp}.\footnote{H. D. Politzer also published this result 
simultaneously \cite{politz}.}  

This result suggests that the symmetry
associated with the quantum
number introduced
by Greenberg, Han and Nambu be used as the 
local symmetry required for a non-abelian quantum
field theory of quark interactions.  
A physical motivation for the introduction
of this new quantum number, called color because there are three
primary colors, 
had been found in
the total rate of $e^+e^-\rightarrow hadrons$, which 
was experimentally seen to be enhanced by a factor of three
over the expectation from only three quarks.  The
rate of the electromagnetic
decay $\pi^0\rightarrow2\gamma$
is similarly enhanced.
This marks the first appearance of the theory of the strong
interactions known today as quantum chromodynamics (QCD), or the
quantum dynamics of the color field.

At this time, it is advantageous for us to leave the historical
treatment of the strong interactions and focus on QCD.  
Modern QCD views hadrons as highly relativistic bound states
of nearly massless quarks and massless 
spin-1 objects known as gluons.\footnote{These are 
the analogue of QED's photons.}
The strong interactions between hadrons are viewed as  
residual interactions, much like the molecular forces
between neutral atoms.
Obviously, this aspect of the
modern theory of strong interactions complicates matters
greatly - especially since we still do not fully understand
the way it works in QED.
Quarks interact by exchanging gluons in exactly the same way that
electrons and protons interact by exchanging photons.  However, in
QCD gluons also exchange gluons.  This new property comes from the
fact that QCD is a non-abelian theory as opposed to QED, an 
abelian theory.  This will be explained in detail in 
Chapter \ref{basics}.  For now, 
we explore some of the qualitative aspects of QCD.

The most important consequence of the non-abelian nature
of QCD is the phenomenon of asymptotic freedom which 
led to its discovery.  This property causes QCD to become
a weakly interacting theory at short distances.  Looking
at this the other way, we see that QCD becomes strongly interacting
at long distances.  Since we don't know how to solve a strongly
interacting quantum field theory, we cannot work out what this
implies quantitatively.  However, we can qualitatively say that
if the interaction becomes stronger and stronger as color charges are
separated, we can never completely liberate a color charge. 
Since a consistent quantum field theory has no choice but to allow
the vacuum to create pairs of particles and antiparticles from 
ambient energy, the energy associated with separation of color charges
will at some point become great enough to permit the creation
of a pair of opposite color charges from the vacuum.  At this point,
the charges we were separating have happily combined with
the new vacuum charges to form color singlets, 
and we find ourselves with two hadrons where 
we once had only one.  This property of QCD is called
color confinement : the color is physically confined in singlet 
arrangements.  Of course, once the coupling gets large 
enough, the calculations required to show asymptotic freedom
are no longer valid.  It is for this reason that we
cannot {\it prove} the existence of color confinement in QCD
and it remains one of the most poorly understood aspects of
the strong interactions, and, more generally, the 
entire SM.

\begin{table}
\begin{tabular}{|c|r|r|}
\hline
{\rm Quark}&{\rm Constituent Mass}($m_e$)&{\rm Current
Mass}($m_e$)\\\hline
u&550-710&3-10\\
d&640-710&6-18\\
s&1000-1050&120-330\\\hline
\end{tabular}
\vspace{0.5in}
\caption{Constituent versus current quark masses for the
three lightest quarks.}
\label{qmass}
\end{table}

One of the biggest mysteries of contemporary QCD is the 
approximate validity of naive quark models.
Several phenomenological
models of quarks were employed in the 1960's to 
get the observed hadron spectrum.  If one assumes that a proton
is made of two $u$ quarks and a $d$ quark, a $\Lambda$ of
a $u$, $d$, and an $s$, a $K^+$ of a $u$ and
an $\overline s$, et cetera, one can make naive predictions
of the masses of the hadrons in terms of the quark masses.
A spin-spin interaction proportional to the inverse
masses of the quarks can account for the observed splitting
between the spin-1/2 and spin-3/2 baryons and give us even
more experimental data with which to compare.  Once we have
fit the few parameters of our phenomenological theory 
to data, we can predict the rest of the observed mass spectrum.
These predictions are rather accurate, to the credit of our naive
models.
The fits also determine effective masses for the
quarks.\footnote{The best fits require one
to use different quark masses for mesons and baryons.
The values obtained from meson data are always smaller than
those from baryons, and the fits are somewhat better in the 
baryon sector ($\sim 1\%$) than in the meson (as bad as 
$\sim10\%$).}  In addition, one can obtain the masses of
quarks in QCD from the current algebra approach mentioned above.
These two methods give very different results, as shown
in Table \ref{qmass}.  

Another red flag seen in this approach is that 
it has completely ignored gluons.  Gluons would certainly 
be expected to contribute to the energy of these highly 
relativistic bound states.  Furthermore, the wavefunctions
used to calculate the splitting between hadrons of different 
spin have assumed that the spin of hadrons comes only from 
quarks.  However, since gluons are vector particles they 
are expected to contribute to the spin of the hadrons as well.
In fact, asymptotically one expects approximately
50\% of the spin of the proton to be carried by gluons.  
These models have also ignored the expected contribution 
from the so-called `sea' quarks.  It is well-known
that relativistic quantum theories allow the vacuum to
require create pairs of particles and antiparticles from the vacuum.
It would certainly be naive of us to imagine that the 
highly relativistic strongly bound hadronic states consist
only of three quarks or a quark and an antiquark.  The simple
fact that these systems are so strongly bound gives the 
system quite enough ambient energy to create many quark-antiquark
pairs out of the vacuum.  Such vacuum fluctuations are called
the `sea of quarks' in the hadron.  Any realistic hadronic model
must take this sea into account as well.  

After all of
this analysis, we are faced with the question of why these
models work so well.  It seems that somehow the 
interactions of QCD arrange the quarks, antiquarks, and gluons
in such a way that a description of hadrons in terms of 
naive valence quarks is appropriate for certain quantities.
These effective degrees of freedom are commonly called 
{\it constituent} quarks.  The standard belief is that while these
constituent quarks have the effective masses listed in the first column
of Table \ref{qmass}, the true quarks of QCD have the much smaller `current'
masses in the second column.  Of course, due to the effect
of confinement, the masses of quarks and
gluons are not well-defined within quantum field theory.
Strictly speaking, the mass of an object 
is defined only as that object travels asymptotically far
away from any other object.  Since colored objects cannot
do this, the `mass' of a colored object has no meaning in this
sense.\footnote{This is actually a property of any unbroken non-abelian
gauge symmetry, whether or not it is asymptotically free.}

One of the major difficulties with QCD is the fact that its 
degrees of freedom are not the same as those we see in the laboratory.
This is an immediate consequence of confinement.  If we
would like to understand {\it any} strong interaction scattering
process, we must first understand the complicated bound states 
involved in the process.  This is an extremely difficult
path to take; we still do not fully understand the scattering of 
complicated atoms.  Nuclear bound states are even more difficult.
Using QCD to understand even simple nuclei is undescribably 
complicated.  For this reason, physicists have worked to find a theory
which is expressed in terms of the low energy degrees of freedom.
Since such a theory has composite objects as its fundamental fields,
it cannot be expected to make sense for arbitrarily large momenta.
However, if the momenta involved are small,
the effects of the internal structure of the fundamental fields 
can be absorbed into unknown constants and fit to the data.
Such theories are required to have the same symmetries as the underlying
theory, but are otherwise unconstrained.  This naturally leads
to a tremendously complicated theory whose lagrangian has infinitely
many terms.  The hope is that there is some small parameter 
in the theory which allows one to calculate only to finite
order.  

There are several theories which have achieved a 
limited degree of success in the past, all of which are derived 
from
chiral symmetry.  Chiral symmetry is an isospin-like symmetry
possessed by QCD in the limit of small quark mass.  An effective theory
based on this symmetry ends up being an expansion in the ratio of
external momentum or the pion mass to a scale which turns out to be 
near the proton mass.  This means the expansion parameter in 
chiral perturbation theory is about 1/7.  One interesting thing
about Chiral Perturbation Theory ($\chi$PT) is that it restores the 
pions to their original place as carriers of the strong interaction.
In $\chi$PT, one sees neutrons and protons exchanging
pions in a way equivalent to that originally envisioned by
Yukawa.  However, this theory is much more complicated.  
The infinitely many terms in its lagrangian make power counting - 
the systematic truncation of the lagrangian - the most important game in town.
In all expansions, one hopes that the next (uncalculated) coefficient
is not unnaturally large.  Even with a very small expansion
parameter, this can lead to trouble.  Unfortunately, many
quantities which have been calculated in $\chi$PT are plagued with
large coefficients.  This fact of life has made low energy 
strong interaction physics very difficult.  One wonders 
whether or not a sensible effective theory exists for this 
domain of nuclear physics.

Fortunately, the high energy side of QCD does not have
these difficulties.  Asymptotic freedom assures us that 
at high energy the QCD coupling is small and we may use perturbation
theory as we do in QED.  On the other hand, confinement
tells us that we cannot escape the low energy physics we do not 
understand.  Any physical process involving the strong interaction
must necessarily also involve complicated bound states since they
are the only physically acceptable external states of the theory. 
For example, DIS can be considered as the collision of a virtual
photon and a proton.  
If we consider low energy scattering,
the proton collectively absorbs the photon and recoils.  At very high energy,
the photon does not see the proton as a whole.  Rather, it 
resolves the bound state into its constituents and is absorbed
by one of the quarks.  
To properly calculate the amplitude for this process, we would
have to consider high energy scattering on an arbitrary parton
in the proton and then average over the existing species of parton.
High energy scattering on arbitrary partons can 
be done in QCD perturbation theory.  However, the 
averaging process 
requires knowledge of the elusive proton wavefunction.
Although we cannot calculate it, this object
appears in many experiments.  Since a proton is a proton
regardless of what experiment we decide to subject it to, this 
information could in principle be used to predict the outcome of
other experiments.  More importantly, if we can use the 
perturbative calculation of the scattering to extract this
information from experiment, we will have quantities which
relate the physical degrees of freedom to those of our microscopic
theory.  This somewhat roundabout way of using perturbation
theory to extract hadronic structural information from experiment is
the topic of this dissertation.  

In Chapter \ref{basics}, I give a somewhat detailed
introduction to the basics of perturbative quantum field theory
as applied to QCD.  This chapter is meant to
provide a mathematical foundation for the rest
of the thesis, as well as familiarize the reader
with some of my conventions.

Chapter \ref{dis} is devoted to the aforementioned
process of DIS.  First considered in the late 1960's, this process
provides the perfect first step for a look at factorization theorems,
parton distribution functions and operator product expansions.
It is here that the theoretical groundwork for 
my original research on deeply virtual Compton scattering (DVCS)
is laid.  

Chapter \ref{dvcs} encompasses the process of DVCS.
Introduced by X. Ji because 
of the structural information it probes
\cite{jispin}, DVCS can be used in conjunction with 
polarized DIS to extract the contribution to
proton spin due to quark {\it orbital} angular momentum.
This quantity cannot be isolated by any other experiment
proposed in the literature.
Using the treatment of Chapter \ref{dis} as a model,
we calculate the one-loop QCD corrections to 
DVCS at the leading twist level and use the results
to generalize Wilson's operator product expansion of two 
electromagnetic currents.  More importantly,
we show that this process is factorizable at this level
to all orders in QCD perturbation theory.  The material
presented in this chapter was first published 
by X. Ji and I in \cite{xdjjon}.

Chapter \ref{htwist} concerns the one-loop evolution
of the twist-3 structure function $f_T(x)$
measured in DIS with transverse polarization. 
After a brief review of some of the complications
that arise in transverse processes,
we discuss the problem of its horrendous
anomalous dimension matrix.  The fact that
this distribution mixes freely
with the more general two-parton correlators
$G_1(x,y)$ and $G_2(x,y)$ causes immense complications
in the analysis of experimental data.
These other
distributions are known to decouple
in the limit of large $N_c$
at the one-loop level \cite{ali}, leaving
only the simple evolution experienced by
twist-2 distributions.  This result is
studied from a new standpoint in which the 
reason for the simplification becomes obvious.
Our technique is extended to predict the 
outcome of the two-loop large-$N_c$ result 
for twist-3 distributions and the one-loop
twist-4.  Unfortunately, this
analysis reveals contributions which will
surely destroy the simplification in these sectors. 
The material presented in this chapter
is based on a paper published last year by
X. Ji and I in \cite{metwist}.

\chapter{The Basics of QCD}
\label{basics}

In the introduction, I have outlined QCD qualitatively as a 
renormalizable quantum
field theory describing the interactions of spin-1/2 quarks
and spin-1 gluons.  The form of this theory is governed by a
gauge principle based on a local $SU(3)$ symmetry.  This chapter is
dedicated to explaining quantitatively what that means.

\section{The Lagrangian}
\label{thelagrangian}

To build the lagrangian of QCD here, I will employ 
certain naturalness arguments which I believe to be a nice
way to look at the SM.  These arguments are certainly
not intended as proofs that this is the only way the
theory of strong interactions can be formulated.  The validity
of the theory is placed solely in the hands of 
experiment.  However, it is interesting to note that 
one can build QCD in this way given only a knowledge of
quantum field theory and an exact $SU(3)$ symmetry 
for some Dirac fields.

We begin with a theory of free spin-1/2 quarks.  Assuming
$n_f$ different types (flavors) of colored quarks, we write our
free lagrangian density as
\begin{equation}
{\cal L} = \sum_{f=1}^{n_f}
{\overline\psi_f}(i\!\not\!\partial-m_f)\psi_f\; .
\end{equation}  
$\psi_f$ denotes the color triplet of Dirac fields corresponding
to quarks of flavor $f$.  We note that quarks of different 
flavor are uncoupled.  
Our implicit assumption that color is
entirely an internal symmetry is reflected in the fact that the 
quark mass does not depend on the color.  The fact that 
this symmetry is internal means that there is no difference 
between quark colors and our theory should be invariant
under arbitrary reparameterizations of color.  Since reparametrizations
must not change the normalization of our fields, we restrict
transformations to $3\times 3$ unitary rotations, i.e. 
$\psi_f\rightarrow U\psi_f$ with $U$ an arbitrary unitary 
$3\times3$ matrix.  If the SM consisted only of this sector, these
transformations could be made separately for each flavor.
However, the weak sector breaks this larger symmetry group
by coupling quarks of different flavors.  In some sense, the weak
sector forces us to choose the same color convention for all flavors;  the
remaining symmetry is concerned with which convention was chosen.

The group of unitary $3\times 3$ matrices is called $U(3)$.  
Certain elements of $U(3)$ do not distinguish between
different colors; they merely give an overall phase to the 
entire triplet.  Such transformations are not associated 
with color reparametrization invariance\footnote{
These are are precisely the transformations which lead one to 
electromagnetic interactions.},  so
we exclude them from our group.  The remaining 
objects form the group of {\it special} unitary
$3\times 3$ matrices, $SU(3)$.
Many of the mathematical properties of $SU(3)$ are discussed in Appendix 
\ref{sun}.  At present we need only the form
of a general\footnote{The colorblind $SU(3)$ 
transformations $e^{2i\pi/3}$ and $e^{4i\pi/3}$
cannot be written in this form.  However, since these transformations are
indeed colorblind,
there are rotations in the other sectors which will compensate for them.
For this reason, we may completely ignore these troublesome phases with the 
rest of the electroweak sector of the SM.} 
$SU(3)$ matrix,
\begin{equation}
U=e^{i\theta^a t^a}\; .
\label{gtrans} 
\end{equation}
The $t^a$, $a$=1 to 8, are the $SU(3)$ generators
introduced in the appendix.  For now, their explicit form is unimportant.
We need only know that they are hermitian and traceless.  
The $\theta^a$ are arbitrary real numbers which we will take to be
small for most of our discussion.  

When we rotate
our quark fields, we do so in the same way at every
point in spacetime.  From the standpoint
of special relativity, this seems like a strange procedure.
We don't transmit any information
during this process, so it does not violate causality.  
However, it still is not a natural thing to do.  Ideally, we would like
to define conventions separately at each point in spacetime
so that the physics we describe is local.  
This means we wish our lagrangian to be invariant 
under arbitrary {\it spacetime dependent} $SU(3)$ transformations.
Such local transformations are called gauge transformations.

Since a derivative compares field values at two infinitesimally
close, but separate, spacetime points without taking into account
a difference in convention, our present lagrangian does not admit
this local symmetry.  However, if we decide that we really want
this symmetry to be local, we can alter our theory slightly 
to admit it.  The reason why the derivative term is not invariant
under the local symmetry is that $\partial\psi$ does not transform
in the same way as $\psi$.  There is an extra additive term
here which reflects the spacetime change in the transformation itself.
If we add an extra term to the derivative to cancel
this change in the transformation, our theory will be invariant.
With this in mind, we introduce a covariant derivative
${\cal D}\equiv\partial+i g{\cal A}$.  

Hermiticity of the
lagrangian\footnote{up to a total derivative} requires 
$g{\cal A}$ hermitian.
Since we want 
${\cal D}\psi$ to transform in the same way as $\psi$ itself, 
we require ${\cal D}\rightarrow U{\cal D}U^\dag$.  This implies that
\begin{equation}
{\cal A}\rightarrow U{\cal A}U^\dag-{i\over g}\;U\partial U^\dag\; .
\label{atrans}
\end{equation}
Any matrix ${\cal A}$ with a nonzero trace can be written as the
sum of a traceless matrix and a multiple of the identity.  As mentioned
above, the identity is colorblind and contributes only to the
electroweak sector of the SM.  For this reason, we consider
${\cal A}$ traceless.\footnote{
This property is not altered by the 
gauge transformation (\ref{atrans}),
as can be seen by inspection.}
All traceless hermitian $3\times 3$ matrices can be written
as a linear combination of the eight $SU(3)$ generators, so
${\cal A}(x)$ contains eight hermitian vector fields - 
one for each generator of $SU(3)$.  Objects which are related
to the generators of a group in this way are said to be in the
{\it adjoint} representation of the group.  

We have promoted our $SU(3)$ color symmetry to 
a local symmetry at the cost of introducing new vector fields.
Since these fields transform under the symmetry,
it is natural to consider them as dynamical fields in our
theory.  In order to do this, we must endow them with the
ability to propagate.  In other words, we must give
them a kinetic term.  
It is immediately obvious that the standard kinetic term,
$1/2\, \partial_\mu{\cal A}_\nu\partial^\mu{\cal A}^\nu$, is not invariant
under the local $SU(3)$ symmetry we have worked so hard 
to install.  The nonlinear transformation (\ref{atrans})
makes it difficult to construct an invariant 
kinetic term for ${\cal A}$.
However, since ${\cal D}$ contains ${\cal A}$ and 
transforms linearly under the action of the 
gauge group, we can use it to form a covariant
object which involves derivatives of ${\cal A}$:
\begin{equation}
{\cal F}^{\mu\nu}\equiv -{i\over g}
\left[{\cal D}^\mu,{\cal D}^\nu\right]\; .
\end{equation}
The scalar ${\cal F}^{\mu\nu}{\cal F}_{\mu\nu}$ contains
the appropriate terms for a vector boson kinetic
energy.  It's trace is also 
invariant under the gauge symmetry.
Actually, this term and the quark piece above are the {\it only} gauge-invariant
terms of dimension four or smaller\footnote{This is required for 
renormalizability, as will be mentioned in Chapter 4.} involving only these
fields.\footnote{There is one other term which is thought to 
contribute to physics beyond perturbation theory.  
Its inclusion violates the discrete spacetime symmetries
of parity and time-reversal, and its 
coefficient is experimentally seen to be 
extremely small.}  In particular, there can be no term that corresponds
to a mass for the ${\cal A}$ fields.  

Collecting these pieces and
normalizing the kinetic term of our vector bosons,
we write the full QCD lagrangian density as
\begin{equation}
{\cal L} = \sum_{f=1}^{n_f} {\overline\psi_f}
(i\,{\cal\not\!\! D}-m_f)\psi_f
-{1\over 2}{\rm Tr}\, {\cal F}^{\mu\nu}{\cal F}_{\mu\nu}\; .
\label{unlag}
\end{equation}
This expression looks nearly identical to that of QED.  
In fact, the only difference is in the gauge group.  
Let us calculate ${\cal F}$ : 
\begin{equation}
{\cal F}^{\mu\nu}=\partial^\mu{\cal A}^\nu-\partial^\nu{\cal A}^\mu
         +ig[{\cal A}^\mu,{\cal A}^\nu]\; .
\end{equation}
The first two terms are familiar from QED, while the last is not.  
Since we can express the gauge field ${\cal A}$ as a linear combination 
of generators of the gauge group, the last term
expresses the commutator of two generators of $SU(3)$.  For abelian
groups, where all of the generators commute,
this term is not present.  This
is trivially the case in QED, where there is only one generator.
In non-abelian theories, like QCD, the inclusion of this term is 
unavoidable.
We will see that it leads to self-interactions
among gluons, the quanta of the ${\cal A}$ fields.  
It is these interactions which cause the theory
to be asymptotically free, as shown in Section \ref{renincov}.

\section{The Road to Quantization}
\label{roadtoquant}

Now that we have a lagrangian, we would like to quantize the 
theory.  In scalar field theory, this is implemented merely
by imposing commutation relations on the fields and their
conjugate momenta.  One introduces creation 
and annihilation operators\footnote{See Appendix \ref{canquant}.}
and proceeds with a straightforward evaluation of matrix elements.
In QED, it is not quite that simple.  Gauge 
symmetry requires the introduction of unphysical degrees of freedom.
This is obvious because a gauge transformation is by 
definition an unphysical thing.  It corresponds merely to
a redefinition of convention.  For each configuration
of the photon field, there exist infinitely many 
physically identical configurations which are obtained
by transforming the original.  Taking all of these into account
separately is highly redundant and creates divergences 
which must be removed before quantization 
can make sense.\footnote{This 
is purely an artifact of our mathematical incompetence.
If we were smarter, we would not have to do such violence to
gauge theories in order to make sense of them.} 
In order to make sense of a gauge theory, one must
first choose a gauge so that every physically distinct 
field configuration is taken into account exactly once.
The way this gauge is chosen is arbitrary, but 
certain choices make the resulting formalism more complicated
than others.  

To fix a gauge, one chooses a function
of the gauge field and sets it equal to a certain function of spacetime.
Normally, people choose a linear function 
to simplify the manipulations necessary to fix the gauge.
A standard example is the axial gauge.  This gauge choice is made
by requiring $n\cdot{\cal A}(x)=\omega(x)$ for some function
$\omega$ and some fixed vector $n$.  
If $n^*$ satisfies $n^*\cdot n=1$,
then  
\begin{equation}
{\cal A}(x)\rightarrow {\cal A}(x)
+\partial\int^x (\omega(z)-n\cdot{\cal A}(z))n^*\cdot dz
\end{equation}
is an abelian gauge transformation which takes an arbitrary
field configuration into one that satisfies the gauge condition.
Hence this condition can be satisfied by all physically distinct
configurations.\footnote{In a non-abelian 
gauge theory, the required transformation is
somewhat more involved.  An explicit solution 
is 
\begin{equation}
U=({\rm P}e^{-ig\int^x\,\omega(z)n^*\cdot dz})
({\rm\overline P}e^{ig\int^x\,n\cdot{\cal A}(z)n^*\cdot dz})\;\; ,
\end{equation} 
where (${\rm\overline P}$) P denotes
(anti-)path ordering of the exponential.  
This ordering is constructed so that 
\begin{equation}
n\cdot\partial U=ig(Un\cdot{\cal A}-\omega U)\;\; .
\end{equation}}
However, this gauge choice does not completely 
specify the gauge.  If $v$ satisfies $v\cdot n=0$, 
the gauge transformation 
\begin{equation}
{\cal A}(x)\rightarrow {\cal A}(x)
+\partial f(v\cdot x) 
\label{resaxi}
\end{equation} 
does not change $n\cdot{\cal A}$.
This is called a residual gauge
transformation.  If one wishes to completely specify the
gauge, one must enforce a further gauge restriction to specify
the residual gauge.  This is not usually necessary.  In many
cases, the ambiguities associated with residual gauge transformations
are not enough to interfere with the quantization process.  With the
choice of axial gauge, the residual degrees of freedom exist
in a special corner of momentum space.  We will see that our gauge-fixed
theory regards this corner as a singular region of momentum
space, and special care must be taken near it.

In the axial gauge QCD and QED are very similar,
but the problems associated with its residual gauge transformations
can be quite severe.  The axial gauge also breaks Lorentz invariance 
explicitly by introducing the special direction $n$ during the 
quantization process.  For these reasons, another gauge-fixing 
function is often employed.  The `covariant gauges' are
obtained by specifying $\partial\cdot{\cal A}(x)=\omega (x)$.
Once again, this gauge can be shown to accommodate all physical 
configurations since the abelian transformation
\begin{equation}
{\cal A}^\mu(x)\rightarrow {\cal A}^\mu(x)+\partial^\mu\int d^{\, 4}z\,
{d^{\, 4}q\over (2\pi)^4}e^{iq\cdot (x-z)}
{\partial_\nu{\cal A}^\nu(z)-\omega(z)\over q^2}
\end{equation}
takes any configuration into one which satisfies 
our gauge condition.\footnote{Once 
again, this transformation is more involved for
non-abelian groups.  In this case, the explicit transformation is
too complicated to present here.} 
This gauge choice also contains an ambiguity since any
transformation of the form
\begin{equation}
{\cal A}^\mu(x)\rightarrow {\cal A}^\mu(x)+\partial^\mu\chi(x)
\label{rescov}
\end{equation}
with $\chi(x)$ harmonic\footnote{I.e. 
$\partial_\mu\partial^\mu\, \chi(x)=0$} leaves $\partial\cdot{\cal A}$
invariant.  Immediately, one can see that the residual transformations 
associated with this gauge choice occupy a much smaller
region of momentum space than those of the axial gauge.  
In the previous case, one has an infinite three-dimensional subspace of 
momentum to contend with.  Here, only the light-cone itself
is a singular region.\footnote{The statement 
$\partial\cdot\partial\chi=0$ reads $q^2=0$ in momentum space.}  This
region is the three-dimensional surface $(q^0)^2=|\,\vec{q}\,|^2$, which
is compact for each value of $|q^0|$.  Furthermore, the 
divergences arising from this ambiguity in axial gauge require
special treatment.  Those appearing in covariant gauges
can be treated along with other divergences which appear naturally
in quantum field theory.  We will see this in detail
in the following sections.

Let us now proceed with the quantization process.  To illustrate the
similarities between QED and QCD, we will not specify the gauge group
immediately.  In my view, the best way to see the process of 
gauge fixing lies in the path integral approach.\footnote{For
a derivation of the path integral and a discussion of its 
technical points and physical meaning, see Ref. \cite{Pathint}.
Essentially, the path integral is an instruction to 
sum the contributions of every field configuration.
This is done by integrating over the value of a 
field $\Phi(x)$ at {\it every} spacetime point $x$.}
In this approach to quantum field theory, vacuum matrix
elements of an operator are understood as weighted
averages of the operator over all possible field 
configurations.  The weight given to a particular configuration
is the exponential of its action, ${\cal S}[\Phi]
\equiv\int d^{\, 4}x\,{\cal L}[\Phi]$ : 
\begin{equation}
\left\langle{\rm T}{\cal O}(x_1,\ldots ,x_n)\right\rangle_{vac}
={\int{\cal D}[\Phi]{\cal O}(x_1,\ldots ,x_n)e^{i{\cal S}[\Phi]}
\over \int{\cal D}[\Phi]e^{i{\cal S}[\Phi]}}\;\;\; ,
\label{exp}
\end{equation}
where $\Phi$ stands for all fields in the theory and ${\rm T}{\cal O}$
is any $n$-point correlation function of any of the fields
$\Phi$ appropriately time-ordered.\footnote{One defines the 
{\it time ordering operator}, T, such that
\begin{equation}
{\rm T}\,\{\Phi(x_1)\Phi(x_2)\}\equiv
\Phi(x_1)\Phi(x_2)\Theta(x_1^0-x_2^0)\pm
\Phi(x_2)\Phi(x_1)\Theta(x_2^0-x_1^0)\;\; .
\end{equation}
The $+$ refers to bosons while the $-$ refers to
fermions.}
The path integral takes
{\it all} field configurations into account, which means that it
is redundantly integrating over a continuous infinity
of physically identical gauge configurations.  The idea behind
gauge fixing is to make this redundant integration explicit
so it will cancel in the ratios appearing in physical
matrix elements.  Let us study only the denominator
of Eq.(\ref{exp}).  Our derivation will follow for 
the numerator as long as ${\cal O}$
is gauge-invariant.
Parameterizing a general gauge transformation by real functions
$\theta^a(x)$ as in Eq.(\ref{gtrans}), we write the identity
\begin{equation}
1=\int{\cal D}\theta^a\delta({\cal G}^a[{\cal A}^\theta]-\omega^a)
\left|{\rm det}\left(
{\delta {\cal G}^a[{\cal A}^\theta]\over\delta\theta^b}
\right)\right|_{\cal A}\;\; .
\label{one}
\end{equation}

This equation says that there exists a unique\footnote{I note here
that neither of the gauge conditions specified above satisfy this
property, as exemplified by Eqs.(\ref{resaxi},\ref{rescov}). 
What we really need is for this expression to be independent
of ${\cal A}$ and $\omega^a$.  Both the axial gauge and the covariant gauge
satisfy this requirement.  The actual number of such 
gauge transformations will cancel in the calculation of 
any physical matrix element as long as it is the {\it same}
for {\it all} configurations.} 
gauge transformation
parameterized by $\theta^a(x)$ which will transform a given arbitrary
field configuration ${\cal A}$ into a new configuration ${\cal A}^\theta$
which satisfies the gauge conditions 
${\cal G}^a[{\cal A}^\theta]=\omega^a$.\footnote{Note 
that we have one gauge condition for each
generator.}
The determinant is the Jacobian of the change of variables between 
${\cal G}$ and $\theta$.  Inserting this into the denominator 
of Eq.(\ref{exp}) and performing the 
inverse gauge transformation parameterized
by $-\theta^a$ on the integration variables,\footnote{
Since this transformation is unitary on 
the fermionic fields and the gauge fields (plus
a trivial constant shift), the measure of the path integral
is invariant.  If it weren't,
the path integral itself wouldn't be gauge invariant.} we obtain
\begin{equation}
\int{\cal D}\theta^a\,{\cal D}[\Phi]\left|{\rm det}
\left({\delta {\cal G}^a[{\cal A}^\theta]\over\delta\theta^b}
\right)\right|_{{\cal A}^{-\theta}}
\delta({\cal G}^a[{\cal A}]-\omega^a)e^{i{\cal S}}\;\; . 
\label{omeg}
\end{equation}
It is easily seen by inspection that the dependence of the 
determinant on $\theta$ is entirely illusory, so the 
integration over $\theta$ factorizes completely and 
cancels in the ratio (\ref{exp}).  This factor directly represents
the redundancy that led to the need for gauge fixing in the 
first place.

Since the $\omega^a(x)$ were introduced as arbitrary functions,
we may consider
a weighted average over all possible $\omega^a$ :
\begin{equation}
{1\over 
\int{\cal D}\omega^a f[\omega^a]}
{\int{\cal D}\omega^a\int{\cal D}[\Phi]
f[\omega^a]
\left|{\rm det}
\left({\delta {\cal G}^a[{\cal A}^\theta]\over\delta\theta^b}
\right)\right|_{{\cal A}^{-\theta}}
\delta({\cal G}^a[{\cal A}]-\omega^a)e^{i{\cal S}}}\;\; .
\end{equation}
The integration over $\omega$ cancels independently
of the form of the functional $f[\omega^a(x)]$
if we use the fact that Eq.(\ref{omeg}) does not actually depend
on $\omega^a$.  On the other hand, we may consider the 
denominator an unimportant (though infinite) constant 
and use the $\delta$-function to do 
the $\omega^a$ integral in the numerator :
\begin{equation}
{1\over \int{\cal D}\omega^a\, f[\omega^a]}\;
\int{\cal D}[\Phi]
f[{\cal G}^a[{\cal A}]]
\left|{\rm det}
\left({\delta {\cal G}^a[{\cal A}^\theta]\over\delta\theta^b}
\right)\right|_{{\cal A}^{-\theta}}e^{i{\cal S}}\; .
\end{equation}
Note that although the constraint is 
no longer explicit, it is obvious from our 
construction that in some sense 
${\cal G}^a[{\cal A}]$ {\it does not depend on} ${\cal A}$.\footnote{
This statement is true in the sense that, once all
physically indistinguishable field configurations
are taken into account, the contribution from each physically
distinct set of configurations is identical.}

All that is left is the calculation of the determinant.
Since this depends nontrivially on the form of the
gauge condition and the specific theory under consideration,
we cannot go any further without specification.

Since QED is less complicated, we will consider it first.
In an axial gauge,
\begin{equation}  
{\cal G}[{\cal A}^\theta]=n\cdot{\cal A}+1/g\, n\cdot\partial\theta\;\; .
\end{equation}
Hence we are asked to evaluate the determinant of
$1/g\, n\cdot\partial$.  The meaning of this object
is not obvious.  If we really wanted to evaluate it, we
would have to consider the actual definition of the path integral.
Such things are considered in \cite{Pathint}.  Here, we do 
not care about this infinite constant.  Since it is independent
of all the relevant fields,
it will cancel in the calculation of physical quantities 
(much like every other constant we have ignored in this derivation).
Hence in this case the determinant may be completely ignored.

In a covariant gauge, the Jacobian,
 ${\rm det}(1/g\,\partial\cdot\partial)$, is
also constant.  This leaves the function $f[\omega]$ as the
sole remnant of our gauge-fixing
in QED.  Since we must
live with it, we intend to use our freedom to choose the form 
of $f[\omega]$ to 
the best of our advantage.  
This is accomplished
by choosing a function which will add a quadratic
term to the lagrangian.  Any other choice would lead to new 
vertices for us to take into account 
in perturbation theory.\footnote{
Actually, any other choice would lead to disaster in the sense
that one still could not define a propagator.  In this way, the 
choice of $f[\omega]$ is made for us by the way we know how
to do quantum field theory.  Of course, in principle the choice is arbitrary.
However, we must know how to live with whatever choice we make.}
The function $f[\omega]=\exp[-i\int\, d^{\, 4}x\,\omega(x)^2/2\xi]$
suits our purpose.  

The only effect of all this work is the simple
shift in the lagrangian density 
\begin{equation}
{\cal L}_{\rm QED}^{axial}={\cal L}_{\rm QED}^{0}-
{(n\cdot{\cal A})^2\over 2\xi}
\end{equation}
for the axial gauge and 
\begin{equation}
{\cal L}_{\rm QED}^{cov}={\cal L}_{\rm QED}^{0}-
{(\partial\cdot{\cal A})^2\over 2\xi}
\end{equation}
for the covariant gauge.  There is one degree of freedom
left from the choice of weighting function - the number $\xi$.  
In principle, this object could be a function of spacetime since the
weighting was {\it completely} arbitrary.  However, again the
limitations of our mathematical abilities force us to be less general
than the derivation requires.  As it is, we take $\xi$ as
an arbitrary constant.  This degree of freedom is useful as a 
check of lengthy calculations since its dependence must cancel
in all gauge-invariant quantities.

We turn now to QCD.  The similarities between QCD and QED
are more apparent in the axial gauge, so we begin with 
this choice.
Our first task is to rigorously define what is meant by
the Jacobian under consideration.  The matrix 
$\delta{\cal G}/\delta\theta$ is unambiguously given by
\begin{equation}
{\cal G}^a[{\cal A}^{\theta(x)+\delta\theta(x)}(x)]
-{\cal G}^a[{\cal A}^{\theta(x)}(x)]=
\int\, d^{\, 4}y\,({\delta{\cal G}/\delta\theta})^{a\, b}
(x,y)\delta\theta^b(y)\; .
\end{equation}
As mentioned above, our Jacobian does not actually depend
on $\theta$.  Hence we may take $\theta=0$ to simplify
our task.  For infinitesimal $\delta\theta$, 
Eq.(\ref{atrans}) reduces to 
\begin{equation}
{\cal A}\rightarrow{\cal A}-{1\over g}[{\cal D},\delta\theta]\; .
\end{equation}
The second term in this expression is the covariant derivative of 
$\delta\theta$ :
\begin{eqnarray}
[{\cal D},\theta]&\equiv& t^a\,{\cal D}^{\,a\, b}\,
\theta^b \;\; ;\nonumber\\
{\cal D}^{\,a\, b}&=&\partial\,\delta^{a\, b}
+g\, f^{a\, b\, c}{\cal A}^c\;\; ,
\end{eqnarray}
where $f^{a\, b\, c}$ are the structure constants of $SU(3)$
defined in Appendix \ref{sun}.
Our matrix is given by 
\begin{equation}
-{1\over g}\, n\cdot{\cal D}_x^{\,a\, b}\delta\theta^b (x)=\int d^{\, 4}y
({\delta{\cal G}/\delta\theta})^{a\, b}(x,y)
\delta\theta^b(y)\;\; ,
\end{equation}
so
\begin{equation}
({\delta{\cal G}/\delta\theta})^{a\, b}(x,y)
=-{1\over g}\, n\cdot{\cal D}_x^{\,a\, b}\delta^{(4)}(x-y)\; .
\end{equation}

This matrix appears to depend on ${\cal A}$ through the covariant
derivative.  However, 
the dependence of the matrix on ${\cal A}$ is of the form
$n\cdot{\cal A}$.  Since we have {\it fixed} this component of 
${\cal A}$, it no longer depends on the physical configuration under
consideration.  This allows us
to ignore the determinant once again,  making the quantization
process in axial gauge identical for QED and QCD.  The only
difference between the two theories appears in the gluonic self-coupling
terms in the kinetic energy.

The situation is not so simple in covariant gauge.  
The derivation of the matrix goes through as before.  We find
\begin{equation}
({\delta{\cal G}/\delta\theta})^{a\, b}(x,y)
=-{1\over g}\, \partial\cdot{\cal D}_x^{\,a\, b}\delta^{(4)}(x-y)\; .
\end{equation}
This time, the dependence on ${\cal A}$ is not illusory.  
Since $\partial$ is a differential operator that
will act on the $\delta$-function as well as the gauge field, 
the determinant
of this matrix will contribute to the dynamics of our gauge-fixed
theory.  We must figure out some way to take it into 
account in our calculations if we want to use this gauge.

By direct analogy to $N$-dimensional integrals, one
can perform
Gaussian path integrals exactly.  
For any nonsingular hermitian matrix ${\cal M}$, we have
\begin{equation}
\int{\cal D}\phi^\dag{\cal D}\phi e^{i\int d^{\, 4}x\, d^{\, 4}y
\;\phi^\dag_i (x){\cal M}_{i\, j}(x,y)\phi(y)_j}
=\left|{\rm det}{\cal M}\right|^{-1} \;
\label{pathbose}
\end{equation}
if the fields $\phi$ and $\phi^\dag$ are bosonic (or commuting) 
fields and 
\begin{equation}
\int{\cal D}{\overline\psi}{\cal D}\psi e^{i\int d^{\, 4}x\, d^{\, 4}y
\;{\overline\psi}_i (x){\cal M}_{i\, j}(x,y)\psi(y)_j}
=\left|{\rm det}{\cal M}\right| \;
\label{pathfermi}
\end{equation}
for fermionic (anticommuting) fields $\psi$, $\overline\psi$.
Hence a simple way to take care of our 
problem is to utilize 
auxiliary anticommuting fields $c$ and $\overline c$  
to exponentiate the determinant.  These fields are {\it not}  physical.
They are only mathematical artifacts we introduce to help us through
the day.  With this little snag out of the way, we write the
QCD lagrangian with covariant gauge fixing as\footnote{The factors of
$i$ and $g$ go into our overall normalization as before.}
\begin{equation}
{\cal L}_{\rm QCD}^{cov}={\cal L}_{\rm QCD}^0
-{(\partial\cdot{\cal A}^a)^2\over 2\xi}
-{\overline c}\,\partial\cdot{\cal D}\,c\; .
\end{equation}

The new anticommuting `ghosts' are very strange objects.  
Contrary to appearances, 
$c$ and $\overline c$ are completely 
unrelated complex fields\footnote{The bar
over $\overline c$ is merely convention.  There is no $\gamma^0$
implied here.}  which behave dynamically
like scalars.  Their sole function is to account for
the Jacobian in Eq.(\ref{one}).  If it weren't 
for the gluon field appearing in the covariant derivative, 
they would not interact with 
any physical field and their fluctuations
would simply shift the energy of the vacuum by a constant amount.
These shifts would then cancel with the fluctuations of the 
denominator of Equation (\ref{exp}) for any properly normalized
physical matrix element.
However, their interaction with $\cal A$ allows an operator 
${\cal O}$ to {\it polarize} these fluctuations 
so that they do not cancel its matrix elements.
We will come back to the story of ghosts later when we consider the 
renormalization of the gluon fields.

At this point, it is convenient to introduce source terms into 
our action in order to define a generating functional
for correlation functions : 
\begin{eqnarray}
{\cal S}[\Phi,{\cal J}]&\equiv & {\cal S}[\Phi]+\int d^{\, 4}x\,
{\cal J}(x)\Phi(x)\nonumber\\
{\cal Z}[{\cal J}]&\equiv &{\int{\cal D}[\Phi]e^{i{\cal S}[\Phi,{\cal J}]}
\over \int{\cal D}[\Phi]e^{i{\cal S}[\Phi,0]}}\; .
\label{defz}
\end{eqnarray}
I have introduced a source for each field in the
theory.
Since taking functional derivatives with respect to the external source
${\cal J}$ and subsequently taking ${\cal J}=0$ 
can give us any correlation function,\footnote{
Of course, here we must worry about the interchange
of functional integration and functional differentiation.
While this interchange is extremely difficult if not impossible
to justify, the theory it leads to is the same as that obtained
by other methods in all cases where there are other methods.
Such technical issues are taken up to some extent in 
\cite{Pathint}.} the only object we need to consider
is ${\cal Z}[{\cal J}]$.  This is by no means a trivial
task.  The difficulties associated with QCD stem almost exclusively
from an inability to calculate this object.  A whole branch of 
QCD is dedicated to calculating this quantity numerically.
Even in this approach there are many problems associated
with discretization, finite lattice size, finite computing
power, etc.  Here, we will approach the problem in
the same way it is approached for QED.  Assuming the coupling
$g$ to be small, we solve the free theory and take interactions
into account perturbatively.  This method is certainly
not without its shortcomings, some of which I will try to explain
along the way.

Since we can solve the free theory using (\ref{pathbose})
and (\ref{pathfermi}), it is
advantageous for us to separate this part from
the interaction part.  We will do this explicitly for the
covariant gauges since they are the most complicated.
The construction is identical in axial gauge (ignoring the ghosts).
\begin{eqnarray}
{\cal L}^{cov}_{\rm QCD}[\Phi,{\cal J}]&=&{\cal L}_{free}[\Phi]+
{\cal L}_{int}[\Phi]+{\cal L}_{source}[\Phi,{\cal J}]\; ;\nonumber\\
{\cal L}_{free}[\Phi]&=&\sum_f{\overline\psi}_f(i\!\not\!\partial
-m_f)\psi_f\nonumber\\
&&-{1\over 4}\, (\partial_\mu{\cal A}_\nu^a-\partial_\nu{\cal A}_\mu^a)
(\partial^\mu{\cal A}^{a\,\nu}-\partial^\nu{\cal A}^{a\,\mu})
-{(\partial\cdot{\cal A}^a)^2\over 2\xi}\nonumber\\
&&-{\overline c}^a \partial^2 c^a\; , \nonumber\\
{\cal L}_{int}[\Phi]&=&-g\sum_f{\overline\psi}_f\, \!\not\!\!\!{\cal A}\,\psi_f
\nonumber\\
&&+{1\over 2}\,
gf^{a\, b\, c}{\cal A}_\mu^a{\cal A}_\nu^b\partial^\mu{\cal A}^{c\,\nu}
-{1\over 4}g^2f^{a\, b\, c}f^{a\, d\, e}{\cal A}^{b\,\mu}{\cal A}^{c\,\nu}
{\cal A}^d_\mu{\cal A}^e_\nu\nonumber\\
&&+gf^{a\, b\, c}{\overline c}^a\partial^\mu{\cal A}_\mu^bc^c\hfil\nonumber\\
{\cal L}_{source}[\Phi,{\cal J}]&=&{\overline\eta}\psi+{\overline\psi}\eta
\nonumber\\
&&+j_\mu^a{\cal A}^{a\,\mu}\nonumber\\
&&+{\overline\zeta}^ac^a+{\overline c}^a\zeta^a\; .
\end{eqnarray}

As advertised, the free part of the lagrangian density is
is quadratic in the fields and the interaction part vanishes with
the coupling $g$.\footnote{Of course, this is the reason $g$ was introduced
in the first place.}  We have argued above that {\it any} correlation
function can be obtained by functional differentiation with respect
to the sources.  This means that we can get {\it any} functional
of the fields under the path integral by judicious use of the
source terms.  In particular, the functional 
$\exp[i{\cal S}_{int}[\Phi]]$ :\footnote{This is why we must fix our gauge
{\it before} introducing sources.  If we did not, we
could only couple sources to gauge-invariant
field combinations.  While this is fine for physical quantities, 
the gauge-dependent separation of the action into 
`free' and `interaction' contributions would prevent us 
from introducing perturbation theory.} :
\begin{equation}
\int{\cal D}[\Phi]e^{i{\cal S}[\Phi,{\cal J}]}=e^{i{\cal S}_{int}
[{1\over i}{\delta\over \delta{\cal J}}]}\int{\cal D}[\Phi]
e^{i({\cal S}_{free}[\Phi]+{\cal S}_{source}[\Phi,{\cal J}])}\; .
\end{equation}

This is as far as we can go without getting our hands dirty.
At this point, we must do the path integral.  Fortunately,
our manipulations have resulted in a problem we can solve.
Our path integral now splits into three integrals - one for each
species of particle in our theory (counting ghosts and antighosts
as the same species).  Since it is only the source dependence
that matters, we can complete the square using\footnote{This
expression is valid for any {\it nonsingular} matrix ${\cal M}$.
This is one reason we must fix our gauge before we quantize.}
\begin{equation}
\Phi^\dag{\cal M}\Phi + {\cal J}^\dag\Phi + \Phi^\dag{\cal J}
=(\Phi^\dag+{\cal J}^\dag{\cal M}^{-1}){\cal M}
(\Phi+{\cal M}^{-1}{\cal J})-{\cal J}^\dag{\cal M}^{-1}{\cal J}\;\; ,
\end{equation}
shift the integration variables, and obtain\footnote{This result
is valid only for complex fields.  For hermitian fields 
we use $1/2\,\Phi{\cal M}\Phi+{\cal J}\Phi=
1/2\,\Phi'{\cal M}\Phi'-1/2\,{\cal J}{\cal M}^{-1}{\cal J}$.}
\begin{eqnarray}
\int{\cal D}[\Phi]e^{i{\cal S}_{free+source}[\Phi,{\cal J}]}
&=&e^{-i\int d^{\, 4}x\,d^{\, 4}y\,{\cal J}^\dag(x)
{\cal M}^{-1}(x,y){\cal J}(y)}\nonumber\\
&&\qquad\qquad\;\times\;\;\int{\cal D}[\Phi]e^{i\int
d^{\, 4}x\,d^{\,4}y\,\Phi^\dag(x){\cal M}(x,y)\Phi(y)}\; .
\end{eqnarray}
Once again, we ignore the infinite shift to the vacuum energy and
obtain 
\begin{equation}
{\cal Z}[{\cal J}]={e^{i{\cal S}_{int}[{1\over i}
{\delta\over\delta{\cal J}}]}e^{-i\int d^{\,4}x\,d^{\,4}y\,
{\cal J}^\dag(x){\cal M}^{-1}(x,y){\cal J}(y)}\over 
e^{i{\cal S}_{int}[{1\over i}
{\delta\over\delta{\cal J}}]}e^{-i\int d^{\,4}x\,d^{\,4}y\,
{\cal J}^\dag(x){\cal M}^{-1}(x,y){\cal J}(y)}\left|_{{\cal J}=0}\right.}
\; .
\end{equation}
All that is left is a calculation of ${\cal M}^{-1}$ for each 
species of particle.

We start with the ghost fields since the inversion
is simplest for them.  The free ghost matrix is written
\begin{equation}
{\cal M}_{ghost}^{a\, b}(x,y)=
-\partial^2\delta^{a\, b}\delta^{\,(4)}(x-y)\; .
\end{equation}
By definition, the inverse of this matrix satisfies
\begin{equation}
\int d^{\, 4}z ({\cal M}^{-1})^{a\,c}(x,z){\cal M}^{c\,b}(z,y)=
\delta^{a\,b}\delta^{\,(4)}(x-y)\; .
\end{equation}
Obviously, the $SU(3)$ inverse is trivial : 
$({\cal M}^{-1})^{a\,b}\sim\delta^{a\,b}$.  It is the spatial
inversion that we must think about. 
Since we have not specified a coordinate system,
${\cal M}^{-1}$ cannot depend on one.  Hence our inverse matrix
is a function only of the difference $x-y$.  Performing
a Fourier transform on this variable, substituting the
Fourier representation for the $\delta$-function in $\cal M$, 
and inverting the transform we obtain
\begin{equation}
(\tilde{\cal M}_{ghost}^{-1})^{a\,b}(k)={\delta^{a\,b}\over k^2}\; ,
\end{equation}
or
\begin{equation}
({\cal M}_{ghost}^{-1})^{a\,b}(x,y)=\int{d^{\,4}k\over(2\pi)^4}
{\delta^{a\,b}\over k^2}e^{-ik\cdot(x-y)}\; .
\end{equation}
In the other two sectors the calculation is similar, but
complicated by the presence of spin indices.\footnote{
It is here that we run into trouble if we allow $\xi$ to
vary with spacetime.  In that case, the Fourier transform 
does not go through in the same way.  Presumably it can
be done, but the calculation will be horribly complicated.}
The matrices
\begin{eqnarray}
({\cal M}_{quark})_{\alpha\beta}(x,y)&=&\left(i\partial_\mu
\gamma^\mu_{\alpha\beta}-m\delta_{\alpha\beta}\right)\delta^{\,(4)}
(x-y)\nonumber\\
({\cal M}_{gluon})^{a\,b}_{\mu\nu}&=&\left[g_{\mu\nu}\partial^2
-\left(1-{1\over\xi}\right)\partial_\mu\partial_\nu\right]
\delta^{a\,b}\delta^{\,(4)}(x-y)
\end{eqnarray}
have inverses \footnote{Here, we have completely suppressed
quark flavor and color indices.  Barring the 
difference in quark mass, these matrices are proportional to
the identity in both spaces.}
\begin{eqnarray}
({\cal M}_{quark}^{-1})_{\alpha\beta}(x,y)&=& 
\int{d^{\,4}p\over(2\pi)^4}{({\not\! p}_{\alpha\beta}+m\delta_{\alpha\beta})
\over p^2-m^2}e^{-ip\cdot (x-y)}\\
({\cal M}_{gluon}^{-1})_{\mu\nu}^{a\,b}(x,y)&=&\int{d^{\,4}q\over(2\pi)^4}
{-g_{\mu\nu}+(1-\xi)(q_\mu q_\nu/ q^2)\over q^2}
\delta^{a\,b}e^{-iq\cdot (x-y)}\; .\nonumber
\end{eqnarray}

The physical meaning of these objects becomes clear when we study
the two-point correlation functions.  For simplicity, we consider
the free theory ($g$=0).  The ghost correlation function
\begin{eqnarray}
\left\langle {\rm T} c^a(x){\overline c}^b(y)\right\rangle_{vac}
&=&{1\over i}{\delta\over\delta{\overline\zeta}^a(x)}
\left(-{1\over i}\right){\delta\over\delta\zeta^b(y)}\nonumber\\
&&\;\;\times\;\;e^{-i\int
d^{\,4}w\,d^{\,4}z\,{\overline\zeta}(w)({\cal M}^{-1}_{ghost})(w,z)
\zeta(z)}\left|_{\zeta={\overline\zeta}=0}\right.\nonumber\\
&=&i({\cal M}^{-1}_{ghost})(x,y)
\end{eqnarray}
is nothing but $i$ times the inverse ghost matrix.\footnote{
The signs in this calculation are somewhat tricky.  
We must remember that the ghost and quark fields and sources
are Grassmann variables, and as such {\it anticommute}
rather than commute.}  This tells us that the amplitude
for a ghost to propagate from $y$ to $x$ is given by 
$i({\cal M}^{-1}_{ghost})(x,y)$.  If $y^0<x^0$, the ghost
is traveling forward in time and should have positive
energy.  To see which energies contribute, we attempt to do the
$k^0$ integral via contour.  The singularities at
$k^0=\pm|\,{\vec k}\,|$ occur on the integration
path so their meaning is ambiguous.  In order to have a well-defined
propagator, we must specify some procedure to handle these singularities.
For $y^0<x^0$, the contour may be closed at infinity in the
lower-half plane and we wish for only positive energies to 
contribute.  This is accomplished by pushing the poles 
$k^0=|\,{\vec k}\,|$ into the lower-half plane and the poles 
$k^0=-|\,{\vec k}\,|$ into the upper-half plane.  
This physical\footnote{
Since the ghosts are not physical particles, the
motivation for the use of this prescription is not clear
in this context.  However, it is 
essential for the ghost propagator to have the same form 
as the gluonic propagator if the ghosts are to do their
job properly.} reasoning leads to the unambiguous propagators
\begin{eqnarray}
\Delta^{a\,b}(k)&=&{i\delta^{a\,b}\over k^2+i\varepsilon}\\
S(p)&=&{i(\not\! p+m)\over p^2-m^2+i\varepsilon}\\
D_{\mu\nu}^{a\,b}(q)&=&{\lbrack -g_{\mu\nu}+
(1-\xi)(q_\mu q_\nu/(q^2+i\varepsilon))\rbrack\over q^2+i\varepsilon}
\;i\delta^{a\,b}
\end{eqnarray}
for ghosts, quarks, and gluons\footnote{Strictly speaking,
the $+i\varepsilon$ in the second term of the gluon propagator
is not required by physics any more than that in the 
ghost propagator.  These prescriptions need only be consistent
with each other to ensure a well-defined gauge theory.  
It is only a matter of aesthetic convention that we treat them
in the same way as the physical propagators.}
of momentum $k$, $p$, and $q$, respectively.

In axial gauges, these arguments go through in exactly the same way.
The quark propagator is unchanged and the gluon propagator 
becomes
\begin{equation}
D_{\mu\nu}^{a\,b}(q)={i\over q^2+i\varepsilon}
\left(-g_{\mu\nu}+
{n_\mu q_\nu+n_\nu q_\mu\over n\cdot q+i \vec n\cdot\vec q\varepsilon}
-{n^2+\xi q^2\over (n\cdot q+i{\vec n}\cdot{\vec q}\varepsilon)^2}
q_\mu q_\nu\right)\delta^{a\,b}\; ,
\label{axgl}
\end{equation}
where I have taken $n^0>0$.  This prescription
is motivated in exactly the same way as above,\footnote{
This prescription is appropriate for ordinary 
time-ordered perturbation theory, in which one imposes
quantization conditions on the fields and their
conjugate momenta at equal times.  In practice, one
often uses the axial gauge with $n^\mu$ light-like
($n^2=0$) and quantizes the theory at equal $n\cdot x$.
In this case, the proper prescription is given by
\begin{equation}
{1\over n\cdot q}\rightarrow{n^*\cdot q\over (n^*\cdot q)
(n\cdot q)+i\varepsilon}\;\; ,
\end{equation}
where $n^{*\mu}$ satisfies $n^*\cdot n=1$.
This rule for handling the singularity, due to 
Mandelstam and Leibbrandt \cite{MLpres}, ensures that
only those particles with positive `light-cone energy', $n\cdot q$,
will propagate forward in `light-cone time', $n\cdot x$.
We will see more of this prescription when we study the 
axial gauge in Section \ref{reninax}.}
but here
we cannot maintain frame independence.  This unpleasing
turn of events is not unexpected since our gauge condition has
already broken Lorentz invariance.  Of course, the terms in
which we need this prescription are merely gauge artifacts and
are not required to satisfy the constraints put on physical
propagation.  Since there are no ghosts in this gauge, there is
nothing we must be consistent with.  For this reason, the complicated
prescription suggested by physical arguments is often ignored.
This leads to singularities of a very special kind 
which are required by gauge invariance to cancel in physical quantities.
If they are handled properly, this does indeed happen.\footnote{
The term `handled properly' is in some sense
defined by this statement.}  However,
the special care one must take in the axial gauge 
can lead to tremendous difficulties.  This is one reason 
the covariant gauges are often employed in spite of the
complications due to ghosts.

Although these $i\varepsilon$'s are quite important in many
aspects of quantum field theory, there are many other 
aspects in which they are not essential.  For this reason,
I will not write them explicitly unless they are expected to contribute.
However, it is understood that they are always there.

\section{Feynman Rules and Perturbation Theory}
\label{feynmanrule}

The free theory is nice, but it cannot be expected to teach us
the true nature of the strong interactions.  At some point,
we must allow our fields to interact with one another.  
Aside from the mathematical difficulties associated with
defining the path integral, everything we have done up to 
now is exact.  In order to go further, we must make some
approximations.  

As it stands, the generating functional
is expressed as a series of derivatives acting on a product
of exponentials.  Since we cannot calculate the infinite number
of derivatives necessary to take the full exponential into 
account, let us assume for the moment that the coupling 
$g$ is sufficiently small to justify an expansion :
\begin{equation}
e^{i{\cal S}_{int}[{1\over i}{\delta\over\delta{\cal J}}]}\sim
1+i{\cal S}_{int}\left[{1\over i}{\delta\over\delta{\cal J}}\right]\; .
\end{equation}
Consider the correlation function\footnote{
$i$ and $j$ refer to quark color while 
$\alpha$ and $\beta$ refer to spin.
Here and in the following, I will ignore the difference 
between the fundamental and anti-fundamental representations
of $SU(3)$.  There are simply too many indices floating
around here for us to be particular about placement.
One can always trace through the derivation to resolve
any confusion.} 
\begin{eqnarray}
&&\left\langle{\rm T}\,{\overline\psi}^i_\alpha(x_1)\psi^j_\beta(x_2){\cal A}_\mu^a(x_3)
\right\rangle_{vac}\nonumber\\
&&\qquad\qquad\qquad=-{1\over i}{\delta\over\delta\eta_i^\alpha(x_1)}\,
{1\over i}{\delta\over\delta{\overline\eta}_j^\beta(x_2)}
{1\over i}{\delta\over\delta{j^\mu_a(x_3)}}{\cal Z}[{\overline\eta},
\eta,j]\Big|_{{\overline\eta}=
\eta=j=0} \; .
\end{eqnarray}
In the free theory, this is zero since the external gluon has nowhere to
propagate.  The way that sources appear in the exponentials implies
that there must be an even number of external gluon fields
and the same number of $\psi$'s as $\overline\psi$'s in order
for a matrix element to be nonzero.  
Taking into account one iteration of the interaction, we see that 
our correlation function becomes 
\begin{eqnarray}
&&\left\langle{\rm T}\,{\overline\psi}^i_\alpha(x_1)\psi^j_\beta(x_2)
{\cal A}_\mu^a(x_3)
\right\rangle_{vac}^{\rm LO}=
\left(-{1\over i}\right){\delta\over\delta\eta_i^\alpha(x_1)}\,
{1\over i}{\delta\over\delta{\overline\eta}_j^\beta(x_2)}
{1\over i}{\delta\over\delta{j^\mu_a(x_3)}}\nonumber\\
&&\qquad\qquad\qquad\times\;\;(-ig)\int d^{\,4}w
\left(-{1\over i}\right)
{\delta\over\delta\eta_\gamma^k(w)}{1\over i}{\delta\over\delta
j^\nu_b(w)}\gamma^\nu_{\gamma\delta}(t^b)_{k\,l}{1\over i}{\delta\over\delta
{\overline\eta}_\delta^l(w)}\nonumber\\
&&\qquad\qquad\qquad\qquad
\times\;\;e^{-i\int d^{\,4}y_1\,d^{\,4}y_2\,
{\overline\eta}(y_1)S(y_1,y_2)\eta(y_2)}\\
&&\qquad\qquad\qquad\qquad\qquad
\times\;\;e^{-i/2\int d^{\,4}z_1\,d^{\,4}z_2\,
j^\mu(z_1)D_{\mu\nu}(z_1,z_2)j^\nu(z_2)}\left|_{{\overline\eta}=\eta=
j=0}\right.  \; .\nonumber
\end{eqnarray}

\begin{figure}
\label{qginteract}
\epsfig{figure=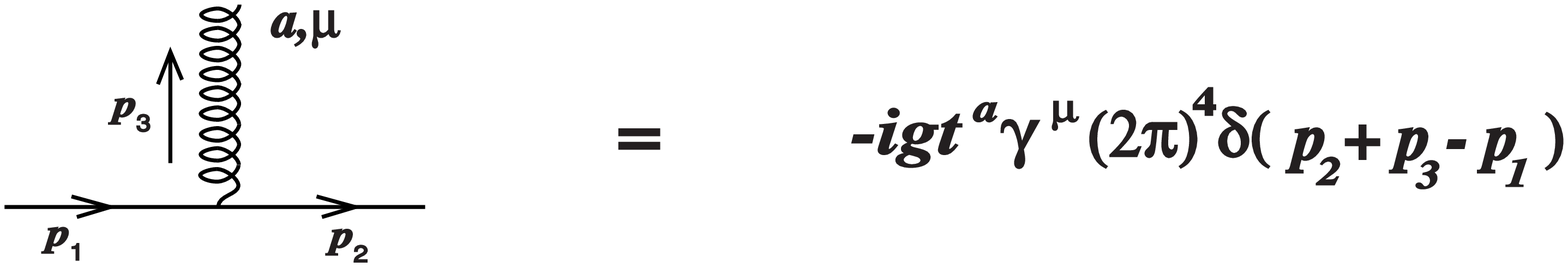,height=2.5cm}
\caption{The fundamental quark-gluon interaction
is represented graphically by a Feynman diagram.
The factor associated with the emission of 
a gluon with polarization $\mu$ and color index $a$
is $-igt^a\gamma^\mu$, along with a $\delta$-function
to ensure momentum conservation.}
\end{figure}

Since the sources are set to zero at the end of the calculation, all 
points must be coupled through propagators.  There are only two 
ways to do this, one of which is zero since the generators of
$SU(3)$ are traceless.  The other gives
\begin{eqnarray}
&&\left\langle{\rm T}\,{\overline\psi}^i_\alpha(x_1)\psi^j_\beta(x_2)
{\cal A}_\mu^a(x_3)
\right\rangle_{vac}^{\rm LO}\nonumber\\
&&\qquad\qquad\qquad=\int d^{\,4}w \,S_{\beta\gamma}^{j\,k}
(x_2,w)\left(-ig\gamma^\nu_{\gamma\delta}(t^b)_{k\,l}\right)
S_{\delta\alpha}^{l\,i}(w,x_1)
D_{\nu\mu}^{b\,a}(w,x_3)\; .
\end{eqnarray}
This expression consists of three propagators, each from an external
point to the common point $w$.  We would like to split it up into
different pieces - one for each propagation and one for the interaction
itself.  However, all of these pieces are at present entangled 
by the integration over $w$.  In momentum space, the separation is
clear :
\begin{eqnarray}
&&\int{d^{\,4}x_1}{d^{\,4}x_2}
{d^{\,4}x_3}e^{-ip_1\cdot x_1}e^{ip_2\cdot x_2}
e^{ip_3\cdot x_3}\left\langle{\rm T}\,{\overline\psi}^i_\alpha(x_1)
\psi^j_\beta(x_2){\cal A}_\mu^a(x_3)
\right\rangle_{vac}^{\rm LO}\nonumber\\
&&\qquad\qquad=S_{\beta\gamma}^{j\,k}(p_2)
[-ig(t^b)_{k\,l}\gamma^\nu_{\gamma\delta}
(2\pi)^4\delta(p_2+p_3-p_1)
]S_{\delta\alpha}^{l\,i}(p_1)
D_{\nu\mu}^{b\,a}(p_3)\; .
\end{eqnarray}
The factor associated with the interaction is just $-igt^b\gamma^\nu$
along with a $\delta$-function enforcing energy-momentum
conservation.
This interaction is represented graphically in Figure 1.1.
The straight lines represent quark propagation and the 
squiggly ones represent gluon propagation.\footnote{On these 
diagrams, ghost propagation is represented by a dashed line.}
Diagrams such
as this one are called Feynman diagrams and are extremely 
useful for organizing perturbation theory.  

\begin{figure}
\label{fig3}
\epsfig{figure=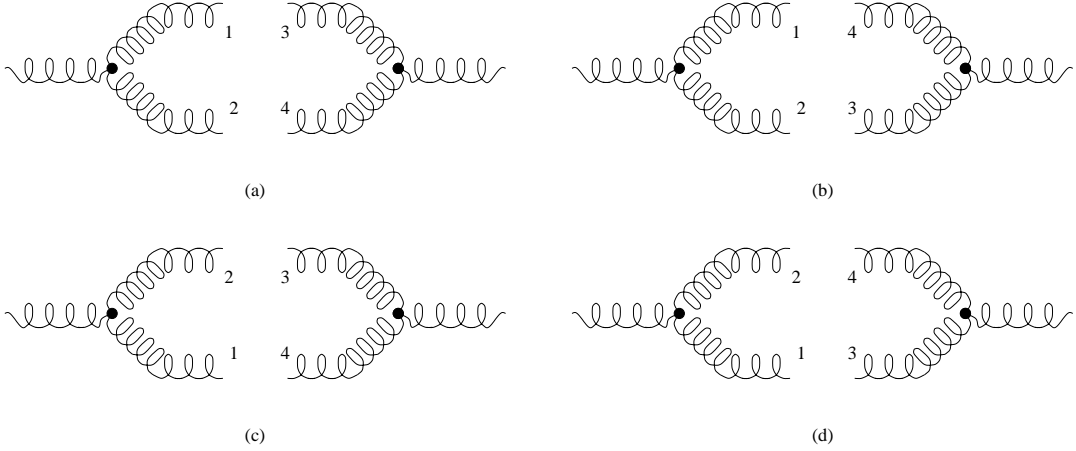,height=2.3in}
\caption{This diagram presents a prime 
example of the reason symmetry factors are sometimes
necessary to avoid overcounting.  The Feynman rule
for the triple gluon vertex counts each of these
contributions as separate.  In reality, however,
only (a) and (b) represent different processes.
In order to obtain the correct result using Feynman rules,
one must include a factor of 1/2.}
\end{figure}
 
The above method can be used to obtain a set of 
rules, called `Feynman rules', for all of the 
terms in the interaction lagrangian.  One simply considers 
a correlation with an 
external field for each particle species in the interaction.
The rules for all of the interactions in QCD with
covariant gauge fixing
are given in Appendix \ref{frulesqcd}.  
Feynman rules for axial gauge calculations are
identical except for the gluon propagator, (\ref{axgl}),
and the fact that there are no ghosts.  With these
rules, one can in principle 
calculate any process to any order in perturbation theory:

1) One draws a propagator for each 
external field in the correlation
and connects them in every way possible 
according to the rules.  

2) Disconnected diagrams
are discarded, as they will cancel with the 
normalization.   

3) Each interaction vertex comes with an associated 
$\delta$-function guaranteeing momentum conservation.  

4) Unconstrained
momenta are integrated over, like every 
other unconstrained 
quantum number in the diagram.  The number 
of such integrations is 
called the number of loops.  

5) The anticommuting nature of 
Grassmann fields generates an extra minus sign for 
each loop associated with
these fields.  

6) Certain diagrams come 
with extra factors because
of their high degree of symmetry.  These 
`symmetry factors' arise because of an inadequacy in our 
Feynman rules.  In the derivation of these rules we assume
that the external fields are all distinct.  While this
is certainly the case if the fields are truly external, 
it need not be true for internal fields.  Consider, for example, the
diagrams of Figure 1.2.  Our rules 
count each separate contraction as a 
distinct contribution to the diagram.  However, there
are really only two distinct contributions;
Figs. 1.2a and d
are identical, as are Figs. 1.2b and c.  Our rules have 
overcounted the contribution of this diagram by a factor of two.
In general, symmetrical diagrams must be divided by the number of 
ways one can interchange propagator lines without changing the 
diagram.  

All of these rules arise naturally from the path integral.
One can always resort to the tedious method above 
to ensure the correct symmetry factor or minus sign.

\section{Renormalization in Covariant Gauges}
\label{renincov}

At this point, it seems like we have all the ingredients necessary for 
a calculation in QCD.  Let's test this hypothesis
by applying our procedure to the quark propagator,
$\langle{\rm T}\,{\overline\psi}(y)\psi(x)\rangle$.
This object receives several kinds of corrections, as shown
in Fig. 1.3.  The diagram in Fig. 1.3a has two disconnected
pieces.  The gluon loop is not associated with the
external sources and therefore cancels with the 
denominator in Eq.(\ref{defz}).  Such disconnected pieces
are called vacuum bubbles and represent shifts in the
vacuum energy.  For our purposes, they can be 
completely ignored.\footnote{Although they will not be useful for us, these
diagrams do contribute to the effective action of the field theory.
Analysis of this object gives insight into the
vacuum structure of a theory.}  Diagrams like the one in
Fig. 1.3b are called one-particle-reducible (1PR) since they can
be broken into two disconnected pieces by cutting only one
propagator.  These diagrams are special because the 
momentum of the `one-particle' propagator is fixed completely
by the conservation law.  This means that the two parts of this
diagram are not coupled by a loop integration and they may
be evaluated separately.  Hence one can consider only
one-particle-irreducible (1PI) diagrams, such as 
those in Fig. 1.3c, and 
string them together to get the effect of the 1PR diagrams.

Calling the 1PI quark self-energy $-i\Sigma$, 
we see that the full correction is\footnote{
Here, we resort to the standard notation
\begin{equation}
i\over\!\not\!p-m
\end{equation} 
for the bare (uncorrected) quark propagator
rather than that found above.  This is understood to mean 
$i(\not\!p-m)^{-1}$.  The $i\varepsilon$ will be ignored for a while,
but it is always understood to be present.}
\begin{eqnarray}
S(p)={i\over\not\! p-m}+{i\over\not\! p-m}\left(-i\Sigma(p)\right)
{i\over\not\! p-m}+\cdots\nonumber\\
={i\over\not\! p-m}\left(
{1\over 1-\Sigma(p)(\not\! p-m)^{-1}}\right)\; ,
\label{1PR}
\end{eqnarray}
where I have summed the geometric series.  In covariant gauge,
Lorentz invariance tells us that $\Sigma$ can depend only on 
$p^\mu$.  Since the strong interactions respect parity
and there is only one vector in the problem, we may 
express $\Sigma$ in terms of two scalar functions of
$p^2$
\begin{equation}
\Sigma(p)=A(p^2)+B(p^2)\not\! p\; .
\end{equation}
The explicit breaking of Lorentz invariance in the axial gauges 
destroys the validity of this expression.  The special techniques necessary
to handle this and the other problems with renormalization in these
gauges will be discussed in Section \ref{reninax}.

\begin{figure}
\epsfig{figure=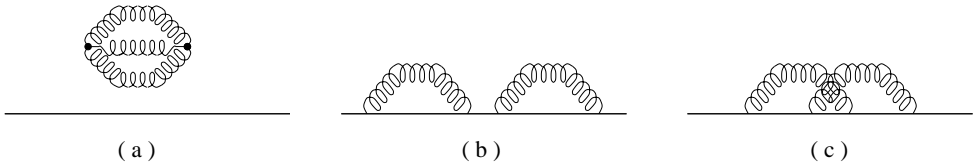,height=0.8in}
\caption{Three different kinds of contributions to 
the quark self-energy.  The gluon loop in 
(a) is uncoupled to the sources and so 
contributes only to the vacuum energy.
The reducible graph in (b) does not
contain any new information; its contributions
can be re-summed with a knowledge of 
irreducible diagrams, like that in (c).}
\label{fig4}
\end{figure}

Substituting this expression for $\Sigma$ into the fully re-summed
propagator, we obtain
\begin{equation}
iS^{-1}(p)=(1-B(p^2))(\not\! p-m)-A(p^2)-mB(p^2)\; .
\end{equation}
Unless $A(p^2)$ and $B(p^2)$ are rather trivial functions near
$p^2=m^2$, the corrected propagator no longer has a pole at this 
position and the residue at the shifted pole is no longer $i$.  
The normalization of the field is a convention, but the fact that it
has changed is disconcerting.  Although arbitrary, this normalization
should be consistent.  The resolution of this apparent difficulty
rests with the fact that we cannot observe the fields present
in our lagrangian.  Turning off the interaction so that we can measure
this quantity $m$ in the lagrangian is 
not an option.\footnote{Of course,
in QCD seeing the objects under consideration {\it at all} is not an
option.  As mentioned in the introduction, defining quark `mass' is itself
difficult.  However, we will ignore these difficulties for the discussion
at hand as they are nonperturbative problems which do not affect
our calculations in perturbation theory.  For the rest of this
chapter, we can ignore confinement completely.  The penalty we
pay for such arrogance will be discussed in the next chapter.}
Hence we are forced to let go of any idea we have that the
parameter $m$ present in the lagrangian is 
a physical observable and that the field $\psi$ is the quark
field we see in experiment.  These quantities are called `bare' or
`unrenormalized' objects and require corrections before we can 
relate them to experiment.

We have the corrected propagator in terms of $A$ and $B$, so we 
can calculate the shift in the mass and the new normalization of
the quark field.  The new pole appears at a renormalized mass
$m_{\rm R}$ such that
\begin{equation}
\left(1-B(m_{\rm R}^2)\right)\left(m_{\rm R}-m\right)-A(m_{\rm R}^2)
-mB(m_{\rm R}^2)=0\; ,
\label{massren}
\end{equation}
and the residue at this pole is given by
\begin{eqnarray}
Z_{\rm F}&=&{1\over i}{\rm Res}[S(p),m_{\rm R}]\nonumber\\
&=&{1\over 1-B(m_{\rm R}^2)
-2m_{\rm R}{\partial}\left(A(p^2)+mB(p^2)\right)/\partial p^2
\left|_{p^2=m_{\rm R}^2}\right.}\;\; .
\end{eqnarray}
Since we have no information on the values of these
bare parameters, it is not useful for us to express predictions in terms
of them.  We would prefer to know the variation of observables with 
physical quantities.  Removing unphysical parameters in favor
of physical quantities is called {\it renormalization}.  It turns out
that every term in our lagrangian requires renormalization.

The idea of renormalization is not new.  Solid state physics has
been using this process for many years.  Electrons in the conduction
band of a metal behave as though their masses are different since
the actual excitations observed involve the background
lattice.  The idea here is the same : we are observing excitations
on top of a background vacuum.  The vacuum in quantum field 
theory is a complicated sea of quantum fluctuations and cannot be
treated as trivial.  The only difference between the solid state
example and renormalization in a quantum field theory is 
that electrons in a metal can be removed from the metal and 
observed free of any background lattice.  They can never be
removed from their electromagnetic field fluctuations and 
observed in their bare form.  This is the underlying reason
bare quantities have no physical meaning.

There are two main ways to view renormalization.  One
way is to simply calculate everything in terms of bare
parameters and substitute known expressions for the 
physical quantities at the end.  This is known as 
unrenormalized perturbation theory.  The unpleasing thing
about unrenormalized perturbation theory is the 
fact that once the calculation is done, one still must 
do the substitution - something which is certainly not always
trivial.  The other main way to renormalize involves a systematic
replacement of bare parameters with renormalized ones at each order in 
perturbation theory.  This renormalized version
has the advantage that unrenormalized
quantities never appear in our expressions.  Although
it will lead to more diagrams, it is generally preferred
to the procrastinator's method. 

Before we can begin the task of renormalization, we must
carefully define our quantities.  We saw above that the 
normalization of the fermion field changes so that the 
residue of the pole becomes $iZ_{\rm F}$.  Consistency
of convention requires the residue for the physical field to
be $i$.  This can be obtained by defining\footnote{
From this point on, I will use the 
subscript $b$ for bare parameters and
leave renormalized parameters with no subscript.}
$\psi=Z_{\rm F}^{-1/2}\psi_b$ since
\begin{eqnarray}
\int d^{\,4}x\,e^{ip\cdot x}\left\langle{\rm T}\,\psi(x){\overline\psi}(0)
\right\rangle_{vac}&&= Z_{\rm F}^{-1}\int d^{\,4}x\,
e^{ip\cdot x}\left\langle{\rm T}\,\psi_b(x){\overline\psi}_b(0)
\right\rangle_{vac}\nonumber\\
&&= i\left\lbrack\not\! p-m-M(p)(\not\!p-m)^2\right\rbrack^{-1}\; ,
\end{eqnarray}
where $M(p)$ is defined to take into account the rest of the 
$p^2$ variation in $A$ and $B$.  $M(p)$
will have nontrivial Dirac structure, but
be regular as $\not\!p\rightarrow m$.  The normalization of the 
two-point correlator is governed by the normalization of the 
kinetic term in the lagrangian, so one might think of this 
renormalization constant as coming from the operator 
${\overline\psi}\,i\!\not\!\!\partial\,\psi$.  
It is also associated
with the field itself.  If we were dealing with a nonlocal 
field theory, this constant is the only renormalization we would need 
since nonlocal products of fields are renormalized simply by products
of $Z_{\rm F}$.  Local field theories are different because the
structure of a local composite operator is quite different than that
of a nonlocal product.  This will become 
quite explicit in our discussion of the coupling.

The gluon and ghost fields also require a field-strength
renormalization.  We define ${\cal A}=Z_{\rm A}^{-1/2}{\cal A}_b$
and $c=Z_{\rm G}^{-1/2}c_b$.\footnote{Although the ghost and anti-ghost
field renormalizations need not be the same since they are unrelated
fields, they will always appear together in our applications.  Hence
for our purposes only the product of their field strengths has any meaning.}
There is also a relationship between the renormalized mass and the 
bare mass.  Since the mass term is the only piece of our lagrangian 
which couples quarks of different chirality, a theory in which
$m_b$=0 is fundamentally different from a theory in which 
$m_b\neq 0$.\footnote{Theories with small fermion masses 
can be approximated by massless theories.  The effects of the
small breaking of this chiral symmetry induced by nonzero 
fermion masses can be taken into account in perturbation theory.
This is the idea behind Chiral Perturbation Theory, the theory 
of pions and nucleons mentioned in the introduction.}
A theory of massless fermions cannot generate a 
fermion mass.  For this reason, the renormalized mass must be proportional
to the bare mass.  We write $m=Z_{\rm m}^{-1}m_b$.  
The renormalization of the coupling must also be multiplicative since
a theory with zero bare coupling is free and certainly will not 
generate a coupling.  Hence we define $g=Z_{\rm g}^{-1}g_b$.
We will see below that even the gauge parameter $\xi$ acquires
multiplicative
renormalization, $\xi=Z_\xi^{-1}\xi_b$.

How does all of this renormalization change our Feynman rules? 
Obviously, it doesn't change them at all in the unrenormalized 
version of perturbation theory since all the action 
happens after the calculation.  However, in order to calculate
quantities in renormalized perturbation theory, we will have
to make some modifications.  We begin by making the trivial observation
that 
\begin{eqnarray}
{\cal L}[\Phi_b,m_b,g_b,\xi_b]&=&{\cal L}[\Phi,m,g,\xi]
\nonumber\\
&&+\sum_f(Z_{\rm F}^f-1){\overline\psi}_f\,i\!\not\!\partial\,\psi_f
\nonumber\\
&&-\sum_fm_f(Z_{\rm m}^fZ_{\rm F}^f-1){\overline\psi}_f\psi_f
\nonumber\\
&&-\sum_fg(Z_{\rm g}Z_{\rm F}Z_{\rm A}^{1/2}-1){\overline\psi}_f
\!\not\!\!\!{\cal A}\,\psi_f
\nonumber\\
&&-{1\over 4}\,(Z_{\rm A}-1)\,
(\partial_\mu{\cal A}_\nu^a-\partial_\nu{\cal A}_\mu^a)
(\partial^\mu{\cal A}^{a\,\nu}-\partial^\nu{\cal A}^{a\,\mu})
\nonumber\\
&&-(Z_{\rm A}Z_\xi^{-1}-1){(\partial\cdot{\cal A}^a)^2\over 2\xi}
\\
&&+{1\over2}g(Z_{\rm g}Z_{\rm A}^{3/2}-1)
f^{a\, b\, c}{\cal A}_\mu^a{\cal A}_\nu^b\partial^\mu{\cal A}^{c\,\nu}
\nonumber\\
&&-{1\over 4}g^2(Z_{\rm g}^2Z_{\rm A}^2-1)
f^{a\, b\, c}f^{a\, d\, e}{\cal A}^{b\,\mu}{\cal A}^{c\,\nu}
{\cal A}^d_\mu{\cal A}^e_\nu
\nonumber\\
&&-(Z_{\rm G}-1){\overline c}^a \partial^2 c^a
\nonumber\\
&&+g(Z_{\rm g}Z_{\rm G}Z_{\rm A}^{1/2}-1)
f^{a\, b\, c}{\overline c}^a\partial^\mu{\cal A}_\mu^bc^c\nonumber\;\; .
\label{Zlag}
\end{eqnarray}
This is nothing more than a simple rearrangement of terms after 
substitution.  The Feynman rules are the same for the first line
of this expression since the lagrangian is the same.  All of the
other terms are intended to be taken into account in perturbation
theory.  These terms are all proportional to $g$ since the free
theory has $Z=1$ for all renormalization constants.  Hence each
of these terms will simply give us another interaction 
vertex to iterate in our Feynman diagrams.  This is 
the tragedy of renormalized perturbation theory.  The 
price of systematically replacing each bare parameter 
with the physical couplings is eight new interaction vertices.  This
is twice as many as we started with!  The extra terms are called
counterterms because they are constructed to cancel certain 
contributions which arise in our calculation.  For the remainder of this
section we will be concerned with calculating their coefficients 
at leading order in perturbation theory.

\begin{figure}
\label{fig5}
\epsfig{figure=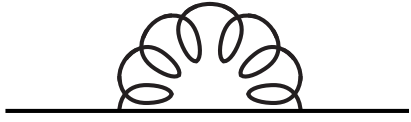,height=1.5cm}
\caption{The one-loop correction to the quark self-energy.}
\end{figure}

Let us begin by actually calculating $A(p^2)$ and $B(p^2)$.  There
is only one diagram which contributes to these quantities 
at next-to-leading order (see Fig. 1.4).
Using the Feynman rules in Appendix \ref{frulesqcd}, 
we find the amplitude for this
diagram to be\footnote{Remember that the external propagators
are not included in the definition of $\Sigma$!}
\begin{equation}
-i\Sigma^{\rm (1-loop)}(p)=\int{d^{\,4}k\over(2\pi)^4}\,(-igt^a\gamma^\mu)
\left({i\over\not\! p-\not\! k-m_b}\right)(-igt^b\gamma^\nu)
D_{\mu\nu}^{a\,b}(k)\; ,
\end{equation}
or\footnote{The casimir  
$C_F\delta^i_j=(t^at^a)^i_j$ is discussed in Appendix \ref{sun}.}
\begin{equation}
\Sigma^{\rm (1-loop)}(p)=ig^2C_F\gamma^\mu\int{d^{\,4}k\over(2\pi)^2}
{(\not\! p-\not\! k+m_b)\lbrack-g_{\mu\nu}+(1-\xi)(k_\mu k_\nu/ k^2)\rbrack
\over k^2[(p-k)^2-m_b^2]}\gamma^\nu\; .
\label{self}
\end{equation}
A closer look reveals this integral to be divergent.  For large $k$,
it behaves as $\int dk/k$.\footnote{Since the vector index must
be carried by the only vector in the problem, $p^\mu$, the linear 
divergence $\int dk$ is thrown away by symmetry.}  This tells
us that extremely large momenta are making contributions to the
process.  The fact that these asymptotically large momenta 
contribute to this process irrespective of the size of the 
characteristic scales in the problem, $p^\mu$ and $m_b$, 
could imply a deep problem with the theory.  It is divergences 
such as this one which turned people away from quantum field theory 
in the end of the 1960's.  We need not give up so quickly, though.
We have just seen that this quantity means nothing by itself.
Only properly normalized physical matrix elements expressed in
terms of physical quantities
are observable.  Hence
in a strict sense it is unimportant that 
this quantity diverges.\footnote{
\label{foot1}In fact, from a theoretical point of view we might 
{\it expect} it to diverge.  If it didn't, the bare parameters of 
the theory would be well-defined finite objects.  However, we can 
think of a bare parameters as the way the particle looks 
extremely close-up, i.e. at asymptotically high energy.  
Since there are always corrections at a higher scale than we 
are observing at, we might not be surprised that there is an infinity
behind them all.  Looking at renormalization in this way, it also
seems quite natural that the divergences from asymptotically large
loop momentum combine with the divergent bare parameters representing
the way the particles look at asymptotically large momentum transfer
to give sensible finite results for physical momentum transfer.}
It implies only that the bare parameters must also diverge.

While the above discussion is fine as a theoretical argument, we
still have not obtained expressions for the quantities $A$ and $B$
(or $Z_{\rm F}$ and $Z_{\rm m}$).  In some sense, we can write 
$A=1+g^2\,\infty_{A}^{(2)}+{\cal O}(g^4\,\infty_A^{(4)})$
with a similar expression for $B$.  However, these 
expressions are not very useful and certainly bring doubts 
about the convergence of our perturbative expansion.
In order to proceed in a constructive way, we must make sense of these
divergences.  We have already argued that the fact that they diverge is 
unimportant, but now we need to find a way to express {\it how} they
diverge.  The important thing here is that we cannot manipulate infinite
expressions.\footnote{
There are several schemes which attempt to
avoid regularization entirely, defining counterterms during the
calculation as sums of divergent integrals in such a way that
the renormalized quantities are well-defined
(see \cite{colren}).
It is not clear that the manipulations used to define these
schemes are applicable to arbitrary order or for an arbitrary process.
At any rate, the scheme introduced below is easier to use
and avoids these problems entirely.}  
It makes no sense at all to write a relation like
$\infty_1-\infty_2=3$.
However, it is perfectly correct and 
meaningful (mathematically) to say $\lim_{x\rightarrow\infty}(x+3\;\; -x)=3$.
What we need to make sense of our theory is a regulator - something to
make the divergences finite until we take a certain limit, which we 
will choose to postpone until the end of the calculation, so that
we can manipulate them.  As we will see, this necessarily introduces
a scale which will act as the separation scale between physics 
we calculate explicitly and physics contained in our renormalized couplings.\footnote{This should 
be obvious from Footnote \ref{foot1}.  We cannot
say that large momentum quantities 
combine to give a moderate momentum
result without referring to a moderate momentum.}  Physical
processes cannot depend on this arbitrary scale of separation, 
but intermediate parameters can and will gain dependence on it.

How can we regularize the theory?  One simple answer is to merely 
introduce a high-momentum cutoff in our integrals.  Instead of 
integrating to infinity, we could integrate only to $\Lambda$.
This procedure has been used before in several theories.  However, 
in gauge theories the price we pay for such savage butchery is high.
If we use this regulator, our theory is no longer renormalizable.
In unrenormalizable theories, divergences appear which cannot be
re-absorbed into the constants we have introduced.  For example, 
if the one-loop quark self-energy in Fig. 1.4 contained a 
divergence proportional to $p^2$, we would require a counterterm
of the form ${\overline\psi}\partial^2\psi$. 
In the absence of such a term, the theory
is not
complete.
Unrenormalizable theories need an infinite number of terms in
order to be complete.  Unless there is a good
perturbative expansion which can be truncated, an infinite number
of experiments need to be done to fix the renormalized coefficients
of each term in the lagrangian.  This is far from practical,
especially since the theory we began with has only
$n_f+1$ parameters - the quark masses and the coupling, $g$.
The price of the simple cutoff regulator is just too high 
for us to pay.  

The reason this cutoff procedure exacts such compensation
is that renormalizability is not an easy thing to come by - especially
in a theory with vector bosons like this one.  In fact,
the only renormalizable theories involving spin-1 fields are 
gauge theories.  The gauge invariance of these theories
plays such a crucial role in the proof \footnote{This proof is the
work that won the Nobel Prize in physics for `t Hooft and Veltman
mentioned in the introduction.  The relevant references are \cite{thooft}.}
that it no longer
works if the method of regulation breaks the 
invariance.\footnote{It can be shown that certain amounts of violence 
can be done to the gauge invariance before the theory 
actually becomes unrenormalizable.  What is required is that
the gauge invariance can be restored by adding a finite number
of new counterterms to the lagrangian.  Even so, a procedure in which
one need not add any terms to the lagrangian (one which
does not break the symmetry) is desirable.}
Our simple cutoff idea butchers the theory with no respect
for either gauge invariance or Lorentz invariance.  
In fact, there are only a few renormalization schemes which 
preserve the gauge invariance of our theory.  The one which
has been proven to be the most useful in many applications, 
including this one, is called Dimensional Regularization (DR).
This scheme was introduced by `t Hooft and Veltman
in 1973 expressly because it does not violate gauge invariance.
Many of the technical details of DR are spelled out in Appendix \ref{dimregapp}.
The basic idea is that the integral in Eq.(\ref{self}) is 
only divergent because we live in 4 dimensions.  If we lived
in 3 dimensions, for example, it would be perfectly 
finite as $k\rightarrow\infty$.  
Furthermore, the theory itself is expected to be well-defined
in {\it any} number of dimensions and its amplitudes are expected to be
{\it meromorphic} functions of the spacetime dimension.
These assertions, along with a set of rules that are outlined in Appendix
\ref{dimregapp}, allow us to calculate matrix elements in an arbitrary 
number of dimensions, $d$.  Our divergences will show up as poles 
in $d-4$.  It is only at the end of the calculation
that we evaluate these (physical) matrix elements at the spacetime 
dimension $d=4$.     

The scale we must introduce along with the 
regularization procedure is mysteriously absent.  This is a good thing 
since the explicit appearance of scales is notorious for breaking
gauge invariance.  However, as argued above, there must be
a scale introduced somewhere.  It turns out that the scale in DR
is hidden in the coupling constant.  To see this, let us analyze 
QCD in $d$ dimensions.

The mass-dimensions of quantum fields are obtained from the requirement
that the action ${\cal S}=\int d^{\,d}x\,{\cal L}$ is dimensionless,
implying that $\cal L$ has mass dimension $d$.  Since derivatives
have mass dimension 1, the kinetic terms in the lagrangian 
fix the dimension of the fields : $(d-2)/2$ for bosons and 
$(d-1)/2$ for fermions.  Applying this result to the 
rest of the lagrangian, we see that the coupling constant
$g$ has mass dimension $(4-d)/2$.  We would rather work with a dimensionless
coupling, so we introduce a scale, $\mu$, to carry the dimension
required.  The bare coupling must also be given dimension, $\mu_0$.
Going back to the bare lagrangian and introducing $Z$-factors again,
we see that $\mu_0$ goes along for the ride; $g$ must be replaced
with $g\mu_0^{(4-d)/2}$ everywhere in the renormalized lagrangian.
The reason that $\mu_0$ appears rather than $\mu$ stems from
the fact that $Z_g$ relates the dimensionless couplings 
$g$ and $g_b$ rather than the dimensionful coefficients
$g\mu^{(4-d)/2}$ and $g_b\mu_0^{(4-d)/2}$.  We will see how
$\mu_0$ is replaced by $\mu$ below.

In principle, we are really
only interested in $d=4$, where $g$ is dimensionless and
the $\mu$ (and $\mu_0$) dependence drops out.  However,
a consistent theory in $d$ dimensions requires the inclusion 
of $\mu$.  Hence all of our final results should be independent of
$\mu$, but its introduction is required by our regulator.  
A study of the $\mu$ dependence of certain intermediate
results leads to a deep understanding of renormalization and
quantum field theory itself.

Armed with a regularization procedure, we are finally ready to 
calculate our $Z$-factors explicitly.  Returning to the
quark self-energy, this time in $d$ dimensions, we write\footnote{
To obtain this result, we have used the 
fact that any term in which the quark
propagator $(p-k)^2-m_b^2$ cancels is zero since it does not 
have any scale dependence.  
This argument is made in
considerably more detail in Appendix \ref{dimregapp}.
The rules for $d$-dimensional $\gamma$ matrix algebra
can be found in Appendix \ref{diralg}.} 
\begin{eqnarray}
\Sigma^{\rm (1-loop)}(p)=ig^2\mu_0^{d-4}C_F\int{d^{\,d}k\over(2\pi)^d}
\left\lbrack{(d-2)(\not\! p-\not\! k)-dm_b\over k^2[(p-k)^2-m_b^2]}\right.
\nonumber\\
\left.+(1-\xi){(p^2-m_b^2)\not\! k-k^2(\not\! p-m_b)\over (k^2)^2[(p-k)^2-m_b^2]}
\right\rbrack\; .
\end{eqnarray}
Using the integration formulae in 
Appendix \ref{covint}, we find\footnote{
Since $g$ appears almost exclusively in the combination $g^2/4\pi$,
I have introduced $\alpha_s\equiv g^2/4\pi$.  This is the true 
expansion parameter of the strong interaction.} 
\begin{eqnarray}
A^{\rm (1-loop)}(p^2)&=&{\alpha_s C_F\over 4\pi}\Gamma\left({\epsilon\over 2}\right)
\left({m_b^2\over 4\pi\mu_0^2}\right)^{-\epsilon/2}\nonumber\\
&&\times\;m_b[d-(1-\xi)]
\int^1_0 dx\,x^{-\epsilon/2}\left[1+(1-x)\left(-{p^2\over m_b^2}\right)\right]
^{-\epsilon/2}\nonumber\\
B^{\rm (1-loop)}(p^2)&=&{\alpha_s C_F\over 4\pi}\Gamma\left({\epsilon\over 2}\right)
\left({m_b^2\over 4\pi\mu_0^2}\right)^{-\epsilon/2}\nonumber\\
&&\times\;\left\lbrace
(2-d)\int_0^1 dx\, (1-x)x^{-\epsilon/2}
\left[1+(1-x)\left(-{p^2\over m_b^2}\right)\right]^{-\epsilon/2}\right.\nonumber\\
&&\;+(1-\xi)\int_0^1 dx\,x^{-\epsilon/2}\left[1+(1-x)\left(-{p^2\over
m^2}\right)\right]
^{-\epsilon/2}\\
&&\left.\;+(1-\xi)\;{\epsilon\over2}\;{p^2-m_b^2\over m_b^2}\int_0^1 dx\,
x^{-\epsilon/2}(1-x)\left[1+(1-x)\left(-{p^2\over m_b^2}\right)\right]
^{-1-\epsilon/2}\right\rbrace\;\; ,\nonumber
\end{eqnarray}
where I have introduced $\epsilon=4-d>0$ for convenience.
At this point, it seems that finding expressions for $Z_{\rm F}$
and $Z_{\rm m}$ is merely a matter of substitution.
In fact, since we only work to a certain order in perturbation theory,
this process is easier than it appears at first. 
Looking at Eq.(\ref{massren}), we see that $m$ is in fact defined
implicitly.  This could present problems - especially in 
view of the involved form of $A(p^2)$ and $B(p^2)$.  At this
order, however, we are not concerned with such complications.
Since $m-m_b$ is necessarily of order $g^2$, we can completely
ignore the factor $B(m^2)$ multiplying it in Eq.(\ref{massren}).  
Furthermore, the difference between evaluating $A(p^2)$ and $B(p^2)$
at $m^2$ and $m_b^2$ is also higher order.\footnote{This is true
if and only if $A(p^2)$ and $B(p^2)$ are analytic functions 
in a region of momentum space which includes both $m^2$ and $m_b^2$.
We have already assumed that this is the case in defining a `residue' 
of the `pole' at $p^2=m^2$ since these concepts do not exist 
unless our denominator is analytic in this region.}
These observations
lead to the trivial relationship
\begin{equation}
m-m_b=A^{\rm (1-loop)}(m_b^2)+mB^{\rm (1-loop)}(m_b^2)+{\cal O}(\alpha_s^2)\, ,
\end{equation}
or\footnote{Here, I use the fact that the last term in 
our expression for $B(p^2)$ does not contribute
at ${\cal O}(\epsilon)$ as $p^2\rightarrow m_b^2$.  Due to the 
singularity in the integral, there will be contributions
at higher orders in the $\epsilon$ expansion.} 
\begin{equation}
Z_{\rm m}=1-{\alpha_s C_F\over 4\pi}\Gamma\left({\epsilon\over2}
\right)\left({m^2\over4\pi\mu_0^2}\right)^{-\epsilon/2}{3-\epsilon\over 2-
\epsilon}+{\cal O}(\alpha_s^2)
\;\; .
\end{equation}

We run into some
difficulty when we attempt to calculate $Z_{\rm F}$.  
One of the derivatives we wish to take 
diverges like $\int_0^1 dx\,x^{-1-\epsilon/2}$.  This is not a 
problem for $\epsilon<0$, but in order to regulate the 
ultraviolet divergences we require $\epsilon>0$.  The origin 
of this problem lies in the fact that the quantities $A$ and $B$
are not analytic for $\epsilon$ strictly positive.  
This divergence comes from the region of small $k$ in the 
integral\footnote{This can be seen from the fact that 
the integrand 
\begin{equation}
\sim {1\over k^2[(p-k)^2-m^2_b]}\;\; ,
\end{equation}
so if
we differentiate with respect to $p^2$ and evaluate at 
$p^2=m_b^2$, we get the expression
\begin{equation}
\sim {1\over k^2(k^2-2p\cdot k)^2}\;\; .
\end{equation}
This integrand does not cause a problem for large $k$, but
can certainly lead to a divergence at small $k$.} 
so it is of a different kind than those
we have seen.  The fact that it appears at all is in fact 
quite disheartening since it signals nonanalytic behavior
of $A$ and $B$ near $m^2_b$.\footnote{Strictly speaking, this is not
necessarily a problem since we really only need analyticity 
near the physical pole $m^2$.  However, we are attempting to calculate
$m-m_b$ as a power series in $\alpha_s$.  Functions which are analytic
at $m$ but not at $m_b$ do not have smooth limits as $\alpha_s\rightarrow 0$,
hence cannot possess a Taylor expansion about $\alpha_s=0$.  
Since Taylor expansions are the only things we know how to calculate
in quantum field theory, we are foiled again.  At this point,
we must either admit defeat and go home or find some way
to understand these divergences without compromising the 
analyticity of $A$ and $B$ near either $m^2$ or $m^2_b$.}
This has the potential to make the theoretical constructs of 
the last few pages collapse.  Fortunately, there is hope for a
solution.  The fundamental principle of dimensional regularization
is the assumption 
that the amplitudes in quantum field theory are {\it meromorphic} functions
of spacetime dimension.  Obviously, $\Sigma(p)$ is not a meromorphic function
of $\epsilon$ in its present form
since we must put restrictions on the sign of $\epsilon$ in
order to make sense of it.  However, this form can be interpreted as
a representation
of the function in which its meromorphic 
nature is not manifest.\footnote{For 
example, the function $\sum_{n=0}^\infty z^n$ is not
a meromorphic function in the complex plane as presented since it
does not converge for $|z|\geq 1$.  However, summing the
series in the region where it converges, we find that it is
a perfectly finite function of $z$ for all $z\neq 1$.}
If we assume that this is the case, we are free to calculate 
our divergent contribution in the region $\epsilon<0$ and continue
the result to $\epsilon>0$ (or any other part of the complex $\epsilon$ plane).
Since this divergence comes from the infrared region of momentum
space, it is sometimes useful
to keep it separate from the ultraviolet divergences.  This is done
via subscript.  Our result for $Z_{\rm F}$ is
\begin{eqnarray}
Z_{\rm F}&=&
1-{\alpha_sC_F\over 4\pi}\left({m^2\over 4\pi\mu_0^2}\right)^{-\epsilon/2}
\left[\Gamma\left(\epsilon_{\rm UV}\over 2\right){1+\epsilon\over1-\epsilon}
\,\xi\right.\nonumber\\
&&\left.\qquad\qquad\qquad\qquad\qquad\qquad+{4\over\epsilon_{\rm IR}}
\Gamma\left(1+{\epsilon\over 2}\right)\right]+{\cal O}(\alpha_s^2) \;\; .
\end{eqnarray}

The {\it scheme} we have used to define 
our renormalized parameters above is called the
{\it onshell} scheme since it defines renormalized quantities at the point
where fields are on their mass shells, i.e. $p^2=m^2$. 
This is not required by the idea of renormalization.  
Only the {\it external} fields
must be normalized properly, and the mass at the pole
will be the physical mass regardless of how we choose to define $m$.  
This fact is exemplified by the existence of unrenormalized perturbation
theory, where we choose not to renormalize at all.  At the end
of the day, we simply eliminate the `bare' parameters in favor of
some arbitrarily defined `renormalized' quantities which will be
measured in experiment.  In two different renormalization schemes
the definitions of the renormalized parameters and the expressions
for physical quantities in terms of these parameters will be different, but
the combination is guaranteed to be the same.  This arbitrariness
allows us to make the choice which will make our calculations
the simplest.  The only residue of the onshell requirements which is
left is the normalization of the {\it external} fields, which {\it must} be 
done in the onshell scheme.  However, since the fundamental fields 
of QCD do
not appear as external fields, even this condition is unimportant for
our purposes.\footnote{In QED one must be careful to perform
this task before comparing with experiment.  Otherwise, inconsistencies
will appear between different experiments.
We will also see that this step is necessary for calculating
`physical' matrix elements involving quarks and 
gluons in the next chapter.}  

One might
say that the easiest thing to do is simply not renormalize.  The problem with
this is that all of our parameters are divergent.  One cannot measure
divergent quantities, so in this case one cannot extract the parameters of
the theory from experiment.  The next best thing is to do the least amount
of work so that our renormalized parameters are finite.  This
scheme is called {\it Minimal Subtraction} (MS).  In MS,
we define our $Z$-factors in such a way that the counterterms
subtract all of the {\it ultraviolet} divergences in our diagrams at 
each order in perturbation theory.  It must be emphasized here that
infrared divergences have nothing to do with renormalization.
As mentioned above, the bare parameters of our theory are
related to the way things look at asymptotically large probing
energies.  Hence they have only to do with ultraviolet divergences.
Infrared divergences must be handled in an entirely different 
way, as will be discussed at length in the next chapter.  
For now, we notice that the infrared divergences in this
calculation occur only when we put the fermion on its mass shell.
Away from the point $p^2=m^2$, the amplitudes are infrared finite.
This is a general phenomenon - offshell matrix elements are always infrared
finite\footnote{This statement is not true for certain gauge choices.
In these cases, the infrared divergences come from singular regions
of `gauge choice space'.  An explicit example of this phenomenon 
is given in Section \ref{reninax}.}
since infrared divergences arise only in singular regions of momentum space.
Since we no longer wish to calculate in the onshell scheme, there is no reason
to consider this singular portion of phase space.

The counterterms which appear at this order in the 
quark correlator shift $A(p^2)$ and $B(p^2)$ to
\begin{eqnarray}
A(p^2)&\rightarrow& A(p^2)+m(Z_{\rm m}Z_{\rm F}-1)\;\; ,\nonumber\\
B(p^2)&\rightarrow& B(p^2)-(Z_{\rm F}-1)\;\; .
\end{eqnarray}
Looking at our expressions for $A$ and $B$, we see that in the 
MS scheme
\begin{eqnarray}
Z_{\rm F}^{\rm MS}=1-{\alpha_s^{\rm MS}C_F\over 4\pi}
{2\over\epsilon}(\xi)\;\; ,\nonumber\\
Z_{\rm m}^{\rm MS}=1-{\alpha_s^{\rm MS}C_F\over 4\pi}{2\over\epsilon}(3)\;\; ,
\end{eqnarray}
where we have explicitly shown the dependence on scheme.
The leftover contributions to the self-energy 
are straightforward to calculate:
\begin{eqnarray}
{\tilde A}^{\rm MS}(p^2)&=&-{\alpha_s^{\rm MS}C_F\over 4\pi}m\left\lbrace
\left(3+\xi\right)\left\lbrack\log{m^2\over\mu_0^2}
+\gamma_{E}-\log(4\pi)\right.\right.\nonumber\\
&&\left.\left.\qquad+\left(1-{m^2\over p^2}\right)\log\left(1-{p^2\over
m^2}\right)\right\rbrack
-2\left(2+\xi\right)\right\rbrace+{\cal O}(\alpha_s^2)\;\; ,\\
{\tilde B}^{\rm MS}(p^2)&=&\;{\alpha_s^{\rm MS}C_F\over 4\pi}\xi\left\lbrack
\log{m^2\over\mu_0^2}+\gamma_{E}-\log(4\pi)\right.\nonumber\\
&&\left.\qquad+\left(1-{m^4\over p^4}\right)
\log\left(1-{p^2\over m^2}\right)-1-{m^2\over p^2}\right\rbrack
+{\cal O}(\alpha_s^2)\;\; .
\end{eqnarray}
These expressions are somewhat messy.  The Euler-Masceroni 
constant $\gamma_{E}$ comes
from the expansion of the $\Gamma$-function and 
the $\log\,4\pi$ comes from the $d$-dimensional
phase space factor $(4\pi)^d$.  A general $\ell$-loop graph will contain 
$\ell$ powers of the coupling and the phase space, so divergences will in general be
expressed as\footnote{Here, I have written the form of the most UV-divergent
part of an $\ell$-loop diagram.  There will, of course, be less divergent pieces.
The $\gamma_{E}$'s and $4\pi$'s in these terms will also arrange
themselves in 
a similar way, although this is more difficult to see.}
\begin{equation}
\Gamma^\ell\left({\epsilon\over 2}\right)
\left({\Lambda^2\over4\pi\mu^2}\right)^{-\ell\epsilon/2}
\sim\left({2\over\epsilon}\right)^\ell \left({\Lambda^2
e^{\gamma_{E}}\over4\pi\mu^2}\right)^{-\ell\epsilon/2}\;\; ,
\end{equation}
where I have used the truncated expansion $\log\;\Gamma(1+\epsilon)\sim
-\gamma_{E}\epsilon$
and the fact that $n\Gamma(n)=\Gamma(n+1)$.  
$\Lambda$ is some mass scale in the process, 
i.e. $\Lambda\sim m$.  Since these factors will always appear with the divergences,
it is convenient to include them in the $Z$-factors.  
Furthermore, we are still plagued with the bare scale $\mu_0$.  A fully 
renormalized quantity should not depend on such an object.  
$Z$-factors relate renormalized and unrenormalized quantities, 
so it is natural to absorb the $\mu_0$ dependence into $Z$.  
Doing this requires the introduction of a scale, which we
choose as $\mu$.  Hence our divergence is written
\begin{equation}
\left\lbrack\left({2\over\epsilon}\right)^\ell \left({\mu^2
e^{\gamma_{E}}\over4\pi\mu_0^2}\right)^{-\ell\epsilon/2}\right\rbrack
\left({\Lambda^2
e^{\gamma_{E}}\over4\pi\mu^2}\right)^{-\ell\epsilon/2}\;\; .
\end{equation}
The quantity in brackets is associated with the divergence (the 
transition from bare to renormalized quantities), and so is
seen as a part of $Z$.  The leftover piece depends only on 
physical parameters and the renormalized scale, and is free to enter
renormalized amplitudes.  

The scheme described above is called 
{\it Modified Minimal Subtraction} ($\overline{\rm MS}$) and 
will be used throughout the rest of this dissertation.
In $\overline{\rm MS}$, $Z_{\rm F}$ and $Z_{\rm m}$ are written
\begin{eqnarray}
Z_{\rm F}^{\overline{\rm MS}}=1-{\alpha_s^{\rm\overline{MS}}(\mu^2)C_F\over
4\pi}{2\over\epsilon}
\left({\mu^2e^{\gamma_{E}}\over4\pi\mu_0^2}\right)^{-\epsilon/2}\;(\xi)
+{\cal O}(\alpha_s^2)\;\; ,\nonumber\\
Z_{\rm m}^{\overline{\rm MS}}=1-{\alpha_s^{\rm\overline{MS}}(\mu^2)C_F\over
4\pi}{2\over\epsilon}
\left({\mu^2e^{\gamma_{E}}\over4\pi\mu_0^2}\right)^{-\epsilon/2}\;(3)
+{\cal O}(\alpha_s^2)\;\; ,
\end{eqnarray}
where we have explicitly shown the dependence of $\alpha_s$ on $\mu^2$.
It is obvious that this object will depend on $\mu^2$ since
$Z_g$ depends on $\mu^2$ and $g_b$ does not.  All
renormalized quantities will depend on $\mu$.  For example,
we see that 
\begin{eqnarray}
\mu{d\,m^{\rm\overline{MS}}\over d\mu}&=&-{1\over 
Z^{\rm\overline{MS}}_{\rm m}}\,
\mu{d\,Z_{\rm m}^{\rm\overline{MS}}\over
d\mu}m^{\rm\overline{MS}}\nonumber\\
&=&-3{\alpha_s^{\rm\overline{MS}}(\mu^2)C_F\over2\pi}m^{\rm\overline{MS}}
+{\cal O}(\alpha_s^2)\;\; .
\end{eqnarray}
This means that QCD runs quark masses down as the probing
scale is increased.
This object is in fact not the physical pole mass
of the quark (even in the absence of confinement).  The physical 
pole mass, as defined in the onshell scheme, must be independent of $\mu$
since it is an observable.  We note here that it is natural for 
these quantities to depend on scale since the scale decides which 
corrections are to be included in renormalized quantities and 
which are to be included explicitly in the amplitudes.

The renormalized quantities $\tilde A$ and $\tilde B$
can also be calculated,
\begin{eqnarray}
{\tilde A}^{\rm\overline{MS}}(p^2)&=&-{\alpha_s^{\rm
\overline {MS}}(\mu^2)C_F\over 4\pi}m\left\lbrace
\left(3+\xi\right)\left\lbrack\log{m^2\over\mu^2}\right.\right.\nonumber\\
&&\left.\left.+\left(1-{m^2\over p^2}\right)\log\left(1-{p^2\over
m^2}\right)\right\rbrack
-2\left(2+\xi\right)\right\rbrace+{\cal O}(\alpha_s^2)\;\; ,\\
{\tilde B}^{\rm\overline{MS}}(p^2)&=&\;
{\alpha_s^{\rm\overline{MS}}(\mu^2)C_F\over 4\pi}\xi\left\lbrack
\log{m^2\over\mu^2}\right.\nonumber\\
&&\left.+\left(1-{m^4\over p^4}\right)
\log\left(1-{p^2\over m^2}\right)-1-{m^2\over p^2}\right\rbrack
+{\cal O}(\alpha_s^2)\;\; ,
\end{eqnarray}
and are somewhat cleaner in this scheme.  
Note the branch cut in our amplitude at $p^2=m^2$.  This
cut leads to the infrared divergence in the onshell scheme.  
It is related to the possibility that the intermediate states
in our calculation are onshell and free to propagate to infinity
(in the absence of confinement, of course).  The condition for this possibility
is $(p-k)^2=m^2$ and $k^2=0$, while $(p-k)^0$
and $k^0$ are positive.  Working in the rest frame of 
$p-k$, one can show that these conditions imply 
$p^2=m^2+2mk^0\ge m^2$.  Having learned our lesson from venturing
into the onshell scheme, we choose to work far away from this kinematical
region and take $p^2<0$.  

We still have one point 
which must be addressed.  What should be used for $\mu$?  If we wanted to calculate
to all orders, this choice would be arbitrary since no physical quantities 
actually depend on $\mu$.  However, we have no intention of calculating {\it any}
physical quantity to all orders and no hope of doing so in a finite amount of time
(at least, not at this stage of the game).  With this in mind, our choice of $\mu$ 
should be dictated by the size of the coefficient of $\alpha_s$. 
For $-p^2<\!\!<m^2$, the leftover terms in $A$ and $B$ are all small except 
possibly the $\mu$-dependent logarithm.  
Choosing $\mu^2=m^2$ eliminates this problem.
However, for $-p^2>\!\!>m^2$ this choice would lead 
to a term $\sim\log(-p^2/m^2)$
in the amplitude, rendering the perturbative expansion useless.  In this
region, it is preferable to choose $\mu^2=-p^2$.  Hence we see that
for the one-loop quark self-energy the natural choice of $\mu^2$ is
the larger of $-p^2$ and $m^2$.  Of course, we have no guarantee that these
choices will not lead to disaster at higher orders in the expansion.  The best
choice of $\mu$ is obviously the one which makes {\it all} corrections 
as small as possible.\footnote{In fact, one way to estimate
the contributions of higher-order effects is to study the dependence
of physical quantities on $\mu^2$.  In regions where
the dependence is slight, our corrections should be small.}

\begin{figure}
\label{fig6}
\epsfig{figure=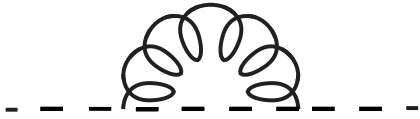,height=1.5cm}
\caption{The one-loop correction to the ghost self-energy.}
\end{figure}

The ghost propagator is renormalized in exactly the same way.
The relevant diagram (Fig. 1.5) is almost identical to the
quark self-energy.  We obtain\footnote{From here on, I will drop
the superscript $\rm\overline{MS}$ on $Z$-factors and the like.  
All results will be given in this scheme unless otherwise specified.} 
\begin{equation}
Z_{\rm G}=1+{\alpha_s(\mu^2)C_A\over16\pi}{2\over\epsilon}
\left({\mu^2e^{\gamma_{E}}\over4\pi\mu_0^2}\right)^{-\epsilon/2}\;(3-\xi)
+{\cal O}(\alpha_s^2)\;\; ,
\end{equation}
where $C_A$ is defined in Appendix \ref{sun}.

The three 
diagrams shown in Fig. 1.6 contribute to the field strength
renormalization of the gluon.  Denoting the 
sum of the amputated 1PI gluon self-energy 
insertions as $-i\Pi^{\mu\nu}$,
the quark loop (Figure 1.6a) contributes\footnote{
The $SU(3)$ generators are normalized by the relation
$T_{\rm F}\delta^{a\,b}={\rm Tr}\,[t^at^b]$.}
\begin{eqnarray}
-i\Pi^{\mu\nu}_{\rm F}(q)&=&-i{2\alpha_s(\mu^2)T_{\rm F}\over\pi}\Gamma
\left({\epsilon\over 2}\right)\sum_f
\left({m_f^2\over 4\pi\mu_0^2}\right)^{-\epsilon/2}\nonumber\\
&\times&\;\;\;\int^1_0dx\,x(1-x)\left[1+x(1-x)\left({-q^2\over m^2}\right)\right]
^{-\epsilon/2}\;\left(q^2g^{\mu\nu}-q^\mu q^\nu\right)\; .
\label{qloop2}
\end{eqnarray}
The transverse structure $q^2g^{\mu\nu}-q^\mu q^\nu$ appears
as a consequence of current conservation, $q_\mu\Pi^{\mu\nu}(q)=0$.
This property of QCD is essential to its renormalizability
\cite{thooft,colren}; 
we will see in Appendix \ref{quantpoint} 
that it persists to all orders.
The gluon and ghost loop contributions 
(Figs. 1.6b and c) do not satisfy
this condition separately; their contributions 
have the UV divergences
\begin{eqnarray}
-i(\Pi^{\mu\nu}_{\rm A})_{\rm UV}(q)&=&i{\alpha_s(\mu^2)
C_A\over 32\pi}{2\over\epsilon}
\left({\mu^2e^{\gamma_{E}}\over4\pi\mu_0^2}
\right)^{-{\epsilon/ 2}}\nonumber\\
&&\qquad\times\;\left\lbrack\left({53\over
3}-4\xi\right)\left(q^2g^{\mu\nu}-q^\mu q^\nu\right)
-\left(q^2g^{\mu\nu}+q^\mu q^\nu\right)\right\rbrack\\
-i(\Pi^{\mu\nu}_{\rm G})_{\rm UV}(q)&=&i{\alpha_s(\mu^2)C_A\over 32\pi}
{2\over\epsilon}
\left({\mu^2e^{\gamma_{E}}\over4\pi\mu_0^2}\right)^
{-{\epsilon\over 2}}\nonumber\\
&&\qquad\times\;\left\lbrack -{1\over 3}\left(q^2g^{\mu\nu}-q^\mu 
q^\nu\right)
+\left(q^2g^{\mu\nu}+q^\mu q^\nu\right)\right\rbrack
\end{eqnarray}
for the gluons and ghosts, respectively.  Note that although each 
contribution separately violates current conservation, the 
sum does not.  This is the reason the ghosts needed to be introduced - they
exist only to cancel the unphysical gluonic degrees of freedom which appear
in covariant gauges.  

\begin{figure}
\label{fig7}
\epsfig{figure=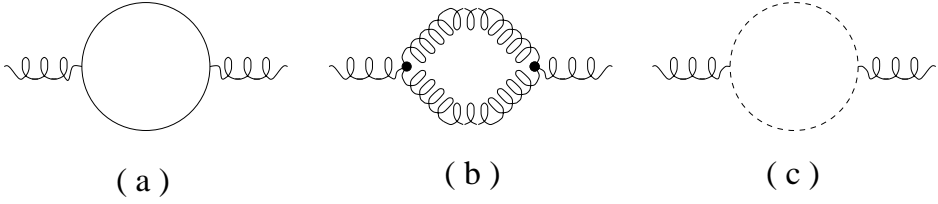,height=1.0in}
\caption{The three contributions to the gluon
field-strength renormalization at 
one-loop order in covariant QCD.}
\end{figure}

$\Pi^{\mu\nu}$ receives two counterterm contributions.  The
field strength contributes $-i(Z_{\rm A}-1)(q^2g^{\mu\nu}-q^\mu q^\nu)$,
while the gauge parameter gives $i(Z_{\rm A}Z_\xi^{-1}-1)q^\mu q^\nu/\xi$.
Since all of our divergences are proportional to the 
transverse structure, we can simply take $Z_{\rm A}=Z_\xi$ 
at this order.  In fact, the counterterm relevant to the
gauge parameter leads to nonconservation of our current.  This
is not surprising since this term came from our gauge fixing, 
so certainly cannot be expected to be gauge invariant.  
However, it can be shown that
even the gauge fixed theory will conserve current
\cite{renward}, which implies
$Z_\xi=Z_{\rm A}$ to all orders.  
We can see this in another way by returning to 
unrenormalized perturbation theory for a moment.
Writing the full 1PI gluon contribution as
\begin{equation}
\label{12345}
-i\Pi_{\mu\nu}^{a\,b}(q^2)=-i\Pi(q^2)(q^\mu q^\nu-q^2g^{\mu\nu})
\delta^{a\,b}\; 
\end{equation}
and re-summing the 1PR contributions, we obtain the full gluon propagator
\begin{equation}
D_{\mu\nu}^{a\,b}(q)={i\left\lbrack-g_{\mu\nu}+\left(1-(1-\Pi(q^2))\xi_b\right)
(q_\mu q_\nu/ q^2)\right\rbrack\over (1-\Pi(q^2))q^2}\delta^{a\,b}\; .
\end{equation}
It is now obvious that the gluon fields and the gauge parameter are
renormalized in the same way to all orders.  This is a direct consequence of
our assumption (\ref{12345}) that the 1PI gluon contribution is transverse.

Defining the other counterterm in such a way that it cancels the 
divergent contributions to $\Pi$, we find
\begin{equation}
Z_{\rm A}=1+{\alpha_s(\mu^2)\over4\pi}
{2\over\epsilon}\left(\mu^2e^{\gamma_{E}}
\over4\pi\mu_0^2\right)^{-\epsilon/2}\left
\lbrack{1\over 2}\left({13\over 3}-\xi\right)C_A-
{2n_f\over3}\left(2T_F\right)\right\rbrack\;\; ,
\label{zaren}
\end{equation}
where $n_f$ is the number of flavors.  

The leftover finite
contribution to $\Pi$ has a serious problem coming from the 
quark loop (Figure 1.6a).  As in the quark self-energy diagram
analyzed above, this expression contains logarithms involving 
the quark masses, the external momentum scale, and the 
renormalization scale.  Once again, these logarithms 
become large unless we choose the renormalization scale near
the larger of the two physical scales.  However,
we cannot just arbitrarily choose $\mu^2$ to
satisfy this one particular diagram; there are
two other diagrams which also contribute.  These other
diagrams demand a renormalization scale comparable to the 
external momentum ($q^2$).  If any of the quarks 
in our theory have masses much larger than this scale,
Figure 1.6a will lead to large finite corrections to the 
amplitude.  

From a physical point of view, this is 
nonsense - extremely heavy quarks should decouple from 
low energy phenomena.  The problem lies in our renormalization procedure
itself.  We are choosing counterterms to
cancel {\it all} ultraviolet divergences, regardless of the scale
at which they become important.  This is not exactly
what we want.  The idea of renormalization is to choose a scale
and calculate explicitly all physics below that scale, while
systematically grouping the physics above it into
constants which renormalize the fields and parameters of the theory.
Since the heavy quark loop has support only for 
loop momenta near or above its mass scale, choosing 
a small renormalization scale should make its contribution 
to the renormalized amplitudes vanish.  What we see is 
quite the contrary : the contribution of the quark loop 
diagram to the renormalized amplitudes is divergent
as the quark mass goes to infinity.  Since this 
diagram has also contributed to the infinite constant
$Z_{\rm A}$, and the strict limit $m_f\rightarrow\infty$
gives no contribution to the amplitude\footnote{This can be seen
explicitly from the expression for the diagram; one takes
$m_f$ to $\infty$ {\it before} doing the integral.}, we can interpret 
the divergence in the renormalized amplitude as an attempt to cancel
the divergence in $Z_{\rm A}$.  This procedure leads us to believe
that we should simply ignore the effects of very heavy 
quarks; the associated physics will be grouped
into renormalization constants with the rest of the high-energy
behavior of the theory.  On the other hand, we cannot expect all of the physics 
associated with a quark to be represented by a simple constant.  
If it could be, QCD would certainly not have such rich structure.
The complete dynamics associated with a heavy quark must 
be represented instead by a series of local operators
which give the heavy quark effects the 
freedom to depend on external momenta.
This is the hallmark of an unrenormalizable theory.
Expanding the integrand of the quark loop contribution
in a power series in $1/m_f$, we see that each term 
is more divergent than the last.\footnote{Of course, this is
because our expansion does not converge for large loop momenta.
These new divergences did not come out of thin air.}
This implies that we will need an infinite number 
of counterterms to cancel them, another indication
that the effective theory is unrenormalizable.  

The explicit
form of this theory can be obtained by simply doing the 
path integral for the heavy quark fields.  The result, up
to a trivial constant factor which cancels with the 
vacuum fluctuations, is 
\begin{equation}
{\rm det}\left(1-{i\not\!\!{\cal D}\over m_f}\right)\;\; .
\end{equation}
This determinant can be exponentiated via the identity 
\begin{equation}
\log{\rm\det}M={\rm Tr}\log M\;\; ,
\end{equation}
valid for any nonsingular matrix $M$, 
and added to the lagrangian of the theory.
The expansion of the logarithm gives the infinite 
series of local operators required to encode the effects of the 
heavy quark.  

However simple this expansion looks, it is certainly
not clear what the trace means in this context.  For this
reason, one usually considers instead 
the equivalent\footnote{again, up 
to a trivial constant} determinant
\begin{equation}
{\rm det}\left(1-(i\!\not\!\partial-m_f)^{-1}g\not\!\!{\cal A}\right)\;\; .
\label{hquarkdet}
\end{equation}
Here, the trace of a particular term in the 
expansion of the logarithm is represented as a heavy quark
loop with the appropriate number 
of external gluon fields.\footnote{One 
simply identifies the factor
$(i\!\not\!\partial-m_f)^{-1}$ with the free propagator
for the heavy quark.}  The contribution 
from each loop is calculated according to the 
above rules, with the gluon fields treated
as external sources.  At this point, there is
no difference between the present calculation
and the one that led to (\ref{qloop2}).
The difference comes in our {\it interpretation}
of the ultraviolet divergences.\footnote{Loops
with five or more gluon fields are ultraviolet
finite.  These contributions are not as subtle as 
those that presently concern us.}
These divergences have the form
\begin{equation}
{2\over\epsilon}\,\left(m_f^2e^{\gamma_E}\over
4\pi\mu_0^2\right)^{-\epsilon/2}\;\; .
\end{equation}
This expression represents a battle between
two different limits.  If $\;\epsilon\,\log(m_f/\mu)<\!\!<1$,
we have an ultraviolet divergence which is handled 
as before by renormalizing the gluon field strength.
On the other hand, if $\;\epsilon\,\log(m_f/\mu)>\!\!>1$,
this term approaches zero.
Since we {\it cannot}
take $\epsilon$ strictly to zero in expressions
of this form, the limit $m_f\rightarrow\infty$
annihilates them.\footnote{One needn't 
worry about this expression for light
quarks.  As we have seen above, the terms reorganize
themselves in this case; $\mu^2$ is compared with
$q^2$ rather than $m_f^2$.}  
Contributions 
which are ultraviolet finite
do not require $\epsilon$ to regulate them.  
Here, we can take $\epsilon$ strictly 
to zero.  Since $m_f$ is not {\it strictly} 
infinity, the factor
\begin{equation}
\left(m_f^2e^{\gamma_E}\over
4\pi\mu_0^2\right)^{-\epsilon/2}\rightarrow 1\;\; .
\end{equation}

Grouping the heavy quark effects into a series 
of local operators is now straightforward.
We simply calculate the contribution
from each {\it nonlocal} term in the expansion
of (\ref{hquarkdet}), throw the ultraviolet
divergence (if any) away, and expand the 
remainder in a series in $1/m_f$.  Each 
term in the expansion is then replaced 
by an effective operator and added to the 
lagrangian.
As an example, we calculate explicitly
the two-gluon contribution :
\begin{eqnarray}
i\int d^{\,4}x\,{\cal O}_2(x)&=&-{g^2\over2}\int\,d^{\,4}x\,d^{\,4}y\,
{\rm Tr}\left\lbrack{\cal M}^{-1}_{quark}
(x,y)\not\!\!\!{\cal A}(y)
{\cal M}^{-1}_{quark}(y,x)\not\!\!\!{\cal A}(x)
\right\rbrack\nonumber\\
&=&-{g^2T_F\over2}\int{d^{\,4}q\over(2\pi)^4}
\tilde{\cal A}^a_\mu(q)\tilde{\cal A}^a_\nu(-q)\nonumber\\
&&\qquad\times\int{d^{\,d}k\over(2\pi)^d}
{{\rm Tr}\,\lbrack(\!\not\!k\,+\!\!\not\!q+m_f)\gamma^\mu
(\!\not\!k\,+m_f)\gamma^\nu\rbrack\over[(k+q)^2-m_f^2]
(k^2-m_f^2)}\nonumber\\
&\rightarrow&-i{\alpha_sT_F\over10\pi}
\int{d^{\,4}q\over(2\pi)^4}
\tilde{\cal A}^a_\mu(q)\tilde{\cal A}^a_\nu(-q)
\left(q^2g^{\mu\nu}-q^\mu q^\nu\right)\left({q^2\over m_f^2}\right)\\
&&\qquad\qquad\qquad\qquad\qquad\qquad\times\left\lbrack 
{1\over3}+{1\over28}\,{q^2\over m_f^2}+\cdots\right\rbrack\nonumber\\
&=&-i{\alpha_sT_F\over10\pi}\int d^{\,4}x
{\cal A}_\mu^a(x)\left(g^{\mu\nu}\partial^2
-\partial^\mu\partial^\nu\right)\left({\partial^2\over m_f^2}\right)\nonumber\\
&&\qquad\qquad\qquad\qquad\qquad\qquad\times
\left\lbrack {1\over3}-{1\over28}\,{\partial^2\over m_f^2}
+\cdots\right\rbrack {\cal A}_\nu^a(x)\;\; .\nonumber
\end{eqnarray}
The final form of the local operator can 
certainly be expressed in a nicer form.
In particular, for an abelian theory it is
entirely equivalent to the gauge-invariant 
combination
\begin{equation}
{\cal O}_2={\alpha_sT_F\over20\pi}{\cal F}^{\mu\nu}
\left({1\over3}-{1\over28}\,{\partial^2\over m_f^2}+\cdots\right)\,
{\partial^2\over m_f^2}\,{\cal F}_{\mu\nu}\;\; .
\end{equation}
For non-abelian theories, it 
is certainly not clear whether 
or not the higher order terms in our
expansion of (\ref{hquarkdet}) will conspire
to form a gauge-invariant combination at the end of the
day.  On one hand, we know that the 
original lagrangian is gauge invariant so we
would expect gauge invariance of our effective lagrangian 
as well.  On the other, the determinant
(\ref{hquarkdet}) is not obviously gauge invariant
due to our cancelation of the infinite 
constant
\begin{equation}
{\det}\left(i\!\not\!\partial-m_f\right)\;\; .
\end{equation}
In any case, we will consider these correction
terms small enough to ignore completely.
This leads us to the result (\ref{zaren})
obtained previously for the gluon field strength 
renormalization, but with $n_f$ interpreted as
the number of {\it active} flavors.

\begin{figure}
\label{fig8}
\epsfig{figure=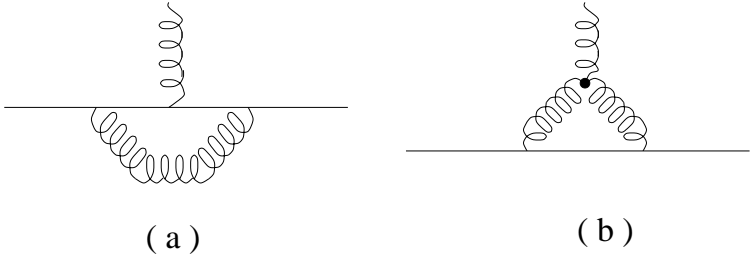,height=1.3in}
\caption{The two vertex correction diagrams at one-loop order in QCD.}
\end{figure}

Now, we need only calculate $Z_g$.  This quantity appears in 
several counterterms.  Unless our gauge symmetry 
is broken by quantum effects, the $g$'s in our lagrangian 
must remain the same after renormalization.  This can be 
checked explicitly by calculating $Z_g$ in all four different 
ways.  However, a general analysis of the ways a symmetry can
be broken by quantum effects tells us that vector symmetries
(symmetries that treat the left- and right-handed fermions in the same 
way) are preserved.\footnote{An exception to this rule
is conformal symmetry, which is broken {\it explicitly}
by our regularization procedure.}   
Trusting this, we calculate the 
one-loop corrections to the quark-gluon vertex shown in 
Figure 1.7.  Since we wish to work in $\overline{\rm MS}$,
only the ultraviolet divergent parts of the amplitudes are
required to deduce $Z_g$.  Diagram 1.7a. contains the 
divergence 
\begin{equation}
{\alpha_s\over4\pi}\left(C_F-
{C_A\over2}\right){2\over\epsilon}\left(\mu^2e^{\gamma_{E}}
\over4\pi\mu_0^2\right)^{-\epsilon/2}\left(\xi\right)\;\; ,
\end{equation}
while 1.7b. contains
\begin{equation}
{\alpha_s\over4\pi}\left(C_A\right){2\over\epsilon}
\left(\mu^2e^{\gamma_{E}}
\over4\pi\mu_0^2\right)^{-\epsilon/2}{3\over4}\left(1+\xi\right)\;\; .
\end{equation}
Here, we have written only the 
constant factor multiplying the leading structure,
$-igt^a\gamma^\mu$.  In this language, the 
counterterm contributes
\begin{equation}
Z_{\rm F}Z_{\rm A}^{1/2}Z_{g}-1\;\; .
\end{equation}

Putting all of this together and requiring the sum
to be ultraviolet finite gives
\begin{equation}
Z_g=1-{\alpha_s(\mu^2)\over4\pi}{2\over\epsilon}\left(\mu^2e^{\gamma_{E}}
\over4\pi\mu_0^2\right)^{-\epsilon/2}\left\lbrack{11\over 6}C_A
-{2T_F\over3}n_f\right\rbrack\;\; .
\end{equation}
As before, $n_f$ is the number of quark flavors whose mass is less 
than the probing scale.  We notice at once that the 
gauge dependence has canceled.  This is necessary
because the gauge coupling is
independent of gauge choice.  It would certainly be
strange if renormalization required this quantity to
acquire gauge dependence.  However, 
the cancelation is highly nontrivial.  Both of the 
diagrams, as well as the field strength renormalizations, are
gauge-dependent.  The gauge dependence
of these diagrams is not truly an artifact of the coupling itself,
but rather is associated with the external fields.

\section{The Beta Function - Asymptotic Freedom}
\label{betaasymp}

As we have seen, the ideas behind renormalization
make it necessary for renormalized parameters
to depend on the scale of separation between 
moderate and ultraviolet momenta.  Although true physical
observables are in fact independent of this scale,
any truncation of the coupling expansion will introduce 
dependence.  Furthermore, our {\it effective} coupling
really does depend on scale.  This scale dependence
governs the region of applicability of perturbation theory,
as we will see below.

The variation of the QCD coupling with scale is 
summarized in the QCD $\beta$-function,
defined via
\begin{equation}
\beta(\alpha_s)\equiv\mu^2{d\alpha_s\over d\mu^2}\;\; .
\label{betafcn}
\end{equation}
Since this dependence comes only from $Z_g$, 
the $\beta$-function will not depend on any scales 
except $\mu^2$.  
Furthermore, the fact that it is dimensionless
implies that even this dependence comes 
exclusively from 
$\alpha_s(\mu^2)$.\footnote{Since the $\beta$-function
is manifestly finite 
{\it when expressed in terms of the renormalized coupling}, 
factors of $(\mu/\mu_0)^{-\epsilon}\rightarrow1$.}
Since 
$g=Z_g^{-1}g_b$ and $g_b$ is independent of $\mu$, we have the relation
\begin{equation}
\beta(\alpha_s)=-2\alpha_s\,Z_g^{-1}\mu^2{dZ_g\over d\mu^2}\;\; .
\end{equation}

The leading order result for $Z_g$ obtained
in the last section gives
\begin{equation}
\beta(\alpha_s)=-\alpha_s^2(\mu^2){1\over4\pi}\left({11\over3}C_A-{4\over3}n_fT_F\right)
+{\cal O}(\alpha_s^3)\;\; .
\label{lobeta}
\end{equation}
A look at the QCD lagrangian, Eq.(\ref{unlag}), 
tells us that the gluon fields are
properly normalized for $T_F=1/2$.\footnote{Appropriately normalized
real vector fields have the kinetic term 
$-1/2\;\partial^\mu{\cal A}^\nu\partial_\mu{\cal A}_\nu$.} 
Since $C_A=3$ for $SU(3)$ with this normalization, the 
leading order $\beta$-function is negative in QCD 
with less than seventeen active flavors.
This means that if we start with a small enough coupling to 
apply perturbation theory, the coupling will decrease as its
scale increases.  A theory whose $\beta$-function exhibits this
behavior is called asymptotically free.

The analysis of the last section tells us that the 
natural choice of $\mu^2$ is the probing
scale relevant to our process.  A different choice
will lead to large coefficients of the expansion
due to leftover logarithms.  In this sense, 
the variation of $\alpha_s$ with scale is a physical phenomenon.
Asymptotic freedom assures us that 
QCD behaves as a weakly coupled theory
when probed at high energy.\footnote{Assuming, of course, 
that there exists some energy at which the coupling
is small enough to apply perturbation theory.  This 
assumption has been thoroughly justified by experiment,
as exemplified by the observation of scaling in DIS.}
On the other hand, our expression for the $\beta$-function
also implies that the effective coupling grows 
we decrease the probing scale.  Truncating
the $\beta$-function at leading order and solving 
(\ref{betafcn}) gives
\begin{equation}
\alpha_s(\mu^2_2)={\alpha_s(\mu^2_1)\over 1+
\beta_0\alpha_s(\mu^2_1)\log\left(\mu_2^2/\mu_1^2\right)}\;\; ,
\end{equation}
where I have introduced $\beta_0\equiv(11-2n_f/3)/4\pi$.
Were this exact, our coupling would diverge at
the scale 
\begin{equation}
\mu_2^2=\mu^2_1e^{-\alpha_s(\mu_1^2)/\beta_0}\;\; .
\end{equation}
The truncation of the perturbative 
expansion will loose its validity long before
the coupling diverges, so this is certainly not
a physical effect.  However, it does imply that there 
exists a scale small enough that QCD is nonperturbative.
Hence we see that asymptotic freedom is a two-edged sword - 
it implies both that perturbation theory will be applicable
at high energy {\it and} that it will not be applicable
at low energy.  Since low energy is related to 
long distance phenomena, this result means that 
the interaction between two quarks will grow as they are
separated until perturbation theory breaks down and their
behavior becomes incalculable.  One can use this as a 
heuristic motivation for confinement, although it
certainly does not constitute a proof.

Although asymptotic freedom tells us that our coupling 
will formally vanish as the energy is increased,
it does not tell us at what energy it will become 
small.  Experimentally, $\alpha_s$ is measured at
the $Z$-boson mass (91.19 GeV) as
\begin{equation}
\alpha_s(M_Z^2)=0.119(2)\;\; .
\end{equation}
While this certainly seems small enough to apply perturbation
theory, higher order effects should be measurable.
For this reason, a systematic treatment of the corrections
to the leading $\beta$-function is desired.  

In general, one writes
\begin{equation}
\beta(\alpha_s)=-\beta_0\alpha_s^2-\beta_1\alpha_s^3-\cdots\;\; .
\end{equation}
Of course, our renormalized coupling will depend on the 
scheme we use to define it.  The relationship between
the renormalized couplings in two different 
schemes can be calculated in perturbation theory :
\begin{equation}
\tilde\alpha_s=\alpha_s+\gamma\alpha_s^2+\delta\alpha_s^3+\cdots\;\; .
\label{schemedep}
\end{equation}
To see how the $\beta$-function
depends on scheme, we write\footnote{We assume here that the 
difference between the two schemes is not scale dependent,
i.e. $\gamma$ and $\delta$ do not depend on scale.}
\begin{equation}
\tilde\beta=\beta+2\gamma\alpha_s\beta+3\delta\alpha_s^2\beta+\cdots\;\; .
\end{equation}
Expanding $\beta$ in terms of $\alpha_s$ and using the 
relations
\begin{eqnarray}
\alpha_s^2&=&\tilde\alpha_s^2-2\gamma\tilde
\alpha_s^3+(5\gamma^2-2\delta)\tilde\alpha_s^4+\cdots\;\; ,\nonumber\\
\alpha_s^3&=&\tilde\alpha_s^3-3\gamma\tilde\alpha_s^4+\cdots\;\; ,\\
\alpha_s^4&=&\tilde\alpha_s^4+\cdots\;\; ,\nonumber
\end{eqnarray}
we find 
\begin{eqnarray}
\tilde \beta_0&=&\beta_0\nonumber\\
\label{req112}
\tilde \beta_1&=&\beta_1\\
\tilde \beta_2&=&\beta_2-\gamma \beta_1-
(\gamma^2-\delta)\beta_0\;\; .\nonumber
\end{eqnarray}
Hence the first two coefficients of the
$\beta$-function are independent of scheme\footnote{This is {\it not}
true if $\gamma$ and $\delta$ are scale dependent.  In that case,
only the first coefficient will be scheme independent.},
but the rest will depend on the conventions used to define 
$\alpha_s$.  

Since these higher-order terms depend on scheme,
they may be chosen in any way we like.  A convenient 
choice {\it defines}
\begin{equation}
\beta_n\equiv\left({\beta_1\over \beta_0}\right)^n\beta_0\;\; ,
\label{req111}
\end{equation}
allowing us to re-sum the $\beta$-function to all
orders in $\alpha_s$:\footnote{It may seem as though
we have just obtained something for nothing.  However,
this manipulation is not a true simplification; it only plays one
on TV.  If we wish to take this scheme seriously, we must
use it consistently in all of our calculations.  One way to do
this is to calculate an amplitude in a known scheme, say $\rm\overline{MS}$,
and re-express the amplitude in terms of the new coupling 
via (\ref{schemedep}).  The constants $\gamma$ and $\delta$ are obtained
from the requirement (\ref{req111}) in conjunction with 
(\ref{req112}).  Even after all of this manipulation,
our amplitude will still be correct only up to the order at which it
was calculated.  On the other hand, this scheme is certainly
not without its merits.  The simple evolution of 
$\alpha_s$ is offers may indeed be worth the effort.}
\begin{equation}
\beta(\alpha_s)=-{\beta_0\alpha_s^2\over1-\beta_1\alpha_s/\beta_0}\;\; .
\label{allorderbeta}
\end{equation}
This function is not Lipshitz continuous near $\alpha_s=\beta_0/\beta_1$,
so we have no guarantee that a solution
to (\ref{betafcn}) exists.  In fact, an analysis of this
equation reveals two distinct branches of solutions.
The region $\alpha_s<\beta_0/\beta_1$ is completely disjoint
from $\alpha_s>\beta_0/\beta_1$.  In the latter case,
the $\beta$-function is {\it not} aymptotically free.  Since
asymptotic freedom is observed in nature, only the 
former region is physically realized.\footnote{Although
the latter region is not physical in the sense
that it is not asymptotically free, it certainly
does exhibit the kind of behavior 
we expect for a theory with well-defined 
infrared behavior.  Unless our coupling
really does diverge in the infrared,
the $\beta$-function must become 
positive as $\alpha_s$ becomes large.  
This, along with the physical requirement of 
Lipshitz continuity, implies the existence
of an infrared fixed point.  These topics are
discussed to some extent in \cite{peskin}.}
However, 
the solution to
Eq.(\ref{betafcn}) in this region,
\begin{equation}
\beta_0\log{\mu_2^2\over\mu_1^2}
={1\over\alpha_s(\mu_2^2)}-{1\over\alpha_s(\mu_1^2)}
+{\beta_1\over\beta_0}\log{\alpha_s(\mu_2^2)\over\alpha_s(\mu_1^2)}\;\; ,
\end{equation}
does not exist for arbitrary $\mu_2^2$.  
It is readily seen that one must take 
\begin{equation}
\mu_2^2\geq
\left({\beta_0\over\alpha_s(\mu_1^2)\beta_1}\right)^{\beta_1/\beta_0^2}\;
{\rm exp}\left\lbrack-{1\over\beta_0}\left({1\over\alpha_s(\mu_1^2)}
-{\beta_1\over\beta_0}\right)\right\rbrack\;\mu_1^2
\end{equation}
to obtain a sensible value for $\alpha_s(\mu_2^2)$.\footnote{
Qualitatively, one can identify the two disjoint
regions at the point $\alpha_s=\beta_0/\beta_1$.
The phase plane portrait of (\ref{betafcn})
then reveals this value as a special kind of 
infrared fixed point.  This analysis does not
survive the transition to quantitative 
reasoning, however, since increasing the scale
then gives ambiguous results.}
The actual value of this bound on $\mu_2^2$
depends on the number of active flavors and the value of
$\mu_1^2$,\footnote{A standard treatment from
the $Z$ mass, changing $n_f$ as
quark masses are crossed, gives a lower bound
of approximately 0.4 GeV$^2$.} but its existence implies that 
we must be more careful when attempting
to work beyond perturbation theory.  
Despite appearances, our expressions are really only
valid in the region of small $\alpha_s$.

One can obtain $\beta_1$ from a two-loop calculation of
any vertex in the theory.  Such a calculation is extremely
complicated, but certain general aspects of it can give
us insight into the behavior of higher order effects.
To begin with, assume 
\begin{equation}
Z_g=\sum_\ell z_\ell\left({\mu^2e^{\gamma_{E}}
\over4\pi\mu_0^2}\right)^{-\ell\epsilon/2}
\left({\alpha_s(\mu^2)\over4\pi}\right)^\ell\;\; .
\end{equation}
We emphasize here that it is the {\it renormalized} coupling which appears
in this expansion.  Keeping all terms up to order $\alpha_s^3$,
we write 
\begin{eqnarray}
\beta(\alpha_s)&=&\alpha_s\left\lbrack z_1\epsilon{\alpha_s\over4\pi}
\left({\mu^2e^{\gamma_{E}}\over4\pi\mu_0^2}\right)^{-\epsilon/2}
-2z_1{\beta\over4\pi}\left({\mu^2e^{\gamma_{\rm
E}}\over4\pi\mu_0^2}\right)^{-\epsilon/2}\right.\nonumber\\
&&\left.\qquad\qquad+(2z_2- z_1^2)
\epsilon\left({\alpha_s\over4\pi}\right)^2
\left({\mu^2e^{\gamma_{\rm
E}}\over4\pi\mu_0^2}\right)^{-2\epsilon/2}\right\rbrack\\
&=&\alpha_s\left\lbrack z_1\epsilon{\alpha_s\over4\pi}
\left({\mu^2e^{\gamma_{E}}\over4\pi\mu_0^2}\right)^{-\epsilon/2}
+(2z_2-3z_1^2)\epsilon\left({\alpha_s\over4\pi}\right)^2
\left({\mu^2e^{\gamma_{\rm
E}}\over4\pi\mu_0^2}\right)^{-2\epsilon/2}\right\rbrack\;\; .\nonumber
\end{eqnarray}
Evidently, the coefficient we are after is 
\begin{equation}
\beta_1={3z_1^2-2z_2\over(4\pi)^2}\epsilon\;\; .
\end{equation}

Since $z_1$ is divergent like $1/\epsilon$, it looks as though the 
second coefficient of the $\beta$-function diverges in the same manner.
This is unacceptable.  The renormalizability of our theory 
guarantees a finite renormalized coupling.  If $g$ is finite, it 
does not have divergent variation with $\mu^2$.  
The only way out of this situation is to 
require a complete cancelation of $z_1^2$, i.e.
if $z_2=a/\epsilon^2+b/\epsilon$ then $a=3(\epsilon z_1)^2/2$.
This is in fact what happens.  The fact that 
$g$ is finite implies that the double poles in 
$z_2$ are not independent of $z_1$.  This behavior is  
expected since double poles
come from regions of momentum space where two loop-momenta
are simultaneously large.  Renormalizability assures us that
all divergences obtained when one loop momentum gets large 
are taken care of by the one-loop counterterms,
so these double poles are naturally related to products of 
the one-loop results.  This is a general aspect of
renormalizable quantum field theories - all higher-order ultraviolet
divergences in an $\ell$-loop diagram are related to the 
divergences in $n<\ell$-loop diagrams.  Only the 
$1/\epsilon$ pole in a diagram gives new information.
Stated compactly, all true $\ell$-loop effects are given by
the residue of the $\ell$-loop amplitude at $\epsilon=0$.
This is a nice result, but difficult to exploit in
practice.  One usually needs to compute the entire divergence
structure of a graph to extract the $1/\epsilon$ pole.
However, it can be an extremely useful check of a calculation.

The result for $\beta_1$ can be found in the literature \cite{ellis}.  
It is
\begin{eqnarray}
\beta_1&=&{1\over(4\pi)^2}{2\over3}\left\lbrack17 
C_A^2-2n_fT_F(3C_F+5C_A)\right\rbrack\nonumber\\
&=&{1\over(4\pi)^2}\left(102-{38\over3}n_f\right)\;\; .
\end{eqnarray}
Since $\beta_1>0$ for $n_f<9$,
this coefficient only serves to hasten the 
asymptotic vanishing of the coupling.

Looking at the diagrams we used to arrive at this result, we
see that the $\beta$-function for QED can be obtained by
taking $C_A=0$ and $T_F=C_F=1$.  In this theory, 
the $\beta$-function is {\it positive} in perturbation theory.
We can understand this in terms of {\it screening} of electric
charge.  A charge sitting in the vacuum polarizes the 
virtual fluctuations.  Since QED causes opposite
charges to attract, this polarization results in a 
screening of the true electric charge.  As we probe
the charge at higher and higher energy, we effectively
penetrate some of this shielding and get a look at the true charge.
Hence, we would expect the QED coupling to increase with scale.
The analogue of this argument in QCD results in 
anti-screening since the gluons themselves participate in the
vacuum polarization.  The effect of gluons is opposite that of
quarks, so acts to increase the charge experienced at great distances.
Although this argument and its extensions cannot rigorously
prove that colored objects are confined, they can certainly
go a long way towards giving us a qualitative understanding of this
phenomenon.

\section{Renormalization in the Axial Gauge -\\ The Price of a 
Ghost-Free Theory}
\label{reninax}

Now that we understand the need for renormalization in 
quantum field theories, we would like to see how it
plays out in another gauge.  As introduced in Section 
\ref{roadtoquant}, the axial gauge seems rather simple.  
There are no ghosts to worry about and the freedom 
to choose $n$ can in principle allow us to 
simplify calculations a great deal.
The only apparent complication is the rather lengthy
gluon propagator, Eq.(\ref{axgl}).  

This gauge
belongs to a set of `physical gauges' since 
contraction of the gauge propagator with its
momentum gives
\begin{equation}
q^\mu{D}_{\mu\nu}=i\left\lbrack {n_\nu\over n\cdot q}-{q_\nu\over(n\cdot q)^2}
\left(n^2+\xi q^2\right)\right\rbrack\;\; ,
\end{equation}
which does not contain a propagator pole at $q^2=0$.  
Hence the unphysical degree of freedom
associated with the longitudinal polarization
state does not propagate in these gauges.\footnote{This is
{\it not} the case in covariant gauges.  There,
contraction gives $-i\xi q_\nu/q^2$.  Hence
in covariant gauges unphysical degrees of freedom 
propagate unless one takes $\xi=0$
(Landau's gauge).}  This is the reason 
there are no ghosts in axial gauges - the 
degree of freedom they live to cancel does not
propagate here.  It seems that axial 
gauge, rather than covariant, is the natural choice 
for gauge theories from this reasoning.

Once we begin to look at renormalization, however, 
this gauge changes its tune.  The new tune is one
of violent attacks on Lorentz invariance and
unregulated divergences.  
To study these effects,
let us consider a simpler situation than the general
case.  Apparently, taking 
$n^2=\xi=0$ simplifies the gluon propagator immensely.
Since we intend to choose $\vec n$ as a special vector 
in our problem, we take it collinear to all 
external momenta in our problem.  Note that this
implies that our use of axial gauge requires 
collinear kinematics.\footnote{One can, of course,
use the axial gauge in situations which are not collinear.
However, the integrals one is faced with in these situations 
are a great deal more complicated than those 
we will encounter.  The choice of axial gauge
is usually made expressly for the reason that the kinematics
of a process favor one direction over all others.
Otherwise, there is no advantage to its use.} 
This space of collinear
vectors is spanned by the basis 
$\lbrace n^\mu=(1,0,0,-1)/2\Lambda, p^\mu=(1,0,0,1)\Lambda\rbrace$,
where $\Lambda$ is an arbitrary scale reflecting the 
residual boost invariance along the 3-direction.\footnote{
This invariance is really an illusion since any specific choice for
$n$ fixes $\Lambda$.}  Our collinear kinematics requirement 
is that all external momenta can be expressed as a linear combination
of $n$ and $p$, i.e. all particles 
move in the 3-direction.\footnote{
This is not as restrictive as it may seem.  In two-particle 
systems, we can always transform into a frame in which the 
momenta are collinear.  In this frame, we are free to quantize the theory
and introduce $n$.  However, once we have introduced $n$
we are no longer free to change frames.  This is 
a consequence of the axial gauges' violence to Lorentz symmetry.}
Note that we have chosen our basis vectors such that $p^2=n^2=0$ and 
$n\cdot p=1$.  

The goal of this section is to obtain the QCD $\beta$-function
in the axial gauge.  Along the way, we will see some major differences
between this gauge choice and the covariant choice.  
As with the covariant gauges, our first task
is to calculate the quark self-energy.
For simplicity, we begin with massless 
quarks.\footnote{Since we are concerned here
only with the ultraviolet behavior of the theory,
this can be done without loss of generality.
We will briefly consider massive quarks
at the end of this section.}
On general grounds, we can decompose the 
1PI contribution as\footnote{We take the quark 
momentum as $q$ here to avoid confusion with the
basis vector $p$.}
\begin{equation}
\Sigma(q)=A\not\! q+B{q^2\over2n\cdot q}\not\! n\;\; .
\end{equation}
The appearance of $\not\!n$ is a direct consequence
of the explicit Lorentz symmetry breaking induced
by our axial gauge choice.  This also leads to the 
fact that $A$ and $B$ are functions of $q^2$ {\it and}
$n\cdot q$.  Our choice $n^2=0$ suppresses their
dependence on $n^2$.  The factor $q^2/2n\cdot q$ multiplying
$B$ causes $A$ and $B$ to have the same dimensionality
(since $n^\mu$ is like an inverse mass) and will prove convenient
in the future.

Iterating the 1PI result to all orders to obtain
the full 1PR propagator as in (\ref{1PR}) gives
\begin{equation}
{i\over\not\!q}\sum_n\left\lbrack
\left(-i\Sigma\right){i\over\not\!q}\right\rbrack^n
={i\over\not\!q}\left\lbrack 
1-A-{1\over2n\cdot q}B\not\!n\not\!q\right\rbrack^{-1}\;\; .
\label{quarkaxial}
\end{equation}
One look at this expression tells us that things
are not as simple as they were in the covariant gauges.
We cannot renormalize the quark fields with a simple constant.
The form of this full propagator implies that our renormalized
fields must be related to the unrenormalized ones
through a rotation in Dirac space;  there is no other
way to remove the $\not\!n$.  Rewriting
(\ref{quarkaxial}) as
\begin{eqnarray}
{1\over1-A}&&\left\lbrack1+{1\over2n\cdot q}
\left(\sqrt{1-B(1-A)^{-1}}-1\right)\not\!q\not\!n\right\rbrack^{-1}\nonumber\\
\label{fullqax}
&&\phantom{\Big\lbrack1+{1\over2n\cdot q}
\Big()\sqrt{1}}\times\;{i\over\not\!q}\;\times\\
&&\left\lbrack1+{1\over2n\cdot q}
\left(\sqrt{1- B(1-A)^{-1}}-1\right)
\not\!n\not\!q\right\rbrack^{-1}\;\; ,\nonumber
\end{eqnarray}
it is obvious that the renormalized fields are 
defined via
\begin{equation}
\psi=\sqrt{1-A}\left\lbrack1+{1\over2n\cdot q}
\left(\sqrt{1- B(1-A)^{-1}}-1\right)\not\!q\not\!n\right\rbrack\psi_b\;\; .
\label{qreninax}
\end{equation}
This transformation is certainly not a constant multiplicative factor.
Its nontrivial dependence on $n\cdot q$ implies that the 
manipulations used in 
Section \ref{renincov} to define renormalized perturbation 
theory will lead to an infinite series of local operators, each of
which renormalizes in its own special way.  Hence
our theory appears unrenormalizable.

What happened?  In covariant gauges, QCD is a perfectly reasonable 
renormalizable field theory.  Since the gauge choice we make is
in principle arbitrary, there should be no real 
difference between the two choices; we certainly should not
have the situation that QCD is renormalizable with one choice
but not the other!  However, we must be much more careful
when discussing renormalizability.  Taken by itself with no gauge-fixing
and no regularization procedure, QCD is a nonsensical field theory.
It is plagued by infinities of many different kinds and 
presents no way for us to extract sensible predictions
from its amplitudes.  Only after we decide on a regularization and
gauge-fixing procedure will QCD mean something.  For this reason,
we certainly cannot expect QCD to be renormalizable for any and every choice
of ultraviolet cut-off and gauge-fixing function.  Renormalizability
is as much a property of these choices as it is of the lagrangian.
As mentioned above, QCD is not renormalizable even in
covariant gauges if the ultraviolet cut-off breaks 
Lorentz invariance.  In the present case, this symmetry is
broken explicitly by the gauge-fixing procedure.  However,
since here it is only the gauge choice that breaks the 
symmetry, we can handle these difficulties
and define a sensible quantum field theory in 
the axial gauge.\footnote{This is valid as long as we choose
an ultraviolet regulator which does not further mutilate
either Lorentz symmetry or gauge invariance.  We will
employ dimensional regularization for this purpose.}

To see how this is done, recall that we are not forced
to use renormalized perturbation theory.  In Section
\ref{renincov} we saw that we could also use unrenormalized 
perturbation theory.  The results are the same as long as
we express the final result in terms of renormalized 
quantities.  In particular, all {\it external} fields in the 
amplitude must be renormalized.  
Writing the full quark propagator as
\begin{equation}
V^{-1}{i\over\not\!q}\gamma^0\left(V^{-1}\right)^\dag\gamma^0\;\; ,
\label{maggie}
\end{equation}
with
\begin{equation}
V\equiv\sqrt{1-A}\left\lbrack1+{1\over2n\cdot q}
\left(\sqrt{1- B(1-A)^{-1}}-1\right)\not\!q\not\!n\right\rbrack\;\; ,
\end{equation}
we see that a properly normalized amputated Greens function
is obtained by simply removing $V^{-1}i/\!\!\!\not\!\!q$ from the
full propagator of all outgoing quark lines
and $i/\!\!\!\not\!\!q\;\gamma^0(V^{-1})^\dag\gamma^0$ from the 
full propagator of all incoming quark lines.  All remaining
divergences\footnote{after a similar manipulation is performed
for all external gluon lines, of course} are 
attributed to the operators themselves.

Now that we understand the basic idea behind
renormalization in the axial gauge,
let us calculate the all-important quantity
$V$.  Since $A$ and $B$ start at ${\cal O}(\alpha_s)$,
a one-loop calculation sees only their leading powers
in $V$ :
\begin{equation}
V\sim\left(1-{1\over2}A-{1\over4n\cdot q}B\not\!q\not\!n\right)\;\; .
\end{equation}
The calculation of $A$ and $B$ is complicated by the presence of 
$1/k\cdot n$ in the gluon propagator.  This term renders the 
integration techniques of Appendix \ref{covint} impotent
since it is not quadratic in loop momenta.\footnote{A variation
of the Feynman parameter approach in Appendix \ref{covint}
based on the fact that
\begin{equation}
{1\over AB}=\int_0^\infty{dz\over\left\lbrack A+zB\right\rbrack^2}\;\; .
\end{equation}
can be employed in this situation.
Obviously, generalizations of this relation can be used to 
combine any number of normal propagators and `light-cone' 
denominators $1/k\cdot n$.  However, the introduction 
of a new Feynman parameter here leads to complications we would rather
avoid.  The techniques introduced below are much better suited 
for our purpose.}  Alternatively, using our special collinear kinematics,
we can decompose the $d$-dimensional integration as\footnote{
Our choice of $p$ and $n$ gives no change in measure.}
\begin{equation}
d^{\,d}k=d(n\cdot k)d(p\cdot k)d^{\,d-2}k_\perp\;\; .
\end{equation}
Since the denominator contains two powers of $p\cdot k$,
this part of the integration may be done by contour\footnote{
The application of contour integration
in this context is spelled out more clearly in 
Sections \ref{pertdis} and \ref{pertdvcs}.}
if no powers
of $p\cdot k$ appear in the numerator. 
Powers that do appear may be removed via the replacement
\begin{equation}
2p\cdot k={n\cdot(q-k)\over(n\cdot q)^2}q^2\;\; ,
\end{equation}
which is valid in the numerator.\footnote{\label{zeromass.}In deriving this
result, I have used
\begin{equation}
\label{onlypv}
\int d(n\cdot k)d(p\cdot k)
f(n\cdot q,n\cdot k){d^{\,d-2}k_\perp\over(2\pi)^{d}}{1\over k^2}
=\int d(n\cdot k)d(p\cdot k)
f(n\cdot q,n\cdot k)
{d^{\,d-2}k_\perp\over(2\pi)^{d}}{1\over (k-q)^2}
=0
\end{equation}
by absence of scale.  This is true regardless of the form of $f$ since
both $n\cdot q$ and $n\cdot k$ are dimensionless.
The validity of this replacement also rests on the fact that 
$q$ has no transverse component.  This is one place where 
complications due to general kinematics rear their ugly heads.}
In this manner, we obtain the result
\begin{eqnarray}
\Sigma&=&-{\alpha_sC_F\over2\pi}\Gamma\left({\epsilon\over2}\right)
\left({-q^2\over4\pi\mu_0^2}\right)^{-\epsilon/2}\int_0^x\,{dy\over x}\left[{y\over x}
\left(1-{y\over x}\right)\right]^{-\epsilon/2}\nonumber\\
&&\qquad\qquad\qquad\qquad\times\;\left\lbrace 
\left(1-{\epsilon\over2}\right)\left(1-{y\over x}\right)\not\!q
+{2 y\over x-y}{q^2\over2x}\not\!n\right\rbrace\;\; ,
\label{sigax}
\end{eqnarray}
where $x\equiv n\cdot q$.
The ultraviolet divergence is contained in the factor $\Gamma(\epsilon/2)$
multiplying our expression.  

In analogy with Section \ref{renincov}, 
we would like 
to include only the ultraviolet divergent part of $\Sigma$
in the definition of $V$.  This procedure leads to an extra divergence
of the form\footnote{For finite
$\epsilon$, this integral is quite finite.  Looking at 
Eq.(\ref{sigax}), we see that the divergence appears as
$\Gamma(-\epsilon/2)$.  However, consistency requires that
only {\it ultraviolet} divergences be included in $V$.
A well-defined separation between infrared
and ultraviolet physics can be obtained only if 
we use different techniques to regularize the two 
types of divergences.}
\begin{equation}
\int_0^x\,{dy\over x}{2x\over x-y}
\label{lcdiv}
\end{equation}
in the second term.  Tracing it back through the calculation,
we see that our new divergence comes directly from the 
factor $1/k\cdot n$ in the gluon propagator.
This behavior is reminiscent of the 
divergence we encountered in Section \ref{renincov}
when we were working in the onshell scheme. 
There, the divergence appeared because of our
choice of external momenta.  Once we decided that
we did not wish to renormalize in the onshell
scheme, it no longer posed a problem for us.
On the other hand, this divergence is present 
regardless of which external states we choose.
Deforming $n^2$ away from zero,
we see that instead of a divergence we obtain terms like
$\log^2(n^2q^2/(q\cdot n)^2)$ in the 
{\it finite} part of our amplitude.  The part of our
amplitude that generates these light-cone
singularities does not
even {\it contribute} to the ultraviolet divergence
for $n^2$ finite.  

To gain insight into the 
meaning of these divergences, it is necessary
to take a closer look at the calculation.
A characteristic integral that leads to divergences
of this sort is
\begin{equation}
\int{d^{\,d}k\over(2\pi)^d}\,{1\over(n\cdot k-\omega)\,k^2\,(q-k)^2}\;\; .
\label{primeint}
\end{equation}
A simple power-counting argument implies immediately
that this integral is not ultraviolet divergent.
However, performing the $k^-$ integral by contour, 
we immediately arrive at the opposite conclusion.
This fact is extremely disconcerting.  Renormalization
proofs rely heavily on power-counting, so the ultraviolet
divergence contained in this one integral has the 
power to destroy the renormalizability of our theory.
Actually performing this integral away from the singular
points in the complex plane of $\omega$ reveals that
it not only diverges, but produces a divergence
which is dependent on {\it logarithms} of the external 
momenta.  Since $\log(x)$ is not analytic at $x=0$,
these divergences cannot even be absorbed by local operators.
If we intend to take this theory seriously,
our lagrangian requires counterterm operators of the form
$\overline\psi\;\log(in\cdot\partial)\,\psi$.
Obviously, this is quite unacceptable.

In order to understand why the integral (\ref{primeint})
does not follow the rules of power-counting, we
must remember how those rules are derived.
Since the square of a Euclidean vector 
is positive-definite, power-counting arguments 
are always valid in Euclidean space.  In order
to use Euclidean arguments on a Minkowskii integral,
our integrand must allow us to perform a rotation
through $\pi/2$ radians in the complex energy plane.
The causal 
prescription for the physical propagator 
poles introduced in Section \ref{roadtoquant}
satisfies this requirement.\footnote{This prescription
forces negative-energy poles to be above the real axis
and positive-energy poles to be below.  Hence the axis can 
be rotated freely in the ultraviolet region 
without encountering any poles.} 
We have not explicitly
chosen a prescription for the pole in the gluon propagator, 
but our manipulations imply that our prescription 
does not depend on any of the other coordinates.
Since $n\cdot k$ is linear in $k^0$, our poles occur
either always above or always below the axis.  No prescription
of this form admits Wick rotations, so we cannot apply 
power-counting techniques.

As mentioned in Section \ref{roadtoquant},
there is a causal prescription for our 
new poles.  Since we intend to perform
our integrals in light-cone coordinates, 
we prefer the Mandelstam-Liebbrandt (ML) prescription \cite{MLpres}
\begin{equation}
{1\over n\cdot k}\rightarrow{p\cdot k\over (p\cdot k)
(n\cdot k)+i\varepsilon}
={2p\cdot k\over k^2+k_\perp^2+i\varepsilon}
\label{mandel}
\end{equation}
over the one involving $\vec k\cdot\vec n$.  
The form of this prescription makes our earlier
mathematical indiscretion apparant.  We were considering
contour integration in the $k^-$ plane, while the poles
in the $k^+$ plane are the ones that truly dictate
the analytic behavior of our integral.  The fact that
we ignored the analytic structure of this pole
before {\it induced} the non-causal principle value (PV) 
prescription\footnote{I should mention here
that we intend to use this prescription
in the most naieve way possible. 
The subtleties of defining products and 
powers of this distribution are considered in \cite{axial}.}
\begin{equation}
{1\over n\cdot k}\rightarrow
\lim_{\delta\rightarrow0}{1\over2}\left({1\over
n\cdot k+i\delta}+{1\over n\cdot k-i\delta}\right)
=\lim_{\delta\rightarrow0}
{n\cdot k\over (n\cdot k)^2+\delta^2}\;\; .
\label{pvpres}
\end{equation}
While this prescription certainly regularizes the 
light-cone divergence, it is not at all 
obvious that it will lead to a renormalizable
quantum field theory.

To illustrate the difference between these two
prescriptions, it is instructive to see how our
calculation works in each.  Since we already have
an expression for the quark self-energy with the 
principle value prescription, we begin with this
choice.  Defining
\begin{equation}
I_0\equiv\lim_{\delta\rightarrow0}
\int_0^1{xdx\over x^2+\delta^2}\sim\int_0^1{dx\over x}\;\; ,
\label{I0def}
\end{equation}
we can extract the ultraviolet divergent parts of
Eq.(\ref{sigax}) :
\begin{eqnarray}
\left[A(x)\right]_{\rm
UV}&=&-{\alpha_sC_F\over4\pi}{2\over\epsilon}\left(\mu^2e^{\gamma_{E}}
\over4\pi\mu_0^2\right)^{-\epsilon/2}\nonumber\\
\left[B(x)\right]_{\rm
UV}&=&-{\alpha_sC_F\over4\pi}{2\over\epsilon}\left(\mu^2e^{\gamma_{E}}
\over4\pi\mu_0^2\right)^{-\epsilon/2}\left(4I_0+4\log x-4\right)\;\; .
\end{eqnarray}
As promised, a dependence on the log of $n\cdot q$ 
has been generated.

In the gluon sector, the situation is analogous.  The fact that 
the axial gauge is `physical' implies that the gluon self-energy
must satisfy current conservation without any help from ghosts.
The most general form for the gluon 1PI self-energy
with our choice of kinematics is
\begin{eqnarray}
\Pi^{\mu\nu}(q)&=&\Pi_1\left(q^\mu q^\nu-q^2g^{\mu\nu}\right)\nonumber\\
&&+\Pi_2\left(q-{q^2\over q\cdot n}\,n\right)^\mu
\left(q-{q^2\over q\cdot n}\,n\right)^\nu\;\; .
\end{eqnarray}
The full gluon propagator, 
\begin{eqnarray}
D^{\mu\nu}_{a\,b}(q)={\delta^{\,a\,b}\over 1-\Pi_1}&&
\left[\delta^\mu_\alpha+\left(\sqrt{1+{\Pi_2\over1-\Pi_1-\Pi_2}}-1\right)
{1\over n\cdot q}n^\mu q_\alpha\right]\nonumber\\
\label{fullgax}
&&\phantom{[\delta^\nu+()}
\times\;{id^{\alpha\beta}(q)\over q^2}\;\times\\
&&\left[\delta^\nu_\beta+\left(\sqrt{1+{\Pi_2\over1-\Pi_1-\Pi_2}}-1\right)
{1\over n\cdot q}q_\beta n^\nu \right]\;\; ,\nonumber
\end{eqnarray}
where $d^{\mu\nu}(q)\equiv-g^{\mu\nu}+(q^\mu n^\nu+q^\nu n^\mu)/q\cdot n$,
is obtained by summing the 1PR contributions.
Inverting the matrix on either side, we see that the 
renormalized gluon field is defined via
\begin{equation}
{\cal A}^\mu=\sqrt{1-\Pi_1}\left[\delta^\mu_\alpha+\left(\sqrt{
1-{\Pi_2\over1-\Pi_1}}-1\right)
{1\over n\cdot q}n^\mu q_\alpha\right]{\cal A}^\alpha_b\;\; .
\label{greninax}
\end{equation}

At the one-loop level, only two diagrams contribute to $\Pi$.
The quark loop is identical to its covariant gauge counterpart
and does not depend on our prescription for the light-cone
singularity.
The other diagram, the gluon loop, can be evaluated using the technique
outlined above.  The full result contains the UV divergence
\begin{eqnarray}
\left[\Pi_1(x)\right]_{\rm
UV}&=&{\alpha_sC_A\over4\pi}{2\over\epsilon}\left(\mu^2e^{\gamma_{E}}
\over4\pi\mu_0^2\right)^{-\epsilon/2}\left({11\over3}
-{4n_fT_F\over 3C_A}-4I_0-4\log x\right)\nonumber\\
\left[\Pi_2(x)\right]_{\rm
UV}&=&{\alpha_sC_A\over4\pi}{2\over\epsilon}\left(\mu^2e^{\gamma_{E}}
\over4\pi\mu_0^2\right)^{-\epsilon/2}\left(-4+4I_0+4\log x\right)\;\; .
\end{eqnarray}

The three-point function that defines the renormalized coupling
receives one-loop corrections from the same diagrams we considered
in covariant gauge.  The integrals are a little more complicated
than in the self-energies,
but can be taken readily
if we are interested only in the ultraviolet
divergent part.  Let us consider a longitudinally
polarized gluon.  These amplitudes are obtained by
contracting the gluon polarization index with $n$.\footnote{
For example, the leading three-point function for 
polarization $\alpha$ is $-igt^a\gamma^\alpha$.  For longitudinal
polarization, we write $-igt^a\!\not\!n$.}
The quark correction (Figure 1.7a) has the divergence
\begin{equation}
\left(-ig_bt^a\!\not\!n\right){\alpha_s\over4\pi}
\left(C_F-{C_A\over2}\right){2\over\epsilon}
\left(\mu^2e^{\gamma_{E}}\over4\pi\mu_0^2\right)^{-\epsilon/2}
\left(4I_0+2\log x_{i}+2\log x_f-3\right)\;\; ,
\end{equation}
while the gluon correction (Figure 1.7b) contains
\begin{equation}
\left(-ig_bt^a\!\not\!n\right){\alpha_sC_A\over8\pi}
{2\over\epsilon}
\left(\mu^2e^{\gamma_{E}}\over4\pi\mu_0^2\right)^{-\epsilon/2}
\left(1+{x_i+x_f\over x_f}\log{x_i\over x_g}
+{x_i+x_f\over x_i}\log{x_f\over x_g}\right)\;\; .
\end{equation}
Here, $x_i$, $x_f$, and $x_g$ are the light-cone momenta
$n\cdot q$ for the initial quark, final quark, and gluon,
respectively.  

The self-energy corrections are performed by tacking 
the full propagators (\ref{fullqax}) and (\ref{fullgax})
on to the leading order diagram, and
removing the middle propagator and the 
external $Z_F$-matrix to renormalize the external 
states.  The quark,
\begin{equation}
\left(-ig_bt^a\!\not\!n\right){\alpha_sC_F\over4\pi}
{2\over\epsilon}
\left(\mu^2e^{\gamma_{E}}\over4\pi\mu_0^2\right)^{-\epsilon/2}
\left(3-4I_0-2\log x_i-2\log x_f\right)\;\; ,
\end{equation}
and gluon,
\begin{equation}
\left(-ig_bt^a\!\not\!n\right){\alpha_sC_A\over8\pi}
{2\over\epsilon}
\left(\mu^2e^{\gamma_{E}}\over4\pi\mu_0^2\right)^{-\epsilon/2}
\left({11\over3}-4-{4n_fT_F\over3C_A}\right)\;\; ,
\end{equation}
combine with the vertex corrections to give the 
full one-loop vertex function,
\begin{eqnarray}
&&\left(-ig_bt^a\!\not\!n\right)\left\lbrack 1+{\alpha_s\over4\pi}
{2\over\epsilon}
\left(\mu^2e^{\gamma_{E}}\over4\pi\mu_0^2\right)^{-\epsilon/2}
\left({11\over6}\,C_A-{2n_fT_F\over3}\right.\right.\nonumber\\
&&\!\!\!\!\left.\left.+{C_A\over 2}
\left({x_i-x_f\over x_f}\log x_i+{x_f-x_i\over x_i}\log x_f
-{(x_i+x_f)^2\over x_i x_f}\log x_g-4I_0\right)\right)\right\rbrack\;\; .
\label{longvert}
\end{eqnarray}
If we simply ignore the dependence on $x_i,x_f$, and $x_g$
for the moment, 
this expression implies that the effective coupling at
scale $\mu$ is given by\footnote{Note the difference between
renormalized and unrenormalized perturbation theory.
In renormalized perturbation theory, we would have 
determined $Z_g$ from a subtraction of ultraviolet divergences.
Here, we evaluate the full contribution without a thought
given to renormalization (except for the external states).
The effective coupling is determined by the requirement
that our matrix element be finite when expressed in terms
of it.}
\begin{equation}
\label{ourbeta}
g(\mu^2)\equiv g_b\left\lbrack 1+{\alpha_s\over4\pi}
{2\over\epsilon}
\left(\mu^2e^{\gamma_{E}}\over4\pi\mu_0^2\right)^{-\epsilon/2}
\left({11\over6}\,C_A-{2n_fT_F\over3}\right)\right\rbrack\;\; ,
\end{equation}
which gives the $\beta$-function found in the last section.

The fact that the logarithms do not cancel implies
that this gauge choice with the noncausal prescription for the 
light-cone singularity is truly not renormalizable, as
predicted.\footnote{In the literature, one can find a 
calculation of the $\beta$-function in this 
gauge which does not contain these
difficulties \cite{cfp}.
Expressions for the momentum-dependent `$Z$-factors' from 
various diagrams are given, with the result that the 
unsightly terms cancel in the renormalized coupling.
Their calculation makes no mention of the gluon
polarization, and the wavefunction renormalization
factors they employ are appropriate for the 
product $\overline\psi\not\!n{\cal A}_\perp\psi$.  Since this
combination does not appear in the lagrangian, it is unclear how
to interpret their result.  In addition, several papers
present calculations of the QCD splitting functions
using PV \cite{Vogel}.  This prescription closely resembles
that used for eikonal propagation introduced in 
Section \ref{partondistdis}, so it may indeed
be appropriate for this purpose.  However, if a
renormalization procedure that produces
sensible results in {\it all} cases exists for
this prescription, it is certainly not straightforward.}
However, the specific form (\ref{pvpres}) of the prescription
is not actually used until the integrals are performed.  If we 
choose instead to allow the light-cone denominators to act only
{\it algebraically}, and simply ignore any terms of the form
$c/k\cdot n$, with $c$ independent of $k$, the 
embarrassing terms never appear.  Unfortunately, 
since the integrals in each diagram contain different 
limits, this `prescription' is hard to quantify mathematically.
In addition, it is entirely unclear how this procedure would
be implemented at higher orders.

The situation for transversely polarized gluons is 
analogous.  Here, even the Dirac structure is nontrivial.
The quark
\begin{equation}
\left(-ig_bt^a\gamma^\perp\right){\alpha_s\over4\pi}
\left(C_F-{C_A\over2}\right){2\over\epsilon}
\left(\mu^2e^{\gamma_{E}}\over4\pi\mu_0^2\right)^{-\epsilon/2}
\left(-1-{x_f\over x_g}\log{x_i\over x_f}\not\!n\not\!p
+{x_i\over x_g}\log{x_f\over x_i}\not\!p\not\!n\right)
\end{equation}
and gluon
\begin{equation}
\left(-ig_bt^a\gamma^\perp\right){\alpha_sC_A\over8\pi}
{2\over\epsilon}
\left(\mu^2e^{\gamma_{E}}\over4\pi\mu_0^2\right)^{-\epsilon/2}
\left(-1+4I_0+4\log x_g
-\log{x_i\over x_f}\not\!n\not\!p-\log{x_f\over x_i}\not\!p\not\!n 
\right)\;\; 
\end{equation}
vertex diagrams combine with the wavefunction renormalization
factors to generate the full vertex
\begin{eqnarray}
(-igt^a\gamma^\perp){\alpha_s\over4\pi}{2\over\epsilon}
\left({\mu^2e^{\gamma_E}\over 4\pi\mu_0^2}\right)^{-\epsilon/2}
&&\!\!\!\!\!\!\left\lbrack
{11\over6} C_A-{2\over3}n_fT_F+{C_A\over 2}\left(
{1\over x_g}\log{x_i\over x_f}\,\not\!n\not\!p_i
-{1\over x_g}\log{x_f\over x_i}\,\not\!p_f\not\!n\right)\right.\nonumber\\
\label{perpvert}
&&+C_F\left({1\over x_g}\log x_f\,\not\!n\not\!p_i
-{1\over x_g}\log x_i\,\not\!p_f\not\!n\right)\\
&&\left.-C_F\left((I_0-{x_i\over x_g}\log x_f)\,\not\!p\not\!n+(I_0
+{x_f\over x_i}\log x_i)\not\!n\not\!p\right)\right\rbrack\;\; .\nonumber
\end{eqnarray}
Once again, we see that the correct one-loop $\beta$-function is 
obtained if the offensive contributions are simply ignored.

Before moving on to the Mandelstam-Liebbrandt prescription,
it is instructive for us to study the way in which the offensive
terms pollute our final results.  In the longitudinal
direction, (\ref{longvert}) shows that the 
unsightly logarithms confine themselves to the 
$C_A$-part of the coupling.  Tacking external states\footnote{
This procedure is somewhat delicate since truly onshell
external states will be annihilated by $\not\!n\not\!p$ whether
or not it is accompanied by $x_i$.  The ambiguity comes
directly from our choice of collinear kinematics.
To resolve it, we could choose to add a small transverse momentum
component to each of our external legs; since this 
pole prescription is flawed in any case, we will not bother.}
onto
(\ref{perpvert}) causes all but the $C_F$-part of these terms to 
vanish.  This fact is intimately related to the gauge choice 
we have made and its implications on the physical states of
our theory.  

Quantizing our theory at equal values of
$n\cdot x\equiv x^+$,\footnote{See Appendix D.1.}
rather than at equal times, we see that
the QCD equations of motion\footnote{See Appendix D.2}
in the gauge $n\cdot{\cal A}=0$ reveal a severe pathology with
our formalism : they do {\it not} dictate the $x^+$ (`time') 
evolution of either ${\cal A}^-$ or  
$\psi_-\equiv{1/2}\not\!n\!\not\!p\,\psi$.
This means that these field operators do {\it not}
propagate freely in this gauge.  Only the transverse
gluon field components and the physical `good' quark field 
$\psi_+\equiv1/2\not\!p\!\not\!n\,\psi$ 
can be directly interpreted
in the perturbative theory we have derived.  The other
components must be re-expressed in terms of the 
freely-propagating ones through their equations 
of motion before they can be considered in our formalism.
In light of this observation, it is not surprising that 
the longitudinal contribution, which is associated with 
${\cal A}^-$, is polluted by $C_A$-type gluon effects and 
the transverse contribution, which contains $\psi_-$, is polluted
by $C_F$-type quark effects.  In some sense, the fact that 
these field components do not propagate freely makes our renormalization
of the external states insufficient to remove all 
associated divergences.  Notice that the residual 
divergence structures depend on {\it all three} momenta,
so the renormalization of these `bad' field components 
actually depends on the diagram they are embedded in, further evidence
of their composite nature.

Now that we have seen the consequences of a noncausal
prescription, let us see what happens when we use
one that is physically motivated.  The most important
practical difference between the Mandelstam-Liebbrandt
prescription (\ref{mandel}) and the PV prescription is 
that ML introduces another vector to the problem.
This effectively adds a new scale and 
invalidates arguments such as that in Footnote \ref{zeromass.}.
However, this added complication is more than compensated by
the fact that ML preserves power counting.\footnote{I should mention
here that the power-counting techniques of covariant calculations
must be slightly modified to obtain the correct results.
Each of our $d$-dimensions can conceivably generate an 
ultraviolet divergence, so we must power count 
each dimension separately.  The integral
\begin{equation}
\int{d^{\,d}k\over(2\pi)^d}\,{1\over (q-k)^2}\,{1\over
(n\cdot k)(n\cdot (k-\ell))(n\cdot(q-k))}
\end{equation}
is divergent even in spite of the fact that its
overall degree of divergence is negative.}
Integrals of the form
\begin{equation}
\int{d^{\,d}k\over(2\pi)^d}\,{1\over n\cdot k (k-q)^2}
\end{equation}
contribute to ultraviolet divergences, but those of the 
form (\ref{primeint}) do not.  This leads to a tremendous simplification.
In particular, the integrals which generate the offensive
contributions in the PV prescription are automatically 
thrown away by ML.

The calculation is straightforward.  One simplifies the 
expression of each diagram, taking care to retain only those
terms which contribute to the UV divergence, and performs
the $k^+$ integration by contour.  The transverse integrations
are now trivial, and one is left with a simple one-dimensional
integral for $k^-$.  With the same conventions as above,
the self-energy diagrams contribute
\begin{eqnarray}
\left[A(x)\right]_{\rm
UV}&=&-{\alpha_sC_F\over4\pi}{2\over\epsilon}\left(\mu^2e^{\gamma_{E}}
\over4\pi\mu_0^2\right)^{-\epsilon/2}\left(1\right)\nonumber\\
\left[B(x)\right]_{\rm
UV}&=&-{\alpha_sC_F\over4\pi}{2\over\epsilon}\left(\mu^2e^{\gamma_{E}}
\over4\pi\mu_0^2\right)^{-\epsilon/2}\left(-4\right)\;\; ;\\
\left[\Pi_1(x)\right]_{\rm
UV}&=&\phantom{-}{\alpha_sC_A\over4\pi}
{2\over\epsilon}\left(\mu^2e^{\gamma_{E}}
\over4\pi\mu_0^2\right)^{-\epsilon/2}\left({11\over3}
-{4n_fT_F\over 3C_A}\right)\nonumber\\
\left[\Pi_2(x)\right]_{\rm
UV}&=&\phantom{-}
{\alpha_sC_A\over4\pi}{2\over\epsilon}\left(\mu^2e^{\gamma_{E}}
\over4\pi\mu_0^2\right)^{-\epsilon/2}\left(-4\right)\;\; ,
\end{eqnarray}
which is certainly quite a bit cleaner than the PV results!
It is interesting to note that the two prescriptions agree
if we ignore the logarithm-type terms.  This behavior
is also seen in the three-point diagrams,
Figs. 1.7(a) and (b) in the longitudinal,
\begin{eqnarray}
\left(-ig_bt^a\!\not\!n\,\right){\alpha_s\over4\pi}
{2\over\epsilon}
\left(\mu^2e^{\gamma_{E}}\over4\pi\mu_0^2\right)^{-\epsilon/2}
\left(C_F-{C_A\over2}\right)
\left(-3\right)\;\; ,\\
\left(-ig_bt^a\!\not\!n\,\right){\alpha_s\over4\pi}
{2\over\epsilon}
\left(\mu^2e^{\gamma_{E}}\over4\pi\mu_0^2\right)^{-\epsilon/2}
\left(\phantom{C_F-}\,\;{C_A\over2}\right)
\left(\phantom{-}1\right)\;\; ,
\end{eqnarray}
and transverse,
\begin{eqnarray}
\left(-ig_bt^a\gamma^\perp\right){\alpha_s\over4\pi}
{2\over\epsilon}
\left(\mu^2e^{\gamma_{E}}\over4\pi\mu_0^2\right)^{-\epsilon/2}
\left(C_F-{C_A\over2}\right)\left(-1\right)\;\; ,\\
\left(-ig_bt^a\gamma^\perp\right){\alpha_s\over4\pi}
{2\over\epsilon}
\left(\mu^2e^{\gamma_{E}}\over4\pi\mu_0^2\right)^{-\epsilon/2}
\left(\phantom{C_F-}\,\;{C_A\over2}\right)\left(-1\right)\;\; ,
\end{eqnarray}
sectors, respectively.  From these results, we easily
arrive at the correct $\beta$-function, (\ref{lobeta}),
pollutant-free.  

In order to understand why ML was able to circumvent the 
problems associated with renormalizing the external
composite `bad' fields and generate the correct $\beta$-function,
we appeal to some of the arguments in the beginning
of this section.  Since the violence to Lorentz symmetry
(which leads directly to the fact that different 
components of the fields behave in different ways) 
occurs only in the choice of gauge, it is 
unphysical.  Its effects will cancel 
in the calculation of gauge-invariant
quantities, giving the same result for all well-defined
gauge choices.  As we have seen, the contribution to the
three-point functionfrom each diagram depends on the 
gluon polarization.  The PV prescription
does not exhibit a cancelation
of this dependence, leading to a directionally-dependent
`renormalized coupling' in direct violation of the 
gauge symmetry.  In ML, the observed cancelation
allows us to define a sensible, directionally-independent
renormalized coupling.  This is direct evidence 
that the causal prescription does not do further 
damage to our theory and invalidate our gauge principle.
Here, the explicit breaking of Lorentz symmetry 
is indeed artificial and constrained to cancel
in a sum over all contributions.

The ML prescription contains one loose end that should be 
taken care of before we conclude this section.
When calculating the self-energy diagrams above,
we implicitly assumed that they have the same form
in both prescriptions.  In light of (\ref{qreninax}) and (\ref{greninax}),
this assumption still leads to renormalization
`constants' which depend on the field momenta.
Since we intend to take ML seriously,
we cannot be satisfied with this behavior.  Looking at these
expressions again, we see that for the collinear 
kinematics we used the replacement $q^\mu/n\cdot q\rightarrow p^\mu$
is valid.  As ML introduces $p^\mu$ {\it independently} of
the external momenta, dependence on $p^\mu$ is perfectly 
acceptable.\footnote{Note that this is {\it not} the case
for PV.  There, dependence on the external momenta is 
unavoidable - even in the absence of the logarithms.}  To test
the validity of this replacement unambiguously, it is necessary
to consider transverse momenta.  We do this explicitly
in the quark sector.

Consider the complete 1PI quark self-energy in ML.  
Its most general form\footnote{Since QCD conserves
parity and the self-energy is hermitian
in the sense that $\Sigma^\dag=\gamma^0\Sigma\gamma^0$, 
terms involving products of two or more $\gamma$-matrices
either vanish or reduce to these structures.} is
\begin{equation}
\Sigma(q)=A\not\!q+B\,p\cdot q\not\!n+C\,n\cdot q\not\!p+Dm_b\;\; .
\end{equation}
For completeness, we work with massive quarks.  Iterating this result
as above, one obtains the full propagator
\begin{equation}
{i\over\not\!q-m_b-\Sigma}\;\; .
\end{equation}
We would like to decompose this into a form reminiscent
of (\ref{maggie}), but with the added requirement that
the matrix $V$ be independent of $q$.  Working through this,
we arrive at the full propagator
\begin{equation}
{1-A-B\over(1-A)^2}\left\lbrack
1-{B\over2(1-A)}\not\!p\not\!n\right\rbrack^{-1}
{i\over\not\!q-m}\left\lbrack
1-{B\over2(1-A)}\not\!n\not\!p\right\rbrack^{-1}\;\; ,
\end{equation}
where the renormalized mass is given by
\begin{equation}
m={1+D\over 1-A}\,m_b\;\; ,
\end{equation}
along with the consistency relation
\begin{equation}
C(1-A-B)+B(1-A)=0\;\; .
\label{consistency}
\end{equation}
This implies that our renormalized fields are given by
\begin{equation}
\psi={1-A\over\sqrt{1-A-B}}\left\lbrack
1-{B\over2(1-A)}\not\!p\not\!n\right\rbrack\psi_b\;\; .
\label{qrenML}
\end{equation}

If $A, B, C$, and $D$ are all independent of $q$
and satisfy (\ref{consistency}), we are free to define
proper renormalized fields in the spirit of Section 1.4
and arrive at a sensible, consistent theory.
At one-loop order, we can calculate these 
quantities explicity; with the common overall factor
\begin{equation}
{\alpha_sC_F\over4\pi}\,{2\over\epsilon}\left({\mu^2e^{\gamma_E}\over
4\pi\mu_0^2}\right)^{-\epsilon/2}\;\; ,
\end{equation}
we have
\begin{eqnarray}
A&=&\phantom{-}1\\
B&=&\phantom{-}2\\
C&=&-2\\
D&=&\phantom{-}2\;\; .
\end{eqnarray}
Note that (\ref{consistency}) is satisfied to the 
order at which we work and the correct gauge-invariant
dependence of $m(\mu)$ on $\mu$,
\begin{equation}
\mu{d\over d\mu} m(\mu)
=-3{\alpha_sC_F\over2\pi}\,m(\mu)\;\; ,
\end{equation}
is reproduced.  

In the collinear limit, $n^\mu$, $k^\mu$, and $p^\mu$
are no longer linearly independent.  Here, we can calculate
only the combined quantities $A_0\equiv A+C$ and $B_0\equiv B-C$.
Fortunately, our consistency relation (\ref{consistency})
allows us to extract $A, B$, and $C$ from these two:
\begin{eqnarray}
A&=&1-\sqrt{1-(2A_0+B_0)+A_0(A_0+B_0)}\;\; ,\\
B&=&{1-A_0-B_0\over 2-A_0-B_0}\,B_0\\
C&=&-{1-A\over2-A_0-B_0}\,B_0\;\; .
\end{eqnarray}
Using these expressions, we see that (\ref{qreninax}) 
is just a re-expression
of (\ref{qrenML}) in terms of $A_0$ and $B_0$.
In this way, we can calculate the correct renormalization
constants using collinear kinematics.

In the gluon sector of the theory, things are similar. 
Here, since there is no mass, we have no need to calculate
explicitly the analogues of $A, B$, and $C$.  The collinear
expressions $\Pi_1$ and $\Pi_2$ presented above
are quite sufficient for all of our calculations.

It is interesting to note that the entire
$\beta$-function is contained in the factor 
$\Pi_1$ of the gluon self-energy.  In our calculation
of the three-point function, all of the other
contributions conspire to cancel 
each other.  This is a general property
of the axial gauges at one-loop order, and 
provides an explicit example of how 
changes in gauge can shift contributions
from graph to graph without changing the 
final result.

In this section, we explored a non-covariant 
gauge choice.  
We have seen how subtle differences in prescription
can make or break renormalizability in 
a gauge which explicitly violates Lorentz symmetry.
If we had broken
the spacetime symmetry more explicitly, 
summing all contributions would not have been enough
to restore its former health.  This is what happens 
when we regularize divergences by a momentum cut-off, 
for example.  There, special counterterms must be added to the 
lagrangian to forcibly restore the symmetry order by order
in perturbation theory.

The axial gauge's problems
are not insurmountable.  As we have shown, 
the theory possesses well-defined matrix 
elements which produce the correct behavior
if implemented properly.  Though some of its aspects
are more involved than those of  
its covariant brother, 
this gauge is not without its charms.
Among other things, it restores a certain degree of 
physical intuition to QCD.  This gauge also
represents tremendous simplifications in some
applications, most notably those involving extremely
high energy processes.  For these reasons, it
is often employed despite its difficulties.

\chapter{Deep-Inelastic Scattering}
\label{dis}

In the last chapter, we saw that asymptotic freedom assures
us QCD perturbation theory is applicable to processes
whose natural scales are large compared to the
typical energy at which the coupling gets large. 
Unfortunately, confinement implies that {\it all}
physical processes are inextricably linked to the 
low energy phenomena that lead to bound states.
In order to calculate even the simplest
of observables, we must first understand the 
complicated QCD bound states that appear in
our experiments.  Since perturbation theory cannot
be applied to these states, the technology developed in the
last chapter is useless for this purpose.  In the absence
of nonperturbative techniques for solving interacting
quantum field theories, we must throw our hands in the
air and give up the task of calculating {\it any}
strong interaction cross-section from scratch.

Although we cannot calculate truly physical 
quantities completely in QCD, there is hope that
we can use experimental results to learn 
about these nonperturbative quantities.
First advanced by Feynman in 1969 \cite{parton},
this idea is based on a separation of scales
in certain processes.  If the high- and low-energy
contributions to a process can systematically be separated,
one can calculate the high-energy physics 
in perturbation theory and use experimental data
to {\it extract} the nonperturbative physics.  
These nonperturbative quantities
can then be used not only to predict the outcome
of other similar experiments, but also to teach us
about the structure of QCD's bound states.

The purpose of this chapter is to describe the
scale-separation process
in the context of inclusive deep-inelastic scattering (DIS).
The formalism developed here can be applied to a
number of physical processes, and represents
the main way QCD is used to predict the outcome
of experiments and give us insight into hadronic
structure.

The process of DIS, along with the relevant
structure functions and kinematics, is
described in Section \ref{diskin}.  It is 
here that we will discuss what is measured in the 
experiment and how the measurements are
related to hadronic physics.

In Section \ref{pertdis}, a qualitative argument
is used to motivate the separation of scales.
We use this argument to perform perturbative
calculations of the hard scattering amplitude
and find that it contains infrared 
divergences.  These divergences signal
soft nonperturbative physics that cannot 
be calculated reliably within perturbation theory.

Section \ref{partondistdis} explains
that these divergences represent contributions
from nonperturbative distribution 
functions.  We carefully define these distributions
and show that our amplitudes are perfectly finite
when expressed in terms of them.  
A measurement of the cross-section, coupled
with knowledge of the perturbative coefficients, 
will now allow us to extract these nonperturbative
quantities.  Some of the structural 
information contained in these distributions
is discussed, as well as certain 
consistency relations they are required to satisfy.

The analysis of Section \ref{factdis}
is devoted to infrared divergences 
and their relation to parton distribution
functions.  It allows us to show that
the soft, nonperturbative contributions
to inclusive DIS can be attributed exclusively
to the parton distribution functions of Section \ref{partondistdis}
to {\it all} orders of QCD perturbation theory.
Such a statement is called a {\it factorization theorem}.
The tools we will use to derive this result were originally
developed by G. Sterman \cite{Stermanpower}, 
S. Libby \cite{Libby}, 
and J. Collins \cite{Collins}
using the results of L. Landau \cite{Landaueqns}.

Since the perturbative physics in our process
is related only to the microscopic degrees of freedom of
QCD, quarks and gluons, all target dependence
is isolated in the nonperturbative distribution functions.
Section \ref{opedis} shows how to use
this fact to arrive at an operator relation between
the nonlocal product of two 
electromagnetic currents and an 
infinite series of local partonic operators.
Such local operator product expansions (OPE's)
were first considered by Wilson in 1969 \cite{ope}.
Since then, they have been used extensively to
study the analytic structure of 
amplitudes and characterize scattering mechanisms.
Using the results of Section \ref{partondistdis},
we derive the next-to-leading order coefficients of
the OPE and discuss the interpretation of the local operators.

Section \ref{sumdis} contains a summary and 
some concluding remarks about inclusive DIS.

\section{Kinematics}
\label{diskin}

Deep-inelastic scattering (DIS) is a process in which a lepton 
is scattered off of a hadron at very high energy.
The final state of this process will be a backscattered
lepton and a complicated hadronic state which we denote
as $|X\rangle$.
We consider an electron scattering via
the exchange of a highly virtual photon off of a proton, but
the theoretical tools developed here also apply to weak exchanges 
and processes involving other hadronic states.
At leading order in the electromagnetic coupling,
the amplitude for this process is given by
\begin{equation}
\left\langle k's', X\left|{\rm T}\left((-i)\int d^{\,4}x\,J^\mu_e(x)A_\mu(x)
(-i)\int d^{\,4}y\,J^\nu_q(y)A_\nu(y)\right)\right|ks, PS\right\rangle\;\; ,
\end{equation}
where I have denoted the electron's initial (final) momentum and spin 
as $k(k')$ and $s(s')$, respectively, and the proton's momentum
and spin by $P$ and $S$, respectively.  The currents
$J^\mu_e=e\,{\overline\psi}_e\gamma^\mu\psi_e$ and 
$J^\mu_q=\sum_fe_f{\overline\psi}_f\gamma^\mu\psi_f$ are the 
usual electron and quark electromagnetic currents, respectively.
Recognizing that there are no photons 
in either external state\footnote{at leading order in the
electromagnetic coupling}
and that the two currents are fundamentally different 
allows us to separate the matrix elements and arrive finally
at the amplitude\footnote{I have used the operator
identity ${\cal O}(x)=e^{i\hat{P}\cdot x}{\cal O}(0)e^{-i\hat{P}\cdot x}$,
where $\hat P$ is the momentum operator introduced in Appendix \ref{unitrepso31}.}
\begin{equation}
{i\over (k'-k)^2}\left\langle k's'\left|J^\mu_e(0)\right|ks\right\rangle
\left\langle X\left|J^\nu_q(0)\right|PS\right\rangle g_{\mu\nu}
(2\pi)^4\delta(p_X+k'-P-k)\;\; ,
\end{equation}
where $p_X$ represents the 4-momentum of the state $|X\rangle$
and I have employed Feynman's gauge ($\xi=1$).

This is an extremely complicated object which is 
different for each possible final hadronic state.
General processes such as this one are difficult to 
pin down theoretically.  However, a great deal of
information is contained in the {\it inclusive} cross-section,
\begin{eqnarray}
\sigma\propto &&{1\over q^4}\left\langle k's'\left|J^\alpha_e(0)
\right|ks\right\rangle
\left\langle ks\left|J^\mu_e(0)\right|k's'\right\rangle g_{\mu\nu}
g_{\alpha\beta}\nonumber\\
&&\times\sum_X \left\langle PS\left|J^\nu_q(0)\right|X\right\rangle
\left\langle X\left|J^\beta_q(0)\right|PS\right\rangle 
(2\pi)^4\delta^{\,(4)}(P+q-p_X)\;\; ,
\label{crosss}
\end{eqnarray}
for DIS.  Here, I have defined $q\equiv k-k'$.  Such
total cross-sections are much cleaner theoretically
and make the relationship between microscopic and
macroscopic degrees of freedom more obvious.

Since the physics 
outside the brackets is  
quite well understood (it can be done within 
the framework of QED),
we ignore it for the time being and focus on the 
hadronic object
\begin{eqnarray}
W^{\mu\nu}&\equiv&{1\over4\pi}\sum_X \left\langle 
PS\left|J^\mu_q(0)\right|X\right\rangle
\left\langle X\left|J^\nu_q(0)\right|PS\right\rangle 
(2\pi)^4\delta(P+q-p_X)\nonumber\\
&=&{1\over4\pi}\int d^{\,4}z\, e^{iq\cdot z}\left\langle 
PS\left|J^\mu(z)J^\nu(0)\right|PS\right\rangle \;\; .
\end{eqnarray}
In deriving the final form I have used the fact that $\sum_X |X\rangle\langle X|=1$ 
for any complete set of orthonormal states.  Since $W^{\mu\nu}$ is
concerned only with quark electromagnetic currents,  
I have dropped the subscript $q$.
The factor $1/4\pi$ is a convention for future convenience.

Let us explore some of the properties of $W^{\mu\nu}$.  Hermiticity of
the currents implies that $(W^{\mu\nu})^*=W^{\nu\mu}$.  $W^{\mu\nu}$ also
conserves parity since all of our interactions and currents do.  
Furthermore,
electromagnetic current conservation implies 
$q_\mu W^{\mu\nu}=q_\nu W^{\mu\nu}=0$.
These properties allow us to deduce the tensor 
structure\footnote{The factors of 2 adorning 
$G_1$ and $G_2$ compensate for the implied factor
of 1/2 in the spin vector $S^\mu$.  Note that this decomposition
does not coincide with that found most often
in the literature.  These structure
functions are proportional to those found in more
standard treatments.}
\begin{eqnarray}
&&W^{\mu\nu}=\phantom{+}\left({q^\mu q^\nu\over q^2}
-g^{\mu\nu}\right)F_1(\nu,q^2)\nonumber\\
&&\phantom{W^{\mu\nu}=}
+\left(P-{\nu\over q^2}q\right)^\mu\left(P-{\nu\over q^2}q\right)^\nu
{1\over\nu}F_2(\nu,q^2)\nonumber\\
\label{Wstruc}
&&\phantom{W^{\mu\nu}=}
-i\epsilon^{\mu\nu\alpha\beta}q_\alpha S_\beta {2M\over\nu}G_1(\nu,q^2)\\
&&\phantom{W^{\mu\nu}=}
-i\epsilon^{\mu\nu\alpha\beta}
q_\alpha\left(\nu S-q\cdot S\, P\right)_\beta
{2M\over \nu^2}G_2(\nu,q^2)\nonumber
\end{eqnarray}
for $W$, where $\nu\equiv q\cdot P$ and 
$M$ is the proton mass.  The totally antisymmetric
Levi-Cevita tensor, $\epsilon^{\mu\nu\alpha\beta}$, is
introduced with the convention $\epsilon^{0123}=+1$ in
Appendix \ref{relativity}.
The real scalar functions $F_{1,2}$ and $G_{1,2}$ are called
{\it structure functions} of the proton.  
Our task is to relate these functions to the 
nonperturbative wavefunction of the proton.
Our
three external vectors define six invariants.  
Since $P^2=M^2$, $S^2=-1/4$, and
$S\cdot P$=0, we are left with two scalars and one pseudoscalar.
The pseudoscalar $q\cdot S$ can contribute only linearly because its
appearance is dictated by the Dirac structure of the matrix element.
In four dimensions, one can reduce any Dirac structure to terms 
independent of or linear in 
$S^\mu$.  Parity invariance forbids 
the latter from appearing anywhere
except in the tensor structures multiplying $G_{1,2}$.  
This leaves us with the
two invariants $q^2$ and $\nu$.

Our analysis up to this point has been quite general.
The decomposition (\ref{Wstruc}) is valid for all\footnote{
Strictly speaking, we must choose physical values for
$q^2$ and $\nu$.  We will see below that this restricts
$q^2\leq0$ and $\nu\geq0$.  However, our structure functions
can be analytically continued to almost every region
of the complex plane of $q^2$ and $\nu$.} 
values of $q^2$ and $\nu$.  However, in order to 
make contact with the fundamental degrees of freedom 
of perturbative QCD, we must specialize to a specific 
region.  In the last chapter, we saw that QCD becomes
nonperturbative at low energies.  This means that
the familiar quarks and gluons of the QCD lagrangian
are no longer the relevant excitations.  At these
low energy scales, we must consider collective excitations 
of quarks and gluons (`quasi-partons') rather than the 
degrees of freedom introduced in the last chapter. 
This behavior is reminiscent of solid state physics,
where strong interactions between electrons and the 
lattice require us to consider `quasi-electron' excitations
rather than the familiar electron-ion excitations 
we see in `free' electrodynamics.  
After some reasonable assumptions,
the lagrangian for these effective degrees of
freedom can be derived directly from the fundamental 
lagrangian.  It can easily be seen that the 
quasi-electrons interact weakly and are therefore
the relevant degrees of freedom.\footnote{Renormalization
can also be viewed in this way : the fundamental 
pointlike fields in our lagrangian interact 
strongly with vacuum fluctuations.  The more appropriate
degrees of freedom are collective excitations of the 
fundamental fields and vacuum fluctuations, the 
renormalized fields.}  Presumably, something similar
is happening in our case.  However, since we are dealing with 
an ultra-relativistic quantum field theory rather than
non-relativistic quantum mechanics,  we can no longer
smoothly make the transition from fundamental to
effective degrees of freedom.\footnote{This is merely
another signal of our incompetence.  Certainly,
a transition can be made; we just don't know where
to begin.}  This leaves us with two options.  We
can either write down a theory which is not\footnote{
directly} motivated
by the fundamental theory, or stick to regions
in which the effective degrees of freedom are 
the quarks and gluons of QCD.  There are many models and 
theories which opt for the former, among them
the nuclear shell model, Yukawa's pion theory,
the constituent quark model, and chiral perturbation theory.
Here, we attempt to learn something about QCD itself 
by taking the latter path.

To this end, we will consider DIS in the {\it Bjorken}
limit, formally defined as 
$Q^2, \nu\rightarrow\infty$ while 
$x_B\equiv Q^2/2\nu$ remains constant.  Here,\footnote{ignoring
the electron mass}
\begin{equation}
Q^2\equiv-q^2=2k^0(k')^0 (1-\cos\theta)
=4k^0(k')^0\sin^2(\theta/2)\geq0\;\; ,
\end{equation}
with $\theta$ the angle between the initial and
final electron spatial momenta.
This special kinematic regime admits quarks
and gluons as appropriate degrees of freedom
since they interact weakly at high energy.
In this limit we can ignore the 
proton mass and all natural scales 
associated with QCD.
A straightforward application of dimensional
analysis\footnote{Our states are normalized
via $\langle\vec p\,'\,|\vec p\,\rangle=
2p^0(2\pi)^3\delta(\vec p\,'\,-
\vec p\,)$, as in Appendix \ref{canquant},
so each state contributes mass dimension
$-1$.} reveals that $W$, and therefore all of 
our structure functions, 
is dimensionless.
In the Bjorken limit, the only relevant
scales in our problem are $Q^2$ and $\nu$.  Hence
our structure functions should depend only on their
ratio, $x_B$.  
While this
is strictly true in the 
formal limit, for any finite $Q^2$ one 
will see violations of this scaling law coming 
from the dependence of the
QCD coupling on scale.  Since QCD is asymptotically 
free, these violations
will get smaller and smaller as $Q^2$ is increased.  This
is one of the main reasons QCD was chosen as the underlying theory
of the strong interactions, as detailed in the Introduction.

The kinematics of the Bjorken limit allow us to write the 
vectors $q$ and $P$ in a very simple way.  Since we have only two
dynamical vectors, we can always choose a frame in which 
their spatial momenta are parallel to the 3-axis.  
In this frame, we can expand $q$ and $P$
in terms of the two basis vectors $p^\mu=\Lambda(1,0,0,1)$ and
$n^\mu=(1,0,0,-1)/2\Lambda$.
$\Lambda$ is an arbitrary scale
reflecting our freedom to boost along the 3-axis.
Note that we have chosen basis vectors which satisfy
$p^2=n^2=0$ and $n\cdot p=1$.  We will call the component 
of a vector which is parallel to $p$ the `$+$' component and
that which is parallel to $n$ the `$-$' component.
This implies
\begin{equation}
k^0=\Lambda k^++k^-/2\Lambda\;\; ;\qquad\qquad
k^3=\Lambda k^+-k^-/2\Lambda\;\; ,
\end{equation} 
for any vector $k$.
Our vectors have the expansion
\begin{eqnarray}
P&=&p+{M^2\over 2}n\nonumber\\
q&=&-\zeta p+{Q^2\over 2\zeta}n\;\; ,
\end{eqnarray}
where 
\begin{equation}
\zeta\equiv{Q^2\over{2x_BM^2}}
\left(-1+\sqrt{1+{4x_B^2M^2\over Q^2}}\right)
\rightarrow x_B
\end{equation}
as $Q^2\rightarrow\infty$.
Since the $-$ component of $P$ cannot form 
large scalars, we will ignore it whenever 
we can and simply write $P=p$.  

Due to the role played by the light cone
in special relativity, truly massless fields are
not associated with spin vectors.  The analogous
degree of freedom for these fields is the 
helicity, $h$.\footnote{I use the convention 
$h=\pm1/2$ rather than $\pm1$.}  
As shown in Appendix \ref{unitrepso31}, the replacement
$MS\rightarrow hP$ is valid as $M\rightarrow0$.  
While this identification gets rid of the 
unsightly factors of $M$ multiplying 
the spin-dependent structure functions,
the tensor structure associated with 
$G_2$ vanishes identically.
In this term, we are certainly not free to 
take the proton mass to zero!  
As a consequence of this,
we will see that the structure 
function $G_2$ is suppressed
in the Bjorken limit.

We are now ready to discuss some of
the contributions to DIS from various structure
functions in the Bjorken limit.  The lepton tensor we ignored above,
\begin{equation}
L^{\mu\nu}\equiv\left\langle k's'\left|J^\mu_e(0)\right|ks\right\rangle
\left\langle ks\left|J^\nu_e(0)\right|k's'\right\rangle\;\; ,
\end{equation}
can be calculated at leading order in QED.  
For massless electrons, 
only helicity eigenstates make sense.  Since these
do not interact with each other, the 
relation
\begin{equation}
u_\alpha(k,\lambda)\overline u_\beta(k,\lambda')=
{1\over2}\delta_{\lambda\lambda'}
\left\lbrack\left(1+2\lambda\gamma_5\right)
\not\!k\right\rbrack_{\alpha\beta}\;\; ,
\end{equation}
where $\lambda,\lambda'=\pm1/2$ label  
helicities, 
holds.  Substitution gives 
\begin{equation}
L^{\mu\nu}=2e^2\left(k^\mu k'^\nu+k^\nu k'^\mu-g^{\mu\nu}k\cdot k'\right)
-4ie^2\epsilon^{\mu\nu\alpha\beta}q_\alpha(\lambda k)_\beta\;\; .
\end{equation}
As required by electromagnetic gauge invariance,
$L$ conserves current if the electrons are both onshell.
This makes the contraction $W^{\mu\nu}L_{\nu\mu}$ especially
painless.  We obtain
\begin{eqnarray}
W^{\mu\nu}L_{\nu\mu}&=&\phantom{+}2e^2Q^2F_1(\nu,Q^2)\nonumber\\
&&+2e^2\left[2P\cdot k'\left(1+{P\cdot k'\over\nu}\right)-x_BM^2\right]
F_2(\nu,Q^2)\nonumber\\
&&+2e^2\lambda{MS\cdot(k+k')\over\nu}Q^2 G_1(\nu,Q^2)\\
&&+4e^2\lambda{MS\cdot(\nu k'-P\cdot k' q)\over\nu^2}Q^2 
G_2(\nu,Q^2)\;\; .\nonumber
\end{eqnarray}
If the initial and final electrons do not have transverse
momentum components, the kinematics of the process require 
$k=q^- n$ and $k'=-q^+ p$.  In this case, the quantity
\begin{equation}
2P\cdot k'=-M^2q^+=x_BM^2+{\cal O}(M^4/Q^2)
\end{equation}
causes the contribution of $F_2$ to be suppressed 
by $M^4/Q^4$ relative to that of $F_1$.  Writing 
\begin{equation}
MS^\mu=s^+p^\mu+M^2s^-n^\mu+MS_\perp^\mu\;\; ,
\end{equation}
on dimensional grounds, we see that $G_1$ 
contributes at the same order as $F_1$ and
$G_2$ is suppressed by $(M^2/Q^2)$.  
Processes with appreciable transverse momentum 
see the much smaller suppression $(k_\perp^2/Q^2)$
of $F_2$ and allow the transverse components 
of $S^\mu$ to contribute as $(MS_\perp\cdot k_\perp/Q^2)$
to $G_{1,2}$, but do not alter the leading behavior.  
We will discuss these contributions in
Chapter \ref{htwist}.  For now, however, we 
ignore the subleading structure functions and focus on
$F_1$ and $G_1$.

\section{Deep-Inelastic Scattering on Partonic Targets}
\label{pertdis}

The Bjorken limit was not introduced merely as a 
way to get rid of two structure functions (although
this aspect is not without its charms).  Rather,
it was alleged to help us make contact with the 
fundamental degrees of freedom in QCD.  To understand
how this happens, let us take a moment to think 
about the physical process at hand.  Our proton
is a conglomerate of partons interacting
strongly with each other.  The process we are interested
in concerns the scattering of an intruder off
this complicated mess.  

Fortunately, some general
arguments allow us to simplify our task.  Causality implies that
our partonic interactions do not happen instantaneously.
Consider, for example, two partons under the influence
of the nonperturbative force that binds them 
together.  Since they feel this strong restoring force, it
stands to reason that they are nearing the end of their
leash.  Their separation distance, $d$, should be of the 
same order as the nucleon size and is related to 
the energy scale, $\Lambda_{\rm QCD}$, at which QCD becomes
nonperturbative :
\begin{equation}
d\sim1/\Lambda_{\rm QCD}\;\; .
\end{equation}
Since they are separated by this distance, the 
interaction requires a time of order $1/\Lambda_{\rm QCD}$
in which to occur.  On the other hand, our virtual 
photon is living on borrowed time.  The form of its propagator
implies that it can only exist for a time of
order $M/Q^2$ in the 
rest frame of the nucleon.\footnote{Its propagator
is of the form 
\begin{equation}
\int {d^{\,4}x\over x^2}\, e^{iq\cdot x}\;\; .
\end{equation}
For large $q\cdot x$, small variations
in $x$ cause wild oscillations
in the integrand which effectively
damp contributions.  Hence dominant contributions
come only from regions where {\it both} $q\cdot x$ {\it and}
$x^2$ are small.  Since $q^\mu$ is not light-like (far from it, in fact),
this implies $x^\mu$ is small.  How small
is dictated by the components of $q^\mu$ in the frame
we are using.  For example, in the rest frame of the 
proton, the time alotted to our virtual photon
is on the order of $1/q^0\sim M/Q^2$.}
If $Q^2/M>\!\!>\Lambda_{\rm QCD}$,
the partons simply do not have time to scatter
nonperturbatively during the intruder's
visit.  Consequently, the intruder sees a frozen
non-dynamic collection of partons rather than the 
complicated interacting soup we do not understand.  
By the time the partons can re-scatter, the process
we are interested in is over and done with.
These final-state interactions cannot affect the total
cross-section for our process\footnote{They can,
however, affect an {\it exclusive} cross-section
like $\gamma^*p\rightarrow n\,2\pi^+\,\pi^-\,\pi^0$.
This process could've begun as 
$\gamma^*p\rightarrow p\,\pi^0$, with re-scattering
effects producing the observed final state
long after the virtual photon scattering.
This is one of the main simplifications
provided by {\it inclusive} DIS.}
and hence are unimportant to our treatment.

Since the partons do not interact with each other
during the scattering, we can treat each parton as though it
were alone.  In this way, the complete scattering
tensor $W^{\mu\nu}$ becomes a weighted average
over `partonic scattering tensors' :
\begin{equation}
W^{\mu\nu}_{T}=\sum_af_{a/ T}w^{\mu\nu}_a\;\; .
\label{neanfact}
\end{equation}
Here, $f_{a/ T}$ is the probability of
finding a parton of type $a$ in our target $T$
and $w_a^{\mu\nu}$ is the scattering tensor 
for a partonic `target' of type $a$.  The sum 
extends over all parton species, spin, and momentum.
I emphasize here that this decomposition can be done
only in cases where the scattering is {\it incoherent},
i.e. the partons do not interact with each other during
the scattering.\footnote{We have shown that for 
$Q^2/M>\!\!>\Lambda_{\rm QCD}$ partons do not interact 
over long distances during the scattering.
Short distance perturbative interactions 
certainly can take place during the scattering.  
However, the exchange of vast amounts of
energy between partons within a nucleon 
is suppressed simply because there is 
not a lot of ambient energy in the medium.
These kinds of parton correlation effects
are suppressed by powers of $Q$.  We will
discuss a certain class of them in the last chapter.}
Measurements of the running of $\alpha_s$
and the size of various hadrons reveals the strong
interaction scale to be of order $300$ MeV; as long
as we have $Q$ of order $2-3$ GeV or more, the 
scattering can reasonably be considered
incoherent.
Some of the approximations of the last section require 
somewhat larger $Q$ since they ignore the nucleon mass
(roughly $1$ GeV).  Ideally, inclusive DIS is considered at energies
of order $10$ GeV or larger.  

Equation (\ref{neanfact}) is a mathematical
expression of the physical separation of
scales in our process.  It states that
the full amplitude can be {\it factorized}
into a product of two parts.
The functions $f_{a/T}$ describing the
probability of finding various partons within
our target, called {\it parton distribution functions}, 
are nonperturbative in nature and 
cannot be calculated using the techniques
considered here.  These objects 
are related to soft, low-energy physics and
do not depend on the specific scattering 
process.
On the other hand, the partonic scattering
tensors $w^{\mu\nu}_a$ are expressed entirely 
in terms of the fundamental degrees of freedom 
in QCD and can in principle be calculated 
in perturbation theory.  They are associated
with the hard, high-energy physics of the 
scattering and 
do not depend on the 
target we consider.
This separation between hard and soft physics,
called {\it factorization}, is indispensable 
to modern QCD.  Many recent results in 
perturbative QCD involve identifying and isolating the 
relevant distributions analogous to
$f_{a/T}$ for a certain process and calculating
their coefficients.  The remainder of this
section is devoted to a calculation
of DIS on partonic targets.  We shall
have a lot more to say about parton distributions
in the next section.

At present, perturbation theory is not readily applicable
to the calculation of $W$ because it does not involve a 
time-ordered product.  We can remedy this in the following
roundabout way.  Consider the tensor
\begin{equation}
\int d^{\,4}z\, e^{iq\cdot z}\left\langle PS\left|J^\nu(0)
J^\mu(z)\right|PS\right\rangle\;\; .
\end{equation}
Introducing a complete set of intermediate states and shifting
the argument of $J^\mu$ to the exponent, we
see that it is equal to
\begin{equation}
\sum_X\left\langle PS\left|J^\nu(0)\right|X\right\rangle\left\langle
X\left|J^\mu(0)\right|PS
\right\rangle (2\pi)^4\delta\left(q+p_X-P\right)\;\; .
\end{equation}
The sum extends over all physical intermediate states with
energy-momentum $p_X=P-q$.  Since we intend to work in the 
region\footnote{The subscript $rf$ reads `in the rest frame of the proton'.
Since the zero component of a four vector is not a scalar,
we must specify a frame to give it meaning.}
$(q^0)_{rf}>\!\!>(P^0)_{rf}=M$, there are no physical states which satisfy the 
$\delta$-function.  Hence we can write
\begin{equation}
W^{\mu\nu}\sim{1\over4\pi}\int d^{\,4}z\, e^{iq\cdot z}\left\langle 
PS\left|[J^\mu(z),J^\nu(0)]\right|PS\right\rangle\;\; ,
\end{equation}
where the `$\sim$' is meant to remind us that this identification is
only valid in the physical region of scattering, i.e. 
$(q^0)_{rf}>0$.\footnote{Since baryon number 
is conserved in all interactions
we will consider, the state $|X\rangle$ must include at least 
one baryon.  Hence its energy must be at least the proton mass.}
We still do not have a time-ordered product, but can use the same
trick on the tensor
\begin{equation}
\label{Tdef}
T^{\mu\nu}\equiv i\int d^{\, 4}z\,e^{iq\cdot z}\left\langle PS\left|
{\rm T}J^{\mu}(z)
J^{\nu}(0)\right|PS\right\rangle\;\; 
\end{equation}
to obtain
\begin{equation}
T^{\mu\nu}\sim i\int d^{\, 4}z\,\Theta(z^0)
\,e^{iq\cdot z}\left\langle PS\left|[J^{\mu}(z),
J^{\nu}(0)]\right|PS\right\rangle\;\; .
\end{equation}
The step-function, $\Theta(z^0)$, comes from the time-ordered product.
The complex conjugate of $T^{\mu\nu}$,
\begin{equation}
\left(T^{\mu\nu}\right)^*\sim-i\int d^{\, 4}z\,\Theta(-z^0)
\,e^{iq\cdot z}\left\langle PS\left|[J^{\nu}(z),
J^{\mu}(0)]\right|PS\right\rangle\;\; ,
\end{equation}
differs from $T^{\nu\mu}$ only in the argument of the $\Theta$-function.
Writing  
\begin{equation}
\Theta(z^0)={1\over2}[1+{\rm sign}(z^0)]\;\; ,
\end{equation}
where ${\rm sign}(x)$ returns the sign of its argument,
it becomes obvious that 
\begin{equation}
\Im m\;T^{(\mu\nu)}\sim2\pi W^{(\mu\nu)}
\end{equation}
for the symmetric parts and 
\begin{equation}
\Re e\;T^{[\mu\nu]}\sim2\pi iW^{[\mu\nu]}
\end{equation}
for the antisymmetric parts.
Decomposing $T^{\mu\nu}$ as above,
\begin{eqnarray}
&&T^{\mu\nu}=\phantom{+}
\left({q^\mu q^\nu\over q^2}-g^{\mu\nu}\right)T_1(\nu,q^2)\nonumber\\
&&\phantom{T^{\mu\nu}=}
+\left(P-{\nu\over q^2}q\right)^\mu\left(P-{\nu\over q^2}q\right)^\nu
{1\over\nu}T_2(\nu,q^2)\nonumber\\
\label{Tdecomp}
&&\phantom{T^{\mu\nu}=}
-i\epsilon^{\mu\nu\alpha\beta}q_\alpha S_\beta{2M\over\nu}S_1(\nu,q^2)\\
&&\phantom{T^{\mu\nu}=}
-i\epsilon^{\mu\nu\alpha\beta}
q_\alpha\left(\nu S-q\cdot S\, P\right)_\beta
{2M\over\nu^2}S_2(\nu,q^2)\nonumber\,\, ,
\end{eqnarray}
we see that the structure functions associated with $W$ 
are just $1/2\pi$ times the imaginary
part of those associated with $T$.  This relation is only true
in the physical region, but we are free to calculate
our structure functions in any way we wish.  Since $T$ involves a 
time-ordered product, it is of exactly the form we need
to apply perturbation theory.

\begin{figure}
\label{fig9}
\epsfig{figure=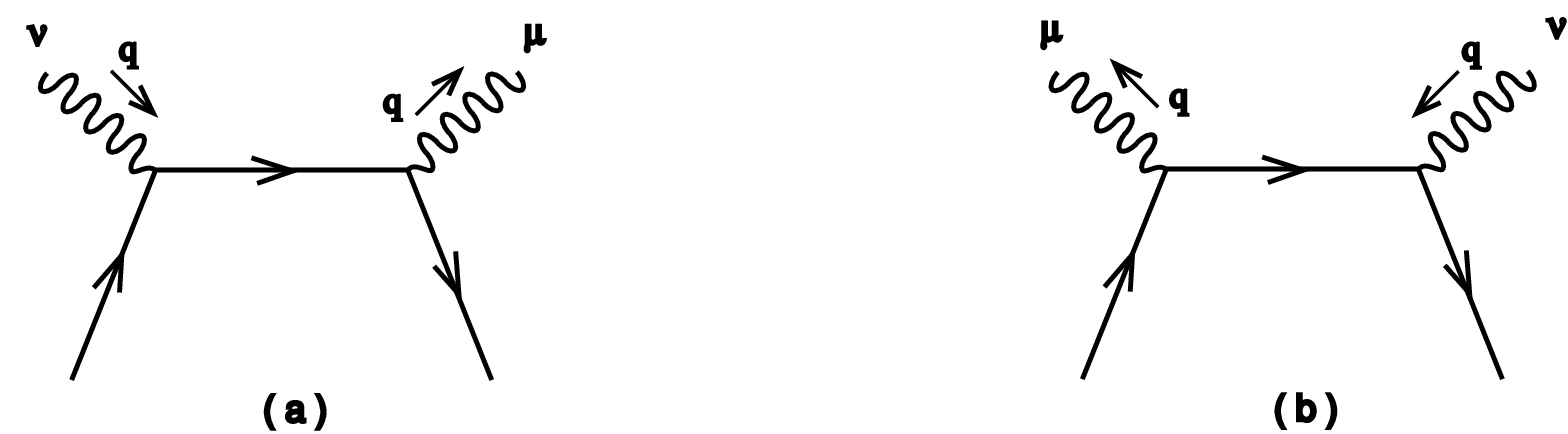,height=1.5in}
\caption{The two leading-order diagrams for
virtual photon-quark scattering.  The crossing symmetry allows
us to obtain one from the other.}
\end{figure}

Let us consider scattering on a massless 
target of momentum $xp$.
The leading order amplitude involves a simple contraction of the 
quark currents with each other and the external states.  The two relevant
diagrams are shown in Figure 2.1.  We need only consider one of
them explicitly because the other can be obtained from 
crossing symmetry, $q\rightarrow-q$ and $\mu\rightarrow\nu$.
This is an exact symmetry of $T$, independent
of the QCD dynamics, as can be seen 
directly from its definition.
Figure 2.1a is trivial to evaluate.  It has the value\footnote{We have 
kept our definition of $\nu=p\cdot q$ rather than changing to 
$xp\cdot q$ since in the end we will be concerned with the 
tensor structures relevant to the true target of momentum
$P$ rather than this fictitious quark target.}
\begin{equation}
{e_q^2\over 2\nu}\;{1\over x_B-x-i\varepsilon}\;{\overline u}(h,xp)
\gamma^\mu(x\!\!\not\!p\,+\!\!\not\!q\,)\gamma^\nu u(h,xp)\;\; ,
\end{equation}
where ${\overline u}(h,xp)$ and $u(h,xp)$ are the Dirac wavefunctions of 
a free quark of momentum $xp$ and helicity $h$ introduced in Appendix 
\ref{dottedandundotted}, and $e_q$ is the charge of
a quark of flavor $q$.\footnote{Here, I resort to the 
standard notation $q$ for quark flavor rather than $f$.}
As above, the Dirac product can be simplified by writing
\begin{equation}
u_\alpha(h,xp)\overline u_\beta(h,xp)=
{1\over2}[(1+2h\gamma_5)x\!\!\not\!p\,]_{\alpha\beta}\;\; ,
\end{equation}
which is valid for massless spinors.  In this way, 
the amplitude reduces to the simple Dirac trace
\begin{equation}
{e_q^2\over 2\nu}\,{x\over x_B-x-i\varepsilon}\,\,{1\over2}{\rm Tr}\left[
\gamma^\mu(x\!\!\not\!p\,+\!\!\not\!q\,)
\gamma^\nu(1+2h\gamma_5)\!\not\!p\,\right]\;\; .
\end{equation}
The `1' contributes only to the symmetric part of $T^{\mu\nu}$,
while $\gamma_5$ contributes only to the antisymmetric part.
Concentrating on the symmetric part for the moment, 
we find
\begin{eqnarray}
T_q^{(\mu\nu)}&=&xe_q^2\left({1\over x_B-x-i\varepsilon}-
{1\over x_B+x-i\varepsilon}\right)
\left({q^\mu q^\nu\over q^2}-g^{\mu\nu}\right)\nonumber\\
&+&2x^2e_q^2\left({1\over x_B-x-i\varepsilon}+
{1\over x_B+x-i\varepsilon}\right){1\over\nu}
\left(p-{\nu\over q^2}q\right)^\mu\left(p-{\nu\over q^2}q\right)^\nu\;\; ,
\end{eqnarray}
where the crossing term, $q\rightarrow-q$ and $\mu\leftrightarrow\nu$,
has been taken into account.\footnote{Note that 
the crossing is not exactly $\mu\leftrightarrow\nu$ and
$q\rightarrow-q$ since the $+i\varepsilon$ must
also change sign.  This fact is related to the
analytic properties of $T^{\mu\nu}$ as a function
of $x_B$.}
The structure functions relevant to our experiment 
are now easy to compute :\footnote{
The $+i\varepsilon$'s introduced in 
Section \ref{roadtoquant} play a pivotal role here.}
\begin{eqnarray}
_qF_1(x,\nu, Q^2)&=& xe_q^2{1\over2}\delta(x_B-x)\nonumber\\
_qF_2(x,\nu, Q^2)&=&x^2e_q^2\delta(x_B-x)=2x_B\;{_qF_1}(x,x_B)\;\; .
\end{eqnarray}

Here, we see explicitly that the structure functions 
scale; they depend only on the ratio $x_B$ between of the 
two large scales $\nu$ and $Q^2$.\footnote{The dependence
of $T_q$ on the $+$ component of the quark momentum
is viewed as dependence on the type of parton.}  Along with 
the Callan-Gross relation, $_qF_2=2x_B\;{_qF_1}$, scaling is a general 
property of this kind of scattering process.
This behavior can be shown quite generally via the 
current algebra techniques of Gell-Mann \cite{curr}.  
The fact that it is broken by quantum effects
was one of the main motivations of D. Gross' search for a proof
that quantum field theory cannot describe experiment.  His subsequent
discovery of asymptotic freedom turned his argument
on its head and led to the realization that only a theory
like QCD can be consistent with experiment.

The antisymmetric part of $T$ is given by the 
$\gamma_5$ term in the trace.
The result is 
\begin{equation}
T^{[\mu\nu]}_q=
-i\epsilon^{\mu\nu\alpha\beta}q_\alpha(h p)_\beta{2xe_q^2\over\nu}
\left({1\over x_B-x-i\varepsilon}+{1\over x_B+x+i\varepsilon}\right)\;\; .
\end{equation}
This implies
\begin{equation}
_qG_1(\nu, Q^2)=xe_q^2{1\over 2}\delta(x_B-x)=\;_qF_1(\nu,Q^2)\;\; .
\end{equation}
Once again, we see scaling.

The presence of $\delta(x_B-x)$ in our structure functions is easy to understand
from the standpoint of `physical' quark scattering.  
At leading order, the only intermediate state accessible to the 
quark-photon system is an excited quark.  If this state is to be physical,
the quark must be onshell.  This implies $2x\nu=Q^2$, or $x=x_B$.  

Our amplitude requires corrections due to virtual gluon exchange, 
as shown in Figure 2.2.  As presented, these diagrams are riddled with
ultraviolet divergences.  The discussion on renormalization
in Section \ref{renincov} tells us how to deal with these divergences.  
It is simplest to use the $\rm \overline{MS}$ scheme in 
Feynman's gauge ($\xi=1$).  In renormalized perturbation theory,
there are three relevant counterterms.  The obvious insertion is
$i(Z_F-1)(x\!\!\not\!\!p\,+\!\!\not\!\!q\,)$ for the intermediate quark 
propagator.  The other two come from the 
electromagnetic sector of the theory.  

Consider for a moment
the quark-photon interaction term in QED.  Its structure
is identical to the quark-gluon interaction with $t^a$ replaced by 1:
\begin{equation}
{\cal L}_{int}^{\rm QED}\sim -Q_qe_b\overline\psi_b\not\!\!A_b\psi_b\;\; ,
\end{equation}
where $Q_q$ is the quark charge,
$e_b$ is the bare electromagnetic coupling,\footnote{
The {\it electromagnetic coupling} is a parameter which
summarizes the strength of the interaction of {\it all}
charged fields with the photon.  It is analogous to the 
QCD coupling $g$.  The {\it charge} of a field describes
the way a field transforms under the gauge group.
It is analogous to the {\bf 3} of $SU(3)$.  For example,
the {\it charge} of an up quark is 2/3 since under an
electromagnetic gauge transformation parameterized by
$\alpha$ its phase rotates by the amount $2\alpha/3$. 
In contrast, the (renormalized) 
{\it electromagnetic coupling}
$e\sim\sqrt{4\pi/137}\sim3/10$
is the same for all electromagnetically charged fields.
When the difference between these two objects is
inessential, it is often ignored by introducing
the `quark electromagnetic charge' 
$e_q\equiv Q_qe$.  However, one must always
remember that they are {\it not} the same.  In particular,
the coupling $e$ is just a number while the charge $Q$ is 
a quantum mechanical operator describing the 
behavior of fields under a gauge transformation.} 
and $A_b$ is the bare photon field.  Defining our $Z$-factors 
in the same way as for QCD, we see that 
this leads to the counterterm
\begin{equation}
-(Z_eZ_FZ_A^{1/2}-1)Q_qe\,\overline\psi\!\not\!\!A\psi
\end{equation}
in the QED lagrangian.
Electromagnetic gauge invariance requires the running
of the coupling to be the same for {\it all} 
electromagnetically charged particles.  Hence the 
QCD ultraviolet divergence structure of the 
quark electromagnetic current must be identical to the 
divergence structure of the quark 
self-energy.  This must be satisfied
{\it diagram by diagram}, i.e. for each QCD diagram the sum of all possible
electromagnetic current insertions must have an ultraviolet divergence which is
identical to (the negative of) that of the corresponding self-energy
diagram.  The cancelation, being a consequence of {\it electromagnetic}
gauge invariance,  is completely independent of the QCD dynamics.

The full counterterm contribution to the amputated
diagrams is taken into
account simply by multiplying the leading order
result by $Z_F$.  While this is all that is necessary
to cancel the ultraviolet divergences, 
it will not result in a properly normalized amplitude.
Since we have not used the onshell renormalization
scheme, the residue of our renormalized propagator
is not $i$.  While this fact is unimportant for intermediate
quark propagation, our external quark wavefunctions must
be correctly normalized.\footnote{Since this calculation
does not represent actual physical scattering,
the difference between properly and improperly
normalized quark fields can be absorbed into
the distribution functions.  However, if we
wish to relate these functions to the effective high-energy
degrees of freedom of QCD, we has better normalize them
properly.}
This is done exactly as in the 
onshell scheme : we multiply the amputated 
Feynman diagrams by the full (1PR) propagators,
(re)renormalize the external states
by dividing by the square root of the 
propagator residue,\footnote{
The word `residue' is used in this context
to mean `mathematical residue divided by $i$', or
`thing that should be 1'.} multiply by the 
inverse propagator, take the 
onshell limit,
and attach external 
free wavefunctions.
This procedure is known as 
Lehmann-Symanzik-Zimmermann (LSZ) 
reduction formula \cite{LSZ}.  It is 
quite general and applies to all quantum field
theories, regardless of whether or not they
are renormalizable.
Since the full propagators contain a 
power of the residue,\footnote{
Actually, since we intend to take the onshell
limit after all of these manipulations, the full
propagator is effectively {\it just} the pole 
contribution.}
the end result is simply multiplying
the amputated graphs by the square root of the residue
for each physical external state.

To see this explicitly, let us examine 
the fully renormalized 
quark propagator in this scheme.  The 1PI graphs
sum to $-i\Sigma=-iB(p^2)\!\not\!\!p$, with a counterterm
contribution $i(Z_F-1)\!\!\not\!p$.  The full propagator
becomes
\begin{equation}
S(p)={i\over(Z_F-B(p^2))\not\!p}\;\; .
\end{equation}
The residue of this object at $\not\!p=0$ 
is $[Z_F-B(0)]^{-1}$.  Since $B(p^2)$ is
dimensionless, it may be written\footnote{As before, 
the $\mu_0$'s appear 
to keep the coupling dimensionless.}
\begin{equation}
B(p^2)=\sum_{n=1}^\infty\alpha_s^nB_n
\left(p^2\over\mu_0^2\right)^{-n\epsilon/2}\;\; .
\end{equation}
Since our renormalized amplitude is free of
ultraviolet divergences, we are free to take 
$\epsilon<0$.\footnote{We {\it cannot} do this
in the unrenormalized amplitude since 
$\epsilon$ is required positive to 
regulate the ultraviolet divergences.
Only in an ultraviolet finite quantity can
we take $\epsilon<0$.} 
In this case, it becomes obvious
that $B(0)=0$.  This implies that the correct
amplitude is obtained by multiplying the 
amputated result
by $Z_F^{-1/2}$ for each 
external quark leg, or $1/Z_F$ in all.
It must be emphasized that this factor
is {\it not} ultraviolet divergent.\footnote{This
is why we never had to take it into account before.  
If we are concerned only with the ultraviolet 
behavior of an operator, it is best not to take the 
onshell limit so this kind of `mixing of
divergences' does not occur.}  All of the
ultraviolet divergences have been removed by renormalization.
This factor contains the {\it infrared}\, divergence generated by
taking the onshell limit.  However, as long as
we recognize that there are no ultraviolet divergences
left in the problem, we may allow this $Z_F$ to cancel
the one from the counterterms.  This means that 
after all is said and done all of the $Z_F$ factors necessary for 
a properly normalized amplitude cancel.  The true
amplitude is simply the sum of the diagrams, without
a thought given to counterterms or proper normalization.
We must keep in mind, though, that all of the
divergences we run into from now on are infrared
and cannot simply be swept into renormalization constants.

In unrenormalized perturbation theory, the story
is somewhat different.  The procedure here
is to calculate {\it all} diagrams, renormalize
{\it external} states, and express the 
result in terms of renormalized
quantities at the end of the day.
The result of these manipulations is guaranteed
to be an ultraviolet finite quantity.
In addition, onshell external states require
infrared (re)renormalization as discussed above.
In our case, the electromagnetic currents do
not require renormalization at this order in 
QED.\footnote{At higher orders, the renormalization
of these currents is reflected in the renormalized 
coupling.}  As mentioned above,
all of their ultraviolet divergences cancel with intermediate
self-energy diagrams.  The external states, on the other
hand, require an overall division by $Z_F$ to cancel
their ultraviolet divergences.  The amplitude
we are left with is guaranteed to be 
free of ultraviolet divergences.  
Once again, we (re)renormalize the onshell
external states through multiplication 
by an infrared divergent $Z_F$.  The end
result is that all of our $Z_F$ factors cancel; 
we need only calculate the amputated diagrams
and attach onshell external wavefunctions.

\begin{figure}
\label{fig10}
\epsfig{figure=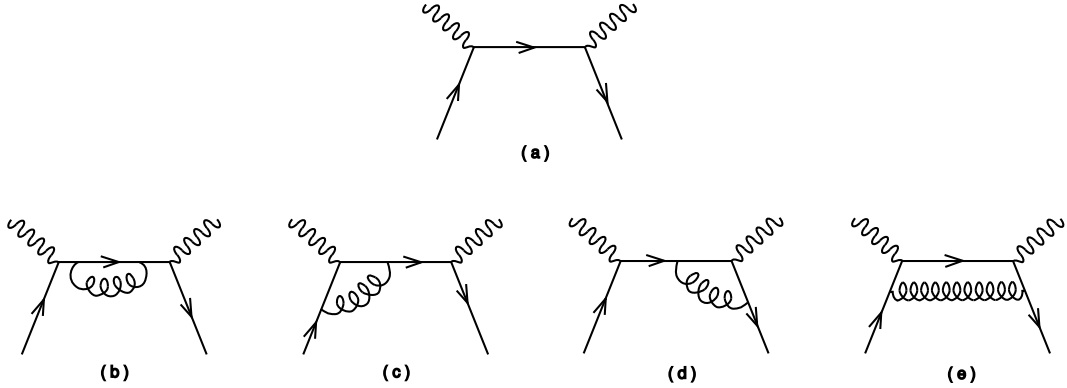,height=2.0in}
\caption{Contributions to virtual photon-quark
scattering at (a) leading and (b)-(e) next-to-leading
order in the QCD coupling.}
\end{figure}

The final step in the LSZ program involves taking the
onshell limit of our amplitudes.  As we have seen,
this leads to infrared divergences in general.
Although they seem like a terrible defect
at first, the appearance of these 
divergences is quite natural.
Their origins and treatment are discussed
in the next section; for now, we simply acknowledge their
existence and proceed. 

There are several ways to do this calculation.
One method is that used for the leading order
amplitude.  The unpolarized (symmetric) part of
$T$ is calculated by averaging over external helicities and 
the polarized (antisymmetric) part is obtained by taking
half the difference of right- and left- handed helicities.
In this way, both parts become a trace which can be simplified
before performing the loop integration. 
Unfortunately, subtleties associated 
with the definition of $\gamma_5$ in 
$d$ dimensions\footnote{see Appendix \ref{dimregapp}.}
make the polarized result ambiguous.
Thus, in order to resolve this ambiguity,
one must choose a consistent $\gamma_5$ scheme.
The most natural thing to do is simply 
perform the loop integration
over the full Dirac structure, leaving the introduction
of $\gamma_5$ for later.  This approach has the advantage
of obtaining both the polarized and unpolarized amplitudes
simultaneously.  
After the loop integration, the only vectors left over
are external and may be taken to be in the first 
four dimensions.
At this point, we are free to use the above method to extract the 
symmetric and antisymmetric parts explicitly.

The intermediate quark self-energy diagram, Figure 2.2b, can
be done using the techniques of Appendix \ref{covint}.
Ignoring terms of order $\epsilon$, the result
is
\begin{eqnarray}
{\alpha_sC_F\over4\pi}\left({Q^2e^{\gamma_{E}}
\over4\pi\mu_0^2}\right)^{-\epsilon/2}\left\lbrack
{2\over\epsilon}+1-\log{x_B-x-i\varepsilon
\over x_B}\right\rbrack
{e_q^2\over2\nu}\,{1\over x-x_B+i\varepsilon}
\gamma^\mu(x\!\!\not\!p\,+\!\!\not\!q\,)\gamma^\nu\nonumber\\
+(\mu\leftrightarrow\nu,q\rightarrow-q)\;\; ,
\end{eqnarray}
where the 
external states have been omitted.  Once again, we
see the addition
$(\mu\leftrightarrow\nu,q\rightarrow-q)$ from the 
photon crossing diagram.
In the interest of 
space, this crossing contribution is implied in all of the 
following results unless indicated otherwise.  
As before, the $+i\varepsilon$ plays
a fundamental role in the extraction of the imaginary
part of the amplitude.  Note that we can ignore
it here simply by endowing $x_B$ with a small negative 
imaginary part, $\tilde x_B\equiv x_B-i\varepsilon$.
This behavior persists in all of our 
diagrams, provided that the crossing is obtained by 
$\tilde x_B\rightarrow-\tilde x_B$.\footnote{One 
can see this by examining each diagram separately,
keeping the $\varepsilon$'s explicitly.  Alternatively,
a study of the amplitude in the complex $x_B$-plane 
reveals $T$ to be an analytic function of $x_B$ with branch 
cuts along the real axis stemming from $\pm x$.  The discontinuity
relevant to our physical structure functions can then be obtained
by approaching the positive real axis from underneath.}
In the following, we drop the tild\'e on $\tilde x_B$ and
simply remember that $x_B$ has a small negative imaginary part.

The calculation of the remaining diagrams is greatly 
simplified by taking $\mu$ and $\nu$ in the transverse directions.
A look at the tensor structures relevant to 
$F_1$ and $G_1$ tells us that this simplification is
free of charge - none of the leading dynamics is affected by it.
Since the external states in our calculation satisfy the 
free equation of motion, 
$\not\!\!p\,u(xp,h)=0$, the intermediate self-energy
diagram may be rewritten for transverse $\mu\nu$ as
\begin{equation}
{\alpha_sC_F\over4\pi}\left({Q^2e^{\gamma_{E}}
\over4\pi\mu_0^2}\right)^{-\epsilon/2}\left\lbrack{2\over\epsilon}+
1-\log{x_B-x\over x_B}\right\rbrack
{e_q^2\over2}\,{1\over x-x_B}\,
\gamma^\mu\!\!\not\!n\,\gamma^\nu\;\; .
\end{equation}
We will see that all of our amplitudes can be expressed in terms 
of this one simple Dirac structure and its hermitian conjugate.

An application of the formulae in 
Appendix \ref{covint} to the
integrals encountered in the remaining diagrams
is possible, but extremely
tedious.  This method of covariant integration does 
not use the special kinematics of our problem.
A superior method\footnote{This method is identical to that
employed in Section \ref{reninax}.  The presence of
a nonzero scale complicates this process,
so we discuss its application explicitly in this case.}
breaks the integration into 
$+$, $-$, and $\perp$ components : 
$d^{\,d}k=dk^+dk^-d^{\,d-2}k_\perp$.  The definitions of 
$+$ and $-$ ensure that the measure is invariant under this
rotation.  Defining $y\equiv k^+$, we may substitute 
$k^\mu=yp^\mu+k^-n^\mu+k_\perp^\mu$ into our expressions.
The fact that $n^2=p^2=0$ and none of our external momenta
have transverse components leads to a tremendous simplification.
In particular, the large $k^-$ behavior of the integral
is never worse than $\int dk^-/(k^-)^2$.  This sets the stage for a 
contour integration in the $k^-$-plane.  Restoring the 
$+i\varepsilon$'s, we can analyze the pole structure of the integrand.
Since $k^-$ is always accompanied by a function of $y$, the position
of its pole will drop from above the real axis ($y<\!\!<0$)
to below the real axis ($y>\!\!>0$) as $y$ makes its trek 
from $-\infty$ to $+\infty$.  The precise position of the 
drop point depends on the pole in question, but for either
very large $y$ or very large $-y$ all of the poles appear 
either below or above.  We are free to close the 
contour in the pole-free half of the plane in these regions, 
giving no contribution to the integral.  For intermediate
$y$, we may close the contour in either the upper- or the 
lower-half-plane; Cauchy's theorem assures us the 
result will be identical.
The transverse dimensions 
now have an integral of the form\footnote{We can always
choose $N^2=0$ by refusing to pick up the 
$Q^2$-dependent pole.  In this case, both integrals  
reduce to
\begin{equation} 
\int {d^{\,d-2}k_\perp\over(2\pi)^{d-2}}{1\over k_\perp^2+M^2}\;\; .
\end{equation}
}
\begin{equation}
\int{d^{\,d-2}k_\perp\over(2\pi)^{d-2}}{\alpha Q^2+
\beta k_\perp^2\over (k_\perp^2+M^2)(k_\perp^2+N^2)}\;\; ,
\end{equation}
where $\alpha$ and $\beta$ are functions of $x$,$x_B$, and $y$,
which can easily be done via the methods of Appendix \ref{dimregapp}.
The final integration can always be done with the formula
\begin{equation}
\int_0^1dy\,y^{\alpha-1}(1-y)^{\beta-1}=
{\Gamma(\alpha)\Gamma(\beta)\over\Gamma(\alpha+\beta)}
\end{equation}
after a suitable change of variables.

This analysis is complicated by the existence
of physical intermediate states for $x_B\geq x$.  
These states lead to branch cuts in our amplitude 
and generally get in the way.  We can avoid them {\it during}
the calculation by working in the unphysical region,
$x_B>\!\!>x$.  At the end of the day, we are
free to analytically continue $x_B$ to the 
physical region of interest.
  
The two vertex diagrams (Figs. 2.2c and d)
are related by hermitian 
conjugation\footnote{This conjugation is to 
be understood as taking place only
in the Dirac sector.  The $-i\varepsilon$ 
associated with $x_B$
does {\it not} change sign.
Since the final state wavefunction comes with a $\gamma^0$,
$\overline u\equiv u^\dag\gamma^0$, the conjugation simply 
reverses the order of the $\gamma$-matrices.}
and $\mu\nu$ exchange, as can be seen directly
from their expressions.  This symmetry leaves us
with only two diagrams to calculate explicitly.
Including both contributions, the vertex gives
\begin{eqnarray}
&&-{\alpha_sC_F\over4\pi}e_q^2\left({Q^2e^{\gamma_E}
\over4\pi\mu_0^2}\right)^{-\epsilon/2}
\left\lbrace{1\over x_B-x}\left(
{2\over\epsilon}+4\right)\right.\nonumber\\
&&\phantom{
-{\alpha_sC_F\over4\pi}e_q^2
}+\left\lbrack\left({1\over x}+{1\over x_B-x}\right)
\left({4\over\epsilon}+3-\log{x_B-x\over x_B}\right)\right.\\
&&\phantom{
-{\alpha_sC_F\over4\pi}e_q^2
\left\lbrack\left({1\over x}+{1\over x_B-x}\right)
\left(\right.\right.}\left.\left.
-{1\over x_B-x}\right\rbrack\log{x_B-x\over x_B}\right\rbrace
\gamma^\mu\!\!\not\!n\,\gamma^\nu\;\; .\nonumber
\end{eqnarray}
The presence
of a squared propagator in the final diagram
(Fig. 2.2e) complicates matters slightly. 
However, since there are still only 
three poles, our method can easily be modified to 
handle this situation.  We obtain
\begin{eqnarray}
&&-{\alpha_sC_F\over4\pi}e_q^2\left({Q^2e^{\gamma_E}
\over4\pi\mu_0^2}\right)^{-\epsilon/2}
\left\lbrace{2\over\epsilon}
{1\over2x}\left(
\gamma^\mu\!\!\not\!n\,\gamma^\nu+\gamma^\nu\!\!\not\!n\,\gamma^\mu\right)
\right.\nonumber\\
&&\phantom{-{\alpha_sC_F\over4\pi}e_q^2\left({Q^2e^{\gamma_E}
\over4\pi\mu_0^2}\right)^{-\epsilon/2}}\;\;\;
+\left\lbrack\left({2\over\epsilon}-
{1\over2}\log{x_B-x\over x_B}\right)\gamma^\mu\!\!\not\!n\,\gamma^\nu
\right.\\
&&\phantom{-{\alpha_sC_F\over4\pi}e_q^2\left({Q^2e^{\gamma_E}
\over4\pi\mu_0^2}\right)^{-\epsilon/2}\;\;\;+[(}
\left.\left.+{5\over2}
\left(\gamma^\mu\!\!\not\!n\,\gamma^\nu
-\gamma^\nu\!\!\not\!n\,\gamma^\mu\right)
\right\rbrack{x_B-x\over x^2}
\log{x_B-x\over x_B}\right\rbrace\;\; .\nonumber
\end{eqnarray}

All of our diagrams contain infrared
divergences\footnote{Strictly speaking, 
the self-energy diagram we calculated first 
does not contain any infrared divergences.  It contains
only ultraviolet divergences which are destined to cancel
with counterterms.  However, since we allowed the 
ultraviolet divergences in our counterterms to
cancel the infrared divergences associated with the
LSZ reduction formula, these ultraviolet divergences
have been transmuted into infrared ones.} 
which do not cancel in the full
amplitude,
\begin{eqnarray}
&&-{\alpha_sC_F\over8\pi}e_q^2\left({Q^2e^{\gamma_E}
\over4\pi\mu_0^2}\right)^{-\epsilon/2}\left\lbrace
\left({2\over\epsilon}+3\right){3\over x_B-x}
\gamma^\mu\!\!\not\!n\,\gamma^\nu\right.\nonumber\\
&&\phantom{-{\alpha_sC_F\over4\pi}e_q^2\left({Q^2e^{\gamma_E}
\over4\pi\mu_0^2}\right)^{-\epsilon/2}[}
+\left\lbrack\left({x_B+x\over x^2}+{2\over x_B-x}\right)
\left({4\over\epsilon}+3-\log{x_B-x\over x_B}\right)\right.\nonumber\\
&&\phantom{-{\alpha_sC_F\over4\pi}e_q^2\left({Q^2e^{\gamma_E}
\over4\pi\mu_0^2}\right)^{-\epsilon/2}[+]}
\left.-3\left({x_B-x\over x^2}
+{1\over x_B-x}\right)\right\rbrack
\log{x_B-x\over x_B}\,\,\gamma^\mu\!\!\not\!n\,\gamma^\nu\\
&&\phantom{-{\alpha_sC_F\over4\pi}e_q^2\left({Q^2e^{\gamma_E}
\over4\pi\mu_0^2}\right)^{-\epsilon/2}[}
+{2\over\epsilon}\;{1\over x}\left(
\gamma^\mu\!\!\not\!n\,\gamma^\nu
+\gamma^\nu\!\!\not\!n\,\gamma^\mu\right)\nonumber\\
&&\phantom{-{\alpha_sC_F\over4\pi}e_q^2\left({Q^2e^{\gamma_E}
\over4\pi\mu_0^2}\right)^{-\epsilon/2}[}
\left.
+5\;{x_B-x\over x^2}
\log{x_B-x\over x_B}\,\,\left(\gamma^\mu\!\!\not\!n\,\gamma^\nu
-\gamma^\nu\!\!\not\!n\,\gamma^\mu\right)\right\rbrace\;\; ,\nonumber
\end{eqnarray}
as promised in the beginning of this chapter.  
At this point, we are free 
to project out 
the structure functions as above.
We obtain
\begin{eqnarray}
&&T^q_1(x,x_B)=-{\alpha_sC_F\over4\pi}e_q^2
\left({Q^2e^{\gamma_E}
\over4\pi\mu_0^2}\right)^{-\epsilon/2}\left\lbrace
\left({2\over\epsilon}+3\right){3x\over x_B-x}+{4\over\epsilon}
\nonumber\right.\\
&&\phantom{T^q_1(,xx_B)=
-{\alpha_sC_F\over4\pi}e_q^2}
+\left\lbrack\left({x_B+x\over x}+{2x\over x_B-x}\right)
\left({4\over\epsilon}+3-\log{x_B-x\over x_B}\right)\right.\\
&&\phantom{T^q_1(,xx_B)=
-{\alpha_sC_F\over4\pi}e_q^2+[]}
\left.\left.
-3\left({x_B-x\over x}+{x\over x_B-x}\right)\right\rbrack
\log{x_B-x\over x_B}\right\rbrace\nonumber\\
&&\phantom{T^q_1(,xx_B)=
-{\alpha_sC_F\over4\pi}e_q^2+[]+++}
+\left(x_B\rightarrow-x_B\right)\nonumber
\end{eqnarray}
for the symmetric part and 
\begin{eqnarray}
&&S^q_1(x,x_B)=
-{\alpha_sC_F\over4\pi}e_q^2
\left({Q^2e^{\gamma_E}
\over4\pi\mu_0^2}\right)^{-\epsilon/2}\left\lbrace
\left({2\over\epsilon}+3\right){3x\over x_B-x}
\nonumber\right.\\
&&\phantom{T^q_1(,xx_B)=
-{\alpha_sC_F\over4\pi}e_q^2}
+\left\lbrack\left({x_B+x\over x}+{2x\over x_B-x}\right)
\left({4\over\epsilon}+3-\log{x_B-x\over x_B}\right)\right.\\
&&\phantom{T^q_1(,xx_B)=
-{\alpha_sC_F\over4\pi}e_q^2[]+}\left.\left.+7\;{x_B-x\over x}-
3\;{x\over x_B-x}\right\rbrack
\log{x_B-x\over x_B}\right\rbrace\nonumber\\ 
&&\phantom{T^q_1(,xx_B)=
-{\alpha_sC_F\over4\pi}e_q^2[]++++}-\left(x_B\rightarrow-x_B\right)\nonumber
\end{eqnarray}
for the antisymmetric.  In these expressions, the 
crossing contribution has been included explicitly.  

\begin{figure}
\label{fig11}
\epsfig{figure=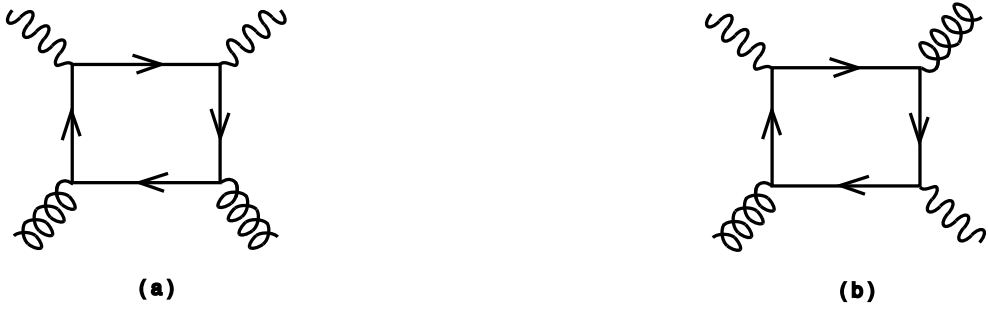,height=4.0cm}
\caption{Leading diagrams for gluon-virtual photon 
scattering.}
\end{figure}

At this order in QCD, virtual photons can also
scatter off gluons in our target 
via the virtual quark loop diagrams
in Figure 2.3.  
These diagrams must be ultraviolet 
finite since there are no counterterms at this order. 
The two gluons have 
transverse polarizations\footnote{As explained in Appendix 
\ref{unitrepso31}, 
massless particles are classified by their helicity.  For 
vector objects, like the gluon, 
the form of the rotation generators
implies that helicity eigenstates are obtained 
only for {\it transverse} polarization.} $i$ and $j$.  
Once again, there are polarized and
unpolarized contributions to the amplitudes.  These can be 
obtained in exactly the same way as for the quarks, by 
performing the integration first and projecting out the 
symmetric and antisymmetric parts of the 
amplitude later.  When averaging over the 
gluon polarizations in $d$ dimensions, 
we divide by $d-2$ rather than 2 because
this is the number of polarization states available
to a gluon in $d$ dimensions.
The result is 
\begin{eqnarray}
&&T^g_1(x,x_B)={\alpha_sT_f\over2\pi}\left(\sum_qe_q^2\right)
\left({Q^2e^{\gamma_E}
\over4\pi\mu_0^2}\right)^{-\epsilon/2}\left\lbrace
-{4\over\epsilon}-2\right.\nonumber\\
&&\qquad\qquad
\quad\left.+\left\lbrack\left(1+2x_B{x_B-x\over x^2}\right)\left(
{4\over\epsilon}+4-\log{x_B-x\over x_B}\right)-2\right
\rbrack \log{x_B-x\over x_B}\right\rbrace\\
&&\qquad\qquad\qquad\qquad\qquad\quad\qquad\qquad\qquad\qquad\qquad\qquad+
\left(x_B\rightarrow-x_B\right)\nonumber
\end{eqnarray}
for the symmetric structure and\footnote{For the 
antisymmetric structure, the easiest way
to get the result is by explicitly taking
the photon and gluon indices in the $1$- and $2$-directions
and interpreting the result as multiplied
by $i\lambda\epsilon^{-+12}$.}  
\begin{eqnarray}
&&S^g_1(x,x_B)={\alpha_sT_f\over2\pi}\left(\sum_qe_q^2\right)
\left({Q^2e^{\gamma_E}
\over4\pi\mu_0^2}\right)^{-\epsilon/2}\nonumber\\
&&\qquad\qquad\quad\times\left
\lbrace\left\lbrack\left(1+2{x_B-x\over x}\right)\left(
{4\over\epsilon}+4-\log{x_B-x\over x_B}\right)-2\right
\rbrack\log{x_B-x\over x_B}\right\rbrace\\
&&\qquad\qquad\qquad\qquad\qquad\qquad\qquad\qquad\qquad\qquad\qquad
-\left(x_B\rightarrow-x_B\right)\nonumber
\end{eqnarray}
for the antisymmetric.  Once again, we see infrared divergences.  These
divergences are intimately related to those in the quark sector,
as we will see in the next section.

\section{Infrared Divergences and Parton Distributions}
\label{partondistdis}

As we have seen, DIS on onshell massless quark and gluon targets 
contains infrared divergences which do not cancel in `physical' amplitudes.
One might say, ``Who cares? These amplitudes are not physical anyway, so
it's not important that they are unavoidably divergent.''.  
However, the physical arguments of last section
indicate that these amplitudes contribute to 
physical scattering processes involving hadrons.
The relation between physical
and partonic amplitudes involves probability
functions which are certainly not expected to diverge.
Furthermore,
since non-abelian nature of QCD contributes only trivially to the 
calculations of the last section, there is no way that our calculation
can `know' that the process isn't physical.  In fact,
this problem also exists in QED where the process {\it is} physical.
How, then, can we make sense of these divergences?

In QED, one argues that the quantity we have calculated is
not really physical.  If we were to consider an actual experiment,
we would have no way to distinguish an electron of momentum 
$k$ from an electron-photon {\it system} 
of momentum $k$ in which the photon
carried either a very small (soft) momentum or moved collinear to
the electron.  These two situations are considered separate 
in the calculation.  If one wants to compare with experiment,
one should add the cross-section for these processes 
to this one.  It is from precisely these regions of 
momentum space that the infrared divergences 
come.\footnote{It is also precisely these regions of 
momentum space that one cannot properly take into
account in QCD perturbation theory.  
The $\beta$-function
of QCD implies that the coupling will {\it grow} 
as $\mu^2$ is {\it decreased} until it no longer makes sense
to use perturbation theory.  Of course, at that 
point we don't know precisely what the behavior of the $\beta$-function
is because we cannot calculate it.  
However, one thing is sure - in the regions associated
with the infrared divergences of the last section 
the dynamics of QCD is nonperturbative.  This 
fact will be addressed below.}  

Since infrared divergences come from regions of
soft or collinear momentum, they are associated
mainly with long time scales.    
The arguments of last section then 
imply that these regions are independent
of the hard scattering.  They correspond
to processes in which, for example, a quark splits into
another quark and a collinear gluon long
before the scattering.  In effect, it
is only the second quark which participates in
the hard scattering.\footnote{This analysis
is complicated by diagrams like Fig. 2.2c.  However, 
as we will see below, its inclusion does little
more than ensure the gauge-invariance of the 
distributions $f_{a/T}$.}  This fact is in direct
conflict with the definition of $w_a^{\mu\nu}$.
These quantities were {\it defined} as that part
of the high-energy scattering which involves a parton
of type $a$.  The above discussion
implies that the object we calculated 
contains a piece in which the parton relevant
to the scattering
is of a different type. 
Apparently, this new type of parton can be found
inside our old parton with a certain probability.
This is reminiscent of the discussion 
of last section.  There, we considered the 
target as a loosely (over the relevant time scale)
bound collection of partons and discovered that
the scattering takes place with each parton
incoherently.  Here, we see that by probing at
a large scale the partons themselves can break up
into other partons.  This suggests that
we consider our partons as targets in the 
sense of Equation (\ref{neanfact}) :
\begin{equation}
T_a^{\mu\nu}=\sum_{a'}\,f_{a'/a}\; t^{\mu\nu}_{a'}\;\; .
\label{partneanfact}
\end{equation}
In the last section, we calculated the left-hand-side of this
equation rather than the coefficients $t_a^{\mu\nu}$
on the right-hand-side.  The nature of these coefficients
implies that they {\it cannot} be calculated directly.  Our
procedure for perturbative calculations necessarily
involves integration over all portions of momentum space,
which will lead unavoidably to contributions from low-energy.
The functions $f_{a'/a}$, on the other hand, involve
no explicit momentum cutoff.  It is merely expected that
they do not receive corrections from asymptotically
large momentum scales.  Unfortunately, the regions
in which they are expected to be generated are nonperturbative
in nature.  We have already included contributions
from these regions in $T^{\mu\nu}_a$.  

Although
these contributions have been calculated incorrectly,
the method we used can be extended {\it consistently}
to incorrect perturbative calculations of other quantities,
$f_{a'/a}$ in particular.  Since the incorrect contributions
are calculated in the same way, we can use Equation (\ref{partneanfact})
to cancel them out and extract correct expressions
for $t^{\mu\nu}_{a}$ order by order in perturbation theory.
It is these quantities (rather, the imaginary parts of
their form factors) which appear in the physical
cross-section and will be free of divergences.
A formal proof of this statement involves 
a complete classification of the regions 
which generate infrared divergences in DIS and 
an understanding of the relationship between 
these regions and the regions which generate the 
parton distributions.  An outline of
such a proof appears in the next section.  For now,
let us see explicitly how it works at one-loop order.
Along the way, we will learn a great deal about the 
parton distributions.

In order to calculate parton distribution
functions in perturbation theory, we must
write them as matrix elements.  The probability 
of finding a quark of flavor $q'$ and momentum
$xp$ in a quark state of flavor $q$ and momentum $p$ is 
the same as the probability that I can {\it remove}
a quark of flavor $q'$ and momentum $xp$\footnote{The 
final expression we will use to define 
$f_{q'h'/qh}(x)$ actually receives contributions from all 
partons whose momentum $k$ satisfies $n\cdot k=x$.  
The other components of $k$ give
subleading contributions 
to the hard scattering.  These kinds of 
corrections are discussed in Chapter \ref{htwist}.} from a 
quark state of flavor $q$ and momentum $p$ and be 
left with a physical\footnote{In calculations
with partonic external states, the word `physical'
is used to mean physical in the absence of confinement.
I.e. a transversely polarized onshell gluon is a `physical'
state while an offshell quark is not.  Since
ghosts are nothing but mathematical artifacts, they
cannot be present in `physical' states.}
state :
\begin{eqnarray}
f_{q'h'/qh}(x)&\sim&\sum_X\;{\left\langle ph\right|\psi_{q'}^\dag(0)
\left|X\right\rangle
\left\langle X\left|\psi_{q'}(0)\right|ph\right\rangle}
\delta(1-x-n\cdot p_X)\nonumber\\
&\sim&\int{d\lambda\over2\pi}e^{-i\lambda x}
\left\langle ph
\right|\psi_{q'}^\dag(\lambda n)\psi_{q'}(0)
\left|ph\right\rangle\;\; .
\label{dumbdist}
\end{eqnarray}
This expression can be viewed as an impressionistic
view of the distributions functions at best since 
its derivation completely ignores the requirement of
gauge invariance and the presence of 
spinor indices on the quark fields.  
The latter
complaint concerns dependence on
$h'$.\footnote{Parity invariance
implies that we can take $h=+1/2$ without
loss of generality.}
Since we wish to consider both values, 
we will leave the spinor indices free for now,
but multiply $\psi^\dag$ by $\gamma^0$ to form a 
true Dirac index.
Achieving gauge invariance is also quite
simple.  According to Appendix \ref{sun},
our distribution is in need of a gauge link
between $0$ and $\lambda n$.  The path 
the gauge link traverses should be along 
$n^\mu$ to minimize its dependence on
$\lambda$.  In this way, we arrive
at the expression
\begin{equation}
\left({\cal M}_{q'/q}\right)_{\alpha\beta}(x)
=\int{d\lambda\over2\pi}e^{-i\lambda x}
\left\langle ph
\left|\overline\psi^{q'}_\beta\left(\lambda n\right)
{\cal G}_n\left(\lambda n,0\right)
\psi^{q'}_\alpha\left(0\right)\right|ph\right\rangle\;\; . 
\label{qdistmat}
\end{equation}

Although $\cal M$ is the minimal gauge invariant extension of 
(\ref{dumbdist}), its interpretation is quite different.
The first manipulation changes it from a 
probability to a
density matrix.  This allows us to consider both 
helicity states simultaneously, much as considering
$f_{a'/a}$ as a function of $x$ allows us to combine
a continuously infinite number of probabilities into
one probability {\it function}.  Our density 
matrix can readily be split into two probability 
functions with the help of some projection operators.
However, the explicit form of the gauge link
\begin{equation}
{\cal G}_n\left(\lambda n,0\right)\equiv{\rm P}\left\lbrace
e^{-ig\int^\lambda_0 n\cdot{\cal A}(\zeta n) d\zeta}\right\rbrace\;\; 
\end{equation}
reveals that the second manipulation has changed the 
{\it nature} of our simple distribution.  Rather than the
removal and subsequent replacement of a quark, 
$\cal M$ describes a correlation between
quark and gluon fields.  This complication is
unavoidable in gauge theories because 
quarks do not make sense by themselves.  The 
fact that a quark field transforms under a gauge
transformation implies that `the probability of
finding a quark' is itself not a gauge invariant concept.
However, once we choose a gauge this statement has
meaning.  In particular, the gauge restriction $n\cdot{\cal A}=0$
makes our gauge link trivial.  In this gauge, we are
free to interpret $\cal M$ as a true quark density 
matrix.  Unfortunately, the difficulties associated
with this singular gauge encourage us to make a less
controversial choice.  Nevertheless, its existence has taught
us that the gluons appearing in the gauge link
are unphysical in the sense that they can be 
removed by a gauge rotation.\footnote{This fact is 
critically dependent on the choice of path 
implicit in the definition of the gauge link.
{\it Any} other choice will necessarily
introduce physical gluons, destroying the 
desired interpretation of $\cal M$.}  Since they
are unphysical, we can view them as gauge 
artifacts and continue to interpret 
$\cal M$ as a quark density matrix.  

The gluons fields in the gauge link
are evaluated at space-time points between
the two quark fields.  These contributions
are generated in DIS by diagrams like
Fig. 2.2c in which a gluon interacts with the 
quark {\it during} the scattering.  Although the arguments
above indicate that there is not enough
time for this to occur, the inclusion of this
graph is required by gauge invariance.  
Since all of our arguments 
concerning factorization have depended 
critically on the idea that no low-energy QCD 
interactions can happen during the scattering,
this process could potentially be very 
dangerous.  The offensive contribution 
is of the form
\begin{equation}
\gamma^\mu{\!\!\not\!p\,+\!\!\not\!q\over(p+q)^2}\gamma^\alpha
{y\!\!\not\!p\,+\!\!\not\!q\over(yp+q)^2}\gamma^\nu
\end{equation}
in the collinear region, where the 
gluon has momentum $(1-y)p$.  In the Bjorken
limit, the leading contribution has Dirac
structure $\gamma^\mu\!\!\not\!n\,\gamma^\alpha\!\!\not\!n\,\gamma^\nu$.
This structure vanishes unless $\alpha=-$, so leading
behavior is associated with the field 
${\cal A}^+=n\cdot\cal A$.\footnote{Remember that 
$\gamma^\alpha$ came from the interaction 
$\not\!\!\!{\cal A}={\cal A}^+\gamma^-+{\cal A}^-\gamma^+
-\vec{\cal A}_\perp\cdot\vec\gamma_\perp$.}
This implies that our troublesome 
graph does not contribute to the leading
behavior in light-cone gauge.  Furthermore, since the 
gluon momentum is proportional to $p^\mu$, 
the leading region is equivalent to one in
which the gluon field has been replaced by
its momentum.  Writing
\begin{equation}
(1-y)p_\alpha\gamma^\mu{\!\!\not\!p\,+\!\!\not\!q
\over(p+q)^2}\gamma^\alpha
{y\!\!\not\!p\,+\!\!\not\!q\over(yp+q)^2}\gamma^\nu
\sim\gamma^\mu{y\!\!\not\!p\,+\!\!\not\!q\over(yp+q)^2}\gamma^\nu
-\gamma^\mu{\!\!\not\!p\,+\!\!\not\!q\over(p+q)^2}\gamma^\nu\;\; ,
\end{equation}
it becomes obvious that this gluon insertion does not
constitute a nontrivial contribution to the scattering
in the infrared region.  In reality,
this graph simply represents a combination of 
scattering off of a quark of momentum $yp$ (just the quark)
and a quark of momentum $p$ (coherent scattering
off of the quark and the gluon combined).
We will see in the next section that this
behavior generalizes to any number of 
gluons to all orders in $\alpha_s$
as a consequence of current conservation
in QCD.  Returning to the parton distribution,
we see that the gluon fields in the gauge link
are generated by graphs like these.  They are
required by gauge invariance, but the same 
principle also forbids them from having a nontrivial
effect on the hard scattering.  Nonetheless,
they are present and must be taken into 
account in any nonsingular gauge.

\begin{figure}
\epsfig{figure=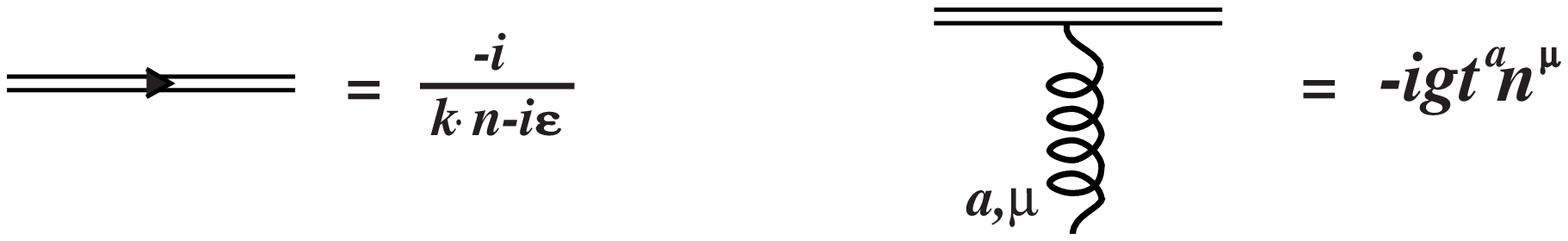,height=2cm}
\caption{Feynman rules for eikonal line 
propagation.}
\label{fig13}
\end{figure}

Now that we have unambiguously defined what
we mean by a parton distribution function,
we would like to calculate it in perturbation
theory.  Once again, we find ourselves in
an awkward situation because of the lack of a 
time-ordered product.  The trick we
used in the last section cannot be applied here
because the term we would like to add is not strictly
zero.  More importantly, the gauge link will not
permit this sleight-of-hand.  However, 
we can massage this quantity into a more useful form 
by writing
\begin{eqnarray}
\Psi(\lambda n)&\equiv&
{\cal G}_n(\infty,\lambda n)\psi(\lambda n)\nonumber\\
{\cal M}_{q'/q}^{\alpha\beta}(x)&=&\int{d\lambda\over2\pi}e^{-i\lambda x}
\left\langle ph\left|\overline\Psi^\beta_{q'}\left(\lambda n\right)
\Psi^\alpha_{q'}\left(0\right)\right|ph\right\rangle\\
&=&\int{d\lambda\over2\pi}e^{-i\lambda (x+n\cdot p_X-1)}\sum_X
\left\langle ph\left|\overline\Psi^\beta_{q'}
\left(0\right)\right|X\right\rangle
\left\langle X\left|\vphantom{\overline\psi^\beta}
\Psi^\alpha_{q'}\left(0\right)
\right|ph\right\rangle\;\; .\nonumber
\end{eqnarray}
The fields $\Psi$ are gauge invariant in the sense that 
gauge transformations at infinity can be taken as trivial in
perturbation theory.\footnote{The nontrivial topology
of our gauge group will cause problems here if we intend to 
do a calculation outside of perturbation theory.}  Since the 
0-component of $n^\mu$ is positive, the path-ordering of the exponential
is equivalent to a time-ordering operation in the matrix element on the 
right.  This exponential represents a coherent
sum of gluon fields propagating from $0$ to
$\infty$.  Coherent propagation of this kind
is known as an {\it eikonal} line and plays 
an important role in proofs of factorization.
The eikonal propagation itself is represented in Feynman 
diagrams as a double line.  The Feynman rules
for its propagation and interaction with gluon fields
are shown in Figure 2.4.  

The calculation of
$\cal M$ is now straightforward :
we simply draw all Feynman diagrams corresponding to the 
matrix element on the right
for a given final state, sum
their contributions, multiply by the complex conjugate\footnote{
To be explicit, I should mention that the complex conjugate
must be multiplied by $\gamma^0$ from the right.},
and sum over final states.
For quark external states, the necessary diagrams up to order 
$\alpha_s$ are shown in 
Figure 2.5.
At this order there are only two final states -
the vacuum and the one gluon state.  The sum over
gluon final states involves an integration\footnote{in 
$d$ dimensions} over
onshell positive-energy momenta and a sum over
polarization states.  This integration is
greatly simplified by the $\delta$-function
constraint on the gluon's $+$ momentum imposed
by the $\lambda$ integration.
The polarization sum in Feynman's 
gauge is given by $-g_{\mu\nu}$.

\begin{figure}
\label{fig14}
\epsfig{figure=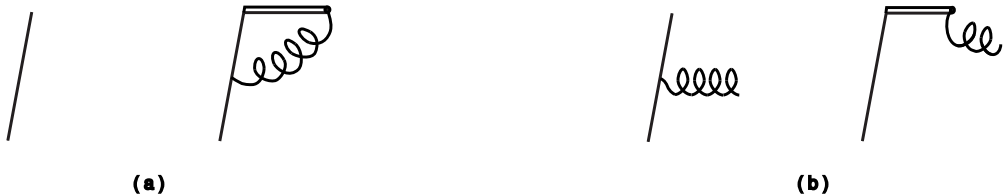,height=2.5cm}
\caption{Contributions to the `quark distribution
in a quark' at (a) leading and (b)-(d) next-to-leading order in QCD.
The final state in (a) and (b) is the vacuum, 
while (c) and (d) represent gluonic external states.}
\end{figure}

The only subtlety left is that all corrections to this
object seem to be exactly zero in dimensional regularization 
due to the lack of scale.  As discussed
in Appendix \ref{dimregapp}, this zero represents a cancelation 
between infrared and ultraviolet divergences.  However,
we really wish to consider a {\it renormalized} matrix element
defined at a certain scale $\mu^2$ to be free of 
ultraviolet divergences.\footnote{Strictly speaking,
we {\it cannot} take $p^2\rightarrow 0$ until we renormalize
our matrix elements because this procedure leads to 
{\it unregularized} infrared divergences.  Until our
distribution is ultraviolet finite, we need
$\epsilon>0$ to regularize ultraviolet divergences.
In this region, $\lim_{p^2\rightarrow0}(p^2)^{-\epsilon/2}$
is certainly {\it not} zero.  However, once the 
ultraviolet divergences have been removed, we
are free to take $\epsilon<0$ and $p^2\rightarrow0$
without ambiguity.}  This object is formally defined 
through its moments, which will be introduced
in Section \ref{opedis}.  The
local operators they are related to can be renormalized
with a $Z$-factor as with the quark-gluon current 
$\overline\psi\!\not\!\!\!{\cal A}\,\psi$ in the 
last chapter.  However, at this order we can 
renormalize simply by subtracting the ultraviolet
divergence in a particular scheme.

To isolate the ultraviolet part of the divergence,
we regularize the infrared 
by taking $p^2$ vanishingly small 
but finite.\footnote{This amounts to 
neglecting it everywhere except in the 
denominators.}  This has the effect of postponing the 
cancelation of infrared and ultraviolet divergences
until the ultraviolet divergence can be identified and 
removed.  In addition to the direct subtraction,
division by $Z_F$ is necessary to
properly normalize the external quark 
states.\footnote{In renormalized
perturbation theory, this factor is required
to renormalize the external states; in unrenormalized
perturbation theory, it renormalizes the 
quark fields.}  
In the $\rm\overline{MS}$ scheme, the full
result is\footnote{The divergent integral here will be 
discussed below.  For now, we just treat it with the 
love and care it requires as an ill-defined mathematical
object.}
\begin{eqnarray}
&&{\cal M}_{q'/q}(x)=
\left\lbrace\delta(1-x)-{\alpha_sC_F\over2\pi}{2\over\epsilon}
\left({\mu^2e^{\gamma_E}\over4\pi\mu_0^2}\right)^{-\epsilon/2}
\right.\nonumber\\
&&\qquad\qquad\qquad\qquad\times\left\lbrack{1+x^2\over1-x}
-\delta(1-x)\int_0^1\,dy
{1+y^2\over1-y}\right\rbrack\\
&&\qquad\qquad\qquad\qquad\qquad\qquad
\left.\vphantom{\left({\mu^2e^{\gamma_E}\over4\pi\mu_0^2}\right)^{-\epsilon/2}}
+{\cal O}(\alpha_s^2)
\right\rbrace \delta_{qq'}u_q(p,h)\overline u_q(p,h)\;\; .\nonumber
\end{eqnarray}
Since 
$u(p,h)\overline u(p,h)=(1+2h\gamma_5)\!\not\!p/2$,
we can split the density matrix into 
two scalar pieces :
\begin{eqnarray}
\label{densitysplit}
\left({\cal M}_{q'h'/qh}\right)_{\alpha\beta}&=&{1\over2}f_{q'/q}(x)
\left(\not\!p\right)_{\alpha\beta}+\tilde f_{q'/q}(x)
\left(\gamma_5h\!\not\!p\right)_{\alpha\beta}\;\; ;\\
\label{unpolsep}
f_{q'/q}(x)&=&{1\over2}\int{d\lambda\over2\pi}e^{-i\lambda x}
\left\langle 
ph\left|\overline\psi^{q'}\!\left(\lambda n\right)
{\cal G}_n\left(\lambda n,0\right)\!\!\not\!n\,
\psi^{q'}\!\left(0\right)\right|ph\right\rangle\\
&=&\delta_{qq'}\left\lbrace\delta(1-x)-
{\alpha_sC_F\over2\pi}{2\over\epsilon}
\left({\mu^2e^{\gamma_E}\over4\pi\mu_0^2}\right)^{-\epsilon/2}
\right.\nonumber\\
\label{unpolpert}
&&\qquad\quad\quad\times\left.\left\lbrack{1+x^2\over1-x}
-\delta(1-x)\int_0^1\,dy
{1+y^2\over1-y}\right\rbrack
+{\cal O}(\alpha_s^2)
\vphantom{\left({\mu^2e^{\gamma_E}\over4\pi\mu_0^2}\right)^{-\epsilon/2}}
\right\rbrace\;\; ;\\
\label{polsep}
\tilde f_{q'/q}(x)&=&{1\over4h}\int{d\lambda\over2\pi}e^{-i\lambda x}
\left\langle ph\left|
\overline\psi^{q'}\left(\lambda n\right)
{\cal G}_n\left(\lambda n,0\right)\!\!\not\!n\,
\psi^{q'}\left(0\right)\right|ph\right\rangle\\
&=&\delta_{qq'}\left\lbrace\delta(1-x)-
{\alpha_sC_F\over2\pi}{2\over\epsilon}
\left({\mu^2e^{\gamma_E}\over4\pi\mu_0^2}\right)^{-\epsilon/2}
\right.\nonumber\\
\label{polpert}
&&\qquad\quad\quad\left.\times\left\lbrack{1+x^2\over1-x}
-\delta(1-x)
\int_0^1\,dy{1+y^2\over1-y}\right\rbrack
+{\cal O}(\alpha_s^2)
\vphantom{\left({\mu^2e^{\gamma_E}\over4\pi\mu_0^2}\right)^{-\epsilon/2}}
\right\rbrace\;\; .
\end{eqnarray}

As mentioned above, the helicity-dependent 
scalar distributions can be projected out
of $\cal M$.  Taking the target 
helicity positive, we have\footnote{The $\not\!n$
here appears simply because it allows us
to project out a scalar distribution
involving single-helicity fields.  Alternatively,
one can simply multiply $\cal M$ by 
$(1\pm\gamma_5)/2$ on the left and work out
the contributing functions.}
\begin{eqnarray}
f_{q'\pm/q+}&=&{1\over2}\;
{\rm Tr}\;\left\lbrack{1\over2}(1\pm\gamma_5){\cal M}\not\!n\,
\right\rbrack\nonumber\\
\label{projectout}
&=&{1\over2}\left(f_{q'/q}(x)\pm\tilde f_{q'/q}(x)\right)\;\; ;\\
\label{unpolquark}
f_{q'/q}(x)&=&f_{q'+/q+}+f_{q'-/q+}\;\; ,\\
\label{polantiquark}
\tilde f_{q'/q}(x)&=&f_{q'+/q+}-f_{q'-/q+}\;\; .
\end{eqnarray}
Hence $f_{q'/q}(x)$ represents the probability
that a quark of flavor $q'$ and momentum $xp$
will be found in a quark of flavor $q$ 
and momentum $p$
regardless of its polarization, and
$\tilde f_{q'/q}(x)$ represents
the {\it difference} between the same helicity
distribution and the opposite 
helicity distribution.  Our result above 
indicates that at one-loop
order the probability of finding a 
quark with opposite helicity is zero.  This
is in agreement with the fact that QCD is a 
vector theory, and as such does not change the 
helicity of its constituents.  However, 
higher order radiative corrections
can generate opposite helicity fermions
through loops.  This simplification is only
an artifact of the low order at which we
work.

Since QCD dynamics are
independent of flavor,\footnote{assuming,
as always, that all quarks are massless}
quarks can only be generated in flavor-singlet
pairs.  Every contribution to $f_{q'/q}(x)$
must be countered by an equal contribution
to an antiquark distribution function
$f_{\overline q'/q}(y)$.  
This suggests that the
integral of the difference between 
quark and antiquark probability distributions
for any given flavor cannot be altered by QCD
dynamics.  This is, in fact, true.
The integral 
\begin{equation}
\int_0^1\,dx\left\lbrack f_{q/T}(x)-f_{\overline q/T}(x)\right\rbrack
\equiv N_{q/T}
\label{qflav}
\end{equation}
is completely independent of QCD dynamics and
returns an integer, the {\it net} number of quarks
of flavor $q$,
for every target $T$.  Of course, in order
to have a rigorous statement, we must first
define the antiquark distribution.
Writing\footnote{Note the position of the gauge 
link.  Its indices are contracted as before.}
\begin{eqnarray}
\label{aqdist}
\left({\cal\overline M}_{\overline q'/q}\right)_{\alpha\beta}(x)
&=&\int{d\lambda\over2\pi}e^{-i\lambda x}
\vphantom{\left\langle ph\left|\overline\psi^{q'}\!\left(\lambda n\right)
{\cal G}_n\left(\lambda n,0\right)\!\!\not\!n\,
\psi^{q'}\!\left(0\right)\right|ph\right\rangle}\left\langle 
ph\left|\psi^{q'}_\alpha\left(\lambda n\right)
\overline\psi^{q'}_\beta\left(0\right)
{\cal G}_n\left(0,\lambda n\right)\right|ph\right\rangle\\
&=&{1\over2}f_{\overline q'/q}(x)
\left(\not\!p\,\right)_{\alpha\beta}+\tilde f_{\overline q'/q}(x)
\left(\gamma_5\,h\!\!\not\!p\,\right)_{\alpha\beta}\;\; ,
\end{eqnarray}
in analogy with Eqs.(\ref{qdistmat}) and 
(\ref{densitysplit}), we see that 
this density matrix
contains the antiquark probability
distributions
\begin{eqnarray}
f_{\overline q'\pm/q+}&=&{1\over2}\;
{\rm Tr}\;\left\lbrack{1\over2}(1\pm\gamma_5)
{\cal\overline M}\not\!n\,\right\rbrack\nonumber\\
\label{helcomp}
&=&{1\over2}\left(f_{\overline q'/q}(x)
\pm\tilde f_{\overline q'/q}(x)\right)\\
f_{\overline q'/q}(x)&=&
f_{\overline q'+/q+}+f_{\overline q'-/q+}\\
\tilde f_{\overline q'/q}(x)
&=&f_{\overline q'+/q+}-f_{\overline q'-/q+}\;\; .
\end{eqnarray}
Here, 
the helicity label does not correspond to the 
helicity of the antiquark itself.  Rather, it
refers to the helicity of the quark related to the 
antiquark.  For example, $f_{\overline q'+/q+}(x)$
represents the probability of finding an 
antiquark of flavor $q'$,
momentum $xp$, and {\it negative} helicity
in a quark of flavor $q$ with momentum
$p$ and positive helicity.  

Since quarks and 
antiquarks are so closely related, it is customary to
combine their contributions into the same 
distribution.  In light of Equation (\ref{aqdist}),
the relationship
\begin{equation}
{\cal M}_{\alpha\beta}(-x)\sim-{\cal\overline M}
_{\alpha\beta}(x)
\end{equation}
is valid in the singular axial
gauge, $n\cdot{\cal A}=0$, provided that we
ignore the anticommutator.\footnote{It can be
shown that this term vanishes 
in a particular quantization scheme.  
See \cite{ColSop}.}  Since we have not at
this point attributed any meaning to the
functions $f_{q'/q}(x)$ on the negative real
axis, we can {\it define}
\begin{eqnarray}
f_{q/T}(-x)\equiv-f_{\overline q/T}(x)\nonumber\\
\label{qaqrelat}
\tilde f_{q/T}(-x)\equiv-\tilde f_{\overline q/T}(x)
\end{eqnarray}
for $x>0$.  This is not in any way a statement 
about the analytic properties of $f_{q'/q}(x)$.\footnote{
The operators which are able to generate {\it connected}
contributions are quite
different in the two cases.}
It is merely a convention to simplify the 
appearance of our equations and acknowledge 
the similarity between quarks and antiquarks.
Using this notation, Eq.(\ref{qflav}) can be written 
in the compact form
\begin{equation}
\int_{-1}^1\,dx\,f_{q/T}(x)=N_{q/T}\;\; .
\label{flavsum}
\end{equation}

The vector nature of QCD allows one to 
construct the sum rule
\begin{equation}
\int_{-1}^1\,dx\,f_{q\pm/T}(x)=N_{q/T}^\pm
\end{equation}
using arguments identical to those that
led to (\ref{flavsum}).  These new quantities
can be used to define the {\it net} number of
positive and negative helicity quarks of 
each flavor in the target $T$.  
According to (\ref{projectout}),
this implies
\begin{eqnarray}
N_{q/T}^++N_{q/T}^-=\int_{-1}^1\,dx\,f_{q/T}(x)
=\phantom{\Delta} N_{q/T}\phantom{\;\; .}\\
\label{deltasigma}
N_{q/T}^+-N_{q/T}^-=\int_{-1}^1\,dx\,
\tilde f_{q/T}(x)=\Delta N_{q/T}\;\; .
\label{ooooo}
\end{eqnarray}
As mentioned above, $N_{q/T}$ is conserved separately
for each flavor.  This fact follows from the 
result\footnote{This identity
is proved in Appendix \ref{quantpoint}; 
it follows from QED gauge invariance.}
\begin{equation}
\partial_\mu\langle{\rm T}\left(\overline\psi_q(x)
\gamma^\mu\psi_q(x)\right)\rangle=0
\end{equation}
for physical matrix elements.

The analogous relation for 
$\Delta N_{q/T}$ is broken by quantum effects 
\cite{anomaly}.  However, these effects
are the same for each species of quark.  
This makes the difference $\Delta N_{q'/T}-\Delta N_{q/T}$
independent of QCD dynamics.  Systematically,
one can define the nonsinglet
distributions
\begin{eqnarray}
f^{i}_{NS/T}(x)&\equiv&\sum_{j=1}^{i}f_{j/T}(x)-if_{i+1/T}\\
\tilde f^{i}_{NS/T}(x)&\equiv&\sum_{j=1}^{i}
\tilde f_{j/T}(x)-i\tilde f_{i+1/T}\;\; .
\end{eqnarray}
In a theory of $n_f$ approximately massless quarks,
the $n_f-1$ nonsinglet distributions
represent purely quark effects.  These distributions
are immune to the dynamical anomaly that
alters $\Delta N_{q/T}$, so the objects
\begin{equation}
\Delta N_{NS/T}^{i}\equiv
\int_{-1}^1\,dx\,\tilde f^{i}_{NS/T}(x)
\end{equation}
are physical quantities which are conserved
to all orders of perturbation theory.\footnote{
This still doesn't put them on the same level as
the $N_{q/T}$.  These quantities are
conserved separately regardless of quark mass
effects, while conservation of the 
nonsinglet $\Delta N$'s relies heavily on 
the assumption of massless quarks.  
Even $N_{q/T}$ is altered
by the flavor-changing weak interactions.
Only the singlet quantity $N_{S/T}$ corresponding
to the total number of quarks in the target 
remains inviolate in the full theory.}
The orthogonal degree of freedom is composed of
the sum of all flavors,
\begin{eqnarray}
f_{S/T}(x)&\equiv&\sum_{j=1}^{n_f}f_{j/T}(x)\\
\tilde f_{S/T}(x)&\equiv&\sum_{j=1}^{n_f}
\tilde f_{j/T}(x)\;\; .
\end{eqnarray} 
This singlet\footnote{`Singlet' and `nonsinglet' 
refer to the transformation properties under the 
flavor group, $SU(n_f)$.  Since the gluon
fields are invariant under the action of this
group and the nonsinglet distributions are
not, these two sectors have no business with each
other.  The singlet distribution, on the other
hand...} distribution is not immune to the
effects of the dynamical anomaly that
alters $\Delta N$.  This implies $\Delta N_{S/T}$,
which represents the difference
between the number of positive and negative helicity
quarks irrespective of flavor in the target,
is dynamically dependent.  This
manifests itself in a dependence on
renormalization scale.  Since it
is not loyal to any 
particular flavor, the effects 
of this distribution cannot be
distinguished from those of the gluon 
field.  As we will see, this fact is 
intimately related to the presence of
infrared divergences in the gluon amplitude.

The soft physics relevant
to gluon contributions to DIS comes in two different
forms.  On one hand, gluons can fluctuate into quark-antiquark
pairs which then participate in the hard scattering. 
In this case, the fluctuation occurs long
before the hard scattering and cannot truly 
be accounted for perturbatively.
This leads to a term of the form
$f_{q/g}\,t^{\mu\nu}_q$ in $T_g^{\mu\nu}$.
The distributions relevant to this contribution 
are contained in Equation (\ref{qdistmat}) 
and its anti-counterpart (\ref{aqdist}), with
gluonic external states.  Defining\footnote{Note
the extra factor of 1/2 accompanying $\tilde f$.
This compensates for the difference in spin between
gluon and quark.}
\begin{eqnarray}
{\cal M}_{q/g\eta}(x)&=&{1\over2}f_{q/g}(x)\not\!p
+{1\over2}\tilde f_{q/g}(x)\gamma_5(\eta\!\!\not\!p\,)\;\; \\
{\cal\overline M}_{\overline q/g\eta}(x)&=&
{1\over2}f_{\overline q/g}(x)\not\!p
+{1\over2}\tilde f_{\overline q/g}(x)
\gamma_5(\eta\!\!\not\!p\,)\;\; ,
\end{eqnarray}
where $\eta=\pm1$ is the gluon helicity,
we have
\begin{eqnarray}
f_{q/g}(x)&=&-{\alpha_sT_F\over2\pi}{2\over\epsilon}
\left({\mu^2e^{\gamma_E}\over4\pi\mu_0^2}\right)^{-\epsilon/2}
\left\lbrack x^2+(1-x)^2\right\rbrack\\
\tilde f_{q/g}(x)&=&-{\alpha_sT_F\over2\pi}{2\over\epsilon}
\left({\mu^2e^{\gamma_E}\over4\pi\mu_0^2}\right)^{-\epsilon/2}
\left\lbrack x^2-(1-x)^2\right\rbrack\\
f_{\overline q/g}(x)&=&\phantom{-}f_{q/g}(x)\\
\tilde f_{\overline q/g}(x)&=&-\tilde f_{q/g}(x)\;\; .
\end{eqnarray}

On the other hand,
gluons can enjoy their own hard scattering
coefficient.  Here, the fluctuation into
a quark-antiquark pair occurs on the same time
scale as the hard interaction, creating an effective
photon-gluon coupling.  This contribution will
have the form $f_{g/g}\,t^{\mu\nu}_g$.  The parton
distribution which appears here must be expressed as
a matrix element of a gluonic operator.  The road
from idea to matrix element is complicated in this case by the
nonlinear transformation law for the gluonic fields.
Here, we need somewhat more than a gauge link to 
form a gauge invariant distribution.  In fact,
the only local object which annihilates the 
one-gluon state and has linear gauge transformation
properties is the field strength ${\cal F}^{\mu\nu}$.
This implies that the matrix element
\begin{eqnarray}
      G^{\mu\nu\alpha\beta}_{g/a}(x)&=&\int{d\lambda\over 
2\pi}e^{-i\lambda x}\left\langle p\eta\left|
{\cal F}^{\mu\nu}\left(\lambda n\right)
{\cal G}_n\left(\lambda n,0\right)
{\cal F}^{\alpha\beta}\left(0\right)
\right|p\eta\right\rangle\nonumber \\
\label{gluondensitymat}
&=& -xf_{g/a}(x)\left(g^{\mu\alpha}p^\nu 
p^\beta -g^{\mu\beta}p^\nu p^\alpha +g^{\nu\beta}
p^\mu p^\alpha -g^{\nu\alpha}p^\mu p^\beta\right) \\ 
&&+i\eta x\tilde f_{g/a}(x)\left(\epsilon^{\mu\alpha\gamma\delta}
p^\nu p^\beta -\epsilon^{\mu\beta\gamma\delta}
p^\nu p^\alpha+\epsilon^{\nu\beta\gamma\delta}
p^\mu p^\alpha -\epsilon^{\nu\alpha\gamma
\delta}p^\mu p^\beta\right)n_\gamma p_\delta\;\; , \nonumber
\end{eqnarray}
contains the vacuum as an intermediate state.
${\cal G}$ is the same gauge link as that in
the quark distribution operator, with 
the generators $T^a$ appearing in the gluon fields 
interpreted in the adjoint representation.\footnote{
See Appendix \ref{sun}.  These indices have been 
suppressed as in the quark distribution operators.}  
The normalization has been chosen 
so that the scalar distributions,\footnote{
The dual field strength is defined by
$\tilde{\cal F}^{\mu\nu}\equiv{1\over2}\,\epsilon^{\mu\nu\alpha\beta}
{\cal F}_{\alpha\beta}$.}
\begin{eqnarray}
      f_{g/a}(x)=-{1\over \;2x\;}
\int {d\lambda\over 2\pi}e^{-i\lambda x}
\langle p\eta| {\cal F}^{\mu\alpha}
\left(\lambda n\right)
{\cal G}_n\left(\lambda n,0\right)
{\cal F}_{\;\;\alpha}^\nu\left(0\right)|
p\eta\rangle n_\mu n_\nu\, \ , \\
    \tilde f_{g/a}(x)=-{i\over 2\eta x}\int{d\lambda\over 2\pi}
e^{-i\lambda x}\langle p\eta| 
{\cal F}^{\mu\alpha}\left(\lambda n\right) 
{\cal G}_n\left(\lambda n,0\right)
\tilde{\cal F}_{\;\;\alpha}^\nu
\left(0\right)|p\eta\rangle n_\mu n_\nu\, \ ,
\end{eqnarray}
which can be isolated by contraction, 
have the leading behavior
\begin{eqnarray}
\label{gluenorm}
f_{g/g}(x)={x\over2}\left\lbrack\delta(1-x)+\delta(1+x)\right\rbrack
+{\cal O}(\alpha_s)\phantom{\;\; .}\nonumber\\
\tilde f_{g/g}(x)={x\over2}\left\lbrack\delta(1-x)
-\delta(1+x)\right\rbrack+{\cal O}(\alpha_s)\;\; .
\end{eqnarray}

As before, 
These distributions can be calculated 
in perturbation theory by introducing new fields
with eikonal propagation and inserting a complete
set of states.\footnote{Since gluons are bosons, the creation and
annihilation operators in each gluon field
are of the same species.  This leads to the 
$\delta(1+x)$ terms above, which do not 
appear in the fermion 
case because the action of antiquark creation
and annihilation operators cannot generate 
connected diagrams at this order.  Note that
this bose symmetry can be taken into account
simply by requiring (anti)symmetry in the argument
$x$ for the (un)polarized distribution.  
This is a general property of the matrix
element (\ref{gluondensitymat}).}
The physical polarization vectors for gluons moving
in the 3-direction are given by $\varepsilon^\mu_\eta
=(0,1,i\eta ,0)/\sqrt{2}$.
Since this distribution appears only with $t^{\mu\nu}_g$,
a quantity that starts with $\alpha_s$,
we do not need the next-to-leading
behavior in our calculation.  This also 
implies that we need not calculate the `gluon
distribution in a quark', $f_{g/q}$, since this
object can only contribute to $T^{\mu\nu}_q$ at order
$\alpha_s^2$. 

To see the connection 
between the 
above distributions and the idea of
removing one gluon, we turn again to the singular
light-cone gauge.  As before, the gauge link
is trivial in this gauge.  Furthermore, 
this choice reduces 
$n_\alpha{\cal F}^{\alpha\mu}$ to $n\cdot\partial{\cal A}^\mu$. 
Introducing a complete set of intermediate states,
we see that $f_{g/a}(x)$ becomes\footnote{neglecting the 
bose symmetry}
\begin{eqnarray}
f_{g/a}(x)&=&-{1\over2x}\int{d\lambda\over2\pi}e^{-i\lambda x}
\sum_X\left\langle ph\left|in\cdot\partial{\cal A}_a^\alpha
\left(\lambda n\right)\right|X\right\rangle
\left\langle X\left|in\cdot\partial{\cal A}^a_\alpha
\left(0\right)\right|ph\right\rangle\nonumber\\
\label{whatwewant}
&=&x\sum_X\delta\left(p_X\cdot n-(1-x)\right)
\left\langle ph\left|{\cal A}_a^\alpha(0)\right|X\right\rangle
\left\langle X\left|{\cal A}_a^\beta(0)\right|ph\right\rangle
\left(-{g_{\alpha\beta}\over2}\right)
\end{eqnarray}
Aside from the factor of $x$, this is exactly what we
would have written down for the unpolarized gluon
distribution if we did not have to worry about
gauge invariance.\footnote{The factor $-g_{\alpha\beta}/2$
averages over transverse gluon polarizations with the
weight $+1$; remember that in axial
gauge ${\cal A}\cdot{\cal A}=
-\vec{\cal A}_\perp\cdot\vec{\cal A}_\perp$.}
However, a proper normalization must take the issue
of gauge invariance into account.
If we wish to normalize all of our distributions
in an unambiguous way, it is necessary to relate them
to a physical observable.  The conventional choice
is the total $+$ momentum carried by a certain parton in
our target.  Interpreting the parton distributions
as probabilities, we would conclude that this
object is simply a sum over the probabilities
weighted by the momentum carried by each parton :
\begin{equation}
n\cdot P_{a/T}=\int_0^1\,dx\,xf_{a/T}(x)\;\; .
\label{partdistnorm}
\end{equation}
On the other hand, the operator representing the momentum 
carried by various parton species can be obtained from the
stress-energy tensor of QCD.  According to the 
analysis of Appendix \ref{quantpoint}, we have
\begin{eqnarray}
\label{momentumenergyconn}
{\hat P}^\mu&=&\int\,d^{\,3}x\,\Theta^{0\mu}(x)\\
\label{QCDemtensor}
\Theta^{\mu\nu}&=&\Theta^{\mu\nu}_q+\Theta^{\mu\nu}_g\\
\label{qenergydis}
\Theta^{\mu\nu}_q&=&\overline\psi\gamma^{(\mu}
\stackrel\leftrightarrow{i{\cal D}}
\vphantom{\gamma}^{\;\nu)}\psi\\
\label{genergydis}
\Theta^{\mu\nu}_g&=&{1\over4}g^{\mu\nu}{\cal F}^2
-{\cal F}^{\mu\alpha}{\cal F}^\nu_{\;\;\alpha}\;\; .
\end{eqnarray}
Lorentz invariance implies
\begin{equation}
\left\langle P\right|\Theta^{\mu\nu}_{q,g}\left|P\right\rangle
=2a_{q,g}P^\mu P^\nu+2b_{q,g}M^2g^{\mu\nu}\;\; .
\label{qgemmatrixele}
\end{equation}
Since $|P\rangle$ is an eigenstate of $\hat P^\mu$,
we have\footnote{Our states are normalized so that
$\langle P'|P\rangle=2P^0(2\pi)^3\delta(\vec P'-\vec P\,)$.}
\begin{eqnarray}
P^\mu\left(2P^0\right)\left(2\pi\right)^3
\delta^{(3)}(0)&=&
\left\langle P\right|\int\,d^{\,3}x\,\Theta^{0\mu}(x)
\left|P\right\rangle\nonumber\\
&=&\left\langle P\right|\Theta^{0\mu}(0)\left|P\right\rangle
\left(2\pi\right)^3\delta^{(3)}(0)\;\; .
\end{eqnarray}
This immediately tells us that 
\begin{equation}
a_q+a_g=1\qquad\qquad b_q+b_g=0\;\; ,
\end{equation}
which promotes the identification of
$a_{q,g}$ with the fraction of momentum carried
by quarks and gluons, respectively.\footnote{The quantity
$b_{q,g}$ should not bother us too much here.
Its presence can be understood in the 
rest frame of the nucleon, where
Equation (\ref{qgemmatrixele}) would otherwise
imply that no species of parton can have nontrivial
momentum dynamics.  These kinds of transverse
effects are suppressed by powers of
$Q^2$ in the Bjorken limit; in any case, the
contribution of $b$ to the $+$ momentum
of the proton for any appreciable value of the 
boost parameter $\Lambda$ is quite negligible.}
These parameters are pure numbers\footnote{ignoring, for the
moment, their dependence on renormalization scale} and
may be obtained in any way we wish.  In particular,
from Equation (\ref{qgemmatrixele}), we see
that
\begin{equation}
a_{q,g}={1\over2}\left\langle P\right|\Theta_{q,g}^{++}
\left|P\right\rangle\;\; .
\end{equation}
Our normalization condition, 
Eq.(\ref{partdistnorm}), then takes the convenient form
\begin{equation}
{1\over2}\left\langle P\right|\Theta^{++}_{q,g}
\left|P\right\rangle
=\int dx\,xf_{q,g/T}(x)\;\; .
\label{normalizationofpartondists}
\end{equation}
It is now easy to see why the `extra' factor
of $x$ adorning the gluon distributions 
is needed.\footnote{A somewhat easier, but less
rigorous way to understand this factor
comes from a different analogy with the quark case.
Above, we have interpreted $\overline\psi$
as the hermitian conjugate to $\psi$.
If instead we think of it as the 
momentum canonically conjugate to $\psi$,
the the analogy with the gluon sector
changes its nature completely.
The canonically conjugate momentum
to the gluon field is $\cal F$, so we
would naively write
\begin{equation}
\langle {\cal F}^{+\alpha}{\cal A}_\alpha\rangle
\end{equation}
for the (unpolarized) gluon distribution.
In order to make this gauge-invariant, we
must eliminate $\cal A$ in favor of $\cal F$.
The `extra' factor of $x$ compensates for the 
necessary derivative.}

Although the correctly normalized gauge-invariant distributions
are given by $f_{g/T}$ and $\tilde f_{g/T}$, 
our perturbative calculation in the last section
was performed with single gluon external 
states.  For these states, distributions of the 
form in Eq.(\ref{whatwewant}) are required.
However, it is the proper distributions
which will survive the transition from
parton external states to physical hadronic ones.
We can account for the difference 
between these two distributions simply
by dividing our perturbative results by
$x$ since
\begin{equation}
\int{d\lambda\over2\pi}e^{-i\lambda x}\left\langle
PS\left|{\cal A}^\mu_\perp(\lambda n){\cal A}^\nu_\perp(0)\right|
PS\right\rangle=\left\lbrack -{2\over d-2}\,{1\over x}\,
g_\perp^{\mu\nu}f_{g/T}(x)+{i\over x}\,
\epsilon^{-+\mu\nu}
\tilde f_{g/T}(x)\right\rbrack
\end{equation}
in the light-cone gauge.

Armed with perturbative expressions for our
parton densities, we are now ready to 
extract the hard scattering coefficients.
Exploiting the fact that we work only
to next-to-leading order, we write
\begin{eqnarray}
T^{\mu\nu}_q&=&\left(T^{\mu\nu}_q\right)^{(0)}+{\alpha_sC_F\over4\pi}
\left\lbrack{2\over\epsilon}\left(
{\mu^2e^{\gamma_E}\over4\pi\mu_0^2}\right)^{-\epsilon/2}
\left(
T^{\mu\nu}_q\right)^{(1)}_{\rm IR}+\left(T^{\mu\nu}_q\right)^{(1)}
\right\rbrack\nonumber\\
f_{q'/q}(x)&=&f^{(0)}_{q'/q}(x)+{\alpha_sC_F\over4\pi}{2\over\epsilon}
\left({\mu^2e^{\gamma_E}\over4\pi\mu_0^2}\right)^{-\epsilon/2}
f^{(1)}_{q'/q}(x)\\
t^{\mu\nu}_{q'}&=&(t^{\mu\nu}_{q'})^{(0)}
+{\alpha_sC_F\over4\pi}(t^{\mu\nu}_{q'})^{(1)}
\end{eqnarray}
for the quark amplitude.
The infrared divergent term in $T_q^{\mu\nu}$
has been designed to include only the infrared divergence
as seen in the $\rm\overline{MS}$ scheme.  For example,
the term 
\begin{equation}
g^{\mu\nu}e_q^2{\alpha_sC_F\over4\pi}\;{3\over x_B-1}\;
{2\over\epsilon}\left(
{Q^2e^{\gamma_E}\over4\pi\mu_0^2}\right)^{-\epsilon/2}
\end{equation}
in $T^{\mu\nu}$ contributes
\begin{equation}
g^{\mu\nu}e_q^2\;{3\over x_B-1}
\end{equation}
to $\left(T^{\mu\nu}_q\right)^{(1)}_{\rm IR}$
and 
\begin{equation}
-g^{\mu\nu}e_q^2\;{3\over x_B-1}\;\log{Q^2\over\mu^2}
\end{equation}
to $\left(T^{\mu\nu}_q\right)^{(1)}$.  
Note that we are considering scattering on
a quark of momentum $p$ rather than $xp$ here.  The 
perturbative coefficient functions are designed 
not to care.  

Equating terms order by order in $\alpha_s$, we see
that Equation (\ref{partneanfact}) implies
\begin{eqnarray}
\label{leadingorder}
\left(T^{\mu\nu}_q\right)^{(0)}=\sum_{q'}\int_{-1}^1 dx\,
f_{q'/q}^{(0)}(x)(t^{\mu\nu}_{q'})^{(0)}(x)\phantom{\;\; .}\\
\label{nexttoleading}
\left(T^{\mu\nu}_q\right)^{(1)}=\sum_{q'}\int_{-1}^1 dx\,
f_{q'/q}^{(0)}(x)(t^{\mu\nu}_{q'})^{(1)}(x)\phantom{\;\; .}\\
\label{requirefact}
\left(T^{\mu\nu}_q\right)^{(1)}_{\rm IR}=\sum_{q'}\int_{-1}^1 dx\,
f_{q'/q}^{(1)}(x)(t^{\mu\nu}_{q'})^{(0)}(x)\;\; .
\end{eqnarray}
Equations (\ref{leadingorder}) and (\ref{nexttoleading})
are used to define $(t^{\mu\nu}_{q'})^{(0)}$
and $(t^{\mu\nu}_{q'})^{(1)}$.
Equation (\ref{requirefact}) is a statement
about the sources of soft physics in our
calculation.  If it is satisfied, all of the 
soft physics relevant to DIS is contained within
the parton distribution functions.  Otherwise, 
other sources of soft physics contribute to this
process.  This situation could imply that one cannot
separate the hard and soft scales.  Fortunately,
this is not the case in inclusive DIS.  Explicit 
substitution of the expressions above into
(\ref{leadingorder}), (\ref{nexttoleading}), and
(\ref{requirefact}) and analogous relations
in the gluon sector, shows that (\ref{requirefact})
is satisfied
for polarized and unpolarized
DIS on quark and gluon targets at order $\alpha_s$.
The results can be expressed for any target
$T$ in the simple form\footnote{The leading
antiquark scattering amplitude is 
just the negative of the leading quark contribution
evaluated at $-x$, so these contributions
can be combined into the same hard scattering
coefficient.} 
\begin{eqnarray}
   T_1(x_B) =\int^1_{-1} {dx\over x} \left[\sum_{q} 
f_{q/T}(x,\mu^2) C_q\left({x\over x_B},{Q^2\over\mu^2}\right)
+f_{g/T}(x,\mu^2) C_g\left({x\over x_B},
{Q^2\over\mu^2}\right) \right]\phantom{\;\; ,} \nonumber \\
S_1(x_B)=  
\int^1_{-1} {dx\over x} \left[\sum_{q}
\tilde f_{q/T}(x,\mu^2) \tilde C_q\left({x\over x_B},{Q^2\over\mu^2}\right)
+\tilde f_{g/T}(x,\mu^2) 
\tilde C_g\left({x\over x_B},{Q^2\over\mu^2}\right) \right]\;\; ,
\label{facdis}
\end{eqnarray}
where\footnote{Note the extra factors of $1/x$
in the gluonic coefficient functions coming
from the difference in definition between the 
properly normalized gluon distributions and 
the distributions relevant to our calculated amplitudes.}
\begin{eqnarray}
C_q\left({x\over x_B},{Q^2\over\mu^2}\right)&=&e_q^2{x\over x_B-x}
-{\alpha_s(\mu^2)C_F\over4\pi}e_q^2
\left\lbrace
\left(3-\log{Q^2\over\mu^2}\right){3x\over x_B-x}-2\log{Q^2\over\mu^2}
\nonumber\right.\nonumber\\
&&\qquad\qquad+\left\lbrack\left({x_B+x\over x}+{2x\over x_B-x}\right)
\left(3-\log{x_B-x\over x_B}-2\log{Q^2\over\mu^2}\right)\right.\nonumber\\
\label{qunpolcoefdis}
&&\qquad\qquad\qquad\left.\left.
-3\left({x_B-x\over x}+{x\over x_B-x}\right)\right\rbrack
\log{x_B-x\over x_B}\right\rbrace\\
&&\qquad\qquad\qquad\qquad\qquad\qquad\qquad
+\left(x_B\rightarrow-x_B\right)\;\; ,\nonumber\\
\tilde C_q\left({x\over x_B},
{Q^2\over\mu^2}\right)&=&e_q^2{x\over x_B-x}
-{\alpha_s(\mu^2)C_F\over4\pi}e_q^2
\left\lbrace
\left(3-\log{Q^2\over\mu^2}\right){3x\over x_B-x}
\nonumber\right.\nonumber\\
&&\qquad\qquad+\left\lbrack\left({x_B+x\over x}+{2x\over x_B-x}\right)
\left(3-\log{x_B-x\over x_B}-2\log{Q^2\over\mu^2}\right)\right.\nonumber\\
\label{qpolcoefdis}
&&\qquad\qquad\qquad\left.\left.+7
\;{x_B-x\over x}-3\;{x\over x_B-x}\right\rbrack
\log{x_B-x\over x_B}\right\rbrace\\ 
&&\qquad\qquad\qquad\qquad\qquad\qquad\qquad
-\left(x_B\rightarrow-x_B\right)\;\; ,\nonumber\\
C_g\left({x\over x_B},{Q^2\over\mu^2}\right)&=&
{\alpha_s(\mu^2)T_f\over2\pi}\left(\sum_qe_q^2\right)
\left\lbrace
2\log{Q^2\over\mu^2}-2\right.\nonumber\\
&&\left.\qquad\qquad+\left\lbrack\left(1+2x_B{x_B-x\over x^2}\right)\left(
4-\log{x_B-x\over x_B}-2\log{Q^2\over\mu^2}\right)\right.\right.\nonumber\\
\label{gunpolcoefdis}
&&\left.\left.\qquad\qquad\qquad\qquad
\vphantom{{Q^2\over\mu^2}}-2\right
\rbrack \log{x_B-x\over x_B}\right\rbrace+
\left(x_B\rightarrow-x_B\right)\\
\tilde C_g\left({x\over x_B},{Q^2\over\mu^2}\right)&=&
{\alpha_s(\mu^2)T_f\over2\pi}\left(\sum_qe_q^2\right)\nonumber\\
&&\qquad\qquad\!\!\times\left\lbrace
\left\lbrack\left(1+2{x_B-x\over x}\right)\left(
4-\log{x_B-x\over x_B}-2\log{Q^2\over\mu^2}\right)\right.\right.\nonumber\\
\label{gpolcoefdis}
&&\left.\left.\vphantom{{Q^2\over\mu^2}}\qquad\qquad\qquad\qquad
-2\right
\rbrack\log{x_B-x\over x_B}\right\rbrace
-\left(x_B\rightarrow-x_B\right)\;\; .
\end{eqnarray}

Renormalization has introduced
violations of the simple scaling behavior 
we saw in the leading order
amplitudes.  The dependence on renormalization
scale is illusory; the coefficient functions
and parton distributions depend separately on
$\mu$ in such a way that the product 
is independent.
However, the dependence of our amplitude
on $Q^2$ is quite real and can be measured
in experiment.  In the absence 
of asymptotic freedom, 
this behavior
would persist even into the Bjorken region so
scaling could never be experimentally observed.
Asymptotic freedom ensures that scaling violations
become smaller as $Q^2$ is
increased.  

Since these scattering coefficients
are completely insensitive to the soft 
infrared physics, (\ref{facdis})
is valid for {\it all} targets.  
In particular, we can derive expressions
for the quantities measured in real
DIS on proton targets simply by taking the 
imaginary part of $T_1$ and $S_1$ as $x_B$
approaches the real axis from below.\footnote{At this
point, we must analytically continue our
expressions to the region $|\,x_B\,|\leq1$.  It
is only in this region that the correspondence
between our amplitude and physical measurements
can be made.}
The parton distribution functions are real, 
so the measured structure functions 
are expressed as convolutions
of the unknown structure functions
with known scattering kernels.  At leading order,
we have simply\footnote{We choose $\mu^2=Q^2$
because this is the most natural value
for it.  The full expression is indeed 
independent - if we calculate to all orders.
Our lack of motivation for this 
task forces us to choose $\mu^2$ in
such a way that corrections are minimized.}
\begin{eqnarray}
F_1(\nu,Q^2)={1\over2}\sum_qe_q^2\left\lbrack 
f_{q/T}(x_B, Q^2)+f_{\overline q/T}(x_B, Q^2)
\right\rbrack\phantom{\;\; .}\\
G_1(\nu,Q^2)={1\over2}\sum_qe_q^2\left\lbrack 
\tilde f_{q/T}(x_B, Q^2)-
\tilde f_{\overline q/T}(x_B, Q^2)\right\rbrack\;\; .
\end{eqnarray}
The interpretation of this result
is clear.  At leading order, the 
photon simply couples to each parton 
completely incoherently.  $F_1$ measures
the total number of charge carriers
with momentum $x_BP$ weighted by the square of the 
charge.  Notice that the 
{\it sum} of the quark and antiquark distributions
is relevant here, rather than the difference,
since both contribute to DIS in the same 
way.  Likewise, $G_1$ measures
the helicity of charged carriers weighted
by squared charge.  Since the 
antiquark distribution has been defined
in such a way that $\tilde f_{\overline q/T}$
measures the net {\it negative} helicity 
of antiquarks, it appears with a sign in this
expression.  The kinematical
restriction $x_B=x$ corresponds to the requirement
that the intermediate quark be `physical'.

A measurement
of the structure functions allows
us to extract the parton distributions
and obtain invaluable information
on nonperturbative hadronic structure.
In particular, integrals
of the structure functions can tell
us how much of a proton's momentum
is carried by quarks or the 
total spin carried by quark helicity
in tritium.  The results of this section allow us to go
even further and predict the 
leading QCD corrections to $F_1$ and 
$G_1$.  All that is needed is the imaginary
parts of the coefficient functions, Eqs.(\ref{qunpolcoefdis}), 
(\ref{qpolcoefdis}), (\ref{gunpolcoefdis}) and
(\ref{gpolcoefdis}).\footnote{
Taking the imaginary part of the logarithms 
in these expressions is somewhat tricky.
For $x<x_B$,
\begin{eqnarray}
\Re e\log{x_B-x\over x_B}&=&\log{x_B-x\over x_B}\nonumber\\
\Im m\log{x_B-x\over x_B}&=&0\;\; ,
\end{eqnarray}
while for $x>x_B$,
\begin{eqnarray}
\Re e\log{x_B-x\over x_B}&=&\log{x-x_B\over x_B}\nonumber\\
\Im m\log{x_B-x\over x_B}&=&-\pi\;\; .
\label{footeqn}
\end{eqnarray}
$-i\pi$ appears rather than $i\pi$ 
because for $x_B$ slightly below the 
real axis, $x_B-x$ is a negative number with a 
negative imaginary part.  Taking the 
branch cut along the negative real axis,
we arrive at (\ref{footeqn}).}

Once the 
parton distributions are known at
a certain scale, their scale evolution is fixed
by perturbative QCD.  This makes the 
study of scaling violations an especially
clean way to view QCD.  In fact, the 
prediction of these violations in DIS  
is one of the most precision tests of 
the theory.

The scale evolution of our distribution
functions is interesting in its
own right.  Physically, it arises 
for the same reason that $\alpha_s(\mu^2)$
runs - things look different at different
scales.  From very far away, I may think that
I am looking at a quark of momentum $p$.
However, if I look closer by increasing my probing
scale, I may find that that quark is really
a quark of momentum $xp$ and a gluon of 
momentum $(1-x)p$.
This line of reasoning leads to the evolution equation
\begin{eqnarray}
\label{primord}
\mu^2{D_a\over D\,\mu^2}f_{a/T}(x,\mu^2)&=&
{\alpha_s(\mu^2)\over2\pi}\sum_{a'}\int_0^1\,dy
f_{a'/T}(y,\mu^2)\int_0^1\,dz P_{a'\rightarrow a}(z)
\delta(x-zy)\\
\label{altpar1}
&=&{\alpha_s(\mu^2)\over2\pi}\sum_{a'}\int_x^1 {dy\over y}P_{a'\rightarrow a}
\left({x\over y}\right)f_{a'/T}(y,\mu^2)\;\; .
\end{eqnarray}
Here, 
$\alpha_sP_{a'\rightarrow a}
(x)d\mu^2/2\pi\mu^2$ represents the 
probability that a parton of type $a'$ and 
momentum $p$ will generate a parton of 
type $a$ and momentum $xp$ due to the infinitesimal
change $d\mu^2$ in scale.  This activity will
add to the distribution $f_{a/T}(x)$.
The `covariant' derivative,
$D$, is designed to include the 
change in $f_{a/T}(x)$ due to partons of 
type $a$ splitting into other partons.
This is known as the endpoint contribution. 
Conceptually, it would seem that this 
term is just be some kind of sum over
probabilities $P_{a\rightarrow a'}$.  However,
classical reasoning does not work here.
In reality, 
we do not have conservation of particles
in a quantum field theory.  The {\it only} conserved
quantities here are those quantum numbers 
associated with some symmetry of our lagrangian.
In this case, the endpoint contributions are fixed by
flavor and momentum conservation.
As mentioned above,\footnote{At this level, it is 
conceptually easier to use the explicit forms of the 
quark and antiquark distributions.  We can always
simplify our expressions later using (\ref{qaqrelat}).}
conservation of flavor implies that
\begin{equation}
\int_0^1dx\,\left\lbrack f_{q/T}(x,\mu^2)-
f_{\overline q/T}(x,\mu^2)\right\rbrack
\end{equation}
is independent of $\mu^2$.  Using
\begin{equation}
\mu^2{D_a\over D\mu^2}\equiv\mu^2{d\over d\mu^2}
+{\alpha_s(\mu^2)\over2\pi}A_a\;\; ,
\end{equation}
we see that this requirement leads to\footnote{In these
kinds of manipulations, it is often easier to
use the primordial form of the evolution equation,
(\ref{primord}), rather than the streamlined
(\ref{altpar1}).  Note that we must be very careful deriving
this result since the integrals involved are singular.
In particular, the form of $P_{q\rightarrow q}(z)$ at leading
order in $\alpha_s(\mu^2)$ leads to an anomalous
dependence of $A_q$ on $x$.}
\begin{equation}
A_q=\int_0^x{dy\over x}\left\lbrack P_{q\rightarrow q}\left({y/x}\right)
-P_{\overline q\rightarrow q}\left({y/x}\right)\right\rbrack\;\; .
\label{flavconsreq}
\end{equation}
Some of the relations
\begin{eqnarray}
P_{q\rightarrow q'}(x)&=&P_{\overline q\rightarrow\overline q'}(x)\\
P_{q\rightarrow\overline q'}(x)&=&P_{\overline q\rightarrow q'}(x)\\
P_{q\rightarrow g}(x)&=&P_{\overline q\rightarrow g}(x)\\
P_{g\rightarrow q}(x)&=&P_{g\rightarrow\overline q}(x)\\
(1-\delta_{qq'})P_{q\rightarrow q'}(x)&=&
(1-\delta_{qq'})P_{q\rightarrow\overline q'}(x)\\
\label{diffflav}
&=&(1-\delta_{qq'})P_{q\rightarrow\overline q}(x)\\
\label{qdiffflav}
P_{q\rightarrow g}(x)&=&P_{q'\rightarrow g}(x)\\
\label{qbdiffflav}
P_{g\rightarrow q}(x)&=&P_{g\rightarrow q'}(x)\\
\label{singletmb}
P_{q\rightarrow q}(x)&=&P_{q'\rightarrow q'}(x)
\end{eqnarray}
are necessary to derive this result.  All
follow from the $CP$ 
and flavor invariance of the QCD lagrangian 
and are valid to all orders in perturbation 
theory.\footnote{Quark mass effects break flavor 
invariance.  As such, they will alter 
relations (\ref{diffflav}), (\ref{qdiffflav}),
(\ref{qbdiffflav}), and (\ref{singletmb}).  
However, this will not
change (\ref{flavconsreq}).}  At the order to which
we work, $P_{q\rightarrow\overline q}$ is zero.  Hence
conservation of flavor requires 
\begin{equation}
A_q=\int_0^x{dy\over x}\,P_{q\rightarrow q}\left({y\over x}\right)\;\; .
\end{equation}

On the other hand, we have calculated $f_{q'/q}(x,\mu^2)$
in perturbation theory.  Manually taking the 
derivative of Eq.(\ref{unpolpert}), we arrive 
at
\begin{eqnarray}
\mu^2{d\over d\mu^2}f_{q/q}(x,\mu^2)&=&{\alpha_s(\mu^2)C_F\over2\pi}
\left\lbrack {1+x^2\over1-x}-
\delta(1-x)\int_0^1dy\,{1+y^2\over1-y}\right\rbrack\nonumber\\
\label{primordpqq}
&=&{\alpha_s(\mu^2)\over2\pi}\int_0^1dydz
\left(C_F\;{1+z^2\over1-z}\right)f_{q/q}(y,\mu^2)\delta(x-zy)\\
&&-{\alpha_s(\mu^2)\over2\pi}\int_0^x{dy\over x^2}\,C_F\;
{x^2+y^2\over x-y}f_{q/q}(x,\mu^2)\;\; \nonumber
\end{eqnarray}
at leading order in $\alpha_s(\mu^2)$.
Comparing with (\ref{primord}), we see that 
\begin{eqnarray}
\label{unpolqsplit}
P_{q\rightarrow q}(x)=C_F\;{1+x^2\over1-x}\;\; ,\\
A_q=C_F\int_0^x{dy\over x^2}\;{x^2+y^2\over x-y}
\end{eqnarray}
at leading order in $\alpha_s$.  In light
of (\ref{flavconsreq}), these two 
quantities are not independent.
Since $P_{q\rightarrow q}(x)$ comes entirely from
gluon intermediate states,
the calculation of vacuum
intermediate states is in some sense 
redundant.  In the light-cone gauge, the absence
of the gauge link causes the diagrammatic calculation
to be incomplete.  In order to obtain the full
result in this gauge, one must 
{\it impose} (\ref{flavconsreq}).\footnote{
This is true for a well-defined, consistent
light-cone gauge choice.  Due
to the similarity between the pole prescription
indicated in Fig. 2.4 and the PV-prescription,
the pathological theory reproduces the 
correct end-point contribution at this order while
ML does not.  However, this does not reconcile
our renormalizability issues with PV.}

We are now in a position to understand the divergent
integral in $A_q$.  Its presence it required to 
{\it cancel} the divergent behavior
of $\int\,P_{q\rightarrow q}$.  To display this
explicitly, one often combines the two terms :
\begin{eqnarray}
\mu^2{d\over d\mu^2}f_{a/T}(x,\mu^2)&=&\sum_{a'}
{\alpha_s(\mu^2)\over2\pi}\int_x^1{dy\over y}\left\lbrack P_{a'\rightarrow a}
\left({x\over y}\right)\right.\nonumber\\
\label{altpar}
&&\left.\qquad\qquad\qquad-A_a\delta_{aa'}
\delta\left(1-{x\over y}\right)\right\rbrack\;f_{a'/T}(y,\mu^2)\\
&=&\sum_{a'}
{\alpha_s(\mu^2)\over2\pi}\int_x^1{dy\over y}\; {\cal P}_{a'\rightarrow a}
\left({x\over y}\right)f_{a'\rightarrow a}(y,\mu^2)\;\; .\nonumber
\end{eqnarray}
In the quark case, we can write explicitly
\begin{equation}
{\cal P}_{q\rightarrow q}(x)=C_F\left\lbrack 
{2\over(1-x)^+}-x-1+{3\over2}\delta(1-x)\right\rbrack\;\; .
\end{equation}
The `+ function', $1/(1-x)^+$, is defined by
\begin{equation}
{1\over(1-x)^+}\equiv\lim_{\varepsilon\rightarrow0}
\left\lbrack {1\over1-x}\Theta(1-x-\varepsilon)-
\delta(1-x)\int_0^{1-\varepsilon}dy\;{1\over1-y}\right\rbrack\;\; .
\end{equation}
Its definition implies
\begin{equation}
\int_0^1dx\;{f(x)\over(1-x)^+}=\int_0^1dx\;{f(x)-f(1)\over1-x}\;\; ,
\end{equation}
which is more useful in practice.

The other conservation law which must be satisfied by
these distributions is that of $+$ momentum.  
We have normalized our distributions in such a 
way that
\begin{equation}
\sum_{a}\int_0^1dx\;x\;f_{a/T}(x,\mu^2)=n\cdot P_T
\end{equation}
is the total $+$ momentum of the target.  Since
this cannot be scale dependent, we have the 
relation
\begin{equation}
\sum_a\int_0^1x\,dx\;A_af_{a/T}(x,\mu^2)=\sum_a\sum_{a'}
\left(\int_0^1z\,dz\;P_{a'\rightarrow a}(z)\right)
\left(\int_0^1y\,dy\;f_{a'/T}(y,\mu^2)\right)\;\; .
\end{equation}
Using the above relations once more, along with the 
fact that $A_q=A_{\overline q}$ is independent of
flavor and of {\it target}, we arrive at
\begin{eqnarray}
\label{gendpoint}
A_g&=&\int_0^x {y\over x}\,{dy\over x}
\;\left\lbrack 2n_fP_{g\rightarrow q}\left({y\over x}\right)
+P_{g\rightarrow g}\left({y\over x}\right)\right\rbrack\\
\label{strictqcond}
A_q&=&\int_0^x {y\over x}\,{dy\over x}
\;\left\lbrack P_{q\rightarrow q}\left({y\over x}\right)
+\left(2n_f-1\right)P_{q\rightarrow\overline q}\left({y\over x}\right)
+P_{q\rightarrow g}\left({y\over x}\right)\right\rbrack\;\; .
\end{eqnarray}
The first of these relations allows us
to deduce $A_g$ without the use of vacuum
insertions (or provides a useful
check of our calculation), while the second
takes the form of an unexpectedly strict
consistency condition on the allowed probabilities
for quark splitting.  This equation can be
used as an extremely strong check
of one's expressions.

Our perturbative calculations of the 
parton distributions allow us to 
extract $P_{g\rightarrow q}(x)$
explicitly by differentiating $f_{q/g}(x)$.
The result,
\begin{equation}
P_{g\rightarrow q}(x)=T_F\left\lbrack x^2+
(1-x)^2 \right\rbrack \;\; ,
\end{equation}
allows us to write the full evolution
equation 
\begin{equation}
\mu^2{D_q\over D\mu^2}f_{q/T}(x,\mu^2)={\alpha_s(\mu^2)\over2\pi}
\int_x^1{dy\over y}\left\lbrack
P_{q\rightarrow q}\left({x/y}\right)f_{q/T}(y,\mu^2)
+P_{g\rightarrow q}\left({x/y}\right)f_{g/T}(y,\mu^2)\right\rbrack\;\; .
\end{equation}
This equation is complicated by the presence
of the gluon distribution.\footnote{Note that the
gluon distribution here has the leading behavior
\begin{equation}
f_{g/g}(x)\sim x\delta(1-x)
\end{equation}
rather than Eq.(\ref{gluenorm}) because we have
chosen to eliminate negative arguments.
In the quark case, this means that we must consider
both a quark and an antiquark distribution.  
Since gluons are their own antiparticles, 
their distributions add.}  As before, we can 
form nonsinglet distributions which decouple
from the gluon sector using the diagonal generators
of $SU(n_f)$.  Due to the symmetry of our theory,
these nonsinglet functions
all satisfy the same evolution equation
\begin{equation}
\mu^2{D_q\over D\mu^2}f_{NS/T}(x,\mu^2)={\alpha_s(\mu^2)\over2\pi}
\int_x^1{dy\over y}\left\lbrack 
P_{q\rightarrow q}\left({x\over y}\right)
-P_{q\rightarrow\overline q}\left({x\over y}\right)
\right\rbrack\,f_{NS/T}(y,\mu^2)
\label{nonsinglet}
\end{equation}
to all orders in perturbation theory.
Its appearance in this evolution equation 
as well as its role in the determination of
$A_q$ make
\begin{equation}
P_{NS}(x)\equiv P_{q\rightarrow q}(x)-P_{q\rightarrow\overline q}(x)
\end{equation}
deserving of the title `nonsinglet splitting function'.

Equation (\ref{nonsinglet}) is also satisfied by the 
nonsinglet antiquark distributions
and the difference between the 
singlet quark and antiquark distributions.
Only the lone full singlet distribution,
\begin{equation}
f_{s/T}(x,\mu^2)\equiv\sum_q\left(f_{q/T}(x,\mu^2)+
f_{\overline q/T}(x,\mu^2)\right)
\end{equation}
can mix with the gluon fields.
Their evolution is coupled according to
\begin{eqnarray}
\mu^2{D_q\over D\mu^2}f_{s/T}(x,\mu^2)&=&
{\alpha_s(\mu^2)\over2\pi}\int_x^1{dy\over y}\left\lbrack
P_s\left({x\over y}\right)f_{s/T}(y,\mu^2)\right.\nonumber\\
&&\left.\phantom{{\alpha_s(\mu^2)\over2\pi}\int_x^1{dy\over y}\lbrack}
+2n_fP_{g\rightarrow q}\left({x\over y}\right)
f_{g/T}(y,\mu^2)\right\rbrack\;\; ,\\
\mu^2{D_g\over D\mu^2}f_{g/T}(x,\mu^2)&=&
{\alpha_s(\mu^2)\over2\pi}\int_x^1{dy\over y}\left\lbrack
P_{q\rightarrow g}\left({x\over y}\right)f_{s/T}(y,\mu^2)\right.\nonumber\\
&&\left.\phantom{{\alpha_s(\mu^2)\over2\pi}\int_x^1{dy\over y}\lbrack}
+P_{g\rightarrow g}\left({x\over y}\right)f_{g/T}(y,\mu^2)\right\rbrack\;\; ,
\end{eqnarray}
where I have defined 
\begin{equation}
P_{s}(x)\equiv 2n_f
P_{q\rightarrow\overline q}(x)+P_{NS}(x)\;\; .
\end{equation}
The coupling of the gluon and singlet 
quark distributions makes data on these objects
extremely difficult to analyze.  Since
one needs to know both distributions entirely
at one scale in order to evolve either to another,
measurements at different scales
are not directly related to one another.  

In the polarized sector, everything works in exactly
the same way.  The nonsinglet distributions
decouple from the gluons and each other
to evolve autonomously, while the fully singlet
distribution mixes with the polarized
gluon distribution.  Here, the definition of the 
antiquark distribution implies that it is the {\it difference}
of the quark and antiquark singlet distributions that forms
the fully singlet distribution, rather than the 
sum.   The endpoint contributions,
being properties of the fields themselves, 
are identical to those in the unpolarized case.  Only
the evolution kernels are different.  
In Appendix \ref{frulesqcd}, the full set of
kernels at leading order in QCD is displayed.
At next-to-leading order, one can find them in \cite{ellis}.
Note that Equation (\ref{strictqcond}) is satisfied
by these functions.

The expressions (\ref{altpar}) are called 
Dokshitzer-Gribov-Lipatov-Altarelli-Parisi
(DGLAP) evolution equations in honor of the 
physicists who first derived them \cite{dglap}.
The functions ${\cal P}_{a'\rightarrow a}(z)$,
known either as {\it splitting functions} or
DGLAP {\it evolution kernels}, allow us to sum
certain effects to all orders in perturbation theory.
For example, although the probability
of finding an antiquark in a quark is zero at
${\cal O}(\alpha_s)$, the nonzero 
splitting functions ${\cal P}_{q\rightarrow g}$
and ${\cal P}_{g\rightarrow\overline q}$
will {\it generate} a nonzero probability 
via the DGLAP equation as the scale is varied.
The distribution generated in this way will
not be the same as that generated by using the 
correct nonzero splitting function 
${\cal P}_{q\rightarrow\overline q}$, but it
will certainly be closer than the direct 
leading order prediction of zero.  
This effect comes from our assumption (\ref{primord})
of the form the evolution will take.  
Once we assert this form, which is motivated
on physical grounds, our equation already knows
{\it part} of the result for the next order.
The use of these
kinds of evolution equations to sum
certain contributions to all orders has been 
developed into a high art form and can be 
extremely useful in analyzing data.  

\section{Factorization to All Orders in DIS}
\label{factdis}

Now that we have seen that the infrared divergences 
present in our one-loop amplitudes can be attributed to  
parton distributions in a 
parton, it is natural to ask if this property
persists to all orders in perturbation theory.
In this section, I will outline a general proof 
of factorization for inclusive DIS.  This proof has been 
undertaken in a rigorous field-theoretic manner 
by several people \cite{Stermanpower,Libby,Collins,factMueller}.
A very instructive recent application of 
the method under consideration here 
can be found in Ref. \cite{jicollins}.

Our proof is structured as follows.  First, we must
identify the infrared-sensitive regions of
our process.  These regions can be characterized
by certain reduced diagrams in which only the 
infrared behavior is emphasized.  Since we work only
at leading order in the Bjorken limit,\footnote{At higher
orders, the arguments of Section \ref{pertdis}
may no longer apply.   Since these corrections
are relevant for smaller values of $Q^2$,
they will involve processes in which the partons
do interact with each other during the scattering.
Depending on the degree of this mixing of scales,
we may or may not have a factorization theorem.
Some of the necessary considerations for 
these power corrections are discussed in 
Chapter \ref{htwist}.  For more detailed 
studies, especially with reference to factorization theorems,
see \cite{htwistfact}.} we will concern ourselves
exclusively with regions that
contribute to the leading behavior of the amplitude.
These regions have special properties which allow
us to show that their contribution does nothing more
than generate the distribution functions introduced in the
last section.  Since all of the infrared-sensitive regions
of our process can be segregated into these nonperturbative
distributions, the remainder of the amplitude is 
due entirely to high energy physics and may be reliably
calculated in perturbation theory.

\begin{figure}
\label{fig15}
\epsfig{figure=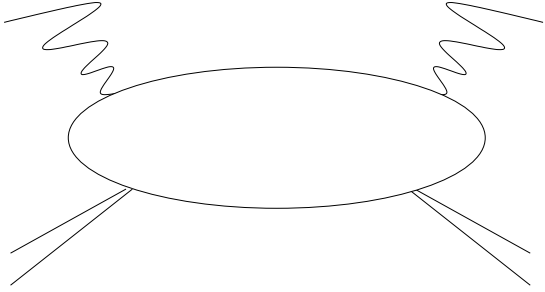,height=1.5in}
\caption{A representation of the all-orders 
scattering process.  The nucleon `blob' is
intended to represent an all-orders 
sum, and the external photons are 
connected in all possible ways to this 
blob.}
\end{figure}

To begin, we must find a way to isolate
the regions of our process which are
sensitive to soft scales.  The easiest and
most effective way to do this is to simply
set all of the soft scales to zero.  Since the 
perturbatively calculable hard scattering regions are
insensitive to these scales, they will be unaffected by this
procedure.  On the other hand, the infrared regions
of interest here will be forced to generate divergences.
It must be emphasized that we are not assuming that the
soft scales in our problem {\it are} zero.  We are setting them
to zero only as a means to {\it identify} the soft physics of our
process.  

To this end, 
we consider a general Feynman diagram
in massless QCD.  
Since we work in principle to 
all orders in perturbation theory, we will consider an external
nucleon state which we draw as a blob (shown in
Figure 2.6).   
The propagator structure of an arbitrary diagram may be
combined into one denominator using
a technique of Feynman's :\footnote{see Appendix \ref{covint}.}
\begin{equation}
\sum_i\alpha_ik^2_i(\ell_j^\mu)+i\varepsilon\;\; .
\label{denominator}
\end{equation}
Here, the $\alpha_i$ are Feynman parameters and 
the $k_i$ are propagator momenta (which are related 
linearly to the loop momenta $\ell_i^\mu$).
Infrared divergences come from regions of the 
joint parameter-momentum space 
where (\ref{denominator}) vanishes.
Not all of these regions can give an infrared divergence.
Consider the integral
\begin{equation}
\int_0^3{dx\over x-1}=\log 2\;\; .
\end{equation}
The integrand surely diverges at $x=1$, 
but the integral itself is finite.\footnote{Mathematically,
this integral is ill-defined.  However, a prescription
such as 
\begin{equation}
\lim_{\delta\rightarrow0}\left\lbrack\int_0^{1-\delta}{dx\over x-1}
+\int_{1+\delta}^3{dx\over x-1}\right\rbrack=\log 2\;\; 
\end{equation}
can be used to define the integral unambiguously.
This is what the $+i\varepsilon$'s in our
propagators do.  In this case, we would obtain an imaginary
part from the prescription :
\begin{equation}
\int_0^3{dx\over x-1+i\varepsilon}=\log2-i\pi\;\; .
\end{equation}}
This
is because the contour of integration can be 
deformed away from the singularity free of charge.
These kinds of integrals can lead to branch cuts
in our amplitudes and contribute to the imaginary
part, but they can never give infrared divergences.

There are only two ways an integral can lead to an
infrared divergence.  The first is if the 
integrand is divergent at an endpoint of the 
integration domain.  In this case, we cannot 
deform the contour away from the singularity.
An example of an
endpoint singularity is given by
\begin{equation}
\int_0^1{dx\over x}\;\; .
\end{equation}
This integral is unavoidable infinite.
Note that this is the kind of integral that
generates the light-cone singularities in
the PV-regulated axial gauge.

The second way to obtain an unavoidably 
infinite integral is by pinching 
the contour so that it cannot be deformed away
from the singularity, for example
\begin{equation}
\lim_{\varepsilon\rightarrow0}\int_{-1}^1{dx\over (x-i\varepsilon)(x+i\varepsilon)}
=\lim_{\varepsilon\rightarrow0}{2\over\varepsilon}\;
{\rm Tan}^{-1}\left({1\over\varepsilon}\right)\;\; .
\end{equation}
This integral is also unavoidable infinite.  However, note that
the expression
\begin{equation}
\lim_{\varepsilon\rightarrow0}\int_{-1}^1{dx\over(x+i\varepsilon)^2}=-2
\label{analcont}
\end{equation}
is {\it not} divergent.  This singularity is not
pinched and therefore the integration contour
may be deformed away from it.\footnote{It may seem 
a little strange that this integral is not infinite.
If we imagine adding up its contributions, 
we would certainly get infinity.  However,
we must not think of our manipulations in this
way.  What we are discussing here is the {\it analytic structure}
of our integrals.  The correct way to interpret this result
is by considering instead 
\begin{equation}
\lim_{a\rightarrow0}\int_{-1}^1{dx\over(x+a)^2}=
\lim_{a\rightarrow0}{2\over a^2-1}=-2\;\; .
\label{realwayhaha}
\end{equation}
In some sense, it is only a numerical accident that 0 lies between 
the two endpoints of integration.  The true singularities in this
family of integrals lie at $a=\pm 1$, the endpoints.  This
may seem like a swindle, but in reality it is the only
way to interpret objects like (\ref{analcont}).  These kinds
of expressions are ill-defined only in specific regions
of our parameter space.  The only sensible thing to
do is evaluate them away from these regions and analytically
continue the result later.  This is exactly what was done in 
Section \ref{pertdis}, where we took $x_B>\!\!>x$ to simplify the 
calculation and analytically continued our amplitudes to 
$x_B\leq x$ at the end.}
These two cases are the only ways a contour can be forced
through a singularity, so these are the only kinds on singularities we need 
consider in our search for infrared divergences.

Since (\ref{denominator}) is linear in
$\lbrace\alpha_j\rbrace$, singularities in this
space {\it cannot} be pinched.  
We can deform the parameter contour to 
avoid all singularities unless either
\begin{equation}
\label{Land1}
\alpha_i=0\quad\quad{\rm or}\quad\quad k_i^2=0
\end{equation}
for {\it each} $i$.  In the first case, the 
singularity occurs at an endpoint so cannot 
be deformed away.\footnote{The other endpoint,
$\alpha_i=1$,
requires all other parameters to be zero
(since Feynman parameters are constrained 
by $\sum_i\alpha_i=1$).  If the denominator 
(\ref{denominator}) vanishes in this case,
$k_i^2=0$.  Hence this also reduces to (\ref{Land1}).}
In the second, (\ref{denominator}) is
independent of $\alpha_i$ so contour deformation
produces no effect.  This region of 
parameter-momentum space will produce an 
infrared divergence if it is pinched in 
momentum space.  Since (\ref{denominator})
is quadratic in $k_i$, there are
two poles for each $k_i$.  These
poles are pinched only if they occur 
at the same point on the real line.
The two zeros of a quadratic form $D(x)$ are 
equal iff $dD(x)/dx=0$ simultaneously with 
$D(x)$.  The structure of Feynman diagrams
is such that one can always choose 
loop momenta to appear with the same sign 
in {\it all} of the propagators.\footnote{This can be seen by inspection.}  
Since propagator momenta always appear quadratically
in the denominator, we can choose this 
sign to be positive.\footnote{All we are saying here
is that one can always choose a momentum routing
such that $k_i=\sum_j\ell_j+q$, where $q$ is some external
momentum and the sum extends over all loops whose
momenta flow through $k_i$.}
With this choice of 
routing, the condition for a pinched singularity 
in $\ell^\mu_j$
becomes
\begin{equation}
\sum_i\alpha_i k^\mu_i=0\;\; .
\label{Land2}
\end{equation}
The sum goes over all propagators through 
which $\ell_j^\mu$ flows.  In order to meet the criteria
for an infrared divergence, Equation (\ref{Land1}) must be
satisfied by each propagator {\it and}
Equation (\ref{Land2}) must be satisfied by each loop
momentum.  The conditions are collectively referred to
as the Landau equations \cite{Landaueqns}.

Equation (\ref{Land2}) has a nice physical interpretation
due to Coleman and Norton \cite{colemannorton}.
Since $\alpha_ik^\mu_i$ is proportional to the 
4-velocity of propagator $i$, we can think of it as
the distance traversed by the particle it
represents in a time $\alpha_ik^0_i$.\footnote{Since the
units of $\alpha_i$ are arbitrary, we can consider
it a frame-independent ratio of time to energy $k^0_i$.}
(\ref{Land2}) states that as we go around the loop 
$\ell_j$, we get back to the point we started from.
Since propagators with $\alpha_i\ne0$ are onshell
($k_i^2=0$), the Landau equations state that 
pinched surfaces come only from {\it classically 
allowed} propagation.\footnote{We can obtain
classically unallowed propagation in quantum mechanics since
momentum eigenstates do not have well-defined spacetime locations.
To see this in another way, recall that momentum conservation
at each vertex is imposed {\it after} integration over the spacetime
location of the vertex.} 
Each solution of the Landau
equations has a corresponding `pinched diagram' in which
all lines representing propagators with $\alpha_i=0$ have
been shrunk to a point\footnote{This is because $\alpha_i=0$ implies
no propagation.  In this region of parameter space,
the integral is independent of $k_i^\mu$ so represents
an position eigenstate.} and all other lines
represent onshell propagation.

\begin{figure}
\label{fig12}
\epsfig{figure=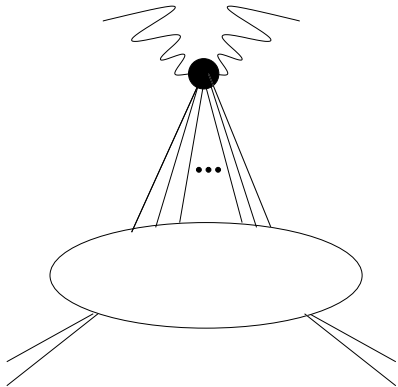,height=2.0in}
\caption{The most general reduced diagram for DIS.  Straight lines
represent collinear (or soft) partons of {\it any} species.}
\end{figure}

For DIS on a hadron target, the most general pinched diagram 
for $T^{\mu\nu}$ is 
shown in Figure 2.7.  Here, we see that any
number of (collinear) parton lines may 
connect the hard scattering to the 
hadron blob.  However, the hard scattering itself must
be considered a point interaction.  Diagrams in which this
is not the case cannot satisfy 
(\ref{Land1}) and (\ref{Land2}) 
simultaneously since two or more onshell propagators
leaving the first photon interaction vertex must be moving
in different directions and therefore can never (classically) recombine
at the other photon vertex.  Shrinking all offshell propagators
to a point thus gives us a pointlike hard interaction.
This implies that summing over all final states in DIS
cancels the infrared divergences associated with final 
state interactions.\footnote{Without the sum over final
states, we could not obtain the cross-section for DIS from
$T^{\mu\nu}$.}  Physically, we can understand this as
a consequence of uniterity.  By definition, final state 
interactions take place on the scale of nonperturbative QCD.
For large $Q^2$, the arguments of Section \ref{pertdis}
imply that these interactions can only
occur long after the hard scattering has taken place.  As such, they
cannot affect the probability that the scattering will 
take place.  By the time final state interactions
come into play, that decision has already been made.
The only thing these interactions can affect is the 
final form of the outgoing physical state.  Therefore, their
effects {\it must} cancel out in the 
inclusive cross-section.\footnote{As $Q^2$ is lowered, the time 
the hard interaction takes
grows until there is no longer a clear separation between 
the (not-so-)hard interaction and final (and initial) state
interactions.  At this point, the above analysis breaks 
down.  These nonperturbative interactions certainly can 
and will affect the total inclusive cross-section.  
In our classification
of reduced diagrams,
the assumption of large $Q^2$ is reflected
in the fact that we take point-like couplings 
for the photon interactions.  Less virtual 
photons have spatially extended couplings
which allow classical propagation between them,
reinstating final state interactions.}

Although all diagrams of the type in Figure 2.7 
can contribute to infrared divergences, they will not 
all contribute at leading order in the Bjorken limit.
The Bjorken suppression of various graphs may be 
deduced from $Q^2$ power counting.  A detailed analysis
\cite{Stermanpower} reveals that only those
diagrams in which the hard scattering is connected to 
the hadron blob by two `physical' parton propagators and any 
number of collinear scalar\footnote{Scalar gluons
can be thought of as gluons whose polarization vectors
are dotted with their momenta.  I.e. the `scalar gluon field'
is written $\partial_\mu{\cal A}^\mu$.}
gluons give leading contributions in the 
Bjorken limit.  The appearance of scalar gluons
can be understood by considering the simplest case of 
one gluon interacting directly with the 
intermediate quark propagation.\footnote{Like Figure 2.2c
in Section \ref{pertdis}.}  
The gluon insertion breaks the quark propagator into 
two, giving a suppression of $Q^2$ unless this
can be canceled by the numerator factors.
Since large terms in the numerator appear only with
$\not\!n$, leading behavior
occurs only when the interaction involves $\not\!p$.
In light-cone gauge, this is not possible since
$\not\!p$ is related to ${\cal A}^+=0$.  In covariant 
gauge, we can obtain $\not\!p$ by contracting the other
side of the propagator with $p^\mu$ :
\begin{equation}
p^\mu\left(-g_{\mu\nu}+(1-\xi){p_\mu p_\nu\over p^2}\right)\gamma^\nu
=-\xi\not\!p\;\; .
\end{equation}
This procedure gives zero for strictly collinear gluons in
light-cone gauge,\footnote{In these manipulations,
we take $p^2=0$ wherever it will not cause a divergence.}
\begin{equation}
p^\mu\left(-g_{\mu\nu}+p_\mu n_\nu+p_\nu n_\mu\right)=0\;\; ,
\end{equation}
so we can say in either case that leading behavior comes
only from scalar gluons.  Generalizing this 
analysis to higher loops and other couplings leads to the 
above result.  

A simple way to understand why these diagrams are favored
over others comes from
light-cone dimensional 
analysis.  Since the dimension of an amplitude is 
fixed, all soft mass dimensions must be 
compensated by the hard scale $Q^2$.
Assuming covariant normalization for the
external states ($\langle p|p\rangle = 2p^0(2\pi)^3
\delta^3(0)$), every external wavefunction
contributes mass dimension $-1$. The collinear quarks 
and gluons in a soft hadron vertex have 
effective mass dimensions depending on their
polarizations. A Dirac field $\psi$ can be written
as a sum of good ($\psi_+$) and bad ($\psi_-$)
components, where $\psi_{\pm} = P_{\pm}\psi$ and
$P_{\pm} = {1\over 2}\gamma^{\mp}\gamma^{\pm}$. 
The good (bad) component has effective 
light-cone mass dimension $1$ ($2$).\footnote{Since we
consider collinear quarks moving in the 3-direction, 
the free quark wavefunction is annihilated by 
$\not\!p$.  This means that the `bad' component
cannot freely propagate.  We will see in Chapter \ref{htwist}
that this part of the quark field really represents a
quark-gluon composite rather
than a simple quark.}  The vector
potential $A^\mu$ has light-cone
components $A^+$, $A^\perp$, and $A^-$, which have
effective mass dimensions 0, 1, and 2, respectively.\footnote{As in
the quark case, this is related to equation of motion constraints.} 
For the reduced diagram shown in Figure 2.7, the only 
soft mass dimension comes from the 
nucleon-quark-gluon blob. 
Using the above 
rule, we find it is $\Sigma m_i-2$.
Here, $m_i$ is the light-cone 
mass dimension of the $i$th collinear parton line.
Our scattering process requires at least two physical
($m\neq0$) partons, so $\Sigma m_i\geq 2$.  Since
physical parton lines add soft dimension to our matrix
element, they lead to suppression in our amplitudes.  Only
the scalar gluon field ${\cal A}^+$ can be added without
penalty.  Hence, we see that the most general 
leading pinched collinear surface is represented 
by Fig. 2.7, with two parton lines with $m=1$ and any
number of scalar gluons connecting the nucleon blob to the
hard scattering.
Noncollinear lines cannot classically reach the hard interaction,
so the only other possibility 
for a pinched singularity involves soft lines.
Since zero is in some sense collinear to everything, 
these lines
are in principle contained in Figure 2.7.
However, we must consider them explicitly as 
they can lead to divergences that will not
factorize into our parton distributions.  

The power counting involving soft quark and gluon lines
is more subtle, and some discussion may be found 
in Ref. \cite{jicollins}.  Essentially, 
one applies a general theorem due to Kinoshita, Lee and
Nauenberg (KLN) \cite{KLN} which states that 
infrared divergences cancel in the 
sum over initial {\it and} final states.
According to the above arguments, 
soft initial state interactions are suppressed
by $Q^2$ while collinear ones only 
serve to enforce the gauge-invariance of our process.
Hence our sum over final state interactions only
is enough to cancel the {\it leading} soft divergences.
Because of this cancelation, we can consider any
reduced diagram with soft lines connecting 
the hard scattering blob to the nucleon jet 
subleading. The
situation here is 
discussed in detail in 
Sterman \cite{sterman}. 

\begin{figure}
\epsfig{figure=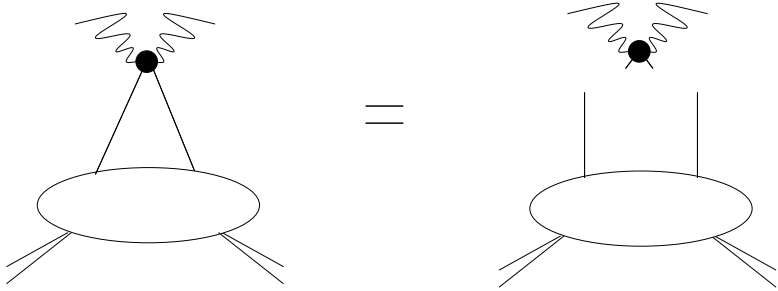,height=1.5in}
\caption{A graphical statement of factorization 
in the axial gauge.}
\label{fig16}
\end{figure}

In the light-cone gauge, where collinear scalar gluons do not 
exist, the jump from this analysis to a factorization
theorem is in some sense obvious.  Here, $\Psi=\psi$
and only those diagrams with two physical partons
connecting the hard and soft physics contribute at leading order.
In this case, factorization is represented graphically by
Figure 2.8.  However, the complications associated with renormalization
in light-cone gauge\footnote{see Section \ref{reninax}} make it desirable
to consider covariant gauges.  In these gauges,
factorization follows from the Ward-Takehashi identities
introduced in Appendix \ref{quantpoint}.  Diagrammatically, 
these identities tell us that the sum of all insertions of a 
scalar gluon is zero.  This is true for all physical matrix elements 
of gauge-invariant operators.  In particular, we can apply it
both to the hard scattering kernel and to the eikonal line.
A graphical illustration of this application is 
shown in Figures 2.9a and b.
Since the right-hand-sides of these graphical equations
are identical, we are free to make the identification 
in 2.9c.  Physically, the scalar gluons are 
{\it coherently} absorbed by the hard jet, which is
represented by an eikonal line.  They do not possess the
immense amounts of energy necessary to resolve the 
individual constituents of the jet, so only the
{\it total} color charge and momentum are relevant.
This kind of coherent absorption cannot affect the 
hard scattering in a nontrivial way; its only effect is to
induce gauge invariance in the final result.  As we saw
in the last section, this is exactly the effect of the 
gauge link in our parton distributions.  Seen from this angle,
it is not surprising that the scalar gluons generate our
distributions.  The proof to all orders is by induction;
applying the Ward identities iteratively,
one easily arrives at the factorization theorem expressed in 
Equation (\ref{facdis}).  

\begin{figure}
\epsfig{figure=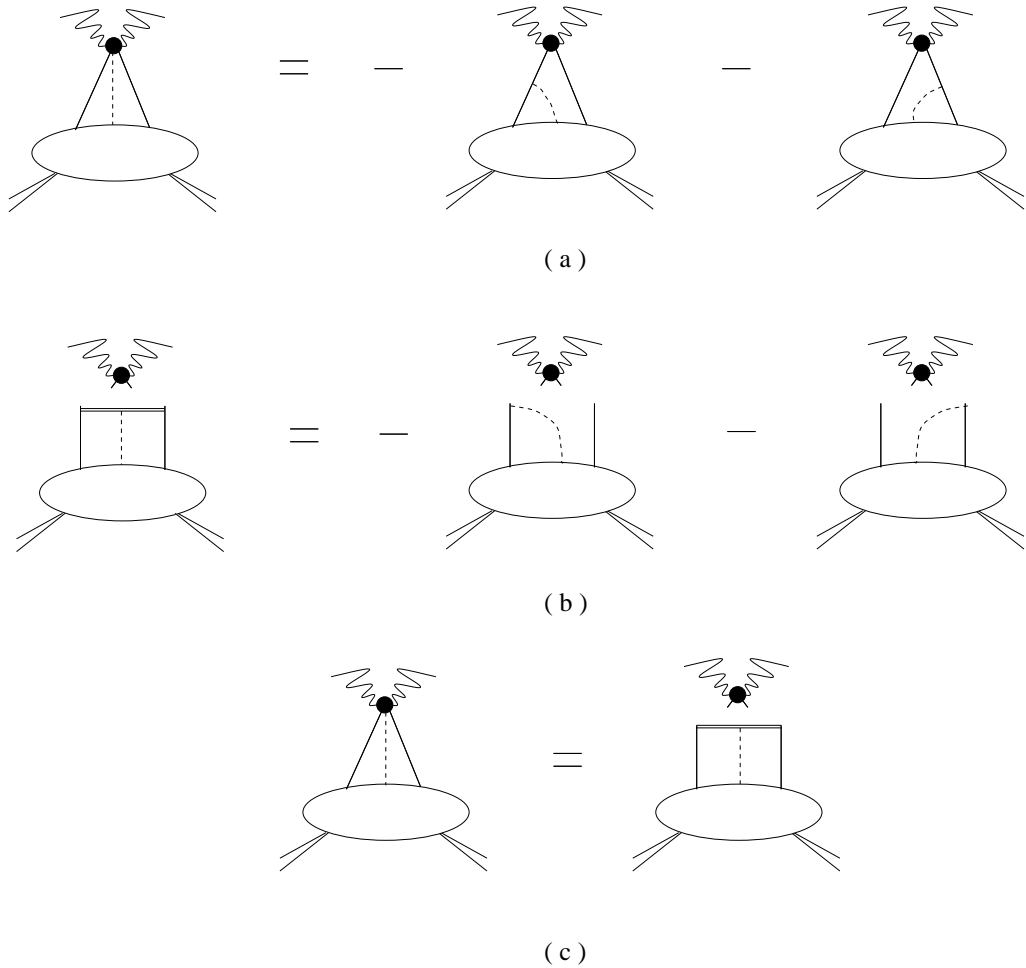,height=5.0in}
\caption{A graphical illustration of the use
of Ward identities to show that scalar gluons
generate the gauge link in our parton distribution
functions.  Dashed lines represent the 
propagation of a scalar gluon.}
\label{fig17}
\end{figure}

\section{The Operator Product Expansion}
\label{opedis}

In the last three sections, we have shown that the 
hard and soft physics of the 
amplitude\footnote{The difference between this
expression and that in Eq.(\ref{Tdef}) is purely
aesthetic, as can be shown immediately by shifting the 
argument and changing $z\rightarrow-z$.}
\begin{equation}
T^{\mu\nu}=i\int d^{\,4}z\,e^{-iq\cdot z}\left\langle PS\left|
{\rm T}\,J^\mu(0)\,J^\nu(z)\right|PS\right\rangle
\label{tdef}
\end{equation}
can be consistently separated from each other 
if the virtuality $Q^2=-q^2>\!\!>M\Lambda_{\rm QCD}$.
Furthermore, we have argued that this separation is
{\it independent} of the target.  
Since the target dependence of this amplitude is contained
entirely within the external states, it is natural to ask whether
or not we can strip these states away and express 
factorization as an operator relation.  This idea was first
introduced by Wilson in 1969 \cite{ope} 
and has been used extensively in deep-inelastic
scattering and other perturbative QCD processes.
The fundamental idea is that for large $Q^2$
the integral in (\ref{tdef}) has support
only in the region of small $z$.  This means that
our amplitude receives contributions only from regions
in which the operator product $J^\mu J^\nu$ is nearly
local,  which suggests a kind of Taylor expansion
of the operator $J^\mu(z)$ about $z=0$.
In this way, we express the operator
\begin{equation}
i\int d^{\,4}z\,e^{-iq\cdot z}\,{\rm T}J^\mu(0)\,J^\nu(z)
\label{opeoperator}
\end{equation}
as an infinite series, known as an Operator Product
Expansion (OPE), of local operators. 
For the product of two currents
separated near the light-cone, the expansion is
threefold. Primarily, it is a twist expansion, 
in which twist-2 contributions are leading whereas
the higher twist terms are suppressed by
powers of $M/Q$.  Each term in the twist expansion
contains an infinite number of local operators of the 
relevant twist. Finally, the coefficients of these operators 
(Wilson coefficients) are themselves expansions 
in the strong coupling constant.  
In this section, I will show how we can convert the results
of Section \ref{partondistdis} into next-to-leading order
results for the Wilson coefficients of the OPE relevant to DIS.
The Wilson coefficients for the unpolarized
DIS process were first calculated at 
order $\alpha_s$ in the $\overline{\rm MS}$
scheme in \cite{bardeen}.  For the polarized 
case, one can find them in \cite{pol1}.  

To get a better handle on the meaning
of the OPE, it is useful to construct the leading
order Wilson coefficients directly from the 
operator, Eq.(\ref{opeoperator}).  Since we take
$z$ near zero, we will be concerned only with 
the most singular terms in this limit.  By inspection,
one can see that singular behavior occurs only
when we contract two quark fields from the different 
currents together to form a propagator.\footnote{This implies 
that we need not consider terms
in the operator product that couple 
different flavors.}  For $z^0<0$,
we have
\begin{eqnarray}
&&i\int d^{\,4}z\,\Theta(-z^0)e^{-iq\cdot z}J^\mu(0)\,J^\nu(z)\nonumber\\
&&\qquad=i\int d^{\,4}z\,\Theta(-z^0)e^{-iq\cdot z}
\int{d^{\,4}k\over(2\pi)^4}e^{ik\cdot z}
\sum_qe_q^2\overline\psi_q(0)\gamma^\mu\;
{i\over\not\!k}\;\gamma^\nu\psi_q(z)\nonumber\\
&&\qquad=i\int d^{\,4}z\,\Theta(-z^0)e^{-iq\cdot z}
\int{d^{\,4}k\over(2\pi)^4}
\sum_qe_q^2\overline\psi_q(0)\gamma^\mu\;\left({i\over-i\!\not\!\partial_z}
e^{ik\cdot z}\right)\;\gamma^\nu\psi_q(z)\nonumber\\
&&\qquad
=-\sum_qe_q^2\overline\psi_q(0)\gamma^\mu\;{1\over i\!\not\!\partial
+\!\!\not\!q}\;\gamma^\nu\psi_q(0)\;\; .
\end{eqnarray}
In the last line, I have integrated by parts\footnote{
The action of the derivative on the $\Theta$-function 
generates a contact term which can be subtracted by 
considering a modified time-ordering operation, T$^*$.
This technical point is discussed in Ref. \cite{itzy}.
Here, we will assume that the appropriate modifications
have been made and ignore this term.}
(ignoring a surface term)
and performed the integrals over $z$ and $k$.  
The result for $z^0>0$ is identical to this one, but
with $q\rightarrow-q$ and $\mu\leftrightarrow\nu$.\footnote{In order
to show this, we have to use the fact that we consider 
only {\it diagonal} matrix elements.  As full-fledged 
operators, these objects differ by a total derivative.}

These results are very similar to those obtained in
Section \ref{pertdis}, as they must be.  We can get those
results
simply by tacking on external
states and allowing the leftover quark fields 
to contract with them.  Alternatively,
using
\begin{equation}
{1 \over\not\!p\,+\!\!\not\!q\,}={\not\!p\,+\!\!\not\!q\,\over q^2}
\;\sum_{n=0}^\infty\left(-{p^2+2p\cdot q\over q^2}\right)^n\;\; ,
\end{equation}
we can turn this expression into the 
expansion
\begin{eqnarray}
i\int d^{\,4}z\,e^{-iq\cdot z}\,{\rm T}J^\mu(0)\,J^\nu(z)&=&
\sum_qe_q^2\,\overline\psi_q\sum_{n=0}^\infty\left\lbrack\gamma^\mu
{i\!\!\not\!\partial\,+\!\!\not\!q\over Q^2}
\left({2q\cdot i\partial-\partial^2\over Q^2}\right)^n
\gamma^\nu\right.\nonumber\\
&&\!\!\!\!\!\!\phantom{\sum_qe_q^2\overline\psi_q}
\left.+\gamma^\nu
{i\!\!\not\!\partial\,-\!\!\not\!q\over Q^2}
\left({-2q\cdot i\partial-\partial^2\over Q^2}\right)^n
\gamma^\mu\right\rbrack\,\psi_q
\end{eqnarray}
involving only local operators.  The operator $\partial^2$ 
cannot hope to compete with $Q^2$ in the Bjorken limit unless
we intend to consider external states with incredible amounts
of ambient energy, so we can safely neglect this term.
On the other hand, if $\partial_\mu$ chooses its orientation
wisely, $q\cdot i\partial$ can survive this limit.  

Simplifying the Dirac structure and cleaning house a little,
we arrive at  
\begin{eqnarray}
&&i\int d^{\,4}z\,e^{-iq\cdot z}{\rm T}J^\mu(0)\,J^\nu(z)\nonumber\\
&&\qquad=\phantom{-}{2\over Q^2}\sum_qe^2_q
\left(g^{\mu\alpha}+{q^\mu q^\alpha\over Q^2}\right)
\left(g^{\nu\beta}+{q^\nu q^\beta\over Q^2}\right)\nonumber\\
\label{F2op}
&&\qquad\qquad\qquad
\times\mathop{{\sum}'}_{\!\!n=0}^{\!\!\infty}\overline\psi_q
\left(i\partial_\alpha\gamma_\beta
+i\partial_\beta\gamma_\alpha\right)
\left({2q\cdot i\partial\over Q^2}\right)^n\psi_q\\
&&\qquad\phantom{=}
-{2\over Q^2}\sum_qe^2_q
\left(g^{\mu\nu}+{q^\mu q^\nu\over Q^2}\right)\nonumber\\
\label{F1op}
&&\qquad\qquad
\qquad\times\mathop{{\sum}'}_{\!\!n=1}^{\!\!\infty}\overline\psi_q\!\!\not\!q\,
\left({2q\cdot i\partial\over Q^2}\right)^n\psi_q\\
&&\qquad\phantom{=}
-i\epsilon^{\mu\nu\alpha\beta}{2\over Q^2}\sum_qe_q^2\nonumber\\
\label{S1op}
&&\qquad\qquad
\qquad\times\mathop{{\sum}'}_{\!\!n=0}^{\!\!\infty}\overline\psi_q
\left({2q\cdot i\partial\over Q^2}i\partial+q\right)_\alpha
\gamma_\beta\gamma_5\left({2q\cdot i\partial\over Q^2}\right)^n\psi_q\\
&&\qquad\phantom{=}
+{2\over Q^2}\sum_qe_q^2\left({q^\mu q^\alpha\over Q^2} g^{\nu\beta}-
{q^\nu q^\beta\over Q^2} g^{\mu\alpha}\right)\nonumber\\
\label{unphop1}
&&\qquad\qquad
\qquad\times\mathop{{\sum}'}_{\!\!n=0}^{\!\!\infty}\overline\psi_q
\left(i\partial_\alpha\gamma_\beta
-i\partial_\beta\gamma_\alpha\right)
\left({2q\cdot i\partial\over Q^2}\right)^n\psi_q\\
\label{unphop2}
&&\qquad\phantom{=}
-{2\over Q^2}g^{\mu\nu}\sum_qe_q^2\mathop{{\sum}'}_{\!\!n=0}^{\!\!\infty}
\overline\psi_qi\!\not\!\partial
\left({2q\cdot i\partial\over Q^2}\right)^n\psi_q\;\; .
\end{eqnarray}
Here, a primed summation indicates that only
even or odd numbers are summed over, depending
on whether the sum starts with 0 or 1.

Contracting with $q_\mu$, we immediately see that
the first two terms automatically conserve current.
However, the last three operators do not
satisfy the Ward identity.  To see why, let us imagine 
taking diagonal
hadronic matrix elements of our expression.
Since the operators on the right-hand-side do 
not have any intrinsic dependence on $q^\mu$,
Lorentz invariance requires all of their vector indices
to be carried by the hadronic momentum and all
of their axial indices to be carried by the spin.
This implies that (\ref{unphop2}) vanishes by the 
equation of motion,\footnote{which, to the order 
at which we work, is simply $i\!\!\not\!\partial\psi=0$.}
and (\ref{unphop1}) does so identically.  In order to have
some other behavior, we need to introduce another 
momentum into the problem - an action that invariably
leads to other contributions which will combine
with these to restore gauge invariance.
The third term, (\ref{S1op}), retains the nonvanishing
gauge dependent piece
\begin{equation}
\epsilon^{\mu\nu\alpha\beta}P_\alpha S_\beta
\end{equation}
after we have taken our matrix element.
In the Bjorken limit, this tensor must
vanish since $S$ is parallel to $P$. 
Although it's more difficult to see,
any deviation from this limit will imply other 
contributions to the amplitude.  We will see in 
Chapter \ref{htwist} that these contributions
do indeed cancel this gauge dependence.
For now, let us consider only leading 
contributions and take this object as zero.

Now that we have taken care of electromagnetic
gauge invariance, it behooves us to consider 
chromodynamic gauge invariance.  As it stands, none
of the operators on the right-hand-side of our 
expansion are gauge-invariant.
The reason for this
is very simple - we are working only to leading order
in QCD.  The culprit is our replacement 
of $\psi(0)\overline\psi(z)$ with 
the Dirac propagator.  To maintain
explicit gauge invariance without calculating
to all orders, we can introduce a gauge link
between the two remaining fields simultaneously
with the replacement.  The effect of this 
procedure is a simple replacement of all
partial derivatives with covariant derivatives.

Stripping away the factors of $q^\mu$,
we see that our expansion involves 
local operators of the form
\begin{equation}
\overline\psi\gamma^{\mu_1}
i{\cal D}^{\mu_2}\cdots i{\cal D}^{\mu_n}\psi\;\; .
\end{equation}
As explained in Appendix \ref{dottedandundotted},
$n$-index tensors such as these
can be separated into several 
disjoint pieces.  Each of these pieces
transforms separately from the others
under the action of the Lorentz group.  
As a symmetry of our theory, Lorentz 
covariance restricts the possible 
divergences that can appear.  For this
reason, operators which begin in the same
irreducible representation of a symmetry 
group will remain together in that representation
after renormalization.  However, Lorentz invariance 
cannot impose relations between two {\it different} 
representations.  The operators that appear
in our expansion actually represent sums 
of operators in different representations, all
of which renormalize independently of
one another.  Rather than this mess, we would prefer
to separate our operators into groups
which stay together after renormalization.
The procedure for this separation is straightforward
and outlined in the appendix.  Here, we mention
only that the tensors required for our purpose 
are those with the highest spin.
All other structures
are subleading.\footnote{This
can be seen simply be taking matrix elements.
The hadron's momentum and helicity
cannot support any antisymmetric tensors and
the traces all lead to powers of
$M^2$.}  In view of this, we define
\begin{eqnarray}
\label{unpolopdef}
{_q\cal O}_n^{\mu_1\mu_2\cdots\mu_n}\equiv\overline\psi_q\gamma^{(\mu_1}
i{\cal D}^{\mu_2}\cdots i{\cal D}^{\mu_n)}\psi_q\phantom{\gamma_5\;\; .}\\
\label{spinopdef}
_q\tilde{\cal O}_n^{\mu_1\mu_2\cdots\mu_n}\equiv\overline\psi_q\gamma^{(\mu_1}
i{\cal D}^{\mu_2}\cdots i{\cal D}^{\mu_n)}\gamma_5\psi_q\;\; .
\end{eqnarray}
In terms of these operators, we can write our 
expansion as
\begin{eqnarray}
&i\int d^{\,4}z\,e^{-iq\cdot z}\,{\rm T}J^\mu(0)J^\nu(z)=&
{2\over Q^2}\sum_qe_q^2\mathop{{\sum}'}_{\!\!n=0}^{\!\!\infty}{(2q_{\mu_1})
(2q_{\mu_2})\cdots(2q_{\mu_n})\over (Q^2)^n}\nonumber\\
\label{lope}
&&\times\left\lbrace\left\lbrack 2\left(\delta^\mu_\alpha
+{q^\mu q_\alpha\over Q^2}\right)\left(\delta^\nu_\beta
+{q^\nu q_\beta\over Q^2}\right)\right.\right.\\
&&\qquad\qquad\left.-2\;{q_\alpha q_\beta\over Q^2}
\left(g^{\mu\nu}+{q^\mu q^\nu\over Q^2}\right)\right\rbrack
{_q\cal O}^{\alpha\beta\mu_1\cdots\mu_n}_{n+2}\nonumber\\
&&\left.\vphantom{\left(\delta^\mu_\alpha
+{q^\mu q_\alpha\over Q^2}\right)}
\qquad\qquad\qquad\qquad-i\epsilon^{\mu\nu
\alpha\beta}q_\alpha g_{\beta\lambda}
\,_q\tilde{\cal
O}^{\lambda\mu_1\cdots\mu_n}_{n+1}\right\rbrace\;\; .\nonumber
\end{eqnarray}

This is the desired 
expansion of the 
product of two electromagnetic currents 
in terms of an infinite series of local
operators.  Each operator appears
with a power of $Q$ to compensate its mass-dimension
and a power of $q_\alpha/Q$ for most of its 
tensor indices.  Removing trivial kinematical 
factors,\footnote{This amounts to ignoring
each index that is `bogged down' in a 
tensor structure involving $\mu\nu$, and
along with it a power of $Q$.  In addition, 
factors of $q_\alpha$ adorning these
tensor structures should also be ignored 
along with a corresponding factor of $Q$.
By taking matrix elements and comparing 
with (\ref{Tdecomp}), one can see that this is the proper
procedure to determine the suppression of
the structure function itself.}
a general operator of spin $s$ and 
mass-dimension $m$ appears in our expansion 
appears as\footnote{This is purely a dimensional
argument.  Note that the mass-dimension
of our operator $\int{\rm T}\,JJ$ is 2.}
\begin{equation}
{(2q_{\mu_1})(2q_{\mu_2})\cdots(2q_{\mu_s})\over (Q^2)^s}
{1\over Q^{m-s-2}}{_q\cal O}^{\mu_1\mu_2\cdots\mu_s}\;\; .
\end{equation}
Taking matrix elements with momentum 
$P$ such that $P\cdot q=\nu$, we see that the
contribution of this operator is suppressed
by $Q^{m-s-2}$.  Since it governs the suppression
of these operators, the {\it twist} 
\begin{equation}
t\equiv m-s
\end{equation}
is of great importance.  
A look at Equations 
(\ref{unpolopdef}) and (\ref{spinopdef})
tells us that the twist of the operators
in our leading expansion is $(n-1)+3-n=2$.
Hence these operators are not suppressed
in the Bjorken limit.  

As mentioned above, operators of different
spin cannot mix under renormalization due to the 
Lorentz invariance of our theory.  
Likewise, the absence of a scale\footnote{The 
scale generated through the coupling 
by dimensional transmutation is of no help here.
Its effect is limited to producing scale variations;
$\mu^2$ can appear only in a ratio.}
in massless
QCD prevents operators with different mass-dimension
from mixing with one-another.  This implies that only
operators of the same twist can mix.  
An exhaustive classification of QCD operators
in terms of twist \cite{xdjhtc} reveals that
the lowest possible twist is twist-2.  Furthermore,
the only other operators of twist-2 in the 
nonsinglet sector,\footnote{Here, we consider only 
operators with nonzero forward matrix elements.
We will meet other twist-2 operators in the 
next chapter.}
\begin{equation}
_q\hat{\cal O}^{\alpha\mu_1\cdots\mu_n}_n=
\overline\psi_q\sigma^{\alpha(\mu_1}i{\cal D}^{\mu_2}
\cdots i{\cal D}^{\mu_n)}\psi_q\;\; ,
\end{equation}
must completely decouple from our operators in 
massless QCD because of their chiral-odd 
structure.\footnote{
These operators relate left- and right-handed quarks.
As such, they cannot contribute perturbatively in massless
QCD.}  Since the operators (\ref{unpolopdef})
and (\ref{spinopdef}) have different transformation properties
under parity, they cannot mix with each other.  This
leads us to the result that all of our operators 
must have diagonal matrix elements which 
evolve multiplicatively.\footnote{
Once gluons are introduced, we will
that the flavor singlet quark
operators mix with the gluonic operators.}
Hence if we
can extract the value of one of these
matrix elements at any scale, we automatically
know it at all (perturbative) 
scales.  This represents
a tremendous simplification 
from the results of Section \ref{partondistdis},
where we found that knowledge of 
the entire parton distribution at one scale
is necessary to evolve any part of it 
to another.  Unfortunately, we will see that this
simplification is not devoid of distasteful attributes.

To see the relation between our expansion (\ref{lope})
and the structure functions of (\ref{Tdecomp}), 
we must take forward matrix elements.
Using Lorentz symmetry, we can express the 
matrix elements of ${_q\cal O}$ and $_q\tilde{\cal O}$
in terms of pure numbers :
\begin{eqnarray}
\label{unpolopscdef}
\left\langle P\left|{_q\cal O}^{\mu_1\cdots\mu_n}_n
\right|P\right\rangle\equiv 2a_n^qP^{(\mu_1}\cdots P^{\mu_n)}
\phantom{M\;\; .}\\
\left\langle P\left|_q\tilde{\cal O}^{\mu_1\cdots\mu_n}_n
\right|P\right\rangle\equiv 4\tilde a_n^qMS^{(\mu_1}\cdots P^{\mu_n)}\;\; .
\label{polopscdef}
\end{eqnarray}
In this way, we arrive at
\begin{eqnarray}
\label{T1oper}
T_1(\nu,Q^2)=2\mathop{{\sum}'}_{\!\!n=2}^{\!\!\infty}
\left(\sum_qe_q^2a_{n}^q\right)\left({1\over x_B}\right)^n
\phantom{_{+1}\;\; .}\\
\label{T2oper}
T_2(\nu,Q^2)=4\mathop{{\sum}'}_{\!\!n=1}^{\!\!\infty}
\left(\sum_qe_q^2a_{n+1}^q\right)\left({1\over x_B}\right)^n
\phantom{\;\; .}\\
\label{T3oper}
S_1(\nu,Q^2)=2\mathop{{\sum}'}_{\!\!n=1}^{\!\!\infty}
\left(\sum_qe_q^2\tilde a_{n}^q\right)\left({1\over x_B}\right)^n
\phantom{_{+1}}\;\; .
\end{eqnarray}
These equations express a direct relation between 
the structure functions of the 
Compton amplitude and the matrix elements of
local QCD operators.  
In particular, barring belligerent behavior on the part 
of $a_n$ and $\tilde a_n$, 
they tell us that our structure functions
are analytic functions of $x_B$ near $x_B=\infty$.
In addition, the hermiticity of ${_q\cal O}$ and $_q\tilde{\cal O}$
implies that our structure functions are real
for $x_B$ on the real axis.  This means that
the true measured structure functions,
$F_1$, $F_2$, and $G_1$, can only be nonzero in 
nonanalytic regions of the $x_B$-plane.  In these 
regions, branch cuts can 
allow discontinuities across the real axis 
which generate an imaginary part.  This fact implies
that our operator product expansion {\it cannot} converge
in the physical region.  However, it can be 
used as a powerful tool to study the analytic
structure of the Compton amplitude.
Moreover, properties such as the 
Callan-Gross relation, $T_2(\nu,Q^2)=2x_BT_1(\nu,Q^2)$,
can be studied more easily from this vantage.

To go further, we must find some way to relate
the scalar matrix elements to one-another.
This is done following a hint from Section \ref{partondistdis}.
There, we found that the integral of $x$ times a structure
function is related to the matrix element of a 
local operator appearing in $\Theta^{\mu\nu}$.
Generalizing this procedure, we write
\begin{eqnarray}
&&\int dx x^{n-1}f_{q/T}(x)\nonumber\\
&&\qquad\qquad={1\over2}\int dx
{d\lambda\over2\pi}e^{i\lambda x}x^{n-1}
\left\langle PS\left|\overline\psi\left(0\right)
{\cal G}_n\left(0,\lambda n\right)
\!\!\not\!n\,\psi\left(\lambda n\right)\right|PS\right\rangle\nonumber\\
&&\qquad\qquad={1\over2}\int dx{d\lambda\over2\pi}\left({1\over i}
{d\over d\lambda}\right)^{n-1}e^{i\lambda x}
\left\langle PS\left|\overline\psi(0)
{\cal G}_n\left(0,\lambda n\right)
\!\!\not\!n\,\psi(\lambda n)\right|PS\right\rangle\nonumber\\
&&\qquad\qquad={1\over2}\int dx{d\lambda\over2\pi}e^{i\lambda x}
\left\langle PS\left|\overline\psi(0)
{\cal G}_n\left(0,\lambda n\right)
\!\!\not\!n\,\left(n\cdot i{\cal D}(\lambda n)\right)^{n-1}
\psi(\lambda n)\right|PS\right\rangle\nonumber\\
&&\qquad\qquad={1\over2}\left\langle PS\left|\overline\psi(0)
\!\!\not\!n\,\left(n\cdot i{\cal D}\right)^{n-1}\psi(0)
\right|PS\right\rangle\;\; ,
\end{eqnarray}
where we have ignored a surface term.
Performing the same manipulations for the polarized
distribution and 
comparing with (\ref{unpolopscdef}) and (\ref{polopscdef}),
we see that 
\begin{eqnarray}
\int dx\,x^{n-1}\,f_{q/T}(x)=a^q_n\phantom{\;\; .}\\
\int dx\,x^{n-1}\,\tilde f_{q/T}(x)=\tilde a_n^q\;\; .
\end{eqnarray}
Integrals of this form are called {\it moments}.
Mathematically, the correspondence between
$a^q_n$ and $f_{q/T}(x)$ is contained in
{\it Mellin} transformation.   
Substituting these expressions into (\ref{T1oper}),
(\ref{T2oper}), and (\ref{T3oper}), switching the 
order of summation and integration,\footnote{I refuse to
justify this mathematically.} and re-summing, we
arrive at
\begin{eqnarray}
T_1(\nu,Q^2)=\int{dx\over x}\;{x^2\over x_B}\;
{1\over x_B-x}\;\sum_qe_q^2\;f_{q/T}(x)+(x_B\rightarrow-x_B)\phantom{\;\; .}\\
T_2(\nu,Q^2)=\int{dx\over x}\;{2x^2}\;
{1\over x_B-x}\;\sum_qe_q^2\;f_{q/T}(x)-(x_B\rightarrow-x_B)\phantom{\;\; .}\\
S_1(\nu,Q^2)=\int{dx\over x}\;\phantom{2}x\;\;
{1\over x_B-x}\;\sum_qe_q^2\;\tilde f_{q/T}(x)-(x_B\rightarrow-x_B)\;\; .
\end{eqnarray}
These expressions are familiar from Section \ref{pertdis},
as they should be.  At this point, we are free to take imaginary
parts and obtain the physical structure functions\footnote{The
mathematical indiscretion above generated these
imaginary parts.  It can be justified in the 
unphysical region where the expansion converges.  Once this
has been done, the re-summed unphysical amplitudes
can be analytically continued to the physical 
region.}
\begin{eqnarray}
F_1(\nu,Q^2)&=&{1\over2}\sum_qe_q^2\left\lbrack f_{q/T}(x_B)
+f_{\overline q/T}(x_B)\right\rbrack\\
F_2(\nu,Q^2)&=&x_B\sum_qe_q^2\left\lbrack f_{q/T}(x_B)
+f_{\overline q/T}(x_B)\right\rbrack\\
G_1(\nu,Q^2)&=&{1\over2}\sum_qe_q^2\lbrack \tilde f_{q/T}(x_B)
-\tilde f_{\overline q/T}(x_B)\rbrack\;\; .
\end{eqnarray}
Note that we have introduced the antiquarks.
Their inclusion does not 
significantly alter
the derivation above.

Since we have only considered the leading order in 
$\alpha_s$, Equation (\ref{lope}) is guaranteed
to have QCD corrections.  A calculation
in the spirit of the above derivation 
would invariably lead to arthritis.  Alternatively,
we can turn the above analysis on its head
and use the results of Section \ref{partondistdis}
to obtain these corrections.  
Our amplitudes have the form 
\begin{equation}
T\sim\int {dx\over x} f(x)C(x,x_B)\;\; .
\end{equation}
In the unphysical
region $x_B>\!\!>x$, we write
\begin{equation}
T\sim\sum_{n=0}^\infty\int {dx\over x} f(x)
c_n\;\left({x\over x_B}\right)^n\;\; .
\end{equation}
Substituting our expressions for the moments,
we obtain an expansion analogous
to (\ref{T1oper}).  Writing
\begin{equation}
\left({1\over x_B}\right)^n={(2q_{\mu_1})
(2q_{\mu_2})\cdots(2q_{\mu_n})\over (Q^2)^n}
P^{\mu_1}P^{\mu_2}\cdots P^{\mu_n}
\end{equation}
and interpreting the resulting expressions
as matrix elements of the local operators
(\ref{unpolopdef}) and (\ref{spinopdef}), we
can strip away the matrix elements and
arrive at the operator product expansion\footnote{The 
factor $e_q^2$ does not appear as the quark coefficient
to all orders.  At higher orders, it is only the 
singlet and nonsinglet distributions
which make sense.  For these distributions, one can 
factor out an electromagnetic charge factor to all orders.
Since the distinction is irrelevant at this
order, I have not made it.}
\begin{eqnarray}
&&i\int d^{\,4}z\,e^{iq\cdot z}\;{\rm T}\,J^\mu(z)J^\nu(0)
\qquad\qquad\qquad\qquad
\qquad\qquad\qquad\qquad\nonumber\\
&&\qquad=\left({q^\mu q^\nu\over q^2}-g^{\mu\nu}\right)
\mathop{{\sum}'}_{\!\!n=2}^{\!\!\infty}{(2q_{\mu_1})
\cdots(2q_{\mu_n})\over (Q^2)^n}\nonumber\\
\label{symmopedis}
&&\qquad\qquad\qquad\qquad\times\left\lbrack \sum_qe_q^2c_n^{\,q}\,\;
{_q\cal O}_n^{\mu_1\cdots\mu_n}
+\left(\sum_qe_q^2\right)
c_n^{\,g}\;\,{_g\cal O}_n^{\mu_1\cdots\mu_n}\right\rbrack\nonumber\\
&&\qquad\phantom{=}-i\epsilon^{\mu\nu\alpha\beta}q_\alpha 
g_{\beta\lambda}{2\over Q^2}
\mathop{{\sum}'}_{\!\!n=1}^{\!\!\infty}{(2q_{\mu_2})
\cdots(2q_{\mu_n})\over (Q^2)^{n-1}}\nonumber\\
\label{asymmopedis}&&\qquad\qquad\qquad\qquad\times
\left\lbrack
\sum_qe_q^2\tilde c_n^{\,q}\;\,_q
\tilde{\cal O}_n^{\lambda\mu_2\cdots\mu_n}
+\left(\sum_qe_q^2\right)
\tilde c_n^{\,g}\;\,_g\tilde{\cal O}_n^{\lambda\mu_2
\cdots\mu_n}\right\rbrack\;\; .
\end{eqnarray}
I have ignored the longitudinal contribution,
although it appears at the twist-2 level, because it is
kinematically suppressed.
The gluonic operators, defined by
\begin{eqnarray}
_g{\cal O}_n^{\mu_1\cdots\mu_n}={\cal F}^{\alpha(\mu_1}
i{\cal D}^{\mu_2}\cdots i{\cal D}^{\mu_{n-1}}
{\cal F}^{\mu_n)}_{\;\;\;\alpha}\phantom{i\;\; .}\\
_g\tilde{\cal O}_n^{\mu_1\cdots\mu_n}={\cal F}^{\alpha(\mu_1}
i{\cal D}^{\mu_2}\cdots i{\cal D}^{\mu_{n-1}}
i\tilde{\cal F}^{\mu_n)}_{\;\;\;\alpha}\;\; ,
\end{eqnarray}
are related to the gluon distributions in
exactly the same way as their quark counterparts.
These operators are also twist-2 and 
mix with the singlet quark operators, as we will see 
below.  The coefficients of the expansion are given by\footnote{
These coefficients do not agree with those in \cite{pol1}
because their choice of $\gamma_5$ scheme is
different from mine.}
\begin{eqnarray}
     c_{n}^q &=& 1-{\alpha_sC_F\over 4\pi}\left\lbrack
             9-{8\over n}+{2\over n+1}+4S_2(n-1)-4T_1^1(n-1)\right.\nonumber\\
&&\left.\phantom{1-{\alpha_sC_F\over 4\pi}[} 
-S_1(n-1)\left(3+{2\over n}+{2\over n+1}\right)\right\rbrack\;\; ,
\\
    \tilde c_{n}^q &=& 1-{\alpha_sC_F\over 4\pi}\left\lbrack 
             9-{8\over n+1}+{2\over n}+4S_2(n-1)
               -4T_1^1(n-1)\right.\nonumber\\ 
&&\left.\phantom{1-{\alpha_sC_F\over 4\pi}[} 
-S_1(n-1)\left(3+{2\over n}+{2\over n+1}\right)\right\rbrack\;\; ,
\\
    c_{n}^g &=&{\alpha_sT_F\over 2\pi}\left\lbrack 
         -{2\over n+2}-2S_1(n-1)\left({1\over n}
 -{2\over n+1}+{2\over n+2}\right)\right\rbrack\;\; ,
\\
      \tilde c_{n}^g  &=&{\alpha_sT_F\over 2\pi}\left
\lbrack 2+2S_1(n-1)\right\rbrack\left({1\over n}-
{2\over n+1}\right)\;\; ,
\label{gluepolcoeff}
\end{eqnarray}
where I have introduced
\begin{eqnarray}
     && S_j(n)\equiv\sum_{i=1}^n{1\over i^j} \ , \nonumber\\
     && T_j^k(n)\equiv\sum_{i=1}^n{S_j(i)\over i^k} \ . 
\end{eqnarray}
These coefficients can be obtained 
simply by expanding the hard scattering kernels
about $x_B=\infty$.  Note the implication that
the kernels are finite in this limit.  This is
one of the requirements
of a local operator product expansion.

Since the moments of the distribution functions
are mathematically equivalent to the functions
themselves,\footnote{The Legendre polynomials
are finite linear combinations of $\{x_i\}$,
so we can always trade Mellin moments for 
Legendre moments.  Since the latter form a complete 
basis of $L_2$ on the interval $[-1,1]$, we can fix the 
distribution functions in $L_2$.  To do better, 
we determine endpoint contributions from the 
conservation requirements, Eqs.(\ref{flavconsreq})
and (\ref{gendpoint}).} it is interesting to 
study their evolution.  According to (\ref{primord}),
we have
\begin{equation}
\mu^2{d\over d\mu^2}a_n^{\,a}(\mu^2)={\alpha_s(\mu^2)
\over2\pi}\sum_{a'}\int_0^1
dz\,z^{n-1}\,{\cal P}_{a'\rightarrow a}(z)\,a_n^{\,a'}(\mu^2)\;\; .
\end{equation}
Hence the evolution of the moments
is governed by the {\it anomalous dimension}
\begin{equation}
\gamma^n_{aa'}\equiv\int_0^1dz\,z^{n-1}\,
{\cal P}_{a'\rightarrow a}(z)\;\; .
\end{equation}
In the nonsinglet sector, we have the simple result 
\begin{eqnarray}
\gamma_{NS}^n=C_F\left\lbrack{3\over2}+{1\over n(n+1)}-2S_1(n)
\right\rbrack\;\; ,\\
a_n^{NS}(\mu^2)\sim\left({\mu^2\over\mu^2_0}
\right)^{\alpha_s\gamma_{NS}^n/2\pi}
a_n^{NS}(\mu_0^2)\;\; .
\end{eqnarray}
This explains why $\gamma$ is called an anomalous dimension.
$a_n$ is dimensionless, but it varies with scale as
though it had dimension $\alpha_s\gamma_{NS}^n/\pi$.\footnote{
If we take the leading variation of 
$\alpha_s(\mu^2)$ with $\mu^2$ into account, 
the factor
\begin{equation}
\left({\mu^2\over\mu_0^2}\right)^{\alpha_s
\gamma_{\rm NS}^n/2\pi}\rightarrow \left(
{\alpha_s(\mu_0^2)\over\alpha_s(\mu^2)}
\right)^{2\gamma_{\rm NS}^n/b_0}\;\; .
\end{equation}
However, since we only use the leading
value of $\gamma_{\rm NS}^n$, this represents
only a partial resummation of higher 
order effects.  This so-called `leading log'
approximation is used extensively in fits to data.
In any case, the two factors coincide in the limit 
of small $\alpha_s$.}
$\gamma$ is related to the renormalization constant 
required to remove its ultraviolet divergences.  
One can think of it as the residue leftover
by renormalization.  Note that $\gamma_{NS}^n\leq 0$
for all values of $n$.  This implies that 
as the scale is raised, less and less of the 
operators in our expansion are important.  It
reflects the fact that as we look closer and closer at
a parton distribution, more and more partons appear.
The more partons there are to share the total momentum
of our target, the less momentum each one has.  Hence
as the scale is increased, the parton distributions 
become sharply peaked at the origin.  Since each higher
moment penalizes small-$x$ contributions more than the 
last, it sees less of the bulk of the distribution.
One more important aspect of our result
is that $\gamma_{NS}^1=0$.  Since the first
moment of the quark distribution is 
conserved, its anomalous dimension is constrained to
be zero.  

In the singlet sector, the quark and gluon
operators mix with each other.  Mathematically,
this is because they contain divergences 
that they can't absorb by themselves.  Physically, it is
because `quark' and `gluon' are not good quantum
numbers in this context.  If we want something
to evolve independently of other things,
it had better be an eigenstate of the 
evolution operator.  Since the gluon and 
singlet quark fields fluctuate into one another,
it doesn't really make sense to separate them
in this way.  To find out what {\it does} make
sense, one can calculate the 
anomalous dimension matrix,
\begin{eqnarray}
\mu^2{d\over d\mu^2}\left(\matrix{a_n^s\cr a_n^g\cr}\right)
={\alpha_s(\mu^2)\over2\pi}\left(\matrix{\gamma^n_{s}&\gamma^n_{sg}\cr
\gamma^n_{gs}&\gamma^n_{g}\cr}\right)
\left(\matrix{a_n^s\cr a_n^g\cr}\right)\;\; ,
\end{eqnarray}
where
\begin{eqnarray}
\gamma^n_{s}&=&C_F\left\lbrack{3\over2}+{1\over n(n+1)}
-2S_1(n)\right\rbrack\\
\gamma^n_{sg}&=&2n_fT_F\left\lbrack{1\over n}
-{2\over(n+1)(n+2)}\right\rbrack\\
\gamma^n_{gs}&=&C_F\left\lbrack {1\over n+1}
+{2\over n(n-1)}\right\rbrack\\
\gamma^n_{g}&=&2C_A\left\lbrack {1\over n(n-1)}+
{1\over(n+1)(n+2)}-S_1(n)\right\rbrack-{4n_fT_F-11C_A\over6}\;\; ,
\end{eqnarray}
and diagonalize it.  The eigenstates of this matrix
evolve independently of one another.
If we now invert the Mellin transform, we will 
find ourselves with two singlet distributions
which make sense (at this order) independently of
one another.  In particular, the evolution of the second
moment is governed by the matrix
\begin{eqnarray}
\gamma^2=\left(\matrix{-{4\over3}\,C_F&\phantom{-}{2\over3}\,n_fT_F\cr
\phantom{-}{4\over3}\,C_F&-{2\over3}\,n_fT_F\cr}\right)\;\; .
\end{eqnarray}
This matrix has the eigenvalues
\begin{equation}
\gamma^2_1=0\qquad\qquad\qquad\gamma^2_2=-{4\over3}
\,C_F-{2\over3}\,n_fT_F\;\; .
\end{equation}
Since the second eigenvalue is negative, 
the associated eigenvector will 
be `run into the ground' as the scale is increased and the
only relevant combination becomes that
associated with the eigenvalue zero.
Since the second moment of our
distributions represents the fraction of 
$+$ momentum carried by quarks and gluons,
this eigenvector tells us these fractions
at asymptotically large momentum scales :
\begin{eqnarray}
P_q(\infty)={2N_cn_fT_F\over2(N_c^2-1)+2N_cn_fT_F}
={3n_f\over16+3n_f}\phantom{\;\; .}\\
P_g(\infty)={2(N_c^2-1)\over2(N_c^2-1)+2N_cn_fT_F}={16\over16+3n_f}\;\; .
\end{eqnarray}
This result was first discovered by 
D. Gross and F. Wilczek in 1974 \cite{grossmom}.
Since there are $(N^2_c-1)$ gluons 
each with two physical polarization states
and $N_cn_f$ quarks each with two polarizations
and `strength' $T_F$, this result has the simple
interpretation of equal momentum 
sharing among all of the relevant degrees of freedom.
Corrections to this result vanish as
\begin{equation}
\left({\mu^2\over\mu_0^2}\right)^{-(16/9+n_f/3)\alpha_s/2\pi}
\end{equation}
as the scale is increased.

The polarized matrix,
\begin{eqnarray}
\tilde\gamma^n&=&\left(\matrix{\tilde\gamma^n_{s}&\tilde\gamma^n_{sg}\cr
\tilde\gamma^n_{gs}&\tilde\gamma^n_{g}\cr}\right)\;\; ,\\
\nonumber\\
\tilde\gamma^n_{s}&=&\tilde\gamma^n_{NS}=\gamma^n_{s}\\
\tilde\gamma^n_{sg}&=&2n_fT_F\left( {2\over n+1}
-{1\over n}\right)\\
\tilde\gamma^n_{gs}&=&C_F\left({2\over n}-{1\over n+1}\right)\\
\tilde\gamma^n_{g}&=&2C_A\left\lbrack{2\over n(n+1)}-S_1(n)\right\rbrack
-{4n_fT_F-11C_A\over6}\;\; ,
\end{eqnarray}
is quite unrelated to $\gamma^n$.  
This is because the separate helicity
of the gluon and singlet quark fields 
also does not make sense by itself;  only the 
combination is important.  Since this is in
a different sector of the theory, there is
no reason that the two eigenstates should be related.
In particular, the unpolarized anomalous dimension matrix
in the singlet sector is divergent for $n=1$ owing
to the fact that the first moment of the gluon
distributions is either {\it not} local\footnote{
Since factors of $x$ turn into derivatives, we can think
of factors of $1/x$ as turning into integrals.}
or {\it not} gauge-invariant.\footnote{Instead of living with
a nonlocal distribution, we could
choose to of cancel the $x$ by 
working in the light-cone gauge.  This process
leads to a local operator, but one that has
no readily accessible physical interpretation.}
However, the polarized matrix
\begin{eqnarray}
\tilde\gamma^1=\left(\matrix{0&0\cr {3\over2}\,
C_F&{1\over6}({11C_A-4n_fT_F})\cr}\right)
\end{eqnarray}
is a singular but perfectly finite object.
This matrix implies that the singlet helicity is
conserved\footnote{The anomaly does not make its
presence known until the next order.} while
the `gluon helicity' {\it grows} with the 
{\it positive} anomalous dimension
\begin{equation}
\label{eq2234}
\tilde\gamma^1_{g}={33-2n_f\over6}\;\; .
\end{equation}
This number is, of course, familiar to us
as the leading coefficient of the $\beta$-function.
It can be shown that this behavior is canceled
by an equally large orbital contribution in
any truly physical quantities\footnote{Note that 
according to (\ref{gluepolcoeff}), 
this first moment of this distribution
does {\it not} appear in polarized DIS.  All higher
moments have {\it negative} anomalous dimensions, 
as they should.} \cite{Hoodbhoyspin}.  The steady-state
solution,
\begin{eqnarray}
\tilde a_1^{s}&=&{4n_fT_F-11C_A\over6}\;\; ,\\
\tilde a_1^g&=&{3\over2}\;C_F\;\; ,
\end{eqnarray}
is not as useful as the unpolarized result
because we have no way to normalize it short
of experiment.  

\section{Summary of DIS}
\label{sumdis}

In this chapter, we have studied one of the 
most important tools of modern particle physics -
deep-inelastic scattering.  
In a special kinematic limit, this process can be 
considered incoherent; one can calculate the 
cross-section independently for each constituent of
a complicated bound state and simply average
over the constituents to obtain the total cross-section.
This is due to a separation of scales 
between the soft physics that dominates the 
bound state interactions and the hard physics
relevant to the virtual photon scattering. 
Using analytic arguments about the 
generation of singularities, one can
show that this {\it factorization} of 
hard and soft scales persists to all orders
in perturbation theory.

The weighting of the average over constituents 
is done via parton distribution functions, which
can be shown within QCD to be related to 
probabilities for finding partons within 
our target hadron.  Although these
distributions are nonperturbative and 
cannot be calculated reliably using
contemporary methods, they are process-independent 
and appear over and over again in different 
experiments.  As such, these distributions
can be measured in one experiment and used to predict
the outcome of another as a test of the theory.
Moreover, the information on nonperturbative
hadronic wavefunctions 
they contain is interesting in itself.
This information
can be used to test the accuracy of models, gain
insight into hadronic structure, and 
help make contact between the infinitesimal 
world of QCD degrees of freedom and the `macroscopic'
world of hadronic interactions.

Scaling, a somewhat robust model-independent
prediction of current algebra, is broken by the evolution
of these parton distributions.  Completely calculable
within perturbation theory, violations of scaling
in deep-inelastic scattering provide a very useful
test of the accuracy of QCD which is 
not obscured by hadronic effects we 
do not understand.
Experimental observation of scaling violations
agrees with the predictions of QCD to better than
10\% over a span in $Q^2$ of four orders of 
magnitude and in $x_B$ of one order of magnitude.

Expanding our amplitudes in the unphysical
region of large $x_B$, we can 
use DIS to derive a local operator expansion
for the product of two electromagnetic currents.
This expansion not only allows us to study
the analytic structure of the amplitude itself, but
also provides a useful playing field on which to test various
assumptions about the behavior of operators in a quantum
field theory.  Furthermore, several of the 
operators that appear in the expansion are important
in their own right as representatives of 
flavor currents, axial currents, and the stress-energy
tensor of QCD.  We will find this aspect
very useful in motivating the topic of the next chapter.

As a whole, deep-inelastic scattering has made tremendous
contributions to many aspects of modern physics.

\chapter{Deeply Virtual Compton Scattering}
\label{dvcs}

In a recent paper, X. Ji introduced deeply virtual Compton 
scattering (DVCS) as a probe to a novel class of 
``off-forward'' parton distributions (OFPD's)
\cite{jispin}. DVCS is a process in which a highly virtual
photon 
scatters on a nucleon target (polarized or unpolarized),
producing an exclusive final state consisting of a
high-energy real photon and a slightly recoiled 
nucleon. 

Several interesting theoretical papers
which study
the DVCS process further have 
since appeared in the literature. In Ref. \cite{ra1}, 
the single-quark scattering was recalculated 
using a different, but equivalent definition of 
the parton distributions.  The evolution equations of the distributions
were derived and some general aspects of 
factorization were discussed.
In Ref. \cite{ji2}, the evolution equations for
OFPD's were derived and the leading-twist DVCS cross sections 
were calculated at order $\alpha_s^0$. Some past and recent 
studies of OFPD's can be found in \cite{ofpd}.
The DVCS process was considered as a limit of unequal mass
Compton scattering, which was studied from the point of view 
of the operator product expansion, in Ref. \cite{chen}. Some early studies 
of unequal mass Compton processes can be found 
in Refs. \cite{wana,muller1}. 
${\cal O}(\alpha_s)$ corrections
to DVCS were studied in Refs. \cite{Belitsky,man,jon}
from different perspectives.

The main motivation for the present study is to see
if the theoretical basis for the DVCS process is up to par 
with other well-known perturbative QCD processes.
More explicitly, we discuss the existence of a factorization 
theorem for this process. For general two
virtual photon processes in the Bjorken limit, 
factorizability is suggested by the analysis of DIS
in the last chapter. In the case of DVCS, where one of the 
photons is onshell, the situation could be different. 
Potential infrared problems can arise 
because of the additional light-like vector 
in this special kinematic limit. However, it is believed 
that these complications will 
not ruin the factorization properties \cite{ji2}.

The material presented in this chapter was originally
published by X. Ji and I in \cite{xdjjon}.  
It is structured as follows.  In the
first section, I will outline one of the
theoretical motivations for the study of DVCS.
The kinematics of this process allow measurement of 
certain matrix elements which are related to the 
{\it total} angular momentum carried by 
quarks in the proton.  Since the 
angular momentum carried by quark {\it spin}
is measured in DIS, DVCS can help 
isolate the contribution due to quark {\it orbital}
angular momentum.  This quantity can give us great insight into
the wavefunction of the proton.  Section \ref{kindvcs}
is devoted to the kinematics of DVCS and the introduction
of OFPD's which embody the nonperturbative physics 
at leading twist in DVCS.

To see factorization at work, it is instructive
to work out one-loop examples. We 
will do this explicitly in Section \ref{pertdvcs}.  As in
Section \ref{pertdis},
we consider the unphysical process
of DVCS on onshell quark and gluon targets.
For completeness we consider both
the symmetric and antisymmetric parts of the amplitudes, 
which are related to helicity-independent and dependent
parton distributions, respectively. The only omission
is the gluon helicity flip amplitude, which 
is discussed in Ref. \cite{hoodbhoy}.  We will 
see that the collinear infrared divergences 
can be interpreted as the one-loop 
perturbative parton distributions in Section 
\ref{factdvcs1lp}. This property is independent of the 
special kinematic limit of DVCS. 

A general proof of factorization in DVCS 
was first given by Radyushkin in his approach
based on the $\alpha$-representation \cite{ra1}.  Here,
we give an alternative proof using 
the tools introduced in the last chapter.
According to these, 
the infrared sensitive contributions 
in a generic Feynman diagram 
can be represented by reduced diagrams. 
In Section \ref{factdvcs}, we will see 
that the leading reduced diagrams for DVCS
are identical to 
those present in DIS. Therefore, the proof of
factorization at this twist level is identical to that
for DIS. 

The factorization properties of general two 
virtual photon processes can be summarized beautifully
in terms of Wilson's operator product expansion. 
This expansion requires operators with total derivatives
to describe the 
off-forward nature of the process \cite{chen,wana,muller1}. 
In Section \ref{opedvcs}, I show how we 
can convert our one-loop results 
into Wilson coefficients of the twist-2 
operators in the $\overline{\rm MS}$ scheme. 
The resulting operator product expansion is
valid for forward as well as a certain class of
off-forward matrix elements.

A summary and conclusion of this chapter appears 
in Section \ref{sumdvcs}.

\section{Motivation}
\label{motdvcs}

One of the most interesting contemporary theoretical
studies in nuclear physics involves the origin
of nucleon spin.  Along with the discovery
that protons are composite objects comes the realization
that their spins must somehow be generated
dynamically by QCD.  As described in the 
Introduction, the earliest quark models saw hadronic
spin as a simple sum of the spins of the constituents.
The three-quark structure of baryons led to 
spins of $1/2$ and $3/2$ for ground states, with orbital
excitations providing higher 
half-integer spins.  However, as viewed from
QCD, these models are fundamentally flawed.  They
make no mention either of the sea quarks, which 
certainly must be taken into account in a full-fledged
quantum field theory, or of the gluons, which
are expected to be responsible for the quark
interactions.  Furthermore, they tacitly assume
that the ground state contains no orbital excitations.  
While this is certainly true in non-relativistic
quantum mechanics, there is no reason to expect it
to be the case in a highly relativistic quantum field theory.
These objections motivate us to take another
look at the origins of nucleon spin, this
time from the standpoint of QCD.

As we have seen in the last chapter, 
the polarized quark density, $\tilde f_{q'/q}(x)$,
represents the difference between the 
probability densities for positive and negative helicity 
quarks carrying
momentum fraction $x$.  This promotes the 
identification
\begin{equation}
s_{q'/T}={1\over2}\int dx\tilde f_{q'/T}(x)
\end{equation}
of the first moment of the polarized quark distribution
with the amount of target spin carried by 
quark helicity.  We will see below that this relation is
indeed correct.  This contribution has been measured in
several experiments, most notably the European 
Muon Collaboration (EMC) in 1987,
with the result that quark spin 
accounts for only $12\pm17\%$ 
of the proton's spin at a scale of approximately
10 GeV$^2$.\footnote{The fit used
to extract this value depends on an assumption
of $SU(3)$ flavor symmetry.   This assumption
accounts for much of the quoted error bar.
A recent global analysis of the relevant data
can be found in Ref. \cite{emcdata}.}
This completely counterintuitive 
result caused the so-called `spin-crisis' and put
to bed any notion that the naive quark model could
be used for anything more than the crudest of
calculations.  Apparently, $\sim85\%$ of nucleon
spin must be attributed to orbital and gluonic 
effects.  To further classify the spin sources,
it was proposed that the first moment of the 
spin-dependent gluon distribution contains information
on the contribution due to gluon spin.  
Unfortunately, this is not the case.  The subtleties
associated with gauge invariance forbid us
from separating the gluon contribution to 
hadronic spin into separate pieces representing
`orbital angular momentum' and `helicity angular momentum'.
These two quantities do not make sense 
separately for gauge fields.\footnote{
This fact is exemplified by the discussion following
Eq.(\ref{eq2234}).  The separation of gluon spin
and orbital angular momentum leads to this positive
anomalous dimension.} 

A correct analysis of nucleon spin
begins with the Pauli-Ljubanskii vector
\begin{equation}
{\hat W}^\mu=-{1\over2}\epsilon^{\mu\nu\alpha\beta}{\hat P}_\nu
{\hat J}_{\alpha\beta}
\end{equation}
introduced in Appendix \ref{unitrepso31}.  
As in the appendix, we classify particle states 
by eigenvalues of the momentum operator $\hat P$ 
and one of the components of the Pauli-Ljubanskii
vector.  In the rest frame of a nucleon,
$p^\mu=(M,0,0,0)$, the 3-component of 
$\hat W$ is $M{\hat J}^{12}$.  The statement that
a proton has spin-1/2 means that there 
are two eigenstates of ${\hat W}^3$ :
\begin{eqnarray} 
{\hat W}^3\left|p^\mu,\uparrow\right\rangle=
\phantom{-}{1\over2}M\left|p^\mu,\uparrow\right\rangle\phantom{\;\; .}\nonumber\\
{\hat W}^3\left|p^\mu,\downarrow\right\rangle=
-{1\over2}M\left|p^\mu,\downarrow\right\rangle\;\; .
\end{eqnarray}
The normalization of our states is such that
\begin{equation}
\left\langle p',h'\left|\right.p,h\right\rangle=2p^0\delta_{hh'}
(2\pi)^3\delta\left(\vec p-\vec p\,'\,\right)\;\; ,
\end{equation}
where $h$ and $h'$ represent $\uparrow$ and $\downarrow$, so
\begin{equation}
\left\langle p,\uparrow\left|{\hat J}^{12}\right|
p,\uparrow\right\rangle={1\over2}\,2M(2\pi)^3\delta^{(3)}(0)\;\; .
\end{equation}
Our choice of the 3-direction was arbitrary, as was
our choice of frame; in
general, for an arbitrary spin direction $s^\mu$
and nucleon momentum $p^\mu$, we have\footnote{
My normalization is such that
$s^2=-1/4$ rather than $-1$.}
\begin{equation}
\left\langle ps\left|{\hat W}^\mu
\right|ps\right\rangle=Ms^\mu\,2p^0(2\pi)^3\delta^{(3)}(0)\;\; .
\label{normalization}
\end{equation}

In order to understand how the spin of a proton is
generated, we need to decompose it into several 
contributions.  As shown in Appendix \ref{quantpoint},
the spatial rotation generators\footnote{As always,
neglected $SU(3)$ and flavor indices are fully contracted.}
\begin{eqnarray}
{\hat J}^{ij}&=&\int\,d^{\,3}x\left({J}_{qs}^{ij}+
{ J}_{qo}^{ij}+{ J}_g^{ij}\right)\\
{ J}_{qs}^{ij}&=&{1\over2}\;\psi^\dag\sigma^{ij}\psi\\
{ J}_{qo}^{ij}&=&\psi^\dag\left(x^i\, i{\cal D}^j
-x^j\, i{\cal D}^i\right)\psi\\
{ J}_{g}^{ij}&=&-\left(x^i{\cal F}^{j\alpha}
-x^j{\cal F}^{i\alpha}\right)
{\cal F}^0_{\,\,\alpha}
\end{eqnarray}
split nicely into three separately gauge-invariant
pieces.  Writing $\hat S\equiv\hat W/M$ and substituting
these expressions into
the definition of $\hat W$,
we find that\footnote{As usual, a vector hat on
a 4-vector denotes the 3-vector formed from
its spatial components - with indices {\it upstairs}.}
\begin{eqnarray}
\label{decomp}
{\hat{\vec S}}\phantom{_{qs}}&=&\int\,d^{\,3}x\left({{\vec S}}_{qs}+
{{\vec S}}_{qo}+{{\vec S}}_g\right)\\
\label{quarkspin}
{{\vec S}}_{qs}&=&{1\over2}\;\overline\psi\;\vec\gamma\;\gamma_5\,\psi\\
\label{quarkorbit}
{{\vec S}}_{qo}&=&\psi^\dag\left(\vec x \times -i\vec{\cal D}\right)\psi\\
\label{gluoncontribution}
{{\vec S}}_{g}\phantom{_s}&=&\vec x\times\left(\vec{\cal E}
\times\vec{\cal B}\,\right)\;\; ,
\end{eqnarray}
where $\cal E$ and $\cal B$ are the chromo-electric
and -magnetic fields, respectively, defined
by ${\cal E}^i\equiv-{\cal F}^{0i}$ and
${\cal B}^i\equiv-\tilde{\cal F}^{0i}$.
Equation (\ref{quarkspin}) has the obvious interpretation
as the contribution from quark intrinsic angular momentum
since it gives a different sign to the left- and right-handed
components of the quark fields.   
Similarly, (\ref{quarkorbit}) is what we expect from an 
orbital contribution to angular momentum once
we recognize $-i{\cal\vec D}$ as the momentum operator
in a gauge theory.  The remaining term is
attributed to gluons. 
The nonlinear behavior of the gauge
field under gauge transformations prevents
and further decomposition.  The canonical
approach to angular momentum discussed in 
Appendix \ref{quantpoint} shows conclusively that
this separation is unavoidably gauge-dependent.
However (\ref{gluoncontribution}) does have a nice
physical interpretation.  Since $\vec{\cal E}\times\vec{\cal B}$
is the Poynting vector representing classical
energy flow, this term has the expected form of
an angular momentum.  In fact, it is exactly
what we would write down for the 
angular momentum of a classical electromagnetic field.

The contribution of each operator
in our decomposition is given 
by the matrix element
\begin{eqnarray}
\left\langle ps\left|\int d^{\,3}x\;
{{\vec S}}_i(x)\right|ps\right\rangle&=&
(2\pi)^3\delta^{(3)}(0)\left\langle ps\left|
{{\vec S}}_i(0)\right|ps\right\rangle\nonumber\\
&=&2Ma_i\vec s\,(2\pi)^3\delta^{(3)}(0)\;\; .
\end{eqnarray}
The last equality follows from rotational invariance and
the fact that QCD conserves parity.\footnote{The left hand
side of this equation is odd (doesn't change sign) 
under parity, so
it must be proportional to a vector which is 
also odd under parity.  The only such vector in this
problem is $\vec s$.}  The constants $a_i$ characterize
the contribution of operator $i$ to the nucleon 
spin.  In view of (\ref{normalization}) and (\ref{decomp}),
we have the sum rule
\begin{equation}
a_{qs}+a_{qo}+a_{g}=1\;\; .
\end{equation}
At this point, we are free to use any means
necessary to extract the $a_i$'s.  
These objects are inherently nonperturbative
and so must be measured in experiment (at least in this
stage of the game).  

As we saw in the last chapter,
certain matrix elements show up naturally in physical
scattering processes.  We have only to 
calculate their coefficients
and perform the measurements.  In particular,
hadronic matrix elements of the operator
$\overline\psi\!\not\!n\,\gamma_5\psi$ can be
measured in polarized DIS.  Exploiting 
Lorentz invariance\footnote{and the fact that QCD conserves
parity}, we can extract $a_{qs}$ from these measurements :
\begin{eqnarray}
\left\langle ps\left|\overline\psi\gamma^\mu\gamma_5\psi
\right|ps\right\rangle&=&4Ma_{qs}s^\mu\nonumber\\
\left\langle ps\left|\overline\psi\!\not\!n\,\gamma_5\psi
\right|ps\right\rangle&=&4ha_{qs}\;\; ,
\end{eqnarray}
where in the last line we have chosen a frame
in which the hadron is moving ultra-relativistically
along the 3-axis.  In light of Equation (\ref{ooooo}),
$a_{qs}$ is just the first moment
of the polarized singlet quark distribution,\footnote{The
antiquark contribution must, of course, be
included.  As expected, it simply extends the 
integration into the negative real axis
according to (\ref{qaqrelat}).}
\begin{equation}
a_{qs}=\int_{-1}^1dx\,\tilde f_{S/T}(x)=\Delta N_{S/T}\;\; .
\end{equation}

As mentioned in Section \ref{partondistdis},
this quantity is not conserved by 
the interactions of QCD.  
It a scale-dependent object, like the 
QCD coupling.  In fact, all three of the operators 
in Equation (\ref{decomp}) evolve 
with scale.  This is a completely natural
phenomenon.  Only the sum of all three 
contributions must be scale-independent.
This means that it makes no sense to ask 
how much spin is carried, for example, by 
quark orbital angular momentum.  Only
when a scale is specified do questions such
as this have meaning.  At high
scales, one can calculate the evolution
of these operators in perturbation theory.
As the scale runs to infinity, only the 
eigenvector of the anomalous dimension matrix
with the smallest (negative) eigenvalue becomes relevant.
In this limit, one can show \cite{jispin,Hoodbhoyspin}
that the percentage of spin carried by quarks and
gluons becomes
\begin{eqnarray}
S_q(\infty)={2N_cn_fT_F\over2(N_c^2-1)+2N_cn_fT_F}
={3n_f\over16+3n_f}\phantom{\;\; .}\\
S_g(\infty)={2(N_c^2-1)\over2(N_c^2-1)+2N_cn_fT_F}=
{16\over16+3n_f}\;\; .
\end{eqnarray}
This result is identical to that found in
Section \ref{opedis} for the distribution of longitudinal 
momentum!  Apparently, it is a somewhat 
general feature of the asymptotic limit.
This is not completely surprising since its
interpretation is simply benevolent sharing
among partons.  However, what this result 
implies about the distribution at moderate
scales is wholly unclear.

As mentioned above, experimental 
results
indicate that the quark spin contribution 
to proton spin is $\sim15\%$ at moderate probing scales
(10 GeV$^2$).  This implies that
a full 85\% of the nucleon spin is attributed
to gluonic and orbital effects at these scales.
Unfortunately, the other operators are not readily available
for comment.  In order to measure $a_{qo}$
directly, one would have to find a process
which is sensitive only to the quark orbital
angular momentum - a formidable, if
not impossible, task.  Luckily, all is
not lost.  The primordial form of 
$\hat J$ involves the energy-momentum tensor
of QCD.  Hopefully, we can entice this
somewhat less exotic operator to help us
extract the information we desire.

We have already studied a process
in which matrix elements of the energy-momentum
tensor appear - unpolarized DIS.  According
to (\ref{QCDemtensor}) and 
(\ref{normalizationofpartondists}), the second moments
of the unpolarized nucleon structure functions
are proportional to its diagonal matrix elements.
Unfortunately, things are not as simple here as
they were in the quark spin sector.  The operator
that we want diagonal matrix elements of is
\begin{equation}
{\hat J}^{ij}=\int\,d^{\,3}x 
\left(x^i\Theta^{0j}-x^j\Theta^{0i}\right)\;\; .
\end{equation}
Writing
\begin{eqnarray}
\langle{\hat J}^{ij}\rangle&=&
\lim_{\Delta^\mu\rightarrow0}\left\langle P+{\Delta\over2}
\left|{\hat J}^{ij}\right|P-{\Delta\over2}\right\rangle\nonumber\\
&=&\lim_{\Delta^\mu\rightarrow0}\int\,d^{\,3}x\,
e^{-i\vec x\cdot\vec\Delta}
\left\langle P+{\Delta\over2}
\right|x^i\Theta^{0j}-x^j\Theta^{0i}
\left|P-{\Delta\over2}\right\rangle\\
&=&\lim_{\Delta^\mu\rightarrow0}\int\,d^{\,3}x\,
\left\langle P+{\Delta\over2}
\left|\left(i{\partial\over\partial\Delta^i}
e^{-i\vec x\cdot\vec\Delta}\right)
\Theta^{0j}-\left(i\leftrightarrow j\right)
\right|P-{\Delta\over2}\right\rangle\nonumber\\
&=&\lim_{\Delta^\mu\rightarrow0}\left\lbrack-i
{\partial\over\partial\Delta^i}\left\langle P+{\Delta\over2}\left|
\Theta^{0j}\right|P-{\Delta\over2}\right\rangle
-\left(i\leftrightarrow j\right)\right\rbrack
\left(2\pi\right)^3\delta\left(\vec\Delta\right)\nonumber\;\; ,
\end{eqnarray}
we see that the spatial coordinates in the definition
of $\hat J$ can be converted into derivatives
on $\Theta$'s {\it off-diagonal} matrix elements.
  
Thanks to the separable form of $\Theta$,
this expression for $\langle\hat J\rangle$ can 
also be split into independent quark and gluon pieces.
Off-diagonal matrix elements of these two pieces
can be separately 
parameterized by form factors :\footnote{
Conservation of the full energy-momentum tensor implies
that $\overline C_q=-\overline C_g$.}
\begin{eqnarray}
\left\langle P+{\Delta\over2}\right|\Theta_{q,g}^{\mu\nu}
\left|P-{\Delta\over2}\right\rangle=
\overline U\left(P+{\Delta\over2}\right)&&\left\lbrack
A_{q,g}(\Delta^2)\gamma^{(\mu}P^{\nu)}\right.\nonumber\\
&&\phantom{[]}+B_{q,g}(\Delta^2){P^{(\mu}i\sigma^{\nu)\alpha}\Delta_\alpha/2M}
\nonumber\\
\label{emformdecomp}
&&\phantom{[]}+C_{q,g}(\Delta^2){\left(\Delta^\mu\Delta^\nu
-g^{\mu\nu}\Delta^2\right)/M}\\
&&\left.\phantom{[]}+\overline C_{q,g}(\Delta^2)Mg^{\mu\nu}
\vphantom{\gamma^{(\mu}P^{\nu)}}\right\rbrack
U\left(P-{\Delta\over2}\right)\;\; .\nonumber
\end{eqnarray}
Here, ($\overline U$) $U$ is the free wavefunction
for an (outgoing) incoming nucleon.
Since we wish to consider a derivative of this
object at $\Delta=0$, only quantities linear in
$\Delta$ can contribute.  Lorentz
invariance\footnote{along with the fact that 
$\Delta\cdot P=0$ since both external states are on the 
same mass-shell} implies that the scalar $\overline U U$
can depend only on $\Delta^2$, so contributions
to $\langle\hat J\rangle$ can come only from
the $A$ and $B$ terms.  Using the {\it Gordon identity},
\begin{eqnarray}
\label{Gidentity}
&&\overline U\left(P+{\Delta\over2}\right)
\gamma^{(\mu}P^{\nu)}U\left(P-{\Delta\over2}\right)
=\overline U\left(P+{\Delta\over2}\right)\left\lbrack 
P^\mu P^\nu/M\right.\\
&&\left.\phantom{\overline U\left(P+{\Delta\over2}\right)
\gamma^{(\mu}P^{\nu)}U\left(P-{\Delta\over2}\right)
=\overline U}
+P^{(\mu}i\sigma^{\nu)\alpha}\Delta_\alpha/2M 
\right\rbrack U\left(P-{\Delta\over2}\right)\;\; ,\nonumber
\end{eqnarray}
it becomes obvious that 
\begin{equation}
\left\langle P\right|{\hat J}_{q,g}^{ij}\left|P\right\rangle
={1\over2}\left\lbrack A_{q,g}(0)+B_{q,g}(0)\right\rbrack
\overline U\left(P\right)\sigma^{ij}U\left(P\right)
\left(2\pi\right)^3
\delta^{(3)}(0)
\end{equation}
in the rest frame of the nucleon.  
Contracting with $\epsilon^{ijk}/2$ and comparing with
(\ref{normalization}) in the rest frame, we
see that the quark and gluon contributions 
to hadronic spin are given by\footnote{
I have used the fact that 
$U^\dag\;\vec\gamma\;\gamma_5 U=4M\vec s$
in the rest frame.}
\begin{equation}
S_{q,g}=A_{q,g}(0)+B_{q,g}(0)\;\;\qquad\qquad ;
\qquad\qquad \sum_{q,g}\left\lbrack A_i(0)+B_i(0)
\right\rbrack=1\;\; .
\label{spindecompse}
\end{equation}
This is an extremely interesting result.  It
states that we can obtain information about
the spin structure of a hadron from {\it unpolarized}
scattering experiments.  This is not a completely
crazy idea since polarization concerns
the {\it alignment} of a particle's spin
rather than the way it is distributed among its
constituents.  Still, it is somewhat unexpected
that we can obtain such useful information on spin
without polarization.  

Although this result still does not tell
us how to measure the quark orbital contribution to 
proton spin directly, knowledge of the above quantities
coupled with the information obtained from polarized DIS 
allows us to extract this information.  At present,
we have very little information
on the nonperturbative 
momentum distribution of partons within hadrons.  
The distributions measured in DIS give us information
only about longitudinal momentum distributions.  
The quark orbital contribution to the proton spin
will give us valuable information on the {\it transverse}
momentum distribution.  In addition, since the 
proton represents the absolute baryonic ground
state, information on its orbital distribution
tells us information about the structure of this global
energy minimum of QCD.

However simple it seems, Equation (\ref{spindecompse})
does have a catch.  As mentioned above,
DIS allows us to probe {\it diagonal} matrix
elements of $\Theta^{\mu\nu}$.  According
to (\ref{emformdecomp}), the form factor
$A(0)$ can be extracted from these measurements.
Unfortunately, the other form factor necessary for
our present purpose does not contribute
to diagonal matrix elements.  In order to extract
$B(0)$, we will have to find some way to probe
slightly off-diagonal matrix elements 
of this operator.  

We discovered in
Section \ref{opedis} that 
the local operators appearing in the 
QCD stress-energy tensor also
contribute to the light-cone
operator product expansion of two electromagnetic
currents.  Although this expansion is
valid only for diagonal matrix elements,
it is certainly plausible to assume that
the operators of interest will also contribute
to a generalized expansion which extends the 
range of validity.  If this is the case, 
then we can extract information
on the off-diagonal matrix elements
of $\Theta$ by studying off-forward two-photon
processes.  In fact, it is easy to show that
{\it all} of Chapter 2 generalizes quite easily to
off-forward processes in which both photons
are deeply virtual (though with different virtualities).
The calculations are almost identical.  However,
there is a major philosophical difference
between these two processes.
Deep-inelastic scattering only becomes
a two-photon process when we sum over final states and 
use the optical
theorem to express its cross-section
as the imaginary part of the Compton
amplitude.  Experimentally, it is 
electron-proton scattering with only 
stray photons in the initial or final 
states.  On the other hand, the general
two-photon processes proposed to help
us extract $B(0)$ are intrinsic
Compton processes.  
Since the initial and final states are necessarily
different, the imaginary parts of these amplitudes
are not related to {\it any} cross-section.  
To get the cross-section for a physical process, we
have to square {\it these} amplitudes!
Furthermore, in order for these processes to be
physical, the initial and final states must be
onshell.  In DIS, the virtual photon
is not a part of the initial or final 
state; it is exchanged between
the electron and the proton during the
process.  We can generate the initial
virtual photon the same way in the present
case, but the deeply virtual photon
in the `final' state presents a 
real problem.  It is simply too much
to ask for the electron to bear the 
burden of both virtual photons.  Unfortunately,
spacelike virtual photons are in no position
to bargain.  If they are not quickly absorbed 
by other particles, they simply will not form
in the first place.  

One way out of
this dilemma is to move into a region where
the final photon is not spacelike, but 
timelike.  Timelike virtual 
photons have many options open to
them.  Depending on their
energy, they can decay into many 
physical final states.
However, the kinematics 
associated with timelike 
virtual photons requires a 
large momentum transfer between 
the initial and final states.  Since we
intend to extrapolate our form factors
to zero momentum transfer, this is
not ideal for our purpose.  

Alternatively,
we can relax the condition that both 
photons have a large virtuality and allow
the final state photon to go onshell.
This process, called deeply virtual Compton scattering (DVCS),
is the subject of the present chapter.
We will find that it is almost ideal for 
our needs in that it probes certain
nonperturbative structure functions, 
the off-forward parton distributions (OFPD's),
whose moments are related to off-diagonal
matrix elements of the stress-energy tensor.
Furthermore, its kinematics permit the 
square of the momentum transfer to vanish 
with the nucleon mass.  This fact allows a 
smooth extrapolation to our region of interest.

\section{Kinematics and Parton distributions}
\label{kindvcs}

Although our ultimate interest is in deeply virtual Compton
scattering, we start by  considering a general Compton
process involving two offshell photons with different 
virtualities.  This and a suitable choice of kinematic 
variables allow us to exploit the full symmetry of 
the problem.  In the general Compton process, a virtual photon 
of momentum $q+\Delta/2$ is absorbed by a hadron of 
momentum $P-\Delta/ 2$, which then emits 
a virtual photon with momentum 
$q-\Delta/ 2$ and recoils with momentum $P+{\Delta /2}$.  
The three independent external momenta can be expanded in 
terms of the light-cone vectors 
\begin{eqnarray}
       p^\mu &=& \left(p^+,0,0,p^+\right) \ ,  \\
      n^\mu &=& {1\over {2 p^+}}\left(1,0,0,-1\right) \ , \nonumber
\end{eqnarray}
where the 3-direction is chosen as the direction of the average hadron 
momentum ($P$), and two transverse vectors.  
As in previous chapters, we call the coefficient of 
$p^\mu$ the + component and that of $n^\mu$ the $-$ component. 
Thus we write 
\begin{eqnarray}
      P^\mu &=& p^\mu+{M^2-{t/4}\over 2} n^\mu \ ,  \nonumber\\
      q^\mu &=& -\zeta  p^\mu+{Q^2\over {2 \zeta}}n^\mu\ ,  \\
     \Delta^\mu &=&-2\xi p^\mu+\xi(M^2-t/4)n^\mu+\Delta_\perp^\mu \ , \nonumber
\end{eqnarray}
where $M$ is the hadron mass (which is taken to be the same for the
initial and final hadrons), $t=\Delta^2$, $Q^2$ is the virtuality of
$q^\mu$, $\xi$ is a measure of the difference of the virtualities of 
the two external photons, $\zeta$ is defined as   
\begin{equation}
\zeta={Q^2\over{2x_B(M^2-t/4)}}
\left(-1+\sqrt{1+{4x_B^2(M^2-t/4)\over Q^2}}\right),
\end{equation}
and $\Delta_\perp^\mu$ is a vector in the transverse directions which has 
squared length $-t\left(1-{\xi^2}\right)-4\xi^2 M^2$.  We have also
introduced $x_B={Q^2/(2 P\cdot q)}$, the analogue of the Bjorken
scaling variable in this off-forward process.  These 
expressions limit the range of $\xi$ to
\begin{equation}
     \xi^2 \leq {-t\over -t+4M^2}\ ,   
\end{equation}
for  fixed $t$, or the range of $t$ to 
\begin{equation}
     -t\geq {4\xi^2M^2\over 1-\xi^2}\ , 
\end{equation}
for  fixed $\xi$.  

According to the last section, our
region of interest is $t=0$.  In order for the final state photon
to be onshell,\footnote{in the Bjorken limit} 
we must have $\xi\ne0$.  In light of the 
above expressions, this is only possible if we take
the limit $M^2\rightarrow0$ {\it before} $t\rightarrow0$.  
Unfortunately,\footnote{Well, maybe fortunately...} 
we do not control the mass of the 
nucleon.  However, it is kinematically allowed to take $\xi$
small so that $-t$ can be much smaller than $M^2$.  Performing
several of these measurements will provide a smooth
extrapolation to the required values.

In the Bjorken limit, these expressions simplify considerably. Since we 
consider only the leading twist in this chapter, we may neglect all but the 
$+$ components of $P^\mu$ and $\Delta^\mu$ (in order to form large 
scalars, one must dot the $+$ component of 
a vector with the $-$ component of $q$).
Hence, in the limit $Q^2\rightarrow\infty$ ($x_B$ remaining finite), we may write
\begin{eqnarray}
     && P^\mu \sim p^\mu \ ,  \nonumber\\
\label{bjdvcs}
     && q^\mu\sim -x_B p^\mu+{Q^2\over 2 x_B} n^\mu \ , \\
     && \Delta^\mu\sim -2\xi p^\mu \ .  \nonumber
\end{eqnarray}     
We note that the external invariants have been reduced from six to 
three by enforcing kinematics and taking the Bjorken limit.  We express these 
three scalars in terms of one mass scale, $Q^2$, and two dimensionless 
parameters, $x_B$ and $\xi$.  Apparently, the limit $\xi\rightarrow0$
recovers the DIS results from the last chapter.  We will find
this to be a useful check of our calculation.  The DVCS limit
is given by $\xi\rightarrow x_B$.  As mentioned above, we cannot 
do DVCS when $\xi=0$.\footnote{Otherwise, we could've just
extracted the required information from DIS measurements!}  However,
measurements at small $x_B$ may allow $-t$ small enough that
the variation to zero holds no surprises.  In particular,
for the relatively large value $x_B\sim0.1$, 
we can have $-t$ as low as $\sim0.04$ GeV$^2$.  
Since typical hadronic scales are $\sim0.1$ GeV$^2$
and larger, this value of $-t$ is actually quite close to zero.

Our goal is to factorize the short and long distance 
physics of the Compton amplitude\footnote{Here, we are actually interested
in this amplitude rather than the analogue of
$W^{\mu\nu}$.}
\begin{equation}
   T^{\mu\nu} = i\int d^{\, 4}z\,
     e^{\,i q\cdot z}\left\langle P+{\Delta\over 2}
      \left|{\rm T} J^\mu\left({z\over 2}
           \right) J^\nu \left(-{z\over 2}\right)
       \right|P-{\Delta\over 2}\right\rangle \ 
\label{comam} 
\end{equation}
in the Bjorken limit,
where $J^\mu = \sum_q e_q \overline{\psi}_q
\gamma^\mu\psi_q$ is the electromagnetic
current and $\psi_q$ is the bare quark field of flavor $q$ and
charge $e_q$.  Note that the operator here is identical to that
considered in DIS.  Only the external states are different.
This immediately implies that all of the diagrams we need to calculate
are topologically identical to those of the last chapter.
The only difference is in the momenta of the 
external states.

As before, the nonperturbative contribution to the Compton amplitude
in Eq.(\ref{comam}) can be expressed in terms of
the off-forward parton distributions contained in 
parton density matrices \cite{ji2}. For quarks, 
we define the light-cone correlation function
\begin{eqnarray}
     {\cal M}^q_{\alpha \beta}\left(x,\xi\right) &=& \int 
{d\lambda \over 2\pi} e^{-i\lambda x} \left\langle 
P+{\Delta\over 2}\left|\overline\psi^q_\beta 
\left({\lambda\over2}n\right){\cal G}_n\left(
{\lambda\over2}n,-{\lambda\over2} n\right)
\psi^q_\alpha\left(-{\lambda\over2} n\right)
\right|P-{\Delta\over 2}\right\rangle \nonumber  \\
 &=& {1\over 2} F_{q/T}(x,\xi,t) {\not\! p}_{\alpha\beta} 
+  \tilde F_{q/T}(x,\xi,t) \left(\gamma_5h\!\not\!p
\right)_{\alpha\beta}+\cdots\;\; ,
\label{qofpd}
\end{eqnarray}
where $h$ is the helicity of the proton.
The ellipses denote contributions  
either of higher twist or chiral-odd structure, 
which do not contribute to the leading 
process under consideration.  The gauge
link, ${\cal G}_n$, is defined as 
in Section \ref{partondistdis}
with a contour that runs along the $n$ direction.
Note that this density matrix is just an off-diagonal
matrix element of the {\it same} field operator
we considered there.\footnote{
The difference in definition between this correlation function
and that in Section \ref{partondistdis} 
(the arguments of the parton fields)
is not important there
since the matrix element is diagonal.  However, here 
the off-diagonal nature of the matrix element creates a 
dependence on the positions of the fields.
I have chosen the symmetric form for later convenience.}

Isolating the scalar distributions, we find
\begin{eqnarray}
     && F_{q/T}(x,\xi,t) \\
&&\quad={1\over 2} \int {d\lambda \over2\pi}
          e^{-i\lambda x}\left\langle P+{\Delta\over 2}\left|
\overline \psi_q\left({\lambda \over 2} n\right)
{\cal G}_n\left({\lambda\over2}n,-{\lambda\over2} n\right)
\not\! n \psi_q\left(-{\lambda \over 2} n\right) 
\right|P-{\Delta\over 2}\right\rangle \ , \nonumber\\  
     &&\tilde F_{q/T}(x,\xi,t) \\
    &&\quad={1\over 4h} \int {d\lambda \over 2\pi}
          e^{-i\lambda x}\left\langle P+{\Delta\over 2}\left|
\overline \psi_q\left({\lambda \over 2} n\right)
{\cal G}_n\left({\lambda\over2}n,-{\lambda\over2} n\right)     
\not\! n\gamma_5 \psi_q\left(-{\lambda \over 2} n\right) 
\right|P-{\Delta\over 2}\right\rangle \;\; .\nonumber
\end{eqnarray}
The ever-present renormalization scale, $\mu$,
has been suppressed here for brevity. 
Dependence on the square of the momentum transfer, $t$,
will also largely be ignored in the following
as it will not affect most of our discussion.
     
Before we go on, let us see how these distributions
are related to the form factors $A(t)$ and $B(t)$
of the stress-energy tensor.  Writing
\begin{eqnarray}
&&\int{d\lambda\over2\pi}
e^{-i\lambda x}\left\langle P+{\Delta\over 2}\left|
\overline \psi_q\left({\lambda \over 2} n\right)
{\cal G}_n\left({\lambda\over2}n,-{\lambda\over2} n\right)
\gamma^\mu \psi_q\left(-{\lambda \over 2} n\right) 
\right|P-{\Delta\over 2}\right\rangle\\
&&\quad=\overline U\left(P+{\Delta\over2}\right)\left\lbrack
\gamma^\mu\;H_{q/T}(x,\xi,t)+i\sigma^{\mu\alpha}\Delta_\alpha/2M
\;E_{q/T}(x,\xi,t)\right\rbrack\,U\left(P-{\Delta\over2}\right)\;\; ,\nonumber
\end{eqnarray}
we see that 
\begin{eqnarray}
&&F_{q/T}(x,\xi,t)={1\over 2}\;
\overline U\left(P+{\Delta\over2}\right)\left\lbrack
\not\!n\,H_{q/T}(x,\xi,t)\right.\nonumber\\
&&\qquad\qquad\qquad\qquad\left.+i\sigma^{\mu\alpha}n_\mu\Delta_\alpha/2M
\;E_{q/T}(x,\xi,t)\right\rbrack\,U\left(P-{\Delta\over2}\right)\;\; .
\end{eqnarray}
On the other hand, the second $x$-moment of 
$F_{q/T}(x,\xi,t)$ is an off-diagonal matrix
element of the quark contribution to the 
stress-energy tensor :\footnote{Note that here
the `$\leftrightarrow$' adorning $\cal D$ is
quite important.  For off-diagonal matrix elements,
it must not be forgotten that $\Theta^{\mu\nu}$ is
a benevolent operator which treats both 
external states with the utmost care and respect.} 
\begin{equation}
\int dx\,x\,F_{q/T}(x,\xi,t)={1\over2}\;
\left\langle P+{\Delta\over 2}\left|
\overline \psi_q\!\not\!n\,n\cdot 
\stackrel{\leftrightarrow}{i\cal D}
\psi_q\right|P-{\Delta\over 2}\right\rangle\;\; .
\end{equation}
Comparison with (\ref{emformdecomp})
gives
\begin{eqnarray}
&&\int dx\,x\,F_{q/T}(x,\xi,t)={1\over 2}\,\overline U
\left(P+{\Delta\over2}\right)\left\lbrack \not\!n\;A_q(t)+
i\sigma^{\mu\nu}n_\mu\Delta_\nu/2M\;B_q(t)\right.\nonumber\\
&&\qquad\qquad\qquad\qquad\qquad\qquad\qquad\qquad
\left.+4\xi^2/M\;C_q(t)\right\rbrack\,U\left(P-{\Delta\over2}\right)\;\; .
\end{eqnarray}
The Gordon identity, Equation (\ref{Gidentity}),
allows us to deduce 
\begin{eqnarray}
\int dx\,x\,H_{q/T}(x,\xi,t)=A_q(t)+4\xi^2\,C_q(t)\;\; ,\\
\int dx\,x\,E_{q/T}(x,\xi,t)=B_q(t)-4\xi^2\,C_q(t)\;\; ,
\end{eqnarray}
from which we immediately see that 
\begin{equation}
S_q=A_q(0)+B_q(0)=\int dx\,x\,\left\lbrack H_{q/T}(x,\xi,0)
+E_{q/T}(x,\xi,0)\right\rbrack\;\; .
\end{equation}
Luckily, the quark contribution to 
angular momentum is not 
sensitive to $\xi$.  The lower
bound on $-t$ causes us to want $\xi$ small 
anyway, but it's certainly nice to know
that we need not perform an 
extrapolation to $\xi=0$; only $t=0$ is
strictly required.  

Many properties of the distributions
$E(x,\xi,t)$ and $H(x,\xi,t)$, as well
as their sister distributions in the 
polarized, chiral-odd, and gluon sector 
are summed up nicely in Ref. \cite{jiofpd}.
There, one can find details of their parton model
interpretation, form factor decompositions
for their moments, models representing their
behavior, and sum rules that constrain
them.  

The contribution of gluons to DVCS requires off-forward distribution
functions.  In analogy with Section \ref{partondistdis}, 
these are written\footnote{Here, $\Delta$ leads to an
additional twist-2 term involving gluon helicity
flip.  This term is considered in \cite{hoodbhoy}.}
\begin{eqnarray}
     && F_{g/T}(x,\xi,t)\\
&&\quad=-{1\over 2x}\int {d\lambda\over 2\pi}e^{-i\lambda x}
\left\langle P+{\Delta\over 2}\left| {\cal F}^{+\alpha}
\left({\lambda\over 2} n\right)
{\cal G}_n\left({\lambda\over2}n,-{\lambda\over2} n\right)
{\cal F}_{\;\;\alpha}^+\left(-{\lambda\over 2}n\right)\right|
P-{\Delta\over 2}\right\rangle\;\;  ,\nonumber\\
   && \tilde F_{g/T}(x,\xi,t)\\
&&\quad=-{i\over 4hx}\int{d\lambda\over 2\pi}
e^{-i\lambda x}\left\langle P+{\Delta\over 2}\left| 
{\cal F}_a^{+\alpha}\left({\lambda\over 2}n\right) 
{\cal G}_n\left({\lambda\over2}n,-{\lambda\over2} n\right)
\tilde{\cal F}_{b\,\,\alpha}^+
\left(-{\lambda\over 2}n\right)\right|P-{\Delta
\over 2}\right\rangle\;\; .\nonumber
\end{eqnarray}

In this chapter, the fact that the external
wavefunctions do not correspond to the same momenta
implies that the identity 
$u\overline u=(1+2h\gamma_5)\!\not\!p/2$
cannot be used.  Instead, we calculate the amplitudes
via a direct substitution of the expansions of the 
parton distributions.  Interpreting these
functions as their leading order values (which
are simple $\delta$-functions for partonic
targets), we
will see that the one-loop calculation generates their
evolution.  For example, the polarized quark 
amplitude is obtained by tracing with $\gamma_5h\!\!\not\!p$
because this is the structure that appears multiplying
$\tilde F_{q/T}$ in (\ref{qofpd}).  The subtleties
associated with defining $\gamma_5$ in $d$ dimensions 
make the polarized result ambiguous.  
To resolve the ambiguity, one must use a well-defined 
$\gamma_5$ scheme.  Several different schemes
can be found in the literature;\footnote{See Appendix \ref{dimregapp}.} 
the prescription used here is equivalent to that
introduced by `t Hooft and Veltman \cite{thooft}.
The fact that this choice favors certain directions
over others makes it necessary to multiply the 
polarized contribution of diagram 3.1e
by the factor $(1+\delta\epsilon)$,\footnote{
This amounts to a redefinition
of $\gamma^\alpha\gamma_5\gamma_\alpha$.}
where $\delta$ is determined by the requirement that the 
nonsinglet axial current is conserved,
whenever one wishes to consider the chiral properties
of a theory \cite{larin2}.  Since we are not concerned with the 
chiral nature of QCD at the moment, we will
ignore this technicality.
        
In the gluon sector,
the fact that the conceptual idea of removing one
gluon and the actual gauge invariant 
distributions that describe this process 
do not coincide leads to extra factors.  In this
case, one can show that 
\begin{eqnarray}
&&\int{d\lambda\over2\pi}e^{-i\lambda x}
\left\langle P+{\Delta\over 2}\right|
{\cal A}^\mu_\perp\left({\lambda\over 2}n\right)
{\cal A}^\nu_\perp\left(-{\lambda\over 2}n\right)\left|P-{\Delta\over2}\right\rangle\nonumber\\
&&\qquad\qquad\qquad=-{x\over x_+x_-}\left\lbrack {2\over d-2}g_\perp^{\mu\nu}F_{g/T}(x,\xi)
-i\epsilon^{\mu\nu\alpha\beta}(2hp)_\alpha n_\beta
\tilde F_{g/T}(x,\xi)\right\rbrack\;\; ,
\end{eqnarray}
where I have defined $x_\pm\equiv x\pm\xi$.  Here, we see 
explicitly the extra factor of $1/(1-\epsilon/2)$ 
multiplying the unpolarized gluon amplitude.

\section{One-Loop Compton Amplitudes \\ on Quark and Gluon Targets}     
\label{pertdvcs}

In this section, we perform a one-loop calculation of the 
general Compton scattering on onshell quark and gluon
targets in the Bjorken limit. The result will be used in 
the next section to show that the infrared divergences present 
in our amplitude are generated exclusively by the 
off-forward parton distributions. 
This property does not change in the limit of 
a real final state photon. 

We begin with an onshell quark target of momentum
$p$.  Here, there are two diagrams at 
leading order (LO) and eight at next-to-leading order 
(NLO).  The renormalization constants necessary 
for this process cancel themselves out in exactly the
same manner we saw in Section \ref{pertdis}.
Half of the diagrams are shown in Fig. 3.1. The
other half will be taken into account by using the 
crossing symmetry, i.e., the simultaneous replacement of $q\rightarrow -q$
and $\mu\leftrightarrow\nu$.  The terms
$\pm(x_B\rightarrow-x_B)$ in the following
formulae reflect this contribution.
Because of time reversal invariance,
the Compton amplitude is also an even function of $\xi$. 
This relates the left and right 
vertex diagrams (Figs. 3.1c 
and 3.1d, respectively) to each other.  
The quark self-energy diagram and the
box diagram are themselves $\xi$-symmetric. 
This symmetry not only 
allows us to reduce the number of graphs at NLO 
from four to three, but also 
becomes a powerful tool which helps us compute 
each amplitude. 

\begin{figure}
\label{jfig2}
\epsfig{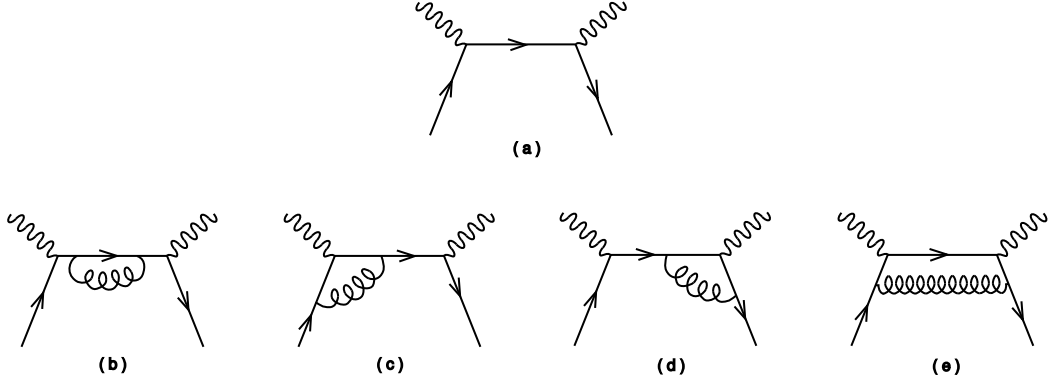}
\caption{Diagrams for Compton scattering on an onshell quark
to order $\alpha_s$.}
\end{figure}           

Examining our graphs, we see that the self-energy diagram 
contains a loop integral with two Feynman denominators, the vertex 
diagram contains one with three, and the box with four. 
A one-loop integral with two propagators is straightforward.  
Difficulties arise, however, with the calculation of three- and 
especially four-propagator integrals.  These difficulties 
may be avoided in this 
calculation because of several simplifications.  
Consider first the box diagram.  The loop integral 
is of the form (with momentum routing as shown in Fig. 3.2)
\begin{equation}
      \int {d^d k\over {(2\pi)}^d} {{{\rm Tr}
\lbrack\gamma^\alpha\left({\not\! k}
-\xi\not\! p \right)\gamma^\mu\left({\not\! k}+{\not\!q}\right)
\gamma^{\nu}\left({\not\! k}+\xi\not\! p \right)\gamma_\alpha
{\not\! p}(\gamma_5)\rbrack}\over
 {(k+\xi p )^2(k-\xi p)^2(k-p)^2(k+q)^2}}\ , 
\end{equation}
where I have replaced $\Delta^\mu$ by $-2\xi p^\mu$.  
In order to simplify the integral, 
we express the trace as a sum of terms which
cancel one of the propagators. This can be done because
both $k^2$ and  $2p\cdot k$ can be written as 
linear combinations of $(k+\xi p)^2$
and $(k-\xi p)^2$, and the trace vanishes 
whenever $k^-$ and $k_\perp^2$ do.  

\begin{figure}
\label{jfig3}
\epsfig{figure=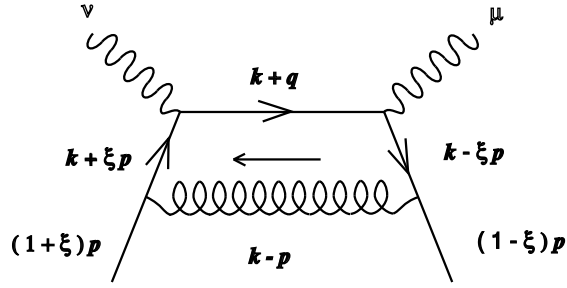,height=4cm}
\caption{The momentum flow in the box diagram}
\end{figure}                
Now that we have shown that a denominator can be canceled, we have
effectively reduced the four-propagator problem to a three-propagator 
one.  Since we have only shown that the numerator will be a 
linear combination of two different denominators, rather that proportional 
to one, the four-propagator integral will in general become 
two three-propagator integrals.  However, we may use 
the $\xi$ symmetry by writing
\begin{eqnarray}
     \phantom{2p\cdot} k^2={1\over 2}\;
\left(k+\xi p\right)^2+(\xi\rightarrow 
-\xi) \ , \nonumber\\
      2p\cdot k\phantom{^2}={1\over 2\xi}\left(k+\xi p \right)^2+
(\xi\rightarrow -\xi) \ .
\end{eqnarray}
     In this way, we consider only the $(k+\xi p )^2$ cancelation
and let the symmetry take care of the rest.  The integral we 
now have is  exactly the same (up to numerator differences) as that 
arising from the right vertex correction.
We will see that this basic integral is the only one we must calculate to 
obtain both the polarized and unpolarized amplitudes for the quark and 
gluon contributions (if we forget, for the moment, the simple self-energy
 diagram).  

The integral has the form
\begin{equation}
      \int {d^dk\over{(2\pi)}^d} {Numerator\over{[(k-p)^2+i\varepsilon]
[(k-{\xi}p)^2+i\varepsilon][(k+q)^2+i\varepsilon]}}\,\,.
\label{int}
\end{equation}
We have found that this integral is easily done 
in light-cone coordinates by expanding $k^\mu$ 
in terms of
the light cone momenta, $p^\mu$ and $n^\mu$.  As in 
Section \ref{pertdis}, we first 
do the $k^-$ integration by contour in the unphysical 
region of large $x_B$ and then the transverse integrations. The
$k^+$ integration is left until the end.  Writing $k^+=yp^+$, 
the value of the integral (\ref{int}) is 
\begin{eqnarray}
   &&    {i\over 16\pi^2}\left({Q^2\over 4\pi}\right)^{-{\epsilon/
2}}{\Gamma({\epsilon/2})\over 1-\xi }
\left\lbrace\left({x_B-a\over x_B}\right)^{-\epsilon/2}\right.\nonumber\\
&&\qquad\qquad\qquad\left.\left.\times\;\int_a^{x_B}
dy\left({{y-a}\over{x_B-a}}\right)^{1-{\epsilon/ 2}}\left(
1-{y-a\over x_B-a}\right)^{-\epsilon/2}
N(a)\right\rbrace\;\right|^{a=1}_{a={\xi}}\;\; ,
\end{eqnarray}
where $N(a)$ is defined by
\begin{eqnarray}
   &&   Numerator\Big|_{k^-={k_\perp^2/ 2p^+(y-a)}}=\alpha+\beta k_\perp^2\;\; ; 
\nonumber\\
   &&\qquad   N(a) =\beta-
{\alpha\over Q^2}{{x_B(x_B-a)}\over{(y-a)(x_B-y)}}\;\; .
\end{eqnarray}
     Doing the $y$-integrals requires some care because a delicate 
cancelation
must occur if one is to get a finite result, but the treatment is 
straightforward.  After the $\xi$ and crossing symmetries are used, 
we find the full NLO result for the symmetric quark amplitude\footnote{
We take $\mu^2=Q^2$ and $\mu\nu$ in the $\perp$
directions as before.}
\begin{eqnarray}
     && T^{(ij)}_q = -g^{ij}\sum_{q'} e_{q'}^2 \delta_{qq'}\left\lbrace
{1\over x_B-1}-{\alpha_s C_F\over 4\pi}\left({Q^2e^{\gamma_E}\over
4\pi\mu_0^2}\right)^{-\epsilon/2}\left\lbrace {3\over
{x_B-1}}\left({2\over\epsilon}+3\right)\right.\right.\nonumber\\
&&\qquad -{1\over\xi}\left\lbrack\left({2\xi\over x_B^2-1}
+{x_B+\xi\over 1-\xi^2}\right)
\left({4\over\epsilon}+3-\log\,\left(1-{\xi\over x_B}\right)
\right)-3\,{x_B-\xi\over 1-\xi^2}\right\rbrack \log\left(1-{\xi\over
x_B}\right)\nonumber \\ 
&&\qquad\qquad \left. 
+\left\lbrack  \left({x_B+1\over 1-\xi^2}+{2\over
x_B-1}\right)  \left({4\over\epsilon}+3-\log\, \left(1-{1\over
x_B}\right)\right)-3\,{x_B-1\over 1-\xi^2} \right.\right. \\
&& \qquad\qquad\qquad\left.\left.\left. - {3\over x_B-1}\right
\rbrack \log\left(1-{1\over x_B}\right)\right\rbrace 
+(x_B\rightarrow -x_B)
\vphantom{\left({Q^2e^{\gamma_E}\over
4\pi\mu_0^2}\right)^{-\epsilon/2}}\right\rbrace\nonumber
\end{eqnarray}                                 
and the antisymmetric amplitude
\begin{eqnarray}
 && T^{[ij]}_q =-i\epsilon^{\alpha\beta ij}n_\alpha
(2h p)_\beta \sum_{q'} 
    e_{q'}^2 \delta_{qq'} \left\lbrace {1\over x_B-1}-{\alpha_s C_F\over
4\pi}\left({Q^2e^{\gamma_E}\over 
4\pi\mu_0^2}\right)^{-\epsilon/2}\left\lbrace {3\over
x_B-1}\left({2\over\epsilon}+3\right)\right.\right.\nonumber\\
&& \qquad\left.\left.-\left\lbrack\left({2x_B\over x_B^2-1}+{x_B+\xi\over
1-\xi^2}\right)\left({4\over\epsilon}+3-\log\,\left(1-{\xi\over
x_B}\right)\right)+7{x_B-\xi\over 1-\xi^2}
\right\rbrack \log\left(1-{\xi\over
x_B}\right)\right.\right.  \nonumber\\
&& \qquad\qquad+\left\lbrack\left({x_B+1\over 1-\xi^2}+{2\over
x_B-1}\right)\left({4\over\epsilon}+3-\log\,
\left(1-{1\over x_B}\right)\right)+7\;{x_B-1\over
1-\xi^2} \right.\\ 
&& \qquad\qquad\qquad\left.\left.\left.
-{3\over x_B-1}\right\rbrack \log\left(1-{1\over
x_B}\right)\right\rbrace-(x_B\rightarrow -x_B)
\vphantom{\left({Q^2e^{\gamma_E}\over
4\pi\mu_0^2}\right)^{-\epsilon/2}}\right\rbrace\nonumber \ . 
\end{eqnarray}   
The divergences in 
these amplitudes are infrared divergences since renormalization
has removed all ultraviolet singularities.  Their 
presence signals the existence of nonperturbative physics 
in the process. As mentioned earlier, they will be 
factorized into nonperturbative matrix elements whose values can 
be extracted from experiment.  We will see this explicitly in
the next section.  For now, we summarize the contributions 
from the gluon 
sector.

\begin{figure}
\label{jfig4}
\epsfig{figure = 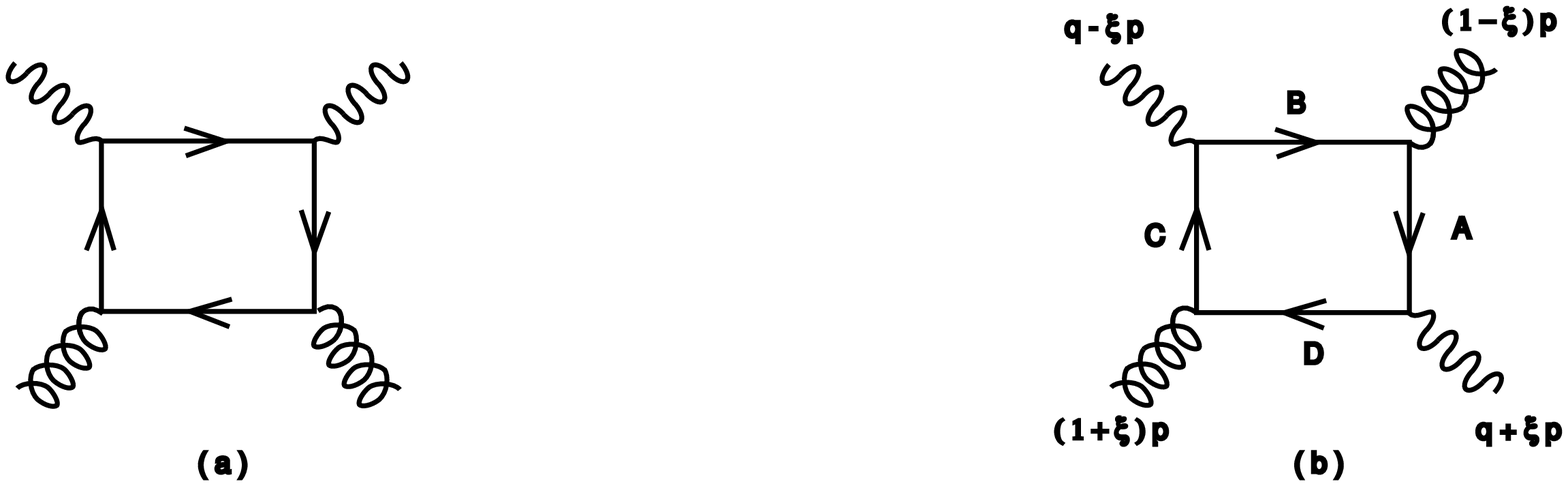, height=4cm}
\caption{Diagrams for gluon Compton scattering at one-loop.}
\end{figure}           
The six graphs that contribute to the LO amplitude for gluon-photon
scattering can be reduced to three by reversing the 
fermion number
flow, and one of these can be eliminated by crossing symmetry.  
The two distinct graphs we are left with are shown in Fig. 3.3.  
The denominator of Fig. 3.3a is
identical to that of the quark box.  Again, the numerator is seen 
to vanish 
whenever $k^2$ and $2p\cdot k$ do, which allows us to cancel one of 
the propagators exactly as above.  Fig. 3.3b is somewhat more tricky.  
This diagram is 
itself symmetric under both crossing and $\xi$ symmetry.  Labeling 
the momenta as shown, we see that under an integral the symmetry 
$q\rightarrow -q$ is 
equivalent to $A\leftrightarrow B$ and $C\leftrightarrow D$ and 
$\xi\rightarrow -\xi$ is equivalent to $B\leftrightarrow D$ and 
$A\leftrightarrow C$.  We also note that here it 
is possible to represent 1 as a linear combination of Feynman  
denominators, which guarantees our ability to cancel one.  
Of course, since 1 does not depend on $q$ or $\xi$, it may be represented 
in the symmetric form
\begin{equation}
      1=-{1\over 2(1-\xi)p\cdot q}A^2+(\xi\rightarrow -\xi)+(q\rightarrow -q).
\end{equation}
    Now it remains only to calculate the trace in a symmetric way and 
substitute the result into the formulae of the quark calculation.  
Averaging the gluon polarization for the symmetric
amplitude in $d$ dimensions, one finds  
\begin{eqnarray}
     &&  T_g^{(ij)}=-{\alpha_sT_F\over 2\pi}g^{ij}\left(\sum_{q}
e_q^2\right)\left({Q^2e^{\gamma_E}\over 4\pi\mu_0^2}\right)^{-\epsilon/2}
{1\over 1-\xi^2}
\nonumber\\ &&\;\;\;\times\;\left\lbrace
-2x_B\left\lbrack\left({x_B\over 1-\xi^2}-{1\over 2\xi}\right)
\left({4\over\epsilon}+4-\log\,\left(1-{\xi\over x_B}\right)\right)+
{1\over\xi}\right\rbrack\left(1-{\xi\over x_B}\right)\log\left(1-
{\xi\over x_B}\right)\right.\nonumber\\
&&\;\;\;\;\;\;\;+\left\lbrack\left(1+2x_B{x_B-1\over 1-\xi^2}\right)\left(
{4\over\epsilon}+4-\log\,\left(1-{1\over x_B}\right)\right)-2\right
\rbrack \log\left(1-{1\over x_B}\right)\\
&&\qquad\qquad\qquad\qquad\qquad\qquad\left. 
\vphantom{\left({Q^2e^{\gamma_E}\over
4\pi\mu_0^2}\right)^{-\epsilon/2}}
+(x_B\rightarrow -x_B)\right\rbrace\nonumber\;\; , 
\end{eqnarray}  
for the symmetric amplitude and 
\begin{eqnarray}
    &&  T_g^{[ij]}=-i{\alpha_sT_F\over 2\pi}\epsilon^{\alpha\beta ij}
n_\alpha (\eta p)_\beta\left(\sum_{q}e_q^2\right)\left({Q^2e^{\gamma_E}\over 4
\pi\mu_0^2}\right)^{-\epsilon/2}{1\over 1-\xi^2} \nonumber\\
&&\;\;\;\times\;\left\lbrace-2x_B\left\lbrack{1\over 1-\xi^2}\left({4\over\epsilon}+4-\log\,
\left(1-{\xi\over x_B}\right)\right)\right\rbrack\left(1-{\xi\over
 x_B}\right)\log\left(1-{\xi\over x_B}\right)\right.\nonumber\\
&&\;\;\;\;\;\;\;+\left.\left\lbrack\left(1+2{x_B-1\over 1-\xi^2}\right)\left(
{4\over\epsilon}+4-\log\,\left(1-{1\over x_B}\right)\right)-2\right
\rbrack \log\left(1-{1\over x_B}\right)\right.\\ 
&&\qquad\qquad\qquad\qquad\qquad\qquad -\left. 
\vphantom{\left({Q^2e^{\gamma_E}\over
4\pi\mu_0^2}\right)^{-\epsilon/2}}
(x_B\rightarrow -x_B)\right\rbrace\nonumber\;\; ,
\end{eqnarray}
where $\eta=\pm 1$ 
is the helicity of the gluon `target',
for the antisymmetric amplitude.  

\section{One-Loop Factorization in DVCS} 
\label{factdvcs1lp}
   
We now turn to the infrared divergences present in all 
of the amplitudes. These divergences arise from 
regions of the loop-momentum integration where some of 
the internal propagators are near their `mass shells'. 
In these regions, perturbative calculations 
are clearly meaningless. The standard procedure of fixing 
this problem is to factorize the amplitudes into their infrared
safe (i.e. devoid of infrared divergences) and infrared
divergent pieces, interpreting the latter as nonperturbative 
QCD quantities. 
The goal of this section 
is to show that all infrared divergences present 
in the Compton amplitudes can be associated with the off-forward
parton distributions.

Since we consider factorization within the framework of perturbation 
theory, we need to compute the parton distributions in quark and gluon 
targets in perturbative QCD. 
At the leading order in $\alpha_s$, one has
\begin{equation}
      F_{q'/q}^{(0)}(x, \xi) = \delta_{qq'}\delta (x-1) \ .
\end{equation}
At next-to-leading order, 
\begin{equation} 
  F^{(1)}_{q'/q}(x,\xi) = 
{\alpha_s(\mu^2)\over 2\pi}\delta_{qq'}
\left(-{2\over \epsilon} + \log (\mu^2 e^{\gamma_E} /4\pi\mu_0^2)\right)
    \left(P_{qq}(x, \xi) + A_q(x,\xi)\delta(x-1) \right)\;\; , 
\label{F1}
\end{equation}
where 
\begin{equation}
A_q(x,\xi) = C_F\left\lbrack{3\over 2} +  \int^x_\xi
{dy\over y-x-i\epsilon}
         +  
\int^x_{-\xi}{dy\over y-x-i\epsilon}\right\rbrack \ 
\end{equation}
can be determined from conditions analogous to 
(\ref{flavconsreq}) and 
\begin{eqnarray}
       P_{qq}(x, \xi) &=& C_F{x^2+1-2\xi^2\over (1-x+i\epsilon)(1-\xi^2)} \ ,
       ~~~~~~~~~~   x >\xi   \nonumber    \\
                 &=& C_F{x+\xi\over 2\xi(1+\xi)}
   \left(1+{2\xi\over 1-x+i\epsilon}\right)\ , ~~~~~ -\xi<x<\xi \nonumber \\
                 &=& 0 \ , ~~~~~~~~~~~~~~~~ x<-\xi \ \
\end{eqnarray}
is calculated in the same way as $P_{q\rightarrow q}(x)$
in Section \ref{partondistdis}.
The quark distributions
contain infrared divergences signaled by the presence of 
the $1/\epsilon$ terms. These divergences reflect the soft 
physics intrinsic to the parton distributions. 
$F^{(1)}_{q'/q}$, calculated for the quark 
target, satisfies the evolution equation derived in Ref. \cite{ji2} :
\begin{equation}
      \mu^2{D_a\, F_{a/T}(x,\xi,\mu^2)\over D\, \mu^2}=
{\alpha_s(\mu^2)\over 2\pi}
\int_x^1 {dy\over y}
P_{ab}\left({x\over y},{\xi\over y}\right)F_{b/T}(y,\xi,\mu^2),
\label{alt}
\end{equation}
where $b$ is summed over all parton species and 
the $P$'s are the 
off-forward Alterelli-Parisi kernels, or splitting functions.  
The `covariant' derivative is defined to include $A_a$, the 
endpoint contribution in Eq.(\ref{F1}).  Note that these
kernels reduce to the standard forward splitting
function (\ref{unpolqsplit}) in the DIS limit
$\xi\rightarrow0$.

We can re-express the symmetric part of the quark Compton amplitude
in terms of the unpolarized, off-forward quark distribution\footnote{
For convenience, we have identified the factorization
scale $\mu$ with $Q$.}
\begin{eqnarray}
    &&  T^{(ij)}_q =\nonumber\\
&&\qquad -g^{ij}\sum_{q'}e_{q'}^2\int
dx F_{q'/q}(x,\xi)\left\lbrace 
{1\over x_B-x}-{\alpha_s(Q^2) C_F\over 4\pi}\left\lbrace {9\over {x_B-x}}
\right.\right.\nonumber\\ &&
\qquad\qquad\left.\left.-{x\over\xi}\left\lbrack\left({2\xi\over x_B^2-x^2}
+{x_B+\xi\over x_+x_-}\right)\left(3-\log\,
\left(1-{\xi\over x_B}\right)\right)-3{x_B-\xi\over x_+x_-}
\right\rbrack \log\left(1-{\xi\over x_B}\right)\right.\right.\nonumber\\
&&\qquad\qquad\qquad +\left.\left.\left\lbrack\left({x_B+x\over x_+x_-}+{2\over x_B-x}
\right)\left(3-\log\,\left(1-{x\over x_B}\right)
\right)\right.\right.\right.\label{qu}\\
&&\qquad\qquad\qquad\qquad\left.\left.\left.-3{x_B-x\over x_+x_-}-{3\over
x_B-x}\right\rbrack \log
\left(1-{x\over x_B}\right)\right\rbrace\right.\nonumber\\ &&
\qquad\qquad\qquad\qquad\qquad\left.\vphantom{9\over x_b}
+(x_B\rightarrow -x_B)\right\rbrace\;\; , \nonumber
\end{eqnarray}
and find that the infrared physics is completely contained
within $F_{q'/q}(x,\xi)$.
Analogously, we find that the antisymmetric 
part of the quark Compton amplitude
can be expressed in terms of the polarized off-forward quark 
distribution
\begin{eqnarray}
    &&  T^{[ij]}_q =\nonumber\\
&&\qquad-i\epsilon^{\alpha\beta ij}n_\alpha (2hp)_\beta
\sum_{q'}{e_{q'}^2} 
\int dx\tilde F_{q'/q}(x,\xi)\left\lbrace {1\over x_B-x}-{\alpha_s 
(Q^2)C_F\over 4\pi}\left\lbrace {9\over x_B-x}\right.\right.  \nonumber\\ &&
\qquad\qquad\left.\left.-\left\lbrack\left({2x_B\over x_B^2-x^2}+{x_B+\xi\over 
x_+x_-}\right)\left(3-\log\,\left(1-{\xi\over x_B}
\right)\right)+7{x_B-\xi\over x_+x_-}\right\rbrack \log\left(1-{\xi
\over x_B}\right)\right.\right.\nonumber\\ &&\label{qp}\qquad\qquad\qquad
+\left.\left.\left\lbrack\left({x_B+x\over x_+x_-}+{2\over x_B-x}
\right)\left(3-\log\,\left(1-{x\over x_B}\right)
\right)\right.\right.\right.\\&&\qquad\qquad\qquad\qquad
\left.\left.\left.+7\;{x_B-x\over x_+x_-}
-{3\over x_B-x}\right\rbrack \log\left(1
-{x\over x_B}\right)\right\rbrace\right.\nonumber\\ &&
\qquad\qquad\qquad\qquad\qquad\left.\vphantom{9\over x_b}
-(x_B\rightarrow -x_B)\right\rbrace \ . \nonumber
\end{eqnarray}

We now turn to Compton scattering on a gluon target. Infrared
divergent contributions come from regions where the quarks 
in the box diagrams are nearly onshell. Therefore, it is natural to 
associate these divergences with the quark  
distributions in a gluon target. 
At order $\alpha_s$, off-forward gluonic matrix elements 
of the quark distribution operators have the value
\begin{equation}
     F^{(1)}_{q/g}(x,\xi) = {\alpha_s(\mu)\over 2\pi}
\left(-{2\over \epsilon} + \log (\mu^2 e^{\gamma_E} /4\pi\mu_0^2)\right)
    P_{qg}(x, \xi) \ , 
\end{equation}
where for $x>\xi$
\begin{equation}
       P_{qg}(x, \xi) = 2T_F {x^2+(1-x)^2-\xi^2 \over 
        (1-\xi^2)^2}\,\, ,
\label{gluesingoffmix}
\end{equation}
and for $-\xi<x<\xi$
\begin{equation}
       P_{qg}(x, \xi) = T_F {(x+\xi)(1-2x+\xi)\over 
              \xi(1+\xi)(1-\xi^2)}\,\, .
\end{equation}
$F^{(1)}_{q/g}(x, \xi)=0$ for $x<-\xi$\footnote{This
accounts for the factor of 2 in Eq.(\ref{gluesingoffmix}).
The antiquark contribution is treated with the 
quarks for simplicity.} 

On the other hand, the finite 
contributions come from regions where large momenta run
through the quark loop.  In these regions, 
the photon has an effective 
pointlike coupling with the gluons in the 
target. At leading order, the off-forward 
gluon distribution in a gluon target is just 
\begin{equation}
       F_{g/g}^{(0)}(x, \xi) = {1\over 2}(\delta(x-1)-\delta(x+1)) \ . 
\end{equation}

Using the above off-forward distributions, we re-express
the symmetric part of the gluon Compton scattering amplitude
\begin{eqnarray}
     &&  T_g^{(ij)} =\nonumber\\&&\qquad -{\alpha_s(Q^2)T_F\over 2\pi}g^{ij}\left(\sum_{q}
e_q^2\right)
\int dx {x\over x_+x_-}F_{g/g}(x,\xi)\nonumber\\ &&\qquad\quad
\times\left\lbrace\left\lbrack\left(1+2x_B{x_B-x\over x_+x_-}\right)\left(
4-\log\,\left(1-{x\over x_B}\right)\right)-2\right
\rbrack \log\left(1-{x\over x_B}\right)\right.\label{gu}\\ &&\qquad
\quad\quad-\left.2x_B\left\lbrack\left({x_B\over x_+x_-}-{1\over 2\xi}\right)
\left(4-\log\,\left(1-{\xi\over x_B}\right)\right)+
{1\over\xi}\right\rbrack\left(1-{\xi\over x_B}\right)\log\left(1-
{\xi\over x_B}\right)\right\rbrace\nonumber\\ &&
\quad\quad\quad\qquad- g^{ij}\sum_{q}{e_{q}^2}\int dx F_{q/g}(x, \xi)
     {1\over x_B-x} \nonumber \\ &&\qquad
\qquad\quad\quad\quad\vphantom{9\over x_b} 
+(x_B\rightarrow -x_B)\nonumber\;\; . 
\end{eqnarray}          
Similarly, we can re-express the antisymmetric part of the gluon
Compton amplitude in terms of helicity-dependent, off-forward
quark and gluon distributions
\begin{eqnarray}
    &&  T_g^{[ij]} = \nonumber\\&&\qquad-i{\alpha_s(Q^2)T_F\over 2\pi}\epsilon^{\alpha\beta ij}
n_\alpha (\lambda p)_\beta\left(\sum_{q}e_q^2\right)
\int^1_{-1} dx {x\over x_+x_-}\tilde F_{g/g}(x,\xi)\nonumber\\ &&\qquad
\quad\times\left\lbrace\left\lbrack\left(1+2x{x_B-x\over x_+x_-}\right)\left(
4-\log\,\left(1-{x\over x_B}\right)\right)-2\right
\rbrack \log\left(1-{x\over x_B}\right)\right.\label{gp}\\ &&\qquad\qquad
\quad-\left.2x_B\left\lbrack{x\over x_+x_-}\left(4-\log\,
\left(1-{\xi\over x_B}\right)\right)\right\rbrack\left(1-{\xi\over
 x_B}\right)\log\left(1-{\xi\over x_B}\right)\right\rbrace\nonumber\\ 
&&\qquad\qquad\qquad
\quad-i\epsilon^{\alpha\beta ij}n_\alpha p_\beta\sum_{q}{e_{q}^2}
\int^1_{-1} dx \tilde F_{q/g}(x, \xi)
     {1\over x_B-x} \nonumber \\     
&&\qquad\qquad\qquad\qquad\quad\vphantom{9\over x_b}
- (x_B\rightarrow -x_B) \nonumber\;\; .
\end{eqnarray}        
The complete set of off-forward 
splitting functions at leading
order can be found in Ref. \cite{ji2}.

The one-loop Compton amplitude on a target $T$
may be summarized 
by the factorization formula\footnote{This form is
correct for spin-1/2 targets.  For
spin-1, the replacement $2h\rightarrow\lambda$ is required.
This convention is fixed by the normalization of the 
distributions.}
\begin{eqnarray}
T^{ij}_T &=& -g^{ij} \int^1_{-1} {dx\over x} \left[\sum_{q} 
F_{q/T}(x, \xi) C_q\left({x\over x_B}, {\xi\over x_B}\right)
+F_{g/T}(x, \xi) C_g\left({x\over x_B}, 
{\xi\over x_B}\right) \right]\qquad \label{fac}\\
&&\!\!\!\!-i\epsilon^{ij\alpha\beta}n_\alpha (2hp)_\beta 
\int^1_{-1} {dx\over x} \left[\sum_{q}
\tilde F_{q/T}(x, \xi) \tilde C_q\left({x\over x_B}, {\xi\over x_B}\right)
+\tilde F_{g/T}(x, \xi) 
\tilde C_g\left({x\over x_B}, {\xi\over x_B}\right) \right]\,\, ,\nonumber
\end{eqnarray}
where $C_q$, $C_g$, $\tilde C_q$ and $\tilde C_g$ are shown to
order $\alpha_s$ in Eqs. (\ref{qu}, \ref{gu}, \ref{qp}, \ref{gp}), 
respectively.\footnote{I emphasize here that 
the contributions from longitudinally polarized photons and 
from photon helicity flip have been neglected. Both effects start at 
order $\alpha_s$, although longitudinal contributions
are kinematically suppressed.}
In the above form, all infrared sensitive contributions
have been isolated in the relevant parton distributions, 
which must be calculated nonperturbatively or 
measured in experiments.  In the DVCS limit
$\xi\rightarrow x_B$, the coefficient functions
remain finite, although they have branch cuts there.
This indicates that factorization holds 
for two-photon amplitudes even when one of the photons
is onshell. We will argue in the next section 
that the above formula, one of the
main results of this thesis, remains valid to 
all orders in perturbation theory. 
     
\section{All Order Factorization of DVCS}
\label{factdvcs}

In this section, we generalize the one-loop result of
the previous section, showing that the 
factorization formula Eq.(\ref{fac}) 
is valid in the DVCS limit
to all orders in perturbation theory.  
The one-loop result indicates
that all soft divergences - those associated with
integration regions where all components of some internal 
momenta are zero - cancel,  
whereas all collinear divergences can be factorized 
into the off-forward parton distributions. To see that 
this happens also at higher orders in perturbation theory, 
it is important to understand how the soft cancelation 
happens in the simplest case.

The self-energy diagram in Fig. 3.1b does not contain any infrared 
divergences because the intermediate quark is far offshell. 
The vertex corrections in Figs. 3.1c and 3.1d potentially 
have infrared divergences, but a simple power counting argument 
indicates that these diagrams are in fact infrared convergent. 
Thus infrared divergences appear only in the box diagram. 
In the region where the gluon momentum $k$ is soft, we can 
approximate the integral as 
\begin{equation}
      \sim  \int {d^4k\over (2\pi)^4} {p'\cdot p}\;
       {1\over p'\cdot k}\; {1\over p\cdot k}\; {1\over k^2} \;\; , 
\end{equation}
where $p$ and $p'$ are the momenta of the two external
quark lines.
On the other hand, the wavefunction renormalization
of an `onshell' quark, $\delta Z$,  also contains infrared divergences. 
Grouping these divergent terms together, we have the entire
soft contribution 
\begin{equation}
   \sim   -{1\over 2}\int {d^4k\over (2\pi)^4}\left({p^\mu\over p\cdot k}
      - {p'^\mu\over p'\cdot k}\right)^2 {1\over k^2} \ . 
\end{equation}
In the collinear approximation, $p'$ is proportional to 
$p$ and thus the above integral vanishes. 

For higher-order Feynman diagrams, a systematic method of
identifying, regrouping and factorizing infrared-sensitive 
contributions has been developed by Sterman, Libby, Collins, 
and others \cite{Stermanpower,Libby,Collins,factMueller}. 
As detailed in Section
\ref{factdis}, this method essentially consists of  
the following steps: 1) simplify the Feynman integrals by 
setting all the soft scales to zero, including the 
quark masses; 2) identify the regions of loop integration 
which give rise to infrared divergences; 3) use 
infrared power counting to find the leading 
infrared-divergent regions; 4) show
that all soft and collinear divergences either cancel 
or factorize into some nonperturbative quantities. 
In the remainder of this section, we examine the validity of
the factorization formula Eq.(\ref{fac}) in the limit
of $x_B=\xi$ following the above steps. 

In DVCS, the leading contributions come essentially
from a massless collinear process in which the external momenta
take the form shown in Eq.(\ref{bjdvcs}). In this simplified 
kinematic region, all infrared-sensitive contributions
appear as $1/\epsilon$ poles in dimensional regularization. 
If a contribution contains no infrared divergences, 
it comes from regions of loop momenta comparable to 
the hard scale $Q^2$, and thus is insensitive to the 
soft scales. An infrared-divergent contribution
must come from the integration regions 
where some internal propagators are near their mass shells. 
The parts of these regions which will actually
generate infrared divergences
can be identified from 
the Landau equations, as we have seen, and characterized by
reduced diagrams.
We will argue below that the general {\it leading} reduced 
diagram for DVCS is the one shown in Fig. 3.4a. Here, an
incoming virtual photon 
and an outgoing real one are attached to the hard 
interaction blob, which in turn is connected
to the forward nucleon jet with two 
collinear quark lines or two physically polarized 
gluon lines.  An arbitrary number of longitudinally polarized 
collinear gluon lines can also connect the hard 
diagram to the soft one.  This diagram is identical to that
of DIS.  Indeed, the only real difference between 
DVCS and DIS lies in the fact that the final state photon
is real.  

\begin{figure}
\label{jfig5}
\epsfig{figure=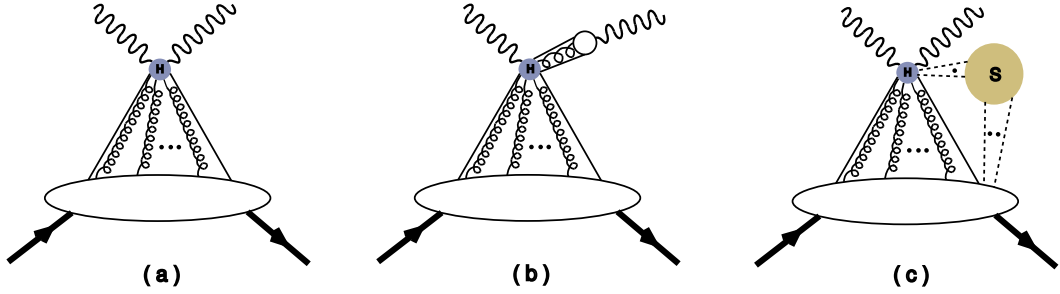,height=3.8cm}
\caption{General reduced diagrams for the DVCS process.}
\end{figure}           

A real photon can have a
pointlike coupling to quarks as well as an
extended coupling via its soft wavefunction. 
In this case, one 
has an additional reduced diagram in which a jet 
of quarks and gluons emerges in the $n^\mu$ direction
and combines into a real photon long after
the hard scattering (Fig. 3.4b). Such a reduced 
diagram has already been considered in 
Ref. \cite{jicollins} and is ${\cal O}(Q^{-1})$ by
infrared power counting.  Indeed, according to the
discussion in Section \ref{factdis}, the photon
wavefunction vertex has a soft mass 
dimension $1 = 2$ (quark lines) $- 1$ (photon state).
A negative hard power $(Q^{-1})$ is needed 
in $T^{\mu\nu}$ to balance it out. 
Hence the soft coupling of quarks to
photons will appear only at next-to-leading twist.

The only remaining difficulties lie in the soft divergences
shown in Figure 3.4c.  However, since we have
already shown that real-photon effects are sub-leading,
one can use the same arguments here as for DIS.  The differences
appear only at higher orders in our twist expansion.

At this point, the only leading reduced diagram is
identical to that in DIS.  The arguments presented in 
Section \ref{factdis} may also be applied here 
in almost exactly the same form.\footnote{The only difference
is in the off-forward nature of the distributions
relevant to this case.}
The eikonal line together with the physical quarks and gluons and 
the nucleon jets form the off-forward parton
distributions defined in Section \ref{kindvcs}. 
In the hard scattering, only the total momentum and 
charge supplied by collinear partons are important.
Thus, one can calculate it with 
incoming physical partons carrying the total momentum
and charge of all the collinear longitudinally-polarized
gluons. In this way, we have a complete 
factorization of the soft and hard physics
in the DVCS process.

\section{Generalized Operator Product Expansion \\ 
and Wilson's Coefficients to the NLO Order}
\label{opedvcs}

As we saw in Section \ref{opedis}, the calculation
of the Compton amplitude can be exploited
to derive an expansion for the product
of two electromagnetic current operators
in terms of local QCD operators.  Since the kinematics
of that chapter require identical initial and 
final states, the OPE we derived there
is valid only for diagonal matrix elements.
Here, however, we have considered a somewhat more general
case in which certain off-diagonal effects have been 
included.  It stands to reason that we can 
use these results to derive a {\it generalized} OPE
which is valid for these special off-forward
matrix elements.

In our operator derivation of the leading 
OPE, Equation (\ref{lope}), we ignored several
`total derivative' operators which cannot
contribute to forward matrix elements.
By treating the initial and final states
symmetrically, systematically using 
$\stackrel{\leftrightarrow}{\cal D}$ in the 
place of $\cal D$, we will see that these
total derivative operators still do not
contribute to the leading expansion. 
However, at next-to-leading order, they
become indispensable as generators of the $\xi$-dependence
of our amplitude.
In the remainder of this section, we
will recast our factorization formula in its generalized
OPE form. In the process, we identify these total derivative 
operators and obtain their Wilson coefficients 
to next-to-leading order in $\alpha_s$. 
The final result agrees with the known OPE, Eq.(\ref{symmopedis}), 
when diagonal matrix elements are considered.  

To derive the generalized OPE, 
we expand $T^{ij}$ in a power 
series about $x_B=\infty$.  In this way, we 
can express the amplitude in terms of
moments of the parton distributions rather than 
the distributions themselves.  Eventually, we will 
relate the moments of parton distributions
to the matrix elements of local operators.
After the aforementioned expansion, we have
\begin{eqnarray}
     && T^{ij}_T=-g^{ij}T_S-i
\epsilon^{\alpha\beta ij}n_\alpha (2hp)_\beta
T_A\,\,\, ;\nonumber\\
     && T_S = 2 \mathop{{\sum}'}_{\!\!n=2}^{\!\!\infty}\,\,
\mathop{{\sum}'}_{\!\!m=0}^{\!\!\infty}\int{dx\over x}{\left({x\over x_B}
\right)}^n{\left({\xi\over x_B}\right)}^m 
\sum_{a=q,g} c^{\,a}_{nm}F_{a/T}(x,\xi)\;\; ,
\\
     && T_A = 2 \mathop{{\sum}'}_{\!\!n=1}^{\!\!\infty}\,\,
\mathop{{\sum}'}_{\!\!m=0}^{\!\!\infty}\int{dx\over x}\left({x\over x_B}\right)^n
\left({\xi\over x_B}\right)^m\sum_{a=q,g} \tilde c^a_{nm}
\tilde F_{a/T}(x,\xi)\;\; .
\label{OPE}
\end{eqnarray} 
The coefficients for the moments
of the quark distributions in the expansion are
\begin{eqnarray}
     c_{nm}^q &=& \delta_{m0}-{\alpha_sC_F\over 4\pi}\left\lbrace
\left\lbrack 9-{8\over n}+{2\over n+1}+4S_2(n-1)-4T_1^1(n-1)\right.
\right.\nonumber \\ 
&& \qquad\left. -S_1(n-1)\left(3+{2\over n}+{2\over n+1}\right)\right]
\delta_{m0}\nonumber\\
&&\qquad\qquad + \left[ {6n\over m(m+n)}-{2\over(n+m)(n+m+1)}+ {4\over m}S_1(m-1)\right.
\\ &&\quad\qquad\qquad \left.\left.
-2S_1(n+m-1)\left({1\over n+m}+{1\over n+m+1}
\right)\right](1-\delta_{m0})\right\}\ , \nonumber\\
    \tilde c_{nm}^q &=& \delta_{m0}-{\alpha_sC_F\over 4\pi}\left
\lbrace\left\lbrack 9-{8\over n+1}+{2\over n}+4S_2(n-1)-4T_1^1(n-1) 
     \right.\right.\nonumber \\
   &&\qquad \left. -S_1(n-1)\left(3+{2\over n}+{2\over n+1}
     \right)\right]\delta_{m0}\nonumber\\
 &&\qquad\qquad  +\left[ {6n\over m(n+m)}+{8\over(n+m)(n+m+1)}+{4\over m}S_1(m-1)\right.
    \\ 
    &&\qquad\qquad \left.\left.
      -2S_1(n+m-1)\left({1\over n+m}+{1\over n+m+1}\right)\right]
    \left(1-\delta_{m0}\right)\right\} \;\; ,\nonumber   
\end{eqnarray}
where I have introduced
\begin{eqnarray}
     S_j(n)&\equiv&\sum_{i=1}^n{1\over i^j} \;\; ,  \nonumber\\
     T_j^k(n)&\equiv&\sum_{i=1}^n{S_j(i)\over i^k} \;\; ,
\end{eqnarray}
as in Section \ref{opedis}.
The above expansion contains only positive powers
of $x$ and $\xi$. This result
is not immediately obvious because of the $x_+x_-$ denominators 
in the amplitudes. In the case of the gluon distribution 
functions, we have an additional factor of $x_+x_-$ in 
the denominator. Since the final OPE contains only local operators,
these factors have to be canceled in the process of 
expansion. This is indeed the case; 
the coefficients of the positive
moments of the gluon distributions are
\begin{eqnarray}
      c_{nm}^g &=&{\alpha_sT_F\over 2\pi}\left\lbrack 
{m\over n+m}-{m+2\over n+m+2}\right.\nonumber\\&&\qquad\qquad\left.-2S_1(n+m-1)\left({1\over n+m}
-{m+2\over n+m+1}+{m+2\over n+m+2}\right)\right\rbrack\;\; ,\nonumber\\
      \tilde c_{nm}^g  &=&{\alpha_sT_F\over 2\pi}\left
\lbrack 2+2S_1(n+m-1)\right\rbrack\left({m+1\over n+m}-
{m+2\over n+m+1}\right)\;\; . 
\end{eqnarray}

Having obtained an expansion involving the moments of the 
distributions, we move toward a general form of
the OPE.  To this end, we consider the moments 
of the parton distributions.  As before, these moments can
be made into local operators via an appropriate use
of integration by parts.  The result is\footnote{As 
mentioned above, the 
off-diagonal nature of these matrix elements complicates
this process slightly.  The derivatives that actually 
appear are $\stackrel\leftrightarrow{\cal D}\,
\equiv\,1/2(\stackrel\rightarrow
{\cal D}-\stackrel\leftarrow{\cal D})$, where the 
arrow indicates the direction the derivative is meant
to operate.  In DIS, this distinction
is unnecessary since $\stackrel\leftrightarrow{\cal D}\,=\,
\stackrel\rightarrow{\cal D}$ in this case.}
\begin{equation}
     \int^1_{-1} dx x^{n-1}F_q(x,\xi)={1\over 2}
\left\langle P_f\left|\overline\psi_q(0)
\stackrel\leftrightarrow{i\cal D}_{(\mu_1}\cdots 
\stackrel\leftrightarrow{i\cal D}_{\mu_{n-1}}\gamma_{\mu_n)}\psi_q(0)
\right|P_i\right\rangle n^{\mu_1}\cdots n^{\mu_n}\;\; .
\end{equation}
Defining 
\begin{equation}
      _q{\cal O}_{\mu_1\mu_2\cdots\mu_n}^n=\overline 
\psi_q(0)\stackrel\leftrightarrow{i\cal D}_{(\mu_1}\cdots \stackrel
\leftrightarrow{i\cal D}_{\mu_{n-1}}\gamma_{\mu_n)}\psi_q(0),
\end{equation}
where $(\cdots )$ signifies that 
the indices are symmetrized and the trace has been removed,
we see that the replacement
\begin{equation}
      (n\cdot i\partial)\,\, _q{\cal O}_n^{+\cdots +} 
  \rightarrow 2\xi{\cal O}_n^{+\cdots +} 
\end{equation}
is valid for the matrix elements we consider.
This prompts us to define
\begin{eqnarray}
      _q{\cal O}^{n,m}_{\mu_1\mu_2\cdots\mu_n} & = & i\partial_{\,(\mu_1}
\cdots i\partial_{\,\mu_m}\overline\psi(0)\stackrel\leftrightarrow
{i\cal D}_{\mu_{m+1}}\cdots\stackrel\leftrightarrow{i\cal 
D}_{\mu_{n-1}}\gamma_{\mu_n)}\psi(0)\ , \nonumber \\ 
      _q{\tilde {\cal O}}^{n,m}_{\mu_1\mu_2\cdots\mu_n} & = & 
i\partial_{\,(\mu_1}
\cdots i\partial_{\,\mu_m}\overline\psi(0)\stackrel\leftrightarrow
{i\cal D}_{\mu_{m+1}}\cdots\stackrel\leftrightarrow{i\cal 
D}_{\mu_{n-1}}\gamma_{\mu_n)}\gamma_5\psi(0)\ , \nonumber \\
     _g{\cal O}^{n,m}_{\mu_1\mu_2\cdots\mu_n} & = &i\partial_{\,(\mu_1}
\cdots i\partial_{\,\mu_m}F_{\mu_{m+1}\alpha}(0)\stackrel\leftrightarrow
{i\cal D}_{\mu_{m+2}}\cdots\stackrel\leftrightarrow{i\cal
D}_{\mu_{n-1}}F^\alpha_{~\mu_n)}(0)\ , \nonumber \\ 
     _g{\tilde {\cal O}}^{n,m}_{\mu_1\mu_2\cdots\mu_n} & 
   = &i\partial_{\,(\mu_1}
\cdots i\partial_{\,\mu_m}F_{\mu_{m+1}\alpha}(0)\stackrel\leftrightarrow
{i\cal D}_{\mu_{m+2}}\cdots\stackrel\leftrightarrow{i\cal
D}_{\mu_{n-1}}i\tilde F^\alpha_{~\mu_n)}(0)\ . 
\end{eqnarray}
After replacing the moments of 
parton distributions in Eq.(\ref{OPE})
with matrix elements of these operators and 
interpreting the result as an operator relation, 
we find the following
generalized OPE,
\begin{eqnarray}
     &&~~~ i\int d^{\,4}z~e^{iq\cdot z}~
    TJ^{\, \mu}\left({z\over 2}\right)J^{\, \nu} \left(-{z\over 2}\right)
   \nonumber \\ 
&=& \left(-g^{\mu\nu}+{q^\mu q^\nu\over q^2}\right)
\mathop{{\sum}'}_{\!\!n =2}^{\!\!\infty}\,\,
\mathop{{\sum}'}_{\!\!m=0}^{\!\!n}
\left({2^{n-m}q_{\mu_1}\cdots q_{\mu_n}\over (Q^2)^n}\right)
 \sum_{a=q,g} c^a_{n-m,m}\,\,_a{\cal O}_{n,m}^{\mu_1
\cdots \mu_n}  \nonumber \\ 
&& - i\epsilon^{\mu\nu \alpha\beta} q_\alpha\,g_{\beta\lambda} 
\mathop{{\sum}'}_{\!\!n = 1}^{\!\!\infty}\,\,
\mathop{{\sum}'}_{\!\!m=0}^{\!\!n}
\left({2^{n-m}q_{\mu_2}\cdots q_{\mu_n}\over 
(Q^2)^n}\right)  \sum_{a=q,g} \tilde c^a_{n-m,m}\,
\,_a\tilde{\cal O}_{n,m}^{\beta \mu_2\cdots \mu_n} \;\; ,
\label{ope2}
\end{eqnarray}
where the primed summations once again imply 
even or odd values only, depending on the initial
value.  It must be pointed out that the generalized OPE does not 
have a unique form. One can define $x_B$ as any nontrivial dimensionless 
invariant formed from the external momenta which remains finite
in the Bjorken limit and expand the amplitude in inverse powers
of this variable.  This will lead to
a different set of coefficient functions, but the physical 
content is the same.  The choice of which OPE to use is determined
by the specifics of the problem at hand.

Since the operators ${\cal O}_{n,m}$ are 
symmetrized and traceless, their rank is $n$.  
The mass dimension of these operators is just the 
dimension of the fields plus that 
of the derivatives, or $3+n-1=4+n-2$.  
Hence these operators are all twist 2.  As we saw in
Section \ref{opedis}, the small number of 
possible twist-2 operators in QCD led to 
autonomous evolution of the diagonal
matrix elements of our operators in the nonsinglet sector.
Here, we see that for a given spin $n$
there are in fact $n$ independent nonsinglet twist-2
operators with the same quantum numbers
that contribute to our generalized OPE.
Since they are not constrained by symmetry
to evolve autonomously, the off-diagonal matrix
elements of these operators will mix
as the scale is varied.  It is this behavior that
leads to the $\xi$ dependence of the off-forward
splitting kernels above.  

Although Lorentz invariance
cannot distinguish between these operators, 
the larger symmetry group of conformal invariance
can.  It is the irreducible representations
of {\it this} symmetry group, which are definite
combinations of our operators ${\cal O}_{n,m}$,
that diagonalize the leading order anomalous dimension
matrix.  Using this larger symmetry group, one can 
determine {\it all} of the above generalized Wilson
coefficients {\it from the results of 
Section \ref{opedis}} \cite{Belitsky}.
Unfortunately, conformal invariance is broken 
by a quantum anomaly\footnote{Conformal invariance
is based on the fact that the (massless) QCD
lagrangian contains no scale.  However,
dimensional transmutation
generates a scale ($\mu^2$)
on which parameters can depend.  
Stated succinctly, the conformal invariance
of QCD is broken by the $\beta$-function.}
beyond this order and
can no longer constrain the form of 
the OPE.  

The above expression contains only the contributions
at leading order in $1/Q^2$.  At the next order in $1/Q$
one has to consider operators of higher twist.  Some of these
operators are discussed in the next chapter.  

\section{Summary of DVCS}
\label{sumdvcs}

In this chapter, we have studied the QCD factorization for 
deeply virtual Compton scattering explicitly at one loop and
then to all orders in perturbation theory. Our conclusion 
is that DVCS is factorizable in perturbation theory at leading twist. 
This statement has the same level of rigor as 
that for DIS studied in Chapter \ref{dis}.
The OFPD's probed in this process are not only 
interesting themselves, but
also can be manipulated to yield 
information on the total spin carried by quarks in the 
proton.  In conjunction with DIS data, we can use
this to obtain the quark orbital contribution to 
proton spin.  This quantity has tremendous implications on
the nature of the proton wavefunction and is not readily
obtainable from any other known process.
This makes DVCS a prime candidate for experimental study in the next 
few years.  Indeed, several experiments are already proposed at
Jefferson Lab.

Expanding our one-loop result for general off-forward Compton
scattering about $x_B=\infty$, we were able to obtain
a generalization of Wilson's OPE.  Our expression is 
valid at leading twist for off-diagonal as well as diagonal matrix
elements.  This generalization requires the introduction 
of total derivative operators whose forward matrix elements vanish.
Our one-loop result gives the coefficients of these operators
to NLO in QCD.  A full generalization would also require
an analysis of longitudinal photon scattering, which has been done
in Ref. \cite{man}, and the photon-helicity flip 
amplitude \cite{hoodbhoy}. 

The scale 
evolution of total derivative operators can best be
studied using conformally-symmetric operators \cite{Belitsky}. 
In fact, it has been known for a long time that 
at the leading-log level, operators of the same twist and
dimension evolve multiplicatively in Gegenbauer 
polynomial combinations. It is a simple exercise to 
transform Eq.(\ref{ope2}) into this basis.
This fact is a consequence of conformal symmetry in massless QCD.
Although this symmetry is broken by a quantum anomaly \cite{confano},
it still has implications
on the behavior of the OPE.  One can actually obtain 
the one-loop generalized OPE coefficients from the 
OPE introduced in Section \ref{opedis} by exploiting 
the constraints from conformal symmetry.  Unfortunately,
the quantum anomaly causes this method to become uneconomical
for higher-order calculations.

Measurements of the DVCS amplitude are made
difficult by the fact that it competes
with the simple bremsstrahlung process 
in which the electron deposits the full momentum
transfer $\Delta$ with one virtual photon, 
emitting the final state photon on its own.
Unfortunately, due to the propagator of the 
virtual photon in this process, it 
wins the competition in the region of 
small $-t$ relevant to spin physics. 
However, this competition is a mixed blessing
since the interference between the two amplitudes
allows a measurement of the phase associated 
with DVCS.  Furthermore, in regions where the 
direct emission dominates, the interference
actually {\it enhances} the effect.
The Bethe-Heitler amplitude for direct emission
is quite well-known.  Since the only hadronic
physics
that enters it are standard electromagnetic
form factors, there should be no difficulty 
in performing a direct subtraction to extract the DVCS 
contribution.  Some estimates
of the cross-section in several different 
kinematical regions is presented in Ref. \cite{ji2}.

\chapter{The Structure Function $f_T(x,\mu^2)$ - \\
Higher Twist Evolution}
\label{htwist}

Until this point, we have been studying only the leading contributions
to scattering processes in the Bjorken limit.  This led
to distribution functions whose moments could be classified 
as twist-2.  As we have seen, partons described by
twist-2 distributions
move collinear to the 
parent hadron.  This is what we expect since 
transverse behavior is suppressed in the Bjorken 
limit.  However, a complete understanding of hadrons
requires information that is not contained in these collinear 
distribution functions.  In particular, transverse momentum
effects are intimately related to partonic correlations
within hadrons.  

A simple example of coherent 
parton scattering is the interference
of a single quark with a
quark {\it and} a gluon in a nucleon target. 
To describe this 
phenomenon, it is necessary to introduce
a three-parton light-cone correlation function
\begin{equation}
 {\cal M}^\alpha (x,y,\mu^2) = \int {d\lambda\over 2\pi} {d\mu\over 2\pi}
     e^{-i\lambda y} e^{-i\mu(x-y)}
   \langle PS| \overline \psi(\lambda)iD^\alpha(\mu n)\psi(0)
     |PS\rangle\;\; , 
\end{equation}
rather than the usual two-parton functions
of the last two chapters.
General parton correlations involve
more than one Feynman variable, and hence
their scale evolution 
is more complicated than those we
have studied.  
  
Experimental study of parton correlations is 
challenging for a number of reasons. One  
is the lack of processes in which all Feynman variables in
a parton correlation can be kinematically controlled. 
For instance, in lepton-nucleon deep-inelastic 
scattering with trasverse polarization, one can measure the 
structure function $f_T(x,\mu^2)$. In 
the Bjorken limit, $f_T(x, \mu^2)$ 
is related to a $y$-moment of the above 
correlation function. Since a moment of $M^\alpha(x,y, \mu^2)$
does not evolve autonomously, knowing the entire
$f_T(x, \mu^2)$ at one scale is not 
sufficient to calculate it at another. This makes
an analysis of $f_T(x,\mu^2)$ data at 
different scales difficult. 

Several years ago, Ali, Braun, and Hiller (ABH) \cite{ali}
made a remarkable discovery that in 
the limit of large number of colors, $N_c$, 
the twist-3 part of $f_T(x, \mu^2)$ does evolve autonomously 
at the one-loop level in the nonsinglet sector. 
Their result has since been widely
used in model calculations and analyses of 
experimental data \cite{use}. More recently, similar
results have been found for the evolutions of other twist-3
functions, $h_L(x, \mu^2)$ and $e(x, \mu^2)$ \cite{other}. 
Given the practical importance of the ABH result, a deeper 
understanding of the large-$N_c$ simplification is 
clearly desirable. Moreover, it is interesting
to investigate the possibility of a similar 
simplification 
at two or more loops and for analogous
twist-4 correlations. 

In this chapter, I present a 
direct calculation of the large-$N_c$
evolution of $f_T(x, \mu^2)$ in the light-cone gauge.
This calculation is based on original research by
X. Ji and I published in \cite{metwist}.
Section \ref{transdis} is devoted to a discussion
of the kinematics and relevant correlations
in our process.  
In Section \ref{evolution},
the details of the calculation are discussed.
Here, we will see that the autonomy of the twist-3 evolution
arises from a special property of one particular 
Feynman diagram.  Since this property is independent of 
the $\gamma$-matrix structure 
of the composite operators inserted, 
the ABH result generalizes immediately to the 
twist-3 parts of $h_L(x, \mu^2)$ and $e(x, \mu^2)$.
Unfortunately, for various reasons, it seems  
there is no similar large-$N_c$ simplification 
for twist-4 functions nor for $f_T(x, \mu^2)$ 
beyond one loop. 

A summary of this chapter appears in Section \ref{sumhtwist}.

\section{DIS With Transverse Polarization}
\label{transdis}

The process we wish to consider here has the same kinematics as
that of Chapter \ref{dis}, but here we take the proton spin 
along a transverse direction (either the 1- or 2-direction).
Of course, this means that we cannot take the proton
mass strictly to zero because massless fermions do not
have transverse spin components.  However, we will ignore this
quantity as small whenever it is subleading.

Since $q\cdot S=0$, the spin-dependent part of $T^{\mu\nu}$
can be written 
\begin{equation}
T^{[\mu\nu]}=-i\epsilon^{\mu\nu\alpha\beta}q_\alpha S_\beta{2M\over\nu}
\left(S_1(\nu,Q^2)+S_2(\nu,Q^2)\right)\;\; .
\end{equation}
Assuming a factorization theorem for this process,
we can express its amplitude as a convolution
of a perturbative scattering coefficient function
with certain nonperturbative distributions.
As before, we write the quark density matrix\footnote{
Here and for the remainder of this chapter, 
we work in the light-cone gauge $n\cdot{\cal A}=0$.}
\begin{equation}
{\cal M}_{\alpha\beta}(x)=
\int{d\lambda\over2\pi}e^{-i\lambda x}\left\langle PS\right|
\overline\psi_\beta(\lambda n)\psi_\alpha(0)
\left|PS\right\rangle={1\over2}f(x)(\not\!P)_{\alpha\beta}
+f_T(x)M(\gamma_5\!\not\!S)_{\alpha\beta}\;\; .
\end{equation}
The first term is just the spin-averaged quark
distribution function of Chapter 2.  Since this
term is spin-independent, the distinction 
between transversely polarized targets and 
longitudinally polarized targets is meaningless to it.

In the limit $M\rightarrow0$, the second structure
reduces to $h\!\not\!\!p\,\,\gamma_5$ as in Section \ref{partondistdis}.  
However, its coefficient contains more than just the 
function we met there.  The easiest 
way to see this is by examining its
moments :
\begin{eqnarray}
\label{fTdef}
f_T(x)&=&-{1\over M}S_\alpha\int{d\lambda\over2\pi}
e^{-i\lambda x}\left\langle PS\right|\overline\psi(\lambda n)
\gamma^\alpha\gamma_5\psi(0)\left|PS\right\rangle\;\; ,\\
\int\,dx\,x^{n-1}f_T(x)&=&-{1\over M}S_\alpha\left\langle
PS\right|\overline\psi\gamma^\alpha\gamma_5\left(i\partial^+\right)^{n-1}
\psi\left|PS\right\rangle\;\; .
\end{eqnarray}
The local operators associated to the moments
of $f_T(x)$ are very similar to those 
related to $\tilde f(x)$.  In fact, using Lorentz
invariance we can write
\begin{eqnarray}
\left\langle PS\right|\overline\psi\gamma^{(\mu_1}
i{\cal D}^{\mu_2}\cdots i{\cal D}^{\mu_n)}\gamma_5\psi\left|PS
\right\rangle&=&2\tilde a_nMS^{(\mu_1}P^{\mu_2}\cdots P^{\mu_n)}\;\; ,\nonumber\\
\left\langle PS\right|\overline\psi\gamma^{+}
i{\cal D}^{+}\cdots i{\cal D}^{+}\gamma_5\psi\left|PS
\right\rangle&=&2\tilde a_nMS^{+}P^{+}\cdots P^{+}\;\; .
\end{eqnarray}
Here, we have symmetrized all of the indices
and removed all possible traces to form
a twist-2 operator.  Working with the kinematics
of Chapter 2, we can extract the value
of $\tilde a_n$ in polarized DIS :
\begin{equation}
\tilde a_n=\int\,dx\,x^{n-1}\tilde f(x)\;\; .
\end{equation}
Hence only the twist-2 piece of $f_T(x)$ is 
known from DIS.  The remaining twist-3
portion contains new physics.
Let us see how $f_T$ contributes to
$T^{[\mu\nu]}$ at leading order.

\begin{figure}
\epsfig{figure=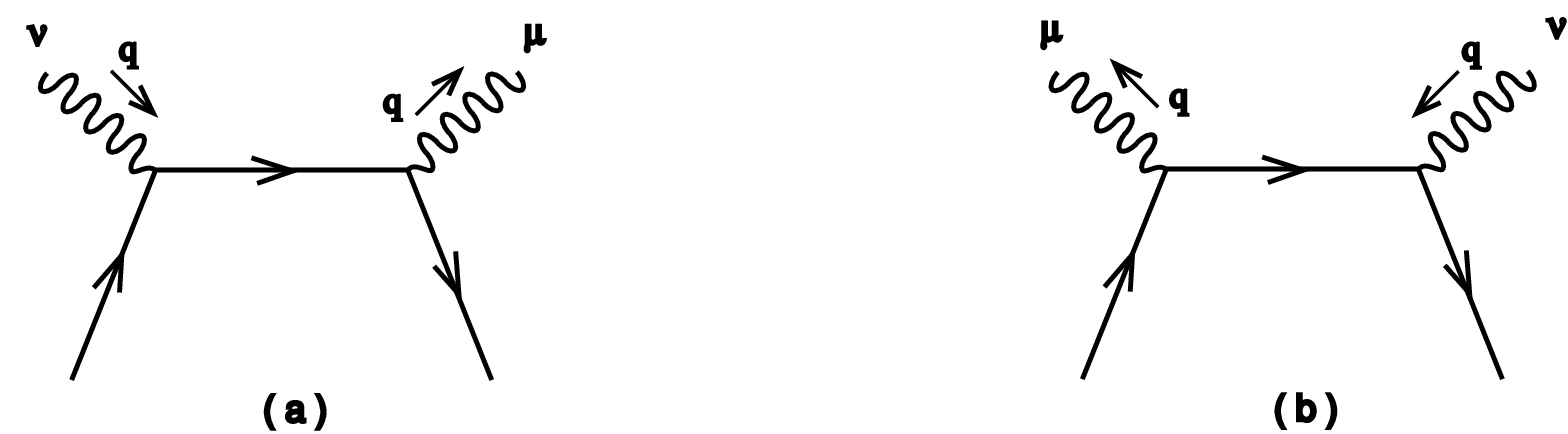,height=1.5in}
\caption{Leading diagrams for `naive'
scattering with transverse polarization.}
\label{fig9g}
\end{figure}

The relevant diagrams for quark scattering 
at leading order are
shown in Figure 4.1.  The   
result,
\begin{equation}
T^{[\mu\nu]}\sim i\epsilon^{\mu\nu\alpha\beta}
 (MS)_\beta\;{2e_q^2\over\nu}\;
\int\,dx\,f_T(x)\,{1\over x-x_B}\,(xp+q)_\alpha\;\; ,
\label{wouldntitbenice}
\end{equation}
does not satisfy the QED Ward identity, $q_\mu T^{\mu\nu}=0$,
unless $S$ and $p$ are parallel.  As usual,
the failure of an amplitude 
to satisfy Ward identities indicates missing
contributions.  Here, we are being reprimanded 
for neglecting the contribution due
to transverse quark momentum.  Our quarks 
are massless, so their spins are aligned with their
momenta.  To support
the transverse
polarization of the nucleon, quarks must
have a transverse component to their momenta.

Revisiting the diagram of Figure 4.1, we take the 
incoming quark to have momentum $xp+k_\perp$ and 
expand the amplitude about $k_\perp=0$.  
The expansion
of the intermediate propagator is 
\begin{equation}
{i\over x\!\not\!p\,+\!\not\!q\,+\!\not\!k_\perp}=
{i\over x\!\not\!p\,+\!\not\!q}+{i\over x\!\not\!p\,
+\!\not\!q}\;{i\!\not\!k_\perp}\;
{i\over x\!\not\!p\,+\!\not\!q}
+{\cal O}(k_\perp^2/Q^2)\;\; , 
\end{equation}
so the leading transverse momentum 
correction takes the form
\begin{equation}
ie_q^2\int\,dx\int d^{\,2}\,k_\perp
\;{\rm Tr}\left\lbrack\gamma^\mu\;{i\over x\!\not\!p\,
+\!\not\!q}\;{i\!\not\!k_\perp}\;
{i\over x\!\not\!p\,+\!\not\!q}
\gamma^\nu{\cal M}(x,k_\perp)\right\rbrack\;\; .
\end{equation}
The transverse momentum distribution, 
\begin{eqnarray}
{\cal M}_{\alpha\beta}(x,k_\perp)&\equiv&\sum_X\left\langle PS\right|
\overline\psi_\beta\left|X\right\rangle\left\langle X\right|
\psi_\alpha\left|PS\right\rangle\delta(n\cdot p_X+x-1)
\delta(p_{\perp X}+k_\perp)\nonumber\\
&=&\int{d\lambda\over2\pi}{d^{\,2}z_\perp\over(2\pi)^2}
e^{-i\lambda x}e^{iz_\perp\cdot k_\perp}
\left\langle PS\right|\overline\psi_\beta(\lambda n)
\psi_\alpha(z_\perp)\left| PS\right\rangle\;\; ,
\end{eqnarray}
represents the probability of finding
a quark with momentum $xp+k_\perp$ in a nucleon
of momentum $P$ and polarization $S$.  
Turning the factor of $k_\perp$ in our amplitude into
a derivative via partial integration, we arrive at\footnote{
The $\perp$ on $\gamma_\alpha^\perp$ is to remind us that the 
sum over $\alpha$ includes only transverse dimensions.}
\begin{equation}
ie_q^2\int\,dx\;{\rm Tr}\left\lbrack\gamma^\mu\;
{i\over x\!\not\!p\,+\!\not\!q}\;{i\gamma^\perp_\alpha}\;
{i\over x\!\not\!p\,+\!\not\!q}
\gamma^\nu{\cal M}^\alpha(x)\right\rbrack\;\; .
\label{gaugedeppiece}
\end{equation}
${\cal M}^\alpha(x)$ is a new distribution representing
the momentum asymmetry induced by the hadron polarization.

\begin{figure}
\epsfig{figure=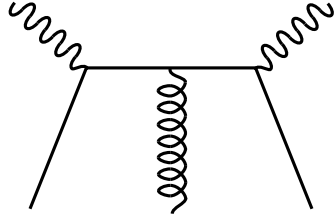,height=3cm}
\label{fig19}
\caption{The gluon correction required for QCD gauge-invariance
in DIS with transverse polarization.}
\end{figure}

In its present form, ${\cal M}^\alpha(x)$ is not gauge-invariant.
Since transverse momentum cannot be separated from
transversely polarized gauge fields, 
we must include these effects to achieve 
gauge invariance.  This implies that the contribution
represented by the 
diagram in Figure 4.2 is necessary.
In terms of the distribution
\begin{eqnarray}
{\cal M}^\alpha(x,y)&\equiv&
\int{d\lambda\over2\pi}\,{d\mu\over2\pi}\;
e^{-i\mu(x-y)}\,e^{-i\lambda y}\left\langle PS\left|
\overline\psi(\lambda n)i{\cal D}^\alpha(\mu n)
\psi(0)\right|PS\right\rangle\nonumber\\
\label{tw3decomp}
\phantom{{\cal M}^\alpha(x,y)}&=&G_1(x,y)MS^\alpha\;\gamma_5\!\!\not\!p
+G_2(x,y)MiT^\alpha\not\!p\;\; ;\\
\,G_1(x,y)&=&-{S_\sigma\over M}
\int{d\lambda\over2\pi}\,{d\mu\over2\pi}\;
e^{-i\mu(x-y)}\,e^{-i\lambda y}\left\langle PS\left|
\overline\psi(\lambda n)\!\!\not\!n\,\gamma_5i{\cal D}^\sigma(\mu n)
\psi(0)\right|PS\right\rangle\\
\,G_2(x,y)&=&i{T_\sigma\over M}
\int{d\lambda\over2\pi}\,{d\mu\over2\pi}\;
e^{-i\mu(x-y)}\,e^{-i\lambda y}\left\langle PS\left|
\overline\psi(\lambda n)\!\!\not\!n\,
i{\cal D}^\sigma(\mu n)
\psi(0)\right|PS\right\rangle\;\; ,
\end{eqnarray}
where I have defined
\begin{equation}
T^\mu\equiv\epsilon^{\mu\nu\lambda\rho}S_\nu P_\lambda n_\rho
\end{equation}
to introduce the other transverse direction, and
\begin{equation}
{\cal D}^\alpha(\mu n)\equiv\partial^\alpha
+igA^\alpha(\mu n)\;\; ,
\end{equation}
our full correction can be neatly summed up as
\begin{equation}
ie_q^2\int\,dx\,dy\;{\rm Tr}\left\lbrack\gamma^\mu\;
{i\over y\!\not\!p\,+\!\not\!q}\;{i\gamma^\perp_\alpha}\;
{i\over x\!\not\!p\,+\!\not\!q}
\gamma^\nu{\cal M}^\alpha(x,y)\right\rbrack\;\; .
\end{equation}

Using the fact that $p^2=n^2=0$, this expression
can be further simplified to
\begin{equation}
{e_q^2\over4\nu}\int {dx\over x-x_B}\;
{\rm Tr}\left\lbrack\gamma^\mu
\!\not\!p\,\gamma_\alpha^\perp\!\not\!n\,\gamma^\nu
\int dy{\cal M}^\alpha(x,y)+\gamma^\mu
\!\not\!n\,\gamma_\alpha^\perp\!\not\!p\,\gamma^\nu
\int dy{\cal M}^\alpha(y,x)\right\rbrack\;\; .
\end{equation}
Substituting (\ref{tw3decomp}) and using the 
relations 
\begin{eqnarray}
G_1(x,y)=\phantom{-}G_1(y,x)\phantom{\;\; .}\\
G_2(x,y)=-G_2(y,x)\;\; ,
\end{eqnarray}
which follow from the reality of our distributions,
leads to
\begin{equation}
-i\epsilon^{\mu\nu\alpha\beta}p_\alpha\left(MS\right)_\beta
{2e_q^2\over\nu}\int dx\left\lbrack\int dy\,\left(
G_1(x,y)+G_2(x,y)\right)\right\rbrack\;{1\over x-x_B}\;\; .
\label{othcontr}
\end{equation}

Apparently, only the
first moments of $G_{1,2}(x,y)$
with respect to $y$ enter our process.
From their definitions, one can 
show that 
\begin{eqnarray}
\int dy\left\lbrack G_1(x,y)+G_2(x,y)\right\rbrack
\qquad\qquad\qquad\qquad\qquad\qquad\qquad\qquad\qquad\nonumber\\
\qquad\qquad={1\over M}\;\int{d\lambda\over2\pi}\left\langle PS\left|
\overline\psi(\lambda n)i{\cal D}^\sigma_\perp(\lambda n)
\left(iT_\sigma\!\!\not\!n\,-S_\sigma\!\!\not\!n\,\gamma_5\right)
\psi(0)\right|PS\right\rangle\;\; .
\end{eqnarray}
Since the expression in parentheses is just
one of the many ways to say 
`$\gamma^\sigma\!\!\not\!n\,\!\!\not\!S\,\gamma_5$',
the equation of motion in this gauge,
\begin{equation}
\not\!n\,\!\!\not\!p\,i\partial^+\psi=-\!\!\not\!n\,
i\!\not\!\!{\cal D}_\perp\psi
\end{equation}
allows us to write finally\footnote{In order
to get rid of the structure $\not\!p\!\!\not\!n\,
-\!\!\not\!n\!\!\not\!p$, we must assert the 
reality of our distributions one more time.}
\begin{equation}
\int dy\left\lbrack G_1(x,y)+G_2(x,y)\right\rbrack
=x\,f_T(x)\;\; .
\end{equation}
Adding (\ref{othcontr})
to (\ref{wouldntitbenice}), we see that 
this contribution does indeed cancel the gauge
dependence of our initial result.  The 
complete amplitude,\footnote{we can't forget the crossing contribution!}
\begin{equation}
T^{[\mu\nu]}=i\epsilon^{\mu\nu\alpha\beta}
q_\alpha (MS)_\beta\;{2e_q^2\over\nu}\;
\int\,dx\,f_T(x)\,\left({1\over x-x_B}-{1\over x+x_B}\right)\;\; ,
\end{equation}
is fully gauge-invariant.  In fact, the only
difference between transverse and longitudinal 
polarization appears to be in the relevant
distribution function.  From the point
of view of the operator product expansion,
this result is obvious.  Since the 
OPE makes no reference to the 
polarization of the target, the coefficient of the twist-2 
contribution must be the same for both processes.
However, the twist-3 contributions are not 
constrained in this way.  In fact, from the analysis
above it seems like a miracle that only the simple
distribution $f_T(x)$ appears in the final 
results rather than $G_1(x,y)$ and $G_2(x,y)$.  
We certainly cannot expect this behavior to
persist in the next order.

The calculation above is extremely disconcerting. 
It seems very strange that we had
to consider {\it gluon} scattering and transverse
momentum effects to obtain {\it QED} gauge invariance.
In a fundamental sense, it seems that this result should've
been automatic.  However, the Feynman 
diagram in Figure 4.1 implicitly assumes that
we can interpret the scattering within
the parton model.  The trouble with a 
parton model interpretation of 
$f_{T}(x)$ can easily be seen by quantizing
the theory on the light front.  
Rather than the treatment discussed in Appendix \ref{canquant},
one can impose canonical commutation relations
on the fields and their conjugate momenta
at equal values of $x^+$ \cite{lfquant}.  In this formalism, 
the `$+$' direction becomes the new `time' variable.
For massless field quanta moving in the $+$
direction, this is a very logical choice for
the time axis.  In light-front quantization,
the quark field contains two very different 
components.  Defining
\begin{equation}
\psi_{\pm}\equiv{1\over2}\gamma^\mp\gamma^\pm\psi\;\; ,
\end{equation}
we see that the equation of motion constrains
$\psi_-$ :
\begin{equation}
   \psi_-(\lambda n) =- {1\over 2} {1\over in\cdot \partial}
     \not\! n i\!\not\!\! {\cal D}_\perp (\lambda n)\psi_+(\lambda n) \ . 
\label{eom}
\end{equation}
Physically, this means that only $\psi_+$ 
contains freely propagating\footnote{in the 
$+$ direction} quark 
quanta.  The $\gamma^-$
in the definition of $\psi_-$ has annihilated
all of the free portions of the quark field
since these satisfy $\not\!pu(xp)=0$.  
Looking at $f_T(x)$, 
\begin{eqnarray}
f_T(x)=-{1\over M}S_\alpha\int{d\lambda\over2\pi}
e^{-i\lambda x}\Big\lbrack\phantom{+}\left\langle PS\right|
\overline\psi_-(\lambda n)
\gamma^\alpha\gamma_5\psi_+(0)\left|PS\right\rangle
\phantom{\big\rbrack\;\; .}\nonumber\\
\phantom{f_T(x)=-{1\over M}S_\alpha\int{d\lambda\over2\pi}
e^{-i\lambda x}\big\lbrack}+\left\langle PS\right|
\overline\psi_+(\lambda n)
\gamma^\alpha\gamma_5\psi_-
(0)\left|PS\right\rangle
\Big\rbrack\;\; ,
\label{better}
\end{eqnarray}
we see that it contains both the `good' component
$\psi_+$ and the `bad' one $\psi_-$.  

\begin{figure}
\epsfig{figure=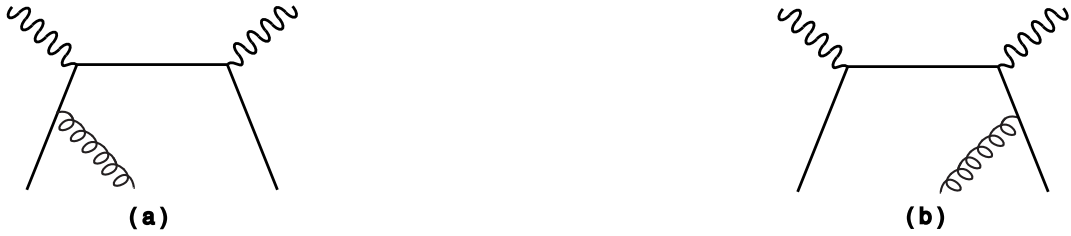,height=3cm}
\caption{These one-particle reducible diagrams represent
the true nature of figure 4.1.  They are obtained by 
eliminating the `bad' component of $\psi$ via the light-cone
equations of motion.}
\label{fig18}
\end{figure}

The 
external lines in our Feynman diagram represent
{\it free} quark fields;\footnote{Our Feynman rules
have been derived in this way; see 
Section \ref{roadtoquant}.} in order
to accurately represent $f_T(x)$ with a diagram, 
we must first eliminate all of the bad quark 
components in favor of the good ones.\footnote{Note 
that this requirement is quite independent
of QCD.  We would have to perform this manipulation
even in a theory without local gauge invariance,
where it would lead only to the transverse momentum
correction.}  This process turns Figure 4.1 into 
Figure 4.3.  An explicit calculation shows that
the contribution from the graphs of Figure 4.3
is identical to (\ref{wouldntitbenice}).\footnote{
Strictly speaking, these gluons must be combined
with the related transverse momentum insertion
to reproduce (\ref{wouldntitbenice}).  For this
reason, it is often advisable to begin with 
bad components and eliminate them via (\ref{eom}).}
However, from Fig. 4.3 it is quite clear
that the contributions of Figure 4.2 
are necessary to ensure 
QED gauge invariance.  
The point here is not that
(\ref{wouldntitbenice}) is wrong, just incomplete.
The form of (\ref{fTdef}) misleads us into 
thinking that it is the only contribution;
the more honest form (\ref{better}) 
reveals the true nature of transverse scattering
to be inextricably coupled with multiparton correlation
effects.  We will see that the effect of this coupling
has consequences far greater than the inclusion
of a few more diagrams.

\section{The Evolution Equation}
\label{evolution}

Now that we have seen that transverse
polarization cannot be decoupled from
coherent partonic correlations, it is natural
to ask what kind of effect this has
on the analysis of the previous chapters.
There, we were able to factorize the 
DIS cross-section into soft and hard sectors
due to the {\it incoherence} of the dominant
scattering mechanism.  It is clear that the
introduction of correlation effects will have
some kind of an impact on this analysis, 
but its full consequences are not obvious.

One immediate effect of coherent scattering
is a complication of the soft sector of 
our amplitude.  If a factorization theorem
does indeed exist at this level, the matrix
elements representing the soft physics 
certainly must encode a great deal more information
than those previously studied.  The distribution
in Equation (\ref{fTdef}) does not look
much different from those we met in Section 
\ref{partondistdis},
but we have just seen that this form
can be quite misleading.  

A reliable way to see how 
much information a distribution contains is by
studying its scale evolution.  For a self-contained
distribution which evolves autonomously, 
only one measurement is enough to determine
the complete behavior.  On the other hand, 
distributions which mix with each other
require external inputs for a study of their
scale dependence.  As we have seen, mixing is 
an indication that a distribution does not contain
all of the relevant physics.  

Since operators of different twist do not 
belong to the same representation of the Lorentz
group, we must separate the twist-2 and -3 
contributions to $f_T(x,\mu^2)$ in order to study
its evolution effectively.  This is most easily
done with the local operators that constitute
its moments rather than the function itself.
The $(n+1)$-st moment of $f_T(x,\mu^2)$ is written
\begin{equation}
   \int dx\, x^{n}\, f_{T}(x, \mu^2) 
   = {1\over 2 M}n_{\mu_1}\cdots n_{\mu_n}
    \left\langle PS\left|\theta^{\perp(\mu_1\cdots\mu_{n})}_n
     \right|PS\right\rangle\;\; ,
\label{mto}
\end{equation}
where 
\begin{equation}
\theta^{\sigma(\mu_1\cdots\mu_{n})}_n
= \overline \psi\gamma^\sigma\gamma_5
i{\cal D}^{(\mu_1}\cdots i{\cal D}^{\mu_{n})}\psi\;\; ,
\end{equation}
with $(\mu_1\cdots\mu_{n})$ indicating symmetrization 
of the indices and removal of the traces. 
The twist-2 $\theta^{(\sigma\mu_1\cdots\mu_{n})}$
(totally symmetric and traceless) and twist-3 
$\theta^{[\sigma(\mu_1]\mu_2\cdots\mu_{n})}$ (mixed symmetric
and traceless) contributions, where $[\sigma \mu_1]$ 
denotes antisymmetrization, can now be separated quite easily
via
\begin{equation}
\theta^{\sigma(\mu_1\cdots\mu_n)}=
\theta^{(\sigma\mu_1\mu_2\cdots\mu_n)}_{n2}
+{2n\over n+1}\theta^{[\sigma(\mu_1]\mu_2
\cdots\mu_n)}_{n3} +\cdots\;\; ,
\end{equation}
where the ellipses denote terms of higher twist.

As mentioned above, the twist-2 part of $\theta_n$
is identical to the operator related to the $(n+1)$st
moment of $\tilde f(x)$.  Its evolution 
has already been studied in Section \ref{opedis}.
We argued in that section that the twist-2 nonsinglet
quark operators must have autonomous evolution because
there are simply no operators with which they can mix.
This statement is not true at the twist-3 level.  
A complete 
basis of twist-3 quark operators 
was first identified in \cite{shuryak},
\begin{eqnarray}
 R_i^n = \overline \psi i{\cal D}^{(\mu_1} 
  \cdots i{\cal D}^{\mu_{i-1}} (-ig){\cal F}^{\sigma \mu_i} 
i{\cal D}^{\mu_{i+1}}
  \cdots i{\cal D}^{\mu_{n-1}} \gamma^{\mu_{n})}\gamma_5\psi
  \phantom{\;\; .}\nonumber \\
 S_i^n = \overline \psi i{\cal D}^{(\mu_1} 
  \cdots i{\cal D}^{\mu_{i-1}} \phantom{(-i)}
g\tilde {\cal F}^{\sigma \mu_i} i{\cal D}^{\mu_{i+1}}
  \cdots i{\cal D}^{\mu_{n-1}} \gamma^{\mu_{n})}
\phantom{\gamma_5}\psi\;\; , 
\end{eqnarray}
where $i=1, ..., n-1$. 
The operator $\theta^{[\sigma(\mu_1]\mu_2\cdots\mu_{n})}_{n3}$
is just a special linear combination of them:
\begin{equation}
  \theta^{[\sigma(\mu_1]\mu_2\cdots\mu_{n})}_{n3} 
={1\over 4n} \sum_{i=1}^{n-1} 
   (n-i)(R_i^n -R_{n-i}^n+S_i^n+S_{n-i}^n) \;\; . 
\label{relation}
\end{equation}
The anomalous dimension matrix in the above operator basis was
first worked out by Bukhvostov et al. and 
later reproduced by a number of authors with different
methods \cite{matrix}. The result is what one would generally
expect.  To evolve the matrix element of 
$\theta^{[\sigma(\mu_1]\mu_2\cdots\mu_n)}$, it is 
not enough just to know it at an initial scale - one must know 
all the matrix elements of $W_i^n=R_i^n-R_{n-i}^n
+S_i^n+S_{n-i}^n$ there.  Important physical
information pertaining to the soft physics 
of our process is simply not probed by the experiment.
This is actually quite clear from the discussion in the
last section.  There, we saw that the 
distributions $G_1(x,y)$ and $G_2(x,y)$ 
are in some sense more fundamental than 
$f_T(x)$.  The local operators $R_i^n$ and $S_i^n$
are related to the moments of {\it these} 
distributions.  

By studying the anomalous dimension matrix 
in the large-$N_c$ limit,\footnote{Large-$N_c$ physics is
discussed in Appendix \ref{largenc}.} 
Ali, Braun and Hiller found that
the eigenvector corresponding to the lowest
eigenvalue is just the linear combination
of twist-3 operators on the right-hand side
of Eq.(\ref{relation}). In other words,  
the twist-3 part of $f_{T}(x, \mu^2)$ evolves 
autonomously in this limit.  
To better understand ABH's result, we 
calculate the large-$N_c$ evolution of 
$f_{T}(x, \mu^2)$ directly. Starting
with the mixed-twist operator
$\theta^{\sigma(\mu_1\mu_2\cdots\mu_{n})}_n$ 
in Eq.(\ref{mto}), we look for possible divergences when 
inserted in multi-point Greens functions. 
To reduce the number of Feynman diagrams, we choose the 
light-cone gauge ${\cal A}^+=0$ and take the $\perp +\cdots +$
component of the $\theta$-operator.  The resulting
operator, 
\begin{equation}
\theta_n \equiv \theta_n^{\perp +\cdots +}
=\overline \psi \gamma^\perp\gamma_5 
(i\partial^+)^{n}\psi\;\; ,
\end{equation}
has the simple 
Feynman rule $\gamma^\perp\gamma_5(k^+)^{n}$, 
where $k$ is the momentum 
of the incoming quark. 

By light-cone power counting, we need only 
consider two- and three-point functions. 
Since the external lines carry color, 
we must ask what type of diagrams 
dominate the large-$N_c$ limit.  As shown in 
Appendix \ref{largenc}, the simple rule 
is that when all external lines 
are drawn to one point (infinity) 
the planer diagrams are leading.
All one-particle-irreducible
(1PI) leading two- and three-point diagrams with one 
$\theta$ insertion are shown in Fig. 4.4. 

The ultraviolet divergences in the two point 
Greens function can obviously be subtracted 
with the matrix element of $\theta_n$ itself. 
The only diagram in which the divergences may 
not be subtracted by $\theta_n$
is Fig. 4.4b. An explicit calculation 
shows that the ultraviolet divergences here
require a subtraction of the local operator
\begin{eqnarray}
   -{1\over 2}C_A {g^2\over 8\pi^2}{2\over\epsilon}
\left({\mu^2e^{\gamma_E}\over4\pi\mu_0^2}\right)^{-\epsilon/2}
   \left[ -{1\over (n+2)} \sum_{i=0}^{n-1}
    \overline \psi \not\! n \gamma_5 (i\partial^+)^i i{\cal D}^\perp
        (i\partial^+)^{n-1-i} \psi \right.\nonumber \qquad\qquad\qquad
  \\
  \quad\quad  +  \left.\left(\sum_{i=1}^{n+1}{1\over i}-{1\over 2(n+1)}\right)
     \left(\overline \psi \;i\not\!\!{\cal D}_\perp\!\not\! 
   n \gamma^\perp\gamma_5(i\partial^+)^{n-1}
   \psi  + \overline \psi 
   (i\partial^+)^{n-1}\gamma^\perp
\gamma_5 \not\! n\; i\not\!\! {\cal D}_\perp 
   \psi\right)\right]\;\; , 
\label{result}
\end{eqnarray}
where I have neglected the contributions of light-cone 
singularities which will be canceled eventually. 
Notice that the first term is present in the  
twist-2 operator 
\begin{equation}
    \theta_{n2}  = {1\over n+1}\left(
    \overline \psi \gamma^\perp \gamma_5(i\partial^+)^{n}\psi
     + \sum_{i=0}^{n-1}
    \overline \psi \gamma^+\gamma_5 (i\partial^+)^ii{\cal D}^\perp
      (i\partial^+)^{n-i-1}\psi\right) \ ,
\end{equation}
\begin{figure}
\label{largefig1}
\epsfig{figure=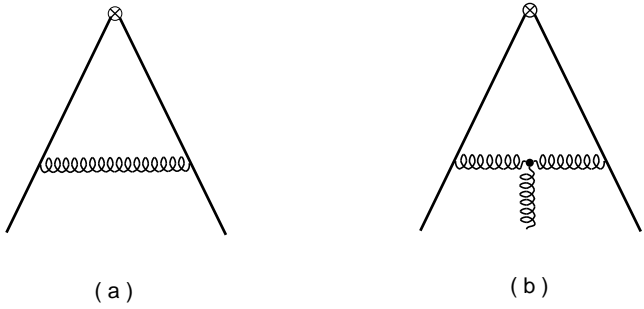,height=4cm}
\caption{The (a) two- and (b) three-point 1PI Feynman diagrams contributing
to the evolution of $\theta_n$ in the large-$N_c$ limit.}
\end{figure}    
\noindent
and the remaining two terms
can be converted to $\theta_n$ by using the 
equation of motion (\ref{eom}). Thus
we easily arrive at the ABH conclusion that $\theta_{n3}$ 
evolves autonomously in the large-$N_c$ limit. 

Including the contribution from Fig. 4.4a as well as
the wavefunction renormalization,
we obtain the evolution equation
\begin{equation}
  \mu^2{ d\theta_n\over d \mu^2}
  = {\alpha_s(\mu^2) \over 2\pi}{ N_c\over 2}
\left[{n+1\over n+2}\theta_{n2} + 
   \left (-2\sum_{i=1}^{n+1}{1\over i} + {1\over n+1} + {1\over 2}\right)
    \theta_n\right] \;\; . 
\end{equation}
Separating out the twist-2 and twist-3 parts, 
we not only recover the well-known twist-2 evolution,\footnote{
Note that this is the evolution of the $(n+1)$st moment
of the distribution rather than the $n$th.}
\begin{equation}
\mu^2{d\theta_{n2}\over d\mu^2} ={\alpha_s(\mu^2)
\over2\pi}\;{N_c\over 2}
\left( {3\over2}
+{1\over (n+1)(n+2)}-2\sum_{i=1}^{n+1}
   {1\over i}\right)\;\; ,
\end{equation}
but also the twist-3 result
\begin{equation}
    \mu^2{ d\theta_{n3}\over d\mu^2}
   ={ \alpha_s(\mu^2)\over 2\pi}\;{N_c\over 2}\left(-2\sum_{i=1}^{n+1}
   {1\over i}+ {1\over n+1} + {1\over 2}\right)\theta_{n3}\;\; ,
\end{equation}
which is identical to the result in Ref. \cite{ali}. 

It is quite clear that the $i$-independence of 
the coefficients in the sum in Eq.(\ref{result})
is the key for the autonomous evolution of $\theta_{n3}$. 
This property is not totally unexpected 
if one inspects Fig. 4.4b more closely. Interpreting 
this diagram in coordinate space, we see that the internal
gluon propagates {\it homogeneously} from one quark to the other. 
By homogeneously, we mean that at
any point along the path of the propagation,
the gluon behaves exactly the same way, 
except, of course, at the points where the gluon 
and quarks interact. The spatial location 
of the interaction with the external gluon 
determines the number of derivatives before 
and after the gluon field in the subtraction 
operator. Since the internal gluon propagation is 
homogeneous, different locations of the triple-gluon 
vertex produce similar physical effects.  
Therefore, the coefficients of the
subtraction operators $\overline \psi \!\not\!\! n 
\gamma_5 (i\partial^+)^i iD^\perp (i\partial^+)^{n-1-i}
\psi$ will be independent of $i$. 
The two extra terms in Eq.(\ref{result})
correspond to the triple-gluon vertex just next to 
the external quark lines, where the homogeneity 
is lost.  However,
these diagrams 
cannot break the symmetry since they can 
always be discarded in favor of the `bad' 
quark component
via the equation of motion (\ref{eom}).

Since the homogeneous property  of the internal
gluon line is independent of the $\gamma$-matrix structure 
of the operator inserted, we conclude that the other 
twist-3 distributions
$e(x, \mu^2)$ and $h_L(x, \mu^2)$ \footnote{These distributions
are defined in \cite{xdjhtc}.} 
also evolve autonomously in the
large-$N_c$ limit. A quick calculation confirms
the evolution equations found in Ref. \cite{other}.

Encouraged by the success of the above approach, 
we apply it to the analogous twist-4 evolution.
In Ref. \cite{jaffe}, the three one-variable
distributions $f_4(x, \mu^2)$, $g_3(x, \mu^2)$ and $h_3(x, \mu^2)$ 
are shown to contain twist-4. 
For example, $f_4(x)$ is defined as
\begin{equation}
      f_4(x) = {1\over M^2}
   \int {d\lambda \over 2\pi} \langle P |\overline \psi(0)
    \gamma^- \psi(\lambda n) |P\rangle \ .
\end{equation}
It was shown in Ref. \cite{ji} that $f_4(x, Q^2)$ 
contributes to the $1/Q^2$ term of the longitudinal 
scaling function $F_L$  of the nucleon
\begin{equation}
    F_L(x_B, Q^2) ={ 2x^2_B M^2\over Q^2}  \sum_a 
       e_a^2 f_{4a}(x_B, Q^2)\ ,  
\end{equation}
where we have neglected higher-order 
radiative corrections. Here, autonomous
evolution of $f_4(x, \mu^2)$ would  
simplify the analysis of 
$F_L$ data immensely. 

\begin{figure}
\label{largefig2}
\epsfig{figure=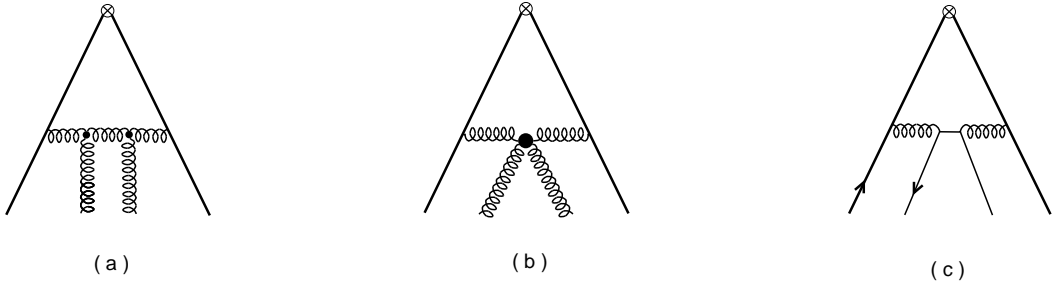,height=3.7cm}
\caption{Four-point 1PI Feynman diagrams contributing
to the evolution of $\hat O$ in the large-$N_c$ limit.}
\end{figure}    

 In the large-$N_c$ limit, we consider one insertion
of the operator $\hat O =\overline \psi\gamma^-(i\partial^+)^{n}\psi$ 
into two-, three- and four-point Green's functions. At one-loop 
order, the 1PI two- and three-point diagrams 
are identical to those in Fig. 4.4 and the 1PI 
four-point diagrams are shown in Fig. 4.5. Only the 
three and four point diagrams can potentially 
destroy the autonomous evolution of $\hat O$. 
Let us start with Fig. 4.5a. One of its divergent 
contributions 
introduces
the local subtraction
\begin{equation}
  \sum_i \overline \psi \,i\!\not\!\! {\cal D}_\perp \!\!\not \! n 
  (i\partial^+)^i i\!\not\!\! {\cal D}_\perp (i\partial^+)^{n-i-2}\psi
   + {\rm h. c.}\;\; ,
\end{equation}
where the coefficients are independent of $i$
because of the homogeneity of the gluon 
propagator. Using the equation of motion, we can 
write this as
\begin{equation}
    \sum_i \overline\psi 
         (i\partial^+)^i i\not\!\! {\cal D}_\perp (i\partial^+)^{n-i-2}\psi
   + {\rm h. c.}
\end{equation}
Since this operator cannot be reduced to either the twist-2 or 
twist-4 part of $\hat O$, the evolution
of the latter cannot be autonomous unless this contribution
is canceled by other diagrams. The only other diagram 
containing the same divergence structure is Fig. 4.4b.
Unfortunately, our explicit calculation did not 
produce this cancelation.  Here, it is not the 
inhomogeneity of gluon
propagation which destroys the autonomous evolution.
In the twist-4 sector, there are simply more operators 
which do not distinguish between quark fields (are derived
from homogeneous propagation) than appear in the 
decomposition of $\hat O$ into twist-2 and -4 contributions.
The same phenomenon
occurs for the twist-4
part of $g_3(x, \mu^2)$ and $h_3(x, \mu^2)$. 

Thus, the large-$N_c$ simplification seems to happen
only for the evolution of the 
twist-3 part of $f_T(x,\mu^2)$, $h_L(x,\mu^2)$ 
and $e(x, \mu^2)$.  Does it happen for them 
at two and higher loops?
In Fig. 4.6, we see two examples of Feynman 
diagrams suspected to break 
the autonomy of the $\theta_{n3}$ evolution, i.e.
they may contain divergences that cannot be
subtracted by $\theta_{n2}$ and $\theta_{n3}$ only. 
The inhomogeneity of
the gluon propagator leads to this suspicion. 
The internal gluon that 
propagates from one quark to another has different 
wavelengths in the different parts of the 
propagation. Its interaction with the 
external gluon will therefore be different at different 
spatial locations. Thus the subtraction operators
have different coefficients depending on 
the number of derivatives before and
after the external gluon field. An explicit
calculation of Fig. 4.6a confirms this suspicion.

\begin{figure}
\label{largefig3}
\epsfig{figure=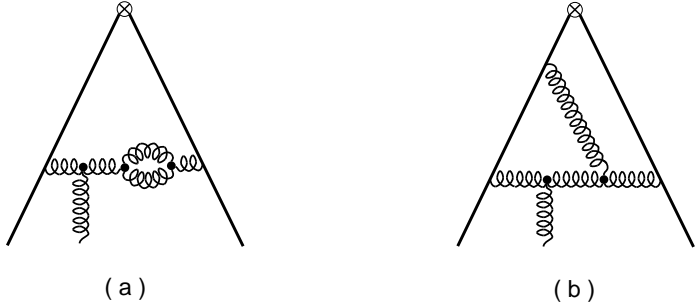,height=4cm}
\caption{Some two-loop 1PI Feynman diagrams that might break
the autonomy of the $\theta_n$ in the large-$N_c$ limit.}
\end{figure}    

This leaves us with only one possibility for autonomous  
two-loop evolution of $\theta_{n3}$: the unwanted structures
cancel in the sum of all large-$N_c$ two-loop 
diagrams.
Calculating all those diagrams is a big 
task. However, even without
an explicit calculation, we do not expect 
the cancelation to happen.
The fundamental reason is that the large-$N_c$ limit
represents only a selection of a subset
of Feynman diagrams, whereas the result 
of an individual diagram is independent 
of this limit. Cancelations 
of a structure do not generally happen among 
Feynman diagrams unless there is a 
symmetry.

\section{Conclusions}
\label{sumhtwist}

Although deep-inelastic scattering with 
transversely polarized targets seems
quite similar to that with longitudinal 
polarization, a closer look reveals
it to be quite deeply entangled with coherent
scattering effects.  This entanglement stems
from the fact that transverse polarization
{\it induces} a transverse momentum distribution
which {\it cannot} be decoupled from the 
effects of physical gluons.  

A study of the evolution of the
distribution function relevant to this
process unearths a large number of 
coupled operators which have been obscured
from view by the equations of motion.
These operators produce an extremely complicated
evolution for the twist-3 part of the measured
structure function, with the effect
that scale variation of the distribution
function {\it cannot} be performed until
one knows the full functional form of the 
more fundamental distributions $G_1(x,y)$ and
$G_2(x,y)$.  Unfortunately, these distributions
are not readily extracted separately in experiment.
Only the combination appearing in $f_T(x)$
shows up at leading order in DIS.
 
A calculation of Ali, Braun, and Hiller shows that the 
anomalous dimension matrix of these operators 
simplifies considerably in the limit of large number
of colors.  The result of this simplification is
that the twist-3 part of $f_T(x)$ evolves
autonomously in one-loop after all.  Corrections to this
evolution appear at ${\cal O}(1/N_c^2)$,
so it should be trusted to approximately 
$10\%$.  This amazing result represents an
incredible simplification of the data analysis
on polarized DIS and allows us to extract valuable
information on gluon-quark correlations
within the proton.

This simplification naturally leads us to ask
if there are any other cases in which it
occurs.  Later calculations have shown that
similar simplifications do indeed occur
for the twist-3 parts of other distributions
relevant to experiment.  An extension of the method
used by Ali, Braun, and Hiller to twist-4 
and higher or to two or more loops 
would be a heroic undertaking.
Here, we have shown 
how to avoid a calculation of the entire anomalous
dimension matrix and get their result directly.
Furthermore, using this method, one can see 
clearly {\it why} the simplification occurs.

We have seen that the large-$N_c$ simplification
happens because this limit, quite accidentally it
seems, happens to throw out exactly the diagrams
which induce mixing.  
In the light-cone gauge, the simplification 
can easily be traced to a special 
property of Fig. 4.4b. 
Unfortunately, this special property
is not enough to ensure autonomous evolution of 
the analogous twist-4 distributions at one loop, 
and does not persist to higher orders in the twist-3
case.  Nonetheless, the discovery
of Ali, Braun, and Hiller remains as a significant
step forward in the study of the $f_T(x, \mu^2)$
structure function. Without the autonomous
one-loop evolution, an analysis of experimental data
on the twist-3 contribution
would be severely constrained. 
 
\appendix
\chapter{Standard Conventions and Formulae}
\label{conventions}

This Appendix is meant to provide a table of standard formulae
and conventions used in the text.  

We work in `God-given' units, where $c=\hbar=1$.  

Throughout, I use the metric
\begin{eqnarray}
g_{\mu\nu}=\left(\matrix{1&0&0&0\cr0&-1&0&0\cr0&0&-1&0\cr0&0&0&-1\cr}\right)
\end{eqnarray}
of signature -2.

The Levi-Cevita symbol is defined via $\epsilon^{0123}=+1$.

When written explicitly, I use the convention that the 
time component of a 4-vector goes first, followed by space.  For
example, $p^\mu=(p^0,p^1,p^2,p^3)$.

Indices in Greek at the beginning of the alphabet ($\alpha,\beta,\ldots$) are
usually Dirac indices.  Exception : $\alpha$ and $\beta$
are fair game for use as Lorentz indices.

Indices in Greek in the middle of the alphabet ($\mu, \nu,\ldots$) are usually
Lorentz indices.  

Indices in Latin at the beginning of the alphabet ($a,b,\ldots$) are usually
in the adjoint representation of $SU(3)$.  

Indices in Latin in the middle of the alphabet ($i,j,\ldots$) are usually
either spatial indices or in the fundamental or conjugate representation
of $SU(3)$.

Context is really the final word here.  If it {\it should be} a 
Lorentz index, it probably is...

The summation convention is everywhere in effect (unless specified otherwise).
This means that any repeated index in the same term of an expression 
is meant to be summed over all possible values of that index.  In certain
spaces, invariant summing requires one contravariant (up) and one
covariant (down) index.  

Suppression of indices implies contraction in the most obvious
way.  Either last index with first index or covariant index
with contravariant (read from left to right).  Exception : 
for double contraction on two indices my convention
is $\theta_{\mu\nu}J^{\mu\nu}\equiv \theta J$.

The difference between the index $i$ and the $\sqrt{-1}=i$
should be clear from context.  I have tried to avoid
grossly misleading presentations, but borderline
cases may be present.

$[,]$ denotes the commutator, defined by
$[A,B]\equiv AB-BA$.

$\{,\}$ denotes the anticommutator, defined by
$\{A,B\}\equiv AB+BA$.

The symbol $\sim$ is used to indicate either 
approximate equality or equality in all important respects.
In most cases, it can be read, ``is essentially the
same as''.

The conjugation operators I employ are \\
\indent\indent\indent$T^*$ for complex\\
\indent\indent\indent$T^\dag$ for Hermitian\\
\indent\indent\indent$\overline T\phantom{^\dag}$ for Dirac. 

The notation $\not\!a$ is shorthand for $a_\mu\gamma^\mu$.

The absence of an explicit
matrix structure where one should be indicates the presence of the
identity.  

The absence of integration limits implies integration over
the {\it entire} relevant space.  Sometimes, integration
is arrested by the support of its integrand.

I do not differentiate explicitly between scalars and 
4-vectors.  This should be clear from the context.  
Spatial 3-vectors {\it always} appear with vector hats, 
i.e. $\vec v$.

I do not specify the number of $\delta$-functions appearing in an expression
explicitly unless this is not obvious from the context.  For example,
$\delta(\,\vec p-\vec p\,'\,)\equiv\delta(p^1-p'\,^1)
\delta(p^2-p'\,^2 )\delta(p^3-p'\,^3)$ .
One important exception is the volume of spacetime,
$(2\pi)^3\,\delta^{(3)}(0)$.

\section{Dirac Algebra}
\label{diralg}

The $\gamma$-matrices are $d$ 
$2^{d/2}\times 2^{d/2}$-dimensional
operators on the $2^{d/2}$-dimensional Dirac spinor
space defined by the {\it Clifford} algebra
\begin{equation}
\left\lbrace \gamma^\mu,\gamma^\nu\right\rbrace=2g^{\mu\nu}{\bf 1}\;\; ,
\end{equation}
where 
$g^{\mu\nu}$ is the metric tensor of our
space.  In four dimensions, one defines
\begin{equation}
\gamma_5\equiv i\gamma^0\gamma^1\gamma^2\gamma^3
=-{i\over4!}\epsilon^{\mu\nu\lambda\rho}\gamma_\mu
\gamma_\nu\gamma_\lambda\gamma_\rho
\end{equation}
so that the projection operators 
\begin{equation}
P_R\equiv {1\over 2}\left(1+\gamma_5\right)\qquad\qquad\qquad\qquad
P_L\equiv {1\over2}\left(1-\gamma_5\right)
\end{equation}
project out positive (right-handed) and negative (left-handed)
helicity states, respectively.  The {\it Weyl} basis 
is defined by the choice
\begin{eqnarray}
\gamma^\mu=\left(\matrix{0&\sigma^\mu\cr
\overline\sigma^\mu&0\cr}\right)\;\; ,
\end{eqnarray}
where $\sigma^i$ are the Pauli matrices,
\begin{eqnarray}
\sigma^1\equiv\left(\matrix{0&1\cr1&0\cr}\right)\;\; ,\qquad\qquad
\sigma^2\equiv\left(\matrix{0&-i\cr i&0\cr}\right)\;\; ,\qquad\qquad
\sigma^3\equiv\left(\matrix{1&0\cr0&-1\cr}\right)\;\; ,
\end{eqnarray}
$\sigma^0$ is the $2\times 2$ identity, and 
\begin{equation}
\overline\sigma^\mu\equiv\sigma_\mu\;\; .
\end{equation}
In this basis, $\gamma_5$ takes the form
\begin{eqnarray}
\gamma_5=\left(\matrix{-{\bf 1}&0\cr0&{\bf 1}\cr}\right)\;\; ,
\end{eqnarray}
where ${\bf 1}$ represents the $2\times 2$ identity. 
The generators of Lorentz transformations,
\begin{equation}
\sigma^{\mu\nu}\equiv {i\over2}\left\lbrack\gamma^\mu
,\gamma^\nu\right\rbrack\;\; ,
\end{equation}
take the form
\begin{eqnarray}
\sigma^{0i}=\phantom{\epsilon^{ijk}}
-i\left(\matrix{\sigma^i&0\cr0&-\sigma^i\cr}\right)\phantom{\;\; .}\\
\sigma^{ij}=\phantom{-i}\epsilon^{ijk}\left(\matrix{\sigma^k&0\cr0&
\phantom{-}\sigma^k}\right)\;\; .
\end{eqnarray}
Under Hermitian conjugation, one has
\begin{eqnarray}
\left(\gamma^\mu\right)^\dag&=&\gamma^0\gamma^\mu\gamma^0\\
\left(\sigma^{\mu\nu}\right)^\dag&=&
\gamma^0\sigma^{\mu\nu}\gamma^0\\
\gamma_5^\dag&=&\gamma_5=-\gamma^0\gamma_5\gamma^0\;\; .
\end{eqnarray}

These matrices are linearly independent and,
along with the identity and $\gamma^\mu\gamma_5$, form 
a closed set under multiplication in $d=4$.
The
relations
\begin{eqnarray}
\gamma_5^2&=&1\\
\sigma^{\mu\nu}\gamma_5&=&{1\over2}\,i\,\epsilon^{\mu\nu\lambda\rho}
\sigma_{\lambda\rho}\\
\gamma^\mu\gamma^\nu&=&g^{\mu\nu}-i\sigma^{\mu\nu}\\
\gamma^\mu\gamma^\nu\gamma^\lambda&=&
g^{\mu\nu}\gamma^\lambda-g^{\mu\lambda}\gamma^\nu
+g^{\nu\lambda}\gamma^\mu+i\epsilon^{\mu\nu\lambda\rho}
\gamma_\rho\gamma_5\\
\gamma^\mu\gamma^\nu\gamma^\lambda\gamma^\rho&=&
g^{\mu\nu}g^{\lambda\rho}-
g^{\mu\lambda}g^{\nu\rho}+
g^{\nu\lambda}g^{\mu\rho}\nonumber\\
&&-i\left(\phantom{+}
g^{\mu\nu}\sigma^{\lambda\rho}-
g^{\mu\lambda}\sigma^{\nu\rho}+
g^{\nu\lambda}\sigma^{\mu\rho}\right.\\
&&\phantom{-i(}\!+\left.
g^{\lambda\rho}\sigma^{\mu\nu}-
g^{\nu\rho}\sigma^{\mu\lambda}+
g^{\mu\rho}\sigma^{\nu\lambda}\right)
-i\epsilon^{\mu\nu\lambda\rho}\gamma_5\nonumber\;\; 
\end{eqnarray}
simplify the decomposition.
In $d$ dimensions, the identities
\begin{eqnarray}
\gamma^\lambda\gamma^\mu\gamma_\lambda&=&(2-d)\gamma^\mu\\
\gamma^\lambda\gamma^\mu\gamma^\nu\gamma_\lambda
&=&2\gamma^\nu\gamma^\mu+(d-2)\gamma^\mu\gamma^\nu
=4g^{\mu\nu}+(d-4)\gamma^\mu\gamma^\nu\\
\gamma^\lambda\sigma^{\mu\nu}\gamma_\lambda&=&(d-4)\sigma^{\mu\nu}\\
\gamma^\lambda\gamma^\mu\gamma^\nu\gamma^\rho\gamma_\lambda&=&
-2\gamma^\rho\gamma^\nu\gamma^\mu+(4-d)\gamma^\mu\gamma^\nu\gamma^\rho\\
\gamma^\lambda\sigma^{\mu\nu}\gamma^\alpha\gamma_\lambda&=&
2\gamma^\alpha\sigma^{\mu\nu}+(4-d)\sigma^{\mu\nu}\gamma^\alpha\\
\gamma^\lambda\gamma^{\mu_1}\gamma^{\mu_2}
\cdots\gamma^{\mu_n}\gamma_\lambda&=&
2\gamma^{\mu_n}\gamma^{\mu_1}\gamma^{\mu_2}
\cdots\gamma^{\mu_{n-1}}\nonumber\\
&&+(-1)^n\left\lbrack 2\gamma^{\mu_3}\gamma^{\mu_2}\gamma^{\mu_1}
\gamma^{\mu_4}\cdots\gamma^{\mu_n}
+(4-d)\gamma^{\mu_1}\gamma^{\mu_2}\cdots\gamma^{\mu_n}\right\rbrack\\
&&+2\sum_{i=1}^{n-3}(-1)^i\gamma^{\mu_{n-i}}\gamma^{\mu_1}
\cdots\gamma^{\mu_{n-i-1}}\gamma^{\mu_{n-i+1}}\cdots
\gamma^{\mu_n}\;\; ;\qquad n>3\nonumber
\end{eqnarray}
prove useful in reducing products of 
$\gamma$-matrices.  

In Feynman diagrams, one is often faced with 
traces of $\gamma$-matrices.  Due to the 
fact that the identity is the {\it only}
matrix in our space whose trace is
nonzero, a nonzero traces implies that 
for each matrix $\gamma^\alpha$ in the trace
there is another matrix $\gamma^\beta$
such that $\gamma^\alpha\gamma^\beta$ is
proportional to the identity.
With this in mind, on can show that the trace formulae
\begin{eqnarray}
{\rm Tr}\,\left\lbrack\gamma^\mu\gamma^\nu\right\rbrack&=&4g^{\mu\nu}\\
{\rm Tr}\,\left\lbrack\gamma^\mu
\gamma^\nu\gamma^\lambda\gamma^\rho\right\rbrack
&=&4\left(g^{\mu\nu}g^{\lambda\rho}-g^{\mu\lambda}g^{\nu\rho}
+g^{\nu\lambda}g^{\mu\rho}\right)\\
{\rm Tr}\,\left\lbrack\gamma^{\mu_1}\cdots\gamma^{\mu_n}\right\rbrack
&=&0\qquad\qquad\qquad n\;\;{\rm odd}\\
{\rm Tr}\left\lbrack\gamma^{\mu_1}\cdots\gamma^{\mu_{2n}}\right\rbrack
&=&{\rm Tr}\left\lbrack\gamma^{\mu_{2n}}\cdots
\gamma^{\mu_{1}}\right\rbrack\\
{\rm Tr}\,\left\lbrack\not\!a\,
\gamma^\alpha\!\!\not\!b\,\gamma^\beta
\!\!\not\!c\,\gamma_\alpha\!\!\not\!d\,\gamma_\beta\right\rbrack
&=&(d-2)\left\lbrace(4-d){\rm Tr}\,\left\lbrack
\not\!a\,\!\!\not\!b\,\!\!\not\!c\,\!\!\not\!d\,
\right\rbrack
-16(a\cdot c)\,(b\cdot d)
\right\rbrace\\
&=&4(2-d)\left\lbrace(8-d)(a\cdot c)\,(b\cdot d)\right.\nonumber\\
&&\qquad\qquad+(d-4)[(a\cdot b)\,(c\cdot d)\\
&&\qquad\qquad\phantom{+(d-4)}
\left.+(a\cdot d)\,(b\cdot c)]\right\rbrace\nonumber\\
{\rm Tr}\,\left\lbrack\not\!a\!\not\!b\!\not\!c\,
\gamma^\alpha\!\!\not\!p\,\gamma^\beta
\!\!\not\!k\,\gamma_\alpha\!\!\not\!q\,\gamma_\beta\right\rbrack
&=&(d-4)(6-d){\rm Tr}\,\left\lbrack
\not\!a\,\!\!\not\!b\,\!\!\not\!c\,\!\!\not\!p\,
\!\!\not\!k\,\!\!\not\!q\,
\right\rbrack\nonumber\\
&&\qquad+4{\rm Tr}\left\lbrace
\not\!a\,\!\!\not\!b\,\!\!\not\!c\,\left\lbrack
(4-d)\left(p\cdot k\not\!q\,+k\cdot q\not\!p\,\right)\right.\right.\\
&&\left.\left.\qquad\qquad\qquad\qquad
-2p\cdot q\not\!k\,
\right\rbrack\right\rbrace\nonumber
\end{eqnarray}
are valid in $d$ dimensions.\footnote{Since the trace
of the identity is an overall factor, we have taken
it as 4 in $d$ dimensions.}  In general,
the trace of an even number of $\gamma$-matrices
is 4 times the sum over all different pairings
of indices multiplied by the sign of the 
permutation.  Written in this way, 
the trace of $2n$ $\gamma$-matrices has
$(2n)!/2^nn!$ terms.

The trace formulae
\begin{eqnarray}
{\rm Tr}\,\left\lbrack\gamma_5\right\rbrack&=&0\\
{\rm Tr}\,\left\lbrack\gamma_5\gamma^\mu\gamma^\nu\right\rbrack&=&0\\
{\rm Tr}\,\left\lbrack \gamma_5\gamma^\mu\gamma^\nu\gamma^\lambda
\gamma^\rho\right\rbrack&=&-4i\epsilon^{\mu\nu\lambda\rho}
\end{eqnarray}
have meaning only in 4 dimensions.

\section{$SU(N)$ Algebra} 
\label{sunforms}

The $(N^2-1)$ generators $t^a$ of the special unitary group in $N$
dimensions normalized to
\begin{equation}
{\rm Tr}\,\left(t^at^b\right)={1\over2}\delta^{ab}
\end{equation}
satisfy the Lie algebra
\begin{equation}
\left\lbrack t^a,t^b\right\rbrack=if^{abc}t^c\;\; ,
\label{iwant1}
\end{equation}
which define the totally antisymmetric 
structure constants $f^{abc}$
of the group.  
The generators also satisfy the Fiertz identity
\begin{equation}
(t^a)^i_j(t^a)^k_l={1\over2}\left(\delta^i_l\delta^k_j
-{1\over N}\delta^i_j\delta^k_l\right)\;\; .
\end{equation}

For $N=3$, we can use the representation
\begin{eqnarray}
t^1={1\over2}\left(\matrix{\phantom{-}0&\phantom{-}1&\phantom{-}0
\phantom{-}\cr\phantom{-}1&\phantom{-}0&\phantom{-}0\phantom{-}\cr\phantom{-}0
&\phantom{-}0&\phantom{-}0\phantom{-}\cr}\right)\;\; ,
\qquad\qquad 
t^2={1\over2}\phantom{1\over\sqrt{3}}
\left(\matrix{\phantom{-}0&-i&\phantom{-}0\phantom{-}\cr 
\phantom{-}i&\phantom{-}0&\phantom{-}0\phantom{-}\cr\phantom{-}0&\phantom{-}0&\phantom{-}0\phantom{-}\cr}\right)\;\; ,\\
\nonumber\\
t^3={1\over2}\left(\matrix{\phantom{-}1&\phantom{-}0&\phantom{-}0
\phantom{-}\cr\phantom{-} 0&-1&\phantom{-}0\phantom{-}\cr\phantom{-}0&\phantom{-}0&\phantom{-}0
\phantom{-}\cr}\right)\;\; ,
\qquad\qquad 
t^4={1\over2}\phantom{1\over\sqrt{3}}
\left(\matrix{\phantom{-}0&\phantom{-}0&\phantom{-}1
\phantom{-}\cr \phantom{-}0&\phantom{-}0&\phantom{-}0
\phantom{-}\cr\phantom{-}1&\phantom{-}0&\phantom{-}0\phantom{-}\cr}\right)\;\; ,\\
\nonumber\\
t^5={1\over2}\left(\matrix{\phantom{-}0&\phantom{-}0&-i
\phantom{-}\cr \phantom{-}0&\phantom{-}0&\phantom{-}0\phantom{-}\cr \phantom{-}i&\phantom{-}0&\phantom{-}0
\phantom{-}\cr}\right)\;\; ,
\qquad\qquad 
t^6={1\over2}\phantom{1\over\sqrt{3}}
\left(\matrix{\phantom{-}0&\phantom{-}0&\phantom{-}0\phantom{-}\cr \phantom{-}0&
\phantom{-}0&\phantom{-}1\phantom{-}\cr\phantom{-}0&\phantom{-}1&\phantom{-}0
\phantom{-}\cr}\right)\;\; ,\\
\nonumber\\
t^7={1\over2}\left(\matrix{\phantom{-}0&\phantom{-}0&\phantom{-}0
\phantom{-}\cr\phantom{-} 0&\phantom{-}0&-i\phantom{-}\cr\phantom{-} 0&\phantom{-}i&\phantom{-}0
\phantom{-}\cr}\right)\;\; ,
\qquad\qquad 
t^8={1\over2\sqrt{3}}\left(\matrix{\phantom{-}1&\phantom{-}0&\phantom{-}0
\phantom{-}\cr \phantom{-}0&\phantom{-}1&\phantom{-}0\phantom{-}\cr\phantom{-}0&\phantom{-}0&-2
\phantom{-}\cr}\right)\;\; ,
\end{eqnarray}
in which the nonzero structure constants are
\begin{eqnarray}
\matrix{\underline a&\underline b&\underline c&&
\underline{f^{abc}}\cr\cr1&2&3&&\phantom{-\sqrt{3}}\,1\cr
1&4&7&&\phantom{-\sqrt3}\,{1\over2}\cr
1&5&6&&\phantom{\sqrt3}-{1\over2}\cr
2&4&6&&\phantom{-\sqrt3}\,{1\over2}\cr
2&5&7&&\phantom{-\sqrt3}\,{1\over2}\cr
3&4&5&&\phantom{-\sqrt3}\,{1\over2}\cr
3&6&7&&\phantom{\sqrt3}-{1\over2}\cr
4&5&8&&\phantom{-\sqrt{3}}{\sqrt{3}\over2}\cr
6&7&8&&\phantom{-\sqrt{3}}{\sqrt{3}\over2}\cr}
\end{eqnarray}

The symmetric structure constants of the 
group, $d^{\,abc}$, are defined by
\begin{equation}
\{t^a,t^b\}={1\over N}\delta^{ab}+d^{\,abc}t^c\;\; .
\label{iwant2}
\end{equation}
In terms of these constants, one can write
\begin{equation}
t^at^b={1\over2}\left({1\over N}\, \delta^{ab}+if^{abc}t^c
+d^{\,abc}t^c\right)\;\; .
\end{equation}

The Jacobi identities,
\begin{eqnarray}
\left\lbrack {\bf F}^a,{\bf F}^b\right\rbrack=
if^{abc}{\bf F}^c\;\; ,\\
\left\lbrack {\bf F}^a,{\bf D}^b\right\rbrack=if^{abc}{\bf D}^c\;\; ,
\end{eqnarray}
where $({\bf F}^a)^{b\,c}\equiv-if^{abc}$
and $({\bf D}^a)^{b\,c}\equiv d^{\,a\,b\,c}$, 
and the relation
\begin{equation}
f^{abe}f^{cde}={2\over N}\left(\delta^{ac}\delta^{bd}-\delta^{ad}\delta^{bc}\right)
+\left(d^{\,ace}d^{\,bde}-d^{\,ade}d^{\,bce}\right)\;\; 
\end{equation}
follow from repeated use of (\ref{iwant1}) and (\ref{iwant2}).
These expressions can be used to show the validity of 
\begin{eqnarray}
t^a\,t^a&=&C_F={N\over2}-{1\over2N}\\
t^b\,t^a\,t^b&=&\left(C_F-{C_A\over 2}\right)t^a=-{1\over2N}\,t^a\\
{\rm Tr}\,[\,t^a\,t^b\,t^c\,]&=&{1\over4}\left(d^{abc}+if^{abc}\right)\\
{\rm Tr}\,[{\bf F}^a{\bf D}^b]&=&0\\
{\rm Tr}\,[{\bf F}^a{\bf F}^b]&=&C_A\delta^{ab}\\
{\rm Tr}\,[{\bf D}^a{\bf D}^b]&=&\left(N-{4\over N}\right)\delta^{ab}\\
{\rm Tr}\,[{\bf F}^a{\bf D}^b{\bf D}^c]&=&\left({N\over2}-{2\over N}\right)if^{abc}\\
{\rm Tr}\,[{\bf F}^a{\bf F}^b{\bf D}^c]&=&{N\over2}\,d^{\,abc}\\
{\rm Tr}\,[{\bf F}^a{\bf F}^b{\bf F}^c]&=&{N\over 2}\,if^{abc}\\
{\rm Tr}\,[{\bf D}^a{\bf D}^b{\bf D}^c]&=&\left(
{N\over2}-{6\over N}\right)\,d^{\,abc}\;\; .
\end{eqnarray}

\section{Covariant integration}
\label{covint}

The standard integrals one encounters
in a dimensionally regularized quantum field theory
with covariant gauge-fixing
are of the form
\begin{equation}
\int\prod_{i=1}^\ell{d^{\,d}l_i\over(2\pi)^d}\,
N(\{l_i\},\{p_j\},\{m_k\})\,\prod_{i=1}^{N_p}\,{1\over k_i^2-m_i^2}\;\; ,
\end{equation}
where $\{l_i\}$ are the loop momenta, $\{k_i\}$
the propagator momenta, $\{m_i\}$ the relevant masses,
and $\{p_i\}$ external momenta.  $N$ is a definite
function of the loop momenta (or, if one prefers, the 
propagator momenta), the external momenta, and 
the masses.  The propagator momenta are necessarily
linear in the loop momenta and can be chosen to
have coefficient 1, so the technique of Feynman 
parameters (along with the use of Lorentz symmetry), 
can always reduce this integral to the 
form of (\ref{dimregint}) at the expense of
introducing $N_p$ new integration 
parameters.  His technique is based
on the identity
\begin{equation}
{1\over AB}=\int_0^1 dx\,dy\,\delta(1-x-y)\,{1\over[xA+yB]^2}\;\; .
\end{equation}
Differentiating with respect to 
$B$ gives 
\begin{equation}
{1\over AB^2}=\int_0^1 dx\,dy\,\delta(1-x-y)\,{2y\over[xA+yB]^3}\;\; ,
\end{equation}
which immediately allows one to derive
\begin{eqnarray}
{1\over ABC}&=&\int_0^1 dw\,dx\,dy\,dz\,\delta(1-x-y)\,
\delta(1-z-w)\,{2w\over[wxA+wyB+zC]^3}\nonumber\\
&=&\int_0^1 dx\,dy\,dz\,\delta(1-x-y-z)\,{2\over[xA+yB+zC]^3}\;\; .
\end{eqnarray}
This process continues; it is easy to show by induction
that the formulae 
\begin{eqnarray}
\label{coupled}
{1\over A_1\cdots A_n}&=&\int_0^1 dx_1\cdots dx_n\,
\delta\left(\sum x_i-1\right)\,{\Gamma\left(n\right)\over\left\lbrack
\sum x_iA_i\right\rbrack^{n}}\\
\nonumber\\
{1\over A_1^{m_1}\cdots A_n^{m_n}}&=&\int_0^1 dx_1\cdots dx_n\,
\delta\left(\sum x_i-1\right)\,{\prod x_i^{m_i-1}\over\left\lbrack
\sum x_iA_i\right\rbrack^{\sum m_i}}\,{\Gamma\left(\sum m_i\right)
\over\prod\Gamma\left(m_i\right)}
\end{eqnarray}
hold for integer $m_i$.  Furthermore, 
the right hand side is an analytic function
of $m_i$ that can be continued throughout the 
complex plane.  
I find the decoupled 
formulae
\begin{eqnarray}
\phantom{i}{1\over A_1\cdots A_{n+1}}\phantom{i}=\int_0^1 dx_1\cdots dx_n\,
{\Gamma(n+1)\prod x_i^{n-i}\over\left\lbrack
\sum A_i(1-x_i)\prod^{i-1}x_j\right\rbrack^{n+1}}\qquad
\qquad\qquad\quad\qquad\qquad\\
{1\over A_1^{m_1}\cdots A_{n+1}^{m_{n+1}}}=\int_0^1 dx_1\cdots dx_n\,
{\left[\prod x_i^{n-i}\right]\prod \left[(1-x_i)\prod^{i-1}
x_j\right]^{m_i-1}\over\left\lbrack
\sum A_i(1-x_i)\prod^{i-1}x_j\right\rbrack^{\sum m_i}}\,{\Gamma\left(\sum m_i\right)
\over\prod\Gamma\left(m_i\right)}\;\; ,\qquad
\end{eqnarray}
where I have defined $x_{n+1}\equiv0$ and $\prod^0\equiv1$,
much more useful in practical calculations.

Once one has combined all of the denominators
in a Feynman integral, one can shift the integration variable
to get rid of all of the directional dependence
of the resulting `monster denominator' via\footnote{This 
expression assumes the use of the coupled 
parameter formula (\ref{coupled}) and ignores masses.
The sums run from 1 to $n$, with $x_{n+1}$ determined by
the $\delta$-function in (\ref{coupled}).}
\begin{equation}
l^2(l-p_1)^2(l-p_2)^2\cdots(l-p_n)^2
\rightarrow\left\lbrack \left(l-\sum x_ip_i\right)^2-\left(\sum x_ip_i\right)^2
+\sum x_ip_i^2\right\rbrack^{n+1}\;\; .
\end{equation}
Any directional dependence in the numerator
is illusory, as one can always remove
external vectors from the integral, 
leaving a free index,\footnote{In multi-loop calculations,
one may find one's self with fractional powers
in the numerator.  These powers are to be considered
part of the denominator and combined via the 
above approach.  One must be careful in these
situations since fractional powers can easily lead
to ambiguities in the complex plane.  
To treat these integrals consistently, one always
performs Wick rotations
{\it before} combining denominators and integrating.
In this way, there will always be an {\it integer}
number of signs.}  so we are left with a 
directionally independent integral.  Using the symmetry of the 
problem, one can now eliminate all 
free indices via
\begin{eqnarray}
\int{d^{\,d}l\over(2\pi)^d}\,{l^\mu (l^2)^\alpha\over(l^2+M^2)^\beta}&=&0\\
\int{d^{\,d}l\over(2\pi)^d}\,{l^\mu l^\nu 
(l^2)^\alpha\over(l^2+M^2)^\beta}&=&{g^{\mu\nu}\over d}\;
\int{d^{\,d}l\over(2\pi)^d}\,{(l^2)^{\alpha+1}\over(l^2+M^2)^\beta}\;\; ,
\end{eqnarray}
or the higher-rank equivalent.  The remaining 
$d$-dimensional integral
is evaluated via
\begin{equation}
\int{d^{\,d}l\over(2\pi)^d}\,{(l^2)^\alpha\over(l^2+M^2)^\beta}
={1\over(4\pi)^{d/2}}\left(M^2\right)^{d/2+\alpha-\beta}
{\Gamma(d/2+\alpha)\Gamma(\beta-\alpha-d/2)\over
\Gamma(d/2)\Gamma(\beta)}\;\; .
\end{equation}
Since this always gives a power of $M^2$, which 
is invariably quadratic in the loop momenta,
this procedure continues until
all of the $d$-dimensional integrals 
have been taken.  One
is then left only with one-dimensional integrals,
which are in principle straightforward.

The following formulae are useful in this
context
\begin{eqnarray}
\Gamma(z)\Gamma(1-z)&=&{\pi\over\sin\pi z}\label{recrelgam}\\
\Gamma\left({z+1\over 2}\right)&=&2^{-z}\sqrt{\pi}
{\Gamma(z+1)\over\Gamma(1+z/2)}\\
\Gamma(2z)&=&2^{2z-1}\pi^{-1/2}\Gamma(z)\Gamma(z+1/2)\\
\log[\Gamma(1+\epsilon)]&=&-\gamma_E\epsilon+\sum_{n=2}^\infty
{(-1)^{n}\over n}\zeta(n)\epsilon^n\\
\int_0^1dx\,x^{\alpha-1}(1-x)^{\beta-1}&=&
{\Gamma(\alpha)\Gamma(\beta)\over\Gamma(\alpha+\beta)}\;\; .
\end{eqnarray}
For more complicated calculations, 
\begin{eqnarray}
&&\int_0^1dx\,dy\,x^{\alpha-1}(1-x)^{\beta-1}y^{\gamma-1}
(1-y)^{\alpha-\gamma-1}(1-xy)^\mu\nonumber\\&&\qquad\qquad\qquad\qquad\qquad\qquad=
{\Gamma(\beta)\Gamma(\gamma)\Gamma(\alpha+\beta+\mu-\gamma)
\over\Gamma(\alpha+\beta+\mu)\Gamma(\alpha+\beta-\gamma)}\;\; 
\end{eqnarray}
may also be of use.
In addition, the integrals
\begin{eqnarray}
&&\int{d^{\,d}k\over(2\pi)^d}\,{1\over(k^2)^\alpha[(k-q)^2]^\beta}\nonumber\\
&&\qquad\qquad={(-q^2)^{d/2-\alpha-\beta}\over(4\pi)^{d/2}}
\,{\Gamma(d/2-\beta)\Gamma(d/2-\alpha)\Gamma(\alpha+\beta-d/2)\over
\Gamma(\alpha)\Gamma(\beta)\Gamma(d-\alpha-\beta)}\;\; ,\\
\nonumber\\
&&\int{d^{\,d}k\over(2\pi)^d}\,{k^\mu\over(k^2)^\alpha[(k-q)^2]^\beta}\nonumber\\
&&\qquad\qquad={(-q^2)^{d/2-\alpha-\beta}\over(4\pi)^{d/2}}
\,{\Gamma(d/2-\beta)\Gamma(d/2+1-\alpha)\Gamma(\alpha+\beta-d/2)\over
\Gamma(\alpha)\Gamma(\beta)\Gamma(d+1-\alpha-\beta)}\,q^\mu\;\; ,\\
\nonumber\\
&&\int{d^{\,d}k\over(2\pi)^d}\,{k^\mu k^\nu\over(k^2)^\alpha[(k-q)^2]^\beta}\nonumber\\
&&\qquad\qquad={(-q^2)^{d/2-\alpha-\beta}\over(4\pi)^{d/2}}
\,{\Gamma(d/2-\beta)\Gamma(d/2+1-\alpha)\Gamma(\alpha+\beta-d/2)\over
\Gamma(\alpha)\Gamma(\beta)\Gamma(d+2-\alpha-\beta)}\\
&&\qquad\qquad\qquad\qquad\times\left\lbrack{1\over2}\,{d/2-\beta\over
\alpha+\beta-d/2-1}\,
q^2 g^{\mu\nu}+\left({d\over2}+1-\alpha\right)
q^\mu q^\nu\right\rbrack\nonumber\;\; ,\\
\nonumber\\
&&\int{d^{\,d}k\over(2\pi)^d}\,{k^\mu k^\nu k^\lambda
\over(k^2)^\alpha[(k-q)^2]^\beta}\nonumber\\
&&\qquad\qquad={(-q^2)^{d/2-\alpha-\beta}\over(4\pi)^{d/2}}
\,{\Gamma(d/2-\beta)\Gamma(d/2+2-\alpha)\Gamma(\alpha+\beta-d/2)\over
\Gamma(\alpha)\Gamma(\beta)\Gamma(d+3-\alpha-\beta)}\nonumber\\
&&\qquad\qquad\qquad\qquad\times\left\lbrack{1\over2}\,
{d/2-\beta\over\alpha+\beta-1-d/2}\,q^2 
\left(
g^{\mu\nu}q^\lambda+g^{\nu\lambda}q^\mu+g^{\lambda\mu}q^\nu\right)\right.\\
&&\qquad\qquad\qquad\qquad\qquad\qquad
\left.+\left({d\over2}+2-\alpha\right)q^\mu q^\nu q^\lambda\right\rbrack\nonumber\;\; ,
\end{eqnarray}
can be done in general and have simple forms.
These integrals have {\it already} been Wick rotated
to Euclidean space.  Loop momenta with free indices
have been left alone, so one must be careful when 
using these formulae with internal contractions.
 
In all the above formulae, $\Gamma(n)$
is the $\Gamma$-function, defined via
\begin{equation}
\Gamma(n)\equiv\int_0^\infty dxx^{n-1}e^{-x}\;\; 
\end{equation}
for $\Re e\,n>0$ and either the reflection
identity (\ref{recrelgam}) or the recursion relation
\begin{equation}
n\Gamma(n)=\Gamma(n+1)
\end{equation}
elsewhere.  It is a meromorphic function 
of $n$ throughout the complex plane with 
simple poles at the negative integers and
zero, with residues
\begin{equation}
{\rm Res}[\Gamma(z),-n]={(-1)^n\over n!}\;\; .
\label{gamres}
\end{equation}
Note that 
\begin{eqnarray}
\Gamma(n+1)&=&n!\\
\Gamma\left({1\over2}\right)&=&\sqrt{\pi}\;\; .
\end{eqnarray}
Its expansion involves the Euler-Masceroni constant
\begin{equation}
\gamma_E\equiv\lim_{n\rightarrow\infty}\left\lbrack
\sum_{i=1}^n{1\over i}-\log n\right\rbrack
=\sum_{n=2}^\infty{(-1)^n\over n}\,\zeta(n)
\sim0.577\;215\;664\;901\;532\;861
\end{equation}
and the Riemann Zeta function,
\begin{equation}
\zeta(s)=\sum_{i=1}^\infty{1\over i^s}
\label{zetadef}
\end{equation}
for $\Re e\,s>1$.  For other values in the 
complex plane of $s$, it is defined via
\begin{equation}
\zeta(s)=2^s\pi^{s-1}{\rm sin}{\pi s\over2}\;\Gamma(1-s)\zeta(1-s)\;\; ,
\label{zetacont}
\end{equation}
and found to be 
a meromorphic function with a simple pole at
$s=1$ with residue
\begin{equation}
{\rm Res}[\zeta(s),1]=1\;\; .
\end{equation}

Values of $\zeta(s)$ at positive even integer $s$
are known via the expansion
\begin{equation}
{x\over e^x-1}=1-{x\over 2}+\sum_{n=1}^\infty
{(-1)^{n+1}\over 2^{2n-1}\pi^{2n}}\;\zeta(2n) x^{2n}\;\; .
\end{equation}
In particular,
\begin{eqnarray}
\zeta(2)&=&{\pi^2\over6}\phantom{\;\; .0}\\
\zeta(4)&=&{\pi^4\over90}\;\; .
\end{eqnarray}
Using these results in conjunction with
(\ref{zetacont}), one can derive
\begin{eqnarray}
\zeta(0)&=&\phantom{}-{1\over2}\\
\zeta(-1)&=&\phantom{}-{1\over12}\\
\zeta(-3)&=&\phantom{-}{1\over120}\;\; .
\end{eqnarray}
In addition, consistency between (\ref{gamres}), (\ref{zetadef}),
and (\ref{zetacont}) implies
\begin{equation}
\lim_{\epsilon\rightarrow0}{\zeta(-2n+\epsilon)\over\epsilon}
=(-1)^n\;{(2n)!\,\zeta(2n+1)\over2^{2n+1}\pi^{2n}}
\end{equation}
for positive integer $n$.

For completeness, I also include the standard 
trancendentals,
\begin{eqnarray}
\pi\sim3.141\;592\;653\;589\;793\;238\phantom{\;\; ,}\\
e\sim2.718\;281\;828\;459\;045\;235\;\; ,
\end{eqnarray}
that appear in our relations.

\section{QCD and Feynman's Rules of Perturbation Theory}
\label{frulesqcd}

The lagrangian density of massless QCD is written
\begin{equation}
{\cal L}=\overline\psi\,i\!\not\!\!{\cal D}\,\psi
-{1\over4}{\cal F}^{\mu\nu}_a{\cal F}_{\mu\nu}^a\;\; ,
\end{equation}
where $\psi$ are the quark fields and ${\cal A}$
are the gluon fields.  The covariant derivative ${\cal D}$
is given by 
\begin{eqnarray}
\stackrel\rightarrow{\cal D}&\equiv&
\stackrel\rightarrow\partial+ig{\cal A}\;\; ,\\
\stackrel\leftarrow{\cal D}&\equiv&
\stackrel\leftarrow\partial-ig{\cal A}\;\; ,\\
{\cal D}^{ab}&\equiv&\delta^{ab}\partial+gf^{abc}{\cal A}_c\;\; ,
\end{eqnarray}
and the gluon field strength, ${\cal F}$, is
\begin{equation}
{\cal F}^{\mu\nu}_a\equiv-{i\over g}\left\lbrack 
{\cal D}^\mu,{\cal D}^\nu\right\rbrack_a=\partial^\mu{\cal A}^\nu_a
-\partial^\nu{\cal A}^\mu_a-gf^{abc}{\cal A}^\mu_b{\cal A}^\nu_c\;\; .
\end{equation}
The Feynman rules for covariant gauge-fixing are shown
in Figure A.1.  For the use of these rules, see Section
1.3.

\begin{figure}
\epsfig{figure=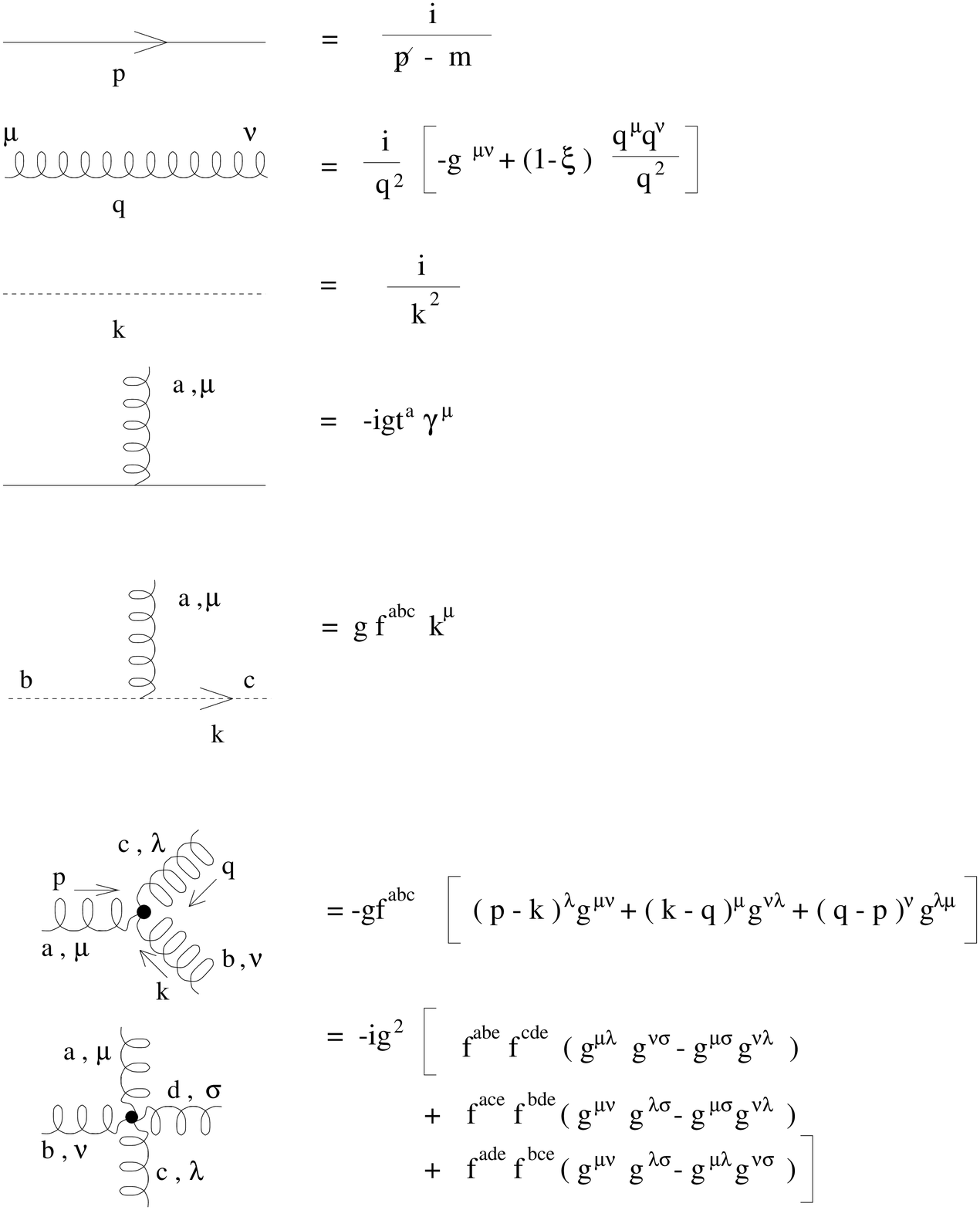,height=6.0in}
\caption{Feynman's rules for a non-abelian 
gauge theory with covariant gauge-fixing.  The 
straight line represents quark propagation,
while the squiggly line represents gluons and 
the dashed ghosts.  In order to avoid index pollution,
most of the indices here have been suppressed.
The proper pole-prescription is given by adding
$i\varepsilon$ to each denominator.  Along with each
vertex, a $\delta$-function ensuring energy-momentum 
conservation is implied.}
\label{frules}
\end{figure}

This lagrangian leads to the 
equations of motion
\begin{eqnarray}
\stackrel\rightarrow{\not\!\!\cal D}\psi&=&0\\
\overline\psi\stackrel\leftarrow{\not\!\!\cal D}&=&0\\
{\cal D}^{ab}_\nu{\cal F}_b^{\nu\mu}&=&g\overline\psi t^a\gamma^\mu\psi
\end{eqnarray}
and the energy-momentum tensor
density\footnote{This expression is also valid for
{\it massive} quarks, in which case I must specify
that the symmetrization of indices in the 
first term does {\it not} indicate a removal of the trace;
the trace of $\Theta^{\mu\nu}$ is given by
\begin{equation}
\Theta^\mu_{\;\;\mu}=m\overline\psi\psi+
{\beta(\alpha_s)\over8\pi\alpha_s^2}\,
{\cal F}_a^{\mu\nu}{\cal F}^a_{\mu\nu}\;\; .
\end{equation}
}
\begin{equation}
\Theta^{\mu\nu}=\overline\psi\gamma^{(\mu}i\!
\stackrel\leftrightarrow{\cal D}\,\!^{\nu)}\psi
-{\cal F}^{\mu\alpha}_a{\cal F}^{\nu}_{a\;\;\alpha}
+{1\over4}g^{\mu\nu}{\cal F}^{\alpha\beta}_a
{\cal F}_{\alpha\beta}^a\;\; ,
\end{equation}
where $\stackrel\leftrightarrow{\cal D}\equiv{1\over2}\left(
\stackrel\rightarrow{\cal D}-\stackrel\leftarrow{\cal D}\right)$.

The coupling of QCD enjoys the $\beta$-function
\begin{eqnarray}
\mu^2{d\alpha_s(\mu^2)\over d\mu^2}&\equiv&\beta(\alpha_s(\mu^2))
=-{1\over4\pi}\,b_0\alpha_s(\mu^2)^2
-{1\over(4\pi)^2}\,b_1\alpha_s(\mu^2)^3-\cdots\;\; ;\\
b_0&=&{11\over3}\,C_A-{4\over3}\,n_fT_F\\
b_1&=&{2\over3}\left\lbrack 17C_A^2-2n_fT_F
\left(3C_F+5C_A\right)\right\rbrack\;\; ,
\end{eqnarray}
where $\alpha_s=g^2/4\pi$, $C_A=3$, $C_F=4/3$, $T_F=1/2$, and
$n_f$ is the number of active (light) quark flavors.

The fundamental Ward identity of QCD can be written
\begin{equation}
g\;\partial_\mu\left\langle{\rm T}\, 
j^\mu_a(x)\,{\cal O}(x_1,\ldots,x_n)
\right\rangle=-i\partial_\mu\left\langle{\rm T}\;{\delta{\cal O}
(x_1,\ldots,x_n)\over
\delta{\cal A}_\mu^a(x)}\right\rangle\;\; ,
\end{equation}
where 
\begin{equation}
j^\mu_a=\overline\psi\gamma^\mu t^a\psi
+f^{abc}{\cal F}^{\mu\alpha}_b{\cal A}_\alpha^c
\end{equation}
is the chromodynamic current.  This identity is 
valid for any gauge-invariant operator $\cal O$,
though in covariant gauges one must add a ghost contribution
to $j^\mu_a$.

To compare fields that are separated 
in spacetime, it is necessary to introduce the gauge link
\begin{equation}
{\cal G}_C(x,y)={\rm P}e^{-ig\int_0^1 A_\mu\left(z(\tau)
\right)(dz^\mu/d\tau)\;d\tau}\;\; ,
\end{equation}
which transforms as
\begin{equation}
{\cal G}_C(x,y)\rightarrow U(x){\cal G}(x,y)U^\dag(y)
\end{equation}
under a gauge transformation.

The one-loop splitting functions for unpolarized
\begin{eqnarray}
P_{NS}(x)&=&C_F\;{1+x^2\over1-x}=P_s(x)\\
P_{sg}(x)&=&T_F\;\left\lbrack x^2+(1-x)^2\right\rbrack\\
P_{gs}(x)&=&C_F\;{1+(1-x)^2\over x}\\
P_{g}(x)&=&2C_A\left\lbrack {x\over1-x}+{1-x\over x}
+x(1-x)\right\rbrack
\end{eqnarray}
and polarized
\begin{eqnarray}
\tilde P_{NS}(x)&=&\tilde P_{s}(x)=P_{NS}(x)\\
\tilde P_{sg}(x)&=&T_F\;
\left\lbrack x^2-(1-x)^2\right\rbrack\\
\tilde P_{gs}(x)&=&C_F\;{1-(1-x)^2\over x}\\
\tilde P_{g}(x)&=&2C_A\left\lbrack 
1-2x+{1\over1-x}\right\rbrack
\end{eqnarray}
parton densities, coupled with the endpoint
contributions for quarks 
\begin{equation}
A_q=C_F\int_0^x {dy\over x^2}\;{x^2+y^2\over x-y}
\end{equation}
and gluons
\begin{equation}
A_g={4n_fT_F-11C_A\over6}+2C_A\int_0^x{dy\over x-y}
\end{equation}
generate the DGLAP evolution equation in the 
nonsinglet
\begin{equation}
\mu^2{D_q\over D\mu^2}f_{NS/T}(x,\mu^2)={\alpha_s(\mu^2)\over2\pi}
\int_x^1{dy\over y}P_{NS}\left({x\over y}\right)
f_{NS/T}(y,\mu^2)
\end{equation}
and singlet
\begin{eqnarray}
\mu^2{D_q\over D\mu^2}f_{s/T}(x,\mu^2)=
{\alpha_s(\mu^2)\over2\pi}\int_x^1{dy\over y}\left\lbrack
P_s\left({x\over y}\right)f_{s/T}(y,\mu^2)
+2n_fP_{sg}\left({x\over y}\right)
f_{g/T}(y,\mu^2)\right\rbrack\;\; \\
\mu^2{D_g\over D\mu^2}f_{g/T}(x,\mu^2)=
{\alpha_s(\mu^2)\over2\pi}\int_x^1{dy\over y}\left\lbrack
P_{gs}\left({x\over y}\right)f_{s/T}(y,\mu^2)
+\phantom{2n_f}P_{g}\left({x\over y}\right)
f_{g/T}(y,\mu^2)\right\rbrack\;\; 
\end{eqnarray}
sectors.  The polarized evolution equations
are obtained simply be adorning these with 
tild\'es.\footnote{As long as one is careful with the definitions
of antiquark distributions.  See Section \ref{partondistdis}.}

\section{Other Useful Relations}
 
Here, I present some miscellaneous relations 
which prove useful in the 
context of quantum field theory :
\begin{eqnarray}
[AB,CD]&=&A[B,C]D+[A,C]BD+C[A,D]B+CA[B,D]\\
&=&A\{B,C\}D-\{A,C\}BD-C\{A,D\}B+CA\{B,D\}\\
\epsilon^{\mu\nu\alpha\beta}\epsilon_{\mu\nu\alpha\beta}&=&-24\\
\epsilon^{\mu\nu\alpha\beta}\epsilon_{\lambda\nu\alpha\beta}
&=&-6\delta^\mu_\lambda\\
\epsilon^{\mu\nu\alpha\beta}\epsilon_{\lambda\rho\alpha\beta}
&=&-2\left(\delta^\mu_\lambda\delta^\nu_\rho
-\delta^\nu_\lambda\delta^\mu_\rho\right)\\
\epsilon^{\mu\nu\alpha\beta}\epsilon_{\lambda\rho\sigma\beta}
&=&-\left(\phantom{+}\delta^\mu_\lambda\delta^\nu_\rho\delta^\alpha_\sigma
-\delta^\alpha_\lambda\delta^\nu_\rho\delta^\mu_\sigma\right.\nonumber\\
&&\phantom{-(}+\delta^\nu_\lambda\delta^\alpha_\rho\delta^\mu_\sigma
-\delta^\mu_\lambda\delta^\alpha_\rho\delta^\nu_\sigma\\
&&\left.\phantom{-(}+\delta^\alpha_\lambda\delta^\mu_\rho\delta^\nu_\sigma
-\delta^\nu_\lambda\delta^\mu_\rho\delta^\alpha_\sigma\right)\nonumber\\
\epsilon^{\mu\nu\alpha\beta}\epsilon_{\lambda\rho\sigma\tau}
&=&-\left(\delta^\mu_\lambda\delta^\nu_\rho\delta^\alpha_\sigma\delta^\beta_\tau
+{\rm 23\;permutations}\right)\\
g^{\mu\nu}\epsilon^{\lambda\rho\sigma\tau}
&=&g^{\mu\lambda}\epsilon^{\nu\rho\sigma\tau}+
g^{\mu\rho}\epsilon^{\lambda\nu\sigma\tau}+
g^{\mu\sigma}\epsilon^{\lambda\rho\nu\tau}+
g^{\mu\tau}\epsilon^{\lambda\rho\sigma\nu}\\
{1\over(x+i\varepsilon)^n}&=&{\rm P} {1\over x^n}+
(-1)^n\,{i\pi\over\Gamma(n)}\,\left({d\over dx}\right)^{n-1}
\delta(x)\\
\int\,dx\,e^{ikx}&=&2\pi\delta(k)
\end{eqnarray}
and generally :
\begin{eqnarray}
\sum_{i=1}^n&=&n\\
\sum_{i=1}^ni&=&{n(n+1)\over2}\\
\sum_{i=1}^ni^2&=&{n(n+1)(2n+1)\over6}\\
\sum_{i=1}^ni^3&=&{n^2(n+1)^2\over4}\\
\sum_{i=0}^ni^k&=&k!\,\lim_{t\rightarrow0}
{d^k\over dt^k}\,{e^{(n+1)t}-1\over e^t-1}\\
\sum_{i=1}^n(-1)^{n-i}&=&{1-(-1)^n\over2}\\
\sum_{i=1}^n(-1)^{n-i}\,i&=&{(n+1)\over2}-{1+(-1)^n\over4}\\
\sum_{i=1}^n(-1)^{n-i}\,i^2&=&{n(n+1)\over2}\\
\sum_{i=1}^n(-1)^{n-i}\,i^3&=&{(n+1)^2(2n-1)\over4}+{1+(-1)^n\over8}\\
\sum_{i=0}^n(-1)^{n-i}\,i^k&=&\,\lim_{t\rightarrow0}
{d^k\over dt^k}\,{e^{(n+1)t}+(-1)^n\over e^t+1}\\
\sum_{i=0}^nx^i&=&{1-x^{n+1}\over1-x}\\
{x^{n-l}y^l-x^{l+j}y^{n-l-j}\over x-y}&=&\left\lbrack \Theta(n-2l-j-1)
\sum_{i=l+j}^{n-l-1}\right.\nonumber\\
&&\qquad\quad\quad\left.-\Theta(2l+j-n-1)\sum_{i=n-l}^{l+j-1}\;\right\rbrack
x^i y^{n-i-1}\\
{1\over1-x}&=&\sum_{i=0}^\infty x^i\\
\log(1-x)&=&-\sum_{i=1}^\infty {x^i\over i}\\
{1\over1-x}\,\log(1-x)&=&-\sum_{i=1}^\infty {x^i}\sum_{j=1}^{i}{1\over j}\\
\log^2(1-x)&=&2\sum_{i=2}^\infty {x^i\over i}\sum_{j=1}^{i-1}{1\over j}\\
(x+y)^n&=&\sum_{i=0}^n{n\choose i} x^i y^{n-i}\\
{n\choose i}&=& {n!\over i!(n-i)!}\\
{n-1\choose i-1}+{n-1\choose i}
&=&{n\choose i}\\
\sum_{i=0}^n{n\choose i}
&=&2^n\\
\sum_{i=j}^n{i\choose j}&=&{n+1\choose j+1}\\
\sum_{i=0}^j{n+i\choose i}&=&{n+j+1\choose j}\\
\sum_{i=j}^n{n\choose i}{i\choose j}x^{n-i} y^{i-j}
&=&{n\choose j}(x+y)^{n-j}\\
\sum_{i=0}^n(-1)^i{n\choose i}&=&\delta_{n0}\\
\sum_{i=0}^j(-1)^i{n\choose i}
&=&(-1)^j{n-1\choose j}\\
\sum_{i=j}^{n}(-1)^i{n\choose i}{i\choose j}&=&(-1)^n\delta_{nj}\\
\sum_{i=j}^n(-1)^i{n\choose i-1}{i\choose j}&=&(-1)^n{n+1\choose j}\\
\sum_{i=j}^n(-1)^i {n+1\choose i+1}{i\choose j}&=&(-1)^j\\
\sum_{i=0}^j{(-1)^i\over{n\choose i}}&=&{n+1\over n+2}\left\lbrack
1+{(-1)^j\over {n+1\choose j+1}}\right\rbrack\;\; .
\end{eqnarray}

\eject

\section{General References}

This section is meant to provide some
standard reference material for further reading
on the subjects presented in this thesis.

\vspace{\baselineskip}

\noindent For a general overview of Quantum Field Theory,
in order of sophistication :
\vspace{\baselineskip}

I. J. R. Aitchison and A. J. G. Hey, {\it Gauge Theories
in Particle Physics}, Institute of Physics Publishing, 
London, 1989.

M. E. Peskin and D. V. Schroder, {\it An Introduction
to Quantum Field Theory}, Addison-Wesley, Reading, MA, 1995.

S. Weinberg, {The Theory of Quantum Fields} Vol. I \& II,
Cambridge University, Cambridge, 1995.

C. Itzykson and J. B. Zuber, {\it Quantum Field Theory},
McGraw-Hill Inc., New York, NY, 1980.
\vspace{\baselineskip}

\noindent For a detailed introduction to perturbative QCD :
\vspace{\baselineskip}

R. D. Field, {\it Applications of Perturbative QCD}, Addison-Wesley,
Reading, MA, 1989.
\vspace{\baselineskip}

\noindent For general renormalizability requirements
and renormalization theorems :
\vspace{\baselineskip}

J. C. Collins, {\it Renormalization}, Cambridge University,
Cambridge, 1984.
\vspace{\baselineskip}

\noindent For a rigorous study of infrared divergences and
factorization theorems :
\vspace{\baselineskip}

G. Sterman, {\it An Introduction Quantum Field Theory,} Cambridge
University, Cambridge, 1993.

A. H. Mueller, ed., {\it Perturbative Quantum
Chromodynamics}, World Scientific, Singapore, 1989.
\vspace{\baselineskip}

\noindent For path integral techniques :
\vspace{\baselineskip}

R. J. Rivers, {\it Path Integral Methods in Quantum Field Theory},
Cambridge University, Cambridge, MA, 1987.
\vspace{\baselineskip}

\noindent For spinors :
\vspace{\baselineskip}

V. B. Berestetskii, E. M. Lifshitz and L. P. Pitaevskii, 
{\it Quantum Electrodynamics}, Pergamon Press, New York, NY, 1980.

H. J. W. M\"uller-Kirsten and A Wiedemann, {\it Supersymmetry :
An Introduction with Conceptual and Calculational Details}, World Scientific,
Singapore, 1987.
\vspace{\baselineskip}

\noindent For group theory :
\vspace{\baselineskip}

M. Hamermesh, {\it Group Theory and Its Application
to Physical Problems}, Addison-Wesley Publishing Co., Reading, MA,
1962.

H. Georgi, {\it Lie Algebras in Particle Physics}, Perseus Books,
Reading, MA, 1982.
\vspace{\baselineskip}

\noindent For integral tables and special mathematical functions :
\vspace{\baselineskip}

G. B. Arfken and H. J. Weber, {\it Mathematical Methods
for Physicists}, Academic Press, San Diego, CA, 1995.

I. S. Gradshteyn and I. M. Ryzhik, {\it Table of Integrals,
Series, and Products}, Academic Press, San Diego, CA, 1994.

M. Abramowitz and I. A. Stegun, eds., {\it Handbook of 
Mathematical Functions}, Dover Publications, New York, NY, 1965.

\chapter{The Lorentz Group and its Representations}
\label{lorentzgroup}

In this appendix, we will discuss some of the more
mathematical properties of the spacetime translation and rotation group.
First, I define some terms.

A {\it group} is a collection of objects which satisfy properties 

1) Any two elements, $x,y$, in the group may be 
composed with each other to 
from another element of the group, $xy$ or $yx$.  This operation
is called {\it multiplication}.  

2) Group multiplication must be 
{\it associative}, so that $\forall x,y,z\in g$ 
\begin{equation}
x(yz)=(xy)z\;\; .
\end{equation}

3) $g$ also must possess 
an {\it identity}, {\bf 1}, such that $\forall x\in g$
\begin{equation}
x{\bf 1}={\bf 1}x=x\;\; .
\end{equation}

4) Each element of the group must have an {\it inverse}, $x^{-1}$,
associated with it such that 
\begin{equation}
xx^{-1}=x^{-1}x={\bf 1}\;\; .
\end{equation}

Group multiplication does {\it not} 
have to be commutative.  In fact, 
it usually isn't.  A group whose multiplication operation is 
commutative is called {\it abelian}.

A {\it representation} of a group is a collection of objects,
one for each element of the group, which satisfy the group
multiplication law.  It is called a representation because it represents the 
abstract group in some concrete way.  For example, the matrices
\begin{eqnarray}
{\bf 1}=\left(\matrix{1&0\cr0&1}\right)\;\;\;\;\;\; 
{\bf 2}=\left(\matrix{1&0\cr0&-1}\right)\;\; ,
\end{eqnarray}
with ordinary matrix multiplication,
represent the group ${\bf Z}_2$.  Here, ${\bf 2}={\bf 2}^{-1}$.  The 
objects $\lbrace 1,0\rbrace$, with the multiplication operation identified
as addition and $1+1$ defined to be $0$, also form a representation of
${\bf Z}_2$.  In this representation, ${\bf 1}=0$.  

A representation ${\bf R}$ is called {\it irreducible} if for 
every $x,y\in{\bf R}$ there exists an element $z\in{\bf R}$ such
that $yz=x$.  
For example, complex 2-vectors with unit norm from an
irreducible representation of $SU(2)$.  Choosing a complex 2-vector 
with unit norm at random to associate with the identity, say $(1,0)$,
we see that any other vector in the representation can be obtained 
via an $SU(2)$ transformation :
\begin{eqnarray}
\left(\matrix{\alpha\cr\beta\cr}\right)=
\left(\matrix{\alpha&-\beta^*\cr\beta&\alpha^*\cr}\right)
\left(\matrix{1\cr0\cr}\right)\;\; .
\end{eqnarray}
Associating the vector $(\alpha,\beta)$ with the transformation
necessary here gives us the representation.  On the other hand,
the representation of $SU(2)$ formed by unitary hermitian 
$2\times2$ matrices is {\it not} irreducible.
Indeed, performing unitary $SU(2)$ transformations on
the unitary hermitian matrix
\begin{eqnarray}
{\bf 1}=\left(\matrix{1&0\cr0&1\cr}\right)
\end{eqnarray}
will certainly not get us to any other element
of our representation.  This element forms a representation
all by itself.  However, once we remove it, we 
are left with an irreducible representation
which any of the three Pauli matrices will
be happy to span for us.  

This appendix is
concerned with the irreducible 
representations of the spacetime symmetry group, $SO(3,1)$.
We will find it to be a continuous
group that relates different inertial frames of
reference to each other.  Since it 
is a fundamental symmetry of nature, 
physical particles and states will form representations
of it.  This is the foothold which we use to formalize a 
theory of objects that we cannot directly interact with
(on their level), and whose interactions do not
always follow our intuition.

Much of the analysis of Section B.2 is based on that found in
S. Weinberg \cite{weinbergbook}. 

\section{Relativity}
\label{relativity}

The contemporary understanding of special relativity is based on the
assumption that the speed of light {\it in a vacuum}, $c$, is 
the same in all reference frames.
To see what this implies, let us
suppose that a beam of light travels in vacuum from spacetime point $0$ to spacetime 
point $P$.  The coordinates of $P$ can be written in a compact 
notation as $x^\mu$, 
where $\mu$ runs from 0 to 3.  $x^0$ is the time coordinate and 
$x^i$, $i=1-3$, are
the spatial components of this `4-vector'.  Since light goes from 
$0$ to $P$ in vacua, 
\begin{equation}
c^2 (x^0)^2=(x^1)^2+(x^2)^2+(x^3)^2\;\; ,
\end{equation}
or\footnote{The component $x^2$ and the 4-vector square 
$x^2=c^2 (x^0)^2-(x^1)^2-(x^2)^2-(x^3)^2$
are obviously quite different.  However, since the 2-component of a vector 
is seldom singled out, confusion is rare.}  
\begin{equation}
c^2 (x^0)^2-(x^1)^2-(x^2)^2-(x^3)^2\equiv x^2=0\;\; .
\label{lc}
\end{equation}
In another reference frame whose origin coincides with this 
one, the final destination of the
light beam will be $P'$.  However, since 
$c$ is the same in all reference frames,
the new coordinates must also satisfy (\ref{lc}).  Hence 
the quantity $x^2$ will be
zero either in {\it all} reference frames or none.

Let us study an example where $x^2\ne 0$.  We consider a laser on a train
pointed at
a mirror on the ceiling.  The train has height $h$ and is moving at a speed 
$v$ with respect to the ground in the 1-direction.
Taking the origin to be the time and place 
the laser beam is turned on, we study the final destination of the `front' of the 
laser beam in two different frames.  Observer A seated in the train
measures the final destination $x_A^\mu=
(2h/c,0,0,0)$.\footnote{Here, as always, component listing 
is read $x^\mu=(x^0,x^1,x^2,x^2)$.}
Observer B standing on the 
ground measures $x_B^\mu=2h/c\,\gamma(1,v,0,0)$, where $\gamma^2=1/(1-v^2/c^2)$.
Hence $x_A^\mu\ne x_B^\mu$, but $x_A^2=x_B^2$.  It can be shown that this
is the case for any process in special relativity.  The assumption that the
speed of light is the same in all reference frames then implies that the
line element
\begin{equation}
ds^2=-g_{\mu\nu}dx^\mu\,dx^\nu\;\; ,
\end{equation}
where 
\begin{equation}
g_{\mu\nu}=\left(\matrix{1&0&0&0\cr 0&-1&0&0\cr 0&0&-1&0 
\cr 0&0&0&-1\cr}\right)\;\; ,
\label{mink}
\end{equation}
is invariant under all changes of frame.  In this expression we
have implied summation over $\mu$ and $\nu$.  Any time an index is repeated 
summation over that index is implied unless 
explicitly stated otherwise.\footnote{This is known as 
Einstein's `summation convention'.  Note that
in order to sum properly, one index must be `up' and the other `down'.
We will see why this is the case below.}  
For simplicity, we have chosen units in which\footnote{
This means if you ask how far away something is, I simply 
tell you how long it takes light
to get there.}  
\begin{equation}
c=1\;\; .
\end{equation}

With these conventions, we may write $x^2=g_{\mu\nu}x^\mu x^\nu$.
The matrix $g_{\mu\nu}$ is called the {\it metric tensor}.  It defines the
geometry of the spacetime we work on.  The specific form (\ref{mink})
is called the Minkowskii metric and is relevant for the flat spacetime 
considered in this work.  

Physical constraints require our changes in frame to be linear\footnote{If
you don't believe me, imagine how a walk to the grocery store then to the post
office looks in two reference frames.  The grocery store has the two
coordinates $G_A$ and $G_B$ while the post office has the two coordinates
$P_A$ and $P_B$.  The vector from the grocery store to the post office
is by definition $P_A-G_A$ in the first frame and $P_B-G_B$ in the second.
If the transformation were nonlinear, then this vector would not 
transform into its
image in the other frame.}, so we can write a general transformation as
\begin{equation}
x^\mu\rightarrow x'^\mu=\Lambda^\mu_\nu\, x^\nu\; .
\label{tform}
\end{equation}
$\Lambda$ is our transformation matrix, known in flat space
as a Lorentz transform.\footnote{By convention, the {\it first} index of 
$\Lambda$ is `up' while the second index is `down'.  It really should be written
$\Lambda^\mu_{\;\;\nu}$.  Indices on matrices are {\it always} understood to
be this way, unless otherwise emphasized.}
Invariance of the line
element implies
\begin{equation}
g_{\alpha\beta}=\Lambda^\mu_\alpha\Lambda^\nu_\beta g_{\mu\nu}\;\; .
\label{Leq}
\end{equation}
Since the metric behaves in this way, we see that the product of any two vectors
$x$ and $p$, $x\cdot p\equiv g_{\mu\nu}x^\mu p^\nu$, is also invariant under 
changes of frame.\footnote{Now we see why the notation $x^2$ was used for the 
strange object in Eq.(\ref{lc}): $g_{\mu\nu}x^\mu x^\nu=x\cdot x=x^2$.}

We can think of Eq.(\ref{tform}) as the way an upper index transforms.
Defining $x_\mu \equiv g_{\mu\nu}x^\nu$, 
we see that a vector with a lowered index
transforms like $x'_\mu=g_{\mu\alpha}\Lambda^\alpha_\nu x^\nu$.  Defining 
$g^{\mu\nu}$ such that $g^{\mu\alpha}g_{\alpha\nu}=\delta^\mu_\nu$, with 
$\delta^\mu_\nu=0$ if $\mu\ne\nu$ and $=1$ if $\mu=\nu$, we can put this
transformation into the form
\begin{equation}
x_\mu\rightarrow x'_\mu=(\Lambda^{-1})^\nu_\mu x_\nu\;\; .
\end{equation}
It is now obvious why $g_{\mu\nu}x^\mu p^\nu=x_\mu p^\mu$ is invariant under
transformations - one transforms like $\Lambda$ and the other like $\Lambda^{-1}$.

Normal vectors, like position and momentum, naturally have an upstairs index.
These are called {\it contravariant} vectors since they transform in a different 
way than the axes (rotating the axes one way produces the same effect as rotating vectors
the opposite way).  Vectors which naturally have a downstairs index,
like the derivative, are called {\it covariant} vectors since they transform in the 
same way as the axes.  To see this, let us examine the derivative operator :
\begin{equation}
{\partial\over\partial x^\mu}f(x)\rightarrow{\partial\over\partial x'^\mu}f(x)
={\partial\over\partial x'^\mu}f(\Lambda^{-1}x')
=(\Lambda^{-1})^\nu_\mu{\partial\over\partial x^\nu}f(x)\;\; .
\end{equation}
Hence, $\partial/\partial x^\mu\equiv\partial_\mu$ does indeed transform 
opposite to $x^\mu$ (or in the same way as the axes).\footnote{When 
indices are suppressed in this kind of calculation, confusion as
to the specific way they are contracted can arise.  The standard
convention is that a proper sum goes from covariant to contravariant.
For example, $(\Lambda x)^\mu=\Lambda^\mu_{\;\;\nu}x^\nu$, but
$\Lambda_\mu^{\;\;\nu}x_\nu=(\Lambda^{-1})^\nu_\mu x_\nu=(x\Lambda^{-1})_\mu$.
Note that in the second example the placement of the indices is
crucial to a correct understanding of the product.  The first
expression is not a proper sum since it goes from contravariant
to covariant.  In order to write it in an unambiguous way,
we had to emphasize the order of the indices.}

Tensors behave in exactly the same way. 
A tensor 
with $n$ contravariant and $m$ covariant indices transforms like
\begin{equation}
T^{\mu_1\cdots\mu_n}_{\nu_1\cdots\nu_m}\rightarrow
(\Lambda)^{\mu_1}_{\alpha_1}\cdots(\Lambda)^{\mu_n}_{\alpha_n}
(\Lambda^{-1})_{\nu_1}^{\beta_1}\cdots(\Lambda^{-1})_{\nu_m}^{\beta_m}
T^{\alpha_1\cdots\alpha_n}_{\beta_1\cdots\beta_m}\;\; .
\end{equation}
Often, we wish to emphasize certain symmetries possessed by tensors.  This is done 
via parentheses and brackets in the following way
\footnote{Sometimes, as we will see below, 
parentheses also imply tracelessness.  It should be clear from the context which
is meant.}
\begin{eqnarray}
T^{(\alpha\beta)\mu_1\cdots\mu_n}&\equiv&{1\over 2}\left(T^{\alpha\beta\mu_1\cdots\mu_n}
+T^{\beta\alpha\mu_1\cdots\mu_n}\right)\nonumber\\
T^{[\alpha\beta]\mu_1\cdots\mu_n}&\equiv&{1\over 2}\left(T^{\alpha\beta\mu_1\cdots\mu_n}
-T^{\beta\alpha\mu_1\cdots\mu_n}\right)\;\; .
\end{eqnarray}

Invariance under coordinate transformations plays an important role in 
both General Relativity and the Standard Model.  In General 
Relativity, this invariance plays a central role and severely constrains the form
of theory.  In the SM, we are concerned mainly with flat space so the constraints
imposed are not as harsh, but the theory is still constrained in
much the same way as rotational invariance constrains classical mechanics.
For example, suppose we wish to calculate a vector quantity which depends only
on a certain momentum, $p$.  A priori, we must calculate 4 different quantities.  
However, the requirement of Lorentz covariance forces this object to transform
like a vector under Lorentz transformations.  Since $p$ is the only vector
in the problem, our quantity must be proportional to $p$.  We need
only calculate the constant of proportionality (which will in general depend
on all invariants in the problem, i.e. masses and $p^2$).  With two indices, 
the problem is a little more complicated.  Suppose we have a two-index tensor which
depends on no momenta.  The argument above would lead to 0 for an answer since there is
`nothing to carry the indices'.  What this means is that there is no reference
to frame in this problem.  Our quantity must be frame-independent.  However, this
does not imply that it is zero since the tensor $g^{\mu\nu}$ satisfies our
requirement, as exemplified by (\ref{Leq}).  In this case, Lorentz invariance
forces our tensor to be proportional to $g^{\mu\nu}$.  

There is only one
other tensor-like object which is invariant under Lorentz transformations.
The Levi-Cevita\footnote{Cevita is pronounced `ch{\"a}v{\=e}t{\u a}', while 
Levi is like `Drove my Chevy to the {\it levi}...'.} tensor is defined
as a 4-index tensor which is completely antisymmetric in all four indices.
Since there are only four possible values for each of the indices, 
the Levi-Cevita tensor has only one independent component.  This 
component is chosen to be 1, i.e. 
\begin{equation}
\epsilon^{0123}=+1\;\; .
\end{equation}
All other components are obtained from the requirement of total antisymmetry.
To see that the Levi-Cevita tensor is invariant, we transform it :
\begin{equation}
\epsilon^{\mu_1\mu_2\mu_3\mu_4}\rightarrow
\Lambda^{\mu_1}_{\nu_1}\Lambda^{\mu_2}_{\nu_2}
\Lambda^{\mu_3}_{\nu_3}\Lambda^{\mu_4}_{\nu_4}
\epsilon^{\nu_1\nu_2\nu_3\nu_4}={\rm det}(\Lambda)
\epsilon^{\mu_1\mu_2\mu_3\mu_4}\;\; ,
\label{ep}
\end{equation}
where the last equality comes from the definition of the determinant,
\begin{equation}
{\rm det}(\Lambda)\equiv {1\over 4!}\epsilon_{\mu_1\mu_2\mu_3\mu_4}
\epsilon^{\nu_1\nu_2\nu_3\nu_4}\Lambda^{\mu_1}_{\nu_1}\Lambda^{\mu_2}_{\nu_2}
\Lambda^{\mu_3}_{\nu_3}\Lambda^{\mu_4}_{\nu_4}\;\; .
\end{equation}
(\ref{Leq}) implies that the determinant of $\Lambda$ is either $+1$ or $-1$.  The 
structure of the Lorentz group is such that all transformations
with determinant $-1$ can be obtained from determinant $+1$ transformations 
and the two discreet operations of parity and time-reversal.  We see from
(\ref{ep}) that these two operations change the sign of the Levi-Cevita
symbol.  This fact causes $\epsilon$ to play an important
role in the spin structure of hadrons in QCD.

Up to now, our analysis has been quite general.  None of the
above manipulations have required the use of the Minkowskii metric
(\ref{mink}).  At this point, we specialize to flat-space and discuss the 
Lorentz transformations explicitly.  (\ref{Leq}) implies that 
$\Lambda$ is orthogonal, i.e. $\Lambda^{\rm T}\Lambda=1$.  In view
of the above discussion on the operations of parity and time-reversal,
we will consider only those transformations with $\det\Lambda=+1$.  The
group of orthogonal transformations in Minkowskii space with unit determinant
is called $SO(3,1)$ (the {\it special orthogonal} group in 3 spacelike and 1 timelike
dimensions).  Any real orthogonal matrix can be written as $e^M$, with $M=-M^{\rm T}$
a real skew-symmetric matrix.  A general skew-symmetric real $4\times 4$ matrix 
has six independent components, and hence the space of these objects can
be spanned by six linearly independent matrices.  The matrices we choose to
form this basis are called the {\it generators} of $SO(3,1)$ since they 
{\it generate} the group.  The algebra satisfied by the generators
is called the {\it Lie}\footnote{As in `... the {\it lie} of the stone.'} algebra of the
group.  Any member of the group is uniquely determined 
by the six coefficients of the generators.  The fact that there is a one-to-one
mapping between these coefficients and the elements of the group means
that the coefficients form a representation of the group,
the {\it adjoint} representation.  

The six generators of $SO(3,1)$ can be conveniently placed into an antisymmetric 
4-tensor, i.e.\footnote{The factor of 1/2 corrects the overcounting of the implied sum
on $\mu\nu$.}
\begin{equation}
M^\alpha_\beta={1\over2}\theta_{\mu\nu} (J^{\mu\nu})^\alpha_\beta\;\; .
\end{equation}
In principle, we are now free to choose whatever form we like for the 
generators $J^{\mu\nu}$, as long as we end up with six linearly independent 
skew-symmetric real matrices.  The most simple choice is also the most natural :
\begin{equation}
(J^{\mu\nu})_{\alpha\beta}=\delta^\mu_\alpha\delta^\nu_\beta
-\delta^\nu_\alpha\delta^\mu_\beta\;\; .
\label{Jdef}
\end{equation}
Working out the commutator of the generators, 
\begin{equation}
\left\lbrack J^{\mu\nu},J^{\alpha\beta}\right\rbrack=g^{\mu\beta}J^{\nu\alpha}
-g^{\nu\beta}J^{\mu\alpha}-g^{\mu\alpha}J^{\nu\beta}+g^{\nu\alpha}J^{\mu\beta}\;\; ,
\label{lieso}
\end{equation}
we find that $SO(3,1)$ is a non-abelian group.  This means that we must
be careful with the order of transformations.  In particular, $\theta$'s for 
successive transformations do {\it not} add.  

Since three-dimensional rotations are among the possible
Lorentz transformations, our generators should include
those for ordinary $SO(3)$ rotations.  This is indeed the case, as we
can see by examining any generator 
with both $\mu$ and $\nu$ spacelike.
For example,
\begin{eqnarray}
(J^{12})^\alpha_\beta=\left(\matrix{0&0&0&0\cr 0&0&-1&0
\cr0&1&0&0\cr0&0&0&0\cr}\right)\;\; .
\end{eqnarray}
It is easy to see that this matrix has four different powers, only
two of which are linearly independent.  Hence the $SO(3,1)$ rotation
it generates is
\begin{eqnarray}
(e^{\theta J^{12}})^\alpha_\beta=\left(\matrix{1&0&0&0\cr 0&\cos\theta&-\sin\theta&0
\cr0&\sin\theta&\cos\theta&0\cr0&0&0&1\cr}\right)\;\; ,
\end{eqnarray}
which is a rotation of $\theta$ radians about the 3-axis.  Hence 
we identify $J^{12}=L^3$.  
For general rotations, the identification\footnote{
It is a standard convention to use Greek letters for general
indices and Roman letters when one wants to specify a spatial index. 
The $\epsilon$ tensor here is the Levi-Cevita tensor for $SO(3)$, with the
convention $\epsilon^{123}=+1$.  Since the metric relevant to $SO(3)$ is trivial,
we make no distinction between contravariant and covariant indices.  However,
a few well-placed minus signs are required when going from a general 
expression to one which specifies spatial indices since $g_{ij}=-\delta_{ij}$.}
$\epsilon^{ijk}J^{ij}=2L^k$ is appropriate.  (\ref{lieso}) then 
implies 
\begin{equation}
\left\lbrack L^i,L^j\right\rbrack =\epsilon^{ijk}L^k\;\; ,
\end{equation}
as expected for generators of $SO(3)$ rotations.
Note that pure rotations form a subgroup of $SO(3,1)$, that
is $SO(3)\subset SO(3,1)$.

The other three generators have one time index and one spatial index.
These objects will generate Lorentz boosts, as can be seen by 
examining $J^{01}$ :
\begin{eqnarray}
(J^{01})^\alpha_\beta=\left(\matrix{0&1&0&0\cr1&0&0&0\cr0&0&0&0
\cr0&0&0&0\cr}\right)\;\; .
\end{eqnarray}
Once again, the powers of this matrix are cyclic so exponentiation is
trivial :
\begin{eqnarray}
 (e^{\theta J^{01}})^\alpha_\beta=\left(\matrix{\cosh\theta&\sinh\theta&0&0
\cr\sinh\theta&\cosh\theta&0&0\cr0&0&1&0\cr0&0&0&1\cr}\right)\;\; .
\end{eqnarray}
Under the action of this transformation,\footnote{We will use the
notation $x^0=t$, $x^1=x$, $x^2=y$, and $x^3=z$ interchangeably.
Confusion between $x^1=x$ and $x^\mu=x$ should be rare.} 
\begin{eqnarray}
t&\rightarrow&t\cosh\theta+x\sinh\theta\nonumber\\
x&\rightarrow&x\cosh\theta+t\sinh\theta\nonumber\\
y&\rightarrow&y\\
z&\rightarrow&z\;\; .\nonumber
\end{eqnarray}
It is obvious from the infinitesimal form of this transformation that we
have given the vector $x$ a velocity $\theta$ along the 1-direction.  
For asymptotically large values of $\theta$, $x$ and $t$ 
approach the same value $(t+x)\cosh\theta$, making the 
vector $x$ approximately lightlike.\footnote{Of course, this vector
does not {\it actually} become lightlike since its invariant 
square is invariant.  However, in our new frame $x^2$ will in fact
be much smaller than the square of its $t$ and $x$ coordinates.  This
makes the vector look lightlike in this frame.}  Hence 
asymptotically large $\theta$ boost the vector $x$ approximately
to the speed of light.  This argument promotes the identification
$v=\tanh\theta$.  We can now eliminate $\theta$ in favor of $v$ to obtain
\begin{eqnarray}
t&\rightarrow&\gamma(t+vx)\nonumber\\
x&\rightarrow&\gamma(x+vt)\nonumber\\
y&\rightarrow&y\\
z&\rightarrow&z\;\; ,\nonumber
\end{eqnarray}
the new coordinates of a vector $x$ which has been endowed with 
a speed $v$ along the 1-direction.  Again, we have 
used $\gamma\equiv(1-v^2)^{-1/2}$.
This transformation reduces to the familiar Galilean transformation
if we reinstate $c$ and take $v<\!\!< c$ (unless we take $xv/c^2$ comparable
to $t$).  

The transformations $J^{0i}\equiv K^i$ do not form a subgroup 
of $SO(3,1)$ since (\ref{lieso}) implies 
\begin{equation}
\left\lbrack K^i,K^j\right\rbrack=-\epsilon^{ijk}L^k\;\; .
\end{equation}
The commutator of $L$ with $K$ is 
\begin{equation}
\left\lbrack L^i, K^j\right\rbrack=\epsilon^{ijk}K^k\;\; ,
\end{equation}
which closes the algebra of the group.  

\section{Unitary Representations of $SO(3,1)$}
\label{unitrepso31}

In this section, we will study how the Lorentz group acts on
fields.  The first thing we must do is expand our
group to include translations.  Under a translation, the
vector $x$ is shifted to $x+a$.  This transformation
is {\it not} linear since $(x+y)+a\ne(x+a)+(y+a)$.
However, a {\it function}
of the vector $x$ will transform linearly under a 
translation :
\begin{equation}
T(a)f(x)=f(x-a)\;\; .
\end{equation}
Note that we have implicitly defined the transformation
$T(a)$ in such a way that it shifts the {\it function}
it acts on forward an amount $a$ rather than the axes.
Hence the value of the un-shifted function at $x$ is the 
same as the value of the shifted function at $x-a$.  
This extension of the Lorentz group is 
called the Poincar\'e\footnote{like {\it s'il vous plait}}
group, and its representations form the 
basis for quantum field theory.

Since we do not want to alter the normalizations of our
states through these transformations, we will be concerned with 
{\it unitary} representations of the Poincar\'e group.
Elements of these representations will be related to the 
coordinate transformations $\Lambda$ and $a$, i.e. $U=U(\Lambda, a)$.
The group multiplication rule, 
\begin{equation}
U(\Lambda_2, a_2)U(\Lambda_1, a_1)=U(\Lambda_2\Lambda_1, a_2+\Lambda_2a_1)\;\; ,
\end{equation}
is governed by these transformations
as well.  This rule implies $U^{-1}(\Lambda, a)=U^\dag(\Lambda,a)
=U(\Lambda^{-1},-\Lambda^{-1}a)$.
$U$ is unitary, so it can be represented as the 
exponential of an anti-hermitian matrix.  The translations
add four generators to the Lorentz group, one for each 
dimension of spacetime, so we have the form
\begin{equation}
U(1+{1\over 2}\theta_{\alpha\beta}J^{\alpha\beta},\epsilon^\mu)
=1-i{1\over 2}\theta_{\alpha\beta}{\hat J}^{\alpha\beta}+i\epsilon^\mu{\hat P}_\mu
\end{equation}
for arbitrary infinitesimal $\epsilon$ and $\theta$.  Note that 
$J$ are the matrices from the last section; they act on 
physical spacetime vectors.  $\hat J$ and $\hat P$ are 
hermitian operators that act on the space in which the fields we are concerned
with live.  
Consider the transformation
\begin{equation}
U(\Lambda,a)U(1+{1\over2}\theta J, \epsilon)U^{-1}(\Lambda,a)=
U(1+{1\over2}\Lambda\theta J\Lambda^{-1}, \Lambda\epsilon-{1\over2}\Lambda\theta J\Lambda^{-1}a)\;\; .
\end{equation}
Expanding in $\theta$ and $\epsilon$, substituting the 
explicit expression for $J$, and noting that each component 
of $\theta$ and $\epsilon$ may be varied independently, we
arrive at
\begin{eqnarray}
\label{Jcov}
U(\Lambda,a){\hat J}^{\mu\nu}U^{-1}(\Lambda,a)&=&(\Lambda^{-1})^\mu_\alpha
(\Lambda^{-1})^\nu_\beta\left({\hat J}^{\alpha\beta}-a^\alpha{\hat P}^\beta
+a^\beta{\hat P}^\alpha\right)\;\; ,\\
\label{Pcov}
U(\Lambda,a){\hat P}^\mu U^{-1}(\Lambda,a)&=&(\Lambda^{-1})^\mu_\nu {\hat P}^\nu\;\; .
\end{eqnarray}
Equation (\ref{Jcov}) states that for pure rotations ($a=0$), transforming the operator 
$\hat J$ is the same as performing the {\it inverse} transformation on the 
coordinate indices of $\hat J$.  The inverse transformation appears here
for the same reason that $x-a$ appears as the argument of $T(a)f(x)$.  
When translations are included, we see that $\hat J$ transforms in exactly
the way we expect an angular momentum to transform under a change of origin.
(\ref{Pcov}) tells us that the operator $\hat P$ transforms as a vector under
Lorentz transformations, but is invariant under translation.
We can obtain the Lie algebra of this group by considering (\ref{Jcov}) and 
(\ref{Pcov}) for infinitesimal transformations $\Lambda$ and $a$.  
Performing the same manipulations as above, we see that
\begin{eqnarray}
\label{Plie1}
-i\left\lbrack\hat J^{\mu\nu},\hat J^{\alpha\beta}
\right\rbrack&=&g^{\mu\beta}\hat J^{\nu\alpha}
-g^{\nu\beta}\hat J^{\mu\alpha}-g^{\mu\alpha}\hat J^{\nu\beta}
+g^{\nu\alpha}\hat J^{\mu\beta}\;\; ,\\
\label{Plie2}
-i\left\lbrack\hat P^\mu, 
\hat J^{\alpha\beta}\right\rbrack&=&g^{\mu\alpha}\hat P^\beta
-g^{\mu\beta}\hat P^\alpha\;\; ,\\
\label{Plie3}
-i\left\lbrack\hat P^\mu,\hat P^\nu\right\rbrack&=&0\;\; .
\end{eqnarray}
The last identity implies that 
the subgroup of pure translations is an abelian group.
The algebra of the Lorentz subgroup is the
same as we found above, as it should be.  However, it must be emphasized
that there we were discussing 4$\times$4 matrices rather than the 
infinite dimensional quantum mechanical operators here.

$\hat P$ and $\hat J$ have a very concrete physical
meaning in quantum mechanics.  To see this, consider a 
system which is translationally invariant.  Since the
phase of an isolated\footnote{If it wasn't isolated, it could not
be translationally invariant.  If state vectors interact, we must translate them all
simultaneously.  This is exactly the same as in classical mechanics,
where we must consider the {\it total} momentum
of a system to obtain conservation.} 
quantum mechanical system is not observable,
the state may change at most by the overall phase
\begin{equation}
T(a)\left|S\right\rangle=e^{i\delta(a)}\left|S\right\rangle\;\; ,
\end{equation}
where $\delta(a)$ is some real function of $a$.  In light of the above 
discussion, we see that $\delta(a)$ should be 
identified with $\hat P\cdot a$.
This implies that $|S\rangle$ is an eigenstate of $\hat P$.  
According to (\ref{Plie3}), this eigenvalue is conserved
under translation.
In classical mechanics,
the conserved quantity related to translation 
invariance is the momentum.\footnote{Remember,
momentum is conserved in the absence of 
external forces, i.e. forces which do not belong to the system we consider.
These forces would break translational invariance, and thus momentum conservation.} 
Hence, we identify the conserved quantity here with the momentum of the 
system.\footnote{This really implies that the momentum of the 
system is {\it proportional} to the eigenvalue.  If the momentum
of the system is $p$ and the eigenvalue is $k$, 
we say $p=\hbar k$.  $p$ has the dimensions of 
momentum (Energy/Speed=Energy since we work in units where 
speed is dimensionless) and $k$ has the dimensions 
of an inverse time (since the exponential of $\hat P\cdot a$ 
makes sense; remember that since speed is dimensionless
length and time are the same unit), so $\hbar$ is Energy$\times$Time.
Once again, we can adjust our units so that $\hbar=1$.
In other units, we can actually measure its value;  
in SI units, $\hbar=1.05457266(63)\times 10^{-34}J\cdot s$.}
Since the momentum of the system is an eigenvalue of $\hat P$, 
we call $\hat P$ the momentum operator.  

Attempting to apply the 
same analysis to the operator $\hat J$, we run into a little snag.
The fact that the $\hat J$'s do not commute with each other
implies that $|S\rangle$ cannot simultaneously be an eigenstate 
of all of them unless most of its eigenvalues are zero.\footnote{
We can see this as follows.  Suppose 
$\hat J^{\mu\nu}|S\rangle=j^{\mu\nu}|S\rangle$.  Now,
$[\hat J^{\mu\nu},\hat J^{\lambda\rho}]|S\rangle=0$ since 
the numbers $j^{\mu\nu}$ and $j^{\lambda\rho}$ commute.  This implies that
the commutator has eigenvalue zero.}
When discussing $\hat J$, it is convenient to define 
$\hat L^i\equiv1/2\epsilon^{ijk}\hat J^{jk}$ and $\hat K^i\equiv J^{0i}$
in analogy with the last section.
The commutation relations (\ref{Plie1}) imply 
\begin{eqnarray}
\label{Ang}
[\hat L^i,\hat L^j]&=&i\epsilon^{ijk}\hat L^k\;\; ,\\
\label{Boost}
[\hat K^i,\hat K^j]&=&-i\epsilon^{ijk}\hat L^k\;\; ,\\
\label{Mix}
[\hat L^i,\hat K^j]&=&i\epsilon^{ijk}\hat K^k\;\; .
\end{eqnarray}
These relations have the interesting consequence that
the operators
\begin{eqnarray}
\hat A^i={1\over2}\left(\hat L^i+i\hat K^i\right)\nonumber\\
\hat B^i={1\over2}\left(\hat L^i-i\hat K^i\right)
\label{ABdef}
\end{eqnarray}
generate commuting subgroups of $SO(3,1)$,
\begin{eqnarray}
\lbrack\hat A^i,\hat A^j\rbrack&=&i\epsilon^{ijk}\hat A^k\nonumber\\
\label{ABcom}
\lbrack\hat B^i,\hat B^j\rbrack&=&i\epsilon^{ijk}\hat B^k\\
\lbrack\hat A^i,\hat B^j\rbrack&=&0\;\; .\nonumber
\end{eqnarray}
Furthermore, the commutation relations satisfied 
by these generators identify the subgroups with 
$SU(2)$.  Hence the group of Lorentz transformations, $SO(3,1)$,
is seen to be nothing more than the direct product
of two commuting $SU(2)$'s.  This will be seen even 
more explicitly when we study the fundamental 
representation of $SL(2,C)$ and its 
relation to $SO(3,1)$ in the next section.

In non-relativistic physics, we concern ourselves
with time evolution and consider a quantity to be {\it conserved}
if it does not change with time.  The Schr\"odinger approach to
quantum mechanics considers state vectors that change with 
time,
\begin{equation}
\left|S(t)\right\rangle=e^{-it\hat H}\left|S(0)\right\rangle\;\; .
\end{equation}
The time dependence is governed by the time-translation operator, or
{\it Hamiltonian}, $\hat H\equiv\hat P^0$.  
The action of an operator on a state will change with time 
because the state itself changes with time.  
In any physical matrix element, we can use this relation
to display the time dependence explicitly :
\begin{equation}
\left\langle S'(t)\right|{\cal O}\left|S(t)\right\rangle
=\left\langle S'(0)\right|e^{it\hat H}{\cal O}
e^{-it\hat H}\left|S(0)\right\rangle\;\; .
\end{equation}
This promotes the identification 
${\cal O}(t)\equiv e^{it\hat H}{\cal O}e^{-it\hat H}$,
which puts the time-dependence of the matrix element into the
operator.  This is Heisenberg's view of quantum mechanics.  
Here, the action of an operator on a state will change with time
because the operator changes with time.  If the 
operator we are concerned with commutes with the Hamiltonian,
it will not depend on time.  Hence {\it all} of its matrix
elements will be conserved.  
Looking at Eqs.(\ref{Plie2}) and (\ref{Plie3}),
we see that $\hat P^i$ and $\hat L^i$ 
are conserved operators.\footnote{
In non-relativistic quantum mechanics,
one usually has some sort of space-dependent potential 
that completely destroys translation invariance.  
These potentials act as static momentum sources which
effectively generate nonzero commutators of the Hamiltonian
with momentum and angular momentum operators.}
Since $\hat L^i$ is a conserved operator associated with 
spatial rotation, we identify it with the angular momentum operator.
If a system is invariant under rotation about a certain axis, say $\hat n$,
the argument given above for $\hat P$ implies that it is 
and eigenstate of $\hat n\cdot\hat L$.
On the other hand, the commutator
\begin{equation}
[\hat H,\hat K^i]=i\hat P^i
\end{equation}
implies that $\hat K^i$ will depend on time.  

In relativistic
quantum mechanics, we are concerned mostly with operators; the external
states can all be formed from the vacuum by the action of various 
operators.  For this reason, it is Heisenberg's approach that is
most useful here.  The fact that time and space mix under Poincar\'e 
transformations makes it natural for us to consider
them together.  In a translationally invariant theory, 
the space dependence of the external states can be removed in
exactly the same way as the time dependence.  For any
spatial 4-vector $x^\mu$ and any quantum operator ${\cal O}$, 
we have the operator identity
\begin{equation}
{\cal O}(x)=e^{ix\cdot \hat P}{\cal O}(0)e^{-ix\cdot\hat P}\;\; .
\end{equation}
Operators that commute with all of the $\hat P^\mu$ do not
depend on coordinates at all.\footnote{This is what
we mean by `translationally invariant'!}

We now turn to a classification of the states in our theory.
Suppose we have a state which represents an electron 
with momentum $p^\mu$ and spin polarization $s^\mu$.
Poincar\'e transformations can change the electron's 
momentum and spin, but certainly cannot change the 
fact that it is an electron.  This means that 
general properties attributed to electrons 
must be associated with operators that commute with 
{\it all} of the generators of the group.  Operators
with this property are called {\it Casimirs} of the 
group.  It is easily shown that the 
mass operator\footnote{Remember that in relativity the invariant square
of the momentum of a system is the square of the 
mass of that system.}
$\hat P\cdot\hat P=\hat P^2$ is a casimir 
of the Poincar\'e group.  The other casimir
has to do with spin.  We know that an electron 
has spin-1/2 regardless of where it is or how we
rotate it.  There must be an operator that 
expresses that.  By inspection, one can see
that the Pauli-Ljubanskii vector
\begin{equation}
\hat W^\mu\equiv-{1\over 2}\epsilon^{\mu\nu\alpha\beta}\hat P_\nu\hat J_{\alpha\beta}
\end{equation}
is exactly what is needed.  Under parity, $\hat W^\mu$ behaves exactly 
as a spin should since $\epsilon$ changes sign.  
$\hat W$ is also orthogonal to $\hat P$ \footnote{
Under the influence of the Levi-Cevita symbol, the operators
$\hat P$ and $\hat J$ commute.} and reduces to a spatial angular momentum 
operator in the rest frame, $\hat P=(M,0,0,0)$.
The commutation relations of this vector with the generators
of the Poincar\'e group are identical with those of 
$\hat P^\mu$ :
\begin{eqnarray}
-i[\hat W^\mu,\hat P^\nu]&=&0\nonumber\\
-i[\hat W^\mu,\hat J^{\alpha\beta}]&=&g^{\mu\alpha}\hat W^\beta
-g^{\mu\beta}\hat W^\alpha\;\; .
\end{eqnarray}
This makes it obvious that its square 
\begin{equation}
\hat W^2=-{1\over2}\hat J_{\mu\nu}\hat J^{\mu\nu}\hat P^2
+\hat J_{\lambda\rho}\hat J^{\sigma\rho}\hat P^\lambda\hat P_\sigma
\end{equation}
is a casimir.\footnote{
This is obvious once one knows that $\hat W$ is a vector
that commutes with $\hat P$.  Any Lorentz scalar will commute with the
rotation generators.  It is the fact the $\hat W$ is translation
invariant that makes it special.}

To classify the irreducible representations
of the Poincar\'e group, we appeal to the idea
of Wigner's `little group' \cite{litgr}.  To begin with,
we diagonalize $\hat P^\mu$ :
\begin{equation}
\hat P^\mu\left|p\right\rangle=p^\mu\left|p\right\rangle\;\; .
\end{equation}
Since $p^\mu$ is a vector, it will transform under Lorentz 
transformations as
\begin{equation}
p^\mu\rightarrow\Lambda^\mu_\nu p^\nu\;\; .
\end{equation}
We can use this to transform our state to a convenient frame
with a representative momentum $\tilde p^\mu$.
In this frame, there will be residual transformations which 
do not change $\tilde p^\mu$, but may act on our state.
This group is called the little group associated with
the reference momentum $\tilde p^\mu$.
The representations of the little group will
tell us how many different states of 
momentum $\tilde p^\mu$ are related to our state 
through the Poincar\'e group.  Since $\hat W$ commutes
with $\hat P$, we can diagonalize one of its components
to break the degeneracy associated with this subgroup.
We will consider two examples of reference momentum 
$\tilde p^\mu$ since these are the only two which actually
occur in nature (to our knowledge). 

As our first example, we take states for which $p^0>0$ and 
$p^2=M^2>0$.  A look at the Lorentz transformations 
tells us that if $p^0>0$ in one frame, it will 
be in all frames, provided $p^2\ge0$.\footnote{
We consider here only the proper Lorentz 
group.  Parity and time-reversal can be taken into account 
separately later.}
Furthermore, there is always a {\it rest frame}, defined by
$\tilde p^\mu=(M,0,0,0)$.  The little 
group associated with this momentum is the group
of spatial rotations, $SO(3)$.  In this frame, 
we have $\hat W^0=0$ and $\hat W^i=M\hat L_i$.  
Since $\hat L^2=\hat L_i\hat L_i$ commutes
with $\hat L_i$, we can simultaneously 
diagonalize $\hat L^2$ and one of the 
$\hat L_i$ :
\begin{eqnarray}
\hat L^2\left|\ell m\right\rangle&=&\ell(\ell+1)\left|\ell m\right\rangle\nonumber\\
\hat L_3\left|\ell m\right\rangle&=&m\left|\ell m\right\rangle\;\; .
\end{eqnarray}
Since $\hat L^2$ is positive definite, we can take 
$\ell\ge0$.  The reason
for the special form of its eigenvalue will become clear shortly.
The hermiticity of $\hat L^2$ and $\hat L_3$ allows us to 
choose orthonormal eigenstates.
Defining 
\begin{equation}
\hat L_{\pm}\equiv \hat L_1\pm i\hat L_2\;\; ,
\end{equation}
we have
\begin{equation}
[\hat L_3,\hat L_\pm]=\pm\hat L_\pm\;\; .
\end{equation}
Hence 
\begin{equation}
\hat L_3\hat L_\pm\left|\ell m\right\rangle=(m\pm 1)\hat L_\pm\left|\ell m\right\rangle\;\; ,
\end{equation}
making $\hat L_\pm$ the {\it raising} and {\it lowering} operators 
associated with $\hat L_3$.  If this process were to go on
indefinitely, we would at some point have a state for which 
$m>\ell$.  However, 
\begin{equation}
\ell(\ell+1)=\left\langle\ell m\right|\hat L^2\left|\ell m\right\rangle
=\left\langle\ell m\right|\hat L_1^2+\hat L_2^2\left|\ell m\right\rangle+m^2
\end{equation}
implies that $m^2\le\ell(\ell+1)$.  This means that there must
be some state $|\ell m_\pm\rangle$ that truncates the process,
\begin{equation}
\hat L_\pm\left|\ell m_\pm\right\rangle=0\;\; .
\end{equation}
Writing $\hat L^2=\hat L_\pm\hat L_\mp+\hat L_3^2\mp\hat L_3$,
it is easy to show that $m_\pm=\pm\ell$.
Starting with the `top' state, $|\ell \ell\rangle$,
we can apply $\hat L_-$ to successively lower $m$ by one
unit each time.  At some point, we must reach the `bottom'
state $|\ell (-\ell)\rangle$ else suffer the consequences of 
inconsistency.  Since we cannot apply $L_-$ a fraction of a time,
this implies that $\ell-n=-\ell$ for some whole number $n$.
Hence $\ell$ must be either integer or half integer.\footnote{This is
the reason we assumed the form $\ell(\ell+1)$ for the eigenvalue
of $\hat L^2$.}  For each $\ell$, we can define all of the
states in our representation from the top state via 
successive application of $\hat L_-$.  The appropriate 
normalization is 
\begin{equation}
\hat L_\pm\left|\ell m\right\rangle=\sqrt{(\ell\mp m)(\ell\pm m+1)}
\left|\ell\;\; m\pm1\right\rangle\;\; ,
\end{equation}
as can be obtained from the requirement that $\ell(\ell+1)=
\langle\ell m|\hat L^2|\ell m\rangle$.\footnote{We also use the 
fact that $(\hat L_\pm|\ell m\rangle)^\dag=\langle\ell m|\hat L_\mp$ .}

In the rest frame of a massive system, we have shown that the
spectrum of $\hat W^2$ is $(-M^2\ell(\ell+1))$, where $\ell\ge0$ is
either integer or half-integer.  Since this object is a casimir
of the Poincar\'e group, different values of $\ell$ correspond to different 
irreducible representations.  For fixed $\ell$, there are $2\ell+1$
different states in the representation.  
Under the action of the rotation group,
these states behave as though they have different angular momentum,
so we say that in
the rest frame the state has angular momentum (spin)
$\ell$.  The spin vector is defined by
$\hat S^\mu\equiv\hat W^\mu/M$.
States in other frames can be obtained by boosting the 
rest frame states.  

The other kind of states which appear in nature have
$p^2=0$, $p^0>0$.  In this case, 
we take $\tilde p^\mu=(\omega,0,0,\omega)$
for our reference momentum.
The little group for this momentum is
generated by the components of $\hat W$ :
\begin{eqnarray}
\hat W^0&=&\omega\hat L_3\nonumber\\
\hat W^1&=&\omega(\hat L_1+\hat K_2)\nonumber\\
\hat W^2&=&\omega(\hat L_2-\hat K_1)\\
\hat W^3&=&\omega\hat L_3\;\; .\nonumber
\end{eqnarray}
These generators satisfy the commutation relations
\begin{eqnarray}
-i[\hat W^1,\hat W^3]&=&-\omega\hat W^2\nonumber\\
-i[\hat W^2,\hat W^3]&=&\phantom{-}\omega\hat W^1\\ 
-i[\hat W^1,\hat W^2]&=&\phantom{-\omega}\,0\;\; ,
\end{eqnarray}
so we can simultaneously diagonalize $\hat W^1$ and 
$\hat W^2$.  However, using the commutation
relations, one can show that the existence of 
one state with finite eigenvalues $w_1$ and $w_2$ for
$\hat W^1$ and $\hat W^2$ implies the existence of 
a continuous infinity of such states :
\begin{equation}
e^{i\theta\hat L_3}\left|w_1,w_2\right\rangle=\left|w_1\cos\theta+w_2\sin\theta,
w_2\cos\theta-w_1\sin\theta\right\rangle\;\; .
\end{equation}
Since we observe no massless states with a continuous degree of freedom
like this, we have no choice but to conclude that 
$\hat W^1$ and $\hat W^2$ annihilate our states.
In this subspace, $\hat W^1$ and $\hat W^2$ commute with $\hat W^3$
so we are free to diagonalize it.  This leaves us 
with no generators to connect different eigenstates 
of $\hat W^3$.  If $\hat W^3$ has more than one eigenvalue, these
states must be in different irreducible representations of the 
Poincar\'e group since they are 
not linked by any of its generators.\footnote{
The other generators are of no use to us here since they
will change the momentum of the state.}  

Massless states of 
3-momentum $\vec p$ are characterized by the action of
the {\it helicity operator}, $\vec p\cdot\hat{\vec L}/p^0$.  The eigenvalue
of this operator, $\lambda$, is called the {\it helicity} of the state.
Written in this way, the helicity of a system is 
invariant under the action of the Poincar\'e group.  
Although the proper Lorentz group cannot connect massless states with
different helicity, the action of parity relates helicity $+\lambda$ 
to helicity $-\lambda$.\footnote{This is because 
parity changes the sign of $\vec p$ while leaving $\hat{\vec L}$ invariant.}
Topological considerations require\footnote{The Lorentz group is not
simply connected, so the identification of a 
$2\pi$ rotation with the identity is misleading.  However, 
going around the group {\it twice} will always 
get us back to the same element.  One way to see this
is through the rotation subgroup, $SO(3)$.
Rotation through an angle $\theta$ off the $z$-axis and $\phi$ around
it (which is represented $(\theta,\phi)$) is physically identical
to $(2\pi-\theta,\pi+\phi)$.  These are two {\it distinct} group elements
which are {\it identified} in this representation.  This
degeneracy is only twofold, so going around twice gets one back to the
same group element.}
\begin{equation}
e^{4\pi i\hat L_3}=e^{4\pi i\lambda}=1\;\; ,
\end{equation}
or $\lambda$ is either integer or half-integer.  
The fact that $\lambda$ is the eigenvalue of an
angular momentum operator tempts us to call it the 
spin projection of the system.  However,
since the square of the Pauli-Ljubanskii vector is 0 regardless
of the value of $\lambda$, we must be careful with such 
terms.  

Explicit forms for the quantum mechanical operators
$\hat P$ and $\hat J$ will in general depend on the specific
theory.  These operators can be obtained from the lagrangian
using general techniques outlined in Section \ref{quantpoint}.

\section{The Spinor Representation of $SO(3,1)$}
\label{dottedandundotted}

In this section, we consider the smallest representation of $SO(3,1)$.
From ordinary quantum mechanics, we know that the group $SO(3)$ is 
locally isomorphic to $SU(2)$ (it satisfies the same Lie algebra).
This led us to the fact that the fundamental representation
of $SO(3)$ can be realized in terms of quantum mechanical 2-spinors.
Now, we are faced with the problem of finding the fundamental representation
of the somewhat larger group, $SO(3,1)$.  
A simple way of going from a certain group
to a larger one is to relax some of the constraints defining the 
smaller group.  Relaxing the determinant condition in $SU(2)$ only yields a 
trivial phase which commutes with everything.  One look at Eq.(\ref{Plie1})
tells us that this is not good enough.  Let us see what happens if 
we relax the uniterity condition. 

The group of $2\times2$ complex matrices with unit determinant
is a representation of $SL(2,C)$, the special linear group
in two complex dimensions.  A general element of $SL(2,C)$ has
four complex entries with one complex condition.  Hence it is composed
of six independent elements.  This certainly is a good sign.
Let us take $\xi^\alpha$ to be a two component object which transforms
under $SL(2,C)$ as
\begin{equation}
\xi^\alpha\rightarrow(T\xi)^\alpha\;\; ,
\end{equation}
and $\eta_\alpha$ to be one which transforms as
\begin{equation}
\eta_\alpha\rightarrow(\eta T^{-1})_\alpha\;\; .
\end{equation}
Obviously, because of the way we have defined $\xi$ and $\eta$, 
$\eta_\alpha\xi^\alpha$ is invariant under $SL(2,C)$.
Our experience with $SO(3,1)$ has taught us that 
conditions on matrices lead to invariants of the group.  There,
we saw that the condition ${\rm det}\Lambda=+1$ led to the 
invariance of the Levi-Cevita symbol.  Here, we define
$\epsilon_{\alpha\beta}$ to be the totally antisymmetric 
tensor of $SL(2,C)$.  We use the convention $\epsilon_{12}=+1$.
Since $\epsilon$ is invariant, the product $\epsilon_{\alpha\beta}\xi^\alpha\eta^\beta$
is also invariant.  This prompts us to identify
\begin{equation}
\eta_\alpha\equiv\epsilon_{\alpha\beta}\eta^\beta\;\; ,
\end{equation}
making $\epsilon$ the metric tensor in this space.  Note that our metric
is {\it anti}symmetric.  This means that we must be extremely careful
with the order of indices and the proper way to sum here.  
For example, $\epsilon_{\alpha\beta}\eta^\alpha=-\eta_\beta\ne\eta_\beta$.
We can define a raising operator such that 
\begin{equation}
\epsilon^{\alpha\beta}\epsilon_{\beta\gamma}=\delta^\alpha_\gamma\;\; .
\end{equation}
Note that this implies that $\epsilon^{\alpha\beta}=-\epsilon_{\alpha\beta}$.

The complex conjugate of $\xi^\alpha$ transforms under $SL(2,C)$ like
$T^*\ne T$.  Hence it forms a separate space of transformations.
Denoting $(\xi^\alpha)^*\equiv\xi^{\dot\alpha}$, we can play the whole game 
over again.  The index $\dot\alpha$ is called a `dotted' index
and lives in a different space than $\alpha$.  Since the 
two indices transform in completely separate ways, there is no metric which
connects them.

As with any group representation, 
products of spinors give us 
higher representations.  For example, the product
\begin{equation}
\zeta^{\alpha\beta}=\xi^\alpha\eta^\beta
\end{equation}
is a spinor tensor of rank two (or a {\it bispinor}).  This representation is reducible,
as we can see by separating it into two independent pieces :
\begin{equation}
\zeta^{\alpha\beta}={1\over 2}\left(\zeta^{\alpha\beta}+\zeta^{\beta\alpha}\right)
+{1\over2}\left(\zeta^{\alpha\beta}-\zeta^{\beta\alpha}\right)\;\; .
\end{equation}
The second term is proportional to the metric $\epsilon^{\alpha\beta}$ since it
is antisymmetric in $\alpha$ and $\beta$.  Hence it is invariant under
$SL(2,C)$.  The first term is symmetric in $\alpha$ and $\beta$.  
We denote this as $\zeta^{(\alpha\beta)}$.  Since it is
symmetric, it only has three independent components.
Note that the total number of components of
$\zeta^{\alpha\beta}$, 4, is conserved in the process of
breaking it up into irreducible representations.
This is a good check that we have done the separation correctly.  
This equation may be written as ${\bf 2}\times{\bf 2}={\bf 3}+{\bf 1}$
since we began with the direct product of two spinor representations (${\bf 2}$)
and ended with the sum of a singlet (${\bf 1}$) and a triplet
(${\bf 3}$) representation.

In general,
a spinor tensor with $2s$ undotted indices may be completely symmetrized
by a systematic removal of all of its `traces' (contractions with the 
metric tensor).  The remaining object transforms as spin $s$ and the 
traces we removed 
transform as spin $s-1, s-2,\ldots,0$ or $1/2$, depending 
on whether $2s$ is even or odd.  These statements can be used as a definition
of `spin' if one so chooses.  The motivation comes from the eigenstates
of the angular momentum operator, which we have not yet discussed in this 
context.  However, the spectrum of eigenvalues of $J^2$ and $J_z$ depends
only on the commutation relations of the group $SO(3,1)$.  Since 
these relations are the same as those of $SL(2,C)$, as we will show below,
the angular momentum operators in our spinor space will have the 
same spectrum we discussed above.  An irreducible representation 
of $SL(2,C)$ with three independent components, like $\zeta^{(\alpha\beta)}$,
will have eigenvalues of $\hat J_z=0,\pm1$ and $\hat J^2=2$.
Thus is behaves like it has angular momentum, or spin, 1.

Mixed combinations of spinors are of a completely another form entirely.
Since dotted and undotted indices live in two different spaces, it
makes no sense to symmetrize or antisymmetrize them.  The
product of a spinor (${\bf 2}$) and an antispinor ($\overline{\bf 2}$)
representation is itself irreducible (${\bf 2}\times\overline{\bf 2}={\bf
4}$).
A mixed bispinor will also be real in the sense that 
\begin{equation}
(\zeta^{\alpha\dot\beta})^*=\zeta^{\dot\alpha\beta}=\zeta^{\beta\dot\alpha}\;\; ,
\label{hermsosl}
\end{equation}
where the last equality comes from the fact that the order of 
dotted and undotted indices is unimportant; these indices are in
two different spaces.  Mixed bispinors form an irreducible
representation of $SL(2,C)$ with four real entries.  If
representations of $SL(2,C)$ can indeed be identified
with representations of $SO(3,1)$, then a relation between 
this irreducible 4-dimensional representation and 
the representation $x^\mu$ from Section \ref{relativity} is possible.

Writing $\zeta^{\alpha\dot\beta}$ as a $2\times2$ matrix,
we see that the relation (\ref{hermsosl}) translates into
a hermiticity condition on $\zeta$.\footnote{The transposition comes from
the convention that undotted indices should label 
rows and the dotted ones columns.  Of course, this convention
is arbitrary, but one must be consistent with whatever convention is chosen.}
Any hermitian $2\times2$ matrix may be expanded in terms of the identity and the
Pauli matrices,
\begin{eqnarray}
\sigma^1\equiv\left(\matrix{0&1\cr1&0}\right)&
\sigma^2\equiv\left(\matrix{0&-i\cr i&0}\right)&
\sigma^3\equiv\left(\matrix{1&0\cr0&-1}\right)\;\; .
\end{eqnarray}
Writing $\overline\sigma^\mu\equiv(1,-\vec\sigma\,)$, we make 
the identification
\begin{equation}
\zeta^{\alpha\dot\beta}=(x^\mu\overline\sigma_\mu)^{\alpha\dot\beta}\;\; .
\label{slsorel}
\end{equation}
We would like for $\zeta$ to behave under $SL(2,C)$ rotations in such a way
that this relation is preserved.  An infinitesimal $SL(2,C)$ 
transformation has six independent
parameters, and so can be expressed
\begin{equation}
(T)^\alpha_{\;\,\beta}=\delta^\alpha_\beta-{i\over 2}\theta_{\mu\nu}
\left(\Sigma^{\mu\nu}\right)^\alpha_{\;\,\beta}\;\; ,
\end{equation}
where the $\Sigma$ are linearly independent $2\times2$ matrices.
Note that we have emphasized the fact that $\alpha$ is the first
index on $T$ and $\beta$ the second.  The transformation for 
$\zeta$ is 
\begin{eqnarray}
(T\zeta)^{\alpha\dot\beta}&=&\zeta^{\alpha\dot\beta}-{i\over2}\theta_{\mu\nu}
\left(\Sigma^{\mu\nu}\right)^\alpha_{\;\,\gamma}\zeta^{\gamma\dot\beta}
+{i\over2}\theta_{\mu\nu}\left(\Sigma^{\mu\nu}\,^*
\right)^{\dot\beta}_{\;\,\dot\gamma}
\zeta^{\alpha\dot\gamma}\nonumber\\
&=&\zeta^{\alpha\dot\beta}-
{i\over2}\theta_{\mu\nu}\left(\Sigma^{\mu\nu}\zeta-\zeta
(\Sigma^{\mu\nu})^\dag\right)
^{\alpha\dot\beta}\;\; ,
\label{ztrans}
\end{eqnarray}
where in the second line we have suppressed the indices. 
The transposition comes from the convention of matrix multiplication.
In another frame, the new $\zeta$ should be related to the new
$x$ in the same way as the old ones for {\it any} choice
of $x^\mu$.  According to (\ref{slsorel}), (\ref{ztrans}),
and (\ref{Jdef}), we require
\begin{equation}
-i\Sigma^{\mu\nu}\overline\sigma^\lambda+i\overline\sigma^\lambda
\left(\Sigma^{\mu\nu}\right)^\dag=g^{\nu\lambda}\overline\sigma^\mu-g^{\mu\lambda}
\overline\sigma^\nu\;\; .
\end{equation}
If we can solve this equation for $\Sigma$, we have found a covariant relationship
between the $\bf 4$ of $SL(2,C)$ and the $\bf 4$ of $SO(3,1)$.

This equation is not as complicated as it seems; the easiest way
to solve it is to recognize that each element is just a $2\times2$ matrix.
{\it Any} $2\times2$ complex matrix can be written as a linear combination of the
Pauli matrices and the identity, if we allow complex coefficients. 
Taking special values for $\mu,\nu$, and $\lambda$ then allows us to deduce
the coefficients.  Defining 
\begin{equation}
\left(\sigma^\mu\right)_{\dot\alpha\beta}\equiv
\epsilon_{\dot\alpha\dot\beta}\epsilon_{\beta\alpha}
\left(\overline\sigma^\mu\right)^{\alpha\dot\beta}\;\; ,
\end{equation}
we have 
\begin{equation}
\left(\Sigma^{\mu\nu}\right)^\alpha_{\;\,\beta}=\left(\overline\sigma^{\mu\nu}\right)
^\alpha_{\;\,\beta}\equiv{i\over 4}\left(\overline\sigma^\mu\sigma^\nu
-\overline\sigma^\nu\sigma^\mu\right)^\alpha_{\;\,\beta}\;\; .
\label{sigup}
\end{equation}
It is easy to show that these objects satisfy the 
commutation relations of $SO(3,1)$, Eq.(\ref{Plie1}).

The relationship between vector indices and 
spinor indices allows us to understand the 
construction of $SO(3,1)$'s irreducible tensor
representations.  According to Eq.(\ref{slsorel}), we
have the correspondence
\begin{equation}
T^{\mu_1\cdots\mu_n}\sim \tau^{\alpha_1\cdots\alpha_n
\dot\beta_1\cdots\dot\beta_n}
\end{equation}
for an arbitrary Lorentz tensor $T$.
We can form irreducible representations
of $SL(2,C)$ by systematically removing
the traces of $\tau$ with respect to 
like indices.  Studying the symmetry of 
different representations under
the interchange of two {\it Lorentz} indices,
i.e. the simultaneous replacement 
$(\alpha_i\dot\beta_i,\alpha_j\dot\beta_j)\rightarrow
(\alpha_j\dot\beta_j,\alpha_i\dot\beta_i)$,
we see that they can be constructed to be
either symmetric or antisymmetric.\footnote{In reality,
the construction is not quite this simple.  The
natural irreducible representations of $SL(2,C)$
do not lend themselves to interpretation in
$SO(3,1)$.  However, certain linear combinations
of them have symmetry properties which are readily
understood within $SO(3,1)$.}

If antisymmetric, one can either have 
\begin{equation}
\epsilon^{\alpha_i\alpha_j}\zeta^{(\dot\beta_i\dot\beta_j)}\phantom{\;\; .}
\end{equation}
or 
\begin{equation}
\epsilon^{\dot\beta_i\dot\beta_j}\zeta^{(\alpha_i\alpha_j)}\;\; .
\end{equation}
Neither of these is hermitian, but the 
sum and the difference (divided by $i$)
both are.  These combinations will be
related to irreducible tensor representations
of the Lorentz group which are antisymmetric 
under the interchange of two indices.

If symmetric, we have two possibilities.  
First, all four indices could be contained within
metric tensors :
\begin{equation}
\epsilon^{\alpha_i\alpha_j}\epsilon^{\dot\beta_i\dot\beta_j}\;\; .
\end{equation}
Inspection of the relationship (\ref{slsorel})
between $SO(3,1)$ and $SL(2,C)$
reveals that this object represents the 
metric tensor $g^{\mu\nu}$ of $SO(3,1)$.
Hence this possibility leads to a trace 
in Lorentz space.  Alternatively,
the structure
\begin{equation}
\zeta^{(\alpha_i\alpha_j)(\dot\beta_i\dot\beta_j)}
\end{equation}
satisfies our requirement.  This 
object is obviously related to a 
symmetric Lorentz tensor which 
does not have a trace.  
This simple argument can be used to
construct the irreducible breakdowns
of {\it any} tensor in $SO(3,1)$.
In particular, it implies that the largest
representation possible is the 
fully symmetric and traceless combination
\begin{equation}
T^{(\mu_1\cdots\mu_n)}\;\; .
\end{equation}
Since this tensor corresponds to a 
spinor tensor with $n$ symmetrized dotted
indices and $n$ symmetrized undotted indices,
its spin is $s=2n/2=n$.  

To see how this works explicitly, 
let us study the decomposition of
the tensor
\begin{equation}
T^{\mu_1\mu_2\mu_3}=\tau^{\alpha_1\alpha_2\alpha_3
\dot\beta_1\dot\beta_2\dot\beta_3}\;\; .
\end{equation}
In order to avoid overcounting, we
must proceed as systematically as possible.
With this in mind, we leave the dotted indices
alone for the moment and work only
with the undotted.
The possible structures are
\begin{eqnarray}
\label{blahS}
\tau_s&=&\tau^{(\alpha_1\alpha_2\alpha_3)
\dot\beta_1\dot\beta_2\dot\beta_3}\\
\tau_{t2}&=&\epsilon^{\alpha_1\alpha_2}\zeta^{\alpha_3
\dot\beta_1\dot\beta_2\dot\beta_3}\\
\tau_{t3}&=&\epsilon^{\alpha_1\alpha_3}\zeta^{\alpha_2
\dot\beta_1\dot\beta_2\dot\beta_3}\;\; .
\end{eqnarray}
Note that we did not take all three
possible traces as they are not all independent.
Traces in $SL(2,C)$ are related to 
antisymmetric structures, so the symmetric 
combination 
\begin{equation}
\epsilon^{\alpha\beta}\zeta^\gamma+
\epsilon^{\beta\gamma}\zeta^\alpha+
\epsilon^{\gamma\alpha}\zeta^\beta=0
\label{consistent}
\end{equation}
is constrained to vanish.  This can be seen
by explicitly checking all three components.
In general, the procedure is
to {\it choose one index} and systematically
remove all of the traces involving {\it that index}.
Other traces are constrained by consistency
relations like (\ref{consistent}).
The dotted indices decompose for {\it each}
of the above structures in the same way.

Equation (\ref{blahS}) gives the irreducible representations
\begin{eqnarray}
\label{blahSS}
\tau_{ss}&=&\tau^{(\alpha_1\alpha_2\alpha_3)
(\dot\beta_1\dot\beta_2\dot\beta_3)}\\
\label{blahST2}
\tau_{st2}&=&\epsilon^{\dot\beta_1\dot\beta_2}
\zeta^{(\alpha_1\alpha_2\alpha_3)
\dot\beta_3}\\
\label{blahST3}
\tau_{st3}&=&\epsilon^{\dot\beta_1\dot\beta_3}
\zeta^{(\alpha_1\alpha_2\alpha_3)
\dot\beta_2}\;\; .
\end{eqnarray}
The first of these is the fully symmetric
spin-3 contribution.  It corresponds to the 
fully symmetric and traceless
\begin{equation}
T^{(\mu_1\mu_2\mu_3)}\;\; .
\label{symm3tens}
\end{equation}
Since three fully symmetrized
$SL(2,C)$ indices have four components :
all the same (2), one different (2), 
this representation has $4\cdot 4=16$ components.

Turning to the trace terms, we see that 
neither (\ref{blahST2}) nor (\ref{blahST3})
is hermitian, so neither can directly 
be related to representations of $SO(3,1)$.
However, their complex conjugates are contained
in\footnote{My notation reads `symmetric',
`trace with $\alpha_2$', and `trace with $\alpha_3$'.
The order refers undotted versus dotted indices.}
$\tau_{t2s}$ and $\tau_{t3s}$.  Representations
of $SO(3,1)$ can only be formed with these
`mixed' spinor tensors via hermitian combinations.
Since we cannot separate these 
from each other while at the same time retaining
hermiticity and completeness of the
representation, $\tau_{st2}$ and 
$\tau_{t2s}$ must {\it combine}
to form {\it one} irreducible 
representation of $SO(3,1)$,
\begin{equation}
T^{[\mu_1(\mu_2]\mu_3)}\;\; .
\label{mixedsymm}
\end{equation}
The fact that 
$SO(3,1)$ possesses these two invariant subgroups is
already obvious from the behavior of its generators
in Eqs.(\ref{ABdef}) and (\ref{ABcom}).  The number of components 
the tensor in (\ref{blahST2}) contains is
$4\cdot 2=8$; (\ref{mixedsymm})
contains 16 components, as does its friend
\begin{equation}
T^{[\mu_1(\mu_3]\mu_2)}\;\; .
\end{equation}
Note that in order to actually form representations
with the indicated symmetry structure, 
we would need to involve the tensor
\begin{equation}
\epsilon^{\alpha_2\alpha_3}\zeta^{\alpha_1(\dot\beta_1
\dot\beta_2\dot\beta_3)}
\end{equation}
and its hermitian conjugate.  
With these six tensors, we can form 
three hermitian Lorentz structures.  However,
in light of the above discussion, only 
two of them are independent.  

We are left with 
\begin{eqnarray}
\tau_{t2t3}&=&\epsilon^{\alpha_1\alpha_2}\epsilon^{\dot\beta_1\dot\beta_3}
\zeta^{\alpha_3\dot\beta_2}\\
\tau_{t3t2}&=&\epsilon^{\alpha_1\alpha_3}\epsilon^{\dot\beta_1\dot\beta_2}
\zeta^{\alpha_2\dot\beta_3}\\
\tau_{t2t2}&=&\epsilon^{\alpha_1\alpha_2}\epsilon^{\dot\beta_1\dot\beta_2}
\zeta^{\alpha_3\dot\beta_3}\\
\tau_{t3t3}&=&\epsilon^{\alpha_1\alpha_3}\epsilon^{\dot\beta_1\dot\beta_3}
\zeta^{\alpha_2\dot\beta_2}\;\; .
\end{eqnarray}
The first two of these expressions
must be combined to form a representation
of $SO(3,1)$, as before.  In this case, 
the presence of two metric tensors 
allows us to construct {\it separate}
hermitian representations.  Involving
the other four mixed-symmetric two-trace 
tensor structures allows us to construct
a fully symmetric\footnote{
under the exchange of {\it pairs} of
indices, i.e. $(\alpha_i\dot\beta_i,\alpha_j\dot\beta_j)\rightarrow
(\alpha_j\dot\beta_j,\alpha_i\dot\beta_i)$.}
tensor which can be identified as the 
trace we removed from (\ref{symm3tens})
and a fully antisymmetric tensor 
which is identified with 
\begin{equation}
T^{[\mu_1\mu_2\mu_3]}\;\; .
\end{equation}
Both of these tensors have 4 components.
The remaining two structures, $\tau_{t2t2}$ 
and $\tau_{T3T3}$, are readily identified 
in $SO(3,1)$ as the other two traces :
\begin{eqnarray}
\tau_{t2t2}\sim g^{\mu_1\mu_2}P^{\mu_3}\phantom{\;\; .}\\
\tau_{t3t3}\sim g^{\mu_1\mu_3}P^{\mu_2}\;\; .
\end{eqnarray}
Hence in $SL(2,C)$ we have the decomposition
\begin{equation}
{\bf 2\times\overline 2\times
2\times\overline 2\times
2\times\overline 2}={\bf 16+}({\bf 8+\overline8}){\bf +}
({\bf 8+\overline8}){\bf +}({\bf 4+\overline4}){\bf+4+4}\;\; ,
\end{equation}
while in $SO(3,1)$,
\begin{equation}
{\bf4\times4\times4}={\bf16}_S{\bf +}{\bf16}_M{\bf+}{\bf16}_M
{\bf+4}_S{\bf+4}_M{\bf+4}_M{\bf+4}_A
\end{equation}
is induced.  Note that in $SO(3,1)$ all three
possible traces show up in the 
final decomposition, the third being formed
from the combination of $SL(2,C)$'s $\bf4+\overline4$
which is orthogonal to the fully antisymmetric
tensor.  This is because traces in $SO(3,1)$
are unconstrained.

This procedure can be extended to tensors of
higher rank without incident.  
While somewhat
tedious, it certainly beats counting components
and enforcing constraint relations to avoid
overcounting.  An equivalent procedure one
can use to directly obtain the representations 
of $SO(3,1)$ is offered by Young's Tableaux.  The
most useful reference that I have come across on this
method is \cite{hammermesh}.

At some point, we would like to use all of this formalism to do physics.
Consider a classical spinor field, $\xi^\alpha(x)$.
Following the discussion of the last section on irreducible representations
of the Poincar\'e group, we wish to diagonalize the momentum
operator $\hat P^\mu$.  This operator has two spinor representations,
$(\hat P\cdot\sigma)_{\dot\alpha\beta}$ and 
$(\hat P\cdot\overline\sigma)^{\alpha\dot\beta}$. 
Each representation necessarily has one dotted and one undotted index in 
$SL(2,C)$, so the object $\hat P\xi$ cannot transform in the same way as
$\xi$ itself.  This leaves us with two options : we can either 
require $\hat P$ to have eigenvalue zero or allow it to change our spinor
into another kind of spinor.  The former option leads to the massless
Weyl spinors that describe neutrinos, and the latter 
leads to the massive Dirac spinors appropriate for electrons.

Weyl spinors are obtained simply from the requirement 
\begin{equation}
(\hat P\cdot\sigma)_{\dot\alpha\beta}\xi^\beta=0\;\; .
\label{weylspin}
\end{equation}
Contracting with $(\hat P\cdot\overline\sigma)^{\gamma\dot\alpha}$ and 
using the identity 
\begin{equation}
\left(\overline\sigma^\mu\sigma^\nu\right)^{\alpha}_{\;\,\beta}=
g^{\mu\nu}\delta^{\alpha}_{\beta}
-2i\left(\overline\sigma^{\mu\nu}\right)^{\alpha}_{\;\,\beta}\;\; ,
\label{anticommute}
\end{equation}
it is easily seen that the casimir $\hat P^2$ has eigenvalue zero so these
spinors can only represent massless states.
Within the proper Lorentz group, this spinor 
stands alone;  it forms its own irreducible representation.
However, under parity our spinor field transforms as\footnote{
This can be seen in the following way :
${\cal P}\xi^\alpha$ cannot transform like 
$\xi^\alpha$ because $\cal P$ does not commute with all 
Lorentz transformations.  However, since the spatial
part of $\cal P$ is proportional to the identity,
it will commute with spatial rotations.
For spatial rotations, our $SL(2,C)$ matrix is unitary
so $(T^{-1})^\dag=T$.  This implies that 
${\cal P}\xi^\alpha$ transforms like $\eta_{\dot\alpha}$.} 
\begin{equation}
{\cal P}\xi^\alpha=\eta_{\dot\alpha}\;\; .
\end{equation}
Hence if we wish to consider representations of the full Lorentz group,
we must include the field $\eta_{\dot\alpha}$ with $\xi^\alpha$ in
our massless representation.\footnote{This is a specific example
of the general argument given in the last section that parity 
relates the two helicities $\pm\lambda$ to each other.
We will see below that $\xi^\alpha$ corresponds to massless
states with helicity $+1/2$ while $\eta_{\dot\alpha}$ 
describes states with helicity $-1/2$.}  This field satisfies
\begin{equation}
(\hat P\cdot\overline\sigma)
^{\alpha\dot\beta}\eta_{\dot\beta}=0\;\; .
\end{equation}
Since we wish to consider these fields together, we write
\begin{eqnarray}
\psi=\left(\matrix{\eta_{\dot\alpha}\cr\xi^\alpha\cr}\right)\;\; ,
\end{eqnarray}
for which the statement of translation invariance becomes
\begin{eqnarray}
\hat P_\mu\left(\matrix{0&\left(\sigma^\mu\right)_{\dot\alpha\beta}\cr
\left(\overline\sigma^\mu\right)^{\alpha\dot\beta}&0\cr}\right)\psi=0\;\; .
\end{eqnarray}

The 4-component spinor $\psi$ is a very strange object.  It mixes 
dotted and undotted indices in a way we have not yet encountered.
This mixing has been forced upon us by parity.  Systems
in which parity is not conserved, like the weak interactions, 
do not require us to consider both spinors together.  In fact,
it has long been thought that the analogue of $\xi^\alpha$ for
neutrinos simply does not exist.  Here we are
concerned with the strong and electromagnetic interactions,
both of which conserve parity, so the existence of one 
helicity necessarily implies the existence of the other.
The $4\times4$ matrices in the parentheses come up over
and over again.  We will call them $\gamma^\mu$.  We also introduce
$\not\!\!\hat P$ as shorthand for $\hat P_\mu\gamma^\mu$.  Using this
compact notation, we write the `equation of motion' for massless
spinor fields as
\begin{equation}
\not\!\!\hat P\psi=0\;\; .
\end{equation}

For a massless spinor moving in the 3-direction,
our equation of motion reduces to
\begin{eqnarray}
2p^0\left(\matrix{0&0&0&0\cr0&0&0&1\cr1&0&0&0\cr0&0&0&0\cr}\right)\psi=0\;\; ,
\end{eqnarray}
which has two linearly independent solutions,
\begin{eqnarray}
u(p,+)=\sqrt{2p^0}\left(\matrix{0\cr0\cr1\cr0\cr}\right)\;\;\;\;{\rm and}
\;\;\;\;u(p,-)=\sqrt{2p^0}\left(\matrix{0\cr1\cr0\cr0\cr}\right)\;\; ,
\end{eqnarray}
which we normalize in such a way that the following relations
hold
\begin{eqnarray}
\label{normmass}
u^\dag(p,+)u(p,+)&=&u^\dag(p,-)u(p,-)=2p^0\\
\label{nullmass}
{\overline u}(p,+)u(p,+)&=&{\overline u}(p,-)u(p,-)=0\\
\label{nullmass2}
{\overline u}(p,+)u(p,-)&=&{\overline u}(p,-)u(p,+)=0\\
\label{compmass}
\sum_{h=\pm} u(p,h){\overline u}(p,h)&=&\not\!p\;\; ,
\end{eqnarray}
where I have defined $\overline u\equiv u^\dag\gamma^0$.
Note that (\ref{normmass}) is not Lorentz invariant 
since $\psi^\dag\psi$ involves a sum over dotted and undotted indices.
However, 
\begin{equation}
\overline\psi\psi=\eta_\alpha\xi^{\alpha}
+\xi^{\dot\alpha}\eta_{\dot\alpha}
\end{equation}
is invariant since it involves contractions with like indices
only.  (\ref{nullmass}) is simply the statement that for massless
spinors the upper and lower components are uncoupled.
This is not the case for massive spinors, where the momentum operator
itself couples dotted and undotted indices.  

Since $\overline\psi\psi$ is invariant and $\gamma^\mu\psi$
transforms in $SL(2,C)$ like $\psi$ itself, the combination
$\overline\psi\gamma^\mu\psi$ transforms under the action of the Lorentz 
group as a vector (it is invariant under $SL(2,C)$).  
This means that $\psi^\dag\psi=\overline\psi\gamma^0\psi$
is just the time component of a 4-vector.  
The last relation
is a statement about the completeness of our solutions.
It is covariant in the sense that both sides transform like the 
$SL(2,C)$ indices on a $\gamma$-matrix.  

Under Lorentz transformations, the lower components of 
$\psi$ transform according to (\ref{sigup}).  The upper components
transform like
\begin{equation}
\left(\Sigma^{\mu\nu}\right)^{\;\,\dot\beta}_{\dot\alpha}
=\left(\overline\sigma^{\mu\nu\;\dag}\right)_{\dot\alpha}^{\;\,\dot\beta}
=\left(\sigma^{\mu\nu}\right)_{\dot\alpha}^{\;\,\dot\beta}
\equiv{i\over 4}\left(\sigma^\mu\overline\sigma^\nu
-\sigma^\nu\overline\sigma^\mu\right)_{\dot\alpha}^{\;\,\dot\beta}\;\; ,
\end{equation}
so $\psi$ transforms according to\footnote{Since the $(2\times2)$
$\sigma^{\mu\nu}$ has dotted and undotted $SL(2,C)$ indices
and this new $(4\times4)$ 
$\sigma^{\mu\nu}$ has only the mixed `Dirac' indices,
confusion should be rare.  Throughout the text, the 
$(2\times2)$ $\sigma^{\mu\nu}$ does not make an appearence.}
\begin{eqnarray}
\Sigma^{\mu\nu}=\left(\matrix{\sigma^{\mu\nu}&0\cr0&
\overline\sigma^{\mu\nu}\cr}\right)
={i\over4}\left[\gamma^\mu,\gamma^\nu\right]\equiv{1\over2}\sigma^{\mu\nu}\;\; .
\label{Dtransform}
\end{eqnarray}
Although this is not a hermitian operator and certainly cannot be
identified with $\hat J$,\footnote{We will see how this 
object is related to $\hat J$ in Section \ref{quantpoint}.} 
it does describe the action of the 
Lorentz group on classical fields $\psi$.  
For a {\it classical}
massless particle moving in the 3-direction, the relevant
helicity operator is 
\begin{eqnarray}
\Sigma^{12}=\left(\matrix{\sigma^3/2&0\cr0&\sigma^3/2\cr}\right)\;\; .
\end{eqnarray}
The action of this operator on our $u$'s tells us that 
$u(p,+)$ represents a particle with helicity $+1/2$ and 
$u(p,-)$ represents a particle with helicity $-1/2$.
In this sense, massless spinors can be said to have
spin-1/2.

Dirac spinors are obtained by writing 
\begin{equation}
(\hat P\cdot\sigma)_{\dot\alpha\beta}\xi^\beta=m\,\eta_{\dot\alpha}\;\; ,
\end{equation}
with $m$ some real constant.  Since $\hat P^2$ is a casimir of the Poincar\'e
group, consistency requires the relation\footnote{We could have 
a different eigenvalue.  In that case, the mass of the 
particle we are describing would be $\sqrt{mm'}$.
However, we can always re-scale the fields to obtain the same 
eigenvalue.}  
\begin{equation}
(\hat P\cdot\overline\sigma)
^{\alpha\dot\beta}\eta_{\dot\beta}=m\xi^\alpha\;\; .
\end{equation}
Once again, we combine the two spinor fields into one 
4-component spinor.  Using the same notation, we write the 
equation of motion for massive spinors as 
\begin{equation}
(\not\!\!\hat P-m)\psi=0\;\; .
\end{equation}
For a spinor at rest, this equation
reduces to 
\begin{eqnarray}
\left(\matrix{-m&0&m&0\cr0&-m&0&m\cr m&0&-m&0\cr0&m&0&-m\cr}\right)\psi=0\;\; .
\end{eqnarray}
Once again, we have two linearly independent solutions :
\begin{eqnarray}
u(m,\uparrow)=\sqrt{m}\left(\matrix{1\cr0\cr1\cr0\cr}\right)\;\;\;\;{\rm and}
\;\;\;\;u(m,\downarrow)=\sqrt{m}\left(\matrix{0\cr1\cr0\cr1\cr}\right)\;\; .
\end{eqnarray}
The analogues of (\ref{normmass}), (\ref{nullmass}), 
(\ref{nullmass2}), and (\ref{compmass})
are
\begin{eqnarray}
\label{normmassive}
u^\dag(p,\uparrow)u(p,\uparrow)&=&u^\dag(p,\downarrow)u(p,\downarrow)=2p^0\\
\label{nullmassive}
{\overline u}(p,\uparrow)u(p,\uparrow)&=&{\overline u}
(p,\downarrow)u(p,\downarrow)=2m\\
{\overline u}(p,\uparrow)u(p,\downarrow)&=&{\overline u}
(p,\downarrow)u(p,\uparrow)=0\\
\label{compmassive}
\sum_{s=\updownarrow} u(p,s){\overline u}(p,s)&=&\not\!p+m\;\; .
\end{eqnarray}
Here, we have used the fact that (\ref{normmassive}) is actually the 
time component of a vector and (\ref{nullmassive}) is invariant under
the Lorentz group to write the appropriate expressions on the 
right-hand-sides of these relations.  Substitution of
our expressions into (\ref{compmassive}) actually gives
$m\gamma^0+m1$, where 1 is the identity in Dirac space.
Since the left-hand-side has the $SL(2,C)$ indices of a $\gamma$-matrix
as before, we deduce that the correct extension of $m\gamma^0+m$
to other frames is $\not\!\!p+m$.  This can be checked by transforming the 
solutions to other frames and explicitly calculating the 
completeness sum.  

Under Lorentz transformations, massive spinors transform 
via the same generators as massless spinors.  The spin
vector introduced in Section \ref{unitrepso31} is given by the spatial components
of $\Sigma$.  Application of these matrices reveals that 
$u(m,\uparrow)$ and $u(m,\downarrow)$ represent 
particles with spin-1/2 up and down along the $z$-axis, respectively.
Spinors representing particles with spin along other axes can be written as linear
combinations of these solutions.  

The $\gamma$-matrices have many nice properties which make them
easy to work with.  The anticommutation relation,
\begin{equation}
\left\lbrace\gamma^\mu,\gamma^\nu\right\rbrace=2g^{\mu\nu}\;\; ,
\end{equation}
can easily be proven using (\ref{anticommute}).  
This relation
tells us that to span the space of possible Dirac matrices
we need only consider antisymmetric products. 
Our explicit expression for the antisymmetric 
combination of two $\gamma$-matrices, $\sigma^{\mu\nu}$, 
tells us that this object represents 
six Dirac matrices that are linearly independent
of each other.
Since there are only four gamma matrices, the antisymmetric 
combination of three will have one missing.  Multiplying
by the square of the missing matrix (which is proportional to 
the identity), we see that this combination can be written
$\gamma^\mu\gamma^0\gamma^1\gamma^2\gamma^3$.  
The antisymmetric product of four matrices must include
all of them.  Defining
\begin{equation}
\gamma_5\equiv i\gamma^0\gamma^1\gamma^2\gamma^3
=-{i\over4!}\epsilon^{\mu\nu\alpha\beta}\gamma_\mu\gamma_\nu
\gamma_\alpha\gamma_\beta\;\; ,
\end{equation}
we see that the space of Dirac matrices is spanned by the 
linearly independent set
\begin{eqnarray}
1&&{\rm Scalar}\nonumber\\
\gamma^\mu&&{\rm Vector}\nonumber\\
\sigma^{\mu\nu}&&{\rm Tensor}\\
\gamma^\mu\gamma_5&&{\rm Pseudovector}\nonumber\\
\gamma_5&&{\rm Pseudoscalar}\nonumber\;\; .
\end{eqnarray}
After each matrix, I have noted the Lorentz transformation
properties of the bilinear $\overline\psi\,\Gamma\,\psi$.
The prefix `pseudo' refers to the transformation properties under
parity.  The presence of $\epsilon$ in the definition of
$\gamma_5$ means that it has an extra sign change under parity.
The properties 
\begin{eqnarray}
\left(\gamma^\mu\right)^\dag&=&\gamma^0\gamma^\mu\gamma^0\\
\left(\sigma^{\mu\nu}\right)^\dag&=&\gamma^0\sigma^{\mu\nu}\gamma^0\\
\left(\gamma_5\right)^\dag&=&\gamma_5\\
\left\lbrace\gamma^\mu,\gamma_5\right\rbrace&=&0
\end{eqnarray}
imply that all these bilinears are real except the pseudoscalar,
which is purely imaginary.  More relations between 
the $\gamma$-matrices can be found 
in Appendix \ref{diralg}.  

The matrix $\gamma_5$ is very useful in forming 
spin projection operators.  The fact that $\gamma_5^2=1$
along with its explicit form in the Weyl basis,
\begin{eqnarray}
\gamma_5=\left(\matrix{-{\bf1}&0\cr0&{\bf 1}\cr}\right)\;\; ,
\end{eqnarray}
imply that the operators
\begin{equation}
P_R\equiv{1\over2}\left(1+\gamma_5\right)\qquad\qquad
P_L\equiv{1\over2}\left(1-\gamma_5\right)
\end{equation}
project out the undotted and dotted, or positive and
negative helicity states, respectively.  Furthermore,
it can easily be checked that the bilinear
\begin{equation}
\overline u(p,s)\gamma^\mu\gamma_5u(p,s)=4ms^\mu
\end{equation}
returns the spin vector for massive particles.  
This formula is also valid for massless 
fermions, provided the replacement $ms^\mu\rightarrow
hp^\mu$ is understood.  For massless particles,
the projection operators are extremely useful.
According to (\ref{compmass}), the bilinear
\begin{equation}
u(p,h)\overline u(p,h)={1\over2}\left(1+2h\gamma_5\right)\not\!p\;\; .
\end{equation}
Here, $h$ is the {\it true} helicity of the 
fermion, taking the values $\pm1/2$.  

Up until this point, we have focused on
{\it positive} energy solutions, i.e. $p^0>0$.
However, for reasons which will become clear in 
Appendix \ref{canquant}, we must also consider
the case in which $p^0<0$.  Here, 
we define $k^\mu\equiv-p^\mu$ and write
\begin{equation}
\not\!\!\hat P\,v(k,h)=\not\!p\,v(k,h)=-\not\!k\,v(k,h)=0
\label{masslessanti}
\end{equation}
for massless fermions and 
\begin{equation}
\not\!\!\hat P\,v(k,s)=\not\!p\,v(k,s)=-\not\!k\,v(k,s)=mv(k,s)
\end{equation}
for massive.  From Equation (\ref{masslessanti}),
it is clear that the solutions $v(k,h)$ for massless fermions
correspond exactly to the positive energy solutions :
\begin{equation}
v(\vec p,h)=u(-\vec p,-h)\;\; .
\end{equation}
Calling the objects represented by
$u$ fermions and those represented by $v$
antifermions, this relation implies that 
an antifermion with positive helicity 
moving along the positive $z$-axis is 
equivalent to a fermion with negative helicity
moving along the negative $z$-axis.  

For massive fermions, the correspondence is
not so simple.  In this case, the Dirac equation
becomes
\begin{eqnarray}
\left(\matrix{m&0&m&0\cr0&m&0&m\cr m&0&m&0\cr0&m&0&m\cr}\right)v(k,s)=0
\end{eqnarray}
in the antifermion rest frame.  Once again, we
have two linearly independent solutions
which we can take as
\begin{eqnarray}
v(k,\downarrow)=\sqrt{m}\left(\matrix{1\cr0\cr-1\cr0\cr}\right)
\;\;\;\;{\rm and}\;\;\;\;
v(k,\uparrow)=\sqrt{m}\left(\matrix{0\cr1\cr0\cr-1\cr}\right)\;\; .
\end{eqnarray}
These solutions also refer to a spin-1/2 object,
but the spin-up and spin-down solutions have
changed roles owing to an extra sign in
the Pauli-Ljubanskii vector.  Here, we
have the relations
\begin{eqnarray}
v^\dag(k,\uparrow)v(k,\uparrow)&=&v^\dag(k,\downarrow)v(k,\downarrow)=2k^0\\
{\overline v}(k,\uparrow)v(k,\uparrow)&=&{\overline v}
(k,\downarrow)v(k,\downarrow)=-2m\\
{\overline v}(k,\uparrow)v(k,\downarrow)&=&{\overline v}
(k,\downarrow)v(k,\uparrow)=0\\
{\overline v}(k,s)u(p,s')&=&{\overline u}(p,s)v(k,s')=0\\
\sum_{s=\updownarrow} v(k,s)
{\overline v}(k,s)&=&-(\not\!p+m)=\not\!k-m\;\; .
\label{acompmass}
\end{eqnarray}

Explicit forms can be found for both $u(p,s)$ and $v(p,s)$
by boosting the rest frame solutions via (\ref{Dtransform}).
The result is
\begin{eqnarray}
u(p,s)=\left(\matrix{\sqrt{p\cdot\sigma}\,\xi_s\cr
\sqrt{p\cdot\overline\sigma}\,\xi_s\cr}\right)\;\;\;\;{\rm and}
\;\;\;\;v(k,s)=\left(\matrix{\sqrt{k\cdot\sigma}\,\xi_{-s}\cr
-\sqrt{k\cdot\overline\sigma}\,\xi_{-s}\cr}\right)\;\; ,
\label{diracsoluns}
\end{eqnarray}
where $\xi_s$ is a 2-component spinor aligned with 
the direction $\vec s$,
\begin{equation}
\vec s\cdot\vec\sigma\,\xi_s=\xi_s\;\; .
\end{equation}
These solutions are constructed so that in the rest 
frame $\vec s\cdot\vec W$ is diagonal.
The square-root of a matrix is understood as an instruction
to take the {\it positive} square root of each eigenvalue.

\chapter{$SU(N)$}
\label{sun}

The special unitary group in $N$ dimensions, $SU(N)$, 
has many applications
throughout physics.  For $N=2$, we have a 
group structure which is homomorphic
to the rotation group, $SO(3)$.  Its 
representations house non-relativistic fields 
with intrinsic angular momentum, or spin,
as well as the orbital excitations
of molecules.  Taking 
$N$ as the number of quark flavors
allows us to represent the approximate 
flavor invariance of the QCD lagrangian.  This 
symmetry was the first true indication that
hadrons are composite.
$SU(2n_f)$ describes the approximate
spin-flavor symmetry of a theory with
$n_f$ extremely massive quarks.
Most recently, $SU(2)$ and $SU(3)$
along with the sister group $U(1)$ 
have been useful in describing 
the full symmetry group of the Standard Model.
Although this thesis is most concerned with the 
group $SU(3)$ and its role in strong interaction
physics, we will find it useful to familiarize
ourselves with the general structure of this
family of groups.

\section{General Properties}

In this section, we will 
discuss the definition and 
some of the mathematical properties of $SU(N)$.
Since we are concerned mainly with the fundamental 
representation of $SU(N)$, we
begin with a discussion in this context.  The 
fundamental representation
of $SU(N)$ can be realized by the group of unitary $N\times N$ matrices 
with determinant $+1$.  
Suppose the vector $q^i$, $i=1$ to $N$, transforms 
in the fundamental
representation of $SU(N)$.  Then 
\begin{equation}
q^i\rightarrow (Uq)^i\;\; .
\end{equation}
This implies that the complex conjugate transforms as
\begin{equation}
(q^i)^*\rightarrow ((Uq)^i)^*=(U^*q^*)^i=((U^\dag)^{\rm T} q^*)^i=(q^*U^\dag)^i\;\; .
\end{equation}
Since $U$ is unitary, the product $q^*q$ is invariant under the transformation.
In analogy with the Lorentz group, let us denote $(q^i)^*\equiv q_i$.  
Objects which transform as $U$ are said to be in the $\bf N$ of $SU(N)$,
while their conjugates are in the $\bf\overline N$.  Since 
$q^*q=q_i q^i=q_i q^j\delta^i_j$
is invariant, we have an invariant 
Kr\"onekker delta.  The invariance of this object 
is directly related to the uniterity
of the elements of $SU(N)$, which prompts us to look for an invariant
related to the determinant of $U$.  Just as in the case with the Lorentz group,
the totally antisymmetric tensor $\epsilon^{i_1\cdots i_N}$ does the trick.
These are the only two invariants in the fundamental representation
of $SU(N)$.

Now that we have discussed the objects these transformations
act on, let us parameterize a general element of the group in 
this representation.  Any unitary matrix
can be written 
\begin{equation}
\label{gen}
U=e^{iH}\;\; ,
\end{equation}
where $H$ is a hermitian matrix.  Using the matrix identity
\footnote{
This identity can 
be proven for any matrix which is the exponential
of a complex multiple of a hermitian matrix 
as follows.  Given
\begin{eqnarray}
{\cal M}&=&e^{\alpha H}\;\; ;\nonumber\\
U^\dag HU&=&\tilde H\;\; ,
\end{eqnarray}
with $\tilde H$ diagonal and $U$ unitary,
we see immediately that
\begin{equation}
{\rm det}\left(U^\dag{\cal M}U\right)
={\rm det}\,{\cal M}={\rm det}\,e^{\alpha\tilde H}
=e^{\alpha\,{\rm Tr}\tilde H}=e^{\alpha\,{\rm Tr}H}
=e^{{\rm Tr\,log}{\cal M}}\;\; .
\end{equation}
The validity of this identity rests mainly on two facts :
that hermitian matrices can be diagonalized by a unitary 
transformation and that the relation is trivially true for 
diagonal matrices.}
$\log\,\det U={\rm Tr}\,\log U$,
we find that $\det U=1\leftrightarrow {\rm Tr}\,H=2n\pi$, with $n\in{\bf Z}$.  
Any $N\times N$ matrix can be decomposed into traceless part and a 
a trace :
\begin{equation}
H=\left(H-{1\over N}{\rm Tr}\,H\right)+{1\over N}{\rm Tr}\, H\;\; .
\end{equation}
Hence our unitary matrix $U$ is the product of the exponential of a traceless
matrix and a phase, $e^{2in\pi/N}$.  Since this phase 
can only attain certain discreet
values, one cannot move continuously from one value of $n$ to another.  
Furthermore, only the set with $n=0$ is a subgroup since it alone contains
its inverses and the identity.  I will call this subgroup the proper 
$SU(N)$ transformations and subsequently 
omit the `proper'.  The whole group can be 
considered (when it is absolutely necessary
to do so) as the product of this $SU(N)$ with the discreet $U(1)$ group 
$e^{2in\pi/N}$, $n\in{\bf Z}$.  Some of the consequences of this 
phase to the global properties
of the SM are discussed in \cite{Gaugetheorypaperback}.

We are now left with the task of parameterizing a traceless 
$N\times N$ hermitian matrix.  Such a matrix has $(N^2-N)$ complex off-diagonal
elements constrained by $(N^2-N)/2$ complex hermiticity constraints and
$N$ real diagonal elements constrained by 1 condition on the trace.
In other words, they have $(N^2-1)$ independent elements.  Any set of
$(N^2-1)$ linearly independent traceless hermitian $N\times N$ matrices will
act as a basis for this space.  Since they exhaust the space of $H$,
they generate all of the elements of $SU(N)$ through (\ref{gen}).
The most natural way to choose these generators
is\footnote{Since $q^i\rightarrow (Uq)^i=U^i_j q^j$, our matrices 
must have one index in the fundamental and one in the conjugate representation.}
\begin{eqnarray}
(H)^i_j&=&A^l_k (M^k_l)^i_j\;\; ;\nonumber\\
(M^k_l)^i_j &=& \delta^k_j\delta^i_l-{1\over N}\delta^k_l\delta^i_j\;\; ,
\label{nat}
\end{eqnarray}
since this choice does not introduce any new objects.  All we are using to construct
our generators are the invariants we have from the above discussion.  
Different choices of the parameters $A^l_k$ will lead to different elements
of the group.  

Although there are in principle $N^2$ matrices $M$, they are not all
linearly independent; they are related by $\delta^i_j(M^k_l)^j_i=0$.
This little inconvenience leads to several problems, not the least of
which is that the $A$'s contain more degrees of freedom than required.
Another difficulty is that our generators, as it stands, 
are not hermitian.  Consider the hermitian conjugate
to $(M^k_l)$ \footnote{We use $\sim$ rather than = here because
the indices $i$ and $j$ are in different spaces on either side of the equation.}
\begin{equation}
((M^k_l)^\dag)^i_j\sim(M^k_l)^j_i\;\; .
\end{equation}
This matrix has the same entries (if we use the convention that `upper' indices
label rows while `lower' indices label columns) as $(M^l_k)^i_j$.  Hence
the hermitian matrices $(M^k_l+M^l_k)$ and $(M^k_l-M^l_k)/i$ should
in principle be used as generators rather than $M^k_l$ and $M^l_k$.  
These generators will have coefficients which are strictly real, while
our original `natural' choice will have complex coefficients 
constrained by the condition
$(A^i_j)^*=A^j_i$.  Although this is just the statement
that complex conjugation interchanges the roles of the fundamental and
complex representations, it is not easy to implement in practice.  
Since the sum $M^k_l+M^l_k$ makes sense only for specific values
of $k$ and $l$,\footnote{We cannot have $k$ or $l$ in 
both the fundamental {\it and} the conjugate representation!} 
these indices can no longer be thought of
as truly in {\it any} representation of $SU(N)$.  This fact leads
to many complications if one wishes to consider $N$ large.
However, in practice we seldom need to consider $N>3$ so 
the hermitian choice is often made.  

Let us consider generators which are all hermitian and which 
do not overcount the relevant degrees of 
freedom.  Taking our new generators to be
$t^a$, $a=1$ to $(N^2-1)$, we write
\begin{equation}
U=e^{i\theta^at^a}\;\; .
\end{equation}
The $\theta^a$'s are in a new $(N^2-1)$-dimensional space.  As with
the Lorentz group, these objects transform in the {\it adjoint}
representation of $SU(N)$.  Note that in the adjoint representation
there are no `upper' and `lower' indices since the $\theta$'s are
real.

Since we are working within a group, products of transformations are themselves
transformations.  In particular,\footnote{
I have suppressed the adjoint
indices via $\theta\equiv\theta^at^a$.} 
\begin{eqnarray}
UVU^\dag V^\dag&=&e^{i\lambda\theta}e^{i\lambda\omega}
e^{-i\lambda\theta}e^{-i\lambda\omega}\nonumber\\
&=&1+\lambda^2[\omega,\theta]+{\cal O}(\lambda^3)
\end{eqnarray}
is an element of the group and hence is expressible in terms
of the generators.  This implies for small $\lambda$ that
\begin{equation}
\omega^a\theta^b[t^a,t^b]=i\eta^ct^c
\end{equation}
for some $\eta$.  Differentiating with respect to $\omega$ and $\theta$, we obtain
\begin{equation}
[t^a,t^b]=if^{a\,b\,c}t^c\;\; ,
\label{liea}
\end{equation}
where I have introduced the $f^{a\,b\,c}$'s as the relevant derivatives of
$\eta$.  The $f^{a\,b\,c}$'s are called the 
{\it structure constants} of the group,
and Equation (\ref{liea}) is said to define the {\it Lie algebra} of the group.
The structure constants are obviously antisymmetric in their
first two indices, but we can make them totally antisymmetric by
choosing the $t^a$'s to be orthogonal in the sense that
\footnote{The normalization is not important here.  It need only
be the same for all generators to induce fully antisymmetric structure 
constants.  In some sense, choosing the same normalization for all of our 
generators restores the symmetry we had to 
break to obtain a basis which does not
overcount the relevant degrees of freedom.  Any other choice would have
a favored direction in adjoint space.} 
\begin{equation}
{\rm Tr}\, t^at^b=T_F\delta^{a\,b}\;\; ,
\end{equation}
where $\delta^{a\,b}$ is the Kr\"onekker delta in adjoint space and
$T_F$ is some positive normalization.
With this choice, 
\begin{equation}
f^{a\,b\,c}=-{i\over T_F}{\rm Tr}\,[t^a,t^b]t^c\;\; .
\label{stcon}
\end{equation}
Total antisymmetry follows from this identity and 
the cyclicity of the trace operation.

Since the Jacobi identity
\begin{equation}
[t^a,[t^b,t^c]]+[t^b,[t^c,t^a]]+[t^c,[t^a,t^b]]=0
\end{equation}
is trivially true, we have a consistency relation for the $f^{a\,b\,c}$'s
\begin{equation}
f^{a\,b\,e}f^{c\,d\,e}+f^{b\,c\,e}f^{a\,d\,e}+f^{c\,a\,e}f^{b\,d\,e}=0\;\; .
\label{jac}
\end{equation}

Let us calculate the transformation law in the adjoint representation.  Since
the $\theta^a$'s and the $A^i_j$'s play the same 
role, under an $SU(N)$ transformation
$\theta\rightarrow U\theta U^\dag$.  For infinitesimal
transformations, we have
\begin{equation}
\theta\rightarrow\theta+i[\omega,\theta]\;\; ,
\end{equation}
or
\begin{equation}
\theta^a\rightarrow\theta^a-f^{a\,b\,c}\omega^b\theta^c\;\; .
\end{equation}
Since this is an $SU(N)$ transformation, it can be written 
$\theta^a\rightarrow U^{a\,b}\theta^b$, with $U$ given by $e^{i\omega^aT^a_A}$,
exactly in the same way as before.  The only difference is that now the 
generators are $(N^2-1)\times(N^2-1)$ dimensional matrices.  This
implies the identification
\begin{equation}
(T^a_A)^{b\,c}=-if^{a\,b\,c}\;\; ,
\end{equation}
from which standpoint the Jacobi identity, (\ref{jac}),
is simply a re-expression of (\ref{liea}) in the adjoint representation.

Since the Kr\"onekker delta and the structure constants seem to be properties of
the group itself rather than a specific transformation, we expect them
to be invariant under transformations.  We can check this by explicit
calculation :
\begin{eqnarray}
\delta^{a\,b}&\rightarrow&\delta^{a\,b}+f^{a\,c\,d}\theta^d\delta^{c\,b}
+f^{b\,c\,d}\theta^d\delta^{a\,c}\nonumber\\
&=&\delta^{a\,b}\nonumber\\
f^{a\,b\,c}&\rightarrow&f^{a\,b\,c}-(f^{b\,c\,e}f^{a\,d\,e}
+f^{c\,a\,e}f^{b\,d\,e}+f^{a\,b\,e}f^{c\,d\,e})\theta^d\\
&=&f^{a\,b\,c}\;\; ,
\end{eqnarray}
where the last equality follows from the Jacobi identity, (\ref{jac}).
In the adjoint representation the generators are the
structure constants, implying that the generators are invariant 
under the action of the group if we rotate {\it all} of their indices.  This
is obvious with the natural choice for the generators, (\ref{nat}), since 
there they are expressed solely in terms of the invariant $\delta^i_j$.   

There is one other invariant in the adjoint representation
which will be useful to us.  This object 
appears because of the fact that the anticommutator of two
generators is a hermitian $(N\times N)$ matrix and so can be represented
as a linear combination of the generators themselves and 
the identity.  We write
\begin{equation}
\left\lbrace t^a,t^b\right\rbrace=\alpha^{a\,b}+d^{\,a\,b\,c}t^c\;\; .
\end{equation}
Taking the trace of both sides, we find that $\alpha^{a\,b}=2T_F/N\,\delta^{a\,b}$.
The $d^{\,a\,b\,c}$'s are, with our choice of normalization, fully
symmetric since
\begin{equation}
d^{\,a\,b\,c}={1\over T_F}{\rm Tr}\,\left\lbrace t^a,t^b\right\rbrace t^c\;\; .
\end{equation}
These structure constants transform according to
\begin{equation}
d^{\,a\,b\,c}\rightarrow d^{\,a\,b\,c}-(f^{a\,d\,e}d^{\,b\,c\,e}+f^{b\,d\,e}d^{\,c\,a\,e}
+f^{c\,d\,e}d^{\,a\,b\,e})\theta^d\;\; .
\end{equation}
An identity similar to Jacobi's,
\begin{equation}
[t^a,\lbrace t^b,t^c\rbrace]+[t^c,\lbrace t^a,t^b\rbrace]
+[t^b,\lbrace t^c,t^a\rbrace]=0\;\; ,
\end{equation}
can be employed to show that the quantity in parentheses is in fact zero.
These definitions allow us to write 
\begin{equation}
t^at^b={1\over 2}\left([t^a,t^b]+\lbrace t^a,t^b\rbrace\right)={1\over 2}
\left({2T_F\over N}\delta^{a\,b}+d^{\,a\,b\,c}t^c+if^{a\,b\,c}t^c\right)\;\; ,
\end{equation}
a very useful result.

Since we have chosen the same normalization for all of our generators, there
is no preferred direction in adjoint space.  Any quantity which does not
depend on an external adjoint quantity must be expressed in terms of 
invariants, in exactly the same way that Lorentz covariant quantities which
do not depend on external vectors must be expressed in terms of
the metric and the Levi-Cevita tensor.  This is a {\it representation independent}
result since it has only to do with the way we have normalized our generators.  Since
all other representations can be built from the fundamental representation,
normalizing the $t^a$ induces a normalization on $T^a_R$ for any representation $R$. 
This normalization is given by 
\begin{equation}
{\rm Tr}_R\, T^a_RT^b_R=T_R\delta^{a\,b}\;\; ,
\end{equation}
where we have used the fact that the left-hand-side is independent of any
external adjoints and hence must be expressed in 
terms of the invariant $\delta$.  

The value $T_R$ depends only on the representation
the trace is performed in.  
For example, let us calculate
$T_A$\footnote{Confusion between the generators 
$T_R^a$ and their normalization $T_R$ should be rare.}
\begin{equation}
{\rm Tr}_A\, T^a_AT^b_A= f^{a\,c\,d}f^{b\,c\,d}=T_A\delta^{a\,b}\;\; .
\end{equation}
The sum on the $f$'s can be expressed in terms of the $t$'s using (\ref{stcon}) :
\begin{equation}
f^{a\,c\,d}f^{b\,c\,d}=-{1\over T_F^2}{\rm Tr}
[t^a,t^c]t^d\;{\rm Tr}[t^b,t^c]t^d\;\; .
\end{equation}
$(t^a)^i_j(t^a)^k_l$ is an object with two fundamental
indices and two conjugate indices
which does not depend on an external vector,
hence it can be expressed in terms of the 
available invariants, $\delta^i_j\delta^k_l$
and $\delta^i_l\delta^k_j$.  Since both matrices are traceless,
we write
\begin{equation}
(t^a)^i_j(t^a)^k_l\propto \delta^i_l\delta^k_j-{1\over N}\delta^i_j\delta^k_l\;\; .
\end{equation}
Contracting with $\delta^j_k\delta^l_i$ gives ${\rm Tr}
\;t^at^a=T_F\delta^{a\,b}\delta^{a\,b}
=(N^2-1)T_F$, so
\begin{equation}
(t^a)^i_j(t^a)^k_l=T_F\left(\delta^i_l\delta^k_j-{1\over N}
\delta^i_j\delta^k_l\right)\;\; .
\end{equation}
This useful relation is known as a {\it Fiertz}\,\footnote{pronounced
almost like {\it fierce}} identity.  Using it we can expand the trace
and do the sum, obtaining
\begin{equation}
T_A=2NT_F\;\; .
\end{equation}

The operator $(T_R^aT_R^a)_{\alpha\beta}$ 
has two indices in the representation $R$ and no adjoint
indices.  Hence it must be proportional to the identity in $R$,
i.e.\footnote{The fact that I have
written both indices downstairs here has no meaning.  What is required is 
the same index structure as the generators have in the representation $R$.}
\begin{equation}
(T_R^aT_R^a)_{\alpha\beta}=C_R\delta_{\alpha\beta}\;\; .
\end{equation}
It is obvious from the Fiertz identity that $C_F=T_F(N^2-1)/N$ and 
$C_A=T_A$.  As with the operators $\hat P^2$ and $\hat W^2$ in $SO(3,1)$,
this operator is a casimir for $SU(N)$.
From here, one can calculate many group structures.  
I list a few useful relations 
in Appendix \ref{sunforms}
for reference.  

Explicit forms for the structure constants can be obtained
for general $SU(N)$ if one uses the `natural' choice for the 
generators, $M^i_j$.  It is straightforward to 
check that
\begin{equation}
(M^i_j)^k_p(M^n_m)^p_l=
\delta^k_j\delta^i_m\delta^n_l-
{1\over N}\delta^i_j\delta^k_m\delta^n_l-
{1\over N}\delta^k_j\delta^n_m\delta^i_l+
{1\over N^2}\delta^i_j\delta^n_m\delta^k_l\;\; .
\label{prodofnat}
\end{equation}
The normalization of these generators is
obtained via
\begin{equation}
(M^i_j)^k_p(M^n_m)^p_l\delta^l_k=
(M^i_j)^n_m\;\; .
\label{TFnat}
\end{equation}
The interpretation of this relation is 
obvious once one considers the symmetry
of the situation.  Above, we normalized our
generators according to ${\rm Tr}\,t^at^b=T_F\delta^{ab}$.
Here, it is not yet clear what form $\delta^{ab}$
will take.  In the most fundamental sense, 
$\delta^{ab}$ is simply an invariant tensor
with two indices in the adjoint representation.
By inspection, one can see that in this
representation for the generators adjoint indices
consist of traceless pairs of fundamental 
and anti-fundamental
indices.  The {\it only} such invariant
object at our disposal is $(M^i_j)^n_m$ itself.  Indeed,
these two conditions were the reason this form was chosen 
for $M$ in the first place!  Hence we 
are led to the identification
\begin{equation}
\delta^{ab}\rightarrow\left(M^i_j\right)^n_m\;\; .
\end{equation}
Proper normalization is obtained 
by the requirement that
\begin{equation}
N^2-1=(M^i_j)^n_m(M^j_i)^m_n\;\; ,
\end{equation}
which can be checked explicitly.
With this in mind, (\ref{TFnat}) implies that
are generators are normalized according to
$T_F=1$.  

From here, we are free to calculate any structures
we wish.  For example,
\begin{equation}
C_F\delta^i_j=\left(M^k_l\right)^i_p
\left(M^n_m\right)^p_j\left(M^l_k\right)^m_n
=\left(N-{1\over N}\right)\delta^i_j\;\; .
\end{equation}
Furthermore, the explicit form of the 
product (\ref{prodofnat}) allows a direct
calculation of the antisymmetric
\begin{eqnarray}
\label{commfdefnat}
\left\lbrack M^i_l,M^j_m\right\rbrack&=&if^{ijk}_{lmn}\,M^n_k\;\; ;\nonumber\\
f^{ijk}_{lmn}&=&-i\left(\delta^i_m\delta^j_n\delta^k_l
-\delta^k_m\delta^j_l\delta^i_n\right)
\end{eqnarray}
and symmetric
\begin{eqnarray}
\label{anticommddefnat}
\left\lbrace M^i_l,M^j_m\right\rbrace&=&{2\over N}
\left(M^i_l\right)^j_m+d^{\,ijk}_{\,lmn}\,M^n_k\;\; ;\nonumber\\
d^{ijk}_{lmn}&=&\delta^i_m\delta^j_n\delta^k_l
+\delta^k_m\delta^j_l\delta^i_n\nonumber\\
&&-{2\over N}\left(
\delta^i_l\delta^j_n\delta^k_m+
\delta^j_m\delta^k_l\delta^i_n+
\delta^k_n\delta^i_m\delta^j_l\right)\\
&&+{4\over N^2}\delta^i_l\delta^j_m\delta^k_n\;\; .\nonumber
\end{eqnarray}
structure constants.  Note that these 
constants have the stated symmetry properties only
under the exchange of {\it pairs} of 
indices.  Note also that $d^{ijk}_{lmn}$ was
{\it chosen} to have this symmetry.  Two of its
terms cannot contribute to (\ref{anticommddefnat}).
Armed with these expressions for general $N$, 
we can directly verify any of the
expressions in Appendix \ref{sunforms}.  Proper normalization
to any value of $T_F$ is trivially performed 
by rescaling the generators.\footnote{As far
as normalization is concerned, $f$ and $d$ count
as generators.}

As a last topic for this section, let us consider 
the problem of finding irreducible representations
of $SU(N)$.  Suppose we wish to consider
$(\bf N\times N)$, 
\begin{equation}
T^{ij}\;\; .
\end{equation}
Since there are no invariants with two fundamental 
indices, the decomposition of this tensor
is trivial.  We simply symmetrize and antisymmetrize
its indices :
\begin{equation}
T^{ij}={1\over2}\left(T^{ij}+T^{ji}\right)
+{1\over2}\left(T^{ij}-T^{ji}\right)\;\; .
\label{twoindex}
\end{equation}
The symmetric representation has $N(N-1)/2+N=N(N+1)/2$
components while the antisymmetric has $N(N-1)/2$.
In this way, we arrive at
\begin{equation}
{\bf N\times N}={\bf {N(N+1)\over 2}+{N(N-1)\over 2}}\;\; .
\end{equation}
Note that we don't need to subtract a trace here;
this concept has no meaning in our space.
Larger numbers of indices in one
representation are handled in an 
analogous way.  One simply adds indices
successively to the decomposition in
(\ref{twoindex}), symmetrizing and
antisymmetrizing as one goes.
For example, $(\bf N\times N\times N)$ 
is decomposed as
\begin{eqnarray}
T^{ijk}&\sim&T^{(ij)k}+T^{[ij]k}\nonumber\\
&\sim&T^{(ijk)}+T^{(i[j)k]}+T^{[i(j]k)}+T^{[ijk]}\;\; .
\end{eqnarray}
For $SU(3)$, this decomposition corresponds to
\begin{equation}
{\bf3\times3\times3}={\bf10+8+8+1}\;\; .
\end{equation}
This result is relevant to the study of 
baryons both in $SU(3)_f$, where the $\bf 8$ corresponds
to the octet of spin-1/2 and the $\bf10$ corresponds
to the decouplet of spin-3/2, and 
the color $SU(3)$ gauge group, where
the $\bf 1$ is the colorless singlet required
for a physical state.  In $SU(3)$, the 
totally antisymmetric combination $T^{[ijk]}$
is obviously invariant since it can be written
in terms of $\epsilon^{ijk}$.
This correspondence allows us to clearly
see the fact that 
\begin{equation}
T^{[i_1\cdots i_{N-1}]}=\epsilon^{i_1\cdots i_N}\tilde T_{i_N}
\end{equation}
transforms as an $\bf\overline N$ 
rather than an $\bf N$.

The decomposition of
$\bf N\times\overline N$ is quite different.
Here, we have the invariant $\delta^i_j$ which
leads to a trace subtraction.
In this case, it is the concept of 
symmetrization and antisymmetrization which 
does not make sense.  This leads to the 
decomposition
\begin{equation}
T^i_j=\left(T^i_j-{1\over N}\delta^i_j\,T^k_k\right)
+{1\over N}\delta^i_j\,T^k_k\;\; .
\end{equation}
The trace is invariant and forms the 
representation {\bf 1} of $SU(N)$, while the 
remainder transforms in the {\it adjoint}
and forms the $({\bf N^2-1})$.  Hence
\begin{equation}
{\bf N\times\overline N}=\left({\bf N^2-1}\right){\bf+1}\;\; .
\end{equation}
This decomposition is relevant to meson physics,
where the $SU(3)_f$ group tells us we will have
an octet (the ${\bf 8}$) and a singlet (the ${\bf 1}$)
and the $SU(3)$ color group tells us that the 
combination of a quark and an antiquark can 
produce a physical color singlet state.

\section{Local $SU(N)$}

As detailed in Chapter \ref{basics}, QCD is a quantum
field theory based on a local $SU(3)$ symmetry.  One 
considers several flavors of quark triplets in the {\bf 3}
and antiquark triplets in the $\overline{\bf 3}$, and requires
the lagrangian to be invariant under local $SU(3)$ transformations.
This leads to the introduction of the {\it covariant} derivative,
$\cal D$.  If we just compared the field at two different points
(using the ordinary derivative), differences in convention 
between the two points would lead to physically unimportant
contributions.  The covariant derivative takes all of these
changes in convention into account and asks for the difference 
between {\it physical} configurations.  However, it can only do this 
for two infinitesimally close points.  

Suppose I want to compare
field values at two points which are a finite distance apart.  
Since these objects are defined with different conventions,
we must first change the convention of one of them to
match the other convention.  This operation is
called {\it parallel transport}.  The changes in convention
occur locally, so we will need to know the convention at
every point on a path between the two points 
to perform this transformation.  The covariant derivative
tells us that local changes in convention are accounted
for by the gauge field $\cal A$.  Hence we expect 
parallel transport to involve some sort
of nonlocal combination of gauge fields.  
The infinite number of paths connecting two points in
4-dimensional spacetime make it necessary for us to 
choose a curve, $C$, to define a unique parallel transport operator.
Defining $C$ by points $z(\tau)$ such that $z(0)=y$ and 
$z(1)=x$, it can be shown that the operator
\begin{equation}
{\cal G}_C(x,y)={\rm P}e^{-ig\int_0^1 A_\mu\left(z(\tau)
\right)(dz^\mu/d\tau)\;d\tau}
\label{gaugelinkdef}
\end{equation}
satisfies the requirements for parallel transport from
$y$ to $x$ along $C$.  The $\rm P$-symbol orders the $\cal A$ fields such that
those corresponding to larger $\tau$ come first, i.e.\footnote{The 
$1/n!$ in the expansion of an exponential appears naturally 
here if the path-ordering is trivial since 
$\int_0^1 d\tau\int_0^\tau d\tau' f(\tau)f(\tau')=1/2
\left(\int_0^1 d\tau f(\tau)\right)^2$
if $f(\tau)$ commutes with itself for different values of $\tau$.}
\begin{eqnarray}
{\rm P}\left\lbrace e^{\int_0^1 M(\tau)d\tau}\right\rbrace&\equiv&
1+\int_0^1 d\tau M(\tau)+\int_0^1 d\tau M(\tau)\int_0^\tau d\tau' M(\tau')
\nonumber\\
&&\;\;\;+\int_0^1 d\tau M(\tau)\int_0^\tau 
d\tau' M(\tau')\int_0^{\tau'} d\tau'' M(\tau'')+\cdots\;\; .
\end{eqnarray}

It may be checked explicitly for infinitesimal gauge 
transformations that 
\begin{equation}
{\cal G}_C(x,y)\rightarrow U(x){\cal G}_C(x,y)U^\dag(y)
\end{equation}
for any choice of the path $C$.
Since all gauge transformations we consider can be obtained 
from repeated application of infinitesimal ones, 
this is all we need.\footnote{As stated before, other transformations can 
be accounted for by rotations in the other sectors of the $SM$.
Topologically nontrivial gauge configurations are associated 
with behavior at infinity and do not concern us here.}
Qualitatively, we can think of (\ref{gaugelinkdef}) as 
iteratively taking changes in convention into account.
The first term corresponds to no change in convention,
the second takes a change in convention into account
at one point along $C$ and sums over all possible 
places the convention could be changed, the third
changes conventions twice and sums over the positions of 
those changes, etc.  The path ordering assures us that 
changes in convention are made in the correct order 
as we traverse $C$.  

The comparison of field values at two different points
in spacetime may now be written
\begin{equation}
\left(\Delta_C\psi\right)(x,y)=\psi(x)-{\cal G}_C(x,y)\psi(y)\;\; .
\end{equation}
From the transformation properties of $\cal G$,
it is obvious that $(\Delta\psi)(x,y)$ transforms like $\psi(x)$.
Of course, we may define this difference at any point in 
spacetime by appropriate use of $\cal G$ :
\begin{equation}
\left(\Delta_{C,C'}\psi\right)(x,y)\left|_z\right.={\cal G}_{C'}(z,x)\psi(x)-{\cal
G}_{C+C'}(z,y)\psi(y)\;\; ,
\end{equation}
where $C'$ is some path connecting $x$ and $z$.  The 
parallel transport operator $\cal G$ is commonly called a 
{\it gauge link} and is very useful in defining
nonlocal distributions, as in Section \ref{partondistdis}.

Different paths used to define ${\cal G}_C$ are {\it not}
equivalent.  The difference between them 
can be characterized by the 
{\it Wilson loop},
\begin{equation}
{\cal U}_{\ell}(x)\equiv{\cal G}_{C'}(x,y){\cal G}_{C}(y,x)\;\; .
\end{equation}
The loop $\ell$ forms a closed curve containing the points
$x$ and $y$ and defined by the paths $C$ and $C'$. 
Expanding this object about $g=0$, we see that its terms
arrange themselves into path-ordered loop
integrals :
\begin{eqnarray}
&&{\cal U}_{\ell}(x)\sim 1-ig\oint_{\ell}{\cal A}(z)\cdot dz\nonumber\\
&&\qquad-ig^2f^{abc}t^c\int_1^2{\cal A}^a(z(\tau))\cdot(dz/d\tau)d\tau\int_0^1
{\cal A}^b(z(\tau'))\cdot(dz/d\tau')d\tau'\nonumber\\
&&\qquad\qquad-g^2\int_0^2{\cal A}(z(\tau))\cdot (dz/d\tau)d\tau
\int_0^\tau{\cal A}(z(\tau'))\cdot (dz/d\tau')d\tau'\,+{\cal O}(g^3)\;\; .
\label{primstok}
\end{eqnarray}
Here, $z^\mu(\tau)$ traces out $C$ for $\tau\in[0,1)$
and $C'$ for $\tau\in(1,2]$.

In abelian theories, one can use Stokes' theorem to convert these
integrals into surface integrals over the 
area $\ell$ bounds.  The same spirit is
present in non-abelian theories, although the 
difficulties associated with noncommutativity
make it less transparent.  To see this, we consider
an infinitesimal loop.  Taking $y=x+dx^\mu+dx^\nu$, 
where $\mu$ and $\nu$ are for the moment {\it fixed}
directions, we can evaluate the 
difference in choice between a path that goes first 
along $dx^\mu$ then along $dx^\nu$ and a path that
does the opposite by approximating the integrals :
\begin{eqnarray}
\oint_\ell {\cal A}(z)\cdot dz
&\sim&\phantom{-}{\cal A}^\mu(x)dx_\mu
+{\cal A}^\nu(x+dx^\mu)dx_\nu\nonumber\\
&&-{\cal A}^\nu(x)dx_\nu-{\cal A}^\mu(x+dx^\nu)dx_\mu\\
&=&\left(\partial^\mu{\cal A}^\nu-
\partial^\nu{\cal A}^\mu\right)dx_\mu dx_\nu\;\; ,
\end{eqnarray}
where I have suspended the Einstein summation convention
for the moment.  Similarly, the 
non-abelian commutator term in (\ref{primstok})
is approximated as
\begin{eqnarray}
&&-ig^2f^{abc}t^c\int_1^2{\cal A}^a(z(\tau))
\cdot(dz/d\tau)d\tau\int_0^1
{\cal A}^b(z(\tau'))\cdot(dz/d\tau')d\tau'\qquad\qquad\qquad\qquad
\qquad\nonumber\\
&&\qquad\qquad\qquad\qquad\qquad
\qquad\qquad\qquad\qquad\sim ig^2f^{abc}t^c{\cal A}_a^\nu(x)
{\cal A}^\mu_b(x)\,dx^\nu dx^\mu\;\; .
\end{eqnarray}
Combining these two contributions,
the integral is expressed
\begin{equation}
-ig{\cal F}^{\mu\nu}(x)\,dx_\mu dx_\nu\;\; .
\end{equation}
The antisymmetric combination of the 
differentials gives an infinitesimal area
element - the area of the bounded region.
This statement is valid for any choices
of $\mu$ and $\nu$, allowing us to 
apply it to the discretization of {\it any}
spacetime region bounded by 
curves $C$ and $C'$.  
Furthermore,
this analysis can be extended to higher orders
in the coupling.

The fact that $\cal F$ itself
is not gauge invariant makes it
difficult to compare infinitesimal
contributions from different regions.
However, the gauge-invariant 
{\it trace} of $\cal U$ admits
a full generalization
of Stokes' theorem :
\begin{equation}
{\rm Tr}\;{\cal U}_\ell={\rm Tr}\,e^{-ig\int_S{\cal 
F}^{\mu\nu}\,d\sigma_{\mu\nu}}\;\; ,
\end{equation}
where $S$ is any surface bounded by $\ell$ and
$d\sigma^{\mu\nu}$ is the four-dimensional 
area element.
This object can be used as a {\it gauge-invariant}
measure of the strength of the gauge field 
in a region.  It is also one of the fundamental building
blocks of the discretized theory of lattice QCD.

\section{The Limit of Large $N_c$}
\label{largenc}

Some very useful results have come from considering
a local gauge theory based on the group $SU(N_c)$
in the limit of large $N_c$.  In this section, I will 
briefly overview some of the main results and 
simplifying features of this limit.  Some useful 
references on this subject are \cite{largennn}.

The lagrangian of large-$N_c$ QCD is identical to
that of normal QCD, except that here there 
are $N_c$ quark colors and ($N_c^2-1$) gluons.
As we have seen above, adjoint objects such as
gluons can be represented as the traceless
combination of fundamental and an anti-fundamental
indices.  Since we intend to take
$N_c$ to infinity, the fact that the
trace degree of freedom should be 
subtracted out is of no great importance.  
Hence, with corrections of order $1/N_c^2$,
we can consider a gluon as a quark-antiquark
combination.\footnote{This is to be
understood as taking place exclusively in the 
$SU(N_c)$ sector of the theory.}  The replacement 
of all gluon propagators with double quark propagators, 
with fermion number flow in opposite directions,
makes it easy to count powers of 
$N_c$.  Each unconstrained loop gives a factor of 
$N_c$ when its colors are summed over.
For example, Figure C.1a comes with the 
factor $N_c^2$ because there are two 
unconstrained `fermion' loops.  
Unless we
can regulate this behavior in some way,
it quickly becomes obvious that each higher
order of perturbation theory will contain
diagrams with more powers of $N_c$.  This is
certainly not the hallmark of a theory 
with a well-defined limit as $N_c\rightarrow\infty$.

\begin{figure}
\epsfig{figure=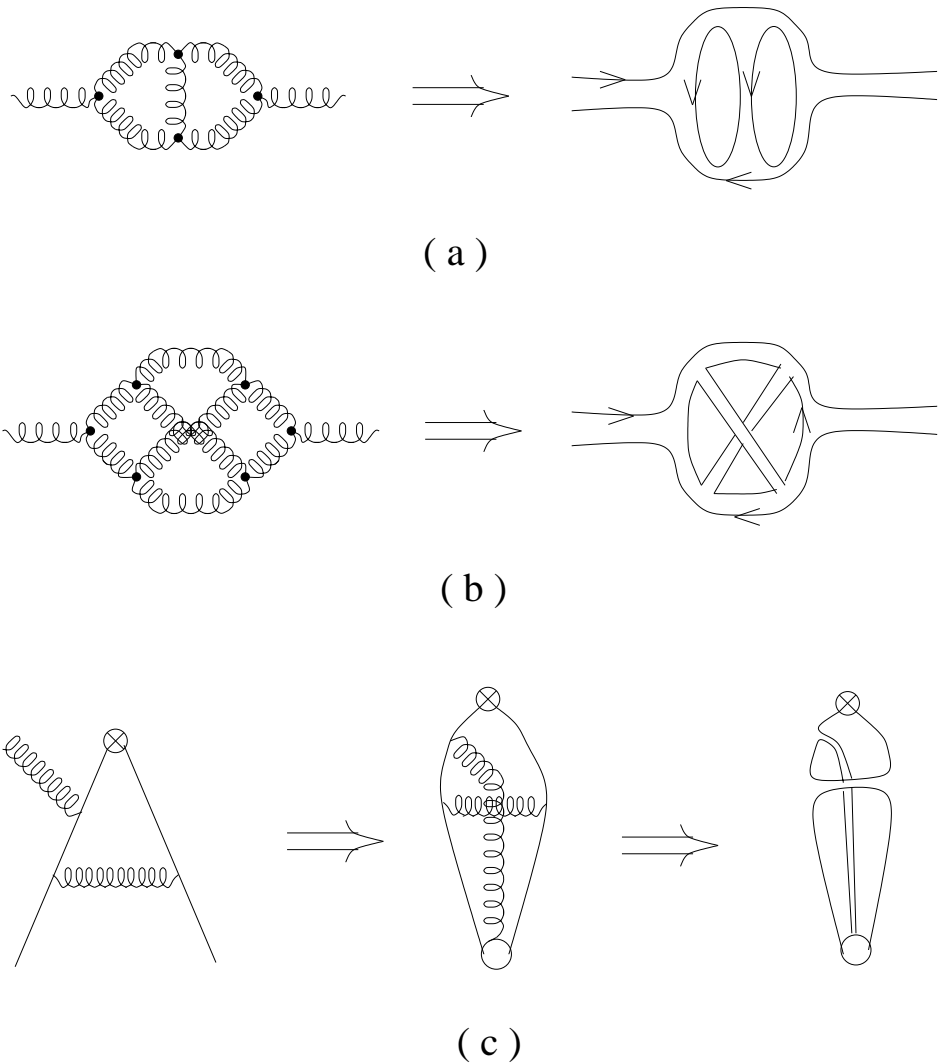,height=5.5in}
\caption{An illustration of the large-$N_c$ 
counting rules.  (a) is enhanced by a power of 
$N_c^2$ because of the unconstrained color loops 
within the bubble.  Since it is a two-loop
graph, it will contribute to the leading 
large-$N_c$ behavior.  On the other hand, (b)
contains only one unconstrained color loop. 
As a three-loop diagram, its contribution 
comes only at ${\cal O}(1/N_c^2)$.  Notice that
(a) is a planar diagram, while (b) must
contain a self-intersection when forced into 
a plane.
As it stands, the 
counting rules for more complicated
correlation functions, like that represented in (c), are 
ambiguous.  At first, we may think that it is planar.  However,
drawing all external lines {\it directly} to the 
open dot (representing infinity), reveals its true nature.}
\label{fig21}
\end{figure}

By inspection, one can see that factors of 
$N_c$ are {\it always} accompanied with factors of
$g_{N_c}^2$, where $g_{N_c}$ is the coupling constant
of our theory.  Since this is {\it our} theory,
we can force $g$ to scale with $N_c$ in any way 
we want.  Comparison with the real 
world requires only that $g_3$ has the value 
measured in experiment.  With this in mind,
we see that our problem can be solved 
by taking
\begin{equation}
g_{N_c}\equiv {g\over\sqrt{N_c}}\;\; .
\end{equation}
This definition causes all of the {\it leading}
powers of $N_c$ at each order in perturbation theory
to contribute at the {\it same} power of 
$N_c$ to the final amplitude under consideration.

Now that we have a well-defined limit, it behooves
us to find out which diagrams contribute in this
limit and which do not.  Since each power
of the coupling contributes a $1/\sqrt{N_c}$
to the amplitude, leading diagrams have
the property that each coupling generates
`half' of an unconstrained color loop.
Real quarks have only one color line, so
quark loops cannot give rise to factors of $N_c$
even though they {\it are} associated with factors
of $g_{N_c}^2$.  This implies that {\it all}
diagrams that involve quark loops are
suppressed by $N_c$.  This is easy to understand
because there are simply more gluons around than
quarks.  Gluons\footnote{In covariant
gauges, one must also include ghosts.  Since 
they appear as adjoints along with the
gluons, their contributions can also
give leading effects.} will not always lead to the 
maximal power of $N_c$ either.  By inspection, it
can easily be seen that diagrams 
like that in Figure C.1b in which 
the color loops are entangled are subleading.
This can be extended to a rule - all
diagrams which possess self-intersecting 
color lines when embedded in a plane 
are subleading in the large-$N_c$ limit.
Since these diagrams cannot be topologically
embedded in a plane, they are referred
to as `nonplanar'.  This rule works
for loops, but must be amended to include
correlation functions with several external
particles.  Here, there exist diagrams
which can be drawn such that they appear
either planar or nonplanar depending
on preference.  However, when all external
lines are drawn {\it directly} to the {\it same} 
point at infinity, only the planar diagrams
will contribute to the leading large-$N_c$ behavior.
This rule is illustrated in Figure C.1c.

Beyond our rule for leading diagrammatic
contributions, not many results
of large-$N_c$ physics affect the topics
presented here.  However, in hadronic physics
this limit represents an extremely useful
simplification.  Among other things, it 
constrains meson-baryon couplings via uniterity 
relations, allows one to justify certain 
hadronic models, like the Skyrme model, 
and makes it easier to relate the macroscopic
world of hadrons to the microscopic one of QCD.
A new symmetry between the light quark
flavor and spin 
becomes manifest in this limit.  This 
`spin-flavor' symmetry arises because
baryons composed of $N_c\rightarrow\infty$ quarks
do not care if you flip the spin of one or
two.  This leads to a symmetry between
the low-lying baryonic spin excitations \cite{Manohar}.  
In particular, it relates the 
$\Delta$ resonance to the nucleon.
Corrections to this result can 
also be systematically calculated, a 
situation hard to come by in hadronic physics.
Furthermore, an even larger symmetry 
group encompasses the baryons if
we simultaneously take some of the 
quark masses to infinity \cite{oktom}.

\chapter{Quantum Mechanics}
\label{qmapp}

The fundamental postulate of quantum mechanics,
\begin{equation}
[x,p]=i\hbar\;\; ,
\label{fundpost}
\end{equation}
is interpreted in quantum field theory as a 
statement about the commutator of a field operator with its
conjugate momentum.  The process known in the sixties as
`second quantization' involves promoting the wavefunctions
of quantum mechanics to operators and imposing the
`canonical commutator' (\ref{fundpost}) on these 
wavefunctions and their conjugate momenta.
The purpose of this appendix is to quickly review
this process and obtain explicit expressions 
for the generators of the Poincar\'e group 
in a quantum field theory.

The analysis of Section D.2 closely follows that in 
S. Weinberg \cite{weinbergbook}.

\section{Canonical Quantization}
\label{canquant}

In ordinary quantum mechanics, the Schr\"odinger
equation,
\begin{equation}
i\hbar{\partial\over\partial t}\phi(\,\vec x\,,t)
=-{\hbar^2\over 2m}\,\nabla^2\phi(\,\vec x\,,t)+V(\,\vec x\,)\phi(\,\vec x\,,t)
\label{schrodingeeq}
\end{equation}
is just the statement that 
\begin{equation}
E={\vec p\,^2\over2m}+V(\,\vec x\,)\;\; ,
\label{classical}
\end{equation}
along with the added understanding that
the quantities $E$ and $\vec p$ are 
operators on the function
space that satisfy the commutation relations
\begin{equation}
\left\lbrack x^\mu,p^\nu\right\rbrack = -i\hbar g^{\mu\nu}\;\; .
\label{commofxt}
\end{equation}
Equation (\ref{schrodingeeq}) follows from the solution
\begin{equation}
p^\mu=i\partial^\mu
\label{momopdef}
\end{equation}
of (\ref{commofxt}).  

Unfortunately, a direct application of (\ref{momopdef})
to the relativistic version of (\ref{classical}),
\begin{equation}
p^2=m^2\;\; ,
\label{relativistic}
\end{equation}
leads to a Hamiltonian whose spectrum is not
bounded from below.  Hence our theory does not possess 
any stable configurations.  The solution to this
dilemma has to do with interpretation.  In non-relativistic
quantum mechanics, the dynamical variables are $\vec x$
and $t$.  One requires $\phi$ to be an eigenfunction
of the energy operator, which depends on the position $\vec x$.
In quantum field theory, the dynamical 
variables are the fields $\phi$ themselves.
Since $t$ is no longer a dynamical variable,
we no longer have any need to bound its 
Fourier conjugate from below.  
This is the fundamental distinction between 
non-relativistic quantum mechanics and 
quantum field theory. 

The dynamical variables of {\it any} quantum
system satisfy canonical commutation relations
with their conjugate momenta.  In order to 
define this statement, we must introduce a lagrangian
for our system.  A lagrangian is dictated by the 
equations of motion it generates, so the first question
we must ask ourselves is what kind of equation of motion
we would like our fields to satisfy.  At some point, 
we intend to consider freely propagating massive 
particles, so a logical choice for the equation of 
motion is (\ref{relativistic}),
\begin{equation}
\left(-\partial^2-m^2\right)\phi=0\;\; .
\label{KGeq}
\end{equation}
This equation can be seen to follow from the 
classical lagrangian density\footnote{The method of
going from lagrangian densities to equations of 
motion is discussed in the next section.}
\begin{equation}
{\cal L}={1\over2}\partial^\mu\phi\partial_\mu\phi-{1\over2}m^2\phi^2\;\; ,
\end{equation}
from which we obtain the conjugate momentum
\begin{equation}
\pi={\delta{\cal L}\over\delta\partial_0\phi}=\partial^0\phi\;\; .
\end{equation}
The canonical commutation relation\footnote{
$\hbar$ was reinstated at the beginning of the Appendix for
nostalgic purposes; from now on, it will be given the value 1.}
\begin{equation}
\left\lbrack\phi(x),\pi(y)\right\rbrack
\Big|_{x^0=y^0}=i\delta(\,\vec x-\,\vec y\,)
\label{KGcanoncomm}
\end{equation}
can be solved by expanding $\phi(x)$ in a complete
basis of solutions to (\ref{KGeq}) with operator coefficients :\footnote{
The hermiticity of $\phi$ relates the two operator coefficients,
as shown.  The factor of $\sqrt{2p^0}$ is for later
convenience.  It belongs to the operators $a_{\,\vec p}$
and $a_{\,\vec p}^\dag$, as the combination is invariant.}
\begin{equation}
\phi(x)=\int{d^4p\over(2\pi)^4}\,\left\lbrack
a_{\vec p}\,e^{-ip\cdot x}+a^\dag_{\vec p}\,e^{ip\cdot x}
\right\rbrack\sqrt{2p^0}\Theta(p^0)(2\pi)\delta\left(p^2-m^2\right)\;\; .
\label{KGfield}
\end{equation}
Consistency with (\ref{KGcanoncomm}) requires
\begin{equation}
\left\lbrack a_{\vec p},a^\dag_{\vec p\,'}\right\rbrack
=(2\pi)^3\delta\left(\,\vec p-\,\vec p\,'\right)\;\; .
\end{equation}

Several comments are in order here.  First, the 
canonical commutation relation (\ref{KGcanoncomm})
must be imposed only at equal times; any other 
choice would violate causality.\footnote{Actually,
this statement is not strictly true.  Lorentz
invariance implies that we can impose this 
relation for equal coordinate along any
{\it timelike} vector we so choose.  In the limit
of large boosts, we can even quantize on the 
light cone.  However, once we have chosen 
a direction and imposed an equal `time' commutation
relation, commutation relations
at different times are fixed and 
can no longer be imposed.}  The presence of the 
$\delta$-function is understood immediately once
we realize that we are quantizing a continuously
infinite number of field {\it densities}
$\phi(\,\vec x\,)$.  Our commutation relations
must be such that the integral over a 
small region about $\vec x$ gives one.
Second, the form of equation (\ref{KGfield})
is dictated by the reality of $\phi(x)$.
Complex fields will in general have different 
operators multiplying their positive- and
negative-frequency solutions.  Note that 
I have included the negative frequency solutions
with the positive ones by changing the 
sign in the exponential.

The field $\phi$, along with 
\begin{equation}
\pi(x)=-i\int{d^4p\over(2\pi)^4}\,\left\lbrack
a_{\vec p}\,e^{-ip\cdot x}-a^\dag_{\vec p}\,e^{ip\cdot x}
\right\rbrack p^0\sqrt{2p^0}\Theta(p^0)(2\pi)\delta\left(p^2-m^2\right)\;\; ,
\end{equation}
can be used to form the {\it true} Hamiltonian
of the system,
\begin{eqnarray}
H&=&\int d^3x\left\lbrack\pi(x)\partial^0\phi(x)
-{\cal L}\right\rbrack\nonumber\\
&=&\int{d^3p\over(2\pi)^3}\,\sqrt{\vec p\,^2+m^2}\left\lbrack
a_{\vec p}^\dag\,a_{\,\vec p}+{1\over2}(2\pi)^3
\delta^{(3)}(0)\right\rbrack\;\; .
\end{eqnarray}
The second term in this expression represents the
sum of the zero-point energy of 
a continuously infinite number of quantum
harmonic oscillators.  This contribution to the 
energy is completely independent of 
the state we consider, and as such cannot 
contribute to any physical energy {\it differences}.
For this reason, we will ignore it completely.
The rest of the Hamiltonian
involves the {\it energy} of a particle of momentum 
$\vec p$,
\begin{equation}
E_{\,\vec p}\equiv \sqrt{\vec p\,^2+m^2}\;\; ,
\end{equation}
and 
satisfies
\begin{eqnarray}
\left\lbrack H, a_{\,\vec p}\vphantom{^\dag}\right\rbrack
=-E_{\,\vec p}\,a_{\,\vec p}\phantom{\;\; .}\\
\left\lbrack H, a^\dag_{\,\vec p}\right\rbrack
=\phantom{-}E_{\,\vec p}\,a^\dag_{\,\vec p}\;\; .
\end{eqnarray}
Assuming a {\it ground state} $|0\rangle$
that satisfies
\begin{equation}
a_{\,\vec p}\left|0\right\rangle=0\;\; ,
\end{equation}
we can construct a whole tower of 
states
\begin{equation}
\left|n_1\vec p_1,n_2\vec p_2,\cdots, n_N\vec p_N\right\rangle
\equiv\prod_{i=1}^N\left(2E_{\vec p_i}\right)^{n_i/2}
{\left(a^\dag_{\,\vec
p_i}\right)^{n_i}\over\sqrt{n_i!}}\left|0\right\rangle
\end{equation}
which have the {\it positive} energy
\begin{equation}
H\left|n_1\vec p_1,n_2\vec p_2,\cdots, n_N\vec p_N\right\rangle
=\sum_{i=1}^Nn_iE_{\,\vec p_i}
\left|n_1\vec p_1,n_2\vec p_2,\cdots, n_N\vec p_N\right\rangle\;\; .
\label{hamiltonian}
\end{equation}
Our states are constructed to have
the covariant normalization
\begin{equation}
\left\langle \vec p\,|\vec p\,'\right\rangle
=2E_{\,\vec p}(2\pi)^3\,\delta\left(\,\vec p-\,\vec p\,'\right)
\end{equation}
for each particle.\footnote{For states with more
than one particle of the same momentum, this
normalization will lead to divergent $\delta$-functions.
However, any truly physical process will involve
wavepackets rather than real momentum eigenstates.
The averaging involved here will smear the $\delta$-function.}

These states have exactly the properties we
expect of a state of $n_i$ free particles 
of mass $m$ and momentum $\vec p_i$.  
Apparently, we are encouraged to 
interpret the operators $a_{\,\vec p}^\dag$ 
and $a_{\,\vec p}$ as {\it creating} and {\it destroying}
a particle, or {\it field quantum}, of momentum $\vec p$.  
Quite unexpectedly, we have come up
with a theory which does {\it not} have a fixed
particle content.  This fact, which stems {\it directly}
from the change in wavefunction interpretation from
innocent state vector to dynamical variable, 
represents a major shift in the way we view physics
as a whole.

The field $\phi$, known as the {\it Klein-Gordon} \cite{KG} field,
is a scalar field because it is invariant
under the action of the Lorentz group.
Although this makes it a good starting place for a
discussion on the quantization of field theories,
its appearence in physical theories 
is rare at best.\footnote{The closest 
friend of the real Klein-Gordon field 
present in the Standard Model is the 
complex Higgs field.  This is the only
field in the Standard Model whose quanta 
have not yet (2000) been discovered experimentally
(unless, of course, one counts the longitudinal
modes of the massive gauge bosons...).}  
Physical theories involve spinor and vector
fields almost exclusively.  For this reason,
we now turn to the Dirac field.

According to the presentation of Section 
\ref{dottedandundotted},
Dirac fields satisfy the equation of motion
\begin{equation}
\not\!p\,\psi(x)=i\!\!\not\!\partial\,\psi(x)=m\psi(x)\;\; .
\label{diraceom}
\end{equation}
This equation follows from the lagrangian density
\begin{equation}
{\cal L}=\overline\psi\left(i\!\!\not\!\partial-m\right)\psi\;\; ,
\label{diraclag}
\end{equation}
from which we can deduce the momentum conjugate to
$\psi$ :
\begin{equation}
{\delta{\cal L}\over\delta\partial_0\psi}=
i\overline\psi\gamma^0=i\psi^\dag\;\; .
\end{equation}
Anticipating a field quantum for $\psi$,
we use the complete basis of solutions to 
(\ref{diraceom}) found in Section \ref{dottedandundotted}
to expand
\begin{equation}
\psi(x)=\int{d^4p\over(2\pi)^4}\sum_{s}\left\lbrack
b_{\,\vec p,s}u(p,s)e^{-ip\cdot x}+d^\dag_{\,\vec p,s}v(p,s)e^{ip\cdot
x}\right\rbrack
\sqrt{2p^0}\Theta(p^0)(2\pi)\delta\left(p^2-m^2\right)\;\; .
\end{equation}
We hope to identify the coefficients in this 
expansion 
with particle creation and annihilation
operators.  

The Dirac field is {\it not} hermitian, 
so we have no right to assume a relation between 
the creation and annihilation operators.  However,
as $e^{-ip\cdot x}$($e^{+ip\cdot x}$) carries 
positive (negative) energy\footnote{Here, I mean
energy in the sense of a wave, rather than 
energy in the sense of the Hamiltonian.} $p^0$($-p^0$), 
we associate it with an annihilation (creation)
operator.\footnote{When $\psi$ acts on a physical 
state, we want the creation of a particle
to take free energy away from the system and 
the destruction of a particle to give free energy to the system.
In this sense, a system must use its wavelike (free)
energy to `make' a particle (bound energy).}
With this in mind, we assume the 
commutators 
\begin{equation}
\left\lbrack b_{\,\vec p,s},b^\dag_{\,\vec p\,',s'}\right\rbrack
=\delta_{ss'}(2\pi)^3\delta(\,\vec p-\vec p\,')=
\left\lbrack d_{\,\vec p,s},d^\dag_{\,\vec p\,',s'}\right\rbrack\;\; , 
\label{wrongcomm}
\end{equation}
with all others vanishing,\footnote{In the free theory, 
these are all uncoupled harmonic oscillators; 
operators from different oscillators should commute.}
and check the canonical commutator
\begin{equation}
\left\lbrack\psi(x),\psi^\dag(y)\right\rbrack\Big|_{x^0=y^0}
=\delta(\,\vec x-\vec y\,)\;\; .
\label{fermwron}
\end{equation}

Unfortunately, our choice (\ref{wrongcomm})
does not cause this relation to be satisfied.\footnote{One
can check this directly using the completeness
relations (\ref{compmass}) and (\ref{acompmass}).}
The culprit is 
the lack of sign 
between the positive- and negative-frequency 
components of 
\begin{equation}
\psi^\dag(x)=\int{d^4p\over(2\pi)^4}\sum_{s}\left\lbrack
b^\dag_{\,\vec p,s}\,u^\dag(p,s)e^{ip\cdot x}+
d_{\,\vec p,s}\,v^\dag(p,s)e^{-ip\cdot x}\right\rbrack
\sqrt{2p^0}\Theta(p^0)(2\pi)\delta\left(p^2-m^2\right)\;\; .
\end{equation}
This 
stems from the fact that there is only 
one derivative in the 
lagrangian density, (\ref{diraclag}),
which is traced in turn to the form
the translation operator takes in $SL(2,C)$.
Hence this seems to be a fundamental property
of spinor fields. 

One solution to this problem is to simply
generate the sign from the commutation
relations.  If we were to take
\begin{equation}
\left\lbrack d_{\,\vec p,s},d^\dag_{\,\vec p\,',s'}\right\rbrack
=-\delta_{ss'}(2\pi)^3\delta(\,\vec p-\vec p')
\label{wrongcomm2}
\end{equation}
rather than (\ref{wrongcomm}), the fields
would indeed satisfy the commutation relation
(\ref{fermwron}).  However, this solution
is exactly as artificial as it seems.  
It only prolongs our suffering
as it leads to a Hamiltonian the is not
bounded from below.  This fact is intuitively
obvious.  A commutation relation like
(\ref{wrongcomm2}) gives $d_{\,\vec p,s}$
the role of creation operator.  Hence the creation
of a $d$-type quantum {\it gives} free energy to a state.
This implies that all of our states 
will eventually {\it decay} to the 
state of {\it infinite} $d$ quanta.
Our aversion to this state of affairs 
was what led us to consider the fields 
as dynamical variables in the first place,
so this `solution' to our problem is 
clearly unacceptable.

Alternatively, we can {\it demand} that 
$d_{\,\vec p,s}$ destroy quanta and search for other
sources of the sign.  To this end, we consider a 
state $|0\rangle$ that both $b$ and $d$ annihilate.
We would also like $b^\dag$ and $d^\dag$ 
to create quanta, so we require that the state
$dd^\dag|0\rangle$ is proportional to $|0\rangle$.
If the commutator $[d,d^\dag]=1$, then 
we arrive at the old result $dd^\dag|0\rangle=|0\rangle$.
However, the reason these two quantities are equal
is that $d$ annihilates $|0\rangle$.  In a 
fundamental sense, it doesn't matter 
whether we {\it subtract} $d^\dag d$ to form the 
{\it commutator} or {\it add} it to form the 
{\it anticommutator}.  On the other
hand, this difference is essential to the 
cancelation of Dirac structures in the 
canonical commutator.  If we choose instead
to impose a canonical {\it anti\/}commutator, 
we will find that
\begin{equation}
\left\lbrace b_{\,\vec p,s},b^\dag_{\,\vec p\,',s'}\right\rbrace
=\delta_{ss'}(2\pi)^3\delta(\,\vec p-\vec p')=
\left\lbrace d_{\,\vec p,s},d^\dag_{\,\vec p\,',s'}\right\rbrace\;\; ,
\end{equation}
with all other anticommutators zero,
satisfy all of our requirements without compromising 
the integrity of our vacuum.  

The decision to impose canonical anticommutation relations
on spinor fields is not extremely outrageous.
As we have seen, these fields are {\it fundamentally}
different from scalar and vector fields. 
In fact, looking at the classical fields, we
would actually {\it expect} this behavior.
Classical vector fields satisfy
\begin{equation}
{\cal A}^\mu g_{\mu\nu}{\cal B}^\nu={\cal B}^\nu g_{\nu\mu} {\cal A}^\mu\;\; ,
\end{equation}
while for spinor fields we have
\begin{equation}
\xi^\alpha\epsilon_{\alpha\beta}\eta^\beta
=-\eta^\beta\epsilon_{\beta\alpha}\xi^\alpha\;\; .
\end{equation}
Hence in classical theories the vector degrees 
of freedom commute, while spinor degrees anticommute.
From this point of view, it seems very natural
that quantization imposes a commutator 
on vectors and an anticommutator on spinors.
This argument also implies that {\it different}
spinor fields should {\it anticommute} rather than commute,
exactly as we found above.

Whatever the reason, the imposition of 
anticommutators on fermionic 
fields changes the theory drastically.  
In particular, the statement
\begin{equation}
\left\lbrace b^\dag_{\,\vec p,s},b^\dag_{\,\vec p,s}\right\rbrace=0
\end{equation}
implies that the creation operator annihilates the 
one-particle state.  Spinor states are either 
filled or empty; there can be no multiple occupation.
This idea was first postulated by Pauli
as a way to explain the observed atomic spectra.
In non-relativistic quantum mechanics,
it is imposed on the theory as an experimental
input.  In quantum field theory, it is 
promoted to a {\it consistency requirement}.  
This is one of the major triumphs of relativistic
quantum mechanics.  Another is the prediction
of antiparticles.  Since the quanta of the 
$d$ operator are created by $\phi$ while those
of the $b$ operator are destroyed, the absence of 
a `$b$-on' is equivalent in some sense to the
presence of a `$d$-on'.  This relationship 
is dubbed particle-antiparticle duality.  
By convention, we take the quanta 
destroyed by $\psi$ as the particles, while
those it creates the antiparticles.  Thus
the quark field $\psi_q$ destroys quarks and creates
antiquarks.  Antiparticles were first predicted by 
P. A. M. Dirac \cite{Diracanti} in 1931, and discovered in the
form of the positron by C. D. Anderson in 1932 \cite{anderson}.
Dirac shared the 1933 Nobel prize in physics with E. Schr\"odinger
for this advancement of the quantum theory, and Anderson 
won the Nobel Prize in 1936 for its experimental verification.

The Hamiltonian of the (correctly quantized) 
free Dirac field can be guessed from
(\ref{hamiltonian}); it is a weighted sum
over the `occupation number' operators
$b^\dag b$ and $d^\dag d$.
Our particle states
\begin{equation}
\left|\,\vec p,s\right\rangle=\sqrt{2E_{\,\vec p}}
\,b^\dag_{\,\vec p,s}\left|0\right\rangle
\end{equation}
are normalized such that
\begin{equation}
\left\langle\,\vec p\,',s'\big|\,\vec p,s\right\rangle
=2E_{\,\vec p}\,\delta_{ss'}(2\pi)^3\delta(\,\vec p\,'-\vec p\,)\;\; .
\end{equation}
Antiparticle states are normalized in the 
same way.  Note that the action of our fields
gives
\begin{eqnarray}
\psi^\dag(x)|0\rangle&=&\int{d^3p\over(2\pi)^3}\,{1\over2E_{\,\vec p}}\,
\sum_s u^\dag(p,s)e^{ip\cdot x}\,|\,\vec p,s\rangle\;\; ,\\
\psi(x)|\,\vec p,s\rangle&=&u(p,s)e^{-ip\cdot x}|0\rangle+\cdots\;\; ,
\end{eqnarray}
so $\psi^\dag(x)$ creates a particle at spacetime point
$x$ and $\psi$ changes the one-particle state into
the vacuum while providing a wavefunction and some free energy.

Canonical quantization of ordinary vector fields goes
through in the same way as for scalars.  The only
difference is the existence of a polarization
vector $\varepsilon^\mu$ analogous to the 
wavefunction $u(p,s)$ for fermions.  Gauge bosons, on the 
other hand, are much more nontrivial to quantize
using the method presented above.  In fact, 
quantization of non-abelian
gauge bosons in covariant gauges requires the introduction
of ghost fields, as discussed in Section \ref{roadtoquant}.
For these kinds of theories, it is much 
clearer to use the path integral approach presented 
there. 

\section{Spacetime Symmetries -\\ The Generators of the Poincar\'e Group}
\label{quantpoint}

In this section, we would like to study the implications
of Poincar\'e symmetry on a quantum field theory.  In
all of the following, we will assume that the 
action relevant to our theory is invariant under 
Poincar\'e transformations of the fields, i.e.
\begin{equation}
{\cal S}[\Phi',(\partial_\mu\Phi)']=
{\cal S}[\Phi,\partial_\mu\Phi]
\end{equation}
for any set of fields $\Phi'$ related to $\Phi$ by
a Poincar\'e transformation.  This is true
of any theory whose lagrangian is a Lorentz 
scalar and does not depend explicitly
on the spacetime coordinates $x^\mu$. 

In a classical theory, we expect such symmetries to
lead to currents and charges which are conserved
by physical processes.  As mentioned
in Section \ref{unitrepso31}, invariance under translation leads to 
conservation of momentum and that under Lorentz 
transformations leads to conservation of angular momentum.
Let us see how this comes about.
To begin with, we must define 
what we mean by a physical process. 
Classically, a physical process is one which
makes the action stationary.  Under an 
arbitrary infinitesimal variation of the fields, 
the action changes by the amount\footnote{
These manipulations can be thought of in
the following way.  We first consider the
action as a function of the {\it independent}
fields $\Phi$ and $\Phi_\mu$.  No one can 
prevent us from varying these fields independently.
However, in the end we are interested only in the
class of fields that satisfy the constraint 
$\Phi_\mu=\partial_\mu\Phi$.  If we 
make independent variations, this constraint
will not be preserved.  On the other hand, if
we correlate the field variations
by imposing $\delta\Phi_\mu=\partial_\mu\delta\Phi$,
field configurations which satisfy $\Phi_\mu=\partial_\mu\Phi$
will continue to satisfy it after the transformation.
In this way, we can differentiate the 
lagrangian with respect to the fields 
$\Phi$ and $\partial_\mu\Phi$ as though
they were independent and still obtain the 
proper constrained variations.  Of course, this
procedure is equivalent to the more orthodox method of 
Lagrange multipliers.}
\begin{eqnarray}
\label{variation}
\delta{\cal S}&=&\int\,d^{\,4}x\left\lbrack
{\delta{\cal L}\over\delta\Phi}\delta\Phi+
{\delta{\cal L}\over\delta\partial_\mu\Phi}\delta\partial_\mu\Phi
\right\rbrack\\
&=&\int\,d^{\,4}x\left\lbrack {\delta{\cal L}\over\delta\Phi}
\delta\Phi+{\delta{\cal L}\over\delta\partial_\mu\Phi}
\partial_\mu\delta\Phi\right\rbrack\\
&=&\int\,d^{\,4}x\left\lbrack {\delta{\cal L}\over\delta\Phi}
-\partial_\mu{\delta{\cal L}\over\delta\partial_\mu\Phi}\right\rbrack
\delta\Phi\;\; .
\end{eqnarray}
In the last line, I have ignored a surface term.  
This contribution does become important 
in certain theories, most notably gravitational
theories and topological models, but it
will not bother us here.  
From the last line, it is apparent that the 
requirement of a stationary action for
{\it arbitrary} infinitesimal field variations
implies the Euler-Lagrange equation
\begin{equation}
{\delta{\cal L}\over\delta\Phi}
-\partial_\mu{\delta{\cal L}\over\delta\partial_\mu\Phi}=0\;\; .
\label{geneqmot}
\end{equation}
This expression is know as the equation of
motion of the theory.  Each independent
field in the theory has an associated equation
of motion which is satisfied by all physical 
processes.  

If we now specialize our variation to a translation
of spacetime coordinates, we can work out
the implications of translation invariance.  
Under a translation, we have the 
variations\footnote{The fact that the 
Pauli-Ljubanskii vector commutes with the generators
of translation implies that this operation
will not mix different elements of the 
same representation of the Poincar\'e group.
In other words, we can translate each field
independently.  This will not be true of Lorentz rotations.} 
\begin{eqnarray}
\delta\Phi(x)&=&\epsilon^\mu(x)\partial_\mu\Phi(x)\nonumber\\
\delta\partial_\mu\Phi(x)&=&\partial_\mu
\epsilon^\nu(x)\partial_\nu\Phi(x)\;\; .
\end{eqnarray}
Substituting these expressions into (\ref{variation}),
we see that $\cal S$ changes by the amount
\begin{equation}
\delta{\cal S}=\int\,d^{\,4}x\left\lbrack\epsilon^\nu(x)\left(
{\delta{\cal L}\over\delta\Phi}\partial_\nu\Phi
+{\delta{\cal L}\over\delta\partial_\mu\Phi}
\partial_\nu\partial_\mu\Phi\right)+
\left({\delta{\cal L}\over\delta\partial_\mu\Phi}\partial_\nu\Phi\right)
\partial_\mu\epsilon^\nu(x)\right\rbrack\;\; .
\end{equation}
The quantity multiplying $\epsilon^\nu$ 
combines to form $\partial_\nu{\cal L}$, as
can be seen via the chain rule, so we are left\footnote{
after ignoring another surface term}
with
\begin{equation}
\delta{\cal S}=\int\,d^{\,4}x\,\partial_\mu\epsilon_\nu(x)
\left\lbrack{\delta{\cal L}\over\delta\partial_\mu\Phi}\partial^\nu\Phi
-g^{\mu\nu}{\cal L}\right\rbrack\;\; .
\end{equation}
Calling the quantity in brackets ${\cal T}^{\mu\nu}$,
we see that invariance of the action
under translation implies that
\begin{equation}
\partial_\mu {\cal T}^{\mu\nu}=0\;\; .
\end{equation}
One often refers to this expression as
the `conservation' of ${\cal T}^{\mu\nu}$
since it implies that the vector
$\int\,d^{\,3}\vec x\,{\cal T}^{0\nu}(x)$ does not depend on time :
\begin{eqnarray}
{d\over dt}\int\,d^{\,3}\vec x\,{\cal T}^{0\nu}(x)&=&
\int\,d^{\,3}\vec x\,{\partial\over\partial x^0}{\cal T}^{0\nu}(x)\nonumber\\
\label{deriveconserve}
&=&-\int\,d^{\,3}\vec x\,{\partial\over\partial x^i}{\cal T}^{i\nu}(x)\\
&=&-\oint\,dS_i\,{\cal T}^{i\nu}(x)\nonumber\\
&=&0\nonumber\;\; .
\end{eqnarray}
The first equality requires the volume of 
integration to be independent of time.  In the second,
we have used the fact that the divergence of $\cal T$ is
zero.  The third is an application of the 
divergence theorem.  The integration
takes place on the boundary of the volume 
and $dS_i$ represents an outward unit normal.  In 
the fourth, we have assumed that this surface integral
vanishes.  If it does not, Equation (\ref{deriveconserve})
states that the change in this quantity 
with time is due only to the value of 
${\cal T}^{i\nu}$ on the boundary.  Physically, 
we can think of ${\cal T}^{i\nu}$ as the `flow' 
of a `charge density' ${\cal T}^{0\nu}$.  This
interpretation makes it easy to understand
(\ref{deriveconserve}).  It says that the 
change in total charge in a volume is equal 
to the amount that came in across the boundary.
This is conservation in its most fundamental
form.  Since the conserved `charge' here is due
to translation invariance, we will call it momentum.
Thus ${\cal T}^{0\nu}$ can be identified with a 
momentum density and ${\cal T}^{i\nu}$ with a momentum
flow.  The integrated charge,
${\cal P}^\mu\equiv\int\,d^{\,3}x{\cal T}^{0\mu}(x)$,
is the conserved total momentum of the system.
For these reasons, $\cal T$ is called the 
{\it canonical energy-momentum tensor} of our
theory.

However nice this derivation is, 
it carries with a few problems.  First, it
is not explicitly covariant.  This problem
stems from the fact that we must choose a 
frame before we ask if a quantity
changes with time.  However, looking at
the derivation, we certainly never used any
specific features of the frame we were in.
In fact, the whole derivation can be 
performed with any timelike vector $v^\mu$
replacing the coordinate $t$.  The statement
then becomes
\begin{equation}
{d\over d\lambda}\int\epsilon_{\alpha\beta\gamma\delta}
\;d\Sigma^{\alpha\beta\gamma}v^\delta v^*_\mu {\cal T}^{\mu\nu}
=-\oint\epsilon_{\alpha\beta\gamma\delta}\;d\sigma^{\alpha\beta}v^\gamma
{\cal T}^{\delta\nu}\;\; ,
\end{equation}
where $d\lambda$ represents a change along the $v^\mu$
direction, $v^{*\,\mu}$ is a vector complementing $v$ such that 
$v^*\cdot v=1$, 
$d\Sigma^{\mu\nu\alpha}$ and $d\sigma^{\mu\nu}$
represent volume and area integration measures,\footnote{
These quantities are best described
in the language of 3- and 2-forms, 
respectively.  In both cases, the orientation
of integration is contained in their definition.
Here, the orientation of the volume integration
must also be specified.  It is timelike.} respectively,
and the integration region cannot change as
one traverses the $v^\mu$ direction.  The physical
content of this equation is identical to that of
(\ref{deriveconserve}).  In fact, due to the 
covariance of the result, we can always choose a 
frame in which $v^\mu=v^{*\,\mu}=(1,0,0,0)$ and use
(\ref{deriveconserve}).\footnote{unless we choose 
$v^\mu$ lightlike, as is done in light-cone 
perturbation theory \cite{ColSop}.}

A more fundamental difficulty with this tensor
is that it is not always symmetric.  It turns
out that defining a conserved angular momentum
density requires a symmetric energy-momentum
tensor.  Non-relativistically, we would
expect the angular momentum density
to have the form
\begin{equation}
\vec L=\vec x\times\vec P\;\; .
\end{equation}
Generalizing this equation to 
four dimensions, we write
\begin{equation}
J^{\mu\nu}=x^\mu P^\nu-x^\nu P^\mu\;\; .
\end{equation}
Interpreting $P$ as a momentum density,
we see that
this is just the time component of the 
tensor
\begin{equation}
{\cal M}^{\alpha\mu\nu}=x^\mu{\cal T}^{\alpha\nu}
-x^\nu{\cal T}^{\alpha\mu}\;\; .
\end{equation}
Taking the divergence of this object 
on $\alpha$, we find that
\begin{equation}
\partial_\alpha{\cal M}^{\alpha\mu\nu}=
{\cal T}^{\mu\nu}-{\cal T}^{\nu\mu}\;\; ,
\end{equation}
effectively ending all of our dreams of a conserved angular
momentum.  

This is not necessarily
a failure.  We have not yet told our calculations
that the lagrangian for our theory is
Lorentz invariant, and we cannot expect
to get angular momentum conservation for free.  
If ${\cal T}^{\mu\nu}$ were manifestly symmetric,
we would have the result that translationally
invariant theories are automatically 
Lorentz invariant and possess conserved 
angular momenta.  Lorentz invariant
scalar theories do indeed possess symmetric
canonical energy-momentum tensors.  This is
due to the simple fact that no antisymmetric 
two-index structure can be constructed in these theories.\footnote{
The structure $\partial^\mu\phi\,\partial^\nu\chi$
can always be transformed away by defining new
fields.}
However, somewhat more input is required in
theories involving fields with nonzero
intrinsic angular momentum.

The biggest problem with ${\cal T}^{\mu\nu}$
occurs in gauge theories like QCD : it is
not gauge invariant.  This fact destroys
any hope we ever had of interpreting any of its
components as physical objects.  $\cal T$ may be conserved
in a gauge theory, but it makes no sense.
It seems that this devastating blow renders all of
our work useless.  However, all is not lost.  
Above, we argued
that the canonical momentum
density cannot be used to locally construct a
conserved angular momentum
density 
for fields with nontrivial spin.
We have also seen that imposing local gauge 
invariance requires the introduction of a gauge field, which  
necessarily transforms like a vector\footnote{This is
true for local {\it internal} symmetries.  Gauge
fields associated with {\it external} symmetries
may have different transformation properties.
For example, the gauge field associated with 
local Lorentz invariance transforms like a 
traceless rank-2 symmetric tensor.  This causes
the associated gauge boson, the graviton, to have
spin-2.} 
since it is on the same footing as $\partial_\mu$.  
Hence the inconsistency of the canonical
energy-momentum tensor in any gauge
theory is already expected on
angular momentum grounds.  This suggests
that the gauge dependent part of $\cal T$ is
related to its antisymmetric part.

If our theory truly is Lorentz invariant,
we must be able to define a conserved angular momentum.
There are several ways to do this.  A straightforward
approach is simply to substitute the infinitesimal 
form of a spacetime dependent Lorentz transformation
into Eq.(\ref{variation}) and follow one's
nose.  This will certainly lead to a conserved
total angular momentum, but the connection between 
angular momentum and momentum is completely 
obscured.  As we saw above, any local 
correspondence between these two quantities
is doomed unless the canonical energy-momentum
tensor is symmetric.  Furthermore, local 
angular momentum densities and currents
defined in this way will not be gauge-invariant 
and therefore cannot have a physical
interpretation.  Alternatively, if we
can somehow define a `new and improved' symmetric energy-momentum
tensor, we can make the 
connection between momentum and angular momentum
explicit by constructing the angular momentum
density locally.  We will see that this construction
also leads naturally to gauge-invariant local densities.

Looking back on our
argument that the total momentum is
conserved, we see that the only crucial property of
$\cal T$ is that its divergence equals zero.
The addition of any divergence-free tensor 
to $\cal T$ will yield another tensor from 
which conserved quantities can be derived.  
However, this argument does not show that the
new quantities are the momenta of the system. 
The conservation of $\cal P$ is a {\it direct} consequence
of translation invariance, as we saw above.  This
more than anything else promotes its identification
with the momentum of the system.  
Alternative tensors may change the 
momentum density, ${\cal T}^{0\mu}$, 
allowing for a conserved angular momentum density,
but must not change the integrated quantity
$\cal P$.  Both of these requirements
are satisfied by the {\it Belinfant\'e} transformation,
\begin{equation}
{\cal T}^{\mu\nu}\rightarrow
{\cal T}^{\mu\nu}+\partial_\alpha\chi^{[\alpha\mu]\nu}\;\; .
\label{bellinfonte}
\end{equation}
The antisymmetry of $\chi$ under
$\mu\alpha$ exchange guarantees the 
new tensor to be divergence-free.  Furthermore,
the difference between the new conserved 
quantities and $\cal P$ is just the surface
integral
\begin{equation}
\int\,d^{\,3}x\,
\partial_\alpha\chi^{[\alpha0]\nu}=\int\,d^{\,3}x
\,\partial_i\chi^{[i0]\nu}=\oint dS_i\,\chi^{[i0]\nu}\;\; ,
\end{equation}
which can certainly be chosen to vanish
over a large enough region.\footnote{barring topological
effects...}
If there exists a $\chi$ such that the remaining
tensor is symmetric, this tensor can be used
to define angular momentum densities
whose integrals represent the 
conserved total angular momentum of
the system.  Let us see how 
the assumption of Lorentz invariance leads
to the existence of such a $\chi$.

Under the action of the Lorentz group,
a scalar lagrangian transforms only because 
its argument does :
\begin{equation}
{\cal L}(x)\rightarrow{\cal L}(\Lambda x)={\cal L}(x)+
{1\over2}\left(\theta_{\mu\nu}J^{\mu\nu}\right)_{\alpha\beta}
x^\beta\partial^\alpha{\cal L}(x)+{\cal O}(\theta^2)\;\; .
\end{equation}
This is a translation of the same 
form as that considered above.  Since
we have already taken the translation
invariance of the action into account,
we will 
ignore the change in spatial 
coordinate and focus on the 
other transformations induced
by the Lorentz group.  

The rotation generators
do not commute with the 
Pauli-Ljubanskii vector, 
so they will in general mix different elements
of irreducible representations of the Poincar\'e group;
under an infinitesimal global Lorentz transformation, 
our fields transform as\footnote{Note that 
the index on the derivative is treated like
a vector index here.}
\begin{eqnarray}
\Phi_m&\rightarrow&\Phi'_m=\Phi_m-{i\over2}
\left(\theta_{\mu\nu}\Sigma^{\mu\nu}\right)_{mn}\Phi_n\nonumber\\
\partial_\alpha\Phi_m&\rightarrow&\partial_\alpha\Phi_m
-{1\over2}\left(\theta_{\mu\nu}
J^{\mu\nu}\right)_{\alpha\beta}\partial^\beta\Phi_m
-{i\over2}\left(\theta_{\mu\nu}\Sigma^{\mu\nu}\right)_{mn}
\partial_\alpha\Phi_n
\end{eqnarray}
Here, $m$ and $n$ label different fields in
our theory.  $\Sigma$ will depend on the 
representation the fields belong to (or, equivalently, the 
spin) and $J$ is the generator of 
coordinate rotations, 
\begin{equation}
\left(J^{\mu\nu}\right)_{\alpha\beta}=\left(
\delta^\mu_\alpha\delta^\nu_\beta-
\delta^\nu_\alpha\delta^\mu_\beta\right)\;\; .
\end{equation}
Under this transformation, the lagrangian itself
must be invariant since we are ignoring its variation
due to the coordinate translation.  Hence
Lorentz invariance implies
\begin{eqnarray}
0&=&-{i\over2}{\delta{\cal L}\over\delta\Phi}\Sigma^{\mu\nu}\Phi
-{i\over2}{\delta{\cal L}\over\delta\partial_\alpha\Phi}
\Sigma^{\mu\nu}\partial_\alpha\Phi-{1\over2}
{\delta{\cal L}\over\delta\partial_\alpha\Phi}
\left(J^{\mu\nu}\right)_{\alpha\beta}\partial^\beta\Phi\nonumber\\
&=&{\delta{\cal L}\over\delta\partial_\mu\Phi}\partial^\nu\Phi
-{\delta{\cal L}\over\delta\partial_\nu\Phi}\partial^\mu\Phi\\
&&-i\left({\delta{\cal L}\over\delta\Phi}\Sigma^{\mu\nu}\Phi
+{\delta{\cal L}\over\delta\partial_\alpha\Phi}
\Sigma^{\mu\nu}\partial_\alpha\Phi\right)\nonumber\\
&=&{\cal T}^{\mu\nu}-{\cal T}^{\nu\mu}-
i\partial_\alpha\left({\delta{\cal L}\over
\delta\partial_\alpha\Phi}\Sigma^{\mu\nu}\Phi\right)\;\; ,\nonumber
\end{eqnarray}
where in the last line I have used the equation
of motion, (\ref{geneqmot}), and the fact that
$\Sigma^{\mu\nu}$ does not depend on $x^\mu$. 

This equation is rather interesting.  Rather
than giving us a divergence-free current
from which we can construct a conserved 
angular momentum density,
Lorentz invariance relates the 
antisymmetric part of 
$\cal T$ to a total divergence.  We can use
this to immediately construct the canonical angular momentum
current density
\begin{equation}
\tilde{\cal M}^{\alpha\mu\nu}\equiv x^\mu{\cal T}^{\alpha\nu}-
x^\nu {\cal T}^{\alpha\mu}
-i{\delta{\cal L}\over\delta\partial_\alpha\Phi}
\Sigma^{\mu\nu}\Phi
\end{equation}
which follows directly from Lorentz invariance.  
As is usually the case with canonical quantities, this
object is easy to interpret : the first two
terms are due to orbital angular momentum 
while the last is due entirely to intrinsic spin
angular momentum.  Unfortunately, $\tilde{\cal M}$
also displays one of the less favorable attributes
of canonical quantities - it is gauge-dependent.

Alternatively, notice that the 
symmetric part of $\cal T$
can be written as
\begin{equation}
{\cal T}^{(\mu\nu)}={\cal T}^{\mu\nu}
-{i\over2}\partial_\alpha\left({\delta{\cal L}\over
\delta\partial_\alpha\Phi}\Sigma^{\mu\nu}\Phi\right)\;\; .
\end{equation}
This is reminiscent of a Belinfant\'e transformation,
(\ref{bellinfonte}), but our version of $\chi$ is
not antisymmetric in $\alpha\mu$.  We can
fix this by subtracting a term in which
$\alpha\leftrightarrow\mu$, but the tensor we
are left with is no longer symmetric.  
Fortunately, the term we need to 
add to restore the symmetry is automatically
antisymmetric in $\mu\alpha$ due to the 
antisymmetry of $\Sigma$; the tensor
\begin{eqnarray}
\Theta^{\mu\nu}&\equiv&{\cal T}^{\mu\nu}-{i\over2}\partial_\alpha\left\lbrack
{\delta{\cal L}\over
\delta\partial_\alpha\Phi}\Sigma^{\mu\nu}\Phi\right.\nonumber\\
&&\qquad\qquad\qquad\qquad\left.-{\delta{\cal L}\over
\delta\partial_\mu\Phi}\Sigma^{\alpha\nu}\Phi
-{\delta{\cal L}\over
\delta\partial_\nu\Phi}\Sigma^{\alpha\mu}\Phi\right\rbrack
\end{eqnarray}
is both symmetric and physically indistinguishable
from $\cal T$.  In gauge theories,
we will also find it to be gauge-invariant.
This is the tensor which can truly 
be interpreted as the momentum
current density of our system.
According to the above discussion, we can use $\Theta$
to form the angular momentum current density
\begin{equation}
{\cal M}^{\alpha\mu\nu}=x^\mu\Theta^{\alpha\nu}
-x^\nu\Theta^{\alpha\mu}\;\; .
\end{equation}
This improved Belinfant\'e form 
differs from $\tilde{\cal M}$ 
locally, but one can show by direct substitution that the conserved
total angular momentum 
\begin{equation}
{\cal J}^{\mu\nu}=\int\,d^{\,3}x\,{\cal M}^{0\mu\nu}(x)
\end{equation}
is unaffected.\footnote{once again,
ignoring surface terms.}

While the above discussion is perfectly
fine in classical theories, some modification 
is clearly needed before these results
can be applied to quantized fields.
In the path-integral approach, the transition
is quite clean.  We begin with the
vacuum matrix element
\begin{equation}
\left\langle{\rm T}\hat{\cal O}(x_1,\ldots,x_n)\right\rangle
={1\over{\cal Z}[0]}\int{\cal D}[\Phi]{\cal O}(x_1,\ldots,x_n)
e^{i{\cal S}[\Phi]}\;\; ,
\end{equation}
where $\hat{\cal O}$ is an arbitrary\footnote{In a 
gauge theory, we will require it to be gauge invariant.} 
collection of field operators depending on the 
spacetime coordinates $x_1,\ldots x_n$.  
Let us ignore the normalization factor ${\cal Z}[0]$
for the moment and focus on the path integral. 
Since $\Phi$ is nothing but a dummy 
variable of integration, it is clear that
\begin{equation}
\int{\cal D}[\Phi]{\cal O}(x_1,\ldots,x_n)
e^{i{\cal S}[\Phi]}=\int{\cal D}[\Phi']{\cal O}'(x_1,\ldots,x_n)
e^{i{\cal S}[\Phi']}\;\; ,
\label{1234}
\end{equation}
where ${\cal O}'$ is the functional $\cal O$ evaluated 
at the field configuration $\Phi'$.
Taking $\Phi'(x)=\Phi(x)+\delta\Phi(x)$,
with $\delta\Phi$ infinitesimal but otherwise 
unconstrained,
we rewrite (\ref{1234}) as\footnote{Recall that 
repeated indices imply summation.  Here, the 
repeated `index' is the spacetime variable $x$.
Also note that the determinant is inverted for
Grassmann fields.} 
\begin{eqnarray}
&&\int{\cal D}[\Phi]{\cal O}(x_1,\ldots,x_n)
e^{i{\cal S}[\Phi]}=\int{\cal D}[\Phi]\left|\det\left(
{\delta\Phi'(x)\over\delta\Phi(y)}\right)\right|\nonumber\\
&&\qquad\quad\times\left\lbrack{\cal O}(x_1,\ldots,x_n)
+{\delta{\cal O}\over\delta\Phi(x)}\delta\Phi(x)
+i{\delta{\cal S}[\Phi]\over\delta\Phi(x)}\delta\Phi(x)
{\cal O}(x_1,\ldots,x_n)\right\rbrack
e^{i{\cal S}[\Phi]}\;\; .
\end{eqnarray}

Let us consider the arbitrary variation 
\begin{equation}
\delta\Phi(x)=\epsilon(x)\;\; .
\end{equation}
In this case, our variation is merely a 
field-independent shift of
the integration variable.  Hence
the Jacobian is
trivial. 
This immediately leads to the relation
\begin{equation}
\int{\cal D}[\Phi]
\left\lbrack{\delta{\cal O}\over\delta\Phi(x)}\delta\Phi(x)
+i{\delta{\cal S}[\Phi]\over\delta\Phi(x)}\delta\Phi(x)
{\cal O}(x_1,\ldots,x_n)\right\rbrack
e^{i{\cal S}[\Phi]}=0\;\; .
\end{equation}
Writing
\begin{equation}
{\delta{\cal S}[\Phi]\over\delta\Phi(x)}\delta\Phi(x)
=\left\lbrack{\delta{\cal L}(x)\over\delta\Phi(x)}
-\partial_\mu{\delta{\cal L}(x)\over\delta\partial_\mu\Phi(x)}
\right\rbrack\epsilon(x)\;\; ,
\end{equation}
and asserting the arbitrariness of 
$\epsilon(x)$, we
see that
\begin{equation}
\left\langle{\rm T}\left\lbrace\left\lbrack{\delta
{\cal L}(x)\over\delta\Phi(x)}
-\partial_\mu{\delta{\cal L}(x)\over\delta\partial_\mu\Phi(x)}
\right\rbrack
{\cal O}(x_1,\ldots,x_n)\right\rbrace\right\rangle
=i\left\langle{\rm T}{\delta{\cal O}
(x_1,\ldots,x_n)\over\delta\Phi(x)}\right\rangle\;\; .
\label{quanteom}
\end{equation}
The right-hand-side of this equation is a series of
{\it contact terms} which involve $\delta(x-x_i)$ for some
$i\in\{1,n\}$, while the left-hand-side is a Green's
function involving the equation of motion 
operator
\begin{equation}
{\delta
{\cal L}\over\delta\Phi}
-\partial_\mu{\delta{\cal L}
\over\delta\partial_\mu\Phi}\;\; .
\label{eommmm}
\end{equation}
Equation (\ref{quanteom}) is an extremely
useful relation.  In particular, it tells us
that all physical matrix elements of (\ref{eommmm})
vanish since in that case the operator $\cal O$ 
represents asymptotic fields and will
have arguments only at infinity.

At this point, it is obvious how to
proceed with other variations of the fields $\Phi$.
All we need is a transformation which leaves the 
measure of the path integral invariant\footnote{Actually,
useful identities can be derived even when the 
measure is not left invariant.  One extremely well-known
example is the derivation of the 
Adler-Bell-Jackiw anomaly \cite{Fujikawa}.}
to arrive at an identity of the form
\begin{equation}
\left\langle{\rm T}\left\lbrace {\cal C}(x)
{\cal O}(x_1,\ldots,x_n)\right\rbrace\right\rangle
=i\left\langle{\rm T}{\delta{\cal O}(x_1,\ldots,x_n)
\over\delta\Phi(x)}\right\rangle\;\; ,
\label{fundamental}
\end{equation}
where ${\cal C}$ is a collection of fields which is 
classically zero.
In particular, the classical relations
involving conservation of energy-momentum and 
angular momentum can easily be cast into this 
form since the transformations of 
translation and Lorentz rotation 
leave the measure invariant.\footnote{For fermions, 
we can see this as a consequence of the invariance
of the product $\overline\psi\psi$, while 
for vector bosons it is due to the fact that 
$\det\Lambda=\pm1$ for the Lorentz transformations
introduced in Section \ref{relativity}.}

The lagrangian density of massless QCD is 
\begin{equation}
{\cal L}=\overline\psi\, i\!\not\!\!{\cal D}\,\psi
-{1\over4}{\cal F}^2\;\; ,
\end{equation}
from which we derive\footnote{These relations 
are {\it not} unique.  In particular, 
the replacement $\stackrel{\rightarrow}{i\cal D}
\rightarrow\,\stackrel{\leftrightarrow}{i\cal D}$
will affect the first four.  However, the equations of 
motion and physical tensor densities are not affected 
by this ambiguity.
We also note that the above derivation assumes commuting,
rather than Grassmann, fields.  To generalize our
expressions, it is sufficient to consider Grassmann
derivatives as coming {\it from the direction of the 
variation}.  For example,
\begin{eqnarray}
{\delta{\cal L}\over\delta\Phi}\,\delta\Phi&\rightarrow&
\delta\overline\psi\,{\stackrel\rightarrow\delta\over\delta
\overline\psi}\,{\cal L}\nonumber\\
&\rightarrow&{\cal L}\,{\stackrel\leftarrow\delta\over
\delta\psi}\,\delta\psi\;\; .
\end{eqnarray}}
\begin{eqnarray}
{\delta{\cal L}\over\delta\psi}&=&-g\overline\psi\,\!\!\not\!\!{\cal A}\\
{\delta{\cal L}\over\delta\partial_\mu\psi}&=&\overline\psi\, i\gamma^\mu\\
{\delta{\cal L}\over\delta\overline\psi}&=&i\!\not\!\!{\cal D}\,\psi\\
{\delta{\cal L}\over\delta\partial_\mu\overline\psi}&=&0\\
{\delta{\cal L}\over\delta{\cal A}_\mu^a}&=&
-g\overline\psi\gamma^\mu t^a\psi-gf^{abc}
{\cal F}^{\mu\alpha}_b{\cal A}^c_\alpha\\
{\delta{\cal L}\over\delta\partial_\mu{\cal A}_\nu^a}
&=&-{\cal F}^{\mu\nu}_a\;\; .
\end{eqnarray}
These relations, along with the 
Lorentz transformation properties
\begin{eqnarray}
\Sigma^{\mu\nu}_{q}={1\over2}\sigma^{\mu\nu}\phantom{\;\; .}\\
\Sigma^{\mu\nu}_{g}=-iJ^{\mu\nu}\;\; ,
\end{eqnarray}
contain the QCD equations of motion,
energy-momentum tensor, angular momentum current density,
and any other symmetry currents of\footnote{unrenormalized and
un-fixed} QCD.  Some useful relations are summarized
in Section \ref{frulesqcd}.

One very useful identity that can easily be
proved at this point is the fundamental {\it Ward
identity}
of QCD.  Taking
\begin{equation}
\delta{\cal A}_a^\mu(x)=\partial^\mu\epsilon_a(x)\;\; ,
\end{equation}
(\ref{fundamental}) becomes\footnote{
Derivatives {\it always} belong {\it outside} the
matrix element.  The easiest way to see this involves the 
sources in the path integral approach.  Since our 
sources bring down powers of the fields rather than their
derivatives, we must differentiate {\it externally}.
Among other things, this implies that derivatives
will always generate contributions from the 
time-ordered product.  These form the 
contact terms on the right-hand-side of our
matrix element identities.  As mentioned in Section
2.5, these contact terms can systematically be removed
by considering the modified T$^*$ product.  See \cite{itzy}.}
\begin{equation}
\partial_\mu\left\langle{\rm T}\left\lbrace
\left\lbrack g\overline\psi\gamma^\mu t^a\psi
+gf^{abc}{\cal F}^{\mu\alpha}_b{\cal A}^c_\alpha\right\rbrack
{\cal O}(x_1,\ldots,x_n)\right\rbrace\right\rangle
=-i\partial_\mu\left\langle{\rm T}{\delta{\cal O}(x_1,\ldots,x_n)
\over\delta{\cal A}_\mu^a(x)}\right\rangle\;\; .
\end{equation}
The expression in brackets
is constructed to be the current the gluon field
couples to, so this equation tells us that Green's
functions involving a gluon current contracted with its
momentum are reduced to contact terms.  In particular,
it implies transversality of the 
gluon propagator and conservation of the color charge.  In this
sense, the terms on the right-hand-side act as
{\it sources} for the color current.  Nonzero contact
terms imply a color source, which will affect the 
total charge.  Note that this identity is also 
obvious from the classical equation of motion of the gluon field,
\begin{equation}
{\cal D}^{ab}_\mu{\cal F}_b^{\mu\alpha}=
g\overline\psi\gamma^\alpha t^a\psi\;\; ,
\end{equation}
and the antisymmetry of the field strength.

An analogous identity implies the conservation of 
quark flavor (or, more generally, baryon and lepton number).
Taking 
\begin{equation}
\delta\psi(x)=i\alpha(x)\psi(x)\;\; ,
\;\;\;\;\delta\overline\psi(x)=-i\alpha(x)\overline\psi(x)
\end{equation}
gives\footnote{
Once again, we must take care with these anticommuting
derivatives.
}
\begin{equation}
\partial_\mu\left\langle{\rm T}\left\lbrace
\overline\psi(x)\gamma^\mu\psi(x)
{\cal O}(x_1,\ldots,x_n)\right\rbrace\right\rangle
=\left\langle{\rm T}\left\lbrace\overline\psi(x)
{\delta{\cal O}(x_1,\ldots,x_n)
\over\delta{\overline\psi}(x)}-{\delta{\cal O}(x_1,\ldots,x_n)
\over\delta{\psi}(x)}\psi(x)\right\rbrace\right\rangle\;\; .
\end{equation}
This relation also finds use as the fundamental Ward 
identity of QED.  

These identities have all been derived ignoring 
the necessary steps of renormalization and gauge-fixing.
The fact that these unrenormalized and un-fixed
Ward identities imply similar relations 
among the fully renormalized gauge-fixed 
Green's functions is nontrivial to show.  
Nonetheless, it has been done in the covariant
gauges \cite{renward} and 
the light-cone gauge \cite{axial}.
Once these renormalized and gauge-fixed 
Ward identities are proven, the renormalizability
of the whole theory is just a step away.  Thanks 
to these identities, all of the ultraviolet
divergences present in the theory are constrained
to take forms which respect our symmetries. 
Since we have chosen the most general lagrangian 
consistent with the symmetry, this implies
that all ultraviolet divergences can be
absorbed into counterterms, which in turn
implies renormalizability.

\chapter{Dimensional Regularization}
\label{dimregapp}

This Appendix is meant to familiarize the reader with the
standard regularization procedure used in quantum field theory.
Introduced by G. `t Hooft and M. Veltman in 1972 
\cite{thooft}, this procedure exploits the analytic 
behavior of integrals as functions of their dimensionality.
Since the ultraviolet (and infrared) divergences
we meet in the text are in some sense accidents of the 
fact that we live in four dimensions, these integrals 
prove entirely convergent when 
evaluated in less (or more)
spacetime dimensions.  The subtleties 
associated with the actual meaning of a 3.5-dimensional
space are completely avoided here; some discussion
can be found in \cite{colren}.  We will view dimensional
regularization exclusively as a consistent 
{\it regularization procedure} for our 
amplitudes.  

It can be shown quite generally
that the coordinates 
\begin{eqnarray}
x_1&=&\rho\, \cos\theta_1\nonumber\\
x_2&=&\rho\, \sin\theta_1\,\cos\theta_2\nonumber\\
x_3&=&\rho\, \sin\theta_1\,\sin\theta_2\,\cos\theta_3\nonumber\\
\vdots&=&\vdots\nonumber\\
x_i&=&\rho\,\prod_{j=1}^{i-1}\sin\theta_{j}\,\cos\theta_{i}\\
\vdots&=&\vdots\nonumber\\
x_{(n-1)}&=&\rho\, \prod_{j=1}^{n-2}\sin\theta_{j}\,\cos\theta_{(n-1)}
\nonumber\\
x_n&=&\rho\, \prod_{j=1}^{n-1}\sin\theta_{j}
\end{eqnarray}
span a Euclidean $n$-dimensional space for integer
$n>1$.  The angles $\theta_i$ all take values
from 0 to $\pi$ except $\theta_{(n-1)}$, which 
takes values on $[0,2\pi)$.  
The 2-norm of the vector $\bf x$ is given 
by $\rho$ and the $(n-1)$ angles specify a location
on the $(n-1)$-dimensional sphere 
defined by $\rho$.  The line element in these
coordinates
\begin{equation}
ds^2=d\rho^2+\rho^2\,d\theta_1^2+
\rho^2\sum_{i=2}^{n-1}
\left(\prod_{j=1}^{i-1}\sin^2\theta_{j}\right)d\theta_i^2
\end{equation}
defines the volume element
\begin{equation}
d\,{\bf x}=\sqrt{\det\,g_{ij}}\,d\rho\,\prod_{i=1}^{n-1}d\theta_i
=\rho^{n-1}\,\left(\prod_{i=1}^{n-1}\sin^{n-i-1}\theta_i\right)
\;d\rho\,\prod_{j=1}^{n-1}d\theta_j\;\; .
\end{equation}
The hyper-area of our $(n-1)$-dimensional sphere 
then becomes
\begin{equation}
\Omega_{(n-1)}=2\pi\prod_{i=1}^{n-2}
\int_0^\pi \sin^{n-i-1}\theta_i\,d\theta_i\;\; ,
\end{equation}
which upon using 
\begin{equation}
\int_0^\pi\sin^n\theta d\theta=2^{n}\,{\Gamma^2
\left({n+1\over2}\right)\over\Gamma(n+1)}
=\sqrt{\pi}\,{\Gamma\left({n+1\over2}\right)
\over\Gamma\left({n\over2}+1\right)}
\end{equation}
reduces to
\begin{equation}
\Omega_{(n-1)}={2\pi^{n/2}\over\Gamma\left(n\over2\right)}\;\; .
\end{equation}

For any function $f(\,{\bf x}\,)$ that depends
only on the magnitude $\rho$ of $\bf x$
we have
\begin{equation}
\int d\,{\bf x}\,f(\,{\bf x}\,)=\Omega_{(n-1)}\int d\rho
\,\rho^{n-1}f(\rho)\;\; .
\label{omegint}
\end{equation}
This expression is correct for any positive integer 
dimension $n$.  Furthermore, these 
integrals respect shift invariance
\begin{equation}
\int d\,{\bf x}\,f(\,{\bf x+ y}\,)=\int d\,{\bf x}\,f(\,\bf x\,)\;\; ,
\end{equation}
for any constant vector $\bf y$,
and the dilation identity
\begin{equation}
\int d\,{\bf x}\,f(a\,{\bf x}\,)
={1\over|a|^{n}}\int d\,{\bf x}\,f(\,{\bf x}\,)\;\; ,
\label{dilation}
\end{equation}
where $a$ is an arbitrary real constant.

We can {\it define} the functional
\begin{equation}
{\cal I}_d[f](\,{\bf y}_1,\ldots,{\bf y}_N)
\end{equation}
as a {\it meromorphic} function of $d$ that
satisfies
\begin{eqnarray}
\label{shiftinv}
{\cal I}_{d}[f](\,{\bf y}_1+{\bf y},\ldots,{\bf y}_N+{\bf y})
&=&{\cal I}_{d}[f](\,{\bf y}_1,\ldots,{\bf y}_N)\\
{\cal I}_{d}[f](\,a{\bf y}_1,\ldots,a{\bf y}_N)&=&
|a|^da^{\phi_f}{\cal I}_d[f]({\bf y}_1,\ldots,{\bf y}_N)
\label{dilinv}
\end{eqnarray}
and takes the value
\begin{equation}
{\cal I}_n[f](\,{\bf y}_1,\ldots,{\bf y}_N)
=\int d\,{\bf x}f(\,{\bf x},{\bf y}_1,\ldots,{\bf y}_N)
\label{constraintondim}
\end{equation}
for $n\in{\cal Z}_+$.\footnote{unless this integral is
not well-defined.  We will see that these cases
are handled quite nicely by the other 
properties of ${\cal I}_d$.}  The constant 
$\phi_f$ is a scale factor which depends only
on $f$.
This does {\it not}
fix our functional uniquely.  One can add
{\it any} analytic function of $d$ which 
vanishes at all positive integers to ${\cal I}_d$
to obtain an equally acceptable functional.  This 
ambiguity will be resolved shortly.

The vectors $\{\,{\bf y}_i\,\}$ define at most $N$ directions,
so in $M>N$ integer dimensions ${\cal I}_M$ contains
at most $N$ directionally-dependent integrals.
Doing these integrals explicitly
gives
\begin{equation}
{\cal I}_{M}[f](\,{\bf y}_1,\ldots,{\bf y}_N)
={\cal I}_{M-N}\left\lbrack
{\cal I}_{N}[f]\right\rbrack(\,{\bf y}_1,\ldots,{\bf y}_N)\;\; ,
\end{equation}
where the remaining $(M-N)$ integrals are
direction {\it independent}, i.e. of the 
form of (\ref{omegint}).
If ${\cal I}_d$ is to be a meromorphic
function of $d$ constrained by (\ref{constraintondim}), 
it must also satisfy this
reduction identity :
\begin{equation}
{\cal I}_{d}[f](\,{\bf y}_1,\ldots,{\bf y}_N)
={\cal I}_{d-N}\left\lbrack
{\cal I}_{N}[f]\right\rbrack(\,{\bf y}_1,\ldots,{\bf y}_N)\;\; .
\end{equation}
This immediately implies that we need not
consider directionally-dependent 
functions $f$.  These can always be reduced
to directionally-independent functions
by doing an integer number of 
integrations.\footnote{although this is 
not always the most practical way to 
extract the value.}

With this in mind, we can use (\ref{omegint})
as a guideline to define a {\it unique}
functional.  For any given directionally-independent 
function $f(\rho)$, we define
\begin{equation}
{\cal I}_d[f]\equiv\Omega_{(d-1)}\int\,d\rho\,\rho^{d-1}f(\rho)\;\; ,
\end{equation}
where the integral on the right is to be evaluated
in a suitable region\footnote{This is taken
to mean anywhere the integral is well-defined.} 
of the complex plane of 
$d$ and analytically continued elsewhere.
Since it has been constructed to 
look like an integral, one often uses the 
notation
\begin{equation}
{\cal I}_d[f](\,{\bf y}_1,\cdots,{\bf y}_N)
=\int d\,{\bf x}\,f(\,{\bf x}\, ,{\bf y}_1,\cdots,{\bf y}_N)\;\; ,
\end{equation}
but it must always be understood that this
is simply a convenient way to write things.
In this form, the shift and dilation invariance
imply that one can manipulate the variable of
integration as in a normal integral.  However, nonlinear
changes of variables require care; 
the interested reader is referred to \cite{colren}.

Now that we know what a `$d$-dimensional integral'
means, let's compute some.
First and foremost, {\it any}
functional ${\cal I}_d[f]$ whose arguments
$\{\,{\bf y}_i\}$ are all zero 
is identically zero.\footnote{We need all scalar
arguments to be zero as well; see below.}
This follows 
from the dilation
property,
\begin{equation}
{\cal I}_d[f]({\bf 0},\ldots,{\bf 0})
=|a|^da^{\phi_f}{\cal I}_d[f]({\bf 0},\ldots,{\bf 0})\;\; ,
\label{diii}
\end{equation}
and the analytic structure of ${\cal I}_d$
as a function of $d$.  This result
is somewhat striking since it implies
that the integrals
\begin{equation}
\int d\,{\bf x}\, (\rho^2)^n
\end{equation}
all take the value zero.  However, 
as a consequence of (\ref{dilinv}), it is 
unalterable.\footnote{Integrals in positive integer
dimensions do not satisfy this equation.  As
a consequence, they are ill-defined.}
Using (\ref{diii}) in conjunction with 
(\ref{shiftinv}), one can also 
easily show that
\begin{equation}
{\cal I}_d[f](\,{\bf y},\ldots,{\bf y}\,)=0\;\; .
\end{equation}
Apparently, ${\cal I}_d$ must have {\it two} vectors
on which to depend if it is to be nonzero.\footnote{
Note that these vectors can be taken as $\bf y$ and $\bf 0$
if one wishes...}  

Throughout the above, I have ignored the dependence
of ${\cal I}_d$ on scalar quantities.
This dependence must, of course, be taken into
account if one is to have a general functional.
In particular, the direction-independent integrals
we have asserted to be the only necessary
consideration depend {\it exclusively} on 
scalar parameters.
Since scalar quantities are not affected by
shifts in the integration variable,
there is no analogue of (\ref{shiftinv})
for these quantities.  However, these parameters
{\it are} affected by dilation :
\begin{equation}
{\cal I}_d[f](a\lambda_1,\ldots,a\lambda_N)
=|a|^{d}a^{\phi_f}{\cal I}_d[f](\lambda_1,\ldots,\lambda_N)\;\; .
\end{equation}
This relation again implies that
parameterless integrals are necessarily 
zero.  However, the lack of 
shift invariance means that we only need one
parameter to obtain
a finite result.  In quantum field theory, 
we will find that the only truly $d$-dimensional
integral one needs to compute is
\begin{equation}
\int{d^{\,d}k\over(2\pi)^d}{\left(k^2\right)^\alpha\over
\left(k^2+M^2\right)^\beta}={1\over(4\pi)^{d/2}}\,
{2\over\Gamma\left({d\over2}\right)}\,\int_0^\infty
dk\;{\left(k^2\right)^{\alpha+(d-1)/2}\over
\left(k^2+M^2\right)^\beta}\;\; .
\label{dimregint}
\end{equation}
For sufficiently small odd $d$ and integer $\alpha$, $\beta$, 
one can extend the integration region to 
negative infinity and take the integral
by contour.  If we are very strong,
we can show
that this gives\footnote{A much
more elegant derivation is given in \cite{colren}.}
\begin{equation}
\label{sss}
{1\over(4\pi)^{d/2}}\left(M^2\right)^{d/2+\alpha-\beta}
\;{\Gamma(d/2+\alpha)\Gamma(\beta-\alpha-d/2)\over
\Gamma(d/2)\Gamma(\beta)}\;\; .
\end{equation}
Analytically continuing this result 
to the full complex plane of $\alpha$, $\beta$, and $d$
gives the required general integral.

Dimensional regularization is a procedure in
which one systematically replaces all four-dimensional
loop-momentum integrals by $d$-dimensional ones.
In this way, the ultraviolet divergences present
in our theory are represented by poles
at $d=4$.  As mentioned above this process
is viewed exclusively as a way to exhibit
ultraviolet (and infrared) divergences 
explicitly as well-defined quantities.  
From this perspective, one would expect no
changes in the theory to be necessary.
Unfortunately, small things have a way
of infiltrating every aspect of a 
quantum field theory.  Interpreting 
a loop-momentum integral as $d$-dimensional
implies in turn that the loop-momenta are themselves 
$d$-dimensional, which implies that the metric
used to form invariant products between them 
is $d$-dimensional.  This fact changes 
the whole structure
of the spacetime symmetry group of our
theory.  

Looking back at Section \ref{relativity}, 
we see that things are not so bad.  
Almost all of the constructs of that 
section can be accommodated in $d$ dimensions.\footnote{
To minimize the change in structure to
our group, we add only spacelike coordinates.}
We have to write 
\begin{equation}
g_{\mu\nu}g^{\mu\nu}=d
\label{gind}
\end{equation}
rather than four and we find ourselves with
$d(d-1)/2$ rotation generators rather
than six.  We can easily keep track of
such factors.  The only change in the above
formalism is that $f$ must now be allowed to 
depend on $d$, which we would've had to consider
at higher orders anyway.\footnote{This does
cause a slight problem with our derivation
of (\ref{diii}) in the case $\phi=-d$.  The 
result of this is that the integral
\begin{equation}
\int{d^{\,d}k\over (k^2)^{d/2}}
\end{equation}
is not uniquely defined.  We cannot take
its integer-dimension value 
since this is also not well-defined, yet
we cannot prove that it is zero using the 
scaling argument.  Even (\ref{sss}) is poorly defined
in this case.
However, since
it is not determined, we
are free to {\it define} it as zero.}
These modifications will affect physical 
amplitudes when multiplied by poles at $d=4$.  
Renormalized quantities are guaranteed to be
finite, but the divergences are certainly not 
guaranteed to leave no trace.  

The intimate relation between
$SO(3,1)$ and $SL(2,C)$ requires us
to continue our Dirac space into
$d$ dimensions as well.  The smallest dimensionality
in which we can find $d$ $\gamma$-matrices which
satisfy the 
Clifford algebra
\begin{equation}
\left\lbrace \gamma^\mu,\gamma^\nu\right\rbrace=2g^{\mu\nu}
\end{equation}
is $2^{d/2}\times2^{d/2}$, so we should
modify the normalizations
\begin{eqnarray}
\gamma^\mu\gamma_\mu&=&d\\
{\rm Tr}\;{\bf 1}&=&2^{d/2}\;\; .
\label{tracenorm}
\end{eqnarray}
Since the trace appears in all amplitudes,
it constitutes an overall factor.  As such, it can
be {\it chosen} not to leave a residue 
by including its effects with the ultraviolet
divergence (along with the $\gamma_E$ and $\log4\pi$
in the $\rm\overline{MS}$ scheme, for example).  For this
reason, people usually just take ${\rm Tr}\;{\bf 1}=4$.

In addition to the above changes,
we must somehow make a connection between 
the $d$-dimensional {\it Euclidean}
space our integrals are defined in and the 
$d$-dimensional {\it Minkowskii}
space in which our theory is defined.
The simple solution to this is 
to rotate our energy integration axis
90$^\circ$ in its complex plane.  This has
the effect
\begin{eqnarray}
k^0\rightarrow ik^0_E\\
k^2\rightarrow-k_E^2\;\; .
\end{eqnarray}
Since we have constructed
our propagators to have a well-defined pole
structure, 
this rotation can be justified for our Feynman integrals
by Cauchy's theorem.\footnote{We can perform
this shift provided we don't cross any poles in the
process.  One can check explicitly that this
leads to no problems here, unless one quantizes
the theory in axial gauge with a pathological
prescription for the spurious singularities.  See
Section 1.6.}  All we must do is
keep track of the sign and the factor of $i$.
At the end of the calculation, we simply rotate
back to obtain amplitudes with physical momenta.

Unfortunately, all of the sectors of our
theory are not so readily expressed in 
$d$ dimensions.  One can trace all of the 
difficulties to the Levi-Cevita tensor,
$\epsilon^{\mu\nu\alpha\beta}$.
Looking back at its definition (\ref{ep}),
we see that this object appears because
of the relation between the determinant of 
$\Lambda$ and its components.  A similar object 
can be defined in $n$ dimensions
as a totally antisymmetric $n$-index tensor.
Its invariance is based on its equal treatment
of {\it all} of the dimensions of our
space.  We have no analogue of this
in $d$ dimensions.  Mathematicians tell
us that if we {\it really} want to take
a non-integer-dimensional space seriously,
its vectors will have an infinite number
of components,\footnote{some discussion 
can be found in \cite{colren}.} an infinite
number of which vanish as $d$ 
approaches a positive integer.
Since we do not intend to introduce 
tensors with an infinite number of
indices into our calculations (and have no idea
how we would if we did intend!), we cannot
determine a definition for $\epsilon$ 
in our dimensionally regularized theory.
As before, this means that we are 
free to define one however we wish, {\it provided that
it is consistent}.\footnote{It also should not do
irreparable damage to Lorentz invariance.}
  Our studies of renormalization
in Section \ref{renincov} imply that we
can choose whatever regularization scheme 
we want, as long as we stick with our scheme 
once it is chosen.  The most obvious choice
is simply to keep the four-dimensional definition.  This
carries the price of breaking the full rotational
symmetry of the $d$-dimensional theory, but
it is difficult to imagine a definition that 
will not.  

When products of two $\epsilon$ tensors
appear, we can pretend not to break the symmetry 
by eliminating the $\epsilon$'s via
\begin{equation}
\epsilon^{\mu\nu\alpha\beta}\epsilon_{\lambda\rho\sigma\tau}
=-\delta^{\mu}_\lambda\delta^\nu_\rho\delta^\alpha_\sigma
\delta^\beta_\tau\;\; + \;\; {\rm permutations}\;\; .
\label{redep}
\end{equation}
Once we have everything in terms
of $\delta$'s, (\ref{gind}) makes us feel like we are
truly working in $d$ dimensions.  
However, the reduction identity
(\ref{redep}) is derived in four dimensions
by working out the implications of the statement, ``If the 
product of two epsilons is not zero, then 
each index on one must have a twin on the other.''.  
This statement is obviously not generalizable
to $d$ dimensions.  On the other hand, one can use 
(\ref{redep}) to define an $\epsilon$ scheme.
This option is described extensively in \cite{gamma1}.
Yet another scheme which considers 
$\epsilon$ a totally antisymmetric tensor 
with four indices in the full $d$-dimensional space
has been brought forth in connection with a 
non-associative Dirac-Clifford algebra \cite{okubo}.

In the fermion sector, the Levi-Cevita symbol's
ambassador, $\gamma_5$, poses similar difficulties.
Because of the role $\gamma_5$ plays in fermion
helicity projection operators and the 
prominence of fermions in real world applications,
different choices for $\gamma_5$'s extension 
have been extensively studied in the literature.
The schemes most often seen are
Bardeen's anticommuting $\gamma_5$ \cite{bardanti}
and `t Hooft and Veltman's $\gamma_5$
prescription \cite{thooft}.  The first of these defines
$\gamma_5$ through the relation
\begin{equation}
\left\lbrace \gamma^\mu,\gamma_5\right\rbrace=0\;\; ,
\end{equation}
which is taken to be valid for {\it all}
$\mu$.  Although this is a natural extension,
it is algebraically inconsistent with a nonzero trace.
Beginning with 
\begin{eqnarray}
d\,{\rm Tr}\,\gamma_5&=&{\rm Tr}\,\gamma^\lambda\gamma_\lambda\gamma_5\nonumber\\
&=&-{\rm Tr}\,\gamma^\lambda\gamma_5\gamma_\lambda\\
&=&-d\,{\rm Tr}\,\gamma_5\;\; ,
\end{eqnarray}
one can show that 
\begin{equation}
d(d-2)(d-4)\cdots(d-2n)\,{\rm Tr}\,\gamma_5
\gamma^{\mu_1}\gamma^{\mu_2}\cdots\gamma^{\mu_{2n}}=0
\end{equation}
for any integer $n$.  This is expected from the
relation $\gamma_5$ has with $\epsilon$; the purpose
of $\gamma_5$ is to force the presence of {\it all}
directions.  However, a $\gamma_5$ which does not
give finite traces is useless to us.  It has been suggested that
dropping some other property of the spinor algebra
may lead to a consistent scheme.  Examples are
the algebra's associativity \cite{okubo} and 
the trace's cyclicity \cite{readpt}.

The `t Hooft and Veltman scheme 
is analogous to the scheme introduced above for 
the Levi-Cevita symbol.  One simply clings
to the four-dimensional definition of $\gamma_5$,
\begin{equation}
\gamma_5=i\gamma^0\gamma^1\gamma^2\gamma^3\;\; ,
\end{equation}
in $d$ dimensions.  This scheme has the advantage of
being fully consistent.  However, it explicitly
breaks the full symmetry of our $d$-dimensional
spacetime.  In particular, its commutation relations 
with the $\gamma^\mu$ read
\begin{eqnarray}
\left\lbrace\gamma^\mu,\gamma_5\right\rbrace=0\;\; 
;\;\;\;\mu\in\{0,3\}\phantom{\;\; .}\\
\left\lbrack\gamma^\mu,\gamma_5\right\rbrack=0\;\; 
;\;\;\;\mu\notin\{0,3\}\;\; .
\end{eqnarray}
This reckless destruction is not without
its price.  In order to ensure that the 
nonsinglet axial current is conserved, one
must perform an extra finite renormalization.
However, once this is done, we are guaranteed a consistent
result.  

In some sense, all of this discussion of $\gamma_5$
is academic for application to QCD.  Since $\gamma_5$
does not appear in the QCD lagrangian, there is 
no real reason to introduce it.  In a vector theory,
$\gamma_5$ functions mainly as a projector 
to help simplify calculations.  On the other
hand, a theory like the Standard Model whose gauge 
bosons couple to a partially axial current 
requires a truly consistent way to separate the 
positive- and negative helicity components
of the Dirac field.  Here, a scheme like 
that of `t Hooft and Veltman's may do irreparable
damage to the gauge theory and lead to 
nonrenormalizability.  All of the standard proofs
that the Standard Model is renormalizable 
assume an ultraviolet regulator that does not break
the gauge symmetry.  However, here it is certainly
not clear that dimensional regularization offers
such a regulator.  Studies of the Ward identities in
such theories can be found in \cite{ovrut}.

Aside from the difficulties associated with
parity-odd structures, dimensional regularization
has proven itself an extremely useful tool.  The simplicity
of its integrals coupled with its respect for
symmetry allow one to trust the correctness of one's
calculation.  One other major advantage of this
prescription is the fact that it can be used to 
regulate infrared as well as ultraviolet divergences.
Historically, one has had to keep infrared divergences
at bay by introducing offshellness or gauge boson masses.
Since the former does not allow the calculation
of onshell physical matrix elements while the latter
does violence to non-abelian gauge symmetries,
these options are clearly not acceptable.  
Dimensional regularization allows one to calculate
onshell matrix elements without breaking
the gauge symmetry through gauge masses.  

However, one cannot simply regulate {\it both}
infrared {\it and } ultraviolet divergences
simultaneously.  This is obvious in a massless
theory, where all corrections would give exactly
zero due to lack of scale.  The correct procedure
begins with a calculation of offshell Green's functions.
These will have ultraviolet divergences which 
are to be regulated dimensionally by taking $d<4$.
Through renormalization, we can re-express our
amplitude in terms of renormalized quantities 
which are ultraviolet finite.
At this point, dimensional regularization is
no longer needed to regularize {\it ultraviolet} physics,
so we are free to analytically continue our
amplitude to $d>4$ where it is well-suited to 
handle {\it infrared} physics.  It is 
now safe to take the onshell limit
without worrying about unregularized divergences.

A list of useful formulae for $d$-dimensional integration,
along with some helpful properties of the $\Gamma$-function,
is given in Section \ref{covint}.

\end{document}